\documentclass[11pt]{article}

\usepackage[margin=1in]{geometry}
\usepackage[T1]{fontenc}
\usepackage{lmodern}
\usepackage{microtype}
\usepackage{amsmath,amssymb,amsthm,mathtools,mathrsfs}
\usepackage{bm}
\usepackage{graphicx}
\usepackage{booktabs,multirow,array,threeparttable,adjustbox}
\usepackage{enumitem}
\usepackage{setspace}
\usepackage{natbib}
\usepackage[colorlinks=true,linkcolor=black,citecolor=blue,urlcolor=blue]{hyperref}
\hypersetup{
  pdftitle={Nonparametric Change-Point Detection and Inference for High-Dimensional Distributions},
  pdfauthor={Ping Zhao; Long Feng},
  pdfkeywords={change-point analysis, high-dimensional inference, distributional change, Gaussian copula, weak cross-sectional dependence, marginal distributions}
}
\usepackage[nameinlink,noabbrev]{cleveref}
\usepackage{caption}
\usepackage{algorithm}
\usepackage{algpseudocode}
\usepackage{placeins}
\usepackage{needspace,etoolbox}

\setlist{nosep,leftmargin=2em}

\newtheorem{theorem}{Theorem}[section]
\newtheorem{proposition}[theorem]{Proposition}
\newtheorem{lemma}[theorem]{Lemma}
\newtheorem{corollary}[theorem]{Corollary}
\newtheorem{assumption}[theorem]{Assumption}
\theoremstyle{remark}

\AtBeginEnvironment{theorem}{\Needspace{5\baselineskip}}
\AtBeginEnvironment{proposition}{\Needspace{5\baselineskip}}
\AtBeginEnvironment{lemma}{\Needspace{5\baselineskip}}
\AtBeginEnvironment{corollary}{\Needspace{5\baselineskip}}
\AtBeginEnvironment{assumption}{\Needspace{5\baselineskip}}
\AtBeginEnvironment{remark}{\Needspace{5\baselineskip}}

\newcommand{\R}{\mathbb{R}}
\newcommand{\E}{\mathbb{E}}
\newcommand{\Pp}{\mathbb{P}}
\newcommand{\bX}{\bm X}
\newcommand{\cK}{\mathcal K}
\newcommand{\cZ}{\mathcal Z}
\newcommand{\cD}{\mathcal D}
\newcommand{\phifun}{\varphi}
\newcommand{\comp}{\mathsf{g}}
\newcommand{\Ssum}{\ensuremath{S_{\mathrm{sum},n,p}}}
\newcommand{\Smax}{\ensuremath{S_{\mathrm{max},n,p}}}
\newcommand{\method}[1]{\textsf{#1}}

\newcommand{\Tcauchy}{\ensuremath{T_{\mathrm{C},n,p}}}
\newcommand{\Pcauchy}{\ensuremath{P_{\mathrm{C},n,p}}}
\newcommand{\hattauSum}{\widehat\tau_{\mathrm{sum}}}
\newcommand{\hattauMax}{\widehat\tau_{\mathrm{max}}}

\title{Nonparametric Change-Point Detection and Inference\\for High-Dimensional Distributions}
\author{Ping Zhao\textsuperscript{1}\qquad Long Feng\textsuperscript{2}\\[0.4em]
\normalsize \textsuperscript{1}School of Mathematical Sciences, Tiangong University, Tianjin, China\\[0.3em]
\normalsize \textsuperscript{2}School of Statistics and Data Science, LEBPS, AAIS, KLMDASR and LMPC,\\
\normalsize Nankai University, Tianjin, China}
\date{}

\begin{document}
\maketitle

\begin{abstract}
High-dimensional distributions can change without altering means or covariances, while the number of affected coordinates is often unknown. We propose nonparametric procedures that address both challenges through standardized rank comparisons of marginal distributions. Sum and maximum scans target dense and sparse changes, and a Cauchy combination adapts to unknown sparsity. The procedures require no marginal moment assumptions and extend to multiple-change detection through wild binary segmentation. Under a weakly dependent Gaussian copula model, we establish asymptotic null distributions, asymptotic independence, detection consistency, and localization guarantees. Simulations demonstrate competitive performance for changes in shape and tails, including alternatives that preserve the first two moments. Applications to gene expression and sensor data illustrate the practical benefits of combining dense and sparse evidence.
\end{abstract}


\section{Introduction}\label{sec:introduction}

Change-point analysis concerns the detection and localization of structural changes in an ordered sequence. Its modern development can be traced to the sequential inspection schemes of \citet{Page1954} and the unknown-location testing problem of \citet{Page1955}. Likelihood-based retrospective inference was developed by \citet{Hinkley1970}, while nonparametric procedures based on ranks and empirical processes were studied by \citet{Pettitt1979}, \citet{Carlstein1988}, and \citet{Dumbgen1991}. For changes in an otherwise unspecified distribution, \citet{Inoue2001} constructed empirical-distribution tests for time series, \citet{HolmesKojadinovicQuessy2013} developed multiplier-based empirical-process procedures for potentially multivariate observations, and \citet{ErlemannLockhartYao2022} studied scan and average Cram\'er--von Mises statistics. Related asymptotic theory for time-series changes was developed by \citet{Picard1985}. The classical foundations are synthesized in the monographs of \citet{BrodskyDarkhovsky1993} and \citet{CsorgoHorvath1997}; broader reviews are provided by \citet{AueHorvath2013}, \citet{HorvathRice2014}, \citet{NiuHaoZhang2016}, and \citet{TruongOudreVayatis2020}. This literature established the basic scan-statistic paradigm: compare the observations before and after each candidate split, aggregate a discrepancy over the split location, and use the optimizing location to estimate the change point.

At a fixed split, testing for a distributional change is a two-sample homogeneity problem. Classical multivariate procedures include empirical-distribution and rank tests \citep{Bickel1969,HallTajvidi2002}, nearest-neighbor and graph constructions \citep{FriedmanRafsky1979,Schilling1986,Henze1988,HenzePenrose1999,Rosenbaum2005,ChenFriedman2017}, and interpoint-distance or energy methods \citep{BickelBreiman1983,MaaPearlBartoszynski1996,BartoszynskiPearlLawrence1997,AslanZech2005,SzekelyRizzo2013}. Other influential approaches include the likelihood-ratio-type empirical-distribution statistics of \citet{Zhang2006}, kernel maximum mean discrepancy \citep{GrettonEtAl2012}, and two-sample procedures designed for high-dimensional observations \citep{LiuModarres2011,BiswasGhosh2014}. These methods provide a broad set of omnibus discrepancies that can be embedded in a change-point scan.

For fixed or moderate dimension, distributional change-point procedures have been developed from several such discrepancies. Kernel change-point testing was introduced by \citet{HarchaouiBachMoulines2009}, energy-distance segmentation by \citet{MattesonJames2014}, and a nonparametric likelihood approach to multiple changes by \citet{ZouYinFengWang2014}; the latter was made computationally more efficient by \citet{HaynesFearnheadEckley2017}. Rank-based methods were studied by \citet{LungYutFongLevyLeducCappe2015}, while \citet{BucherKojadinovicRohmerSegers2014} focused on changes in cross-sectional dependence through sequential empirical copulas. Graph-based scans were developed by \citet{ChenZhang2015}, refined to obtain asymptotically distribution-free calibration by \citet{ChuChen2019}, and extended to repeated observations by \citet{SongChen2022}; \citet{ChenChu2023} review this line of work. Kernel segmentation and its computational theory were further developed by \citet{GarreauArlot2018}, \citet{CelisseMarotPierreJeanRigaill2018}, and \citet{ArlotCelisseHarchaoui2019}. Related nonparametric formulations include data-depth ranks \citep{ChenouriMozaffariRice2020}, Fr\'echet mean and variance changes for metric-space data \citep{DubeyMuller2020}, and Kolmogorov--Smirnov localization theory \citep{MadridPadillaYuWangRinaldo2021}.

Recent work has expanded distributional change-point analysis to serial dependence, learned representations, and complex data objects. \citet{McGonigleCho2025} use joint characteristic functions to detect changes in marginal distributions and nonlinear serial dependence. \citet{KanrarJiangCai2025} construct a model-free scan from the area under the curve of a classifier. The rank-transformation framework of \citet{CuiGengWangWangZou2026} yields distribution-free, model-agnostic procedures with finite-sample guarantees, whereas \citet{LiYu2026} learn an integral-probability-metric critic through a neural CUSUM objective. These contributions greatly broaden the observation spaces and distributional features that can be analyzed without committing to a parametric pre- and post-change model.

The literature with theory tailored specifically to a diverging dimension is considerably smaller. \citet{Li2020} gives a distance method for high-dimensional observations, and \citet{GrundyKillickMihaylov2020} propose a geometrically motivated low-dimensional mapping. The data-adaptive $U$-statistic framework of \citet{LiuZhouZhangLiu2020} accommodates several high-dimensional parameters and alternative sparsity patterns. A closely related homogeneity-metric construction for general high-dimensional distributional changes is developed by \citet{ChakrabortyZhang2021}. \citet{MadridPadillaYuWangRinaldo2022} establish minimax localization theory under smooth multivariate density changes. Kernel and $U$-statistic procedures are studied by \citet{WangLiMengLi2023} and \citet{BonieceHorvathJacobs2024}, and \citet{DrikvandiModarres2025} propose a distribution-free distance method for non-sparse high-dimensional changes. A recent preprint by \citet{ChoudhuryXie2026} considers coordinate-averaged angular kernels for marginal distributional shifts in the high-dimension, low-sample-size regime. Collectively, these papers demonstrate the feasibility of nonparametric high-dimensional change-point inference, while also showing that explicit null calibration and power depend critically on how the distributional discrepancy is represented, standardized, and aggregated.

By comparison, a large part of the high-dimensional change-point literature targets changes in low-order moments or other finite-dimensional parameters. For mean vectors, important contributions include simultaneous-sequence methods \citep{ZhangSiegmundJiLi2010}, panel CUSUM tests \citep{HorvathHuskova2012,Jirak2015,Cho2016}, sparse-projection and minimax procedures \citep{WangSamworth2018,EnikeevaHarchaoui2019}, finite-sample identification and missing-data extensions \citep{YuChen2021,FollainWangSamworth2022}, multiscale online detection \citep{ChenWangSamworth2022}, and adaptive combinations of sum- and maximum-type scans \citep{WangFeng2023}. For covariance and dependence structures, representative methods include \citet{AueHormannHorvathReimherr2009}, \citet{ChoFryzlewicz2015}, \citet{BarnettOnnela2016}, \citet{AvanesovBuzun2018}, and \citet{WangYuRinaldo2021}. The review of \citet{LiuZhangLiu2022} documents the breadth of this rapidly developing area. Moment-specific procedures are powerful when their target parameter changes, but a distribution may change in scale, tail thickness, skewness, modality, or another feature while preserving the mean and even the covariance matrix. Such alternatives motivate inference that is sensitive to the marginal distributions themselves.

This paper first studies a single change in the coordinatewise marginal distributions of high-dimensional observations and then extends the resulting local inference to multiple changes. The formulation is nonparametric in the margins: no location--scale family, density model, or moment condition is imposed. At each candidate split, we apply the likelihood-ratio-type two-sample statistic of \citet{Zhang2006} to every coordinate. Under a continuous null margin, the statistic depends only on the pooled ranks. Its exact finite-sample null mean and variance can therefore be computed once for each sample size and split point, independently of the unknown marginal distributions. We use these quantities to form standardized deficits. Their coordinate average produces a dense scan that accumulates many moderate marginal changes, whereas their coordinatewise maximum produces a sparse scan that retains sensitivity when only a few margins change. Both statistics are defined as maxima of standardized evidence and hence reject for large values.

To accommodate alternatives with unknown sparsity, we establish the asymptotic independence of the dense and sparse scan statistics under the null and construct a Cauchy combination test from their corresponding \(p\)-values. This combined procedure integrates evidence from both scans without requiring prior knowledge of the number of affected coordinates. We further develop single-change estimators with explicit localization rates and multiple-change procedures with consistent recovery of the number of changes and vanishing relative location error under method-specific signal conditions. Simulation studies demonstrate the favorable finite-sample performance of the proposed methods across a range of dense and sparse alternatives, while an empirical application illustrates their practical effectiveness in detecting and locating distributional changes.

Our main contributions are as follows.

\begin{enumerate}[label=(\roman*)]
\item We propose three complementary rank-based procedures for detecting high-dimensional marginal distribution changes: a sum-type scan that accumulates moderate signals across many coordinates, a maximum-type scan that targets changes in a few coordinates, and a Cauchy combination test that is robust to unknown sparsity. Our methodological innovation lies in converting coordinatewise likelihood-ratio-type rank statistics into standardized deficits using their exact finite-sample null means and variances, then integrating complementary aggregation strategies to accommodate varying signal sparsity without imposing parametric marginal models or moment conditions.

\item Under a Gaussian copula model with weak cross-sectional dependence, we establish the asymptotic null distributions of SUM and MAX and prove their asymptotic independence. A central contribution is the development of sharp distributional theory for the likelihood-ratio-type rank statistic of \citet{Zhang2006}: despite its strong empirical performance, the precise null-tail and extreme-value characterizations needed for high-dimensional scanning have, to our knowledge, remained unavailable. For MAX, we retain the nonlinear contributions of the extreme ranks and combine conditional Gaussian-bridge approximation with exponential tilting to derive sharp tail probabilities for the original rank statistic. A nonlinear Bellman equation and convergent Bernoulli representations determine the endpoint prefactors, while local exceedance-cluster analysis and a weak-dependence Poisson approximation yield the Gumbel limit with explicit centering and scaling. Separate boundary analysis and an analytic convolution account for the boundary-dominated and critical regimes. For SUM, a row-kernel approximation establishes the Gaussian-process limit, and conditioning on finitely many extreme coordinates yields SUM--MAX asymptotic independence, providing the theoretical foundation for Cauchy combination.

\item We establish explicit localization rates for single-change estimation and separate consistency theorems for SUM-WBS, MAX-WBS, and Cauchy-adaptive WBS. The multiple-change results allow vanishing marginal signals and different active coordinate sets at different changes. They cover every fixed polynomial dimension \(p=n^a\), \(0<a<2\), with arbitrary coordinate copulas for MAX-WBS. SUM-WBS and Cauchy-WBS additionally assume a common Gaussian copula with a correlation row sum growing as a fixed power of \(\log n\). Each method has its own sufficient signal condition; the adaptive condition allows its successful branch to vary between change points. The proofs control the actual interval and component selections and establish correct change counts and vanishing relative location error.

\end{enumerate}

The remainder of the paper is organized as follows. Section~\ref{sec:statistics} introduces the model and the standardized rank scans, establishes their null limits and asymptotic independence, and develops single-change testing and localization theory. Section~\ref{sec:wbs} extends the procedures to multiple changes through componentwise and Cauchy-adaptive wild binary segmentation, with separate conditions for recovering the number and locations of changes. Sections~\ref{sec:simulation} and~\ref{sec:application} present simulations and empirical applications. Section~\ref{sec:conclusion} concludes. Technical derivations and proofs are collected in Appendices~\ref{app:rank}--\ref{app:wbs}; Appendix~\ref{app:analytic-size} assesses the asymptotic null calibration numerically.

\paragraph{Notation.}
Bold italic letters denote vectors, bold upright letters denote matrices,
and $(\cdot)^\top$ denotes transposition.
For a random element $Y$, $\mathcal L(Y)$ denotes its probability law;
in particular, $\mathcal L(\bX_i)$ is the joint distribution of the
random vector $\bX_i$.
The symbols $\overset{d}{=}$, $\xrightarrow{d}$ (or $\Rightarrow$),
and $\xrightarrow{\Pp}$ denote equality in distribution, convergence
in distribution, and convergence in probability, respectively.
For an event $A$, $\mathbf1(A)$ denotes its indicator, whereas $|B|$
denotes the cardinality of a finite set $B$.
For real numbers $a$ and $b$, write
$a\vee b=\max\{a,b\}$, $a\wedge b=\min\{a,b\}$, and
$a_+=\max\{a,0\}$.
The symbols $\lfloor x\rfloor$ and $\lceil x\rceil$ denote the floor
and ceiling of $x$, respectively.
For positive sequences $a_n$ and $b_n$, $a_n\sim b_n$ means
$a_n/b_n\to1$, while $a_n\asymp b_n$ means that their ratio is
bounded above and away from zero for all sufficiently large $n$.
For random variables $Y_n$ and a positive deterministic sequence $a_n$,
$Y_n=O_{\Pp}(a_n)$ means that $Y_n/a_n$ is bounded in probability.
For a bounded function $f$, $\|f\|_\infty$ denotes its supremum norm
over the specified domain.
All logarithms are natural.
Unless otherwise stated, generic positive constants $c$ and $C$
are independent of $n$ and $p$ and may change from line to line;
subscripts indicate dependence on fixed parameters. The symbols $\tau^\star$ and $\lambda^\star=\tau^\star/n$ are reserved for the single-change location and fraction, whereas $\tau_q$ indexes multiple change points. The label $\comp\in\{\mathrm{sum},\mathrm{max}\}$ denotes a component procedure.

\section{High-dimensional model and standardized statistics}\label{sec:statistics}

\subsection{Model and coordinatewise discrepancy}

Let $\bX_1,\ldots,\bX_n\in\R^p$ be independent random vectors, where
$\bX_i=(X_{i1},\ldots,X_{ip})^\top$ and $p=p_n$ may diverge with $n$. We test
\begin{equation*}
\begin{aligned}
H_0:&\quad \mathcal L(\bX_1)=\cdots=\mathcal L(\bX_n),\\
H_1:&\quad \bX_i\sim F^{(1)},\quad i\le\tau^\star,
\qquad
\bX_i\sim F^{(2)},\quad i>\tau^\star,
\end{aligned}
\end{equation*}
where $\tau^\star\in\{1,\ldots,n-1\}$ is unknown. Write $F_j^{(1)}$ and $F_j^{(2)}$ for the two marginal distribution functions of coordinate $j$. The targeted alternative is
\[
\exists j\in\{1,\ldots,p\}:\quad F_j^{(1)}\ne F_j^{(2)}.
\]

Fix $\eta\in(0,1/2)$ and let
\[
m_n=\lceil n\eta\rceil,
\qquad
\cK_n=\{m_n,\ldots,n-m_n\},
\qquad
\mathcal I_\eta=[\eta,1-\eta].
\]
For $k\in\cK_n$, define
\[
\widehat F^{(k)}_{1j}(x)=\frac1k\sum_{i=1}^k\mathbf1(X_{ij}\le x),
\qquad
\widehat F^{(k)}_{2j}(x)=\frac1{n-k}\sum_{i=k+1}^n\mathbf1(X_{ij}\le x).
\]
Let $X_{(1)j}\le\cdots\le X_{(n)j}$ be the pooled order statistics and set
\[
\phifun(u)=u\log u+(1-u)\log(1-u),
\qquad
h(u)=-\phifun(u),
\qquad 0\log0=0.
\]
Following the likelihood-ratio-type two-sample procedure of \citet{Zhang2006}, we compare the two empirical marginal distributions at the pooled order statistics. More precisely, we apply Zhang's two-sample $Z_A$ statistic to coordinate $j$ and candidate split $k$:
\begin{equation}\label{eq:coordinate-stat}
Z_j(k)
=\sum_{\ell=1}^n
\frac{k h\!\left\{\widehat F^{(k)}_{1j}(X_{(\ell)j})\right\}
+(n-k)h\!\left\{\widehat F^{(k)}_{2j}(X_{(\ell)j})\right\}}
{(\ell-1/2)(n-\ell+1/2)}.
\end{equation}
The statistic accumulates likelihood-ratio discrepancies over the pooled order statistics with an Anderson--Darling-type tail weight. Under the present sign convention, a smaller $Z_j(k)$ indicates a larger discrepancy between the two empirical marginal distributions.


Let $\mathfrak S_n$ be the set of permutations of $\{1,\ldots,n\}$ and let
$\bm\Pi=(\Pi_1,\ldots,\Pi_n)$ be uniform on $\mathfrak S_n$. Define
\[
C_{k\ell}(\bm\Pi)=\sum_{i=1}^k\mathbf1(\Pi_i\le\ell),
\qquad
d_{\ell,n}=(\ell-1/2)(n-\ell+1/2),
\]
and
\begin{equation*}
\mathcal Z_{n,k}(\bm\Pi)
=\sum_{\ell=1}^n
\frac{k h\{C_{k\ell}(\bm\Pi)/k\}
+(n-k)h\{(\ell-C_{k\ell}(\bm\Pi))/(n-k)\}}
{d_{\ell,n}}.
\end{equation*}
For a continuous marginal distribution, $Z_j(k)$ has the same null distribution as
$\mathcal Z_{n,k}(\bm\Pi)$. Hence define
\begin{equation*}
\mu_{n,k}=\E_\Pi\{\mathcal Z_{n,k}(\bm\Pi)\},
\qquad
s_{n,k}^2=\operatorname{Var}_\Pi\{\mathcal Z_{n,k}(\bm\Pi)\}.
\end{equation*}
These constants depend only on $(n,k)$.
The deterministic zero-gap value is
\[
 \overline Z_n=\sum_{\ell=1}^{n-1}
        \frac{n h(\ell/n)}{d_{\ell,n}}.
\]
For $0<u<1$, concavity of the logarithm gives
\[
 h(u)=2\{u\log(u^{-1/2})+(1-u)\log((1-u)^{-1/2})\}
 \le2\log(\sqrt u+\sqrt{1-u})
 \le2\sqrt{u(1-u)}.
\]
For a sample with no ties within coordinates, concavity of $h$,
$d_{\ell,n}\ge\ell(n-\ell)/2$, and integral comparison give
\begin{equation}\label{eq:rank-statistic-uniform-range}
 0\le Z_j(k)\le\overline Z_n
 \le4\sum_{\ell=1}^{n-1}\frac1{\sqrt{\ell(n-\ell)}}
 \le4\int_0^1\frac{du}{\sqrt{u(1-u)}}=4\pi<16.
\end{equation}


Define
\[
A_n(k)=\frac1p\sum_{j=1}^p Z_j(k),
\qquad
D_{j,n}(k)=\frac{\mu_{n,k}-Z_j(k)}{s_{n,k}}.
\]
The local dense and sparse score curves are
\[
D_{\mathrm{sum},n,p}(k)
=\frac{\sqrt p\{\mu_{n,k}-A_n(k)\}}{s_{n,k}},
\qquad
D_{\mathrm{max},n,p}(k)=\max_{1\le j\le p}D_{j,n}(k).
\]
For each coordinate, set
\[
\mathcal M_{j,n}=\max_{k\in\cK_n}D_{j,n}(k).
\]
Their global raw scans are
\begin{equation}\label{eq:global-stats}
\Ssum=\max_{k\in\cK_n}D_{\mathrm{sum},n,p}(k),
\qquad
\Smax=\max_{k\in\cK_n}D_{\mathrm{max},n,p}(k)
=\max_{1\le j\le p}\mathcal M_{j,n},
\end{equation}

At each candidate split $k$, the standardized deficit $D_{j,n}(k)$
measures how far $Z_j(k)$ falls below its null mean, in units of its
null standard deviation. A large positive deficit therefore provides
evidence of a difference between the marginal distributions before
and after $k$ in coordinate $j$. The dense score
$D_{\mathrm{sum},n,p}(k)=p^{-1/2}\sum_{j=1}^p D_{j,n}(k)$
aggregates this evidence across coordinates and is designed to detect
changes spread over many margins. In contrast, the sparse score
$D_{\mathrm{max},n,p}(k)$ retains the largest coordinatewise deficit
and targets changes concentrated in a small number of margins.
Maximizing these scores over the candidate splits gives $\Ssum$
and $\Smax$, respectively, and both tests reject for large values.

Although Section~\ref{sec:null} establishes the asymptotic null
distributions of $\Ssum$ and $\Smax$, the numerical assessment in
Appendix~\ref{app:analytic-size} shows that the resulting approximations
can still exhibit noticeable size distortions at moderate sample sizes.
To obtain more reliable finite-sample calibration, we therefore use
whole-vector permutation rather than relying exclusively on the
limiting critical values. Specifically, we randomly reorder the
observation vectors while keeping their coordinates together and
recompute the entire scan for each permutation. This preserves the
observed cross-coordinate dependence and, under exchangeability of
the observation vectors under the null, yields finite-sample valid
$p$-values for the two component tests when the standard plus-one
correction is used; see Appendix~\ref{app:rank}.
The use of permutation calibration alongside asymptotic analysis
is also common in nonparametric change-point detection
\citep{MattesonJames2014,ChakrabortyZhang2021,DrikvandiModarres2025}.

For finite-sample calibration, we permute the observation index while keeping every $p$-variate vector intact. For $b=1,\ldots,B$, independently draw a permutation $\pi_b$ uniformly from $\mathfrak S_n$ and define
\[
\bX_i^{*(b)}=\bX_{\pi_b(i)},
\qquad i=1,\ldots,n.
\]
For every $k\in\cK_n$, split the permuted sequence into
$\{\bX_1^{*(b)},\ldots,\bX_k^{*(b)}\}$ and
$\{\bX_{k+1}^{*(b)},\ldots,\bX_n^{*(b)}\}$. From these two subsamples, recompute the coordinatewise empirical distribution functions and obtain
$Z_j^{*(b)}(k)$ from \cref{eq:coordinate-stat}. Using the same deterministic null constants $\mu_{n,k}$ and $s_{n,k}$ as for the observed data, set
\[
A_n^{*(b)}(k)=\frac1p\sum_{j=1}^p Z_j^{*(b)}(k),
\]
\[
D_{\mathrm{sum},n,p}^{*(b)}(k)
=\frac{\sqrt p\{\mu_{n,k}-A_n^{*(b)}(k)\}}{s_{n,k}},
\qquad
D_{\mathrm{max},n,p}^{*(b)}(k)
=\max_{1\le j\le p}
\frac{\mu_{n,k}-Z_j^{*(b)}(k)}{s_{n,k}},
\]
and
\[
\Ssum^{*(b)}
=\max_{k\in\cK_n}D_{\mathrm{sum},n,p}^{*(b)}(k),
\qquad
\Smax^{*(b)}
=\max_{k\in\cK_n}D_{\mathrm{max},n,p}^{*(b)}(k).
\]
The upper-tail Monte Carlo $p$-values are then
\[
\widehat p_{\mathrm{sum}}
=\frac{1+\sum_{b=1}^B\mathbf1\{\Ssum^{*(b)}\ge\Ssum\}}{B+1},
\qquad
\widehat p_{\mathrm{max}}
=\frac{1+\sum_{b=1}^B\mathbf1\{\Smax^{*(b)}\ge\Smax\}}{B+1}.
\]
Permuting whole vectors, rather than permuting coordinates separately, preserves the contemporaneous cross-coordinate dependence within each observed vector. Under $H_0$, exchangeability of the vector sequence makes the observed and permuted statistics conditionally exchangeable.

For the permutation-calibrated adaptive test, include the observed ordering as $b=0$ and write $S_{\comp,n,p}^{*(0)}=S_{\comp,n,p}$. Rank each component within the same collection of $B+1$ orderings:
\[
 \widehat p_{\comp}^{(b)}
 =\frac1{B+1}\sum_{r=0}^B
   \mathbf1\{S_{\comp,n,p}^{*(r)}\ge S_{\comp,n,p}^{*(b)}\},
 \qquad b=0,\ldots,B,\quad
 \comp\in\{\mathrm{sum},\mathrm{max}\}.
\]
Thus $\widehat p_{\comp}^{(0)}=\widehat p_{\comp}$. To keep the Cauchy transform finite at the endpoints, define
\[
 \delta_{\mathrm{clip}}=8\cdot2^{-52},\qquad
 c_{\mathrm{clip}}(u)=\min\{1-\delta_{\mathrm{clip}},
                              \max(\delta_{\mathrm{clip}},u)\}.
\]
The same clipping and equal-weight combination are applied to every ordering:
\[
 T_{\mathrm C}^{\mathrm{perm},(b)}
 =\frac12\tan\!\left[\pi\left\{\frac12-
                c_{\mathrm{clip}}(\widehat p_{\mathrm{sum}}^{(b)})\right\}\right]
 +\frac12\tan\!\left[\pi\left\{\frac12-
                c_{\mathrm{clip}}(\widehat p_{\mathrm{max}}^{(b)})\right\}\right].
\]
The resulting permutation $p$-value is
\[
 \widehat p_{\mathrm C}
 =\frac1{B+1}\sum_{b=0}^B
   \mathbf1\{T_{\mathrm C}^{\mathrm{perm},(b)}
                      \ge T_{\mathrm C}^{\mathrm{perm},(0)}\}.
\]
This construction calibrates the combined score by its rank among the same observed and permuted orderings. It defines the finite-sample adaptive procedure used in the numerical studies; the analytic Cauchy value in Section~\ref{sec:null} is defined from the limiting component distributions.

\subsection{Asymptotic null distributions}\label{sec:null}

Let $\Phi$ denote the standard normal distribution function.
For a correlation matrix $\mathbf R_p=(\rho_{j\ell})$, define
\[
\mathfrak r_p=\max_{1\le j\le p}\sum_{\ell\ne j}|\rho_{j\ell}|.
\]
For neighborhoods $\mathcal B_{j,p}\subseteq\{1,\ldots,p\}$, require
\[
 j\in\mathcal B_{j,p},\qquad
 \ell\in\mathcal B_{j,p}\ \Longleftrightarrow\ j\in\mathcal B_{\ell,p},
\]
and define
\[
D_p=\max_{1\le j\le p}|\mathcal B_{j,p}|,
\qquad
\delta_p=\max_{1\le j\le p}
\sum_{\ell\notin\mathcal B_{j,p}}|\rho_{j\ell}|.
\]

\begin{assumption}[Approximately sparse Gaussian copula null]\label{ass:copula-null}
Under $H_0$, the common margins $F_1,\ldots,F_p$ are continuous.
The transformed row vectors satisfy
\[
Y_{ij}=\Phi^{-1}\{F_j(X_{ij})\},
\qquad
\bm Y_i=(Y_{i1},\ldots,Y_{ip})^\top
\overset{\mathrm{iid}}{\sim}N_p(\bm0,\mathbf R_p),
\]
The correlation matrix satisfies the following conditions.
\begin{enumerate}[label=(\roman*)]
\item\label{ass:copula-pairwise}
There is a fixed constant $\rho_\star\in[0,1)$ such that
\[
\max_{j\ne\ell}|\rho_{j\ell}|\le\rho_\star.
\]
\item\label{ass:copula-row}
\begin{equation}\label{eq:sum-dependence-rates}
\frac{\mathfrak r_p}{\log n}\longrightarrow0,
\qquad
\frac{\mathfrak r_p^2}{\sqrt p}\longrightarrow0.
\end{equation}
\item\label{ass:copula-neighborhood}
There are neighborhoods of the preceding form such that
\begin{equation}\label{eq:max-dependence-rates}
\frac{\log(D_p\vee2)}{\log p}\longrightarrow0,
\qquad
\delta_p\log p\longrightarrow0.
\end{equation}
\end{enumerate}
\end{assumption}

Assumption~\ref{ass:copula-null} separates the marginal distributions
from the cross-coordinate dependence. The Gaussian copula restriction
is imposed on the transformed vectors rather than on the original
observations, so the continuous margins may be non-Gaussian and
heavy-tailed, without requiring their moments to exist. In particular,
$\rho_{j\ell}$ describes the correlation between the latent Gaussian
coordinates, not necessarily the Pearson correlation between the
original variables. This semiparametric formulation is widely used
in high-dimensional rank-based analysis
\citep{LiuHanYuanLaffertyWasserman2012}.

Assumption~\ref{ass:copula-null}\ref{ass:copula-pairwise} keeps distinct
latent coordinates uniformly away from perfect positive or negative
correlation, while allowing nonvanishing correlations between them.
Related restrictions appear in high-dimensional extreme-value
and change-point theory \citep{Jirak2015,WangFeng2023}.
The quantity $\mathfrak r_p$ measures the largest total absolute
correlation associated with a single coordinate.
Condition~\eqref{eq:sum-dependence-rates} limits the accumulation
of cross-coordinate dependence in the dense scan, ensuring that
the covariance corrections and higher-order remainders arising
in its Gaussian-process approximation are asymptotically negligible.
This condition permits $\mathfrak r_p$ to diverge.

Condition~\eqref{eq:max-dependence-rates} distinguishes relatively
strong local dependence from weak residual dependence.
The requirement $D_p=p^{o(1)}$ permits each coordinate to have
a growing, but subpolynomial, number of neighbors, whereas
$\delta_p\log p\to0$ controls the total absolute correlation
outside these neighborhoods at the scale relevant to extreme
events. The neighborhoods need not follow any prescribed ordering
of the coordinates, and correlations outside them need not be
exactly zero. Together with the pairwise correlation bound,
these restrictions control simultaneous exceedances across
coordinates and support the Poisson approximation underlying
the Gumbel limit for MAX. Combined with the dense-scan conditions,
they also permit the asymptotic separation of the aggregate SUM
fluctuations from the rare coordinates determining MAX.
Related neighborhood and aggregate-correlation conditions are
used in high-dimensional maximum tests and SUM--MAX
asymptotic-independence theory
\citep{WangFeng2023,FengJiangLiLiu2024}; the particular rates imposed
here are tailored to the present rank-based scans.

For a growing-band model with $\rho_{j\ell}=0$ when
$|j-\ell|>m_p$ and $|\rho_{j\ell}|\le\rho_\star$ for $j\ne\ell$, take
$\mathcal B_{j,p}=\{\ell:|j-\ell|\le m_p\}$. Then
\[
D_p\le2m_p+1,
\qquad
\mathfrak r_p\le2\rho_\star m_p,
\qquad
\delta_p=0.
\]
Both dependence-rate conditions hold if
\[
m_p=o(\log n),
\qquad
\frac{\log(m_p\vee2)}{\log p}\longrightarrow0.
\]
Indeed the second relation gives $m_p=p^{o(1)}$ and hence
$m_p^2/\sqrt p\to0$. Uniformly positive definite banded correlation
matrices with these bounds provide one class of examples.

For a causal invertible finite-order ARMA correlation with geometrically decaying coefficients, choose a neighborhood width $m_{n,p}=p^{o(1)}$ whose remaining rowwise correlation tail is $o(1/\log p)$. Then \eqref{eq:max-dependence-rates} holds. In particular, fixed AR(1) and ARMA correlations meet the neighborhood condition even though their matrices are dense.

For $t\in\mathcal I_\eta$, let
\[
k_n(t)=\min\{n-m_n,\max(m_n,\lfloor nt\rfloor)\}.
\]
The dense process is the function
$t\mapsto D_{\mathrm{sum},n,p}\{k_n(t)\}$ on $\mathcal I_\eta$.
Define the standardized Brownian-bridge correlation kernel
\[
\rho_{\mathrm B}(s,t)
=\frac{s\wedge t-st}{\{s(1-s)t(1-t)\}^{1/2}},
\qquad s,t\in\mathcal I_\eta,
\]
and the covariance kernel
\begin{equation}\label{eq:limit-process-kernel}
\mathcal C(s,t)=\rho_{\mathrm B}(s,t)^2,
\qquad s,t\in\mathcal I_\eta.
\end{equation}
Let $\mathbb G=\{\mathbb G(t):t\in\mathcal I_\eta\}$ be the centered Gaussian process with covariance kernel $\mathcal C$. With
\[
\ell(t)=\log\!\left\{\frac{t}{1-t}\right\},
\qquad
L_\eta=\log\!\left(\frac{1-\eta}{\eta}\right),
\]
we have
\[
\{\mathbb G(t):t\in\mathcal I_\eta\}
\overset d=
\{\mathbb U(\ell(t)):t\in\mathcal I_\eta\},
\]
where $\mathbb U$ is a stationary Ornstein--Uhlenbeck process with
$\operatorname{Cov}\{\mathbb U(u),\mathbb U(v)\}=e^{-|u-v|}$.

\begin{theorem}[Gaussian-process limit for the dense statistic]\label{thm:sum-null}
Suppose Assumption~\ref{ass:copula-null}\ref{ass:copula-pairwise}--%
\ref{ass:copula-row} holds, and let $n\to\infty$ and $p=p_n\to\infty$
along any joint sequence satisfying \eqref{eq:sum-dependence-rates}.
Then the dense process converges in the supremum norm on
$\mathcal I_\eta$ to the centered Gaussian process with covariance
$\mathcal C$ in \eqref{eq:limit-process-kernel}.  Consequently
\begin{equation*}
 \Ssum\xrightarrow d\mathcal M_\eta
 :=\sup_{t\in\mathcal I_\eta}\mathbb G(t)
 =\sup_{|u|\le L_\eta}\mathbb U(u).
\end{equation*}
\end{theorem}

For a fixed nominal level $\alpha\in(0,1)$, let
$c_{\mathrm{sum},1-\alpha}(\eta)$ be the $(1-\alpha)$ quantile of
$\mathcal M_\eta$. The dense test is
\begin{equation}\label{eq:sum-rejection-rule}
\phi_{\mathrm{sum},\alpha}
=\mathbf1\{\Ssum>c_{\mathrm{sum},1-\alpha}(\eta)\}.
\end{equation}
The reference distribution is continuous. Theorem~\ref{thm:sum-null} therefore gives asymptotic size $\alpha$ for the dense test.

The maximum of $p$ coordinatewise scans increases with $p$, so its
null calibration requires a centering constant $\mathfrak b_{n,p}$ and
a positive scale $\mathfrak a_{n,p}$. Their purpose is to place the
threshold $\mathfrak b_{n,p}+\mathfrak a_{n,p}x$ at a level where the
expected number of exceeding coordinates approaches $e^{-x}$.
We first identify the leading location and tail slope, then incorporate
the smaller centering correction and establish the Gumbel law.
All these quantities depend
only on $n,p$ and the trimming fraction $\eta$, since the continuous
one-coordinate null rank distribution is distribution-free.

The relevant dimension is $\gamma_n=(\log p)/(\log n)$. As this ratio
increases, exceptionally long runs of rank labels can affect the tail
of the scan. The point at which they do so is determined by one
constant $b_\eta$. For $0<t<1$, $0\le x\le1$, and a measurable
function $f:[0,\infty)\to[0,1]$, define
\[
 h_t(x)=x\log\frac{x}{t}+(1-x)\log\frac{1-x}{1-t},
 \qquad Hf(s)=\frac1s\int_0^s f(z)\,dz\quad(s>0),
\]
with $0\log0=0$. Here $h_t$ is the Bernoulli relative entropy and $Hf$
is the running average of $f$. Define $b_\eta$ as the largest constant
for which
\begin{equation}\label{eq:main-hardy-threshold}
 b_\eta\int_0^\infty h_t(Hf(s))\,ds
 \le \int_0^\infty h_t(f(s))\,ds
\end{equation}
holds for every $t\in[\eta,1-\eta]$ and every measurable
$f:[0,\infty)\to[0,1]$ with a finite, positive right-hand side.
Thus $b_\eta$ compares the entropy of an averaged rank-label profile
with the cost of producing that profile. It satisfies
$0<b_\eta\le1/4$ and is distinct from the centering constant
$\mathfrak b_{n,p}$. Appendix~\ref{app:max-deterministic-inputs} gives
its equivalent differential-equation characterization and the
constants needed for numerical evaluation.

Write $N=\log n$ and define
\[
 q_n=(\sqrt{\gamma_n}+\sqrt{1+\gamma_n})^2,
 \qquad \lambda(z)=\frac{1-\sqrt{1-4z}}2
 \quad(0\le z\le1/4).
\]
The leading raw level is $Ny_n$, where $y_n$ and the tail slope
$\zeta_n$ are
\begin{equation}\label{eq:main-max-leading-level}
 \zeta_n=\min\left\{\frac{1-q_n^{-2}}4,b_\eta\right\},
 \qquad y_n=\frac{\gamma_n+\lambda(\zeta_n)}{\zeta_n}.
\end{equation}
Below the boundary, $y_n=q_n$; after the slope reaches $b_\eta$,
$y_n=\{\gamma_n+\lambda(b_\eta)\}/b_\eta$. Consequently the leading
center is $(y_n-1)\sqrt{N/2}$, and an increase of
$1/(\zeta_n\sqrt{2N})$ in the threshold for a single coordinate scan changes its rare-event
tail probability by an asymptotic factor $e^{-1}$. The moment expansions
underlying this conversion are
\begin{equation}\label{eq:main-null-moment-orders}
 n(\overline Z_n-\mu_{n,k})=\log n+O(1),
 \qquad n^2s_{n,k}^2=2\log n+O(1),
\end{equation}
uniformly over the trimmed grid; their correction terms are developed
in Appendix~\ref{app:moment-correction-definitions}.

The leading center alone does not determine the Gumbel calibration:
the scan over split locations and the exact rank studentization also
contribute a multiplicative tail factor. This factor enters the centering through the construction in
Appendix~\ref{app:max-normalizing-constants}. Specifically, let $C_n>0$
be the subcritical scan coefficient in
\eqref{eq:main-subcritical-normalization}, with its dependence on
$p$ and $\eta$ suppressed in the notation. Let
$\mathcal R_N^{\rm fc}(U)$ be the critical scan-tail approximation
at raw level $U$ given by the positive convolution in
\eqref{eq:main-critical-convolution}, and let $h_n^{\rm B}$ be the
boundary correction in \eqref{eq:main-boundary-normalization}.
These are deterministic quantities specified there by integrals,
convergent series and differential equations; they require no
unknown marginal distribution or fitted copula parameter.

For $b_\eta<1/4$, define the transition point and switching width by
\[
 q_c=(1-4b_\eta)^{-1/2},\qquad
 \gamma_c=\frac{(q_c-1)^2}{4q_c},\qquad
 \epsilon_N=N^{-1/4}.
\]
The three tail factors enter through the single definition
\begin{equation}\label{eq:main-max-tail-factor}
 \mathcal C_{n,p}=
 \begin{cases}
 C_n,&\gamma_n<\gamma_c-\epsilon_N,\\
 p\mathcal R_N^{\rm fc}(Ny_n),
       &|\gamma_n-\gamma_c|\le\epsilon_N,\\
 \exp\{b_\eta h_n^{\rm B}\},
       &\gamma_n>\gamma_c+\epsilon_N.
 \end{cases}
\end{equation}
The first branch applies below the transition, the second covers its
shrinking neighborhood, and the third applies above it. If
$b_\eta=1/4$, set $q_c=\gamma_c=\infty$ and use
$\mathcal C_{n,p}=C_n$ throughout. The critical convolution provides a deterministic normalization
across the transition in the dimension ratio.

The normalizing constants have the common form
\begin{equation}\label{eq:full-max-branch-selection}
 \mathfrak a_{n,p}=\frac1{\zeta_n\sqrt{2N}},\qquad
 \mathfrak b_{n,p}=(y_n-1)\sqrt{N/2}
                       +\mathfrak a_{n,p}\log\mathcal C_{n,p}.
\end{equation}
Thus $\mathfrak a_{n,p}$ converts the local tail slope to the Gumbel
scale, while $\mathfrak b_{n,p}$ adds the log-tail correction to the
leading location. In the one-coordinate scan-tail expansions,
$\mathcal C_{n,p}$ is the prefactor after accounting for $p$
coordinates: a shift of $\mathfrak a_{n,p}\log\mathcal C_{n,p}$
removes it. This yields
$p\Pp\{\mathcal M_{1,n}>\mathfrak b_{n,p}
+\mathfrak a_{n,p}x\}\to e^{-x}$. The three cases in
\eqref{eq:main-max-tail-factor} correspond to the subcritical, critical
and boundary pairs in Appendix~\ref{app:max-normalizing-constants}.
Equation~\eqref{eq:full-max-branch-selection} expresses these pairs
in a common form.

\begin{theorem}
\label{thm:max-full-analytic}
Under Assumption~\ref{ass:copula-null}, suppose $n,p\to\infty$ and
\[
 0<\underline\gamma\le\frac{\log p}{\log n}
                         \le\overline\gamma<\infty.
\]
Then, for every fixed $x\in\mathbb R$,
\begin{equation*}
 \Pp\!\left\{\frac{\Smax-\mathfrak b_{n,p}}
                       {\mathfrak a_{n,p}}\le x\right\}
                    \longrightarrow\exp(-e^{-x}).
\end{equation*}
\end{theorem}

The dimension ratio need not converge for this result. The weak
dependence conditions make simultaneous exceedances sufficiently rare
that the one-coordinate tail calibration gives the Gumbel law for
the maximum across coordinates. Write
$F_{\mathrm G}(x)=\exp(-e^{-x})$ and
$g_{1-\alpha}=-\log\{-\log(1-\alpha)\}$. The analytic sparse
rejection rule is
\begin{equation}\label{eq:max-rejection-rule}
 \phi_{\mathrm{max},\alpha}
 =\mathbf1\{\Smax>\mathfrak b_{n,p}
                       +\mathfrak a_{n,p}g_{1-\alpha}\}.
\end{equation}
Theorem~\ref{thm:max-full-analytic} gives asymptotic size $\alpha$
for every fixed $\alpha\in(0,1)$. For the subsequent joint limit, we use the polynomial-dimensional
regime
\begin{equation}\label{eq:max-polynomial-regime}
 n\to\infty,\qquad p=p_n\to\infty,\qquad
 \gamma_n\to\gamma\in(0,\infty).
\end{equation}

Let $\mathcal G$ have distribution $F_{\mathrm G}$ and write
\[
F_\eta(x)=\Pp(\mathcal M_\eta\le x).
\]
The distribution $F_\eta$ is continuous by the no-atom property for the supremum of a nondegenerate continuous Gaussian process \citep{AzaisWschebor2009} and can be evaluated using the Ornstein--Uhlenbeck representation.

Define the analytic component $p$-values by
\begin{equation}\label{eq:component-analytic-pvalues}
P_{\mathrm{sum},n,p}=1-F_\eta(\Ssum),
\qquad
P_{\mathrm{max},n,p}
=1-F_{\mathrm G}\!\left(\frac{\Smax-\mathfrak b_{n,p}}{\mathfrak a_{n,p}}\right)
\end{equation}

\begin{theorem}[Joint null limit and asymptotic independence]
\label{thm:joint-independence}
Under Assumption~\ref{ass:copula-null} and
\eqref{eq:max-polynomial-regime},
\begin{equation}\label{eq:joint-independent-limit}
 \left(\Ssum,
       \frac{\Smax-\mathfrak b_{n,p}}{\mathfrak a_{n,p}}\right)
 \xrightarrow d(\mathcal M_\eta,\mathcal G),
 \qquad \mathcal M_\eta\perp\mathcal G,
\end{equation}
\end{theorem}

By continuity of the two limiting distribution functions, the
component values in \eqref{eq:component-analytic-pvalues} converge
jointly to independent $\operatorname{Unif}(0,1)$ variables. Building on this result, we propose an adaptive testing procedure.

For a fixed weight $\omega\in(0,1)$, define the adaptive Cauchy statistic
\begin{equation*}
\Tcauchy(\omega)
=\omega\tan\!\left[\pi\left\{\frac12-P_{\mathrm{sum},n,p}\right\}\right]
 +(1-\omega)\tan\!\left[\pi\left\{\frac12-P_{\mathrm{max},n,p}\right\}\right]
\end{equation*}
and its combined $p$-value
\begin{equation*}
\Pcauchy(\omega)
=\frac12-\frac1\pi\arctan\{\Tcauchy(\omega)\}.
\end{equation*}
Equal weights $\omega=1/2$ are used by default.

\begin{corollary}[Cauchy calibration]\label{cor:cauchy-null}
Under the conditions of Theorem~\ref{thm:joint-independence},
for fixed \(\omega,\alpha\in(0,1)\),
\begin{equation}\label{eq:cauchy-null-limit}
 \Tcauchy(\omega)\xrightarrow d\operatorname{Cauchy}(0,1),
 \qquad
 \Pcauchy(\omega)\xrightarrow d\operatorname{Unif}(0,1),
\end{equation}
and
\[
 \Pp\{\Pcauchy(\omega)\le\alpha\}\longrightarrow\alpha.
\]
\end{corollary}

The combined rejection rule is
\begin{equation*}
\Pcauchy(\omega)\le\alpha,
\qquad\text{equivalently}\qquad
\Tcauchy(\omega)\ge\cot(\pi\alpha).
\end{equation*}

\subsection{Consistency under alternatives}\label{sec:power}

Let $F$ and $G$ be continuous univariate distribution functions, let $t\in(0,1)$, and write $H=tF+(1-t)G$. The population counterpart of \cref{eq:coordinate-stat} is
\begin{equation}\label{eq:population-Z}
\cZ_t(F,G)
=-\int
\frac{t\phifun\{F(x)\}+(1-t)\phifun\{G(x)\}}
{H(x)\{1-H(x)\}}\,dH(x).
\end{equation}
Define the weighted Jensen gap
\[
\cD_t(F,G)
=\int
\frac{t\phifun\{F(x)\}+(1-t)\phifun\{G(x)\}-\phifun\{H(x)\}}
{H(x)\{1-H(x)\}}\,dH(x).
\]
The identity \(H=tF+(1-t)G\) gives
\begin{equation}\label{eq:gap}
\cZ_t(F,G)=\frac{\pi^2}{3}-\cD_t(F,G),
\qquad
\cD_t(F,G)\ge0,
\end{equation}
with strict positivity whenever $F$ and $G$ differ on a set of positive $H$-measure.

Let $\lambda^\star=\tau^\star/n$. For a candidate fraction $t=k/n$, define the left and right population marginal distributions
\[
(F^L_{j,t},F^R_{j,t})=
\begin{cases}
\displaystyle\left(F_j^{(1)},
\frac{\lambda^\star-t}{1-t}F_j^{(1)}+
\frac{1-\lambda^\star}{1-t}F_j^{(2)}\right),&t<\lambda^\star,\\[1.1em]
(F_j^{(1)},F_j^{(2)}),&t=\lambda^\star,\\[0.4em]
\displaystyle\left(
\frac{\lambda^\star}{t}F_j^{(1)}+
\frac{t-\lambda^\star}{t}F_j^{(2)},F_j^{(2)}\right),&t>\lambda^\star.
\end{cases}
\]
Let
\[
\Delta_j(t)=\cD_t(F^L_{j,t},F^R_{j,t}),
\]
\[
\Delta_{\mathrm{sum}}(t)=\frac1p\sum_{j=1}^p\Delta_j(t),
\qquad
\Delta_{\mathrm{max}}(t)=\max_{1\le j\le p}\Delta_j(t),
\]
and, for $\comp\in\{\mathrm{sum},\mathrm{max}\}$,
\[
Q_{\comp}(t)=\frac{\pi^2}{3}-\Delta_{\comp}(t).
\]
Each gap curve $\Delta_{\comp}(t)$ is uniquely maximized, equivalently each $Q_{\comp}(t)$ is uniquely minimized, at $t=\lambda^\star$ whenever at least one marginal distribution changes.

At the true change fraction define
\[
\Delta_j=\cD_{\lambda^\star}(F_j^{(1)},F_j^{(2)}),
\qquad
\overline\Delta_n=\frac1p\sum_{j=1}^p\Delta_j,
\qquad
\Delta_{\max,n}=\max_{1\le j\le p}\Delta_j.
\]
The first signal is appropriate for dense alternatives, while the second is appropriate for sparse alternatives.

\begin{assumption}[Marginal regularity under the alternative]\label{ass:regularity}
The true fraction satisfies $\lambda^\star\in\mathcal I_\eta$, the marginal distribution functions $F_j^{(1)}$ and $F_j^{(2)}$ are continuous for every $j$, and
\[
\frac{\log(pn)}{n}\to0.
\]
\end{assumption}

Put
\[
L_{n,p}=\log(16pn^4),
\qquad
\varepsilon_{n,p}=\sqrt{\frac{L_{n,p}}{m_n}}+\frac1n,
\qquad
 a_{n,p}=\varepsilon_{n,p}^{1/2}\log\!\left(\frac{e}{\varepsilon_{n,p}}\right),
\]
and let $a_{n,1}$ denote the same expression with $p=1$.
For a permutation $\pi\in\mathfrak S_n$, let $Z_j^\pi(k)$ denote
\eqref{eq:coordinate-stat} computed from $(\bX_{\pi(1)},\ldots,\bX_{\pi(n)})$.

\begin{proposition}[Uniform approximation from elementary conditions]\label{prop:uniform}
Under Assumption~\ref{ass:regularity}, there is a constant $C_\eta<\infty$, depending only on $\eta$, such that, for all sufficiently large $n$,
\[
\Pp\!\left[
\max_{k\in\cK_n}\max_{1\le j\le p}
\left|Z_j(k)-\left\{\frac{\pi^2}{3}-\Delta_j(k/n)\right\}\right|
>C_\eta a_{n,p}
\right]
\le \frac{1}{2n^3}.
\]
For a uniformly random permutation $\pi$, conditionally on every realized pooled sample having no ties within each coordinate,
\[
\Pp_\pi\!\left[
\max_{k\in\cK_n}\max_{1\le j\le p}
\left|Z_j^\pi(k)-\frac{\pi^2}{3}\right|
>C_\eta a_{n,p}
\,\middle|\,
\bX_1,\ldots,\bX_n
\right]
\le \frac{1}{4n^2}.
\]
\end{proposition}

Let $C_\eta$ be the constant in Proposition~\ref{prop:uniform} and define
\[
\delta_{\mu,n}
=\max_{k\in\cK_n}\left|\mu_{n,k}-\frac{\pi^2}{3}\right|,
\qquad
\varrho_{n,p}=C_\eta a_{n,p}+\delta_{\mu,n}.
\]
The deterministic quantity $\delta_{\mu,n}$ is obtained from $\mu_{n,k}$.
The bounded range $0\le Z_j^\pi(k)\le16$ and Proposition~\ref{prop:uniform}
give
\begin{equation}\label{eq:null-mean-uniform-bound}
 \delta_{\mu,n}
 \le C_\eta a_{n,1}
 +16\sup_k\mathbb P_\pi\{|Z_j^\pi(k)-\pi^2/3|>C_\eta a_{n,1}\}
 \le C_\eta a_{n,1}+4n^{-2}.
\end{equation}
Consequently
\[
\varrho_{n,p}\le C_\eta\{a_{n,p}+a_{n,1}\}+4n^{-2},
\]
after enlarging $C_\eta$ if necessary.

For the power statements use the fixed Gaussian SUM threshold and
the analytic MAX center and positive scale in
\eqref{eq:sum-rejection-rule} and \eqref{eq:max-rejection-rule}.
Define the standardized signal excesses
\[
 B_{\mathrm{sum},n}
 =\frac{\sqrt p\{\overline\Delta_n-\varrho_{n,p}\}}{s_{n,\tau^\star}},
\]
\begin{equation}\label{eq:sparse-power-ld}
 B_{\mathrm{max},n}
 =\frac{\{\Delta_{\max,n}-\varrho_{n,p}\}/s_{n,\tau^\star}
                         -\mathfrak b_{n,p}}{\mathfrak a_{n,p}} .
\end{equation}
These quantities describe rejection under the alternative; null size
is supplied separately by the null distribution theorems.

\begin{theorem}[Consistency of the dense and sparse tests]\label{thm:power}
Under Assumption~\ref{ass:regularity}, with \(s_{n,\tau^\star}>0\)
and \(p\ge2\) eventually, for each
\(\comp\in\{\mathrm{sum},\mathrm{max}\}\),
\[
 B_{\comp,n}\longrightarrow\infty
 \quad\Longrightarrow\quad
 \mathbb P_{H_1}(\phi_{\comp,\alpha}=1)\longrightarrow1 .
\]
\end{theorem}

\begin{corollary}[Consistency of the Cauchy combination]\label{cor:cauchy-power}
Let $P_{\mathrm{sum},n,p}$ and $P_{\mathrm{max},n,p}$ be defined in \cref{eq:component-analytic-pvalues}. If both signal conditions in Theorem~\ref{thm:power} hold, then
\[
\Pcauchy(\omega)\xrightarrow{\Pp}0.
\]
More generally, the same conclusion holds if one component $p$-value converges to zero and the Cauchy transform of the other component is $O_{\Pp}(1)$.
\end{corollary}

By \eqref{eq:main-null-moment-orders}, $s_{n,k}\sim\sqrt{2\log n}/n$ uniformly over the trimmed scan, so the leading standardized dense and sparse signals are
\[
\frac{n\sqrt p\,\overline\Delta_n}{\sqrt{2\log n}}
\qquad\text{and}\qquad
\frac{n\Delta_{\max,n}}{\sqrt{2\log n}}.
\]
The sparse signal is compared with the original-rank center and scale in \cref{eq:sparse-power-ld}. Theorem~\ref{thm:max-full-analytic} specifies these quantities throughout the polynomial dimension range, with the boundary leading center and all critical transitions included. If $p_{\mathrm{chg},n}$ coordinates each contribute a gap of at least $\delta_{\mathrm{chg},n}$, then $\overline\Delta_n\ge(p_{\mathrm{chg},n}/p)\delta_{\mathrm{chg},n}$, which displays the gain of the dense statistic when the change is spread over many coordinates.

The Cauchy combination test exploits the complementary power of the
SUM and MAX procedures. Under dense alternatives, SUM accumulates
moderate changes across many coordinates, whereas under sparse
alternatives, MAX retains strong signals that may be diluted by
aggregation. Since $\cot(\pi p)\sim(\pi p)^{-1}$ as $p\downarrow0$,
a small component $p$-value produces a large positive contribution
to the combined statistic. In particular, the combined test is
consistent if both component $p$-values converge to zero, or if
one converges to zero while the Cauchy transform of the other
remains bounded in probability. The latter condition prevents
an increasingly negative contribution from offsetting the evidence
provided by the powerful component. Under the corresponding signal
conditions, the combined procedure therefore accommodates both
dense and sparse changes without requiring prior knowledge of
the number of affected coordinates.

\subsection{Single change-point estimation}\label{sec:localization}

The local scores entering the two tests are $D_{\mathrm{sum},n,p}(k)$ and $D_{\mathrm{max},n,p}(k)$ in \cref{eq:global-stats}; for $\comp\in\{\mathrm{sum},\mathrm{max}\}$, write $D_{\comp,n,p}(k)$ for the corresponding score. We estimate the change point by the split attaining the corresponding test statistic:
\begin{equation*}
\hattauSum
=\operatorname*{arg\,max}_{k\in\cK_n}D_{\mathrm{sum},n,p}(k),
\qquad
\hattauMax
=\operatorname*{arg\,max}_{k\in\cK_n}D_{\mathrm{max},n,p}(k).
\end{equation*}
The smallest maximizer is selected in the event of a tie. Thus $\hattauSum$ is the split producing $\Ssum$, while $\hattauMax$ is the split producing $\Smax$. The sparse null center $\mathfrak b_{n,p}$ and positive scale $\mathfrak a_{n,p}$ do not depend on the candidate split $k$. Centering and scaling $\Smax$ for null calibration preserve the maximizing split.

Put
\[
w_{\mathrm{sum},p}=\sqrt p,
\qquad
w_{\mathrm{max},p}=1,
\qquad
s_{n,\min}=\min_{k\in\cK_n}s_{n,k},
\qquad
s_{n,\max}=\max_{k\in\cK_n}s_{n,k}.
\]
The population counterpart of the local standardized score is
\begin{equation}\label{eq:population-standardized-location-score}
\Psi_{\comp,n,p}(t)
=w_{\comp,p}\frac{\mu_{n,k_n(t)}-Q_{\comp}(t)}{s_{n,k_n(t)}}
=w_{\comp,p}\frac{\mu_{n,k_n(t)}-\pi^2/3+\Delta_{\comp}(t)}{s_{n,k_n(t)}},
\qquad t\in\mathcal I_\eta.
\end{equation}
For every fixed $\epsilon>0$, define its separation at the true change point by
\begin{equation*}
g_{\comp,n,p}(\epsilon)
=\Psi_{\comp,n,p}(\lambda^\star)
-\max_{\substack{k\in\cK_n:\ |k/n-\lambda^\star|\ge\epsilon}}
\Psi_{\comp,n,p}(k/n).
\end{equation*}

Define
\[
\widetilde g_{\comp,n}(\epsilon)
=\Delta_{\comp}(\lambda^\star)
-\max_{\substack{k\in\cK_n:\ |k/n-\lambda^\star|\ge\epsilon}}
\Delta_{\comp}(k/n),
\]
\[
\kappa_n=\frac{s_{n,\max}}{s_{n,\min}}.
\]

\begin{proposition}[Studentized population separation]\label{prop:studentized-separation}
If $s_{n,\min}>0$, then, for $\comp\in\{\mathrm{sum},\mathrm{max}\}$ and every $\epsilon>0$,
\[
\frac{s_{n,\min}}{w_{\comp,p}}g_{\comp,n,p}(\epsilon)
\ge
\frac{\widetilde g_{\comp,n}(\epsilon)}{\kappa_n}
-\Delta_{\comp}(\lambda^\star)(\kappa_n-1)-2\delta_{\mu,n}.
\]
For fixed $\eta\in(0,1/2)$, \eqref{eq:main-null-moment-orders} gives
\[
\kappa_n=1+O\{(\log n)^{-1}\}.
\]
\end{proposition}

For a quantitative location bound, suppose that
\(\beta_\comp>0\) and \(\underline c_{\comp,n,p}>0\)
satisfy
\begin{equation}\label{eq:local-separation-rate}
 \Psi_{\comp,n,p}(\lambda^\star)-\Psi_{\comp,n,p}(k/n)
 \ge\underline c_{\comp,n,p}
       \min\{|k/n-\lambda^\star|^{\beta_\comp},1\},
 \qquad k\in\cK_n.
\end{equation}
With the uniform-approximation constant \(C_\eta\), define
\[
 r_{\comp,n,p}
 =\left\{\frac{4C_\eta w_{\comp,p}a_{n,p}}
 {s_{n,\min}\underline c_{\comp,n,p}}\right\}^{1/\beta_\comp}.
\]

\begin{theorem}[Consistency and localization rate]\label{thm:location}
Under Assumption~\ref{ass:regularity}, let \(s_{n,\min}>0\).
For either component, the conditions
\begin{equation}\label{eq:separation-location}
 \frac{s_{n,\min}g_{\comp,n,p}(\epsilon)}
             {w_{\comp,p}a_{n,p}}\longrightarrow\infty
 \quad(\forall\epsilon>0)
\end{equation}
imply \(\widehat\tau_\comp/n\xrightarrow{\mathbb P}\lambda^\star\).
Under \eqref{eq:local-separation-rate}, if \(r_{\comp,n,p}\to0\),
then, for all sufficiently large \(n\),
\[
 \mathbb P\{|\widehat\tau_\comp/n-\lambda^\star|
                      >r_{\comp,n,p}\}\le(2n^3)^{-1}.
\]
\end{theorem}

The separation in \cref{eq:separation-location} is stated for the same studentized population score as the estimator and therefore accounts for the finite-sample variation of both $\mu_{n,k}$ and $s_{n,k}$. Proposition~\ref{prop:studentized-separation} gives a directly checkable sufficient condition in terms of the Jensen-gap separation and shows that the deterministic scale variation is $O\{(\log n)^{-1}\}$, while the null-centering variation is measured exactly by $\delta_{\mu,n}$.

\section{Multiple change-point estimation by componentwise and adaptive WBS}\label{sec:wbs}

For an unknown number of changes, let
\[
0=\tau_0<\tau_1<\cdots<\tau_{K_n}<\tau_{K_n+1}=n,
\]
where $K_n\ge1$ is unknown, and suppose that
\begin{equation}\label{eq:multiple-change-model}
\bX_i\sim F^{(q)},
\qquad
\tau_{q-1}<i\le\tau_q,
\qquad q=1,\ldots,K_n+1.
\end{equation}
Write $F_j^{(q)}$ for the $j$th marginal distribution in segment $q$, define
\[
\mathcal T_n=\{\tau_1,\ldots,\tau_{K_n}\},
\]
and let
\begin{equation*}
\mathfrak d_n=\min_{0\le q\le K_n}(\tau_{q+1}-\tau_q)
\end{equation*}
be the minimum spacing. At every $\tau_q$, at least one marginal distribution is assumed to change.

\subsection{Local interval statistics}

For an interval $I=(s,e]$ of length $L=e-s$, define the trimmed local split set
\[
\mathcal K(I)
=\{s+\lceil\eta L\rceil,\ldots,e-\lceil\eta L\rceil\}.
\]
Using only $\bX_{s+1},\ldots,\bX_e$, let $Z_{j,I}(k)$ be the local version of \cref{eq:coordinate-stat}. The exact null constants for this interval are $\mu_{L,k-s}$ and $s_{L,k-s}$. Put
\[
A_I(k)=\frac1p\sum_{j=1}^p Z_{j,I}(k),
\]
\begin{equation}\label{eq:local-wbs-scores}
D_{\mathrm{sum},I}(k)
=\frac{\sqrt p\{\mu_{L,k-s}-A_I(k)\}}{s_{L,k-s}},
\qquad
D_{\mathrm{max},I}(k)
=\max_{1\le j\le p}
\frac{\mu_{L,k-s}-Z_{j,I}(k)}{s_{L,k-s}}.
\end{equation}
The corresponding interval statistics are
\[
S_{\mathrm{sum}}(I)=\max_{k\in\mathcal K(I)}D_{\mathrm{sum},I}(k),
\qquad
\mathcal M_j(I)=\max_{k\in\mathcal K(I)}
\frac{\mu_{L,k-s}-Z_{j,I}(k)}{s_{L,k-s}},
\qquad
S_{\mathrm{max}}(I)=\max_{1\le j\le p}\mathcal M_j(I).
\]
For each interval length $L$, use the deterministic original-rank
constants $\mathfrak a_{L,p},\mathfrak b_{L,p}$ defined by
\eqref{eq:full-max-branch-selection}, together with the exact
null mean $\mu_{L,k-s}$ and standard deviation $s_{L,k-s}$.
The analytic interval values are
\begin{equation}\label{eq:local-component-pvalues}
P_{\mathrm{sum}}(I)=1-F_\eta\{S_{\mathrm{sum}}(I)\},
\qquad
P_{\mathrm{max}}(I)
=1-F_{\mathrm G}\!\left\{
\frac{S_{\mathrm{max}}(I)-\mathfrak b_{L,p}}{\mathfrak a_{L,p}}
\right\}.
\end{equation}
For $\comp\in\{\mathrm{sum},\mathrm{max}\}$, the local component location is
\begin{equation}\label{eq:wbs-component-location}
\widehat k_{\comp}(I)
=\operatorname*{arg\,max}_{k\in\mathcal K(I)}D_{\comp,I}(k),
\end{equation}
where the smallest maximizer is selected in the event of a tie.

Following \citet{Fryzlewicz2014}, independently generate $M_n$ random intervals
\[
\mathcal R_n=\{I_m=(s_m,e_m]:1\le m\le M_n\}
\]
uniformly from
\[
\mathfrak I_{n,\ell_n}
=\{(s,e]:s,e\in\{0,\ldots,n\},\ s<e,\ e-s\ge\ell_n\},
\]
where $\ell_n$ is the minimum interval length. The same interval family can be used for all three WBS procedures below, so their differences are entirely due to the interval evidence and local location rule.

\subsection{SUM-WBS and MAX-WBS}

For a fixed component $\comp\in\{\mathrm{sum},\mathrm{max}\}$, componentwise WBS compares intervals through $P_{\comp}(I)$ and splits the selected interval at $\widehat k_{\comp}(I)$. Taking $\comp=\mathrm{sum}$ gives SUM-WBS, which accumulates dense marginal changes; taking $\comp=\mathrm{max}$ gives MAX-WBS, which targets sparse marginal changes. Ties between intervals are resolved first by shorter length and then by smaller left endpoint.

\begin{algorithm}[!htbp]
\caption{Componentwise wild binary segmentation (SUM-WBS or MAX-WBS)}\label{alg:component-wbs}
\begin{algorithmic}[1]
\Require Data $\bX_1,\ldots,\bX_n$; random intervals $\mathcal R_n$; component $\comp\in\{\mathrm{sum},\mathrm{max}\}$; trimming constant $\eta$; stopping level $q_{\comp,n}$.
\State For every $I\in\mathcal R_n$, compute $P_{\comp}(I)$ from \cref{eq:local-component-pvalues} and $\widehat k_{\comp}(I)$ from \cref{eq:wbs-component-location}.
\State Initialize $\widehat{\mathcal T}_{\comp}=\varnothing$.
\Procedure{ComponentWBS}{$a,b;\comp$}
  \State $\mathcal R_n(a,b)\gets\{(s,e]\in\mathcal R_n:a\le s<e\le b\}$.
  \If{$\mathcal R_n(a,b)=\varnothing$}
    \State \Return
  \EndIf
  \State $I^\star\gets\operatorname*{arg\,min}_{I\in\mathcal R_n(a,b)}P_{\comp}(I)$.
  \If{$P_{\comp}(I^\star)>q_{\comp,n}$}
    \State \Return
  \EndIf
  \State $\widehat\tau\gets\widehat k_{\comp}(I^\star)$ and $\widehat{\mathcal T}_{\comp}\gets\widehat{\mathcal T}_{\comp}\cup\{\widehat\tau\}$.
  \State \Call{ComponentWBS}{$a,\widehat\tau;\comp$} and \Call{ComponentWBS}{$\widehat\tau,b;\comp$}.
\EndProcedure
\State \Call{ComponentWBS}{$0,n;\comp$}; sort $\widehat{\mathcal T}_{\comp}$ and return it.
\end{algorithmic}
\end{algorithm}

For a nominal stopping parameter $\alpha$, one may take
\[
q_{\mathrm{sum},n}=q_{\mathrm{max},n}=\frac{\alpha}{M_n}.
\]
The theorems below use the analytic interval values and
$q_{\comp,n}=\alpha_n/M_n$ with $\alpha_n\to0$.
The simulations use whole-vector permutation thresholds, whose
recursive calibration is separate from this analytic bound.

\subsection{Cauchy-adaptive WBS}

The adaptive interval evidence combines the two component $p$-values.
Let $q_{\mathrm C,n}\downarrow0$ be the Cauchy-WBS stopping level.
To bound the negative contribution of a large component value, define
\[
\widetilde P_{\comp}(I)=\min\{P_{\comp}(I),1-q_{\mathrm C,n}\},
\qquad \comp\in\{\mathrm{sum},\mathrm{max}\},
\]
With equal weights, define
\begin{equation*}
T_{\mathrm C}(I)
=\frac12\tan\!\left[\pi\left\{\frac12-\widetilde P_{\mathrm{sum}}(I)\right\}\right]
+\frac12\tan\!\left[\pi\left\{\frac12-\widetilde P_{\mathrm{max}}(I)\right\}\right],
\end{equation*}
\begin{equation}\label{eq:local-cauchy-wbs-pvalue}
P_{\mathrm C}(I)
=\frac12-\frac1\pi\arctan\{T_{\mathrm C}(I)\}.
\end{equation}
\begin{lemma}[Dominance of one small component in the stabilized Cauchy value]\label{lem:wbs-cauchy-dominance}
For every $0<q_{\mathrm C,n}\le1/4$, if
\[
\min\{P_{\mathrm{sum}}(I),P_{\mathrm{max}}(I)\}\le \frac{q_{\mathrm C,n}}{4},
\]
then the stabilized Cauchy value satisfies
\[
P_{\mathrm C}(I)\le q_{\mathrm C,n}.
\]
Conversely,
\[
P_{\mathrm C}(I)\le q_{\mathrm C,n}
\quad\Longrightarrow\quad
\min\{P_{\mathrm{sum}}(I),P_{\mathrm{max}}(I)\}\le q_{\mathrm C,n}.
\]
\end{lemma}

Lemma~\ref{lem:wbs-cauchy-dominance} relates the component rejection events to rejection by the stabilized Cauchy combination.

The component used to locate the change within $I$ is selected by
\begin{equation*}
\comp_{\mathrm C}(I)=
\begin{cases}
\mathrm{sum},&P_{\mathrm{sum}}(I)\le P_{\mathrm{max}}(I),\\
\mathrm{max},&P_{\mathrm{max}}(I)<P_{\mathrm{sum}}(I),
\end{cases}
\end{equation*}
and
\begin{equation}\label{eq:wbs-local-location}
\widehat k_{\mathrm C}(I)
=\widehat k_{\comp_{\mathrm C}(I)}(I).
\end{equation}
The Cauchy value detects an interval-level change; the component with the smaller $p$-value supplies its location.

For a current segment $(a,b]$, Cauchy-adaptive WBS retains the intervals in $\mathcal R_n$ that are contained in $(a,b]$, chooses the one with the smallest $P_{\mathrm C}(I)$, and splits at $\widehat k_{\mathrm C}(I)$ whenever that minimum does not exceed $q_{\mathrm C,n}$. Ties are resolved first by the shorter interval and then by the smaller left endpoint.

\begin{algorithm}[!htbp]
\caption{Cauchy-adaptive wild binary segmentation}\label{alg:cauchy-wbs}
\begin{algorithmic}[1]
\Require Data $\bX_1,\ldots,\bX_n$; random intervals $\mathcal R_n$; trimming constant $\eta$; stopping level $q_{\mathrm C,n}$.
\State For every $I\in\mathcal R_n$, compute \cref{eq:local-wbs-scores,eq:local-cauchy-wbs-pvalue,eq:wbs-local-location}.
\State Initialize $\widehat{\mathcal T}_{\mathrm C}=\varnothing$.
\Procedure{CWBS}{$a,b$}
  \State $\mathcal R_n(a,b)\gets\{(s,e]\in\mathcal R_n:a\le s<e\le b\}$.
  \If{$\mathcal R_n(a,b)=\varnothing$}
    \State \Return
  \EndIf
  \State $I^\star\gets\operatorname*{arg\,min}_{I\in\mathcal R_n(a,b)}P_{\mathrm C}(I)$.
  \If{$P_{\mathrm C}(I^\star)>q_{\mathrm C,n}$}
    \State \Return
  \EndIf
  \State $\widehat\tau\gets\widehat k_{\mathrm C}(I^\star)$ and $\widehat{\mathcal T}_{\mathrm C}\gets\widehat{\mathcal T}_{\mathrm C}\cup\{\widehat\tau\}$.
  \State \Call{CWBS}{$a,\widehat\tau$} and \Call{CWBS}{$\widehat\tau,b$}.
\EndProcedure
\State \Call{CWBS}{$0,n$}; sort $\widehat{\mathcal T}_{\mathrm C}$ and return it.
\end{algorithmic}
\end{algorithm}

For a nominal stopping parameter $\alpha$, use $q_{\mathrm C,n}=\alpha/(4M_n)$.
The theoretical bounds below use $q_{\mathrm C,n}=\alpha_n/(4M_n)$ with
$\alpha_n\to0$ and the deterministic dominance lemma.

\subsection{Common conditions and method-specific signals}

The three procedures need different signal assumptions. SUM accumulates
marginal changes across coordinates, MAX uses the largest marginal change,
and the Cauchy procedure may use different types of evidence at different
change points. We state separate consistency results while allowing the
set of changing coordinates to vary between adjacent regimes.

For \(I=(s,e]\), \(L=e-s\), and \(k\in\mathcal K(I)\), put
\(t_{I,k}=(k-s)/L\). Write \(F_{ij}=F_j^{(q)}\) on regime \(q\) and define
\[
 F^L_{j,I,k}=\frac1{k-s}\sum_{i=s+1}^kF_{ij},\qquad
 F^R_{j,I,k}=\frac1{e-k}\sum_{i=k+1}^eF_{ij},\qquad
 \Delta_{j,I}(k)=\mathcal D_{t_{I,k}}(F^L_{j,I,k},F^R_{j,I,k}).
\]
The adjacent marginal signals and their supports are
\[
 \begin{split}
 d_{qj,n}&=\inf_{2/5\le t\le3/5}
       \mathcal D_t(F_j^{(q)},F_j^{(q+1)}),\\
 \mathcal S_{q,n}&=\{j:F_j^{(q)}\ne F_j^{(q+1)}\},\qquad
 s_{q,n}=|\mathcal S_{q,n}|,\\
 \Delta_{\Sigma,q,n}&=\sum_{j=1}^p d_{qj,n},\qquad
 \Delta_{\infty,q,n}=\max_{j\le p}d_{qj,n}.
 \end{split}
\]
Thus \(\Delta_{\Sigma,q,n}\) is a total signal, without division by \(p\);
\(s_{q,n}\) counts all changed margins, including arbitrarily weak ones.

Fix constants \(0<a_-<a_+<2\), \(\beta_{\mathrm W}>0\),
\(d_*,c_\ell,c_*,C_*>0\), all independent of \(n\). Let
\[
 N_n^{\mathrm W}=\log(en),\qquad
 \overline B_n=(N_n^{\mathrm W})^{\beta_{\mathrm W}}.
\]
The constants \(c_*,C_*\) are density-ratio bounds.
They permit nonparametric margins with infinite moments.

\begin{assumption}[Common conditions for WBS consistency]\label{ass:wbs-explicit}
In \eqref{eq:multiple-change-model}, the rows are independent,
\(K_n\ge1\), and
\[
 \mathfrak d_n\ge d_*n,\qquad
 c_\ell n\le\ell_n\le2\mathfrak d_n/3,\qquad
 n^{a_-}\le p\le n^{a_+}.
\]
For continuous distribution functions \(H_j\), all regimes satisfy
\(c_*\le dF_j^{(q)}/dH_j\le C_*\), \(H_j\)-almost everywhere.
The density bounds hold uniformly over \(n,q,j\).
The \(M_n\) intervals are drawn independently and uniformly, with
replacement, from \(\mathfrak I_{n,\ell_n}\), independently of the data,
and form the fixed pool used throughout the recursion. Assume
\(M_n\to\infty\) and \(M_n/N_n^{\mathrm W}\to0\).
\end{assumption}

For SUM-WBS and Cauchy-WBS we additionally require a common Gaussian
copula with a controlled correlation row sum. Precisely, writing
\(F_{ij}=F_j^{(q)}\) for \(\tau_{q-1}<i\le\tau_q\), the condition is
\begin{equation}\label{eq:wbs-sum-copula}
 \begin{split}
 \mathbf Y_i&=\bigl(\Phi^{-1}\{F_{i1}(X_{i1})\},\ldots,
                    \Phi^{-1}\{F_{ip}(X_{ip})\}\bigr)^\top
                 \overset{\mathrm{iid}}{\sim}N_p(\bm0,\mathbf R_p),\\
 \mathbf R_p&=(\rho_{j\ell})\succeq0,\qquad
 \rho_{jj}=1,\qquad
 B_p=1+\max_{j\le p}\sum_{\ell\ne j}|\rho_{j\ell}|
          \le\overline B_n .
 \end{split}
\end{equation}
Here \(\Phi\) is the standard normal distribution function; the
matrix \(\mathbf R_p\) may depend on \(n\) but is common across regimes.
Singular correlation matrices are allowed. MAX-WBS permits arbitrary
coordinate copulas.

Choose a fixed trimming fraction \(0<\eta<\eta_0\), where
\(\eta_0>0\) depends only on the density-ratio and spacing constants.
Lemma~\ref{lem:wbs-relative-population} constructs such an \(\eta_0\)
by comparing the scores at trimming endpoints with adjacent marginal
signals. This choice is uniform over the signal strengths.
Define the stopping exponent and levels by
\begin{equation}\label{eq:wbs-weak-stopping}
 \Lambda_n=\overline B_n(N_n^{\mathrm W})^2,\qquad
 \alpha_n=e^{-\Lambda_n},\qquad
 q_{\mathrm{sum},n}=q_{\mathrm{max},n}=\frac{\alpha_n}{M_n},\qquad
 q_{\mathrm C,n}=\frac{\alpha_n}{4M_n}.
\end{equation}
These levels apply to the exact interval statistics and location rules
in Algorithms~\ref{alg:component-wbs} and \ref{alg:cauchy-wbs}.
The deterministic sequences \(\overline B_n\) and \(\Lambda_n\)
specify the thresholds; MAX-WBS permits arbitrary coordinate dependence.
Under the respective conditions in the theorems below,
Lemma~\ref{lem:wbs-weak-null-stopping} shows that the probability of a
spurious split on a homogeneous interval tends to zero. The proof
uses finite-sample rank bounds and deterministic reference functions
at the stated stopping levels.

For \(g\in\{\mathrm{sum},\mathrm{max},\mathrm C\}\), let \(\widehat K_g\)
be the reported number of changes and
\(\widehat\tau_{g,1}<\cdots<\widehat\tau_{g,\widehat K_g}\) their
ordered locations. For \(\epsilon>0\), define the success event
\[
 \mathcal A_{g,n}(\epsilon)=
 \left\{\widehat K_g=K_n,\ 
   \max_{1\le q\le K_n}|\widehat\tau_{g,q}-\tau_q|\le\epsilon n\right\},
\]
where the location comparison is evaluated only when
\(\widehat K_g=K_n\). The signal ratios used below are
\begin{equation}\label{eq:wbs-component-signal-ratios}
 \begin{split}
 R_{\Sigma,q,n}
 &=\frac{n\Delta_{\Sigma,q,n}}
 {\sqrt{pN_n^{\mathrm W}\Lambda_n}+s_{q,n}(N_n^{\mathrm W})^2},\\
 R_{\infty,q,n}
 &=\frac{n\Delta_{\infty,q,n}}{\Lambda_n}.
 \end{split}
\end{equation}
The first denominator contains both the fluctuation of a sum of
homogeneous coordinate scores and the remainder from the changing
coordinates. Under the stated marginal class, the total population gap alone does
not uniformly control the finite-rank bias. For MAX, the coordinatewise error is
already negligible when its signal dominates \(\Lambda_n\).

\begin{theorem}[Consistency of SUM-WBS]\label{thm:wbs-sum-consistency}
Suppose Assumption~\ref{ass:wbs-explicit} and
\eqref{eq:wbs-sum-copula} hold. Run Algorithm~\ref{alg:component-wbs}
with \(\comp=\mathrm{sum}\), a fixed \(0<\eta<\eta_0\), and stopping
level \(q_{\mathrm{sum},n}\) from \eqref{eq:wbs-weak-stopping}. If
\[
 \frac1{M_n^4}\min_{q\le K_n}R_{\Sigma,q,n}\longrightarrow\infty,
\]
then \(\mathbb P\{\mathcal A_{\mathrm{sum},n}(\epsilon)\}\to1\)
for every fixed \(\epsilon>0\).
\end{theorem}

\begin{theorem}[Consistency of MAX-WBS]\label{thm:wbs-max-consistency}
Suppose Assumption~\ref{ass:wbs-explicit} holds. Run
Algorithm~\ref{alg:component-wbs} with \(\comp=\mathrm{max}\), a fixed
\(0<\eta<\eta_0\), and stopping level \(q_{\mathrm{max},n}\) from
\eqref{eq:wbs-weak-stopping}. If
\[
 \frac1{M_n^4}\min_{q\le K_n}R_{\infty,q,n}\longrightarrow\infty,
\]
then \(\mathbb P\{\mathcal A_{\mathrm{max},n}(\epsilon)\}\to1\)
for every fixed \(\epsilon>0\).
\end{theorem}

For Cauchy-WBS, define the signal ratio
\begin{equation}\label{eq:wbs-cauchy-signal-ratio}
 R_{\mathrm C,q,n}=
 \max\left\{
 R_{\Sigma,q,n},\
 \frac{n\Delta_{\infty,q,n}}
 {\Lambda_n+s_{q,n}^2(N_n^{\mathrm W})^3/p}
 \right\}.
\end{equation}
The support term in the sparse branch controls the SUM remainder when
the adaptive procedure is driven by sparse evidence but selects SUM
for localization. Thus the adaptive sufficient condition combines
a dense branch with a sparse branch that also controls the
approximation error of the selected component.

\begin{theorem}[Consistency of Cauchy-adaptive WBS]\label{thm:wbs-cauchy-consistency}
Suppose Assumption~\ref{ass:wbs-explicit} and
\eqref{eq:wbs-sum-copula} hold. Run Algorithm~\ref{alg:cauchy-wbs}
with a fixed \(0<\eta<\eta_0\) and stopping level
\(q_{\mathrm C,n}\) from \eqref{eq:wbs-weak-stopping}. If
\[
 \frac1{M_n^4}\min_{q\le K_n}R_{\mathrm C,q,n}\longrightarrow\infty,
\]
then \(\mathbb P\{\mathcal A_{\mathrm C,n}(\epsilon)\}\to1\)
for every fixed \(\epsilon>0\).
\end{theorem}

The active set and the branch satisfying the adaptive signal condition
may vary with \(q\). The factor \(M_n^4\) provides a sufficient sampling
margin for excluding random interval endpoints from the shrinking
localization neighborhoods. For example, \(M_n=\lceil\log N_n^{\mathrm W}\rceil\)
introduces only an iterated-logarithm margin. Appendix~\ref{app:wbs}
establishes relative location consistency and termination.

All fixed powers \(p=n^a\), \(0<a<2\), are included by choosing
\(a_-<a<a_+\). For SUM and Cauchy, the correlation row sum in
\eqref{eq:wbs-sum-copula} may diverge as any fixed positive power of
\(\log n\); MAX allows arbitrary coordinate copulas, which may differ
between regimes. To illustrate the distinct signal conditions, take
independent coordinates, \(M_n=\lceil\log N_n^{\mathrm W}\rceil\), and
\(p=n^a\). If all \(p\) margins have gaps of order
\((N_n^{\mathrm W})^{2+\beta_{\mathrm W}/2}/n\), the SUM condition
holds, whereas the MAX signal ratio tends to zero. If one margin has
gap of order \((N_n^{\mathrm W})^{\beta_{\mathrm W}+3}/n\), the MAX
condition and the sparse branch of the adaptive condition hold,
whereas the SUM signal ratio tends to zero. Different change points
can have these two forms in the same sequence. Such vanishing gaps
are realized, for example, by margins
\(F_j^{(q)}=H_j+\theta_{qj,n}(H_j^2-H_j)\) with a fixed bound
\(|\theta_{qj,n}|\le\theta_*<1\), for which adjacent gaps are comparable to
\((\theta_{q+1,j,n}-\theta_{qj,n})^2\).

\section{Simulation studies}\label{sec:simulation}

The experiments in this section assess whole-vector permutation calibration, with the orbit-calibrated adaptive value defined in Section~\ref{sec:statistics}. Appendix~\ref{app:analytic-size} separately evaluates the direct asymptotic calibration of SUM, MAX, and their Cauchy combination.

We compare the proposed sum, maximum, and Cauchy-adaptive procedures, denoted by \method{ZAS}, \method{ZAM}, and \method{ZAC}, respectively. The comparison methods are KDist \citep{ChakrabortyZhang2021}, the high-dimensional distance procedure HDD \citep{DrikvandiModarres2025}, E-Divisive \citep{MattesonJames2014}, and the graph scan gSeg \citep{ChenZhang2015}. Whole-vector permutation is used throughout this section so that contemporaneous dependence is preserved. Unless stated otherwise, candidate splits use trimming fraction $\eta=0.1$ and tests are conducted at level $0.05$.

\subsection{Testing}

We first examine finite-sample size under a Gaussian null. The observations are independent over time and satisfy $\bX_i\sim N_p(\bm0,\boldsymbol{\Sigma}_m)$, where $m\in\{\mathrm{AR},\mathrm{CS}\}$, $\rho=0.5$,
\[
(\boldsymbol{\Sigma}_{\mathrm{AR}})_{jk}=\rho^{|j-k|},
\qquad
\boldsymbol{\Sigma}_{\mathrm{CS}}=(1-\rho)\mathbf I_p+\rho\bm1_p\bm1_p^{\top}.
\]
We consider $n\in\{100,200\}$ and $p\in\{100,200,400\}$. Each configuration uses 1,000 Monte Carlo replications and $B=199$ permutations.

\begin{table}[!htbp]
  \centering
  \caption{Empirical size at nominal level $0.05$ under the Gaussian null. Each entry is based on 1,000 Monte Carlo replications and 199 permutations.}
  \label{tab:sim-size}
  \small
  \setlength{\tabcolsep}{4.5pt}
  \begin{tabular}{ccrrrrrrr}
    \toprule
    $n$ & $p$ & \method{ZAC} & \method{ZAM} & \method{ZAS} & KDist & HDD & E-Div. & gSeg \\
    \midrule
    \multicolumn{9}{l}{\textit{Panel A: AR(1), $\rho=0.5$}} \\
    100 & 100 & 0.042 & 0.050 & 0.047 & 0.041 & 0.036 & 0.066 & 0.044 \\
    100 & 200 & 0.046 & 0.048 & 0.045 & 0.050 & 0.041 & 0.048 & 0.040 \\
    100 & 400 & 0.041 & 0.046 & 0.053 & 0.043 & 0.042 & 0.052 & 0.052 \\
    200 & 100 & 0.046 & 0.049 & 0.045 & 0.051 & 0.056 & 0.062 & 0.043 \\
    200 & 200 & 0.058 & 0.047 & 0.058 & 0.051 & 0.048 & 0.053 & 0.042 \\
    200 & 400 & 0.047 & 0.046 & 0.054 & 0.049 & 0.041 & 0.055 & 0.040 \\
    \midrule
    \multicolumn{9}{l}{\textit{Panel B: compound symmetry, $\rho=0.5$}} \\
    100 & 100 & 0.054 & 0.049 & 0.054 & 0.057 & 0.054 & 0.050 & 0.056 \\
    100 & 200 & 0.061 & 0.063 & 0.055 & 0.052 & 0.048 & 0.053 & 0.039 \\
    100 & 400 & 0.051 & 0.053 & 0.056 & 0.049 & 0.045 & 0.059 & 0.049 \\
    200 & 100 & 0.048 & 0.040 & 0.050 & 0.045 & 0.042 & 0.038 & 0.056 \\
    200 & 200 & 0.052 & 0.060 & 0.052 & 0.051 & 0.040 & 0.054 & 0.053 \\
    200 & 400 & 0.047 & 0.052 & 0.046 & 0.051 & 0.038 & 0.055 & 0.057 \\
    \bottomrule
  \end{tabular}
\end{table}

Table~\ref{tab:sim-size} shows that all procedures remain close to the nominal level under both dependence structures. The proposed tests exhibit no systematic size deterioration as either the sample size or dimension increases, and the remaining fluctuations are consistent with Monte Carlo variability.

For power, let $n=p=200$ and place a single change at $\tau^\star=100$. Independent innovations $\bm\varepsilon_i=(\varepsilon_{i1},\ldots,\varepsilon_{ip})^\top$ are transformed according to
\begin{equation}\label{eq:sim-factor-model}
\bX_i=\mathbf L_m\bm\varepsilon_i,
\qquad \mathbf L_m\mathbf L_m^\top=\boldsymbol{\Sigma}_m,
\qquad m\in\{\mathrm{AR},\mathrm{CS}\}.
\end{equation}
Here $\mathbf L_m$ is the lower triangular Cholesky factor with positive diagonal. Before the change, every innovation is standard normal. After the change, the first $s$ innovations follow a standardized Student distribution and the remaining innovations remain normal:
\[
\varepsilon_{ij}\sim
\begin{cases}
N(0,1), & i\le\tau^\star,\ 1\le j\le p,\\
\sqrt{\{\varkappa_m(s)-2\}/\varkappa_m(s)}\,t_{\varkappa_m(s)}, & i>\tau^\star,\ 1\le j\le s,\\
N(0,1), & i>\tau^\star,\ s<j\le p.
\end{cases}
\]
The population mean and covariance are therefore unchanged, and the alternative acts only through marginal shape and tail thickness. The sparsity index $s$ counts changed independent-component margins rather than changed observed coordinates. The grid is
\[
s\in\{1,2,5,10,20,30,40,50,60,80,120,200\}.
\]
For AR(1),
\[
\varkappa_{\mathrm{AR}}(s)=\min\{30,\,2.0001+0.1\log_2s\}.
\]
For compound symmetry, define the transition weight
\[
 w_{\mathrm{CS}}(s)=
 \begin{cases}
 0,&s\le20,\\
 \log_2(s/20)/\log_2(50/20),&20<s<50,\\
 1,&s\ge50,
 \end{cases}
\]
and use
\[
 \varkappa_{\mathrm{CS}}(s)
 =\min\bigl\{30,\,2.0001
       +0.1\{1-w_{\mathrm{CS}}(s)\}\log_2s
       +0.05w_{\mathrm{CS}}(s)\log_2(p/s)\bigr\}.
\]
This retains the AR(1) rule through $s=20$ and reaches the dense-end rule at $s=50$. Two additional AR(1) experiments replace the Student change by skewness changes: the affected post-change innovations follow either $\operatorname{Exp}(1)-1$ or $\{\Gamma(4,1)-4\}/2$. Both distributions have mean zero and variance one. Every power configuration uses 1,000 replications and $B=199$ permutations.

\begin{figure}[!htbp]
  \centering
  \includegraphics[width=0.82\textwidth]{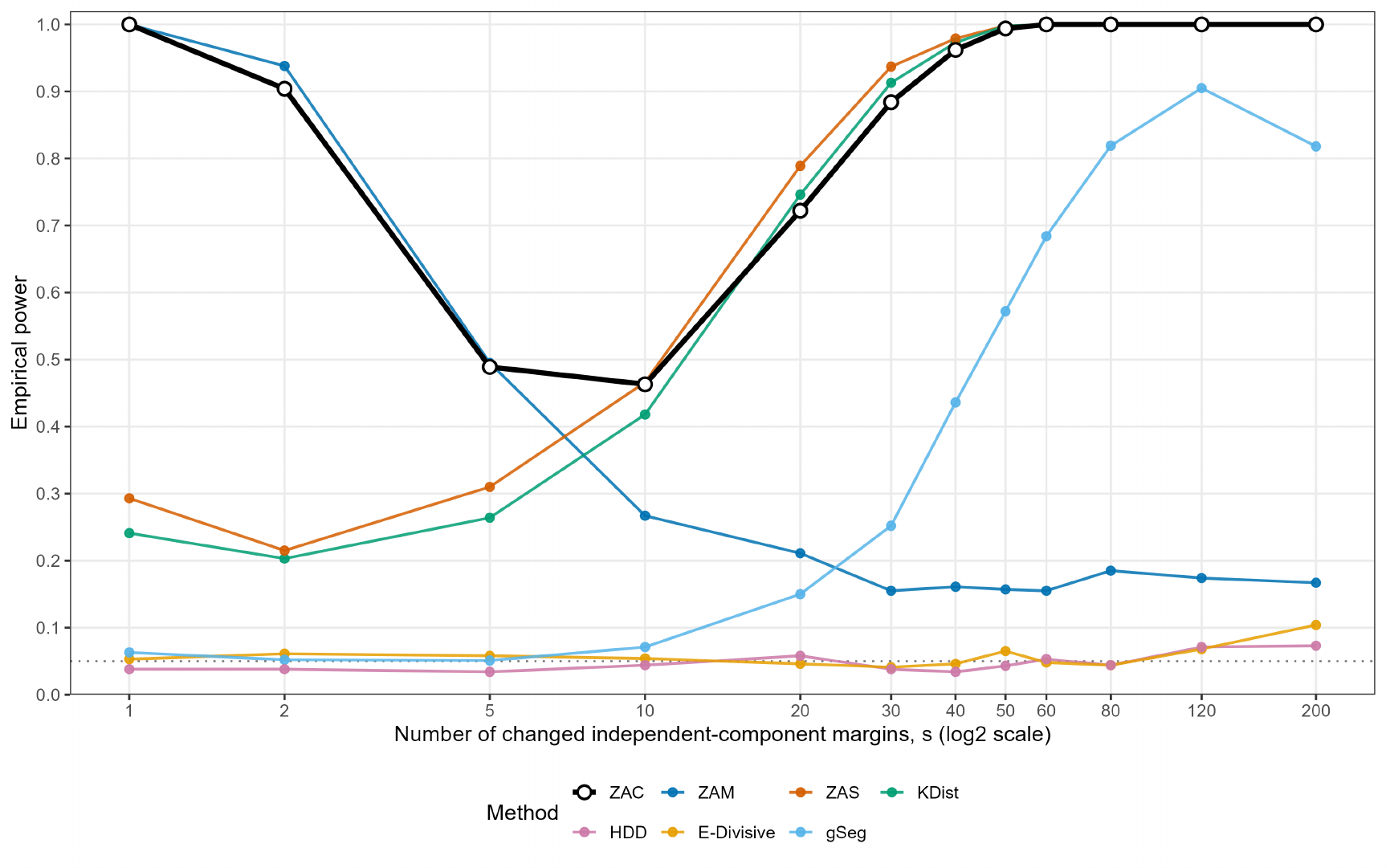}
  \vspace{0.6em}
  \includegraphics[width=0.82\textwidth]{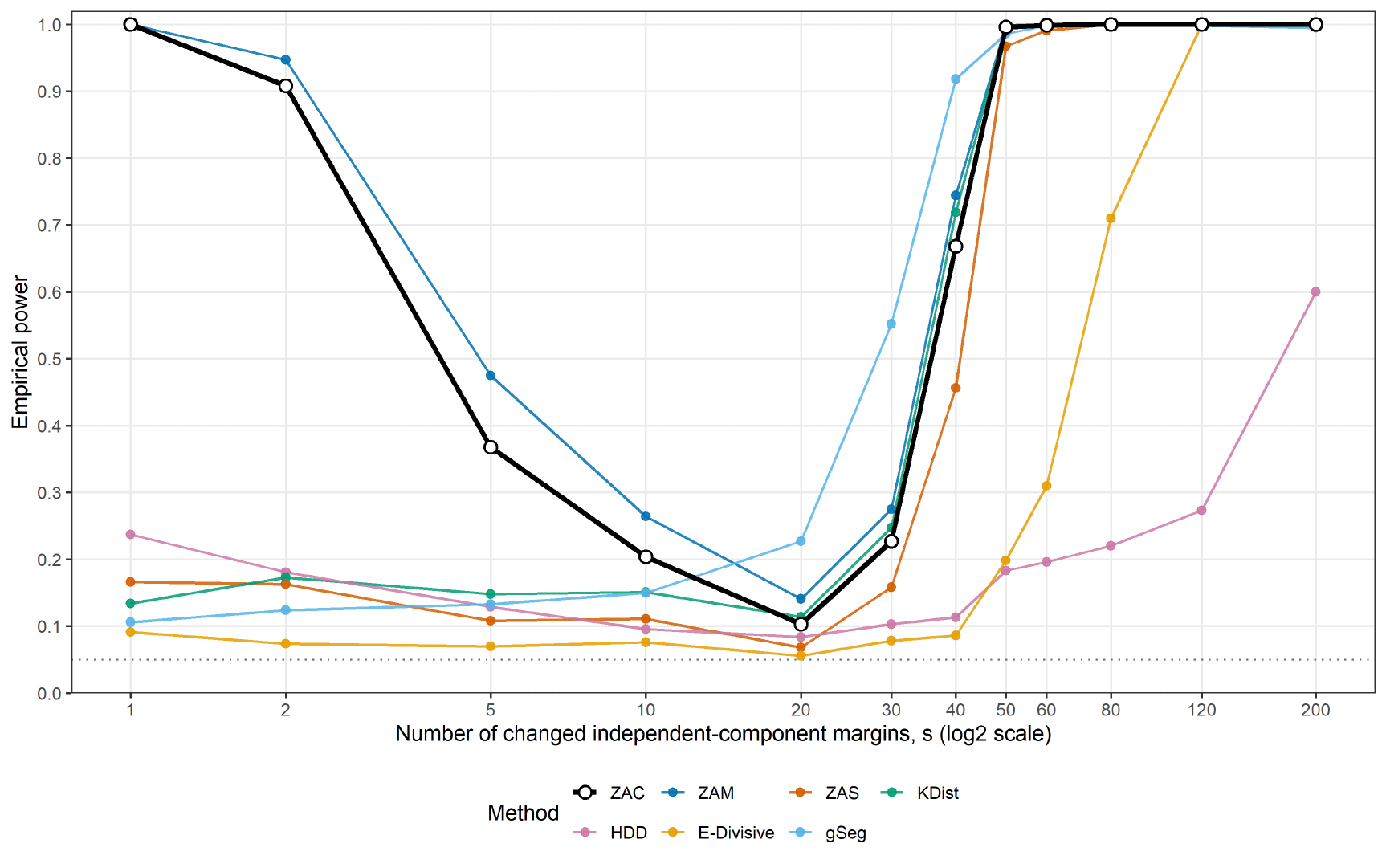}
  \caption{Empirical power for the standardized Student-tail alternatives. The upper panel uses AR(1) dependence and the lower panel uses compound symmetry. The horizontal axis is displayed on a base-2 logarithmic scale and the dotted line marks the nominal level.}
  \label{fig:sim-power-student}
\end{figure}

\begin{figure}[!htbp]
  \centering
  \includegraphics[width=0.72\textwidth]{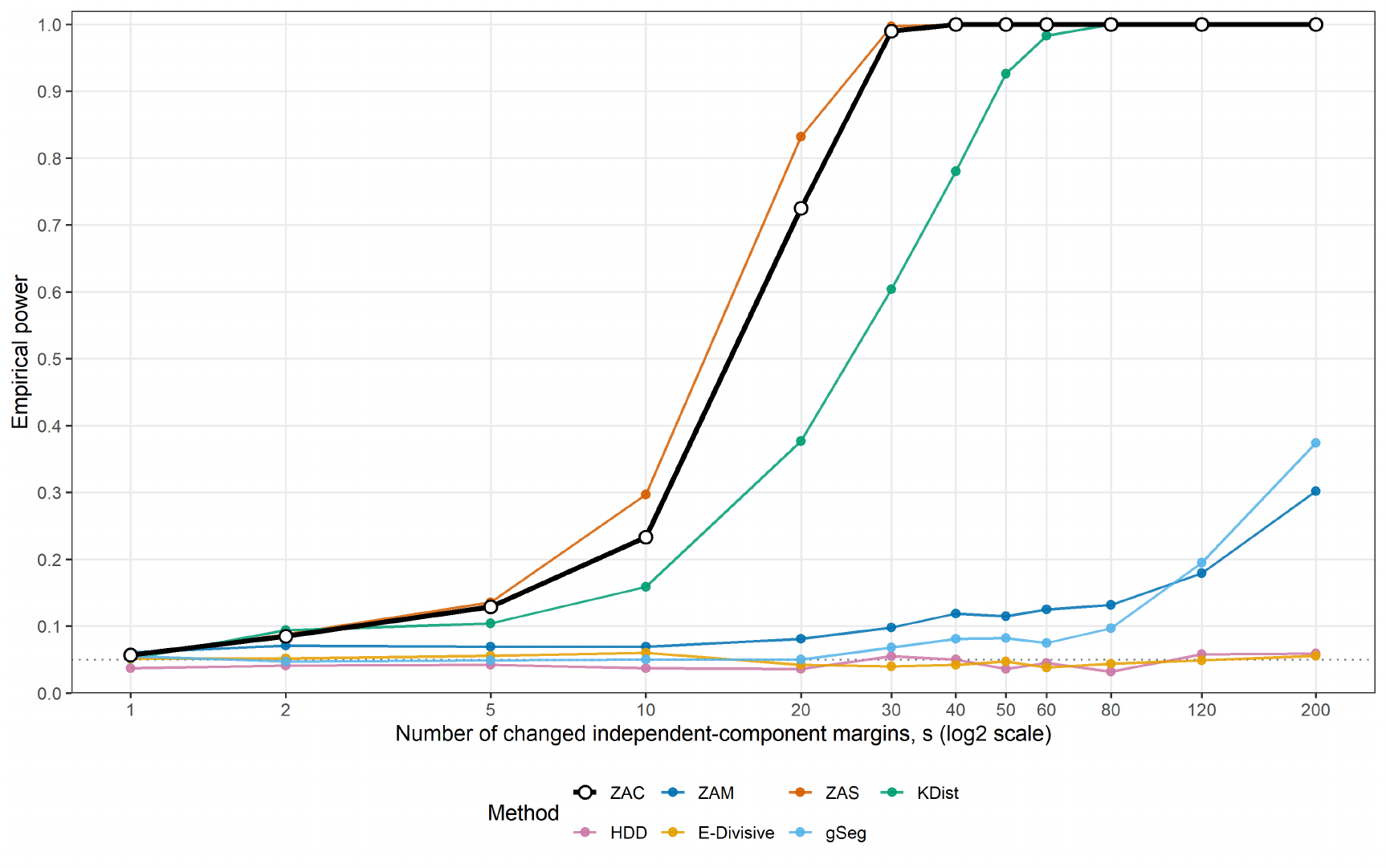}
  \vspace{0.6em}
  \includegraphics[width=0.72\textwidth]{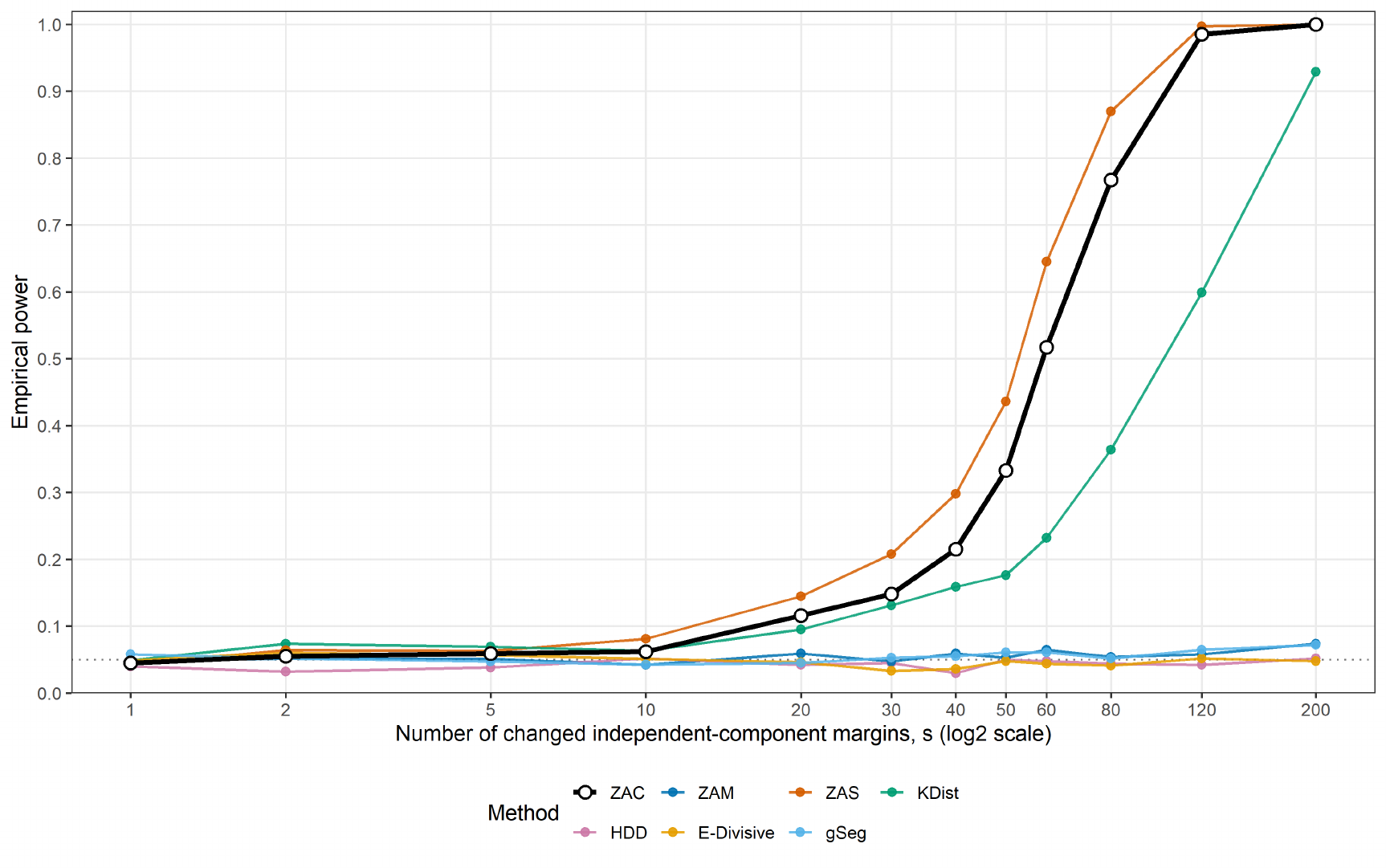}
  \caption{Empirical power under AR(1) skewness changes. The upper panel uses $N(0,1)\to\operatorname{Exp}(1)-1$, and the lower panel uses $N(0,1)\to\{\Gamma(4,1)-4\}/2$.}
  \label{fig:sim-power-skewness}
\end{figure}

Figures~\ref{fig:sim-power-student} and \ref{fig:sim-power-skewness} display a clear sparsity-dependent division of labor. Maximum aggregation is most effective when a change is concentrated in very few margins, whereas sum aggregation becomes preferable when many margins contribute moderate evidence. The Cauchy procedure follows the stronger component closely and therefore avoids committing to a fixed sparsity regime. This transition is stable under both dependence structures. The external distance and graph procedures become competitive for sufficiently dense alternatives, but they are generally less sensitive at the sparse endpoint.

The skewness experiments reinforce the same interpretation from a different distributional direction. Changes in only a few skewed margins are difficult for all methods, while the sum and adaptive procedures improve rapidly as the change becomes widespread. KDist is the strongest external benchmark in these settings, but the proposed dense aggregation is more responsive over much of the moderate-to-dense range. Overall, the testing results support the intended complementarity of the component statistics and the stability of their adaptive combination.

\subsection{Single change-point}

We next evaluate the location estimators under the same AR(1) and compound-symmetry Student-tail alternatives. Each procedure returns one maximizing split $\widehat\tau$, and accuracy is measured by $|\widehat\tau-\tau^\star|/n$. The \method{ZAS} and \method{ZAM} locations maximize their respective scan statistics; the \method{ZAC} location is inherited from the component with the smaller permutation $p$-value. Each configuration again uses 1,000 Monte Carlo replications and $B=199$ permutations.

\begin{figure}[!htbp]
  \centering
  \includegraphics[width=0.74\textwidth]{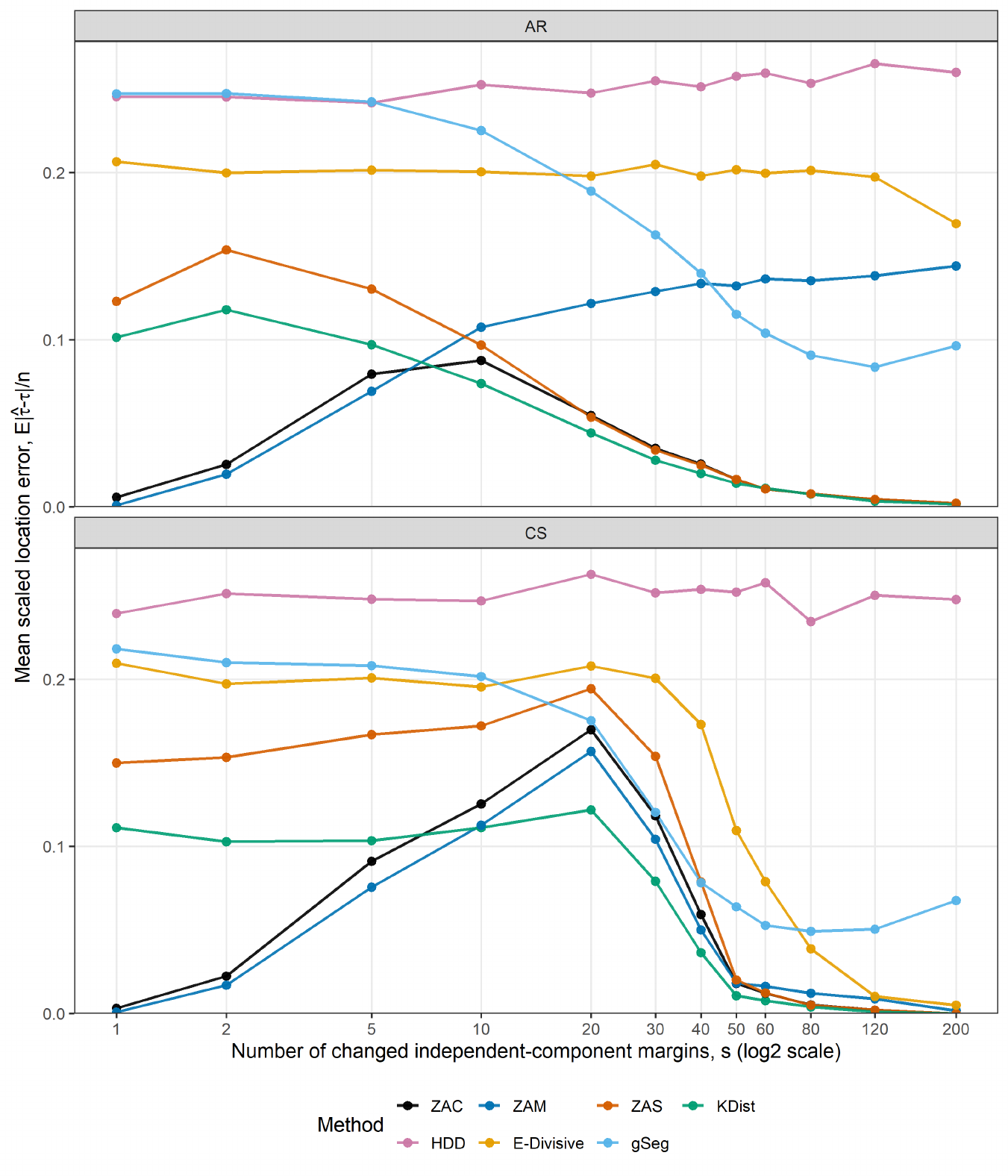}
  \caption{Mean scaled absolute single-change location error under the AR(1) and compound-symmetry Student-tail alternatives. Lower curves indicate more accurate localization.}
  \label{fig:sim-single-location}
\end{figure}

The localization results in Figure~\ref{fig:sim-single-location} closely mirror the testing results. The maximum-based estimator is particularly accurate for concentrated alternatives, while the sum-based estimator improves as the signal becomes dense. The adaptive estimator remains close to the better component across the sparsity grid. Under compound symmetry, moderate sparsity is the most difficult regime because strong equicorrelation and a relatively weak aggregate discrepancy flatten the population objective near the change. Once distributed evidence becomes pronounced, the sum, adaptive, and energy-distance estimators localize the change sharply. These findings confirm that matching the localization criterion to the detection statistic preserves sparse-to-dense adaptivity.

\subsection{Multiple change-points}

The multiple-change experiment uses the AR(1) version of \eqref{eq:sim-factor-model} with $n=p=200$ and $\rho=0.5$. The true change points are $\mathcal T_n=\{0.3n,0.7n\}$. The first $s$ innovations are normal before the first change, standardized Student between the two changes, and normal again after the second change; all remaining innovations are normal throughout. The Student degrees of freedom follow the AR(1) rule above, so the mean and covariance remain constant in all three regimes.

The multiple-change versions of \method{ZAS}, \method{ZAM}, and \method{ZAC} are implemented through SUM-WBS, MAX-WBS, and Cauchy-adaptive WBS. They use a common collection of 50 random intervals together with the full interval, minimum random-interval length 40, local trimming fraction $0.1$, and $B=99$ permutations. KDist uses its native WBS implementation, HDD and gSeg are embedded in recursive segmentation, and E-Divisive uses its native divisive recursion. Each value of $s$ is evaluated over 100 Monte Carlo replications.

Let $\widehat{\mathcal T}$ be the estimated set for the procedure being evaluated and let $L_{\max}$ denote the longest true segment. We report the probability of estimating exactly two changes, the adjusted Rand index for the induced segmentation, and the scaled Hausdorff loss
\begin{equation}\label{eq:sim-scaled-hausdorff}
 d_H(\mathcal T_n,\widehat{\mathcal T})=
 \frac{1}{L_{\max}}
 \max\left\{
   \max_{\tau\in\mathcal T_n}\min_{\widehat\tau\in\widehat{\mathcal T}}|\tau-\widehat\tau|,
   \max_{\widehat\tau\in\widehat{\mathcal T}}\min_{\tau\in\mathcal T_n}|\widehat\tau-\tau|
 \right\},
\end{equation}
For an empty estimated set, define $d_H(\mathcal T_n,\varnothing)=n/L_{\max}$. These criteria distinguish change-count errors, worst-case localization errors, and overall segmentation disagreement.

\begin{figure}[!htbp]
  \centering
  \includegraphics[width=0.76\textwidth]{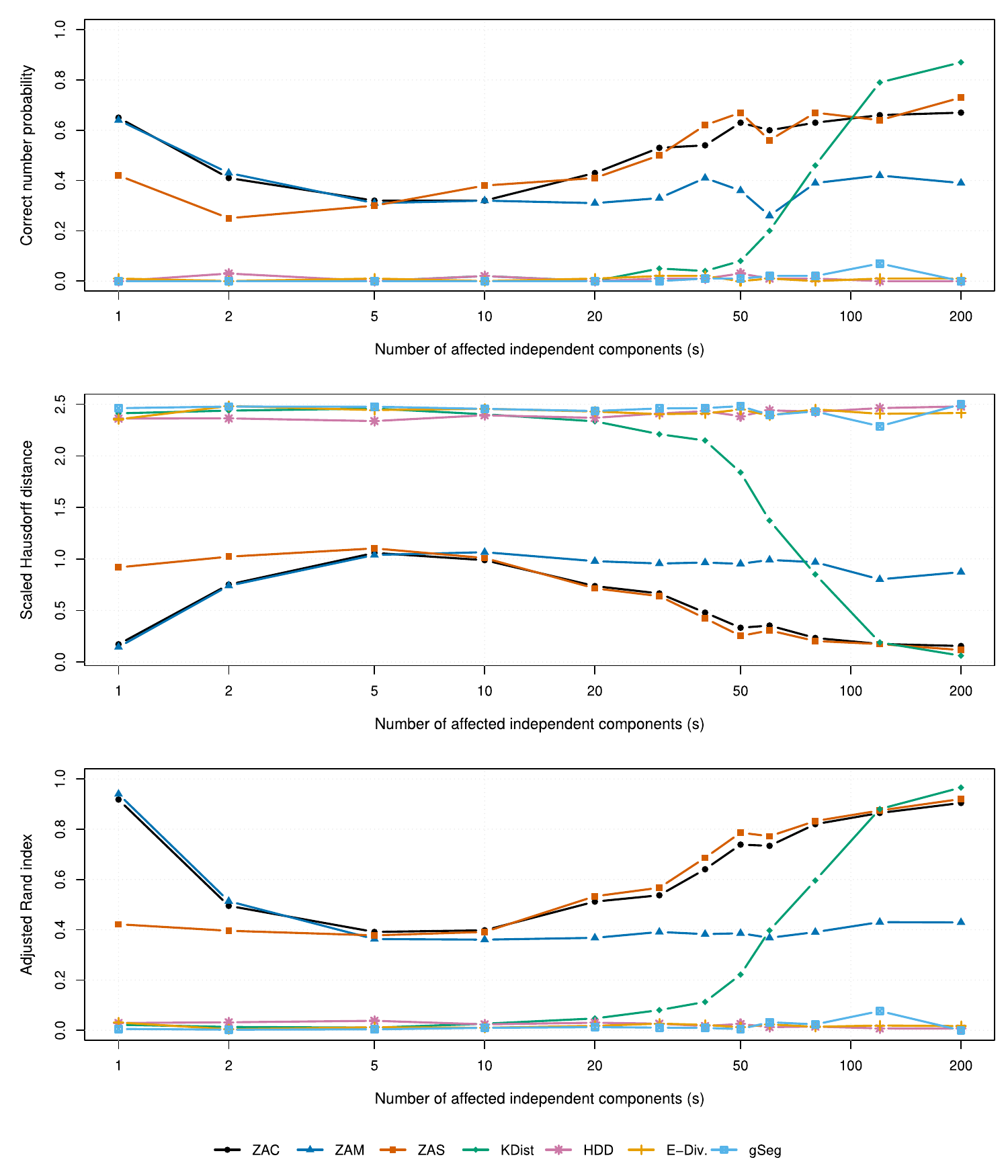}
  \caption{Multiple-change performance under the AR(1) Gaussian--Student--Gaussian alternative. The panels report exact recovery of two changes, mean scaled Hausdorff loss in \eqref{eq:sim-scaled-hausdorff}, and mean adjusted Rand index.}
  \label{fig:sim-multiple}
\end{figure}

Figure~\ref{fig:sim-multiple} shows that the proposed procedures are especially effective for sparse and intermediate marginal-shape changes. MAX-WBS and the adaptive procedure have the clearest advantage when only a few margins change, whereas SUM-WBS becomes more effective as evidence accumulates across coordinates. The middle of the sparsity grid is the most demanding region because the design increases the number of changed margins while weakening each individual tail change. KDist improves markedly at the fully dense endpoint and provides a strong global benchmark there. The three evaluation criteria reveal complementary features: exact-count recovery penalizes any additional split, while the Hausdorff and adjusted-Rand measures can still recognize accurate locations and segmentations when the estimated count is imperfect.

\section{Empirical analysis}\label{sec:application}

\subsection{BrainCloud data}

The first application uses the BrainCloud study of human prefrontal-cortex development and aging \citep{ColantuoniEtAl2011}, available from the NCBI Gene Expression Omnibus under accession GSE30272. We start from the normalized Series Matrix, regress each probe on an intercept and the two surrogate variables supplied in the sample annotation, and do not include age in this nuisance adjustment. Restricting attention to subjects aged at least 20 years gives $n=148$ adult samples. After removing probes with missing or zero-variance values, we retain the 2,000 probes with the largest residual variances, center and scale them robustly, and order the subjects by age. Thus the analysis remains high dimensional, with $p=2000>n$.

All methods are applied to the same age-ordered matrix with trimming fraction $0.1$, nominal level $0.05$, and $B=999$ whole-subject permutations. Let $\operatorname{age}_{(i)}$ be the age of the $i$th ordered subject. For an estimated split $\widehat\tau$, the reported age is the midpoint
\[
 \widehat a=\frac{\operatorname{age}_{(\widehat\tau)}
                    +\operatorname{age}_{(\widehat\tau+1)}}2.
\]

\begin{table}[!htbp]
  \centering
  \caption{BrainCloud global tests and single-change location estimates. All methods use the same $148\times2000$ adult expression matrix and 999 permutations.}
  \label{tab:braincloud-results}
  \small
  \begin{threeparttable}
  \setlength{\tabcolsep}{5pt}
  \begin{tabular}{lrlrrr}
    \toprule
    Method & $p$-value & Decision & $\widehat\tau$ & Bracketing ages & $\widehat a$ (years) \\
    \midrule
    \method{ZAC} & 0.001 & Reject & 56  & 40.847--41.000 & 40.923 \\
    \method{ZAM} & 0.001 & Reject & 56  & 40.847--41.000 & 40.923 \\
    \method{ZAS} & 0.006 & Reject & 56  & 40.847--41.000 & 40.923 \\
    KDist        & 0.002 & Reject & 56  & 40.847--41.000 & 40.923 \\
    HDD          & 0.728 & Do not reject & 116 & 57.482--57.633 & 57.558 \\
    E-Divisive   & 0.063 & Do not reject & 56  & 40.847--41.000 & 40.923\tnote{a} \\
    gSeg         & 0.001 & Reject & 31  & 29.986--30.033 & 30.010 \\
    \bottomrule
  \end{tabular}
  \begin{tablenotes}[flushleft]
    \footnotesize
    \item[a] E-Divisive does not report a location after nonrejection. The displayed value is the maximizer from a fixed-one-change scan and does not alter its global $p$-value or decision.
  \end{tablenotes}
  \end{threeparttable}
\end{table}

The most important feature is the agreement among the proposed procedures. Sparse, dense, and adaptive aggregation all reject homogeneity and select the same midlife boundary, so the empirical conclusion is not driven by a particular sparsity choice. KDist independently supports the same location, while the graph and HDD procedures emphasize different aspects of the high-dimensional geometry and return substantially different maximizers. The result should be interpreted as an age-ordered cross-sectional transition rather than a longitudinal within-person change or a causal threshold for brain aging. Under the common preprocessing and screening scheme, however, agreement across complementary proposed statistics provides strong evidence of a stable distributional transition in adult cortical expression profiles.

\subsection{Gas Sensor Array Drift data}

The second application uses the UCI Gas Sensor Array Drift at Different Concentrations data set \citep{Vergara2012}. The measurements come from a 16-sensor metal-oxide array, with eight features extracted per sensor, giving $p=128$ variables. The data are arranged in ten temporal batches spanning 36 months; the experimental platform and concentration-calibration problem are described by \citet{RodriguezLujanEtAl2014}. We retain the class-1 ethanol observations and adjust each feature for gas concentration through robust standardization, a natural-cubic-spline location regression in log concentration, and a corresponding scale regression. A final median/MAD standardization is applied featurewise.

To balance temporal information without artificial replication, at most 50 observations are sampled without replacement from each batch. The retained batch sizes are
\[
(50,50,50,50,28,50,50,30,50,50),
\]
which gives $n=458$. Within-batch observations are randomly reordered using a fixed seed because exact acquisition times are unavailable, while the order of the ten batches is preserved. The nine reference batch boundaries occur after indices
\[
50,\ 100,\ 150,\ 200,\ 228,\ 278,\ 328,\ 358,\ 408.
\]
They are used as temporal reference transitions rather than imposed stochastic change points.

The global tests use trimming fraction $0.1$, nominal level $0.05$, and $B=999$ whole-vector permutations. All methods except HDD reject temporal homogeneity. For multiple-change estimation, the proposed procedures use 100 WBS intervals and 199 local permutations; the competing procedures use their corresponding recursive implementations.

\begin{figure}[!htbp]
  \centering
  \includegraphics[width=\textwidth]{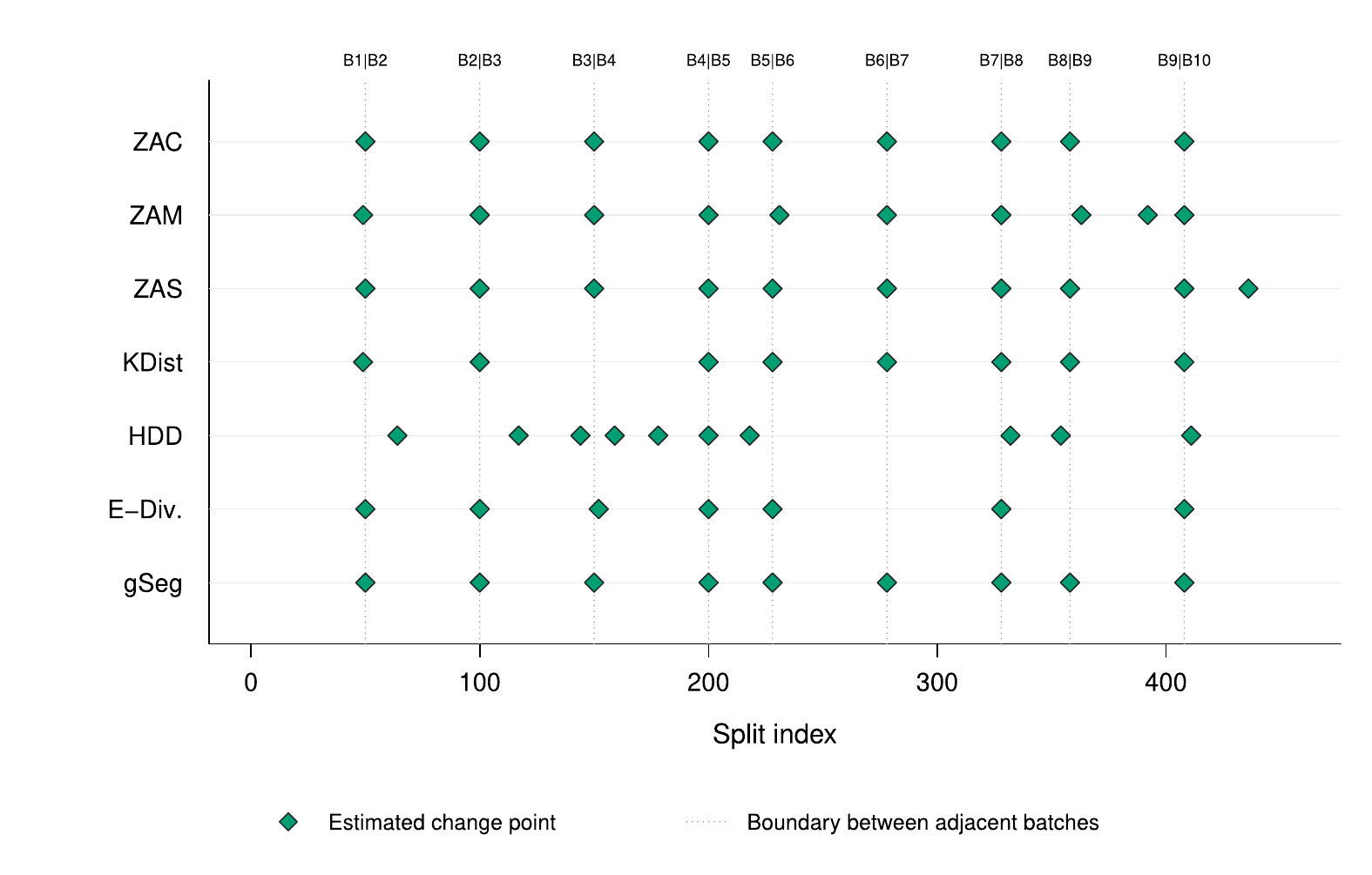}
  \caption{Multiple-change estimates for the concentration-adjusted Gas Sensor Array Drift data. Diamonds indicate estimated changes and dotted vertical lines indicate the nine boundaries between adjacent temporal batches.}
  \label{fig:gas-multiple}
\end{figure}

Figure~\ref{fig:gas-multiple} compares the estimated locations with the batch boundaries.

The adaptive Cauchy procedure recovers the complete sequence of batch transitions without introducing an additional within-batch split. The maximum and sum procedures identify the same core sequence but each adds one local split, indicating mild sensitivity to within-batch heterogeneity when either aggregation rule is used alone. The exact agreement of gSeg with the reference boundaries gives an independent geometric confirmation, whereas KDist and E-Divisive undersegment and HDD returns a less stable collection of locations. These results support pervasive long-term distributional drift after concentration adjustment and illustrate the stabilizing role of adaptive aggregation in multiple-change selection.

The interpretation remains descriptive. Batch labels provide only coarse temporal ordering, exact dates are unavailable within batches, and concentration support differs across batches. The location and scale adjustment reduces the dominant concentration effect but cannot eliminate all confounding between concentration and batch where the supports do not overlap. Accordingly, the estimated transitions should not be interpreted as causal estimates of sensor aging at precise acquisition times.

\FloatBarrier

\section{Conclusion}\label{sec:conclusion}

The standardized rank scans provide complementary tests for dense and sparse marginal changes. The original MAX scan has an analytic Gumbel law across the polynomial dimension range, with distinct subcritical, critical, and boundary normalizations. The SUM process limit and SUM--MAX asymptotic independence justify fixed-level Cauchy calibration. For multiple changes, Theorems~\ref{thm:wbs-sum-consistency}--\ref{thm:wbs-cauchy-consistency} give separate sufficient conditions for correct change counts and vanishing relative location error. They allow weak signals and changing active sets. MAX-WBS allows arbitrary coordinate copulas; SUM-WBS and Cauchy-WBS additionally use the explicit Gaussian-copula condition \eqref{eq:wbs-sum-copula}, with polylogarithmically growing correlation row sums. The SUM and adaptive conditions retain support-dependent remainders needed by the present proof; sharper aggregation bounds under additional control of the marginal signal shape remain open. The minimum segment fraction bounds the number of changes; their locations may vary subject to the spacing condition.

Several questions remain open. The proposed procedures identify a multivariate distributional change through changes in the marginal laws; a change confined to the copula may therefore be weakly detected or missed. Developing joint marginal-and-dependence scans is an important extension. The dependence conditions for the single-change null limits allow slowly increasing rowwise correlation strength and growing approximate neighborhoods, but exclude near-perfect pairwise correlations and dense factor-driven dependence whose aggregate contribution does not vanish. Under such stronger cross-sectional dependence, the Gaussian-process covariance, extreme-value clustering, and asymptotic independence may change and require new normalization. Further work should allow serially dependent observations beyond the independent or exchangeable setting, develop data-driven calibration under complex dependence, and study computationally efficient multiple-change inference when changes are frequent or closely spaced.

\bibliographystyle{apalike}
\bibliography{references}

\appendix

\section{Exact null rank constants and finite-sample calibration}\label{app:rank}

For $|\rho|<1$, let
$\{(Y_i^{(1)}(\rho),Y_i^{(2)}(\rho)):1\le i\le n\}$ be independent bivariate normal vectors with zero means, unit variances, and correlation $\rho$. Let
$\bm\Pi^{(1)}(\rho)$ and $\bm\Pi^{(2)}(\rho)$ be the two rank permutations induced by these samples, and define
\[
\gamma_{n,kr}(\rho)
=\operatorname{Cov}\!\left\{
\mathcal Z_{n,k}\bigl(\bm\Pi^{(1)}(\rho)\bigr),
\mathcal Z_{n,r}\bigl(\bm\Pi^{(2)}(\rho)\bigr)
\right\}.
\]
For $\rho=1$, use the same rank permutation in both arguments, so that
\[
\gamma_{n,kr}(1)
=\operatorname{Cov}_\Pi\{\mathcal Z_{n,k}(\bm\Pi),
\mathcal Z_{n,r}(\bm\Pi)\}.
\]
In particular, $\gamma_{n,kr}(0)=0$.

For compact finite-sample formulas, define
\[
c^-_{k\ell}=\max(0,k+\ell-n),
\qquad
c^+_{k\ell}=\min(k,\ell),
\]
\[
g_{n,k,\ell}(c)
=\frac{k h(c/k)+(n-k)h\{(\ell-c)/(n-k)\}}
{d_{\ell,n}},
\qquad c^-_{k\ell}\le c\le c^+_{k\ell},
\]
and
\[
w_{n,k,\ell}(c)
=\frac{\binom{\ell}{c}\binom{n-\ell}{k-c}}
{\binom nk}.
\]
For $1\le\ell<m\le n$, put
\[
d^-_{k;\ell,m}(c)=\max(c,k+m-n),
\qquad
d^+_{k;\ell,m}(c)=\min(k,c+m-\ell),
\]
\[
w_{n,k;\ell,m}(c,d)
=\frac{\binom{\ell}{c}
\binom{m-\ell}{d-c}
\binom{n-m}{k-d}}
{\binom nk},
\qquad
d^-_{k;\ell,m}(c)\le d\le d^+_{k;\ell,m}(c).
\]

\begin{proposition}[Marginal rank law and Gaussian-copula covariance]
\label{prop:rank-law}
Fix $\eta\in(0,1/2)$ and let $\bX_1,\ldots,\bX_n$ be independent,
identically distributed row vectors with continuous marginal cdfs
$F_1,\ldots,F_p$. No condition is imposed on dependence within a row.
Let $R_{ij}=\sum_{a=1}^n\mathbf1\{X_{aj}\le X_{ij}\}$ and
$\bm\Pi_j=(R_{1j},\ldots,R_{nj})$ be the vector of observed ranks.
Each $\bm\Pi_j$ is marginally uniform on $\mathfrak S_n$, and
\[
 Z_j(k)=\mathcal Z_{n,k}(\bm\Pi_j)\quad\text{a.s.},\qquad
 \E Z_j(k)=\mu_{n,k},\qquad
 \operatorname{Var}\{Z_j(k)\}=s_{n,k}^2,\qquad k\in\mathcal K_n.
\]
For every $\bm\pi\in\mathfrak S_n$ and $k\in\mathcal K_n$,
\[
 0\le\mathcal Z_{n,k}(\bm\pi)\le\overline Z_n\le16.
\]
If, in addition, Assumption~\ref{ass:copula-null} holds, then
\[
 \operatorname{Cov}\{Z_j(k),Z_\ell(r)\}
       =\gamma_{n,kr}(\rho_{j\ell})
       \qquad(1\le j,\ell\le p,\ k,r\in\mathcal K_n).
\]
\end{proposition}

Under Assumption~\ref{ass:copula-null}, each coordinate scan vector
is measurable with respect to its own latent Gaussian column.
The marginal rank law, scalar null moments, and pathwise bound
hold under arbitrary coordinate copulas.

The exact null moments are
\begin{equation}\label{eq:exact-null-mean}
\mu_{n,k}
=\sum_{\ell=1}^n
\sum_{c=c^-_{k\ell}}^{c^+_{k\ell}}
g_{n,k,\ell}(c)w_{n,k,\ell}(c),
\end{equation}
and the exact finite-sample null variance is
\begin{align}
 s_{n,k}^2
&:=\operatorname{Var}_\Pi\{\mathcal Z_{n,k}(\bm\Pi)\}
 =\gamma_{n,kk}(1) \notag\\
&=\sum_{\ell=1}^n
  \sum_{c=c^-_{k\ell}}^{c^+_{k\ell}}
  g_{n,k,\ell}(c)^2w_{n,k,\ell}(c) \notag\\
&\quad
 +2\sum_{1\le\ell<m\le n}
  \sum_{c=c^-_{k\ell}}^{c^+_{k\ell}}
  \sum_{d=d^-_{k;\ell,m}(c)}^{d^+_{k;\ell,m}(c)}
  g_{n,k,\ell}(c)g_{n,k,m}(d)
  w_{n,k;\ell,m}(c,d)
 -\mu_{n,k}^2.
\label{eq:exact-null-variance}
\end{align}

\begin{proof}
Fix a coordinate \(j\). For \(\bm\pi\in\mathfrak S_n\), define
\(E_{\bm\pi,j}=\{X_{\pi_1j}<\cdots<X_{\pi_nj}\}\). Continuity and
exchangeability give
\[
 \Pp(X_{ij}=X_{rj})=0\quad(i\ne r),\qquad
 \sum_{\bm\pi\in\mathfrak S_n}\Pp(E_{\bm\pi,j})=1,\qquad
 \Pp(E_{\bm\pi,j})=\Pp(E_{\bm\pi',j})=\frac1{n!}.
\]
Consequently, with \(R_{ij}=\sum_{a=1}^n\mathbf1(X_{aj}\le X_{ij})\),
\[
 \Pp\{(R_{1j},\ldots,R_{nj})=\bm\pi\}=\frac1{n!},\qquad
 X_{ij}<X_{rj}\ \Longleftrightarrow\
 \Phi^{-1}\{F_j(X_{ij})\}<\Phi^{-1}\{F_j(X_{rj})\}\quad\text{a.s.}
\]
Thus the marginal normal-score transform preserves the rank vector
\(\bm\Pi_j\) under arbitrary coordinate copulas.
For \(C_{k\ell}=\sum_{i\le k}\mathbf1(R_{ij}\le\ell)\), direct
substitution in \eqref{eq:coordinate-stat} yields
\[
 \widehat F^{(k)}_{1j}(X_{(\ell)j})=\frac{C_{k\ell}}k,\qquad
 \widehat F^{(k)}_{2j}(X_{(\ell)j})=\frac{\ell-C_{k\ell}}{n-k},
 \qquad Z_j(k)=\mathcal Z_{n,k}(\bm\Pi_j)\quad\text{a.s.}
\]
Under Assumption~\ref{ass:copula-null}, for \(j\ne\ell\), the
joint Gaussian coordinate blocks satisfy
\[
 \begin{pmatrix}(Y_{ij})_{i\le n}\\(Y_{i\ell})_{i\le n}\end{pmatrix}
 \sim N_{2n}\!\left(
 0,\begin{pmatrix}\mathbf I_n&\rho_{j\ell}\mathbf I_n\\
                  \rho_{j\ell}\mathbf I_n&\mathbf I_n\end{pmatrix}\right),
 \qquad
 \operatorname{Cov}\{Z_j(k),Z_\ell(r)\}
   =\gamma_{n,kr}(\rho_{j\ell}).
\]
For \(j=\ell\), both statistics use the same permutation and the
covariance is \(\gamma_{n,kr}(1)\).
The remaining range and moment calculations use only the uniform
marginal permutation law. The pathwise identity and \eqref{eq:rank-statistic-uniform-range}
give the asserted bound on \(\mathcal Z_{n,k}\).
Let \(S=\{\Pi_1,\ldots,\Pi_k\}\). Counting the selected ranks in
two, and then three, disjoint groups gives
\begin{align*}
 \Pp_\Pi(S=A)&=\binom nk^{-1}\quad(|A|=k),\displaybreak[1]\\
 \Pp_\Pi(C_{k\ell}=c)
 &=\frac{\binom{\ell}{c}\binom{n-\ell}{k-c}}{\binom nk}
   =w_{n,k,\ell}(c),\displaybreak[1]\\
 \Pp_\Pi(C_{k\ell}=c,C_{km}=d)
 &=\frac{\binom{\ell}{c}\binom{m-\ell}{d-c}
                  \binom{n-m}{k-d}}{\binom nk}
   =w_{n,k;\ell,m}(c,d).
\end{align*}
The binomial factors are nonzero precisely on the ranges stated
above. Since \(\mathcal Z_{n,k}=\sum_\ell g_{n,k,\ell}(C_{k\ell})\),
\[
 \begin{aligned}
 \E_\Pi\mathcal Z_{n,k}
 &=\sum_{\ell,c}g_{n,k,\ell}(c)w_{n,k,\ell}(c),\\
 \E_\Pi\mathcal Z_{n,k}^{\,2}
 &=\sum_{\ell,c}g_{n,k,\ell}(c)^2w_{n,k,\ell}(c)
   +2\sum_{\ell<m}\sum_{c,d}
      g_{n,k,\ell}(c)g_{n,k,m}(d)w_{n,k;\ell,m}(c,d),\\
 \operatorname{Var}_\Pi(\mathcal Z_{n,k})
 &=\E_\Pi\mathcal Z_{n,k}^{\,2}
       -(\E_\Pi\mathcal Z_{n,k})^2 .
 \end{aligned}
\]
These are \eqref{eq:exact-null-mean} and
\eqref{eq:exact-null-variance}.
\end{proof}

The exact moments can be evaluated by a recursion over the pooled
ranks. Fix \(k\), let \(S=\{\Pi_1,\ldots,\Pi_k\}\), and write
\[
 T_\ell=\sum_{r=1}^{\ell}g_{n,k,r}(C_{kr}),\qquad
 I_{\ell+1}=\mathbf1\{\ell+1\in S\},\qquad T_0=C_{k0}=0 .
\]
Conditional on the selected labels among the first \(\ell\) ranks,
there are \(k-C_{k\ell}\) selected labels among the remaining
\(n-\ell\). For \(0\le\ell<n\), uniform subset counting gives
\[
 \begin{aligned}
 q_{\ell c}(1)
 &=\frac{\binom{n-\ell-1}{k-c-1}}{\binom{n-\ell}{k-c}}
   =\frac{k-c}{n-\ell},&
 q_{\ell c}(0)
 &=\frac{\binom{n-\ell-1}{k-c}}{\binom{n-\ell}{k-c}}
   =\frac{n-\ell-k+c}{n-\ell},\\
 C_{k,\ell+1}&=C_{k\ell}+I_{\ell+1},&
 T_{\ell+1}&=T_\ell+g_{n,k,\ell+1}(C_{k,\ell+1}).
 \end{aligned}
\]
Define the unnormalized conditional moments
\[
 M_{\ell c}^{(a)}
   =\mathbb E_\Pi[T_\ell^a\mathbf1\{C_{k\ell}=c\}],
 \quad a=0,1,2,\qquad
 (P_{\ell c},L_{\ell c},R_{\ell c})
   =(M_{\ell c}^{(0)},M_{\ell c}^{(1)},M_{\ell c}^{(2)}).
\]
Use \(M_{00}^{(0)}=1\), \(M_{00}^{(1)}=M_{00}^{(2)}=0\);
infeasible states and their transition terms are zero.
Conditioning on \(C_{k\ell}=c'-b\) and expanding the power gives
\[
 \begin{aligned}
 M_{\ell+1,c'}^{(a)}
 &=\sum_{b=0}^1
   \mathbb E_\Pi[
   \{T_\ell+g_{n,k,\ell+1}(c')\}^a
   \mathbf1\{C_{k\ell}=c'-b,I_{\ell+1}=b\}]\\
 &=\sum_{b=0}^1 q_{\ell,c'-b}(b)
      \sum_{r=0}^a\binom ar
       g_{n,k,\ell+1}(c')^{a-r}M_{\ell,c'-b}^{(r)} .
 \end{aligned}
\]
Equivalently, the three moments are updated together:
\begin{equation*}
 \begin{pmatrix}P_{\ell+1,c'}\\L_{\ell+1,c'}\\R_{\ell+1,c'}\end{pmatrix}
 =\sum_{b=0}^1q_{\ell,c'-b}(b)
 \begin{pmatrix}
 1&0&0\\ g&1&0\\g^2&2g&1
 \end{pmatrix}
 \begin{pmatrix}P_{\ell,c'-b}\\L_{\ell,c'-b}\\R_{\ell,c'-b}\end{pmatrix},
 \qquad g=g_{n,k,\ell+1}(c').
\end{equation*}
The terminal identities are
\[
 C_{kn}=k,\qquad T_n=\mathcal Z_{n,k}(\bm\Pi),\qquad
 P_{nk}=1,\quad L_{nk}=\mathbb E_\Pi\mathcal Z_{n,k},\quad
 R_{nk}=\mathbb E_\Pi\mathcal Z_{n,k}^2,
\]
and hence
\begin{equation*}
 \mu_{n,k}=L_{nk},\qquad s_{n,k}=(R_{nk}-L_{nk}^2)^{1/2}.
\end{equation*}
There are at most \((n+1)(k+1)\) states and two transitions per
state, giving \(O(nk)\) arithmetic operations and \(O(k)\) memory
when two consecutive rank layers are retained.
Complementing the uniform subset gives
\[
 S^c\sim\operatorname{Unif}\{A:|A|=n-k\},\qquad
 |S^c\cap\{1,\ldots,\ell\}|=\ell-C_{k\ell},\qquad
 g_{n,n-k,\ell}(\ell-c)=g_{n,k,\ell}(c).
\]
The two entropy terms are interchanged, so
\(\mathcal Z_{n,n-k}(S^c)=\mathcal Z_{n,k}(S)\) pathwise.
Taking its first two moments proves
\begin{equation*}
 \mu_{n,k}=\mu_{n,n-k},\qquad s_{n,k}=s_{n,n-k}.
\end{equation*}

\subsection{Finite-sample permutation validity}

\begin{theorem}[Permutation validity]\label{thm:permutation}
Suppose $\bX_1,\ldots,\bX_n$ are exchangeable under $H_0$, with the same deterministic tie rule for observed and permuted samples. For $\comp\in\{\mathrm{sum},\mathrm{max}\}$ and every $\alpha\in(0,1)$,
\[
\Pp\!\left(\widehat p_{\comp}\le\alpha\right)\le\alpha.
\]
\end{theorem}

When all permutations are enumerated, randomization at the upper rejection boundary attains level $\alpha$ exactly. Use an independent $U\sim\operatorname{Unif}(0,1)$ for that boundary decision.

\begin{proof}
Write \(\bX=(\bX_1,\ldots,\bX_n)\). Let \(\mathcal G=\mathfrak S_n\) act by
\((G\bX)_i=\bX_{G(i)}\). Choose independent uniform
\(G_0,\ldots,G_B\), independently of \(\bX\), and put
\begin{equation*}
 \widetilde{\bX}=G_0\bX,\qquad \pi_b=G_0^{-1}\circ G_b,\quad 1\le b\le B.
\end{equation*}
For arbitrary \(g_0,h_1,\ldots,h_B\in\mathcal G\),
\[
 \begin{aligned}
 &\mathbb P(G_0=g_0,\pi_1=h_1,\ldots,\pi_B=h_B\mid\bX)\\
 &\qquad=\mathbb P(G_0=g_0,G_1=g_0\circ h_1,\ldots,
                           G_B=g_0\circ h_B\mid\bX)
          =|\mathcal G|^{-(B+1)},\\
 &(\pi_b\widetilde{\bX})_i
   =\bX_{G_0(G_0^{-1}(G_b(i)))}=\bX_{G_b(i)},\qquad
 \widetilde{\bX}\overset d=\bX .
 \end{aligned}
\]
Thus the relative permutations are independent of
\(\widetilde{\bX}\), and for either statistic \(T\),
\begin{equation*}
 \bigl(T(\widetilde{\bX}),T(\pi_1\widetilde{\bX}),\ldots,
                         T(\pi_B\widetilde{\bX})\bigr)
   =\bigl(T(G_0\bX),T(G_1\bX),\ldots,T(G_B\bX)\bigr).
\end{equation*}
For arbitrary Borel sets \(A_0,\ldots,A_B\), conditional independence gives
\[
 \begin{aligned}
 &\mathbb P\{T(G_i\bX)\in A_i,\ 0\le i\le B\mid\bX\}\\
 &\qquad=\prod_{i=0}^B
 \left\{\frac1{|\mathcal G|}\sum_{g\in\mathcal G}
                         \mathbf1\{T(g\bX)\in A_i\}\right\}.
 \end{aligned}
\]
The product is invariant under index permutations. By the preceding
identity and independence of the relative permutations, the observed
and simulated statistics have this exchangeable law. Write
\(V_0,\ldots,V_B\) for these variables and set
\[
 R_i=\sum_{b=0}^B\mathbf1\{V_b\ge V_i\},\qquad
 A_q=\{i:R_i\le q\},\quad q\in\{0,\ldots,B+1\}.
\]
If \(A_q\ne\varnothing\), choose \(i_*\in A_q\) of smallest
\(V_i\). Then
\[
 |A_q|\le\sum_{b=0}^B\mathbf1\{V_b\ge V_{i_*}\}
          =R_{i_*}\le q .
\]
This inequality also holds for \(A_q=\varnothing\) and includes ties.
Exchangeability gives
\begin{equation*}
 \mathbb P(R_0\le q)
 =\frac1{B+1}\mathbb E\sum_{i=0}^B\mathbf1\{R_i\le q\}
 \le\frac q{B+1}.
\end{equation*}
Since \(R_0=(B+1)\widehat p_\comp\), taking
\(q=\lfloor\alpha(B+1)\rfloor\) proves
\[
 \mathbb P(\widehat p_\comp\le\alpha)
 \le\frac{\lfloor\alpha(B+1)\rfloor}{B+1}\le\alpha.
\]
For complete enumeration, let \(R\) be the randomized upper rank of
\(T(G_0\bX)\) among all \(T(g\bX)\), \(g\in\mathcal G\). Set \(J=|\mathcal G|\) and
\(\mathcal G_x=\{g:T(g\bX)=x\}\). Independently order the labels
within each tie group uniformly. With
\(a_x=\#\{g:T(g\bX)>x\}\), the randomized upper rank satisfies
\[
 \begin{aligned}
 R\mid\{G_0\in\mathcal G_x,\bX\}
     &\sim\operatorname{Unif}\{a_x+1,\ldots,a_x+|\mathcal G_x|\},\\
 \mathbb P(R=r\mid\bX)
     &=\frac{|\mathcal G_x|}{J}\frac1{|\mathcal G_x|}
       =\frac1J,\quad a_x<r\le a_x+|\mathcal G_x|.
 \end{aligned}
\]
Put \(m=\lfloor\alpha J\rfloor\), \(v=\alpha J-m\), and take an
independent \(U\sim\operatorname{Unif}(0,1)\). Then
\[
 \begin{aligned}
 \mathbb P\{\{R\le m\}\cup\{R=m+1,U\le v\}\mid\bX\}
 &=\frac mJ+\frac vJ=\alpha .
 \end{aligned}
\]
This proves the exact randomized level as well.
\end{proof}

To simulate the dense reference process on
$-L_\eta=u_0<\cdots<u_J=L_\eta$, initialize $\mathbb U(u_0)\sim N(0,1)$ independently of the innovations below.
Use the exact Ornstein--Uhlenbeck recursion
\[
\mathbb U(u_{r+1})
=e^{-(u_{r+1}-u_r)}\mathbb U(u_r)
+\{1-e^{-2(u_{r+1}-u_r)}\}^{1/2}\varepsilon_{r+1},
\qquad \varepsilon_{r+1}\overset{\mathrm{iid}}\sim N(0,1).
\]
The sparse critical value is available directly as
$\mathfrak b_{n,p}+\mathfrak a_{n,p}g_{1-\alpha}$ from \cref{eq:max-rejection-rule}.

\section{Exact null moments and the dense process}\label{app:dense-null}

We derive the exact null moment expansions and then establish the dense process limit.

\subsection{Null-moment correction constants}
\label{app:moment-correction-definitions}

The following deterministic constants refine
\eqref{eq:main-null-moment-orders}. We use the Bernoulli entropy
$h_t$ defined in Section~\ref{sec:null}; the finite sums below
also provide the moment inputs for the MAX tail coefficients.

Write \(z_+=\max\{z,0\}\). We use
\[
 \Gamma(a)=\int_0^\infty z^{a-1}e^{-z}\,dz\quad(a>0),\qquad
 \psi(a)=\frac{\Gamma'(a)}{\Gamma(a)},\qquad
 \gamma_{\mathrm E}=-\psi(1),\qquad
 \zeta(a)=\sum_{j=1}^{\infty}j^{-a}\quad(a>1).
\]

For integers \(\ell\ge1\), let
\[
 p_{0,0}(t)=1,\qquad
 p_{\ell,a}(t)=\binom{\ell}{a}t^a(1-t)^{\ell-a}
       \quad(0\le a\le\ell),\qquad
 g_\ell(t)=\ell\sum_{a=0}^{\ell}p_{\ell,a}(t)h_t(a/\ell).
\]
For \(1\le\ell\le m\), define the deterministic finite sum
\begin{equation}\label{eq:main-moment-covariance}
 \begin{split}
 C_{\ell m}(t)
 =\sum_{a=0}^{\ell}\sum_{d=0}^{m-\ell}
 &p_{\ell,a}(t)p_{m-\ell,d}(t)
       \{\ell h_t(a/\ell)-g_\ell(t)\}\\
 &{}\times\{m h_t((a+d)/m)-g_m(t)\},
 \qquad C_{m\ell}(t)=C_{\ell m}(t),
 \end{split}
\end{equation}
The null-moment corrections are the convergent series
\begin{equation}\label{eq:main-moment-constants}
 \begin{aligned}
 m(t)&=\gamma_{\mathrm E}+2\log2-1+
 \sum_{\ell=1}^{\infty}
 \left(\frac1{\ell-1/2}+\frac1{\ell+1/2}\right)
                          \{g_\ell(t)-1/2\},\\
 v_0(t)&=2\gamma_{\mathrm E}+1-\frac{\pi^2}{6}
 +\sum_{c\in\{-1/2,1/2\}}\sum_{\ell,m\ge1}
 \left\{\frac{C_{\ell m}(t)}{(\ell+c)(m+c)}
                        -\frac1{2\max(\ell,m)^2}\right\}.
 \end{aligned}
\end{equation}
For \(0<t<1\) and \(y>1\), define
\[
 d_0(t,y)=m(t)+\frac{y-1}{4}v_0(t).
\]
The moment constants satisfy, uniformly on the trimmed grid,
\[
 n\{\overline Z_n-\mu_{n,k}\}=\log n+m(k/n)+o(1),\qquad
 n^2s_{n,k}^2=2\log n+v_0(k/n)+o(1).
\]

The proofs begin with the exact rank decomposition.

Put $t=k/n$, $v=t(1-t)$, $u_\ell=\ell/n$,
$w_\ell=u_\ell(1-u_\ell)$, and $b_\ell=\ell\wedge(n-\ell)$.
Throughout this appendix $\eta n\le k\le(1-\eta)n$.  Define
\begin{align*}
 A_\ell&=C_{k\ell}-t\ell,\\
 G_{k\ell}&=nh(u_\ell)-kh(C_{k\ell}/k)
 -(n-k)h\{(\ell-C_{k\ell})/(n-k)\},\\
 q_{k\ell}&=\{2nvw_\ell d_{\ell,n}\}^{-1},\\
 F_{n,k}&=\sum_{\ell=1}^{n-1}G_{k\ell}/d_{\ell,n},\qquad
 Q_{n,k}=\sum_{\ell=1}^{n-1}q_{k\ell}A_\ell^2.
\end{align*}
Then, exactly,
\[
 \mu_{n,k}-\mathcal Z_{n,k}=F_{n,k}-\E F_{n,k},\qquad
 s_{n,k}^2=\operatorname{Var}(F_{n,k}).
\]
Write $D_{n,k}=(F_{n,k}-\E F_{n,k})/s_{n,k}$ for the one-coordinate standardized deficit.
All the constants in this appendix may depend on the fixed trimming fraction.

Set $e_\ell=G_{k\ell}-A_\ell^2/(2nvw_\ell)$.

\begin{lemma}[Entropy remainder under sampling without replacement]
\label{lem:weighted-entropy-projection}
For every fixed integer $r\ge2$,
\[
 \|A_\ell\|_r\le C_r\sqrt{b_\ell},\qquad
 \Pp(|A_\ell|>c_\eta b_\ell)\le2e^{-c'_\eta b_\ell},
 \qquad 0\le G_{k\ell}\le\frac{A_\ell^2}{nvw_\ell}.
\]
Then
\begin{equation}\label{eq:entropy-remainder-correct}
 \|F_{n,k}-Q_{n,k}\|_2\le C_\eta n^{-1},\qquad
 |\operatorname{Cov}(A_\ell,e_\ell)|+
 |\operatorname{Cov}(A_\ell^2,e_\ell)|\le C_\eta.
\end{equation}
\end{lemma}
\begin{proof}
The two expressions for the likelihood-ratio gap of a $2\times2$ table give
\begin{align*}
G_{k\ell}
 &=k\operatorname{KL}(C_{k\ell}/k\Vert u_\ell)
 +(n-k)\operatorname{KL}\{(\ell-C_{k\ell})/(n-k)\Vert u_\ell\}\\
 &=\ell\operatorname{KL}(C_{k\ell}/\ell\Vert t)
 +(n-\ell)\operatorname{KL}\{(k-C_{k\ell})/(n-\ell)\Vert t\}.
\end{align*}
The table identity and the Bernoulli chi-square bound give
\[
 G_{k\ell}\le
 k\frac{(A_\ell/k)^2}{w_\ell}
 +(n-k)\frac{(A_\ell/(n-k))^2}{w_\ell}
 =\frac{A_\ell^2}{nvw_\ell}.
\]
Let \(\xi_1,\ldots,\xi_{b_\ell}\) be independent
Bernoulli\((t)\) variables and set
\(\varepsilon_\ell=1\) for \(\ell\le n/2\), and \(-1\) otherwise.
The convex comparison in Theorem~4 of \citet{Hoeffding1963} yields
\[
 \E e^{zA_\ell}
 \le \E e^{\varepsilon_\ell z\sum_{i=1}^{b_\ell}(\xi_i-t)}
 =e^{b_\ell\psi_t(\varepsilon_\ell z)},\qquad
 \psi_t(z)=\log(1-t+te^z)-tz.
\]
Writing \(t_z=te^z/(1-t+te^z)\), we have
\[
 \psi_t(0)=\psi_t'(0)=0,\qquad
 \psi_t''(z)=t_z(1-t_z)\le\tfrac14,\qquad
 \psi_t(z)=z^2\int_0^1(1-a)\psi_t''(az)\,da\le z^2/8.
\]
Consequently
\[
 \E e^{zA_\ell}\le e^{b_\ell z^2/8},\qquad
 \Pp(|A_\ell|>x)\le2e^{-2x^2/b_\ell},\qquad
 \E|A_\ell|^r
 \le2r\int_0^\infty x^{r-1}e^{-2x^2/b_\ell}\,dx
 \le C_r b_\ell^{r/2}.
\]
These bounds prove the asserted moments and large-deviation
probability.
Define
\[
 a_{3,k\ell}=-\frac{h'''(u_\ell)}6
 \{k^{-2}-(n-k)^{-2}\},\qquad
 \mathcal R_{4,k\ell}=e_\ell-a_{3,k\ell}A_\ell^3.
\]
On \(|A_\ell|\le c_\eta b_\ell\), Taylor's theorem with the
linear terms canceled gives
\begin{equation}\label{eq:entropy-cubic-expansion}
 e_\ell=a_{3,k\ell}A_\ell^3+\mathcal R_{4,k\ell},\qquad
 |a_{3,k\ell}|\le C_\eta b_\ell^{-2},\qquad
 |\mathcal R_{4,k\ell}|\le C_\eta |A_\ell|^4 b_\ell^{-3}.
\end{equation}
On the complementary event, \(|e_\ell|\le C_\eta b_\ell\).
The Taylor expansion and the sixth and eighth count moments give
\[
 \begin{aligned}
 \|e_\ell\|_2
 &\le Cb_\ell^{-2}\|A_\ell\|_6^3
       +Cb_\ell^{-3}\|A_\ell\|_8^4
       +Cb_\ell e^{-cb_\ell}
 \le Cb_\ell^{-1/2},\\
 \left\|\sum_\ell e_\ell/d_{\ell,n}\right\|_2
 &\le\frac Cn\sum_{\ell=1}^{n-1}b_\ell^{-3/2}
 \le Cn^{-1},
 \end{aligned}
\]
where \(d_{\ell,n}\ge nb_\ell/4\).
The covariance calculation uses the signed third and fifth moments of the count.
The factorial identity
\[
 \E(C_{k\ell})_j=\frac{(k)_j(\ell)_j}{(n)_j}
\]
gives
\[
 \E A_\ell^3=
 \frac{\ell v(1-2t)(n-\ell)(n-2\ell)}{(n-1)(n-2)}.
\]
For $n>4$, set $K=k(n-k)$ and $L=\ell(n-\ell)$.  The fifth central
moment is, exactly,
\begin{equation*}
 \E A_\ell^5=\E A_\ell^3\,
 \frac{(10n+24)KL-12n^2(K+L)+n^3(n+5)}
 {n^2(n-3)(n-4)}.
\end{equation*}
Both identities follow by expanding
\((C_{k\ell}-k\ell/n)^j\) into falling factorials for \(j\le5\).
Reflection gives
\[
 |\E A_\ell^3|\le Cb_\ell,\qquad
 |\E A_\ell^5|\le C(b_\ell+b_\ell^2).
\]
The signed moments are used before taking absolute values:
\[
 \begin{aligned}
 |\operatorname{Cov}(A_\ell^2,a_{3,k\ell}A_\ell^3)|
 &\le Cb_\ell^{-2}
    \{|\E A_\ell^5|+\E A_\ell^2|\E A_\ell^3|\}\le C,\\
 |\operatorname{Cov}(A_\ell,a_{3,k\ell}A_\ell^3)|
 &=|a_{3,k\ell}|\E A_\ell^4\le C .
 \end{aligned}
\]
For the fourth-order remainder on the expansion event,
\[
 \begin{aligned}
 |\operatorname{Cov}(A_\ell^2,\mathcal R_{4,k\ell})|
 &\le Cb_\ell^{-3}
   \{\E|A_\ell|^6+\E A_\ell^2\,\E A_\ell^4\}\le C,\\
 |\operatorname{Cov}(A_\ell,\mathcal R_{4,k\ell})|
 &\le Cb_\ell^{-3}
   \{\E|A_\ell|^5+\E|A_\ell|\,\E A_\ell^4\}\le C .
 \end{aligned}
\]
Replacing the truncated odd moments by full moments, and
bounding the original complementary-event terms, costs at most
a polynomial in \(b_\ell\) times \(e^{-cb_\ell}\).
This proves both covariance bounds in
\eqref{eq:entropy-remainder-correct}.
\end{proof}

\begin{proposition}[Uniform null mean and variance]
\label{prop:rank-covariance}
Fix $0<\eta<1/2$ and use the uniform permutation law of
Proposition~\ref{prop:rank-law}, equivalently an iid continuous
one-coordinate null sample. Uniformly for $k\in\mathcal K_n$,
\begin{equation*}
 \E F_{n,k}=\frac{\log n+O_\eta(1)}n,\qquad
 s_{n,k}^2=\frac{2\log n+O_\eta(1)}{n^2}.
\end{equation*}
\end{proposition}
\begin{proof}
The exact variance of a hypergeometric count gives
\[
 \E A_\ell^2=\frac{n^2vw_\ell}{n-1},\qquad
 \E\frac{A_\ell^2}{2nvw_\ell}=\frac{n}{2(n-1)},\qquad
 \sum_{\ell=1}^{n-1}d_{\ell,n}^{-1}
 =\frac{2\log n+O(1)}n.
\]
The first assertion follows from Lemma~\ref{lem:weighted-entropy-projection}.
The variance of the quadratic term follows from the exact permutation moments. Let
\[
 c_{i,\ell}=\mathbf1(i\le\ell)-\ell/n,\quad
 S_{\ell m}=\sum_i c_{i,\ell}c_{i,m}=\ell\wedge m-\ell m/n,
 \quad T_{\ell m}=\sum_i c_{i,\ell}^2c_{i,m}^2.
\]
If $(Z_1,\ldots,Z_n)$ is a permutation of $k$ copies of $1-t$ and
$n-k$ copies of $-t$, then $A_\ell=\sum_i c_{i,\ell}Z_i$.  In particular,
\[
 \operatorname{Cov}(A_\ell,A_m)=\frac{nv}{n-1}S_{\ell m}.
\]
For centered \(X,Y\), use
\[
 \operatorname{cum}(X,X,Y,Y)
   =\E(X^2Y^2)-\E X^2\,\E Y^2-2(\E XY)^2.
\]
Define
\[
 \alpha_{n,k}=-\frac{n^2v(6nv-n-1)}{(n-3)(n-2)(n-1)},\qquad
 \varepsilon_{n,k}=
 \frac{nv(4n^2v-n^2-6nv+2n-1)}{(n-3)(n-2)(n-1)^2}.
\]
Grouping the four indices by equality pattern yields
\begin{equation}\label{eq:rank-fourth-cumulant-correct}
 \operatorname{cum}(A_\ell,A_\ell,A_m,A_m)
 =\alpha_{n,k}T_{\ell m}
 +\varepsilon_{n,k}(S_{\ell\ell}S_{mm}+2S_{\ell m}^2).
\end{equation}
For distinct displayed indices, the five equality patterns have moments
\begin{align*}
\E Z_1^4&=v(1-3v),&
\E Z_1^3Z_2&=-v(1-3v)/(n-1),\\
\E Z_1^2Z_2^2&=\{(n+3)v^2-v\}/(n-1),&
\E Z_1^2Z_2Z_3&=\{2v-(n+6)v^2\}/\{(n-1)(n-2)\},\\
\E Z_1Z_2Z_3Z_4&=-3\E Z_1^2Z_2Z_3/(n-3).
\end{align*}
The five moments follow recursively from
\(Z_i^2=(1-2t)Z_i+v\) and \(\sum_iZ_i=0\).
Substitution of these equality patterns, followed by subtraction
of the three covariance pairings, gives
\eqref{eq:rank-fourth-cumulant-correct}.
Since \(T_{\ell m}\le b_\ell\wedge b_m\),
\(|\alpha_{n,k}|\le C\) and \(|\varepsilon_{n,k}|\le C/n\),
their weighted contributions satisfy
\[
 \begin{aligned}
 \sum_{\ell,m}q_{k\ell}q_{km}
       |\alpha_{n,k}|T_{\ell m}
 &\le\frac C{n^2}
       \sum_{a,b\ge1}\frac{a\wedge b}{a^2b^2}
 \le Cn^{-2},\\
 \sum_{\ell,m}q_{k\ell}q_{km}|\varepsilon_{n,k}|
       (S_{\ell\ell}S_{mm}+2S_{\ell m}^2)
 &\le C(\log n)^2/n^3\le Cn^{-2}.
 \end{aligned}
\]
For the two covariance-pair terms, initially replace $d_{\ell,n}$ by
$\ell(n-\ell)$.  Their sum is
\[
 \frac{n^2}{2(n-1)^2}
 \sum_{\ell,m=1}^{n-1}
 \frac1{\{n-(\ell\wedge m)\}^2(\ell\vee m)^2}
 =\frac{2\log n+O(1)}{n^2}.
\]
To see the logarithmic constant, on $\ell\le m\le n/2$ the sum is
\[
 \sum_{m\le n/2}m^{-2}\sum_{\ell\le m}(n-\ell)^{-2}
 =\sum_{m\le n/2}\{(n^2m)^{-1}+O(n^{-3})\}
 =n^{-2}\{\log n+O(1)\}.
\]
The reflected triangle gives the same logarithmic term,
and exchanging the indices doubles their sum. The remaining
rectangle and the diagonal are \(O(n^{-2})\).
Restoring the half-shifted denominators gives
\[
 \begin{aligned}
 &\left|2\sum_{\ell,m=1}^{n-1}q_{k\ell}q_{km}
       \operatorname{Cov}(A_\ell,A_m)^2
 -\frac{n^2}{2(n-1)^2}\sum_{\ell,m=1}^{n-1}
       \frac1{\{n-(\ell\wedge m)\}^2(\ell\vee m)^2}\right|\\
 &\qquad\le\frac C{n^2}\sum_{a,b\ge1}
          \frac{a^{-1}+b^{-1}}{(a\vee b)^2}
 \le Cn^{-2}.
 \end{aligned}
\]
so \(\operatorname{Var}(Q_{n,k})
 =\{2\log n+O_\eta(1)\}/n^2\).
Define \(a_{\ell\ell}=1\), \(b_{\ell\ell}=0\), and
\[
\begin{array}{lll}
m<\ell:&a_{m\ell}=\dfrac{m(m-1)}{\ell(\ell-1)},&
 b_{m\ell}=\dfrac{m(\ell-m)(1-2t)}{\ell(\ell-1)},\\[2mm]
m>\ell:&a_{m\ell}=\dfrac{(n-m)(n-m-1)}{(n-\ell)(n-\ell-1)},&
 b_{m\ell}=-\dfrac{(m-\ell)(n-m)(1-2t)}{(n-\ell)(n-\ell-1)}.
\end{array}
\]
Put
\[
 c_{m\ell}=\frac{v}{n-1}
       \{m(n-m)-a_{m\ell}\ell(n-\ell)\}.
\]
Conditional hypergeometric sampling gives
\[
 \E(A_m^2\mid A_\ell)
      =a_{m\ell}A_\ell^2+b_{m\ell}A_\ell+c_{m\ell}.
\]  Since $e_\ell$ is a function of $A_\ell$,
\[
 |\operatorname{Cov}(A_m^2,e_\ell)|
 \le C_\eta(|a_{m\ell}|+|b_{m\ell}|).
\]
For $\ell\le n/2$ this is bounded respectively by
$C_\eta m/\ell$ for $m\le\ell$, by $C_\eta$ for
$\ell\le m\le n/2$, and by $C_\eta(n-m)/n$ for $m>n/2$.
Consequently,
\begin{align*}
 &\left|\operatorname{Cov}\left(
 Q_{n,k},\sum_\ell e_\ell/d_{\ell,n}\right)\right|\\
 &\quad\le\frac{C_\eta}{n^2}
 \left\{\sum_{\ell\le n/2}\frac{1+\log\ell}{\ell^2}
       +\frac{(\log n)^2}{n}\right\}
 \le C_\eta n^{-2},
\end{align*}
Reflecting the upper rank half gives the same bound.
With \(R=F_{n,k}-Q_{n,k}\),
\[
 \operatorname{Var}(F_{n,k})
 =\operatorname{Var}(Q_{n,k})
       +2\operatorname{Cov}(Q_{n,k},R)+\operatorname{Var}(R)
 =\frac{2\log n+O_\eta(1)}{n^2}.
\]
\end{proof}

For \(B_\ell\sim\operatorname{Bin}(\ell,t)\), the finite sum defining
\(g_\ell(t)\) in Appendix~\ref{app:moment-correction-definitions} gives
\[
 g_\ell(t)=\mathbb E\{\ell h_t(B_\ell/\ell)\}.
\]
\begin{proposition}[Explicit constant in the null mean]
\label{prop:rank-mean-constant}
The series defining \(m(t)\) in \eqref{eq:main-moment-constants}
is absolutely and locally uniformly convergent. Uniformly over
$k\in\mathcal K_n$,
\[
 n\E F_{n,k}=\log n+m(k/n)+o(1).
\]
In particular $m(t)=m(1-t)$.
\end{proposition}
\begin{proof}
Write $t=k/n$ and
$e_\ell=G_{k\ell}-A_\ell^2/(2nvw_\ell)$ as above.
Using the signed third moment in
\eqref{eq:entropy-cubic-expansion},
\[
 |\E e_\ell|
 \le Cb_\ell^{-2}|\E A_\ell^3|
      +Cb_\ell^{-3}\E A_\ell^4
      +Cb_\ell^Ce^{-cb_\ell}
 \le C_\eta/b_\ell.
\]
The exact quadratic mean is
\[
 \E\{A_\ell^2/(2nvw_\ell)\}=\frac{n}{2(n-1)}.
\]
Let \(I_r\) indicate that pooled rank \(r\) belongs to the first \(k\)
rows. For \(b=(b_1,\ldots,b_\ell)\in\{0,1\}^{\ell}\), define
\(a=\sum_{r=1}^{\ell}b_r\) and
\(\mathcal A_b=\bigcap_{r=1}^{\ell}\{I_r=b_r\}\). Then
\[
 \Pp(\mathcal A_b)
 =\frac{(k)_a(n-k)_{\ell-a}}{(n)_\ell}
 =t^a(1-t)^{\ell-a}+O_\ell(n^{-1}),
 \qquad t\in[\eta,1-\eta].
\]
The other table term and the exact quadratic mean satisfy
\[
 \begin{aligned}
 \E\!\left[(n-\ell)\operatorname{KL}
       \left(t-\frac{A_\ell}{n-\ell}\,\middle\Vert\,t\right)\right]
 &\le \frac{\E A_\ell^2}{(n-\ell)t(1-t)}
 \le C_\ell/n,\\
 \E\frac{A_\ell^2}{2nvw_\ell}&=\frac{n}{2(n-1)},\\
 \E e_\ell&=g_\ell(t)-\tfrac12+O_\ell(n^{-1}).
 \end{aligned}
\]
Complementation gives the same limit for \(\E e_{n-\ell}\);
the two endpoint weights retain their different half-shifts:
\[
 \frac n{d_{\ell,n}}\longrightarrow\frac1{\ell-1/2},
 \qquad
 \frac n{d_{n-\ell,n}}\longrightarrow\frac1{\ell+1/2}.
\]
The domination \(n/d_{\ell,n}\le C/b_\ell\) gives
\[
 \left|n\sum_{b_\ell>M}\frac{\E e_\ell}{d_{\ell,n}}\right|
 \le C_\eta\sum_{b>M}b^{-2}\le C_\eta/M .
\]
For a finite-$n$ bound, couple the first $\ell$ population labels
sampled with and without replacement. Write their sums as
$B_\ell^{\rm bin}$ and $B_\ell^{\rm hyp}$. A mismatch requires a
repeated index in the sample with replacement, whence
\[
 \begin{split}
 P(B_\ell^{\rm bin}\ne B_\ell^{\rm hyp})
 &\le\frac{\ell(\ell-1)}{2n},\qquad
 0\le\ell h_t(x)\le C_\eta\ell\quad(0\le x\le1),\\
 \left|E\{\ell h_t(B_\ell^{\rm hyp}/\ell)\}-g_\ell(t)\right|
 &\le C_\eta\ell^3/n,\\
 E\!\left[(n-\ell)h_t\left(t-\frac{A_\ell}{n-\ell}\right)\right]
 &\le\frac{EA_\ell^2}{(n-\ell)t(1-t)}
 \le C_\eta\ell/n\qquad(\ell\le n/4).
 \end{split}
\]
The binomial Taylor bound gives $|g_\ell(t)-1/2|\le C_\eta/\ell$.
The exact quadratic contribution and the endpoint weight errors are
\[
 \begin{split}
 \frac n{2(n-1)}\sum_{\ell=1}^{n-1}
 \left\{\frac1{\ell-1/2}+\frac1{n-\ell+1/2}\right\}
 &=\log n+\gamma_{\mathrm E}+2\log2-1+O(\log n/n),\\
 \left|\frac n{d_{\ell,n}}-\frac1{\ell-1/2}\right|
 +\left|\frac n{d_{n-\ell,n}}-\frac1{\ell+1/2}\right|
 &\le C/n\qquad(\ell\le n/4).
 \end{split}
\]
Summing the coupling error with weights $C/\ell$, and then using the
preceding $1/M$ tail bound on both halves, proves
\begin{equation}\label{eq:rank-mean-constant-error}
 \sup_{k\in\mathcal K_n}|nEF_{n,k}-\log n-m(k/n)|
 \le C_\eta\{M^{-1}+M^3/n+\log n/n\},
 \qquad 1\le M\le n/4.
\end{equation}
This identifies the series in \eqref{eq:main-moment-constants}, with
absolute and uniform convergence. Replacing $B_\ell$ by
$\ell-B_\ell$ proves symmetry.
\end{proof}

For $t\in(0,1)$ let $B_1,B_2,\ldots$ be independent Bernoulli$(t)$
variables and put
\[
 S_\ell=\sum_{i=1}^\ell(B_i-t),\qquad
 \Phi_\ell(t)=\ell h_t(t+S_\ell/\ell).
\]
For \(\ell\le m\), independence of \(S_m-S_\ell\) and the first
\(\ell\) labels identifies the finite sum in
\eqref{eq:main-moment-covariance}:
\[
 \mathbb E\Phi_\ell(t)=g_\ell(t),\qquad
 \operatorname{Cov}\{\Phi_\ell(t),\Phi_m(t)\}=C_{\ell m}(t).
\]

\Needspace{5\baselineskip}
\begin{proposition}[Explicit constant in the null variance]
\label{prop:rank-variance-constant}
The series defining \(v_0(t)\) in \eqref{eq:main-moment-constants}
is absolutely and locally uniformly convergent. The function $v_0$ is
continuous, satisfies $v_0(t)=v_0(1-t)$, and, uniformly for
$k\in\mathcal K_n$,
\[
 n^2s_{n,k}^2=2\log n+v_0(k/n)+o(1).
\]
\end{proposition}

More specifically, for every fixed trimming fraction there is $C_\eta$
such that, for $1\le M\le n/4$ and all sufficiently large $n$,
\begin{equation}\label{eq:rank-variance-constant-error}
 \sup_{k\in\mathcal K_n}
 \left|n^2s_{n,k}^2-2\log n-v_0(k/n)\right|
 \le C_\eta\left\{
 M^{-1/2}+\frac{1+\log(M+1)}M
 +\frac{M^4}n+\frac{(\log n)^2}n\right\}.
\end{equation}
\begin{proof}
Throughout the proof $t=k/n$, $v=t(1-t)$, and all constants are uniform for $t\in[\eta,1-\eta]$.
First consider the Bernoulli boundary variables. Put
\[
 q_\ell=S_\ell^2/(2\ell v),\qquad e_\ell^\infty=\Phi_\ell-q_\ell,
 \qquad \kappa_4=v(1-6v).
\]
The binomial versions of the signed-moment calculation in
Lemma~\ref{lem:weighted-entropy-projection} give
\[
 \|e_\ell^\infty\|_2\le C_\eta\ell^{-1/2},\qquad
 |\operatorname{Cov}(S_\ell,e_\ell^\infty)|
 +|\operatorname{Cov}(S_\ell^2,e_\ell^\infty)|\le C_\eta .
\]
These are the binomial versions of the signed bounds in
Lemma~\ref{lem:weighted-entropy-projection}. For $\ell\le m$, independent
increments and conditional hypergeometric sampling respectively give
\begin{align*}
 \operatorname{Cov}(q_\ell,q_m)
 &=\frac{\ell}{2m}+\frac{\kappa_4}{4v^2m},\\
 E(S_\ell^2\mid S_m)
 &=\frac{\ell(\ell-1)}{m(m-1)}S_m^2
 +\frac{\ell(m-\ell)(1-2t)}{m(m-1)}S_m
 +\frac{\ell(m-\ell)v}{m-1}.
\end{align*}
For \(m\ge2\), these conditional identities and the signed
bounds imply
\[
 |\operatorname{Cov}(q_\ell,e_m^\infty)|
 +|\operatorname{Cov}(e_\ell^\infty,q_m)|\le C_\eta/m,\qquad
 |\operatorname{Cov}(e_\ell^\infty,e_m^\infty)|
       \le C_\eta/\sqrt{\ell m};
\]
\(m=1\) is immediate. Expanding
\(C_{\ell m}=\operatorname{Cov}(q_\ell+e_\ell^\infty,
q_m+e_m^\infty)\) and using \((\ell+c)^{-1}\le2/\ell\)
gives, for \(\ell\le m\) and \(c\in\{-1/2,1/2\}\),
\[
 \left|\frac{C_{\ell m}(t)}{(\ell+c)(m+c)}-\frac1{2m^2}\right|
 \le C_\eta\left\{\frac1{\ell m^2}
                    +\frac1{\ell^{3/2}m^{3/2}}\right\}.
\]
The shift of the Gaussian weight has the first order because
\[
 \left|\frac1{(\ell+c)(m+c)}-\frac1{\ell m}\right|
 \le C(\ell^{-2}m^{-1}+\ell^{-1}m^{-2}).
\]
Consequently its tail is bounded by
\[
 \sum_{m>M}\sum_{\ell\le m}\frac1{\ell m^2}
       \le C\frac{1+\log(M+1)}M,\qquad
 \sum_{m>M}\sum_{\ell\le m}\frac1{\ell^{3/2}m^{3/2}}
       \le CM^{-1/2}.
\]
Reflection across the diagonal proves
\begin{equation}\label{eq:variance-series-tail}
 \sum_{c\in\{-1/2,1/2\}}\!
 \sum_{\max(\ell,m)>M}
 \left|\frac{C_{\ell m}(t)}{(\ell+c)(m+c)}
             -\frac1{2\max(\ell,m)^2}\right|
 \le C_\eta\left\{M^{-1/2}+\frac{1+\log(M+1)}M\right\}.
\end{equation}
This gives absolute locally uniform convergence, continuity
of the sum, and symmetry by complementing the Bernoulli labels.
We next identify the finite-bridge constant with this series, keeping
an explicit uniform truncation error. Write
\[
 T_n=nF_{n,k}=Q_n^*+R_n^*,\qquad
 Q_n^*=\sum_{\ell=1}^{n-1}a_\ell A_\ell^2,\quad
 R_n^*=\sum_{\ell=1}^{n-1}\omega_{\ell,n} e_\ell,
\]
where $e_\ell$ is the finite-bridge entropy remainder in
Lemma~\ref{lem:weighted-entropy-projection}, and
\[
 a_\ell=\frac{n^2}{2v\ell(n-\ell)d_{\ell,n}},\qquad
 \omega_{\ell,n}=\frac n{d_{\ell,n}},\qquad
 a_\ell^0=\frac{n^2}{2v\ell^2(n-\ell)^2}.
\]
Let \((\mathsf G_1,\ldots,\mathsf G_{n-1})\) be a centered Gaussian vector with the same covariance as
the bridge counts, namely
$\E\mathsf G_\ell\mathsf G_m=nvS_{\ell m}/(n-1)$, and put
\[
 Q_{0,n}^{G}=\sum_{\ell=1}^{n-1}a_\ell^0\mathsf G_\ell^2,\qquad
 V_{0,n}=\operatorname{Var}(Q_{0,n}^{G}).
\]
For any sum restricted to the boundary set
$\mathcal E_M=\{1,\ldots,M\}\cup\{n-M,\ldots,n-1\}$,
use a subscript $M$. Define
\[
 \mathcal H_n=\operatorname{Var}(T_n)-V_{0,n},\qquad
 \mathcal H_{n,M}=\operatorname{Var}(T_{n,M})
                -\operatorname{Var}(Q_{0,n,M}^{G}).
\]
We establish the uniform truncation bound
\begin{equation}\label{eq:finite-variance-truncation}
 |\mathcal H_n-\mathcal H_{n,M}|
 \le C_\eta\left\{M^{-1/2}+\frac{1+\log(M+1)}M
                         +\frac{(\log n)^2}n\right\}.
\end{equation}
To prove the truncation estimate, expand the difference into
four terms:
\[
 \mathcal H_n=\mathcal V_1+\mathcal V_2+\mathcal V_3+\mathcal V_4,
\]
\[
 \begin{aligned}
 \mathcal V_1
 &=2\sum_{\ell,m}(a_\ell a_m-a_\ell^0a_m^0)
                  \operatorname{Cov}(A_\ell,A_m)^2,\\
 \mathcal V_2
 &=\sum_{\ell,m}a_\ell a_m
                  \operatorname{cum}(A_\ell,A_\ell,A_m,A_m),\\
 \mathcal V_3
 &=2\sum_{\ell,m}a_m\omega_{\ell,n}\operatorname{Cov}(A_m^2,e_\ell),
 \qquad \mathcal V_4=\operatorname{Var}(R_n^*).
 \end{aligned}
\]
For each term subtract the same sum restricted to
\(\mathcal E_M^2\).
The relative denominator discrepancy is
\(|a_\ell/a_\ell^0-1|\le C/b_\ell\).
On either common rank half, the unshifted covariance-pair
summand is at most \(C/\max(b_\ell,b_m)^2\). Thus
\[
 |\mathcal V_1-\mathcal V_{1,M}|
 \le C\sum_{\max(a,b)>M}
       \frac1{\min(a,b)\max(a,b)^2}
       +\frac C{n^2}\sum_{\ell,r\le n/2}(\ell^{-1}+r^{-1})
 \le C\frac{1+\log(M+1)}M+\frac{C\log n}{n}.
\]
The last term accounts for opposite halves, where their
unshifted covariance-pair summand is \(O(n^{-2})\).
For \(\mathcal V_2\), use
\eqref{eq:rank-fourth-cumulant-correct},
\(a_\ell\le C/b_\ell^2\), and
\(T_{\ell m}\le b_\ell\wedge b_m\).
The \(\alpha_{n,k}\) term has the same summable majorant.
For the other coefficient,
\[
 \begin{aligned}
 |\varepsilon_{n,k}|
 \sum_{\ell,m}a_\ell a_m
          (S_{\ell\ell}S_{mm}+2S_{\ell m}^2)
 &\le\frac Cn
       \left\{\left(\sum_\ell a_\ell S_{\ell\ell}\right)^2
              +2\sum_{\ell,m}a_\ell a_mS_{\ell m}^2\right\}\\
 &\le C(\log n)^2/n .
 \end{aligned}
\]
Hence
\[
 |\mathcal V_2-\mathcal V_{2,M}|
 \le C\frac{1+\log(M+1)}M+C(\log n)^2/n .
\]
For the mixed term, the conditional count identities in
Proposition~\ref{prop:rank-covariance} imply
\[
 |\operatorname{Cov}(A_m^2,e_\ell)|
 \le C\min(m/\ell,1),\qquad \ell,m\le n/2.
\]
Therefore
\[
 |a_m\omega_{\ell,n}\operatorname{Cov}(A_m^2,e_\ell)|
 \le\frac C{\min(\ell,m)\max(\ell,m)^2}.
\]
The other common half follows by reflection.
On opposite halves its bound is \(C/(n\ell r)\),
where \(r=n-m\). It follows that
\[
 |\mathcal V_3-\mathcal V_{3,M}|
 \le C\frac{1+\log(M+1)}M+\frac Cn
          \left(\sum_{\ell\le n/2}\ell^{-1}\right)^2 .
\]
Finally,
\[
 \|R_n^*\|_2\le C,\qquad
 \|R_n^*-R_{n,M}^*\|_2
       \le C\sum_{b>M}b^{-3/2}\le CM^{-1/2},
\]
and hence
\[
 |\mathcal V_4-\mathcal V_{4,M}|
 \le\|R_n^*-R_{n,M}^*\|_2
       \{\|R_n^*-\E R_n^*\|_2+
          \|R_{n,M}^*-\E R_{n,M}^*\|_2\}
 \le CM^{-1/2}.
\]
Summing the four estimates proves
\eqref{eq:finite-variance-truncation}.
Let \(I_r\) indicate membership of rank \(r\) in the first \(k\)
rows, and put
\[
 \mathbf I_M=(I_1,\ldots,I_M,I_n,\ldots,I_{n-M+1}).
\]
Let \(\{B_r^{(c)}:r\ge1,\ c\in\{-1/2,1/2\}\}\) be independent
Bernoulli\((t)\) variables and let \(\mathbf B_M\) concatenate their
first \(M\) labels in the same order. Define
\[
 d_{n,M}=\sup_{A\subseteq\{0,1\}^{2M}}
       |\Pp(\mathbf I_M\in A)-\Pp(\mathbf B_M\in A)|.
\]
After \(r\) labels are revealed the next success probability differs
from \(t\) by at most \(r/(n-r)\). Sequential maximal coupling gives
\[
 d_{n,M}\le\sum_{r=0}^{2M-1}\frac r{n-r}\le CM^2/n .
\]
For $\ell\le M$, the entropy-table identity and the elementary KL
upper bound imply, deterministically,
\[
 \left|G_{k\ell}-\ell\operatorname{KL}(C_{k\ell}/\ell\Vert t)\right|
 \le C_\eta\ell^2/n.
\]
The analogous bound holds at the upper endpoint in terms of the
number of success labels among the top $\ell$ ranks. Moreover
\[
 \left|\frac n{d_{\ell,n}}-\frac1{\ell-1/2}\right|\le C/n,
 \qquad
 \left|\frac n{d_{n-\ell,n}}-\frac1{\ell+1/2}\right|\le C/n.
\]
Define the independent endpoint functional
\[
 T_M^\infty=\sum_{c\in\{-1/2,1/2\}}\sum_{\ell=1}^M
 \frac{\ell}{\ell+c}
 \operatorname{KL}\!\left(
       \ell^{-1}\sum_{r=1}^{\ell}B_r^{(c)}\,\middle\Vert\,t\right).
\]
Both \(T_{n,M}\) and \(T_M^\infty\) are bounded by \(C_\eta M\).
On the coupled equal-label event their difference is at most
\(C_\eta M^2/n\). Consequently,
\[
 \begin{aligned}
 |\E T_{n,M}-\E T_M^\infty|
     &\le CMd_{n,M}+CM^2/n,\\
 |\E T_{n,M}^2-\E(T_M^\infty)^2|
     &\le CM^2d_{n,M}+CM^3/n\le CM^4/n,\\
 |(\E T_{n,M})^2-(\E T_M^\infty)^2|
     &\le CM|\E T_{n,M}-\E T_M^\infty|\le CM^4/n .
 \end{aligned}
\]
For the Gaussian reference, insert its exact covariance
\(nvS_{\ell m}/(n-1)\) on \(\mathcal E_M^2\); comparison
with the independent endpoint covariances costs at most
\(CM\log(M+1)/n+CM^2/n^2\le CM^4/n\).
Consequently
\begin{equation}\label{eq:finite-boundary-series-comparison}
 \left|\mathcal H_{n,M}
 -\sum_{c\in\{-1/2,1/2\}}\sum_{\ell,m\le M}
 \left\{\frac{C_{\ell m}(t)}{(\ell+c)(m+c)}
                    -\frac1{2\max(\ell,m)^2}\right\}
 \right|\le C_\eta M^4/n.
\end{equation}
The Gaussian variance formula gives the reference term explicitly:
\[
 V_{0,n}=\frac{n^4}{2(n-1)^2}
 \sum_{\ell,m=1}^{n-1}
 \frac1{\{n-\min(\ell,m)\}^2\max(\ell,m)^2}.
\]
Writing its sum as twice the lower triangle minus the diagonal gives
\[
 V_{0,n}=\frac{n^2}{(n-1)^2}(A_n-D_n/2),
\]
where
\[
 A_n=n^2\sum_{m=1}^{n-1}\frac1{m^2}
                   \sum_{j=n-m}^{n-1}\frac1{j^2},\qquad
 D_n=n^2\sum_{m=1}^{n-1}\frac1{m^2(n-m)^2}.
\]
Write \(H_j=\sum_{r\le j}r^{-1}\),
\(H_j^{(2)}=\sum_{r\le j}r^{-2}\), and
\(\mathfrak t_j=\sum_{r\ge j}r^{-2}\). Partial fractions give
\[
 D_n=2H_{n-1}^{(2)}+4H_{n-1}/n,\qquad
 \sum_{a=1}^{j-1}\frac1{a^2(j-a)^2}
   =\frac{2H_{j-1}^{(2)}}{j^2}+\frac{4H_{j-1}}{j^3}.
\]
The triangle is the region \(m,j<n\), \(m+j\ge n\).
Extend this region to positive indices, subtract the two
tails with an index at least \(n\), and add their intersection:
\[
 \begin{aligned}
 A_n
 &=n^2\left\{\sum_{j=n}^\infty
     \left(\frac{2H_{j-1}^{(2)}}{j^2}
                         +\frac{4H_{j-1}}{j^3}\right)
              -2\zeta(2)\mathfrak t_n+\mathfrak t_n^2\right\}\\
 &=n^2\left\{\sum_{j=n}^\infty
       \left(-\frac{2\mathfrak t_j}{j^2}+\frac{4H_{j-1}}{j^3}\right)
              +\mathfrak t_n^2\right\}.
 \end{aligned}
\]
Now
\[
 \mathfrak t_j=j^{-1}+O(j^{-2}),\quad
 H_{j-1}=\log j+\gamma_{\mathrm E}+O(j^{-1}),\quad
 \sum_{j=n}^\infty j^{-3}=\frac1{2n^2}+O(n^{-3}),
\]
and integral comparison gives
\[
 \sum_{j=n}^\infty\frac{\log j}{j^3}
 =\frac{\log n}{2n^2}+\frac1{4n^2}
                     +O(\log n/n^3).
\]
Substitution yields
\[
 A_n=2\log n+2\gamma_{\mathrm E}+1+O(\log n/n),\qquad
 D_n=2\zeta(2)+O(\log n/n).
\]
Thus
\begin{equation}\label{eq:unshifted-gaussian-variance-constant}
 V_{0,n}=2\log n+2\gamma_{\mathrm E}+1-\zeta(2)
                  +O(\log n/n).
\end{equation}
Combining \eqref{eq:variance-series-tail},
\eqref{eq:finite-variance-truncation},
\eqref{eq:finite-boundary-series-comparison}, and
\eqref{eq:unshifted-gaussian-variance-constant} proves
\eqref{eq:rank-variance-constant-error}. Letting first $n\to\infty$
and then $M\to\infty$ proves the uniform expansion. Alternatively
one may choose any $M=M_n\to\infty$ with $M_n^4/n\to0$.
\end{proof}

\subsection{The dense process limit}\label{app:sum-full}

Throughout this subsection put $N=\log n$, $t=k/n$,
$v_t=t(1-t)$, and
\[
 T_j(k)=nF_{n,k,j},\qquad Q_j(k)=nQ_{n,k,j},\qquad
 R_j(k)=T_j(k)-Q_j(k).
\]
Thus $Q_j$ here is $n$ times the quadratic quantity in
Appendix~\ref{app:dense-null}.

All centering constants below are exact.

Define the explicit error quantities
\[
 \epsilon^{\rm app}_{n,p}=\frac{1+\mathfrak r_p}{N},
\]
\[
 \epsilon^{\rm mart}_{n,p}
 =\frac{(1+\mathfrak r_p)^3}{p}
 +\frac{(1+\mathfrak r_p)^2N^2}{n^2}
 +\frac{(1+\mathfrak r_p)^3}{pN^2}
 +\frac{(1+\mathfrak r_p)^2}{n}
 +\frac{(1+\mathfrak r_p)^2}{pN}.
\]
Constants depend only on the fixed trimming fraction and moment
order; process convergence uses the original step interpolation on
\(\mathcal I_\eta\). Let \(\mathscr U\) denote the unordered row vectors.
Exchangeability gives a single permutation, shared by all coordinates,
\[
 \mathbb P(\pi=\sigma\mid\mathscr U)=\frac1{n!},
 \qquad \sigma\in\mathfrak S_n,\qquad
 \pi\perp\!\!\!\perp\mathscr U .
\]
For the pooled ranks \(R_{ij}\), set
\[
 c_{j,\ell,i}=\mathbf1\{R_{ij}\le\ell\}-\ell/n,\qquad
 a_{\ell,n}=\frac{n^2}{d_{\ell,n}\ell(n-\ell)},\qquad
 K_j=\sum_{\ell=1}^{n-1}a_{\ell,n}c_{j,\ell}c_{j,\ell}^{\top}.
\]
If $I_k$ indicates the first $k$ rows of $\pi$, then
\[
 Q_j(k)=\frac{I_k^\top K_jI_k}{2v_t},\qquad K_j\mathbf1=0.
\]
All $K_j$ are permutation conjugates of one positive matrix, and
\begin{equation}\label{eq:sum-full-matrix-bounds}
 \operatorname{tr}K_j=2N+O(1),\quad
 \operatorname{tr}K_j^2=4N+O(1),\quad
 \|K_j\|_{\rm op}\le C,\quad \sum_i(K_j)_{ii}^2\le C.
\end{equation}
The trace identities follow from
Proposition~\ref{prop:rank-covariance}.
For the norm, apply Hardy on the lower half and reflect the upper:
\[
 a_{\ell,n}\le C/\ell^2,\qquad
 \sum_\ell\frac{(\sum_{i\le\ell}z_i)^2}{\ell^2}
          \le4\sum_i z_i^2
 \quad\Longrightarrow\quad z^\top K_jz\le C\|z\|^2 .
\]
At pooled rank \(r\), the diagonal is bounded by
\(C\{r^{-1}+(n-r+1)^{-1}\}\); squaring and summing proves
\(\sum_i(K_j)_{ii}^2\le C\).

For a symmetric matrix $K$ with $K\mathbf1=0$, write $d_i=K_{ii}$ and
$\tau=\operatorname{tr}K$.

Define $A_{ii}=0$ and, for $i\ne h$,
\[
 A_{ih}=K_{ih}+\frac{d_i+d_h}{n-2}
                 -\frac{\tau}{(n-1)(n-2)}.
\]
Since \(K\mathbf1=0\), direct summation gives
\[
 \sum_{h\ne i}A_{ih}
 =-d_i+\frac{(n-2)d_i+\tau}{n-2}
                -\frac{\tau}{n-2}=0.
\]
Put \(U_k(A)=\sum_{a<b\le k}A_{\pi(a),\pi(b)}\),
\(D_k(d)=\sum_{a\le k}d_{\pi(a)}\), and
\(\mathcal F_k=\sigma(\mathscr U,\pi(1),\ldots,\pi(k))\).
Define, for \(0\le k\le n-2\),
\[
 M_k^{(2)}=\frac{U_k(A)}{(n-k)(n-k-1)},\qquad
 M_k^{(1)}=\frac{D_k(d)-k\tau/n}{n-k}.
\]
Substitution of \(A\) into the quadratic form and conditional summation give
\begin{align}
 I_k^\top KI_k-\E_\pi(I_k^\top KI_k)
 &=2U_k(A)+\frac{n-2k}{n-2}\{D_k(d)-k\tau/n\},
 \label{eq:sum-full-harmonic-decomposition}\\
 \E_\pi(M_{k+1}^{(r)}\mid\mathcal F_k)
 &=M_k^{(r)},\qquad r\in\{1,2\},\quad 0\le k\le n-3.
 \label{eq:sum-full-permutation-martingales}
\end{align}
For the first process, zero row sums imply
\[
 \E_\pi(U_{k+1}-U_k\mid\mathcal F_k)
 =\frac1{n-k}\sum_{a\le k}\sum_{i\notin\pi[1:k]}A_{\pi(a),i}
 =-\frac{2U_k}{n-k}.
\]
For the second,
\[
 \E_\pi\{d_{\pi(k+1)}-\tau/n\mid\mathcal F_k\}
 =-\frac{D_k-k\tau/n}{n-k}.
\]
These recursions verify
\eqref{eq:sum-full-permutation-martingales}.

For zero-diagonal matrices $A,B$ with zero row sums,
\begin{align}
 \E_\pi\{U_k(A)U_k(B)\}
 &=\frac{k(k-1)(n-k)(n-k-1)}
 {2n(n-1)(n-2)(n-3)}\langle A,B\rangle_F,
 \label{eq:sum-full-slice-covariance}\\
 \langle A(K),A(H)\rangle_F
 &=\operatorname{tr}(KH)-\frac n{n-2}\sum_iK_{ii}H_{ii}
  +\frac{\operatorname{tr}K\operatorname{tr}H}{(n-1)(n-2)}.
 \notag
\end{align}
In the edge expansion, equal edges, pairs sharing one
endpoint, and disjoint pairs have summed weights in ratio
\(1:-2:1\). Their inclusion probabilities are
\((k)_r/(n)_r\), \(r=2,3,4\), so
\[
 \E_\pi\{U_k(A)U_k(B)\}
 =\frac{\langle A,B\rangle_F}{2}
  \left\{\frac{(k)_2}{(n)_2}
            -2\frac{(k)_3}{(n)_3}
            +\frac{(k)_4}{(n)_4}\right\}.
\]
Factoring the expression in braces gives
\eqref{eq:sum-full-slice-covariance}.
The same zero-row-sum expansion makes \(U_k\) orthogonal
to the degree-one term in
\eqref{eq:sum-full-harmonic-decomposition}.

We also use the following fixed-degree consequence of balanced-slice
hypercontractivity: a harmonic multilinear polynomial $P$ of degree at
most two satisfies
\begin{equation}\label{eq:sum-full-slice-moments}
 \|P\|_{L^r(\pi)}\le C_{\eta,r}\|P\|_{L^2(\pi)},\qquad
 \eta n\le k\le(1-\eta)n,\quad r\ge2.
\end{equation}
Proposition~5.1 and Lemma~5.2 of \citet{FilmusMossel2019}
apply to the two polynomials
\[
 P_1(x)=\sum_i(d_i-\tau/n)x_i,\qquad
 P_2(x)=\sum_{i<h}A_{ih}x_ix_h .
\]
Indeed, their degree, harmonicity, and slice proportion satisfy
\[
 \begin{aligned}
 \deg P_r&\le2,\qquad k/n\in[\eta,1-\eta],\\
 \sum_i\partial_iP_1&=\sum_i(d_i-\tau/n)=0,\\
 \sum_i\partial_iP_2&=\sum_hx_h\sum_{i\ne h}A_{ih}=0 .
 \end{aligned}
\]
For the entropy remainder,
\[
 \|R_j(k)\|_r
 \le C_r\sum_\ell\frac n{d_{\ell,n}}\,b_\ell^{-1/2}
 \le C_r\sum_\ell b_\ell^{-3/2}\le C_r.
\]
Together with slice hypercontractivity,
\(\operatorname{Var}(Q_j)=O(N)\), and exact studentization,
this gives
\begin{equation}\label{eq:sum-full-coordinate-moments}
 \sup_{n,j,k\in\mathcal K_n}\|D_{j,n}(k)\|_r\le C_{\eta,r}.
\end{equation}
\begin{lemma}\label{lem:sum-full-entropy-process}
Under Assumption~\ref{ass:copula-null}\ref{ass:copula-pairwise},
\[
 \E\max_{k\in\mathcal K_n}
 \left|p^{-1/2}\sum_j\{R_j(k)-\E R_j(k)\}\right|^2
 \le C(1+\mathfrak r_p).
\]
\end{lemma}
\begin{proof}
For $\ell\le n/2$ write $A=C-t\ell$ and
\[
 e_\ell(t,A)=
 \ell\operatorname{KL}(t+A/\ell\Vert t)
 +(n-\ell)\operatorname{KL}(t-A/(n-\ell)\Vert t)
 -\frac{A^2}{2v_t}\left(\frac1\ell+\frac1{n-\ell}\right).
\]
On $|A|\le c_\eta\ell$, Taylor expansion, with $A$ held fixed when
differentiating in $t$, gives
\[
 |e_\ell|\le C|A|^3/\ell^2,\quad
 |e_{\ell,A}|\le CA^2/\ell^2,\quad
 |e_{\ell,AA}|\le C|A|/\ell^2,\quad
 |e_{\ell,AAA}|\le C/\ell^2,\quad
 |e_{\ell,t}|\le C|A|^3/\ell^2.
\]
The corresponding mixed derivatives satisfy the same powers of $\ell$.
The next inclusion $\xi$ is Bernoulli with probability
$(\ell-C)/(n-k)$, so
\[
 \Delta A=\xi-\ell/n,\quad
 \E(\Delta A\mid\mathcal F_k)=-\frac{A}{n(1-t)},\quad
 \E\{(\Delta A)^2+|\Delta A|^3\mid\mathcal F_k\}\le C\ell/n.
\]
Choose fixed \(c_\eta^*>0\) and \(\ell_0\) so that
\(\ell\ge\ell_0\), \(|A|\le c_\eta^*\ell\) keep both next states
in the Taylor chart. Define
\[
 \mathcal B_\ell=\{\ell<\ell_0\}\cup\{|A|>c_\eta^*\ell\},
 \qquad
 \Delta e_\ell=e_\ell(t+n^{-1},A+\Delta A)-e_\ell(t,A).
\]
On \(\mathcal B_\ell^c\), conditional Taylor expansion gives
\[
 n|\E(\Delta e_\ell\mid\mathcal F_k)|
 \le C\left\{\frac{|A|^3}{\ell^2}
              +\frac{|A|}{\ell}+\frac1\ell\right\},
 \qquad
 \|n\E(\Delta e_\ell\mid\mathcal F_k)\|_2
          \le C\ell^{-1/2}.
\]
The event \(\mathcal B_\ell\) is \(\mathcal F_k\)-measurable.
Let \(\Delta_a e_\ell\) denote the increment with
inclusion label \(a\in\{0,1\}\). For \(\ell\le\eta n/4\),
\[
 |\Delta_0e_\ell|\le C\ell/n,\qquad
 |\Delta_1e_\ell|\le C\log(\ell+1),\qquad
 \Pp(\xi=1\mid\mathcal F_k)\le C\ell/n.
\]
Here the first bound differentiates the compact complementary
logit; the second uses the finite differences of \(x\log x\).
Hence
\[
 n|\E(\Delta e_\ell\mid\mathcal F_k)|\mathbf1_{\mathcal B_\ell}
 \le C\{\ell+\ell\log(\ell+1)\}\mathbf1_{\mathcal B_\ell}
 \le C\ell^2\mathbf1_{\mathcal B_\ell}.
\]
For \(\ell>\eta n/4\), the same bound follows from
\(n\le C\ell\) and \(|e_\ell|\le C\ell\). The exponential
count bound therefore gives
\[
 \|n\E(\Delta e_\ell\mid\mathcal F_k)
                         \mathbf1_{\mathcal B_\ell}\|_2
 \le C\ell^2\Pp(\mathcal B_\ell)^{1/2}
 \le C\ell^2e^{-c\ell}\le C\ell^{-1/2}.
\]
Reflection gives
\[
 \|e_\ell(k)\|_2+
 \|n\E(\Delta e_\ell(k)\mid\mathcal F_k)\|_2
          \le Cb_\ell^{-1/2}.
\]
After summation with weights \(n/d_{\ell,n}\),
\[
 \|R_j(k)\|_2\le C,\qquad
 \|\E(\Delta R_j(k)\mid\mathcal F_k)\|_2\le C/n .
\]
These predictable quantities are functions of their own
Gaussian column. For centered column functions, Gaussian
chaos expansion at cross covariance \(\rho_{j\ell}I_n\) gives
\[
 |\E fg|\le|\rho_{j\ell}|\|f\|_2\|g\|_2,\qquad
 \left\|p^{-1/2}\sum_j f_j\right\|_2^2
       \le(1+\mathfrak r_p)\max_j\|f_j\|_2^2.
\]
Put \(k_0=\min\mathcal K_n\), \(k_1=\max\mathcal K_n\), and
let \(X_k=p^{-1/2}\sum_j(R_j(k)-\E R_j(k))\), and write
\(d_k=\E(X_{k+1}-X_k\mid\mathcal F_k)\).
Then
\[
 \|X_k\|_2\le C\sqrt{1+\mathfrak r_p},\qquad
 \|d_k\|_2\le C\sqrt{1+\mathfrak r_p}/n.
\]
For the martingale
\(M_k=X_k-X_{k_0}-\sum_{h=k_0}^{k-1}d_h\),
Minkowski gives
\(\|M_{k_1}\|_2\le C\sqrt{1+\mathfrak r_p}\) at the
last trimmed time. Doob and the same drift bound yield
\[
 \|\max_{k_0\le k\le k_1}|X_k|\|_2
 \le\|X_{k_0}\|_2+2\|M_{k_1}\|_2
                  +\sum_{h=k_0}^{k_1-1}\|d_h\|_2
 \le C\sqrt{1+\mathfrak r_p}.
\]
\end{proof}

For \(t\in\mathcal I_\eta\), put
\(\bar t_n(t)=k_n(t)/n\), and define
\[
 a_i(t)=
 \frac{\mathbf1\{i\le k_n(t)\}-\bar t_n(t)}
           {\sqrt{n\bar t_n(t)\{1-\bar t_n(t)\}}}.
\]
For \(0<a<1/2\), define
\begin{equation*}
 \chi_a(y,z)=\int_a^{1-a}
 \frac{\{\mathbf1(y\le\Phi^{-1}u)-u\}
       \{\mathbf1(z\le\Phi^{-1}u)-u\}}{u^2(1-u)^2}\,du.
\end{equation*}
This is a positive, canonical kernel in independent observation rows.

\Needspace{5\baselineskip}
\begin{lemma}\label{lem:sum-full-population-comparison}
With
$Z_{j,a}(k)=\sum_{i<h}a_i(t)a_h(t)\chi_a(Y_{ij},Y_{hj})$,
\[
 \E\max_{k\in\mathcal K_n}
 \left|p^{-1/2}\sum_j
 \{T_j(k)-\E T_j(k)-Z_{j,1/n}(k)\}\right|^2
 \le C(1+\mathfrak r_p).
\]
At each fixed split the individual centered errors in the
entropy, rank-to-population, and population-diagonal comparisons
have uniformly bounded $L^4$ norms.
\end{lemma}
\begin{proof}
Put $P=I-\mathbf1\mathbf1^\top/n$ and, within one coordinate,
\[
 K^{\rm pop}=\frac1nP[\chi_{1/n}(Y_i,Y_h)]_{i,h}P,\qquad
 Q^{\rm pop}(k)=\frac{I_k^\top K^{\rm pop}I_k}{2v_t}
              =\frac12\sum_{i,h}a_i(t)a_h(t)\chi_{1/n}(Y_i,Y_h).
\]
We first prove
\begin{equation}\label{eq:sum-full-order-matrix-error}
 \big\|\|K^{\rm pop}-K\|_F\big\|_4\le C,\qquad
 \|\operatorname{tr}(K^{\rm pop}-K)\|_4\le C.
\end{equation}
Let $U_{(r)}$ be the uniform order statistics.
On the lower half of
the threshold integral, the uncentered population matrix entry whose
maximal rank is $r$ is
\[
 f_n(U_{(r)}),\qquad
 f_n(u)=\frac1n\int_{\max(u,1/n)}^{1/2}
                  \frac{ds}{s^2(1-s)^2},
\]
with value zero for $u>1/2$.
Its rank counterpart is
\(F_n(r)=\sum_{r\le\ell\le\lfloor n/2\rfloor}a_{\ell,n}\).
In rank coordinates define the lower-half matrix discrepancy
\[
 (\Delta K^-)_{ih}
 =f_n(U_{(\max(i,h))})-F_n(\max(i,h)).
\]
Cell integration and the derivative of \(f_n\) give
\[
 |F_n(r)-f_n(r/n)|\le C/r^2,\qquad
 |f_n(u)-f_n(v)|
 \le\frac Cn|\min(u^{-1},n)-\min(v^{-1},n)|.
\]
For $r>m$ the exact beta moments are
\[
 \E U_{(r)}^{-m}=
 \frac{n(n-1)\cdots(n-m+1)}{(r-1)(r-2)\cdots(r-m)}.
\]
Expanding the second and fourth centered powers, for $r\ge5$, gives
\[
 \E|U_{(r)}^{-1}-n/r|^2\le Cn^2/r^3,\qquad
 \E|U_{(r)}^{-1}-n/r|^4\le Cn^4/r^6.
\]
For the first four ranks use the bounded clipped function.
Thus, uniformly in \(r\),
\[
 \|f_n(U_{(r)})-F_n(r)\|_4\le Cr^{-3/2}.
\]
There are \(2r-1\) matrix entries with maximal rank \(r\).
Minkowski applied to their squares gives
\[
 \left\|\|\Delta K^-\|_F\right\|_4^2
 =\left\|\sum_{i,h}|(\Delta K^-)_{ih}|^2\right\|_2
 \le\sum_r(2r-1)
          \|f_n(U_{(r)})-F_n(r)\|_4^2
 \le C\sum_r r^{-2}\le C .
\]
Projection by \(P\) is contractive. Reflecting the upper
threshold half preserves the rank count in this sum.
For the trace write $N_n(u)=\sum_i\mathbf1(U_i\le u)$,
$B_n(u)=N_n(u)-nu$, and $w(u)=u(1-u)$.
Exactly,
\[
 \operatorname{tr}K^{\rm pop}
 =\int_{1/n}^{1-1/n}\frac{du}{w(u)}
  +\frac1n\int_{1/n}^{1-1/n}\frac{(1-2u)B_n(u)}{w(u)^2}\,du
  -\frac1{n^2}\int_{1/n}^{1-1/n}\frac{B_n(u)^2}{w(u)^2}\,du.
\]
Define
\[
 \xi_n(v)=\int_{1/n}^{1-1/n}
       \frac{(1-2u)\{\mathbf1(v\le u)-u\}}{u^2(1-u)^2}\,du.
\]
The linear term is \(n^{-1}\sum_i\xi_n(U_i)\). Its independent
centered summands satisfy
\[
 |\xi_n(u)|\le C\{(u\vee n^{-1})^{-1}
       +((1-u)\vee n^{-1})^{-1}+\log n\},\qquad
 \E\xi_n^2\le Cn,\quad\E\xi_n^4\le Cn^3.
\]
Expansion of the independent fourth moment gives
\[
 \E\left(n^{-1}\sum_i\xi_n(U_i)\right)^4
 =n^{-4}\{n\E\xi_n^4+3n(n-1)(\E\xi_n^2)^2\}\le C.
\]
For the final quadratic term,
\[
 \left\|\frac1{n^2}\int_{1/n}^{1-1/n}
                  \frac{B_n(u)^2}{w(u)^2}\,du\right\|_4
 \le\frac1{n^2}\int_{1/n}^{1-1/n}
                  \frac{\|B_n(u)\|_8^2}{w(u)^2}\,du
 \le C\log n/n .
\]
Indeed, \(\|B_n(u)\|_8^2\le Cnw(u)\), and
\eqref{eq:sum-full-matrix-bounds} gives
\[
 \left|\int_{1/n}^{1-1/n}\frac{du}{w(u)}
                  -\operatorname{tr}K\right|\le C,\qquad
 \|\operatorname{tr}(K^{\rm pop}-K)\|_4
 \le C+C+C\log n/n\le C.
\]
Put \(\Delta K=K^{\rm pop}-K\), \(d_i=(\Delta K)_{ii}\),
and \(\tau=\operatorname{tr}\Delta K\).
Equations~\eqref{eq:sum-full-harmonic-decomposition},
\eqref{eq:sum-full-slice-moments}, and
\eqref{eq:sum-full-order-matrix-error} imply
\[
 \begin{aligned}
 &\|U_k(A(\Delta K))\|_4+
                  \|D_k(d)-k\tau/n\|_4\\
 &\qquad\le C\{\|\|\Delta K\|_F\|_4+\|\tau\|_4\}\le C,\\
 &\left\|\frac{I_k^\top\Delta K I_k
                   -\E_\pi(I_k^\top\Delta K I_k)}{2v_t}\right\|_4
 \le C .
 \end{aligned}
\]
The uncentered quadratic-form difference has conditional mean
\[
 \E_\pi\{Q^{\rm pop}(k)-Q(k)\}
 =\E_\pi\frac{I_k^\top\Delta K I_k}{2v_t}
 =\frac{n\,\operatorname{tr}(K^{\rm pop}-K)}{2(n-1)}.
\]
It is independent of \(k\) and has bounded \(L^4\) norm.
For the aggregate harmonic terms define
\[
 H_{2,k}=p^{-1/2}\sum_jU_k(A(\Delta K_j)),\qquad
 H_{1,k}=p^{-1/2}\sum_j\{D_k(d_j)-k\tau_j/n\}.
\]
Maximal correlation and the preceding coordinate bounds yield
\(\E|H_{r,k}|^2\le C(1+\mathfrak r_p)\), \(r=1,2\).
With \(k_+=\lfloor(1-\eta)n\rfloor\), the martingales in
\eqref{eq:sum-full-permutation-martingales} give
\[
 \begin{aligned}
 M_{2,k}&=\frac{n^2H_{2,k}}{(n-k)(n-k-1)},&
 M_{1,k}&=\frac{nH_{1,k}}{n-k},\\
 \E\sup_{k\in\cK_n}|H_{r,k}|^2
 &\le C\E\max_{k\le k_+}|M_{r,k}|^2
 \le 4C\E|M_{r,k_+}|^2
 \le C(1+\mathfrak r_p).
 \end{aligned}
\]
The centered aggregate conditional mean obeys the same fixed-time
bound and is constant along the scan. For the population diagonal,
direct integration gives
\[
 \chi_{1/n}(y,y)\le
 C\{(\Phi(y)\vee n^{-1})^{-1}
       +((1-\Phi(y))\vee n^{-1})^{-1}\},
\quad
 \E\chi_{1/n}(Y,Y)^2\le Cn,\quad
 \E\chi_{1/n}(Y,Y)^4\le Cn^3.
\]
For $X_i=\chi_{1/n}(Y_i,Y_i)-\E\chi_{1/n}(Y_i,Y_i)$ and \(t=k/n\),
\[
 \frac12\sum_{i=1}^n a_i(t)^2X_i
 =\frac{(1-2t)\sum_{i\le k}X_i+t^2\sum_{i\le n}X_i}{2nv_t}.
\]
Independent fourth-moment expansion bounds this expression
in \(L^4\). For the aggregate let
\(X_i^\Sigma=p^{-1/2}\sum_jX_{ij}\). Row independence and
maximal correlation give
\[
 \operatorname{Var}(X_i^\Sigma)\le Cn(1+\mathfrak r_p),\qquad
 \E\max_{k\le n}\left|\sum_{i\le k}X_i^\Sigma\right|^2
 \le4\sum_{i\le n}\operatorname{Var}(X_i^\Sigma)
 \le Cn^2(1+\mathfrak r_p).
\]
Division by the squared \(n\)-scale in the displayed
diagonal formula proves its aggregate process bound.
Combining this with
Lemma~\ref{lem:sum-full-entropy-process} proves the lemma.
\end{proof}

Let $Y,Y'$ be independent rows with distribution $N_p(0,R_p)$, and set
\[
 h_n(y,z)=\sum_{j=1}^p\chi_{1/n}(y_j,z_j),\qquad
 g_n(y)=\E h_n(y,Y')^2.
\]
The integral operator $\mathscr T_n$ acts on centered
$L^2\{N_p(0,R_p)\}$ by
$(\mathscr T_nf)(y)=\E\{h_n(y,Y')f(Y')\}$.

\begin{lemma}\label{lem:sum-full-row-kernel-bounds}
Under Assumption~\ref{ass:copula-null}\ref{ass:copula-pairwise}--%
\ref{ass:copula-row},
\begin{align}
 \|\mathscr T_n\|_{\rm op}&\le C(1+\mathfrak r_p),&
 \operatorname{tr}\mathscr T_n^2
 &=4pN\{1+O(\epsilon^{\rm app}_{n,p})\},
 \label{eq:sum-full-operator-bounds}\\
 \frac{\E h_n(Y,Y')^4}{n^2(pN)^2}
 &\le C\left\{\frac{(1+\mathfrak r_p)^3}{p}
             +\frac{(1+\mathfrak r_p)^2N^2}{n^2}\right\},&
 \frac{\E g_n(Y)^2}{n(pN)^2}
 &\le C\left\{\frac{(1+\mathfrak r_p)^3}{pN^2}
             +\frac{(1+\mathfrak r_p)^2}{n}\right\},
 \notag\\
 \frac{\operatorname{tr}\mathscr T_n^4}{(pN)^2}
 &\le C\frac{(1+\mathfrak r_p)^2}{pN}.
 \label{eq:sum-full-four-cycle}
\end{align}
\end{lemma}
\begin{proof}
For a centered univariate $f$, put
$F(u)=\int_0^u f(\Phi^{-1}s)\,ds$.
The single-coordinate operator has quadratic form
$\int_{1/n}^{1-1/n}F(u)^2/w(u)^2\,du$.
Hardy's inequality on each half bounds it by $C\|f\|_2^2$.
For centered coordinate functions, maximal correlation gives
\[
 \E\left(\sum_jf_j(Y_j)\right)^2
 \le\sum_j\|f_j\|_2^2+
       \sum_{j\ne l}|\rho_{jl}|\|f_j\|_2\|f_l\|_2
 \le(1+\mathfrak r_p)\sum_j\|f_j\|_2^2.
\]
Factoring through the direct sum of coordinate spaces proves the
operator bound, including singular $R_p$.
For independent standard Gaussian variables \(Y,Y'\),
\[
 \E\{\chi_{1/n}(Y,Y')^2\}
 =\int_{1/n}^{1-1/n}\int_{1/n}^{1-1/n}
 \frac{\{\min(u,v)-uv\}^2}{w(u)^2w(v)^2}\,du\,dv
 =4N+O(1).
\]
For a centered Gaussian pair \((Z,Z')\) with unit variances
and correlation \(\rho\), define
\[
 C_\rho(u,v)=\Pp\{Z\le\Phi^{-1}(u),\,Z'\le\Phi^{-1}(v)\}-uv.
\] Choose
\(\rho_\star<\rho_1<1\) and \(2<q<1+1/\rho_1\).
For \(R_\rho=\left(\begin{smallmatrix}1&\rho\\\rho&1\end{smallmatrix}\right)\),
write \(\E_0\) for expectation under \(N(0,I_2)\). Its density
relative to that law is
\[
 L_\rho(z)=(1-\rho^2)^{-1/2}
       e^{-z^\top(R_\rho^{-1}-I_2)z/2}.
\]
Gaussian integration gives
\[
 \begin{aligned}
 \E_0 L_\rho^q
 &=(1-\rho^2)^{-q/2}
       \det\{qR_\rho^{-1}+(1-q)I_2\}^{-1/2}\\
 &=\frac{(1-\rho^2)^{-(q-1)/2}}
          {\sqrt{1-(q-1)^2\rho^2}}
 \le C_{\rho_1,q},\qquad |\rho|\le\rho_1 .
 \end{aligned}
\]
Set \(a=1-1/q>1/2\), \(u_*=u\wedge(1-u)\), and
\(v_*=v\wedge(1-v)\). Reflection and H\"older's inequality yield
\[
 |C_\rho(u,v)|\le C(u_*v_*)^a,\qquad
 \int_0^1\!\!\int_0^1
       \frac{C_{\rho_1}(u,v)^2}{w(u)^2w(v)^2}\,du\,dv
 \le C\left(\int_0^{1/2}u^{2a-2}\,du\right)^2<\infty.
\]
The two-row kernel is centered in each row. Its Hermite expansion
therefore has no terms of total degree below two; with
\(c_r\) the squared norm of its degree-\(r\) projection,
\[
 c_r\ge0,\qquad
 \E\{\chi_{1/n}(Z_1,Z_2)\chi_{1/n}(Z'_1,Z'_2)\}
       =\sum_{r\ge2}c_r\rho^r,\qquad
 \sum_{r\ge2}c_r\rho_1^r\le C,
\]
where the two Gaussian row pairs have cross covariance
\(\rho I_2\). Thus
\[
 \sum_{r\ge2}c_r|\rho_{j\ell}|^r
 \le\frac{\rho_{j\ell}^2}{\rho_1^2}
                     \sum_{r\ge2}c_r\rho_1^r
 \le C\rho_{j\ell}^2 .
\]
Consequently
\[
 \operatorname{tr}\mathscr T_n^2
 =4pN+O\!\left(p+\sum_{j\ne\ell}\rho_{j\ell}^2\right)
 =4pN+O\{p(1+\mathfrak r_p)\}.
\]
For the fourth moment, the almost-everywhere derivative is
\[
 \partial_y\chi_{1/n}(y,z)=-
 \frac{\phi(y)\{\mathbf1(z\le y)-\Phi(y)\}}
 {\Phi(y)^2\{1-\Phi(y)\}^2}
 \mathbf1\{n^{-1}<\Phi(y)<1-n^{-1}\}.
\]
Let $Z_j=|\partial_y\chi_{1/n}(Y_j,Y'_j)|^2$.
The Gaussian density bound
$\phi(\Phi^{-1}u)\le Cu\sqrt{\log(1/u)}$ and the binary centered
second and fourth moments yield
\begin{align*}
 \E Z_j
 &\le C\int_{1/n}^{1/2}\frac{\log(1/u)}u\,du\le CN^2,\\
 \E Z_j^2
 &\le C\int_{1/n}^{1/2}\frac{\log^2(1/u)}{u^3}\,du
 \le Cn^2N^2.
\end{align*}
The two-dimensional blocks $(Y_j,Y'_j)$ and $(Y_l,Y'_l)$ have
identity marginal covariances and cross covariance $\rho_{jl}I_2$.
Their maximal correlation is $|\rho_{jl}|$.
Consequently
\[
 |\operatorname{Cov}(Z_j,Z_l)|
 \le|\rho_{jl}|\sqrt{\operatorname{Var}(Z_j)
                         \operatorname{Var}(Z_l)}
 \le C|\rho_{jl}|n^2N^2.
\]
Expand the square before summing:
\begin{align*}
 \E\left(\sum_jZ_j\right)^2
 &=\sum_j\operatorname{Var}(Z_j)
   +\sum_{j\ne l}\operatorname{Cov}(Z_j,Z_l)
   +\left(\sum_j\E Z_j\right)^2\\
 &\le C\{p(1+\mathfrak r_p)n^2N^2+p^2N^4\}.
\end{align*}
The other-row derivatives have the same bound, and
Cauchy--Schwarz bounds the product of the two gradient sums.
For standard Gaussian measure, the Poincar\'e inequality
\(\operatorname{Var} g\le\E|\nabla g|^2\), as stated on p.~446 of
\citet{Ledoux1992}, applied to centered \(f\) and to \(f^2\)
gives
\[
 \begin{aligned}
 \E f^4
 &=\operatorname{Var}(f^2)+(\E f^2)^2\\
 &\le4\E(f^2|\nabla f|^2)+(\E|\nabla f|^2)^2\\
 &\le4(\E f^4)^{1/2}(\E|\nabla f|^4)^{1/2}
                              +\E|\nabla f|^4 .
 \end{aligned}
\]
Solving this quadratic inequality gives
\(\|f\|_4\le C\||\nabla f|\|_4\).
For the Gaussian rows with covariance \(R_p\), the linear
change of variables adds
\(\|R_p\|_{\rm op}^{1/2}\le\sqrt{1+\mathfrak r_p}\).
Thus
\[
 \E h_n^4
 \le C(1+\mathfrak r_p)^2
       \{p(1+\mathfrak r_p)n^2N^2+p^2N^4\}.
\]
The displayed weak derivatives belong to \(L^4\) for fixed \(n,p\).
Smooth approximation in Gaussian \(W^{1,4}\), followed by the
preceding inequality, gives
\[
 \begin{aligned}
 \|h_n\|_4
 &\le C\sqrt{1+\mathfrak r_p}\,\||\nabla h_n|\|_4,\\
 \frac{\E h_n^4}{n^2(pN)^2}
 &\le C\left\{\frac{(1+\mathfrak r_p)^3}{p}
       +\frac{(1+\mathfrak r_p)^2N^2}{n^2}\right\}.
 \end{aligned}
\]
For \(d_j(Y)=\chi_{1/n}(Y_j,Y_j)\), the diagonal bounds in
Lemma~\ref{lem:sum-full-population-comparison} give
\[
 \E d_j\le CN,\qquad
 \operatorname{Var}(d_j)\le Cn,\qquad
 |\operatorname{Cov}(d_j,d_l)|
       \le C|\rho_{jl}|n.
\]
Consequently
\[
 \E h_n(Y,Y)^2
 \le C\{p(1+\mathfrak r_p)n+p^2N^2\}.
\]
In the Hilbert space
\(\bigoplus_{j=1}^p L^2([1/n,1-1/n],du/[u^2(1-u)^2])\), put
\[
 \Phi_y=(\mathbf1\{y_j\le\Phi^{-1}(u)\}-u)_{j,u},
 \qquad h_n(y,z)=\langle\Phi_y,\Phi_z\rangle.
\]
Let
\(C_\Phi=\E(\Phi_Y\otimes\Phi_Y)\). Then
\[
 g_n(y)=\langle\Phi_y,C_\Phi\Phi_y\rangle
 \le\|C_\Phi\|_{\rm op}\|\Phi_y\|^2
 =\|\mathscr T_n\|_{\rm op}h_n(y,y).
\]
It follows that
\[
 \E g_n^2
 \le C(1+\mathfrak r_p)^2
             \{p(1+\mathfrak r_p)n+p^2N^2\},
\]
which gives the second bound after division. Finally,
\[
 \operatorname{tr}\mathscr T_n^4
 \le\|\mathscr T_n\|_{\rm op}^2\operatorname{tr}\mathscr T_n^2
 \le C(1+\mathfrak r_p)^2pN,
\]
proving \eqref{eq:sum-full-four-cycle}.
\end{proof}
For \(t\in\mathcal I_\eta\), all time-indexed quadratics mean their
values at \(k_n(t)\). Let \(Y_i,Z_i\) be independent standard normals,
independent across \(i\), and put
\(Y_i^{(\rho)}=\rho Y_i+\sqrt{1-\rho^2}Z_i\).
A superscript \((\rho)\) denotes evaluation on this column.
For \(|\rho|\le\rho_0<1\), decompose the centered population quadratic as
\[
 \begin{aligned}
 Q^{\rm pop}(t)-\E Q^{\rm pop}(t)&=O_t+V_t,\\
 O_t&=\sum_{i<h}a_i(t)a_h(t)\chi_{1/n}(Y_i,Y_h),\\
 V_t&=\tfrac12\sum_i a_i(t)^2
             \{\chi_{1/n}(Y_i,Y_i)-\E\chi_{1/n}(Y_i,Y_i)\}.
 \end{aligned}
\]
For another trimmed time \(s\), row degeneracy,
\(\sum_i a_i(t)^2=1\), and the copula integral just bounded give
\[
 \begin{aligned}
 \operatorname{Cov}(O_t,V_s^{(\rho)})&=
              \operatorname{Cov}(V_t,O_s^{(\rho)})=0,\\
 |\operatorname{Cov}(O_t,O_s^{(\rho)})|
 &\le\left|\sum_{i<h}a_i(t)a_h(t)a_i(s)a_h(s)\right|
        \int_{1/n}^{1-1/n}\!\!\int_{1/n}^{1-1/n}
           \frac{C_\rho(u,v)^2}{w(u)^2w(v)^2}\,du\,dv\le C,\\
 |\operatorname{Cov}(V_t,V_s^{(\rho)})|
 &\le\|V_t\|_2\|V_s\|_2\le C .
 \end{aligned}
\]
Define \(P_\rho f(y)=\E_Z f(\rho y+\sqrt{1-\rho^2}Z)\),
where \(Z\sim N(0,I_n)\). Gaussian conditioning gives
\[
 \|P_\rho(Q^{\rm pop}-\E Q^{\rm pop})\|_2^2
 =\operatorname{Cov}\{Q^{\rm pop}(Y),Q^{\rm pop}(Y^{(\rho^2)})\}\le C .
\]
For the centered error \(E_t=Q(t)-Q^{\rm pop}(t)
-\E\{Q(t)-Q^{\rm pop}(t)\}\), the approximation bound gives
\[
 \begin{aligned}
 |\operatorname{Cov}(E_t,Q^{\rm pop}(s)^{(\rho)})|
 &\le\|E_t\|_2
       \|P_\rho\{Q^{\rm pop}(s)-\E Q^{\rm pop}(s)\}\|_2\le C,\\
 |\operatorname{Cov}(E_t,E_s^{(\rho)})|&\le C .
 \end{aligned}
\]
The same bounds apply to the entropy remainder. If \(F_{t,r}\)
is the degree-\(r\) Hermite projection of a centered rank
quadratic or entropy statistic, comparison at fixed
\(\rho_1\in(\rho_\star,1)\) yields
\[
 \begin{aligned}
 \sum_{r\ge1}\rho_1^r\|F_{t,r}\|_2^2&\le C,\\
 \left|\sum_{r\ge1}\rho^r
                   \langle F_{t,r},F_{s,r}\rangle\right|
 &\le\frac{|\rho|}{\rho_1}
       \left(\sum_{r\ge1}\rho_1^r\|F_{t,r}\|_2^2\right)^{1/2}
       \left(\sum_{r\ge1}\rho_1^r\|F_{s,r}\|_2^2\right)^{1/2}
 \le C|\rho|.
 \end{aligned}
\]
Thus, at any two trimmed times,
\[
 |\operatorname{Cov}\{Q_j(k),Q_\ell(h)\}|
 +|\operatorname{Cov}\{T_j(k),T_\ell(h)\}|
 \le C|\rho_{j\ell}|,\qquad j\ne\ell .
\]
The entropy term is handled by its same bounded-\(L^2\)
remainder and the identical smoothing argument.

Within a coordinate,
\eqref{eq:sum-full-permutation-martingales} and
\eqref{eq:sum-full-slice-covariance} give
$\operatorname{Cov}\{Q_j(k),Q_j(h)\}
=2N\rho_{\rm B}(k/n,h/n)^2+O(1)$.

In particular, for $A_\Sigma=p^{-1/2}\sum_jA(K_j)$,
\[
 \E\|A_\Sigma\|_F^2\le C(N+\mathfrak r_p).
\]
The diagonal contribution in
\eqref{eq:sum-full-harmonic-decomposition}, divided by $\sqrt{2N}$,
has supremum $L^2$ norm at most
$C\sqrt{(1+\mathfrak r_p)/N}$.

The entropy error has the same bound.

Define the exact martingale
\[
 \mathscr M_n(k)=
 \frac{n^2p^{-1/2}\sum_jU_k\{A(K_j)\}}
 {\sqrt{2N}(n-k)(n-k-1)}.
\]
Its terminal second moment and Doob's inequality give
\[
 \E\max_{k\in\mathcal K_n}
 \left|\frac{p^{-1/2}\sum_j\{Q_j(k)-\E Q_j(k)\}}{\sqrt{2N}}\right|^2
 \le C(1+\mathfrak r_p/N).
\]
The entropy process bound extends this inequality to $T_j$.

Since \(n^2s_{n,k}^2=2N+O(1)\),
\[
 \begin{aligned}
 &\E\max_{k\in\mathcal K_n}
 \left|p^{-1/2}\sum_j\{T_j(k)-\E T_j(k)\}
       \left(\frac1{ns_{n,k}}-\frac1{\sqrt{2N}}\right)\right|^2\\
 &\qquad\le\frac C{N^2}
 \E\max_{k\in\mathcal K_n}
 \left|\frac{p^{-1/2}\sum_j\{T_j(k)-\E T_j(k)\}}{\sqrt{2N}}\right|^2
 \le\frac{C(1+\mathfrak r_p/N)}{N^2}.
 \end{aligned}
\]

For fixed $t_1,\ldots,t_d$ and real $\lambda_1,\ldots,\lambda_d$, define
\[
 b_{ih}=\sum_a\lambda_a a_i(t_a)a_h(t_a),\qquad
 S_n=\sum_{i<h}b_{ih}h_n(Y_i,Y_h),\qquad
 V_\lambda=\sum_{a,b}\lambda_a\lambda_b
                \rho_{\rm B}(t_a,t_b)^2.
\]
Grid times are understood in these expressions.

Then $|b_{ih}|\le C/n$
and
\[
 \sum_{i<h}b_{ih}^2=\tfrac12V_\lambda+O(n^{-1}),\qquad
 \operatorname{Var}(S_n)=2pN\{V_\lambda+O_\lambda(\epsilon^{\rm app}_{n,p})\}.
\]
Indeed,
\[
 \sum_{i<h}a_i(s)a_h(s)a_i(t)a_h(t)
 =\frac12\left\{\left(\sum_i a_i(s)a_i(t)\right)^2
                    -\sum_i a_i(s)^2a_i(t)^2\right\}
 =\frac12\rho_{\rm B}(s,t)^2+O(n^{-1}).
\]
The variance formula follows from row degeneracy.
If \(V_\lambda=0\), it gives convergence in \(L^2\).
Otherwise let
\(M_h=\sum_{i<h}b_{ih}h_n(Y_i,Y_h)\).
Conditional on \(Y_h\), these summands are independent and
centered, so
\[
 \E M_h^4
 =\sum_{i<h}b_{ih}^4\E h_n^4
   +6\sum_{i<\ell<h}b_{ih}^2b_{\ell h}^2\E g_n^2 .
\]
Using \(|b_{ih}|\le C/n\) and summing \(h\) gives
\begin{equation*}
 \sum_h\E M_h^4\le C\left\{\frac{\E h_n^4}{n^2}
                                  +\frac{\E g_n^2}{n}\right\}.
\end{equation*}
The predictable quadratic variation splits exactly as
\[
 \begin{aligned}
 \sum_h\E(M_h^2\mid\mathcal F_{h-1})
 &=\sum_i c_i g_n(Y_i)
       +\sum_{i<\ell}d_{i\ell}G_n(Y_i,Y_\ell),\\
 c_i&=\sum_{h>i}b_{ih}^2\le C/n,\qquad
 d_{i\ell}=2\sum_{h>\ell}b_{ih}b_{\ell h}=O(n^{-1}),\\
 G_n(y,z)&=\E\{h_n(y,Y')h_n(z,Y')\}.
 \end{aligned}
\]
The diagonal variance is at most \(C\E g_n^2/n\).
The kernel \(G_n\) is canonical, so distinct unordered pairs
have zero covariance, and
\[
 \operatorname{Var}\!\left(\sum_{i<\ell}d_{i\ell}G_n(Y_i,Y_\ell)\right)
 =\sum_{i<\ell}d_{i\ell}^2\,\E G_n^2
 \le C\operatorname{tr}\mathscr T_n^4 .
\]
Their sum therefore satisfies
\begin{equation*}
 \operatorname{Var}\left\{\sum_h\E(M_h^2\mid\mathcal F_{h-1})\right\}
 \le C\left\{\frac{\E g_n^2}{n}
                         +\operatorname{tr}\mathscr T_n^4\right\}.
\end{equation*}
Divide the two estimates by
\(\operatorname{Var}(S_n)^2\asymp_\lambda(pN)^2\) and apply
Lemma~\ref{lem:sum-full-row-kernel-bounds}. The resulting
normalized bracket variance and fourth-moment sum are bounded
by \(C_\lambda\epsilon^{\rm mart}_{n,p}\).
Every component vanishes under the stated assumptions; for example,
\[
 \frac{\mathfrak r_p^3}{p}
 =\frac{\mathfrak r_p^2}{\sqrt p}
      \frac{\mathfrak r_p}{\sqrt p}\to0,\qquad
 \frac{(1+\mathfrak r_p)^2N^2}{n^2}
 \le CN^4/n^2\to0.
\]
The remaining terms are bounded by these assumptions in
the same direct way.

For real random variables \(X,Y\), write
\(d_{\rm K}(X,Y)=\sup_{x\in\mathbb R}|\Pp(X\le x)-\Pp(Y\le x)|\);
a distribution argument denotes any variable having that law.
For martingale differences \(X_h\) with
\(s^2=\sum_h\E X_h^2>0\) and finite fourth moments, the moment-order-two
inequality in \citet[Theorem~1.1]{Mourrat2013} is
\[
 d_{\rm K}\!\left(\frac{\sum_hX_h}{s},N(0,1)\right)
 \le C\left[
 \E\left\{s^{-2}\sum_h\E(X_h^2\mid\mathcal F_{h-1})-1\right\}^2
 +s^{-4}\sum_h\E X_h^4\right]^{1/5}.
\]
Take \(X_h=M_h\).
The preceding variance and fourth-moment calculations verify
these hypotheses and bound the bracketed quantity by
\(C_\lambda\epsilon^{\rm mart}_{n,p}\).

More explicitly, for $V_\lambda>0$ this gives
\[
 \sup_x\left|
 \Pp\left\{\frac{S_n}{\sqrt{2pN}}\le x\right\}
       -\Phi\left(\frac{x}{\sqrt{V_\lambda}}\right)\right|
 \le C_\lambda\{\{\epsilon^{\rm mart}_{n,p}\}^{1/5}+\epsilon^{\rm app}_{n,p}\}.
\]
The second term includes the explicit variance error.

Combining Lemma~\ref{lem:sum-full-population-comparison} and
$n^2s_{n,k}^2=2N+O(1)$ gives
\[
 \E\max_{k\in\mathcal K_n}
 \left|D_{\mathrm{sum},n,p}(k)
 -\frac{\sum_{i<h}a_i(t)a_h(t)h_n(Y_i,Y_h)}{\sqrt{2pN}}
 \right|^2\le C\epsilon^{\rm app}_{n,p}.
\]
Let \(a=(\epsilon^{\rm app}_{n,p})^{1/3}\) and
\(W_{n,\lambda}=\sum_j\lambda_j
D_{\mathrm{sum},n,p}(k_n(t_j))\).
The process comparison and Chebyshev give
\[
 \Pp\left\{\left|W_{n,\lambda}
             -S_n/\sqrt{2pN}\right|>a\right\}
 \le C_\lambda\epsilon^{\rm app}_{n,p}/a^2 .
\]
Sandwiching its CDF between those of
\(S_n/\sqrt{2pN}\) at \(x-a\) and \(x+a\), and using the
bounded normal density, yields
\[
 \sup_x\left|\Pp(W_{n,\lambda}\le x)
                 -\Phi(x/\sqrt{V_\lambda})\right|
 \le C_\lambda\left\{(\epsilon^{\rm app}_{n,p})^{1/3}
                    +(\epsilon^{\rm mart}_{n,p})^{1/5}\right\}.
\]
Cram\'er--Wold now proves finite-dimensional convergence.

For tightness, the finite-dimensional limit of the exact
martingale is
\[
 \mathscr M(t)=\frac{t}{1-t}\mathbb G(t),\qquad
 \operatorname{Cov}\{\mathscr M(s),\mathscr M(t)\}
       =\{s/(1-s)\}^2,\quad s\le t.
\]
Truncate at the initially measurable event
\(E_{n,K}=\{\|A_\Sigma\|_F^2\le KN\}\).
The matrix second-moment bound gives
\(\Pp(E_{n,K}^c)\le C/K\).
Slice hypercontractivity, the exact covariance identity,
and Doob's inequality give
\[
 \E\left[\sup_k|\mathscr M_n(k)|^6
                       \mathbf1_{E_{n,K}}\right]\le C_K .
\]
Use \(\mathscr M_n(t)=\mathscr M_n(k_n(t))\), and, for a bounded
function \(f\) on \(\mathcal I_\eta\), define
\[
 \operatorname{osc}_\delta(f)
 =\sup_{\substack{s,t\in\mathcal I_\eta\\|s-t|\le\delta}}|f(t)-f(s)|.
\]
For a fixed mesh \(t_i\) of width \(\delta\), put
\(\Delta_i\mathscr M_n=\mathscr M_n(t_{i+1})-\mathscr M_n(t_i)\).
For \(A>0\),
\[
 \E[|\Delta_i\mathscr M_n|^4\mathbf1_{E_{n,K}}]
 \le \E\min\{|\Delta_i\mathscr M_n|^4,A\}
                         +C_KA^{-1/2}.
\]
Finite-dimensional convergence applies to the bounded continuous
term; take \(n\to\infty\), then \(A\to\infty\). The Gaussian
limit has fourth increment moment at most \(C\delta^2\), so
\[
 \limsup_n\E[|\Delta_i\mathscr M_n|^4
                         \mathbf1_{E_{n,K}}]\le C\delta^2 .
\]
Because the truncation is measurable at time zero, the stopped
increment remains a martingale. Doob's fourth-moment inequality
therefore gives
\[
 \Pp\!\left(E_{n,K}\cap
   \left\{\max_{t_i\le k/n\le t_{i+1}}
    |\mathscr M_n(k)-\mathscr M_n(t_i)|>\epsilon\right\}\right)
 \le \frac{C}{\epsilon^4}
     \E[|\Delta_i\mathscr M_n|^4\mathbf1_{E_{n,K}}].
\]
Sum over \(O(\delta^{-1})\) cells, enlarging cells once to
cover adjacent-mesh oscillations:
\[
 \limsup_n\Pp\{\operatorname{osc}_\delta(\mathscr M_n)>\epsilon\}
 \le C/K+C_\epsilon\delta .
\]
Let \(\delta\downarrow0\), then \(K\uparrow\infty\).
This proves tightness with continuous limits and vanishing jumps.

The same martingale estimates for a deterministic set $A$ of $q$
coordinates, with normalization $\sqrt p$ retained, give
\[
 \E\max_{k\in\mathcal K_n}
 \left|p^{-1/2}\sum_{j\in A}D_{j,n}(k)\right|^2
 \le C\frac qp\left\{1+\frac{1+\mathfrak r_p}{N}\right\}.
\]
They also bound the full quadratic SUM supremum in $L^2$.

Consequently the deterministic relative studentization error $O(N^{-1})$
is negligible uniformly, completing the process proof.

The preceding bounds establish the process convergence
\begin{equation}\label{eq:sum-full-functional-limit}
 \{D_{\mathrm{sum},n,p}(k_n(t)):t\in\mathcal I_\eta\}
 \ \Rightarrow\
 \{\mathbb G(t):t\in\mathcal I_\eta\}
\end{equation}
in the supremum norm, along any joint sequence satisfying
Assumption~\ref{ass:copula-null}\ref{ass:copula-pairwise}--\ref{ass:copula-row}.

The continuous mapping theorem proves Theorem~\ref{thm:sum-null}.

To verify continuity of the reference CDF, let \(Z\sim N(0,1)\) and a standard Brownian motion \(B\) be independent,
put \(T=2L_\eta\) and
\(\zeta(r)=\sqrt2\,B((e^{2r}-1)/2)\), and represent the stationary
process as \(e^{-r}\{Z+\zeta(r)\}\). For \(z_2>z_1\),
\[
 e^{-T}(z_2-z_1)
 \le\sup_r e^{-r}\{z_2+\zeta(r)\}
       -\sup_r e^{-r}\{z_1+\zeta(r)\}
 \le z_2-z_1 .
\]
Conditional on \(\zeta\), a level has at most one preimage
in \(Z\), which has probability zero. Hence the supremum
has no atoms.

\subsection{Exponential bounds at a fixed split}
\label{app:sum-exponential-controls}

Use $T_j(k)$, $Q_j(k)$ and $R_j(k)=T_j(k)-Q_j(k)$
as in Appendix~\ref{app:sum-full}. Put
\[
 N=\log n,\qquad B_p=1+\mathfrak r_p,\qquad
 D_{\mathrm{sum},n,p}(k)=p^{-1/2}\sum_{j=1}^p D_{j,n}(k).
\]
The bounds below concern a fixed $k$ and are uniform over
$k\in\mathcal K_n$. Their constants do not identify a sharp
Gaussian tail exponent.

\begin{lemma}\label{lem:sum-rank-exponential-controls}
Fix $\eta\in(0,1/2)$. Under the uniform-permutation law $\mathbb P_0$
of any independent, identically distributed continuous scalar sample,
there are constants $c_\eta,C_\eta>0$ such that, for all sufficiently
large $n$, all $k\in\mathcal K_n$ and all $y\ge0$,
\begin{equation}\label{eq:sum-entropy-exponential-tail}
 \mathbb P_0\{|R_j(k)|>y\}\le C_\eta e^{-c_\eta y},
\end{equation}
and
\begin{equation}\label{eq:rank-raw-bernstein}
 \mathbb P_0\{|T_j(k)-\mathbb E_0T_j(k)|>y\}
 \le C_\eta\exp\{-c_\eta\min(y^2/\log n,y)\}.
\end{equation}
These two bounds apply marginally to each coordinate without any
restriction on dependence between coordinates. The exact scalar
studentization also satisfies
\[
 \log\mathbb E_0 e^{sD_{j,n}(k)}\le C_\eta s^2,
                    \qquad |s|\le c_\eta\sqrt{\log n}.
\]
If, in addition, the normal-score transformed rows are independent
$N_p(0,\mathbf R_p)$ vectors, where $\mathbf R_p=(\rho_{j\ell})$
is positive semidefinite with unit diagonal, put
\[
 B_p=1+\max_j\sum_{\ell\ne j}|\rho_{j\ell}|.
\]
Then
\[
 \log\E e^{sD_{\mathrm{sum},n,p}(k)}\le C_\eta B_ps^2,
             \qquad |s|\le c_\eta\sqrt{p\log n}/B_p,
\]
and, for every $x\ge0$,
\begin{equation}\label{eq:sum-pointwise-bernstein}
 \Pp\{|D_{\mathrm{sum},n,p}(k)|>x\}
 \le2\exp\!\left[
 -c_\eta\min\left\{\frac{x^2}{B_p},
                   \frac{x\sqrt{p\log n}}{B_p}\right\}\right].
\end{equation}
The constants depend only on $\eta$; no restriction on the dimension,
no bound on $B_p$, and no nonsingularity or neighborhood condition
is needed for these inequalities.
\end{lemma}

\begin{proof}
Fix one coordinate and write $t=k/n$, $v=t(1-t)$.
In rank order, let $(B_1,\ldots,B_n)$ be uniform among binary
vectors with exactly $k$ ones, and set
\[
 A_l=\sum_{i=1}^l(B_i-t),\qquad
 d_l=(l-\tfrac12)(n-l+\tfrac12),\qquad
 h_t(x)=\operatorname{KL}(x\Vert t).
\]
The exact entropy and quadratic expressions are
\[
 \begin{split}
 T&=\sum_{l=1}^{n-1}\frac n{d_l}
 \left\{l h_t\!\left(t+\frac{A_l}{l}\right)
 +(n-l)h_t\!\left(t-\frac{A_l}{n-l}\right)\right\},\\
 Q&=\frac1{2v}\sum_{l=1}^{n-1}
       \frac{n^2 A_l^2}{d_l\,l(n-l)}.
 \end{split}
\]
The omitted $l=n$ entropy term is zero.
For \(t+d\in[0,1]\), separate \(|d|\le\eta/2\) from its
complement. On the first set Taylor's theorem gives
\[
 h_t(t)=h_t'(t)=0,\qquad h_t''(t)=v^{-1},\qquad
 h_t'''(x)=(1-x)^{-2}-x^{-2},
\]
\[
 \left|h_t(t+d)-\frac{d^2}{2v}\right|
 \le\frac{|d|^3}{6}
       \sup_{\eta/2\le x\le1-\eta/2}|h_t'''(x)|
 \le\frac4{3\eta^2}|d|^3.
\]
On the second set,
\[
 \left|h_t(t+d)-\frac{d^2}{2v}\right|
 \le\log(1/\eta)+\frac1{2\eta(1-\eta)}
 \le\frac{8|d|^3}{\eta^3}
       \left\{\log(1/\eta)+\frac1{2\eta(1-\eta)}\right\}.
\]
Thus the two bounds, including \(t+d=0,1\), imply
\[
 \left|h_t(t+d)-\frac{d^2}{2v}\right|
 \le C_\eta^{(0)}|d|^3,\qquad
 C_\eta^{(0)}
 =\frac4{3\eta^2}
   +\frac8{\eta^3}
       \left\{\log(1/\eta)+\frac1{2\eta(1-\eta)}\right\}.
\]
For \(b_l=\min(l,n-l)\), substitute the two increments
\(d=A_l/l\) and \(d=-A_l/(n-l)\) and use
\[
 \frac n{d_l}\{l^{-2}+(n-l)^{-2}\}
 \le\frac4{b_l}\frac2{b_l^2}=\frac8{b_l^3}.
\]
\[
 |R|\le8C_\eta^{(0)}
             \sum_{l=1}^{n-1}\frac{|A_l|^3}{b_l^3}.
\]
Let $\mathcal F_l=\sigma(B_1,\ldots,B_l)$. Sampling without
replacement gives
\[
 \begin{split}
 \E(B_{l+1}\mid\mathcal F_l)&=t-\frac{A_l}{n-l},\\
 M_l&=\frac{A_l}{n-l},\\
 M_{l+1}-M_l
 &=\frac{B_{l+1}-\E(B_{l+1}\mid\mathcal F_l)}{n-l-1}.
 \end{split}
\]
For \(l<m\le\lfloor n/2\rfloor\), the increment is
conditionally centered with range length at most \(2/n\).
Conditional Hoeffding's inequality therefore makes
\[
 \mathcal E_l(s)
 =\exp\!\left\{sM_l-\frac{s^2l}{2n^2}\right\},
 \qquad s>0,
\]
a nonnegative supermartingale. With
\(\tau=\min\{\inf\{l:M_l\ge a\},m\}\), optional sampling at
this bounded stopping time gives
\[
 1\ge\E\mathcal E_\tau(s)
 \ge e^{sa-s^2m/(2n^2)}
             \Pp(\max_{l\le m}M_l\ge a).
\]
Consequently,
\[
 \Pp(\max_{l\le m}M_l\ge a)
 \le\inf_{s>0}e^{-sa+s^2m/(2n^2)}
 =e^{-a^2n^2/(2m)}.
\]
Since $A_l\ge z$ implies $M_l\ge z/n$, the two signs give
\[
 \Pp\{\max_{l\le m}|A_l|\ge z\}
 \le2e^{-z^2/(2m)}.
\]
Reversing the label sequence proves the same inequality
for $A_{n-l}$.
For each dyadic $b=2^r\le\lfloor n/2\rfloor$, put
$m_b=\min(2b,\lfloor n/2\rfloor)$ and
\[
 Z_b=b^{-1/2}\max\left\{
         \max_{l\le m_b}|A_l|,
         \max_{l\le m_b}|A_{n-l}|\right\}.
\]
Since \(m_b\le2b\), the two reflected maximal inequalities
give \(\Pp(Z_b>z)\le4e^{-z^2/4}\). Define
\[
 W_b=8C_\eta^{(0)}
       \sum_{b\le b_l<2b}\frac{|A_l|^3}{b_l^3}.
\]
There are at most \(2b\) such ranks, and \(|A_l|\le\sqrt b\,Z_b\), so
\[
 |R|\le\sum_bW_b,\qquad
 W_b\le16C_\eta^{(0)}b^{-1/2}Z_b^3=C_0b^{-1/2}Z_b^3.
\]
The deterministic inequalities \(|A_l|\le b_l<2b\)
also give \(W_b\le D_0b\), where
\(C_0=16C_\eta^{(0)}\) and \(D_0=8C_0\).
For $y\ge8D_0$, choose a dyadic $b_0$ with
$y/(8D_0)<b_0\le y/(4D_0)$. Deterministically,
\[
 \sum_{b<b_0}W_b\le D_0\sum_{b<b_0}b\le y/4.
\]
For \(b=b_02^s\), set
\(a_s=(1-2^{-1/4})2^{-s/4}\), so \(\sum_{s\ge0}a_s=1\).
The remaining blocks satisfy
\[
 \{|R|>y\}\subseteq
 \bigcup_{s\ge0}\{W_{b_02^s}>3ya_s/4\},
\]
\[
 \begin{aligned}
 \Pp(W_{b_02^s}>3ya_s/4)
 &\le \Pp\!\left(
 Z_{b_02^s}>
   \left\{\frac{3ya_s(b_02^s)^{1/2}}{4C_0}\right\}^{1/3}
                 \right)\\
 &\le4\exp\{-c_\eta y^{2/3}b_0^{1/3}2^{s/6}\}.
 \end{aligned}
\]
Because \(b_0>y/(8D_0)\),
\[
 \Pp(|R|>y)
 \le4\sum_{s\ge0}e^{-c_\eta' y2^{s/6}}
 \le C_\eta e^{-c_\eta''y}.
\]
If no block remains, the event is empty.
Increasing $C_\eta$ covers $0\le y<8D_0$ and proves
\eqref{eq:sum-entropy-exponential-tail}.
Integration of the exponential tail yields, after decreasing
\(c_\eta\),
\[
 \E e^{c_\eta|R|}
 =1+c_\eta\int_0^\infty e^{c_\eta y}\Pp(|R|>y)\,dy
 \le C_\eta.
\]
Put \(R^\circ=R-\E R\). Since \(\E R^\circ=0\),
\[
 \E e^{sR^\circ}
 \le1+\frac{s^2}{2}\E\{(R^\circ)^2e^{|sR^\circ|}\}
 \le1+C_\eta s^2,\qquad |s|\le c_\eta,
\]
and hence \(\log\E e^{sR^\circ}\le C_\eta s^2\).
We next bound the quadratic term. Theorem~1 of
\citet{SambaleSinulis2022} applies to multilinear polynomials
on a uniform multislice; for a centered quadratic form with
zero diagonal and matrix $A$, its conclusion is
\[
 \Pp\{|X^\top AX-\E(X^\top AX)|>y\}
 \le2\exp\{-c\min(y^2/\|A\|_{\rm F}^2,y/\|A\|_{\rm op})\}.
\]
The alphabet diameter here is one. For a linear form, the same
theorem gives $2\exp(-cy^2/\|\beta\|_2^2)$.
No independence of the slice entries is required.
In the universal rank basis,
$Q=X^\top KX/(2v)$ with $X_i=B_i-t$.
Equation~\eqref{eq:sum-full-matrix-bounds} gives
\[
 K\mathbf1=0,\qquad
 \|K\|_{\rm op}\le C_\eta,\qquad
 \|K\|_{\rm F}^2\le C_\eta N,\qquad
 \sum_iK_{ii}^2\le C_\eta.
\]
Write \(K^\circ=K-\operatorname{diag}(K_{ii})\).
The identity \(X_i^2=(1-2t)X_i+v\) gives
\[
 Q-\E Q
 =\frac1{2v}
       \{X^\top K^\circ X-\E(X^\top K^\circ X)\}
   +\sum_i\frac{(1-2t)K_{ii}}{2v}X_i .
\]
Here \(\E X=0\), so the quadratic polynomial has zero
expected gradient. Moreover,
\[
 \|K^\circ\|_{\rm F}^2\le C_\eta N,\qquad
 \|K^\circ\|_{\rm op}\le C_\eta,\qquad
 \sum_i\left\{\frac{(1-2t)K_{ii}}{2v}\right\}^2\le C_\eta.
\]
Applying the quadratic and linear multislice inequalities
and splitting the threshold into two halves gives
\[
 \Pp(|Q-\E Q|>y)
 \le C_\eta\exp\{-c_\eta\min(y^2/N,y)\}.
\]
Let \(Q'\) be an independent copy. Jensen's inequality
conditional on \(Q\), followed by the preceding tail bound,
gives
\[
 \E e^{s(Q-\E Q)}\le\E e^{s(Q-Q')},
\]
\[
 \begin{aligned}
 \E|Q-Q'|^{2m}
 &=2m\int_0^\infty y^{2m-1}\Pp(|Q-Q'|>y)\,dy\\
 &\le C_\eta^{2m}\{N^m m!+(2m)!\}\\
 &\le C_\eta^{2m}\{(mN)^m+m^{2m}\},\qquad m\ge1.
 \end{aligned}
\]
The difference is symmetric. Expand its hyperbolic cosine and
use $(2m)!\ge m!m^m$ and $m^{2m}/(2m)!\le C^{2m}$:
\[
 \begin{split}
 \E e^{s(Q-Q')}
 &\le1+\sum_{m\ge1}\frac{(C_\eta s^2N)^m}{m!}
                  +\sum_{m\ge1}(C_\eta|s|)^{2m},\\
 \log\E e^{s(Q-\E Q)}
 &\le C_\eta Ns^2\qquad(|s|\le c_\eta).
 \end{split}
\]
With \(Q^\circ=Q-\E Q\) and \(T^\circ=T-\E T\),
Cauchy--Schwarz combines the two centered estimates:
\[
 \log\E e^{sT^\circ}
 \le\frac12\log\E e^{2sQ^\circ}
       +\frac12\log\E e^{2sR^\circ}
 \le C_\eta Ns^2,\qquad |s|\le c_\eta.
\]
For \(y>0\), take
\(s=\min\{y/(2C_\eta N),c_\eta/2\}\). Then
\[
 \Pp(T^\circ>y)
 \le e^{-sy+C_\eta Ns^2}
 \le C_\eta e^{-c_\eta\min(y^2/N,y)}.
\]
Replacing \(T^\circ\) by \(-T^\circ\) proves
\eqref{eq:rank-raw-bernstein}.
By exact rank studentization,
\[
 D_{j,n}(k)=\frac{T_j(k)-\E_0T_j(k)}
                 {\sqrt{\operatorname{Var}_0(T_j(k))}},\qquad
 \operatorname{Var}_0(T_j(k))=2N+O_\eta(1).
\]
Consequently,
\[
 \log\E_0 e^{sD_{j,n}(k)}
 \le C_\eta s^2,\qquad |s|\le c_\eta\sqrt N .
\]
For the remaining aggregate bounds, impose the additional Gaussian
representation in the lemma and let
$Y_j=(Y_{1j},\ldots,Y_{nj})^\top$ denote its columns.
Their joint covariance satisfies
\[
 \operatorname{Cov}(Y_1,\ldots,Y_p)
       =\mathbf R_p\otimes I_n\le B_p I_{pn}.
\]
Theorem~1(i) of \citet{ChenDafnisPaouris2015} states that
a Gaussian block vector with covariance
$T\le\operatorname{diag}(q_jT_{jj})$ satisfies
\[
 \E\prod_j f_j(Y_j)
 \le\prod_j\{\E f_j(Y_j)^{q_j}\}^{1/q_j}
\]
for nonnegative measurable $f_j$. Singular joint covariance
is allowed. Here $T_{jj}=I_n$ and $q_j=B_p$ verify its matrix
hypothesis.
Apply the Gaussian-block inequality with
\(f_j(Y_j)=\exp\{zD_{j,n}(k)/\sqrt p\}\):
\[
 \begin{aligned}
 \log\E e^{zD_{\mathrm{sum},n,p}(k)}
 &\le \frac1{B_p}
       \sum_{j=1}^p
       \log\E e^{B_pzD_{j,n}(k)/\sqrt p}\\
 &\le C_\eta B_pz^2,\qquad
       |z|\le c_\eta\sqrt{pN}/B_p.
 \end{aligned}
\]
For either sign of the statistic, choose
\[
 z=\min\left\{\frac{x}{2C_\eta B_p},
                 \frac{c_\eta\sqrt{pN}}{2B_p}\right\}.
\]
The bound \(\Pp(\pm D_{\mathrm{sum},n,p}(k)>x)
 \le e^{-zx+C_\eta B_pz^2}\), summed over the two signs,
is \eqref{eq:sum-pointwise-bernstein}.
\end{proof}

\section{Analytic calibration of the MAX statistic}
\label{app:sparse-null}

The Gaussian quadratic transform provides the conditional middle-bridge kernel.
Bellman coefficients and conditional rank-label dynamics then determine the
original statistic's local tails and scan probabilities.

\subsection{Deterministic inputs and analytic normalizations}
\label{app:max-normalizing-constants}
\label{app:max-deterministic-inputs}

This subsection gives the deterministic construction of the three
tail factors in \eqref{eq:main-max-tail-factor}, followed
by the branchwise normalizing pairs. The moment corrections
$m(t),v_0(t),d_0(t,y)$ are defined in
Appendix~\ref{app:moment-correction-definitions}. The entropy Hardy
constant determines the boundary, the Bellman coefficients retain
the endpoint contribution, and the final scan factors account for
maximization over the split location. The entropy Hardy constants
below give the same $b_\eta$ as \eqref{eq:main-hardy-threshold}.

For measurable \(f:[0,\infty)\to[0,1]\), put
\[
 Hf(s)=\frac1s\int_0^s f(z)\,dz\quad(s>0),\qquad
 I_t(f)=\int_0^\infty h_t(f(s))\,ds,\qquad
 J_t(f)=\int_0^\infty h_t(Hf(s))\,ds .
\]
Define the entropy Hardy constants by
\begin{equation*}
 C(t)=\sup_{\substack{f:[0,\infty)\to[0,1]\\0<I_t(f)<\infty}}
               \frac{J_t(f)}{I_t(f)},\qquad
 b(t)=C(t)^{-1},\qquad
 b_\eta=\min_{\eta\le t\le1-\eta}b(t).
\end{equation*}
For fixed \(t\) and \(C>4\), the differential equation
\begin{equation*}
 u_C'(x)=\frac{u_C(x)\{1-u_C(x)\}h_t'(x)}{C\{x-u_C(x)\}},
 \quad u_C(t)=t,\quad
 u_C'(t)=\frac{1-\sqrt{1-4/C}}2
\end{equation*}
specifies the branch through its displayed derivative at the removable
central singularity. Set
\[
 C_+(t)=\inf\{C>4:t<u_C(x)<x\ \forall x\in(t,1)\},
 \qquad b_+(t)=C_+(t)^{-1}.
\]
The variational and differential definitions give
\[
 C(t)=\max\{C_+(t),C_+(1-t)\},\qquad
 4\le C(t)\le\frac2{t(1-t)},\qquad b_\eta=b(\eta).
\]

For \(0\le\vartheta<1/4\), define
\[
 r(\vartheta)=\sqrt{1-4\vartheta},\qquad
 a(\vartheta)=\frac{1-r(\vartheta)}2,\qquad
 \kappa(\vartheta)=\frac{a(\vartheta)}2,\qquad
 \lambda(\vartheta)=a(\vartheta).
\]
For \(0<\vartheta<b(t)\), let \(u_{t,\vartheta}\) be the central branch
\begin{equation*}
 u'=\frac{\vartheta u(1-u)h_t'}{x-u},\qquad
 u(t)=t,\quad u'(t)=a(\vartheta),
 \qquad
 g_{t,\vartheta}(x)=(1-\vartheta)h_t(x)
                         -h_{u_{t,\vartheta}(x)}(x).
\end{equation*}
For \(c>-1\), write \(u=u_{t,\vartheta}\) and \(A=A_{t,\vartheta,c}\).
Define \(A\) by \(A(t)=1\) and
\begin{equation*}
 (u-x)(\log A)'=
 \kappa(\vartheta)+\vartheta c h_t-\vartheta(u-x)h_t'
 -\frac{g_{t,\vartheta}''}{2}\{u(1-u)+(u-x)^2\}.
\end{equation*}
The value at \(x=t\) is given by continuous extension. To specify the
endpoint coefficients by deterministic calculations, let \(m\) be a nonnegative
integer and set
\[
 \mathsf L_{0,c}(0;t,\vartheta)=1,\qquad
 \mathsf L_{m,c}(s;t,\vartheta)=0\quad(s\notin\{0,\ldots,m\}),
\]
and, for \(m\ge0\) and \(s\in\{0,\ldots,m+1\}\), use the recursion
\begin{equation}\label{eq:main-endpoint-recursion}
 \begin{split}
 \mathsf L_{m+1,c}(s;t,\vartheta)
 ={}&\exp\left\{\vartheta\frac{m+1}{m+1+c}
                            h_t\!\left(\frac{s}{m+1}\right)\right\}\\
 &{}\times\{(1-t)\mathsf L_{m,c}(s;t,\vartheta)
                    +t\mathsf L_{m,c}(s-1;t,\vartheta)\}.
 \end{split}
\end{equation}
For \(m\ge1\), put
\[
 k_{m,c}(t,\vartheta)=m^{-\kappa(\vartheta)}
 \sum_{s=0}^m \mathsf L_{m,c}(s;t,\vartheta)
 A_{t,\vartheta,c}(s/m)e^{m g_{t,\vartheta}(s/m)} .
\]
The coefficient and the combined transform amplitude are
\begin{equation*}
 \begin{aligned}
 K_c(t,\vartheta)
 &=k_{1,c}(t,\vartheta)
       +\sum_{m=1}^\infty\{k_{m+1,c}(t,\vartheta)-k_{m,c}(t,\vartheta)\},\\
 \Psi_t(\vartheta)
 &=\sqrt{r(\vartheta)}K_{-1/2}(t,\vartheta)K_{+1/2}(t,\vartheta).
 \end{aligned}
\end{equation*}
The series is absolutely convergent for \(0<\vartheta<b(t)\).
Its terms are finite sums obtained from
\eqref{eq:main-endpoint-recursion}.

Write \(N=\log n\), \(\gamma_n=(\log p)/N\), and
\begin{equation*}
 q_c=(1-4b_\eta)^{-1/2},\qquad
 \gamma_c=\frac{(q_c-1)^2}{4q_c},
 \qquad q_c=\gamma_c=\infty\quad(b_\eta=1/4).
\end{equation*}
The remaining critical and boundary constants are needed only when
\(b=b_\eta<1/4\). The following definitions apply to fixed \(t<1/2\)
with \(b(t)<1/4\), and are evaluated at \(t=\eta\).
Set
\[
 b_t=b(t),\quad a_t=a(b_t),\quad
 h_{1,t}=\log(1/t),\quad H_t=\log\{(1-t)/t\},
 \quad e_t(q)=\{1+\exp(q-H_t)\}^{-1}.
\]
For \(c\in\{-1/2,1/2\}\), define
\begin{equation*}
 \alpha_c(t)=\frac12-\frac{a_t}{2}-b_t c h_{1,t},
 \qquad
 B_c(t)=\frac{1-a_t}{2a_t}-c h_{1,t}.
\end{equation*}
The functions \(u_{t,b_t}\) and \(A_{t,b_t,c}\) are the limits of their
subcritical branches on every compact subinterval of \((0,1)\).
Define
\begin{equation*}
 \begin{aligned}
 A_{*,c}(t)&=\lim_{q\to\infty}
 A_{t,b_t,c}(1-e_t(q))e_t(q)^{1/2}q^{-B_c(t)},\\
 C_U(t)&=\lim_{q\to\infty}q^{-1/b_t}
 \left.\partial_\vartheta u_{t,\vartheta}(1-e_t(q))
                                                    \right|_{\vartheta=b_t-},\\
 D_c(t)&=\frac{b_tA_{*,c}(t)}{\sqrt{2\pi}}
                     \exp\{b_t c h_{1,t}\psi(c+1)\}.
 \end{aligned}
\end{equation*}
When \(\alpha_c(t)<0\), write
\(K_c(t,b_t)=\lim_{\vartheta\uparrow b_t}K_c(t,\vartheta)\).
Define
\[
 \begin{aligned}
 \beta_c(t)&=\tfrac12\mathbf1_{\{\alpha_c(t)>0\}}
                     +\tfrac32\mathbf1_{\{\alpha_c(t)=0\}},\\
 P_c(t)&=
 \begin{cases}
 D_c(t)\Gamma(\alpha_c(t))C_U(t)^{-\alpha_c(t)},&\alpha_c(t)>0,\\
 2D_c(t)/3,&\alpha_c(t)=0,\\
 K_c(t,b_t),&\alpha_c(t)<0,
 \end{cases}\\
 \alpha_t&=\sum_{c\in\{-1/2,1/2\}}(\alpha_c(t))_+,\qquad
 \beta_t=\sum_{c\in\{-1/2,1/2\}}\beta_c(t),\\
 P_t&=\sqrt{r(b_t)}\prod_{c\in\{-1/2,1/2\}}P_c(t).
 \end{aligned}
\]
In particular, \(\alpha_\eta\ge\alpha_{-1/2}(\eta)>0\), and
\[
 \Psi_\eta(b-\Delta)\sim
 P_\eta\Delta^{-\alpha_\eta}\{\log(1/\Delta)\}^{\beta_\eta}
 \quad(\Delta\downarrow0).
\]
The exponents \(\alpha_\eta,\beta_\eta\) are distinct from the test
level \(\alpha\). Put
\[
 v_\eta=\eta(1-\eta),\quad b_1=b_+'(\eta)>0,\quad
 V=2q_c^3,\quad d(t)=d_0(t,q_c).
\]
For \(\alpha_\eta<1\), define the improper integral
\begin{equation*}
 \Theta_0=\lim_{\epsilon\downarrow0}
 \int_{\eta+\epsilon}^{1-\eta-\epsilon}
               \Psi_t(b)e^{-bd(t)}\frac{dt}{t(1-t)}.
\end{equation*}
For the other two cases, set
\[
 P_0=\frac{2P_\eta e^{-bd(\eta)}}{v_\eta b_1(\beta_\eta+1)}
       \quad(\alpha_\eta=1),\qquad
 P_A=\frac{2P_\eta e^{-bd(\eta)}}{v_\eta b_1(\alpha_\eta-1)}
       \quad(\alpha_\eta>1).
\]

The critical profile is also specified by the same differential equation.
For \(t<1/2\) with \(b(t)<1/4\), let
\[
 x_{t,0}=\frac{e\,t}{1-t+e\,t},\qquad
 s_t(x)=\exp\left\{\int_{x_{t,0}}^x
                 \frac{dz}{u_{t,b(t)}(z)-z}\right\},\quad t<x<1.
\]
The decreasing function \(s_t\) maps \((t,1)\) onto \((0,\infty)\).
For \(s>0\), define
\begin{equation*}
 r_t(s)=s_t^{-1}(s),\qquad f_t(s)=u_{t,b(t)}(r_t(s)),\qquad
 \mathcal J_t=\int_0^\infty h_t(r_t(s))\,ds,
\end{equation*}
and
\[
 D_t^+=\int_0^\infty
   \{t(1-r_t(s))h_t'(r_t(s))-(r_t(s)-t)\}\,ds.
\]
For \(y>q_c\), define the boundary scan coefficients
\begin{equation}\label{eq:main-boundary-coefficients}
 \begin{aligned}
 \mathcal H_\eta^{\rm bd}(y)
 &=\frac{b\{q_c-1-(y-q_c)D_\eta^+/\mathcal J_\eta\}}
                  {v_\eta b_1(y-q_c)},\\
 \mathcal C_\eta^{\rm bd}(y)
 &=\frac{2P_\eta}{b\Gamma(\alpha_\eta)}
 (y-q_c)^{\alpha_\eta-1}\mathcal H_\eta^{\rm bd}(y)e^{-bd_0(\eta,y)} .
 \end{aligned}
\end{equation}

Set
\[
 q_n=(\sqrt{\gamma_n}+\sqrt{1+\gamma_n})^2,\qquad
 \vartheta_n=\frac{1-q_n^{-2}}4,\qquad
 \lambda''(\vartheta)=2(1-4\vartheta)^{-3/2}.
\]
When \(\vartheta_n<b_\eta\), define
\begin{equation}\label{eq:main-subcritical-normalization}
 \begin{aligned}
 \Theta_n&=\int_\eta^{1-\eta}\Psi_t(\vartheta_n)
       e^{-\vartheta_n d_0(t,q_n)}\frac{dt}{t(1-t)},\\
 C_n&=\frac{(q_n-1)\sqrt N\,\Theta_n}
                         {\sqrt{2\pi\lambda''(\vartheta_n)}},\\
 A_{n,p}^{\rm S}&=\frac1{\vartheta_n\sqrt{2N}},\qquad
 B_{n,p}^{\rm S}=(q_n-1)\sqrt{N/2}
                         +\frac{\log C_n}{\vartheta_n\sqrt{2N}} .
 \end{aligned}
\end{equation}
This is the subcritical pair.

For the critical pair suppose \(b=b_\eta<1/4\), and put
\begin{equation}\label{eq:main-IG-kernel}
 g_N(v)=\frac{N}{2\sqrt\pi\,v^{3/2}}
              \exp\left\{-\frac{(v-N)^2}{4v}\right\}\quad(v>0),
 \qquad g_N(v)=0\quad(v\le0).
\end{equation}
Define
\[
 c_{\rm tail}=\frac{2P_\eta e^{-bd(\eta)}}
                        {v_\eta b_1\Gamma(\alpha_\eta)},\qquad
 w_\eta^{\rm C}(z)=c_{\rm tail}z^{\alpha_\eta-2}\{\log(e+z)\}^{\beta_\eta},\qquad z>0.
\]
If \(0<\alpha_\eta<1\), set
\[
 h_0=\frac{1-\alpha_\eta}{2},\qquad
 M_{\beta_\eta}=
 \begin{cases}
 \{2\beta_\eta/(eh_0)\}^{\beta_\eta},&\beta_\eta>0,\\
 1,&\beta_\eta=0,
 \end{cases}
\]
and
\[
 B_0=\max\left\{e,
    \left(\frac{2c_{\rm tail}M_{\beta_\eta}}{h_0\Theta_0}\right)^{1/h_0}\right\},
 \qquad m_0=\Theta_0-\int_{B_0}^\infty w_\eta^{\rm C}(z)\,dz .
\]
For \(0<\alpha_\eta<1\), the cutoff gives \(m_0\ge\Theta_0/2>0\). Let \(\delta_0\) denote unit mass at zero and define the positive measure
\begin{equation}\label{eq:main-critical-positive-measure}
 \mu(dz)=
 \begin{cases}
 w_\eta^{\rm C}(z)\,dz,\quad z>0,&\alpha_\eta>1,\\
 \mathbf1_{\{z\ge1\}}w_\eta^{\rm C}(z)\,dz,&\alpha_\eta=1,\\
 m_0\delta_0(dz)+\mathbf1_{\{z\ge B_0\}}w_\eta^{\rm C}(z)\,dz,&0<\alpha_\eta<1.
 \end{cases}
\end{equation}
Put
\begin{equation}\label{eq:main-critical-convolution}
 \mathcal R_N^{\rm fc}(U)=N(q_c-1)
           \int_{[0,\infty)}e^{-bz}g_N(U-z)\,\mu(dz),
\end{equation}
and define
\begin{equation}\label{eq:main-critical-normalization}
 \begin{gathered}
 y_n=
 \begin{cases}
 q_n,&\gamma_n\le\gamma_c,\\
 \{\gamma_n+\lambda(b)\}/b,&\gamma_n>\gamma_c,
 \end{cases}
 \quad U_{0,n}=Ny_n,\quad
 \zeta_n=\min\{(1-y_n^{-2})/4,b\},\\
 H_n=\frac{\log\{p\mathcal R_N^{\rm fc}(U_{0,n})\}}{\zeta_n},
 \qquad A_{n,p}^{\rm C}=\frac1{\zeta_n\sqrt{2N}},\qquad
 B_{n,p}^{\rm C}=\frac{U_{0,n}-N+H_n}{\sqrt{2N}}.
 \end{gathered}
\end{equation}

For the boundary pair, \(\gamma_n>\gamma_c\), put
\(y_n=\{\gamma_n+\lambda(b)\}/b\) and \(E_n=y_n-q_c\). For \(NE_n>1\), define
\begin{equation}\label{eq:main-boundary-normalization}
 \begin{gathered}
 h_n^{\rm B}=\frac{(\alpha_\eta-1)\log N
          +\beta_\eta\log\log(NE_n)+\log\mathcal C_\eta^{\rm bd}(y_n)}b,\\
 A_{n,p}^{\rm B}=\frac1{b\sqrt{2N}},\qquad
 B_{n,p}^{\rm B}=(y_n-1)\sqrt{N/2}+\frac{h_n^{\rm B}}{\sqrt{2N}} .
 \end{gathered}
\end{equation}

The branchwise pairs above are exactly the common pair in
\eqref{eq:full-max-branch-selection}. By
\eqref{eq:main-max-leading-level}, writing
$\vartheta_n=(1-q_n^{-2})/4$,
\[
 \gamma_n=\vartheta_nq_n-\lambda(\vartheta_n),\qquad
 y_n=\begin{cases}
 q_n,&\gamma_n\le\gamma_c,\\
 \{\gamma_n+\lambda(b_\eta)\}/b_\eta,&\gamma_n>\gamma_c.
 \end{cases}
\]
In the subcritical branch, $\zeta_n=\vartheta_n$ and
$\mathcal C_{n,p}=C_n$. In the critical branch,
$\log\mathcal C_{n,p}=\zeta_n H_n$. In the boundary branch,
$\zeta_n=b_\eta$ and
$\log\mathcal C_{n,p}=b_\eta h_n^{\rm B}$.
Substitution gives the respective pairs
$(A_{n,p}^{\rm S},B_{n,p}^{\rm S})$,
$(A_{n,p}^{\rm C},B_{n,p}^{\rm C})$, and
$(A_{n,p}^{\rm B},B_{n,p}^{\rm B})$ without approximation.
The boundary formula requires $N(y_n-q_c)>1$, which holds for
all sufficiently large $n$ under the switching rule in
\eqref{eq:main-max-tail-factor}.

The subcritical normalization is the pair
\((A_{n,p}^{\rm S},B_{n,p}^{\rm S})\) in
\eqref{eq:main-subcritical-normalization}; its coefficient uses the exact
rank moment corrections and the Bellman amplitude \(\Psi_t\).

\begin{theorem}[Original-rank analytic Gumbel law below the boundary threshold]
\label{thm:max-subcritical-gumbel}
Under the pairwise-correlation and neighborhood conditions
of Assumption~\ref{ass:copula-null}, suppose that the limit in
\eqref{eq:max-polynomial-regime} satisfies $0<\gamma<\gamma_c$.
For every real $x$,
\begin{equation}\label{eq:rank-subcritical-gumbel-law}
 P\{(\Smax-B_{n,p}^{\rm S})/A_{n,p}^{\rm S}\le x\}
       \longrightarrow\exp(-e^{-x}).
\end{equation}

\end{theorem}

For $b=b_\eta<1/4$, retain the constants $q_c,V,\alpha_\eta,
\beta_\eta,\Theta_0,P_0,P_A$ defined above in Appendix~\ref{app:max-deterministic-inputs}.
For $a>0$, define
\begin{equation}\label{eq:calibration-fractional-gaussian}
 \mathfrak J_a(x)=\frac1{\Gamma(a)\sqrt{2\pi V}}
     \int_0^\infty z^{a-1}e^{-(x-z)^2/(2V)}\,dz,
 \qquad \mathfrak J_0(x)=\frac{e^{-x^2/(2V)}}{\sqrt{2\pi V}}.
\end{equation}
Set
\begin{equation}\label{eq:calibration-core-amplitude}
 \begin{array}{c|c|c}
 &\mathcal A_{\rm c}(x)&(\nu_*,\ell_*)\\ \hline
 \alpha_\eta<1 &(q_c-1)\Theta_0\mathfrak J_0(x)&(1/2,0)\\[1mm]
 \alpha_\eta=1 &(q_c-1)P_0\,2^{-\beta_\eta-1}\mathfrak J_0(x)&(1/2,\beta_\eta+1)\\[1mm]
 \alpha_\eta>1 &(q_c-1)P_A\,2^{-\beta_\eta}\mathfrak J_{\alpha_\eta-1}(x)&(\alpha_\eta/2,\beta_\eta)
 \end{array}
\end{equation}
and let $\tau_N=(\log p-N\gamma_c)/\sqrt N$ and $x_N=\tau_N/b$.
The bounded critical window uses
\[
 \begin{gathered}
 h_N=\frac{\log p-N\gamma_c+\nu_*\log N+\ell_*\log\log N
                     +\log\mathcal A_{\rm c}(x_N)}b,\\
 B_{n,p}^{\rm c}=(q_c-1)\sqrt{N/2}+\frac{h_N}{\sqrt{2N}},
 \qquad A_n^{\rm c}=\frac1{b\sqrt{2N}}.
 \end{gathered}
\]

\begin{theorem}[Analytic MAX limit in the bounded critical window]
\label{thm:max-critical-core}
Under conditions \ref{ass:copula-pairwise} and
\ref{ass:copula-neighborhood} of Assumption~\ref{ass:copula-null}, suppose
$b(\eta)<1/4$ and $\tau_N$ stays in a fixed compact real interval.
For every fixed $y\in\mathbb R$,
\[
 \Pp\{(\Smax-B_{n,p}^{\rm c})/A_n^{\rm c}\le y\}
                 \longrightarrow\exp\{-e^{-y}\}.
\]
\end{theorem}

The lower critical calibration uses $g_N$ in \eqref{eq:main-IG-kernel}.
For $y>1$ let $\vartheta(y)=(1-y^{-2})/4$, and, for $u/N$ near $q_c$,
write
\[
 \mathcal L_N(u)=
 \log\frac1{(b-\vartheta(u/N))_++N^{-1/2}}.
\]
This logarithm is positive for all sufficiently large $n$ in the
following theorem. Define
\begin{equation}\label{eq:lower-unified-intensity}
 \begin{aligned}
 \mathcal R_N^{\mathrm{lc}}(u)
 &=N(q_c-1)\Theta_0 g_N(u),&&\alpha_\eta<1,\\
 &=N(q_c-1)P_0\{\mathcal L_N(u)\}^{\beta_\eta+1}g_N(u),
       &&\alpha_\eta=1,\\
 &=\frac{N(q_c-1)P_A}{\Gamma(\alpha_\eta-1)}
       \int_0^u w^{\alpha_\eta-2}\{\log(e+w)\}^{\beta_\eta}
                      e^{-bw}g_N(u-w)\,dw,&&\alpha_\eta>1.
 \end{aligned}
\end{equation}
The kernel $g_N$ makes this calibration integral positive.

Define the leading raw level and tilt by
\begin{equation*}
 y_n=
 \begin{cases}
  (\sqrt{\gamma_n}+\sqrt{1+\gamma_n})^2,&\gamma_n\le\gamma_c,\\
  \{\gamma_n+\lambda(b)\}/b,&\gamma_n\ge\gamma_c,
 \end{cases}
 \qquad
 u_{0,n}=Ny_n,\qquad \zeta_n=\min\{\vartheta(y_n),b\}.
\end{equation*}
The definitions agree at \(\gamma_n=\gamma_c\).
Set
\begin{equation}\label{eq:lower-unified-normalization}
 \begin{gathered}
 H_n^{\rm lc}=\frac{\log\{p\mathcal R_N^{\mathrm{lc}}(u_{0,n})\}}{\zeta_n},
 \qquad
 \mathfrak b_{n,p}^{\mathrm{lc}}
       =\frac{u_{0,n}-N+H_n^{\rm lc}}{\sqrt{2N}},\\
 \mathfrak a_{n,p}^{\mathrm{lc}}
       =\frac1{\zeta_n\sqrt{2N}},
 \end{gathered}
\end{equation}

\begin{theorem}[All lower critical approaches and the bounded upper core]
\label{thm:max-critical-lower-unified}
Under the pairwise-correlation and neighborhood conditions of
Assumption~\ref{ass:copula-null}, let $b=b_\eta<1/4$ and suppose
\begin{equation}\label{eq:lower-unified-range}
 \gamma_n\longrightarrow\gamma_c,\qquad
 \limsup_{n\to\infty}\tau_N<\infty .
\end{equation}
Then
\[
 \Pp\left\{\frac{\Smax-\mathfrak b_{n,p}^{\mathrm{lc}}}
                    {\mathfrak a_{n,p}^{\mathrm{lc}}}\le x\right\}
       \longrightarrow\exp(-e^{-x}),\qquad x\in\mathbb R.
\]
\end{theorem}

\begin{theorem}[Original MAX throughout the critical transition]
\label{thm:max-full-critical}
Assume \ref{ass:copula-pairwise} and
\ref{ass:copula-neighborhood} of Assumption~\ref{ass:copula-null},
\(b=b(\eta)<1/4\), and
\[
 \gamma_n=\frac{\log p}{\log n}\longrightarrow\gamma_c.
\]
For each fixed \(x\in\mathbb R\),
\[
 \Pp\left\{\frac{\Smax-B_{n,p}^{\rm C}}
                       {A_{n,p}^{\rm C}}\le x\right\}
       \longrightarrow \exp\{-e^{-x}\}.
\]
\end{theorem}

\begin{theorem}[Analytic MAX limit in the fixed boundary phase]
\label{thm:max-full-boundary}
Assume the pairwise-correlation and neighborhood conditions of
Assumption~\ref{ass:copula-null}, \(b<1/4\),
and a limit \(\gamma>\gamma_c\) in
\eqref{eq:max-polynomial-regime}. Then
\[
 \Pp\!\left\{
  \frac{\Smax-B_{n,p}^{\rm B}}
       {A_{n,p}^{\rm B}}\le x\right\}
       \longrightarrow \exp(-e^{-x}),\qquad x\in\mathbb R .
\]
\end{theorem}
\subsection{Weak-correlation Poisson reduction}

Let $E_{j,n}$ be an event measurable in Gaussian column $j$. The following
result converts its marginal probability into a coordinate-maximum law.
For each positive integer \(r\), define the ordered distinct and separated sets
\[
 \begin{split}
 \mathcal J_r&=\{(j_1,\ldots,j_r)\in\{1,\ldots,p\}^r:
                                      j_a\ne j_b\ (a\ne b)\},\\
 \mathcal S_r&=\{J\in\mathcal J_r:j_a\notin\mathcal B_{j_b,p}
                                      \text{ for every }a\ne b\},\\
 \kappa_*&=(1-\rho_\star)/(1+\rho_\star)>0.
 \end{split}
\]

\begin{proposition}[Poisson reduction for coordinate events]
\label{prop:sparse-factorization}
Under the null Gaussian-copula model in
Assumption~\ref{ass:copula-null}, with the bounds in
\ref{ass:copula-pairwise} and \ref{ass:copula-neighborhood}, let $E_{j,n}$ be identically defined
measurable events in the individual latent Gaussian columns, with common
probability $\pi_n$. If $p\pi_n\to\lambda\in(0,\infty)$, then
\[
 \sum_{j=1}^p\mathbf1(E_{j,n})\ \xrightarrow d\
 \operatorname{Poisson}(\lambda).
\]
For every fixed \(r\),
\[
 \begin{split}
 \left|\sum_{J\in\mathcal S_r}P(\cap_{j\in J}E_{j,n})-(p\pi_n)^r\right|
       &\le C_r(\delta_p\log p+D_p/p),\\
 \sum_{J\in\mathcal J_r\setminus\mathcal S_r}P(\cap_{j\in J}E_{j,n})
       &\le C_rD_p^{r-1}p^{-\kappa_*}e^{C_r\delta_p\log p}.
 \end{split}
\]
\end{proposition}
\begin{proof}
For Gaussian blocks with block covariance $T$ and exponents $a_j>0$, the
Gaussian H\"older inequalities give
\[
 T\preceq\operatorname{diag}(a_j I)
 \ \Longrightarrow\
 \E\prod_j f_j\le\prod_j(\E f_j^{a_j})^{1/a_j},
\]
and, in the reverse order,
\[
 T\succeq\operatorname{diag}(a_jI)
 \ \Longrightarrow\
 \E\prod_jf_j\ge\prod_j(\E f_j^{a_j})^{1/a_j}
\]
\citep[Theorem~1]{ChenDafnisPaouris2015}. Approximation permits
\(f_j=\mathbf1_{E_{j,n}}\). For a separated \(r\)-tuple,
\[
 (1-\delta_p)I_r\preceq R_J\preceq(1+\delta_p)I_r
 \ \Longrightarrow\
 (1-\delta_p)I_{rn}\preceq R_J\otimes I_n
                         \preceq(1+\delta_p)I_{rn}.
\]
The exponents therefore do not depend on the number of rows, and
\begin{equation}\label{eq:tail-separated-factorization}
 \pi_n^{r/(1-\delta_p)}
 \le\Pp\left(\bigcap_{j\in J}E_{j,n}\right)
 \le\pi_n^{r/(1+\delta_p)}.
\end{equation}
For a separated ordered tuple \(J\), this implies
\[
 \left|\log\frac{\Pp(\bigcap_{j\in J}E_{j,n})}{\pi_n^r}\right|
 \le \frac{r\delta_p}{1-\delta_p}\log(1/\pi_n)
 =O_r(\delta_p\log p).
\]
The separated tuple count satisfies
\[
 |\mathcal S_r|=(p)_r+O_r(D_pp^{r-1}),\qquad
 \sum_{J\in\mathcal S_r}\Pp\Bigl(\bigcap_{j\in J}E_{j,n}\Bigr)
 =(p\pi_n)^r\{1+O_r(\delta_p\log p+D_p/p)\}.
\]
For a nonseparated tuple \(J\), join \(j,k\) when
\(j\in\mathcal B_{k,p}\) or \(k\in\mathcal B_{j,p}\). Write
\[
 J=J_1\mathbin{\dot\cup}\cdots\mathbin{\dot\cup}J_c,\qquad
 |J_i|=m_i,\quad c<r,\quad\sum_i m_i=r.
\]
For \(j\in J_i\),
\[
 \sum_{\substack{k\in J_i\\k\ne j}}|\rho_{jk}|
       \le(m_i-1)\rho_\star,
 \qquad
 \sum_{k\in J\setminus J_i}|\rho_{jk}|\le\delta_p.
\]
Set \(D_{jj}=1+(m_i-1)\rho_\star+\delta_p\) on \(J_i\).
For every real vector \(z\),
\[
 z^\top(D-R_J)z
 \ge\sum_j\left(D_{jj}-1-\sum_{k\ne j}|\rho_{jk}|\right)z_j^2
 \ge0.
\]
Gaussian H\"older with these diagonal exponents gives
\[
 \Pp\left(\bigcap_{j\in J}E_{j,n}\right)
 \le\pi_n^{\sum_{i=1}^c m_i/\{1+(m_i-1)\rho_\star+\delta_p\}}.
\]
Choose one root for each component and a spanning tree. This counts at most
$C_rp^cD_p^{r-c}$ tuples of the specified component sizes. For $m\ge2$,
\[
 \frac{m}{1+(m-1)\rho_\star}-1
 =\frac{(m-1)(1-\rho_\star)}{1+(m-1)\rho_\star}\ge\kappa_*.
\]
At least one component is nontrivial. Summing over the finitely many
component-size patterns gives, for \(N_n=\sum_{j=1}^p\mathbf1(E_{j,n})\),
\[
 \begin{split}
 0\le\sum_{J\notin\mathcal S_r}\Pp\Bigl(\bigcap_{j\in J}E_{j,n}\Bigr)
 &\le C_rD_p^{r-1}p^{-\kappa_*}e^{C_r\delta_p\log p}\longrightarrow0,\\
 \big|\E(N_n)_r-(p\pi_n)^r\big|
 &\le C_r(p\pi_n)^r(\delta_p\log p+D_p/p)\\
 &\quad+C_rD_p^{r-1}p^{-\kappa_*}e^{C_r\delta_p\log p}
 \longrightarrow0.
 \end{split}
\]
Thus \(\E(N_n)_r\to\lambda^r\). Ordinary moments are finite linear
combinations of factorial moments, and the Poisson moment generating
function is finite near zero; the method of moments proves the result.
\end{proof}

\subsection{Exact transform and saddlepoint calculation for the Gaussian field}

Let $X(u,v)$ be a centered Gaussian field with covariance
\[
 \E X(u,v)X(u',v')=e^{-|u-u'|/2}e^{-|v-v'|/2},
 \qquad u,u'\in\R,\quad v,v'\ge0,
\]
and put $Q_R(u)=\int_0^R X(u,v)^2\,dv$. The outer coordinate $u$ is logit
time; $v$ is the inner logit coordinate of the rank threshold. The field
\(Q_R\) provides the Gaussian reference transform.

Let \(r_G(z)=\sqrt{1-8z}\) be the principal square root and
\(d_G(z)=\{1-r_G(z)\}/\{1+r_G(z)\}\). Set
\[
 \begin{gathered}
 \Lambda(z)=\frac{1-\sqrt{1-8z}}4,\qquad
 \theta(q)=\frac{1-q^{-2}}8,\qquad
 \Lambda''\{\theta(q)\}=4q^3,\\
 \mathcal C_0\{\theta(q)\}=\frac{2\sqrt q}{q+1},\qquad
 \mathcal I(q)=q\theta(q)-\Lambda\{\theta(q)\}
             =\frac{(q-1)^2}{8q}.
 \end{gathered}
\]

\begin{lemma}[Gaussian transform and fixed-anchor sharp tail]
\label{lem:gaussian-transform-correct}
For \(\operatorname{Re}z<1/8\),
\begin{equation}\label{eq:gaussian-exact-mgf-correct}
 M_R(z):=\E e^{zQ_R(0)}
 =\frac{2\sqrt{r_G(z)}}{1+r_G(z)}
 e^{R\{1-r_G(z)\}/4}
 \{1-d_G(z)^2e^{-r_G(z)R}\}^{-1/2}.
\end{equation}
For \(q\) in a compact subset of \((1,\infty)\),
uniformly as \(R\to\infty\),
\begin{equation*}
 \Pp\{Q_R(0)>Rq\}
 =\frac{\mathcal C_0\{\theta(q)\}}
 {\theta(q)\sqrt{2\pi R\Lambda''\{\theta(q)\}}}
 e^{-R\mathcal I(q)}\{1+o(1)\},
\end{equation*}
\end{lemma}

Differentiating the transform at zero gives
\(\E Q_R=R\) and \(\operatorname{Var}(Q_R)=4(R-1+e^{-R})\).
\begin{proof}
For fixed $u$, write $Z_s=X(u,s)$. Its stochastic differential equation
and generator are
\[
 dZ_s=-\tfrac12 Z_s\,ds+dW_s,\qquad
 \mathcal L f(x)=\tfrac12 f''(x)-\tfrac12 x f'(x).
\]
The Feynman--Kac representation for Markov additive functionals is given
in \citet[pp.~118--119, equations~(7)--(8)]{FitzsimmonsPitman1999}.
A fixed horizon is a hitting time for the space--time process $(s,Z_s)$.
For real $\theta<1/8$, we verify the representation by solving its backward
equation and checking integrability:
\[
 \partial_T f(T,x)=\mathcal L f(T,x)+\theta x^2f(T,x),
 \qquad f(0,x)=1.
\]
Substitution of $f(T,x)=\exp\{a_Tx^2/2+c_T\}$ gives
\[
 a'_T=a_T^2-a_T+2\theta,\qquad
 c'_T=a_T/2,\qquad a_0=c_0=0.
\]
Write $r_G=r_G(\theta)$, $d_G=d_G(\theta)$ and $a_\pm=(1\pm r_G)/2$; then $d_G=a_-/a_+$.
Separating the two roots of the quadratic equation gives
\[
 \frac{a_T-a_-}{a_T-a_+}=d_G e^{-r_G T},
 \qquad
 a_T=a_-\frac{1-e^{-r_G T}}{1-d_G e^{-r_G T}}.
\]
Integration of $c'_T=a_T/2$ then yields
\[
 c_T=\frac{a_-T}{2}
      -\frac12\log\frac{1-d_G e^{-r_G T}}{1-d_G}.
\]
These formulas include $\theta=0$ by continuity. They give
\[
 \begin{cases}
  a_T\le0,\quad c_T\le0,&\theta\le0,\\
  0\le a_T\le a_-<1/2,\quad 0\le c_T\le a_-T/2,
       &0<\theta<1/8.
 \end{cases}
\]
Fix $T<\infty$ and let $\tau_L=\inf\{s:|Z_s|\ge L\}$, where $L>|x|$.
It\^o's formula makes the following stopped process a martingale:
\[
 M_{s\wedge\tau_L}
 =\exp\left\{\theta\int_0^{s\wedge\tau_L}Z_v^2\,dv\right\}
  f(T-s\wedge\tau_L,Z_{s\wedge\tau_L}),
 \qquad 0\le s\le T.
\]
For $\theta\le0$ it lies in $[0,1]$. For $\theta>0$, choose
$\theta_*\in(\theta,1/8)$ and $p_*>1$ sufficiently close to one that
\[
 p_*\theta\le\theta_*,
 \qquad p_*a_-(\theta)\le a_*(\theta_*),
 \qquad a_*(\theta_*):=\frac{1-\sqrt{1-8\theta_*}}2.
\]
The function $\phi_*(x)=e^{a_*(\theta_*)x^2/2}$ satisfies
\[
 (\mathcal L+\theta_*x^2)\phi_*=\frac{a_*(\theta_*)}{2}\phi_*.
\]
Thus It\^o's formula gives a nonnegative local martingale
\[
 L_s=\exp\left\{\theta_*\int_0^sZ_v^2\,dv
                  -\frac{a_*(\theta_*)s}{2}\right\}\phi_*(Z_s),
 \qquad
 \E_x L_{s\wedge\tau_L}\le\phi_*(x).
\]
The preceding inequalities imply the bound
\[
 \sup_{L>|x|}\sup_{0\le s\le T}
 \E_x M_{s\wedge\tau_L}^{\,p_*}
 \le
 \exp\left\{\frac{p_*a_-(\theta)+a_*(\theta_*)}{2}T\right\}\phi_*(x).
\]
Consequently the stopped martingales are uniformly integrable.
Continuity of the Ornstein--Uhlenbeck paths and $L\to\infty$ now give
\[
 f(T,x)=\E_x\exp\left\{\theta\int_0^T Z_s^2\,ds\right\}<\infty.
\]
Under stationarity, $Z_0\sim N(0,1)$. Since $a_T<1$, direct Gaussian
integration gives
\[
 M_T(\theta)
 =\frac{e^{c_T}}{\sqrt{2\pi}}
   \int_{\mathbb R}e^{-(1-a_T)x^2/2}\,dx
 =\frac{e^{c_T}}{\sqrt{1-a_T}}.
\]
Using $a_++a_-=1$, we obtain
\[
 1-a_T
 =a_+\frac{1-d_G^2e^{-r_G T}}{1-d_G e^{-r_G T}},
 \qquad
 \frac{e^{c_T}}{\sqrt{1-a_T}}
 =\frac{2\sqrt{r_G}}{1+r_G}\,
   e^{T(1-r_G)/4}(1-d_G^2e^{-r_G T})^{-1/2}.
\]
This proves \eqref{eq:gaussian-exact-mgf-correct} for real $\theta<1/8$.
For any compact $K\subset\{z:\operatorname{Re}z<1/8\}$, choose
$\theta_*>\max(0,\sup_{z\in K}\operatorname{Re}z)$ with $\theta_*<1/8$.
For every fixed nonnegative integer $h$,
\[
 \sup_{z\in K}Q_R^h|e^{zQ_R}|
 \le C_{K,h}e^{\theta_*Q_R},
 \qquad \E e^{\theta_*Q_R}<\infty.
\]
Hence $M_R$ is holomorphic on that half-plane. On the same domain,
\[
 \operatorname{Re}r_G(z)>0,\qquad
 |d_G(z)|<1,\qquad |d_G(z)^2e^{-r_G(z)R}|<1.
\]
The right side of \eqref{eq:gaussian-exact-mgf-correct} is therefore
holomorphic with its branches fixed at $z=0$. The identity theorem
extends the real formula to the stated complex domain.
Finally, stationarity and the normal fourth-moment identity give
\[
 \E Q_R=R,\qquad
 \operatorname{Var}(Q_R)
 =2\int_0^R\int_0^R e^{-|s-t|}\,ds\,dt
 =4(R-1+e^{-R}).
\]
For real $\theta=\theta(q)$, introduce the exponential tilt
$d\Pp_\theta=e^{\theta Q_R}d\Pp/M_R(\theta)$. The exact formula gives
\[
 (\log M_R)'(\theta)=Rq+O(1),\qquad
 (\log M_R)''(\theta)=4Rq^3+O(1),
\]
uniformly on the stated compact set. Here and below constants can depend on
that compact set.
The saddlepoint calculation requires a local normal approximation. Under the
tilt the characteristic function of $Q_R-Rq$ is
$e^{-itRq}M_R(\theta+it)/M_R(\theta)$. For $|t|\le t_0$, its logarithm is
\[
 -2Rq^3t^2+O(R|t|^3+|t|).
\]
Let \(\varphi_{R,\theta}(t)=e^{-itRq}M_R(\theta+it)/M_R(\theta)\).
The exact square-root formula, including its terminal factor, gives
constants \(t_0,T,c,C,d>0\), uniform over the stated compact set, such that
\[
 |\varphi_{R,\theta}(t)|\le
 \begin{cases}
 C e^{-cRt^2},&|t|\le t_0,\\
 C e^{-cR},&t_0\le|t|\le T,\\
 C(1+|t|)^d e^{-cR\sqrt{|t|}},&|t|\ge T.
 \end{cases}
\]
Consequently
\[
 \int_{\mathbb R}|\varphi_{R,\theta}(t)|\,dt\le C R^{-1/2},
 \qquad
 \varphi_{R,\theta}(w/\sqrt R)\longrightarrow e^{-2q^3w^2}.
\]
Fourier inversion and dominated convergence therefore give a density
\(f_{R,\theta}\) with
\[
 \sup_y f_{R,\theta}(y)\le C/\sqrt R,\qquad
 \sqrt R\, f_{R,\theta}(Rq+y)
 \longrightarrow(8\pi q^3)^{-1/2}
\]
uniformly for bounded $y$. Consequently
\[
 \Pp(Q_R>Rq)=M_R(\theta)e^{-\theta Rq}
 \int_0^\infty e^{-\theta y}f_{R,\theta}(Rq+y)\,dy.
\]
The uniform density bound permits dominated convergence in this integral.
Since $\theta q-\Lambda(\theta)=\mathcal I(q)$, the result follows.
\end{proof}

\subsection{A nonlinear boundary constant and the subcritical fixed-split tail}

We write $T_{n,k}=nF_{n,k}$ and use $\vartheta$ to tilt $T_{n,k}$.
Thus $\vartheta=2\theta(q)$ for the Gaussian parameter above.

This subsection concerns marginal fixed-split laws.

For $t\in(0,1)$ set $h_t(x)=\operatorname{KL}(x\Vert t)$. For a measurable
function $f:[0,\infty)\to[0,1]$, put
\[
 Hf(s)=s^{-1}\int_0^s f(u)\,du,\qquad
 I_t(f)=\int_0^\infty h_t(f(s))\,ds,\qquad
 J_t(f)=\int_0^\infty h_t(Hf(s))\,ds,
\]
and define the modular Hardy constant and its reciprocal by
\begin{equation*}
 C(t)=\sup_{0<I_t(f)<\infty}\frac{J_t(f)}{I_t(f)},\qquad
 b(t)=C(t)^{-1}.
\end{equation*}
The supremum can equivalently be taken over piecewise constant functions
equal to $t$ after a finite time, followed by approximation. The constant is
defined by a deterministic integral variational problem. The quadratic Hardy
inequality and $2(x-t)^2\le h_t(x)\le(x-t)^2/\{t(1-t)\}$ imply
\[
 4\le C(t)\le\frac2{t(1-t)},\qquad 0<b(t)\le\frac14.
\]
The lower bound follows from arbitrarily small perturbations and a quadratic
Hardy extremizing sequence.

For \(C>4\), let \(u_C\) denote the smaller-slope solution of
\begin{equation}\label{eq:hardy-shooting-ode}
 \frac{du_C}{dx}=\frac{u_C(1-u_C)h_t'(x)}{C(x-u_C)},\qquad
 u_C(t)=t,\quad u_C'(t)=\alpha_C,
 \quad \alpha_C=\frac{1-\sqrt{1-4/C}}2.
\end{equation}
Define \(C_+(t)\) as the infimum of \(C>4\) for which
\(t<u_C(x)<x\) throughout \(t<x<1\).
For \(0<\vartheta<b(t)\), put
\[
 r=\sqrt{1-4\vartheta},\qquad
 a=(1-r)/2,\qquad \kappa=a/2.
\]
Let \(u=u_{t,\vartheta}\) denote the central solution of
\begin{equation}\label{eq:bellman-u-ode}
 u'=\frac{\vartheta u(1-u)h_t'}{x-u},\qquad
 u(t)=t,\quad u'(t)=a,
\end{equation}
On its domain define
\begin{equation}\label{eq:bellman-g-definition}
 g_{t,\vartheta}(x)=(1-\vartheta)h_t(x)
                         -\operatorname{KL}(x\Vert u(x))
\end{equation}
The associated Bellman boundary-value problem is
\begin{equation}\label{eq:bernoulli-bellman-equation}
 g-xg'+\log(1-t+t e^{g'})+\vartheta h_t(x)=0,
 \qquad g(t)=g'(t)=0,\quad g''(t)=a/\{t(1-t)\}.
\end{equation}

\begin{lemma}[Hardy ODE and a strict Bellman branch]
\label{lem:hardy-bellman-branch}
\[
 C(t)=\max\{C_+(t),C_+(1-t)\}.
\]
The central branch \(u_{t,\vartheta}\) extends to \([0,1]\),
is bounded away from zero and one, and satisfies \(u(x)<x\)
above \(t\) and \(u(x)>x\) below \(t\).

The function in \eqref{eq:bellman-g-definition} solves
\eqref{eq:bernoulli-bellman-equation}, has bounded first
derivative on \([0,1]\), and is smooth in the interior.
Its second derivative has at most logarithmic endpoint singularities.

Moreover \(g\ge0\) and \(\partial_\vartheta g(x)>0\) for
\(x\ne t\). These properties and parameter analyticity are
uniform on compact strictly subcritical parameter sets.
\end{lemma}

\begin{proof}
The smaller-slope germ of \eqref{eq:hardy-shooting-ode} is
\[
 u_C(t+d)=t+\alpha_Cd+
 \frac{(1-2t)(\alpha_C-1/2)}
 {Ct(1-t)(2-3\alpha_C)}d^2+O(d^3).
\]
The chosen germ removes the central singularity. At a common point,
\[
 \frac{d\alpha_C}{dC}<0,\qquad
 \partial_C\left\{\frac{u(1-u)h_t'(x)}{C(x-u)}\right\}<0
 \quad(t<u<x),
\]
so scalar comparison makes admissibility monotone in \(C\).
To verify existence for large \(C\), set
\[
 K_t=\sup_{t<x\le1}\frac1{x-t}
                \int_t^x\frac{h_t'(y)}{y-t}\,dy<\infty .
\]
The integrand is bounded at \(t\) and has an integrable logarithm at \(1\).
While \(u_C(y)-t\le(y-t)/2\),
\[
 u_C(x)-t\le\frac1{2C}\int_t^x\frac{h_t'(y)}{y-t}\,dy
 \le\frac{K_t}{2C}(x-t)<\frac{x-t}{2}
 \quad(C>\max\{4,K_t\}).
\]
This excludes a first crossing and gives extension to \(1\).
For an admissible branch define
\[
 V_C(x)=C\operatorname{KL}(x\Vert u_C(x))-(C-1)h_t(x),
 \qquad V_C'(x)=-Ch_t'(u_C(x)).
\]
Convexity in \(z\in[0,1]\) now gives
\[
 h_t(x)\le C h_t(z)+V_C(x)+(z-x)V_C'(x),
\]
with equality at \(z=u_C(x)\). For \(x(s)=Hf(s)\),
\(sx'(s)=f(s)-x(s)\), and therefore
\[
 \begin{split}
 \int_\epsilon^M h_t(x(s))\,ds
 &\le C\int_\epsilon^M h_t(f(s))\,ds
      +[sV_C(x(s))]_\epsilon^M,\\
 \epsilon V_C(x(\epsilon))&\longrightarrow0,\qquad
 M V_C(x(M))=O\{M(x(M)-t)^2\}=O(M^{-1})\longrightarrow0.
 \end{split}
\]
The reflected and upper potentials satisfy
\[
 V_C(t-)=V_C(t+)=0,\qquad V_C'(t-)=V_C'(t+)=0.
\]
Their boundary terms therefore cancel at every crossing of \(t\).
For controls equal to \(t\) after a finite time, the preceding estimate gives
\[
 J_t(f)\le CI_t(f)
\]
after \(\epsilon\downarrow0\), \(M\uparrow\infty\); approximation extends it
to the admissible class. Conversely, take
\[
 4<C<C_+(t),\qquad u_C(x_0)=x_0,\quad t<x_0<1,\qquad
 f(s)=x_0\quad(0<s\le1).
\]
For \(s>1\), use the equality trajectory with \(z=\log s\):
\[
 \frac{dx}{dz}=u-x,\qquad
 \frac{du}{dz}=-\frac{u(1-u)}C h_t'(x),\qquad
 x(0)=u(0)=x_0.
\]
On this trajectory,
\[
 |x(s)-t|+|u(s)-t|=O(s^{-k_C}),\qquad
 k_C=\frac{1+\sqrt{1-4/C}}2>\frac12,
\]
so
\[
 \int_1^\infty\{h_t(x(s))+h_t(u(s))\}\,ds
 \le C\int_1^\infty s^{-2k_C}\,ds<\infty.
\]
The prefix also satisfies equality because
\(V_C(x_0)=(1-C)h_t(x_0)\). Thus
\[
 J_t(f)=CI_t(f),\qquad
 \sup_fJ_t(f)/I_t(f)\ge C_+(t)
\]
by \(C\uparrow C_+(t)\); at threshold \(4\), the quadratic lower bound
already matches. Reflection gives the two-sided characterization.
For \(C=1/\vartheta>C(t)\), strict comparison gives
\[
 0<\inf_{[0,1]}u\le\sup_{[0,1]}u<1,\qquad
 g'=\operatorname{logit}(u)-\operatorname{logit}(t).
\]
Substitution proves \eqref{eq:bernoulli-bellman-equation}; the ODE gives
bounded \(g'\) and logarithmic endpoint growth of \(g''\).
For \(q_\vartheta=\partial_\vartheta g\), differentiation gives
\[
 q_\vartheta+(u-x)q_\vartheta'=-h_t(x).
\]
Along $dx/ds=u(x)-x$, starting at $x_0$, the solution tends to $t$ at rate
$1-a>1/2$. Thus
\[
 \partial_\vartheta g(x_0)
 =\int_0^\infty e^s h_t(x(s))\,ds>0\quad(x_0\ne t).
\]
Together with $g_{t,0}=0$ this proves nonnegativity. Local analytic
dependence follows from the removable germ and ordinary parameter dependence
away from the central point; compact strict-subcritical sets stay a positive
distance from contact.
\end{proof}

Let $N_\ell$ be the number of successes in the first $\ell$ members of an
iid Bernoulli$(t)$ sequence. For $c>-1$ put
\[
 T_m^c=\sum_{\ell=1}^m\frac{\ell}{\ell+c}h_t(N_\ell/\ell),
 \qquad v=t(1-t).
\]
For the original statistic, $c=-1/2$ at the lower rank end and $c=+1/2$ at
the upper rank end. Define a positive transport amplitude
$A=A_{t,\vartheta,c}$ by $A(t)=1$ and
\begin{equation}\label{eq:boundary-transport-amplitude}
 (u-x)(\log A)'=
 \kappa+\vartheta c h_t-\vartheta(u-x)h_t'
 -\tfrac12g''\{u(1-u)+(u-x)^2\}.
\end{equation}
The quotient has a removable singularity at $t$: the constant numerator
there is $\kappa-a/2=0$. At the endpoints its derivative is at most
logarithmic. Thus $A$ and $A^{-1}$ are bounded on $[0,1]$. Set
\[
 W_m^c(s;\vartheta)=m^{-\kappa}A(s/m)e^{m g(s/m)}.
\]

\begin{lemma}[Convergent analytic Bernoulli endpoint coefficient]
\label{lem:analytic-boundary-coefficient}
For $0<\vartheta<b(t)$ the limit
\begin{equation}\label{eq:analytic-boundary-coefficient}
 K_c(t,\vartheta)=\lim_{m\to\infty}
 E\{e^{\vartheta T_m^c}W_m^c(N_m;\vartheta)\}
\end{equation}
exists in $(0,\infty)$, with real-parameter error $O(m^{-1})$, uniformly on
compact strict-subcritical parameter sets. It extends analytically to a
complex neighborhood of every real such point, and the convergence is
locally uniform there.
\end{lemma}

\begin{proof}
Write $K_m(\vartheta)$ for the expectation in
\eqref{eq:analytic-boundary-coefficient}. The exact row ratio is
\[
 R_m(s)=\sum_{b=0}^1t^b(1-t)^{1-b}
 \exp\!\left\{\vartheta\frac{m+1}{m+1+c}
 h_t\!\left(\frac{s+b}{m+1}\right)\right\}
 \frac{W_{m+1}^c(s+b)}{W_m^c(s)}.
\]
Set $x=s/m$ and expand $(s+b)/(m+1)=x+(b-x)/(m+1)$. The order-one row sum is
one by the Bellman equation. Under this order-one tilt, $b$ is
Bernoulli$(u(x))$. With $d=u-x$ and $M_2=u(1-u)+d^2$, the coefficient of
$m^{-1}$ is
\[
 -\kappa+\tfrac12g''M_2+d(\log A)'
                 +\vartheta d h_t'-\vartheta c h_t=0.
\]
Hence
\[
 R_m(s)=1+O(m^{-2})\quad(s/m\in[\epsilon,1-\epsilon]),\qquad
 0<c\le R_m(s)\le C
\]
globally, by boundedness of \(g',h_t,A,A^{-1}\).
For the interpolated Bernoulli path put
\[
 \mathcal L=\{f:f(0)=0,\ 0\le f'\le1\},\quad
 \mathcal I(f)=\int_0^1h_t(f')\,ds,\quad
 \mathcal V(f)=\vartheta\int_0^1h_t(f(s)/s)\,ds+g(f(1)).
\]
On the compact Lipschitz path space,
\[
 \sup_{f\in\mathcal L}\int_0^\epsilon h_t(f(s)/s)\,ds
 \le\epsilon\|h_t\|_\infty,\qquad
 |T_m^c/m-T_m^0/m|\le C\log m/m.
\]
Thus \(\mathcal V\) is bounded and continuous. The bounded-increment
Laplace principle of \citet{Mogulskii1977} applies with rate
\(\mathcal I\). Legendre duality and the Bellman equation give
\[
 h_t(f')-\vartheta h_t(f/s)
 \ge g(f/s)+(f'-f/s)g'(f/s)
 =\frac d{ds}\{s g(f(s)/s)\}.
\]
Integration gives
\[
 \mathcal V(f)\le\mathcal I(f),\qquad
 \mathcal V(ts)=\mathcal I(ts)=0,\qquad
 \lim_{m\to\infty}m^{-1}\log K_m(\vartheta)
       =\sup_{\mathcal L}(\mathcal V-\mathcal I)=0 .
\]
Choose \(\vartheta<\vartheta'<b(t)\). If \(H\) is closed and
\(t\notin H\), then
\[
 \delta_H:=\inf_{x\in H}
             \{g_{\vartheta'}(x)-g_\vartheta(x)\}>0,\qquad T_m^c\ge0.
\]
Comparison of the two positive weights gives
\[
 E\{e^{\vartheta T_m^c}W_m^c(N_m;\vartheta)
                  \mathbf1(N_m/m\in H)\}
 \le C m^{\kappa(\vartheta')-\kappa(\vartheta)}
 e^{-m\delta_H}K_m(\vartheta')=e^{-m\delta_H+o(m)}.
\]
The exact identity
\[
 K_{m+1}-K_m
 =E\{e^{\vartheta T_m^c}W_m^c(N_m)(R_m(N_m)-1)\}
\]
therefore gives
\[
 |K_{m+1}-K_m|\le Cm^{-2}K_m+Ce^{-c_0m},\qquad
 K_m\ge A_{\min}m^{-\kappa}.
\]
For sufficiently large \(m_0\), iteration yields
\[
 \begin{split}
 \sup_{m\ge m_0}K_m
 &\le\left(K_{m_0}+C\sum_{j\ge m_0}e^{-c_0j}\right)
             \prod_{j\ge m_0}(1+Cj^{-2})<\infty,\\
 \inf_{m\ge m_0}K_m
 &\ge\prod_{j\ge m_0}(1-Cj^{-2})
       \left(K_{m_0}-C'\sum_{j\ge m_0}e^{-c_0j}\right)>0,\\
 \sum_{j\ge m}|K_{j+1}-K_j|&\le C/m .
 \end{split}
\]
Here \(m_0\) is chosen so that the exponential tail in the second line is
less than \(A_{\min}m_0^{-\kappa}/2\). This proves the asserted positive
limit and its remainder, and gives the explicit series
\[
 K_c=K_1+\sum_{m\ge1}
 E\{e^{\vartheta T_m^c}W_m^c(N_m)(R_m(N_m)-1)\}.
\]
For complex $z$ near a real $\vartheta$, choose real
$\vartheta_2>\vartheta_1>\vartheta$ below the threshold and shrink the
complex neighborhood so that
\[
 \operatorname{Re}z<\vartheta_1,\quad
 \operatorname{Re}g_z\le g_{\vartheta_1},\quad
 0\le\kappa(\vartheta_1)-\operatorname{Re}\kappa(z)<\delta<1.
\]
The central quadratic expansion and strict off-center monotonicity
give, after shrinking the neighborhood,
\[
 \E[e^{\operatorname{Re}zT_m^c}|W_m^c(N_m;z)|]
 \le Cm^{\kappa(\vartheta_1)-\operatorname{Re}\kappa(z)}
          K_m(\vartheta_1)\le Cm^\delta .
\]
For endpoint rows in a fixed closed set \(H\) away from \(t\),
\[
 \inf_{z\in U,\ x\in H}
       \{g_{\vartheta_2}(x)-\operatorname{Re}g_z(x)\}>0 .
\]
The resulting exponential endpoint margin and interior row error give
\[
 \sup_{z\in U}|K_{m+1}(z)-K_m(z)|
 \le Cm^{-2+\delta}+Ce^{-cm},\qquad
 \sum_{m\ge M}\sup_{z\in U}|K_{m+1}(z)-K_m(z)|
 \le C M^{-1+\delta}.
\]
The locally uniformly convergent series of analytic summands defines
an analytic continuation of \(K_c\) to \(U\).
\end{proof}

Put
\[
 \lambda(\vartheta)=\frac{1-\sqrt{1-4\vartheta}}2,\qquad
 \vartheta_q=\frac{1-q^{-2}}4,\qquad
 J(q)=\frac{(q-1)^2}{4q},
\]
and define
\begin{equation*}
 \Psi_t(\vartheta)=\sqrt{r(\vartheta)}
             K_{-1/2}(t,\vartheta)K_{+1/2}(t,\vartheta),
 \qquad r(\vartheta)=\sqrt{1-4\vartheta}.
\end{equation*}

\begin{theorem}[Analytic strict-subcritical fixed-split tail]
\label{thm:rank-fixedsplit-subcritical}
Uniformly for \((t,q)=(k/n,q)\) in compact subsets of
\[
 \{(t,q):0<t<1,\ q>1,\ \vartheta_q<b(t)\},
\]
and \(|y_n|\le h_n\sqrt{\log n}\), where \(h_n\to0\)
is deterministic,
\begin{equation}\label{eq:rank-fixedsplit-subcritical-tail}
 P\{nF_{n,k}>q\log n+y_n\}
 =\frac{\Psi_t(\vartheta_q)}
 {\vartheta_q\sqrt{2\pi\log n\,\lambda''(\vartheta_q)}}
 n^{-J(q)}e^{-\vartheta_q y_n}\{1+o(1)\},
 \qquad \lambda''(\vartheta_q)=2q^3.
\end{equation}
\end{theorem}

\begin{proof}
Set
\[
 L=\log n,\qquad K=8,\qquad m=\lfloor L^K\rfloor,\qquad
 (t,q)\in\mathcal K\Subset
 \{(t,q):t\in(0,1),\ q>1,\ \vartheta_q<b(t)\}.
\]
All constants are uniform on \(\mathcal K\). Reveal the bottom and
top \(m\) labels. For a revealed path with \(b\) successes, the likelihood
relative to \(2m\) independent Bernoulli\((t)\) labels is
\[
 \frac{P\{\operatorname{Bin}(n-2m,t)=k-b\}}
 {P\{\operatorname{Bin}(n,t)=k\}}=1+O(m^2/n),
\]
uniformly over \(0\le b\le2m\). This is the conditional-binomial
identity; Stirling's formula at distance \(O(m)\) from the mean gives its
relative \(O(m^2/n)\) remainder. The exact entropy-table decomposition
also gives
\[
 |T_{\rm lower}+T_{\rm upper}-T_m^{-1/2}-T_m^{+1/2}|
 \le C\sum_{l\le m}\frac ln\le Cm^2/n.
\]
Here the other entropy cell is \(O(l^2/n)\), and its weight is \(O(l^{-1})\);
the limiting-weight correction is \(O(n^{-1})\).
Let \(\mathcal F_{n,m}^{\partial}\) be the sigma-field of these two revealed
paths. Conditional on \(\mathcal F_{n,m}^{\partial}\), the middle endpoints
and success fraction are
\[
 A_m=s_-,\qquad A_{n-m}=-s_+,\qquad
 t_{\rm mid}=t-\frac{s_-+s_+}{n-2m}=t+O(m/n).
\]
The bridge-specific KMT theorem of \citet[Theorem~1.2 and
Remark~1.3]{DimitrovWu2021}, with its constants uniform on compact slopes,
gives a conditional coupling with a Gaussian bridge of these endpoints such
that, for every fixed $A>0$,
\[
 P\{\sup_{m\le l\le n-m}|A_l-G_l|>C_A L
                   \mid\mathcal F_{n,m}^{\partial}\}\le n^{-A}.
\]
Using the same standard bridge to replace
\(t_{\rm mid}(1-t_{\rm mid})\) by \(v=t(1-t)\) costs
\(O(m\sqrt{L/n})\) outside probability \(n^{-A}\).
Put \(b_l=l\wedge(n-l)\). Hypergeometric concentration and the
conditional Gaussian bridge bounds give a common event, of exceptional probability at most \(Cn^{-A}\), on
which
\[
 |A_l|+|G_l|\le B_A\sqrt{b_lL},\qquad
 |A_l-G_l|\le C_A L,\qquad
 \max_l|G_{l+1}-G_l|\le C_A\sqrt L .
\]
On this event, put
\[
 \widehat w_l=\frac{n^2}{d_{l,n}\,l(n-l)},\qquad
 G_{k,l}=l\,h_t(C_l/l)
       +(n-l)h_t\{(k-C_l)/(n-l)\}.
\]
Taylor's formula for these two entropy cells gives
\[
 \begin{split}
 \left|\sum_{l=m}^{n-m}\frac{nG_{k,l}}{d_{l,n}}
       -\frac1{2v}\sum_{l=m}^{n-m}\widehat w_l A_l^2\right|
 &\le C\sum_{l=m}^{n-m}\frac{n}{d_{l,n}}
                         \frac{|A_l|^3}{b_l^2}\\
 &\le CL^{3/2}\sum_{l=m}^{n-m}b_l^{-3/2}
 \le CL^{3/2}/\sqrt m,\\
 \frac1{2v}\left|\sum_{l=m}^{n-m}\widehat w_l(A_l^2-G_l^2)\right|
 &\le C\sum_{l=m}^{n-m}b_l^{-2}|A_l^2-G_l^2|\\
 &\le C\sum_{l=m}^{n-m}
       \{L^{3/2}b_l^{-3/2}+L^2b_l^{-2}\}\\
 &\le C\{L^{3/2}/\sqrt m+L^2/m\}.
 \end{split}
\]
For the continuous Gaussian bridge \(G(x)\), put
\[
 w(x)=\frac{n^2}{x^2(n-x)^2},\qquad
 Z_{\log(x/(n-x))}=\frac{G(x)}{\sqrt{vx(1-x/n)}}.
\]
Define the weighted approximation errors by
\[
 \mathcal E_{\rm shift}
 =\frac1{2v}\left|\sum_{l=m}^{n-m}(\widehat w_l-w(l))G_l^2\right|,
 \quad
 \mathcal E_{\rm Riemann}
 =\frac1{2v}\left|\sum_{l=m}^{n-m}w(l)G_l^2
                         -\int_m^{n-m}w(x)G(x)^2\,dx\right|.
\]
Since \(du/dx=n/[x(n-x)]\), the integral divided by \(2v\) is
\(\frac12\int Z_u^2du\). The Gaussian modulus between successive
integer arguments and the preceding envelope give
\[
 \begin{split}
 \mathcal E_{\rm shift}&\le CL\sum_{l\ge m}l^{-2}\le CL/m,\\
 |\Delta G_l|&\le C\sqrt L,\qquad
 |\log w(l+1)-\log w(l)|\le C/b_l,\\
 \mathcal E_{\rm Riemann}
 &\le C\left\{L\sum_{l\ge m}l^{-3/2}
                 +L\sum_{l\ge m}l^{-2}\right\}
 \le C(L/\sqrt m+L/m).
 \end{split}
\]
Together with the two preceding estimates, these vanish for \(K=8\).
Under the rank-logit transformation the middle has covariance
\[
 \operatorname{Cov}(Z_u,Z_{u'})=e^{-|u-u'|/2}
\]
and OU-bridge interval length
\[
 R_{n,m}=2\log\{(n-m)/m\}=2L-2\log m+O(m/n),
\]
and endpoints $x=s_-/\sqrt{mv(1-m/n)}$, $y=-s_+/\sqrt{mv(1-m/n)}$. Let
$\widetilde T_n$ be the sum of the two independent Bernoulli boundary
functionals and one half of this conditional OU square integral. Write
$Z_m^\pm=s_\pm/\sqrt{mv}$ and restrict to $\mathcal A_n=\{|Z_m^-|,|Z_m^+|\le
B_A\sqrt L\}$. The preceding probability-level construction gives, for every
$z$,
\begin{align}
 (1-o(1))P\{\widetilde T_n>z+\epsilon_n,\mathcal A_n\}-O(n^{-A})
 &\le P\{T_{n,k}>z\}\nonumber\\
 &\le(1+o(1))P\{\widetilde T_n>z-\epsilon_n,\mathcal A_n\}
                                      +O(n^{-A}),\label{eq:rank-hybrid-probability-sandwich}
\end{align}
where
\[
 \epsilon_n=C_A\{L^{3/2}/\sqrt m+L^2/m+m^2/n\}\longrightarrow0.
\]
The \(O(n^{-A})\) events have been removed at probability level before
exponential weighting. For \(w\) near the real saddle use the principal
square root to define
\[
 r(w)=\sqrt{1-4w},\qquad a(w)=\frac{1-r(w)}2,\qquad
 \kappa(w)=\frac{a(w)}2,\qquad\psi_w(x)=e^{a(w)x^2/2}.
\]
The exact conditional OU transform is
\begin{equation}\label{eq:exact-ou-bridge-transform}
 F_R(w;x,y)=e^{\kappa(w)R}\frac{\psi_w(x)}{\psi_w(y)}
                       \frac{p_{r(w)}(R;x,y)}{p_1(R;x,y)},
\end{equation}
where
\[
 p_r(R;x,y)=\sqrt{\frac r{2\pi(1-e^{-rR})}}
 \exp\!\left\{-\frac{r(y-e^{-rR/2}x)^2}{2(1-e^{-rR})}\right\}.
\]
Indeed $(\tfrac12\partial_x^2-\tfrac12x\partial_x+\tfrac w2x^2)
\psi_w=\kappa(w)\psi_w$, and conjugation gives OU drift $-rx/2$. Uniformly
on the endpoint cutoff, the exact kernel is
\[
 F_{R_{n,m}}(w;x,y)=e^{\kappa(w)R_{n,m}}\sqrt{r(w)}
                        \psi_w(x)\psi_w(y)\{1+O(Le^{-cR_{n,m}})\}.
\]
The eigenfunction matches the nonlinear boundary function because
\[
 W_m^c(N_m;w)=m^{-\kappa(w)}\psi_w(Z_m)\{1+O(\eta_n)\},
 \qquad \eta_n=L^{3/2}/\sqrt m+\sqrt{L/m},
\]
uniformly on the cutoff.
For the complex expectation, control the absolute replacement error. Choose a
slightly larger real $\vartheta_1$ still subcritical and shrink the complex
neighborhood so that
\[
 \operatorname{Re}w<\vartheta_1,\quad
 \operatorname{Re}g_w\le g_{\vartheta_1},\quad
 0\le\kappa(\vartheta_1)-\operatorname{Re}\kappa(w)
 \le\delta<\tfrac12-\tfrac3{2K}=\tfrac5{16}.
\]
The boundary lemma gives
\[
 \sup_w E[e^{\operatorname{Re}wT_m^c}|W_m^c(N_m;w)|]\le Cm^\delta,
 \qquad
 \eta_n m^\delta
 =O\{L^{3/2-K/2+K\delta}+L^{1/2-K/2+K\delta}\}\longrightarrow0.
\]
The eigenfunction error is therefore bounded by
\[
 \eta_n\,\E[e^{\operatorname{Re}wT_m^c}|W_m^c(N_m;w)|]
 \le C\eta_nm^\delta\longrightarrow0 .
\]
Choose a further real parameter \(\vartheta_2\). For fixed small \(\epsilon>0\),
\[
 g_{\vartheta_2}(x)-\operatorname{Re}g_w(x)\ge
 \begin{cases}
 c(x-t)^2,&|x-t|\le\epsilon,\\
 c_\epsilon,&|x-t|>\epsilon .
 \end{cases}
\]
Thus the cutoff complement has weighted mass at most
\(m^C(e^{-cB_A^2L}+e^{-cm})\). The remaining absolute errors satisfy
\[
 \begin{split}
 \mathcal E_{\rm middle}&\le CLe^{-cR_{n,m}}m^{2\delta},\\
 \mathcal E_{\rm cutoff}&\le m^C(e^{-cB_A^2L}+e^{-cm}),\\
 \mathcal E_{\rm endpoint}&\le C(1+L)m^{1+2\delta}/n,
 \end{split}
 \qquad
 \mathcal E_{\rm middle}+\mathcal E_{\rm cutoff}
                 +\mathcal E_{\rm endpoint}\longrightarrow0.
\]
The last line includes the endpoint variance \(1-m/n\) and the
\(O(m/n)\) length correction. Local uniform convergence of the analytic
boundary coefficients consequently gives
\begin{equation}\label{eq:truncated-hybrid-transform}
 \mathcal M_n(w):=E[e^{w\widetilde T_n}\mathbf1_{\mathcal A_n}]
 =n^{\lambda(w)}\{\Psi_t(w)+o(1)\},
\end{equation}
locally uniformly near the saddle. The powers cancel as
\[
 e^{\kappa(w)R_{n,m}}m^{2\kappa(w)}
 =n^{2\kappa(w)}\{1+O(m/n)\},\qquad 2\kappa(w)=\lambda(w).
\]
Since \(\Psi_t(\vartheta_q)>0\), a smaller neighborhood has
\(\inf|\Psi_t(w)|>0\) whenever a relative form is used.
For \(s\in\mathbb R\),
\[
 \operatorname{Re}r(\vartheta_q+is)\ge r(\vartheta_q),\qquad
 \operatorname{Re}a(\vartheta_q+is)\le a(\vartheta_q).
\]
The exact kernel \eqref{eq:exact-ou-bridge-transform} gives
\[
 |F_{R_{n,m}}(\vartheta_q+is;x,y)|
 \le C(1+|s|)^{1/4}
 e^{R_{n,m}\operatorname{Re}\kappa(\vartheta_q+is)}
 \psi_{\vartheta_q}(x)\psi_{\vartheta_q}(y).
\]
For \(z=\vartheta_q+is\), \(R=R_{n,m}\), and \(r=r(z)\), the
quadratic exponent left after removing the endpoint eigenfunctions is
\[
 \begin{split}
 \mathfrak e_R(z;x,y)
 &=\log\frac{F_R(z;x,y)}{F_R(z;0,0)}-\frac{a(z)}2(x^2+y^2)\\
 &=-\frac r2\left\{\frac{(y-e^{-rR/2}x)^2}{1-e^{-rR}}-y^2\right\}
   +\frac12\left\{\frac{(y-e^{-R/2}x)^2}{1-e^{-R}}-y^2\right\}.
 \end{split}
\]
Consequently,
\[
 |\mathfrak e_R(z;x,y)|
 \le C\{|r|e^{-\operatorname{Re}rR/2}+e^{-R/2}\}(x^2+y^2).
\]
On the retained endpoints,
\[
 |r|\le\sqrt2\,\operatorname{Re}r,\quad x^2+y^2\le CL,\quad
 \operatorname{Re}r\ge r(\vartheta_q)>0,\quad
 |1-e^{-rR}|\ge1-e^{-r(\vartheta_q)R}.
\]
The residual tends uniformly to zero. The real boundary expectations,
normalized by \(m^\kappa\), are bounded. Dividing by
\eqref{eq:truncated-hybrid-transform} gives
\[
 \left|\frac{\mathcal M_n(\vartheta_q+is)}
                 {\mathcal M_n(\vartheta_q)}\right|
 \le C(1+|s|)^{1/4}
 \exp\{R_{n,m}[\operatorname{Re}\kappa(\vartheta_q+is)
                                  -\kappa(\vartheta_q)]\}.
\]
For some fixed \(s_0<S\), the preceding modulus is bounded by
\[
 \left|\frac{\mathcal M_n(\vartheta_q+is)}
                  {\mathcal M_n(\vartheta_q)}\right|
 \le C(1+|s|)^{1/4}
 \begin{cases}
 e^{-cLs^2},&|s|\le s_0,\\
 e^{-cL},&s_0<|s|\le S,\\
 e^{-cL\sqrt{|s|}},&|s|>S .
 \end{cases}
\]
At \(s=w/\sqrt L\), \eqref{eq:truncated-hybrid-transform} gives
\[
 e^{-iwq\sqrt L}
 \frac{\mathcal M_n(\vartheta_q+iw/\sqrt L)}
      {\mathcal M_n(\vartheta_q)}
 \longrightarrow e^{-\lambda''(\vartheta_q)w^2/2}.
\]
The three bounds give integrable Fourier domination. Under the positive
law proportional to \(e^{\vartheta_q\widetilde T_n}\mathbf1_{\mathcal A_n}\),
Fourier inversion therefore yields
\[
 \sup_z\left|\sqrt L f_{n,\vartheta_q}(Lq+\sqrt Lz)
 -\frac{e^{-z^2/(2\lambda''(\vartheta_q))}}
             {\sqrt{2\pi\lambda''(\vartheta_q)}}\right|\to0.
\]
In particular $\sup f_{n,\vartheta_q}\le C/\sqrt L$. The exact tail-tilt
identity now gives, uniformly for the stated $y_n$,
\begin{align*}
 P\{\widetilde T_n>Lq+y_n,\mathcal A_n\}
 &=\mathcal M_n(\vartheta_q)e^{-\vartheta_q(Lq+y_n)}
   \int_0^\infty e^{-\vartheta_q z}
                    f_{n,\vartheta_q}(Lq+y_n+z)\,dz\\
 &\sim\frac{\Psi_t(\vartheta_q)}
 {\vartheta_q\sqrt{2\pi L\lambda''(\vartheta_q)}}
 e^{-LJ(q)-\vartheta_q y_n}.
\end{align*}
The density bound dominates the tail integral by
\(Ce^{-\vartheta_q z}\). Choose
\[
 A>\sup_{\mathcal K}J(q)+\epsilon,\qquad \epsilon>0,\qquad
 |y_n|\le h_n\sqrt L,\quad h_n\to0 .
\]
Then the additive probability error in
\eqref{eq:rank-hybrid-probability-sandwich} has relative size
\[
 \frac{O(n^{-A})}{L^{-1/2}e^{-LJ(q)-\vartheta_q y_n}}
 \le C\sqrt L\,e^{-\epsilon L+O(h_n\sqrt L)}
 \longrightarrow0,
 \qquad e^{\pm\vartheta_q\epsilon_n}=1+o(1).
\]
The two sandwich bounds consequently prove
\eqref{eq:rank-fixedsplit-subcritical-tail} for the original statistic.
\end{proof}

The coefficient $\Psi_t$ is specified by the Bellman ODE, its
transport equation, and a convergent Bernoulli expectation series.

The transform approximation applies to the truncated hybrid in
the proof. Its exceptional probability is $O(n^{-D})$ for every
fixed $D$; no approximation to the unrestricted rank moment-generating
function is used.

The prefix and cluster estimates below transfer the fixed-split
calculation to the strict scan. The subsequent critical and
boundary sections treat the remaining normalizations.

For the scan, define the prefix law under the Bernoulli endpoint
tilt and Bellman weight by
\[
 d\mu_{m,\vartheta}=K_m(\vartheta)^{-1}
       e^{\vartheta T_m^c}W_m^c(N_m;\vartheta)\,dP_t.
\]
Let $H_{l,m}(s)$ be the unnormalized continuation from state $N_l=s$ to the
terminal Bellman weight at time $m$.

\begin{lemma}[Tilted boundary prefix moments]\label{lem:rank-prefix-tilt}
On every compact strictly subcritical parameter set, there are
$\epsilon,C>0$ such that
\begin{equation}\label{eq:rank-prefix-exponential-moment}
 \sup_{m\ge l\ge1}E_{\mu_{m,\vartheta}}
       \exp\{\epsilon(N_l-lt)^2/l\}\le C.
\end{equation}
and
\begin{equation}\label{eq:rank-prefix-continuation}
 \sup_{s,\,m\ge l}
 \left|H_{l,m}(s)/W_l^c(s;\vartheta)-1\right|
 \le C(1+\log l)^3/l.
\end{equation}
The continuation ratio is bounded above and away from zero for all $1\le
l\le m$, after changing the constants for finitely many small $l$.
\end{lemma}
\begin{proof}
Suppress the parameters in the notation. Differentiating the Bellman and
transport equations, with their removable central singularities, gives
\[
 \begin{aligned}
 |g''(x)|+|(\log A)'(x)|+|h_t'(x)|
       &\le C\{1+|\log(x(1-x))|\},\\
 |g'''(x)|+|(\log A)''(x)|+|h_t''(x)|
       &\le C/[x(1-x)].
 \end{aligned}
\]
The row ratio in the proof of Lemma~\ref{lem:analytic-boundary-coefficient}
therefore satisfies, with $j=\min(s,l-s)$,
\begin{equation}\label{eq:rank-prefix-global-row}
 |R_l(s)-1|\le\frac{C(1+\log l)^2}{l(1+j)}.
\end{equation}
For \(x=s/l\), \(\xi\in\{0,1\}\), put
\[
 \begin{gathered}
 d_\xi=\frac{\xi-x}{l+1},\qquad
 \ell_\xi^0=g(x)+(\xi-x)g'(x)+\vartheta h_t(x),\\
 \ell_\xi=\vartheta\frac{l+1}{l+1+c}h_t(x+d_\xi)
       +\log W_{l+1}(s+\xi)-\log W_l(s).
 \end{gathered}
\]
\[
 A_\xi^{\rm row}=-\kappa+\tfrac12g''(x)(\xi-x)^2
 +(\xi-x)(\log A)'(x)+\vartheta(\xi-x)h_t'(x)-\vartheta c h_t(x),
 \quad r_\xi=\ell_\xi-\ell_\xi^0-A_\xi^{\rm row}/l.
\]
For \(j\ge2\), every \(y\) between \(x\) and \(x+d_\xi\) satisfies
\(y\wedge(1-y)\ge j/(2l+2)\). Taylor's integral remainder gives
\[
 |r_\xi|\le C/(lj)+C(1+\log l)/l^2,\qquad
 |A_\xi^{\rm row}|^2\le C(1+\log l)^2.
\]
The Bellman and transport equations state, respectively,
\[
 \pi_\xi=t^\xi(1-t)^{1-\xi}e^{\ell_\xi^0},\qquad
 \sum_\xi\pi_\xi=1,\quad\sum_\xi\pi_\xi A_\xi^{\rm row}=0,
 \quad
 R_l(s)-1=\sum_\xi\pi_\xi\{e^{A_\xi^{\rm row}/l+r_\xi}-1-A_\xi^{\rm row}/l\}.
\]
For \(j=0,1\), at an endpoint \(x_0\in\{0,1\}\) and an inward
increment \(d\), the same derivative bounds integrate to
\[
 |g(x_0+d)-g(x_0)-d g'(x_0)|
       \le Cd^2(1+|\log|d||),
\]
\[
 |h_t(x_0+d)-h_t(x_0)|
 +|\log A(x_0+d)-\log A(x_0)|
       \le C|d|(1+|\log|d||).
\]
These inequalities, with \(|d|\le2/l\), and the Bellman cancellation
give \eqref{eq:rank-prefix-global-row} also for \(j=0,1\). Dividing the
two row entries by \(R_l(s)\) defines the transitions
\[
 \widetilde P_l(s,s+\xi)=
 \frac{t^\xi(1-t)^{1-\xi}e^{\vartheta V_{l+1}(s+\xi)}W_{l+1}(s+\xi)}
      {W_l(s)R_l(s)},\qquad \xi\in\{0,1\},
\]
where \(V_{l+1}(r)=(l+1)h_t(r/(l+1))/(l+1+c)\).
Let \(N_l\) be the chain with these transitions and
\(\xi_{l+1}=N_{l+1}-N_l\), with its natural filtration \(\mathcal F_l\).
The bounds on
\(g',h_t,A,A^{-1}\) give
\[
 \epsilon_0\le P(\xi_{l+1}=1\mid\mathcal F_l)
                  \le1-\epsilon_0.
\]
Each label count accumulated over \(r\) steps therefore dominates
\(B_r\sim\operatorname{Bin}(r,\epsilon_0)\). No joint independence is needed:
\[
 \frac1{1+\min(a,b)}\le\frac1{1+a}+\frac1{1+b},
 \qquad
 \E\frac1{1+B_r}
 =\frac{1-(1-\epsilon_0)^{r+1}}{(r+1)\epsilon_0}
 \le\frac1{(r+1)\epsilon_0}.
\]
Consequently
\[
 \widetilde E_{l,s}\{1+\min(N_{l+r},l+r-N_{l+r})\}^{-1}
       \le C/(r+1).
\]
Together with \eqref{eq:rank-prefix-global-row}, this yields
\[
 \sup_s\widetilde E_{l,s}\sum_{k=l}^\infty|R_k(N_k)-1|
 \le C\sum_{r\ge0}\frac{(1+\log(l+r))^2}{(l+r)(r+1)}
 \le C(1+\log l)^3/l=:\epsilon_l.
\]
The same conditional bound holds after every stopping time not smaller
than \(l\). Put \(U_l=\sum_{k=l}^\infty|R_k(N_k)-1|\). Expanding into
ordered time indices and conditioning successively gives
\[
 \sup_s\widetilde E_{l,s}U_l^r\le r!(C\epsilon_l)^r,\qquad r\ge1.
\]
For fixed \(q>0\), choose \(l\) so that \(Cq\epsilon_l<1/2\). Then
\[
 \sup_s\widetilde E_{l,s}e^{qU_l}
 \le\sum_{r\ge0}(Cq\epsilon_l)^r
 =\frac1{1-Cq\epsilon_l}=1+O_q(\epsilon_l).
\]
Since the rows tend uniformly to one,
\(|\log R_k|\le2|R_k-1|\) for sufficiently large \(k\); the same bound
therefore controls their logarithmic products. The exact Doob identity
\[
 H_{l,m}(s)/W_l(s)
       =\widetilde E_{l,s}\prod_{k=l}^{m-1}R_k(N_k)
\]
proves \eqref{eq:rank-prefix-continuation}, including its uniform upper and
lower bounds. Finitely many early rows have deterministic positive bounds.
For a fixed compact parameter set \(\mathcal K\), choose \(\Delta>0\) such that
\[
 \vartheta'=\vartheta+\Delta<b(t),\qquad
 (t,\vartheta,c)\in\mathcal K,
 \qquad
 \phi=g_{\vartheta'}-g_\vartheta,\quad
 d_\kappa=\kappa(\vartheta')-\kappa(\vartheta)>0.
\]
Lemma~\ref{lem:hardy-bellman-branch} and
\eqref{eq:bernoulli-bellman-equation} give
\[
 \phi(x)=\int_\vartheta^{\vartheta'}\partial_\nu g_\nu(x)\,d\nu>0
 \quad(x\ne t),\qquad
 \phi(t)=\phi'(t)=0,\qquad
 \lim_{x\to t}\frac{\phi(x)}{(x-t)^2}=\frac{d_\kappa}{t(1-t)}>0.
\]
The amplitude bounds following \eqref{eq:boundary-transport-amplitude}
and compactness provide constants \(c_\phi,C_\phi,a_-,a_+>0\) with
\[
 c_\phi(x-t)^2\le\phi(x)\le C_\phi(x-t)^2,\qquad
 a_-\le\frac{A_{\vartheta',c}(x)}{A_{\vartheta,c}(x)}\le a_+
 \quad(0\le x\le1).
\]
Define
\[
 \begin{aligned}
 L_l(s)&=e^{l\phi(s/l)}
              \frac{A_{\vartheta',c}(s/l)}{A_{\vartheta,c}(s/l)},\\
 Q_l(s)&=\frac{\widetilde E_\vartheta
                [L_{l+1}(N_{l+1})\mid N_l=s]}{L_l(s)}.
 \end{aligned}
\]
Substitution of the two normalized transition probabilities yields the exact identity
\[
 Q_l(s)=\left(1+\frac1l\right)^{d_\kappa}
   \frac{R_l^{\vartheta'}(s)}{R_l^\vartheta(s)}
   \widetilde E_{\vartheta'}
       [e^{-\Delta V_{l+1}(N_{l+1})}\mid N_l=s].
\]
Choose \(\delta>0\) with \([t-2\delta,t+2\delta]\subset(0,1)\)
uniformly on \(\mathcal K\). Denote the preceding row coefficients at
parameter \(\nu\), for \(\xi\in\{0,1\}\), by \(A_{\xi,\nu}^{\rm row},r_\xi^\nu,\pi_\xi^\nu\). For
\(\nu\in\{\vartheta,\vartheta'\}\) and \(|s/l-t|\le\delta\),
\[
 |A_{\xi,\nu}^{\rm row}|+l^2|r_\xi^\nu|\le C,\qquad
 \sum_\xi\pi_\xi^\nu A_{\xi,\nu}^{\rm row}=0,\qquad \sum_\xi\pi_\xi^\nu=1.
\]
Consequently, for sufficiently large \(l\),
\[
 \begin{aligned}
 |R_l^\nu(s)-1|
 &\le\sum_\xi\pi_\xi^\nu\left\{|r_\xi^\nu|
  +\frac12e^{|A_{\xi,\nu}^{\rm row}/l+r_\xi^\nu|}
                       |A_{\xi,\nu}^{\rm row}/l+r_\xi^\nu|^2\right\}\le\frac C{l^2},\\
 \left|\log\frac{R_l^{\vartheta'}(s)}{R_l^\vartheta(s)}\right|
 &\le\frac C{l^2}.
 \end{aligned}
\]
Put \(x=s/l\), \(x_\xi=(lx+\xi)/(l+1)\), \(\xi\in\{0,1\}\).
For \(|x-t|\le\delta\), Taylor's integral formula gives
\[
 \begin{gathered}
 |x_\xi-x|\le\frac1{l+1},\qquad
 |h_t'(x)|\le C|x-t|,\qquad
 \sup_{|y-t|\le2\delta}|h_t''(y)|\le C,\\
 |h_t(x_\xi)-h_t(x)|\le C\left(\frac{|x-t|}{l}+\frac1{l^2}\right),\\
 \left|\frac{l+1}{l+1+c}-1\right|\le\frac Cl,\qquad
 |V_{l+1}(s+\xi)-h_t(x)|
 \le C\left\{\frac{|x-t|}{l}+\frac{h_t(x)}l+\frac1{l^2}\right\}.
 \end{gathered}
\]
Also,
\[
 \begin{gathered}
 h_t''(y)=\frac1{y(1-y)}\ge4,\qquad
 h_t(x)=(x-t)^2\int_0^1(1-r)h_t''(t+r(x-t))\,dr
          \ge2(x-t)^2,\\
 \frac{|x-t|}{l}\le\epsilon(x-t)^2+\frac1{4\epsilon l^2}
 \quad(\epsilon>0).
 \end{gathered}
\]
The identity for \(Q_l\), applied to each of its two possible next states, implies
\[
 \begin{aligned}
 \log Q_l(s)
 &\le d_\kappa\log(1+l^{-1})-\Delta h_t(x)
       +C\left\{\frac{|x-t|}{l}+\frac{h_t(x)}l+\frac1{l^2}\right\}\\
 &\le-\frac\Delta2 h_t(x)+\frac{d_\kappa}{l}+\frac{C_2}{l^2},
 \qquad |x-t|\le\delta .
 \end{aligned}
\]
Here \(\epsilon\) is first chosen small enough and \(l\) then large enough;
\(C_2\) is uniform on \(\mathcal K\). For \(|x-t|>\delta\) and \(l\ge2/\delta\),
\[
 |x_\xi-t|\ge\frac\delta2,\qquad
 \frac{l+1}{l+1+c}\ge\frac12,\qquad
 V_{l+1}(s+\xi)\ge\frac{\delta^2}{4}
\]
for sufficiently large \(l\), including the states \(s=0,l\).
By \eqref{eq:rank-prefix-global-row},
\[
 \log Q_l(s)\le-\frac{\Delta\delta^2}{4}
                +\frac{C(1+\log l)^2}{l}
 \le-\frac{\Delta\delta^2}{8},\qquad
 Q_l(s)\le\rho:=e^{-\Delta\delta^2/8}<1.
\]
Choose fixed constants
\[
 \overline d=\sup_{\mathcal K}d_\kappa<\infty,\qquad
 M\ge\frac{2(\overline d+3)}\Delta,\qquad
 C_M=a_+e^{C_\phi M/2},\qquad
 C_Q=(\overline d+1)e^{\overline d+1}.
\]
Take \(l_0\ge2\) large enough for all preceding bounds and for
\[
 \frac{C_2}{l_0}\le1,\qquad
 \frac M{l_0}<2\delta^2,\qquad
 \rho\le1-\frac1{l_0}.
\]
For \(l\ge l_0\), the two regions satisfy
\[
 \begin{aligned}
 h_t(s/l)\ge M/l,\quad |s/l-t|\le\delta
 &\ \Longrightarrow\ Q_l(s)\le e^{-2/l}\le1-1/l,\\
 |s/l-t|>\delta
 &\ \Longrightarrow\ Q_l(s)\le\rho\le1-1/l,\\
 h_t(s/l)<M/l
 &\ \Longrightarrow\ |s/l-t|<\delta,\quad L_l(s)\le C_M,\\
 h_t(s/l)<M/l
 &\ \Longrightarrow\ Q_l(s)-1
       \le e^{(\overline d+1)/l}-1\le C_Q/l.
 \end{aligned}
\]
With \(C_1=(C_Q+1)C_M\), both cases therefore give
\[
 \widetilde E_\vartheta[L_{l+1}(N_{l+1})\mid N_l=s]
 \le(1-1/l)L_l(s)+C_1/l,\qquad l\ge l_0.
\]
Initialize the normalized chain by
\[
 \widetilde P_\vartheta(N_1=b)
 =\frac{t^b(1-t)^{1-b}e^{\vartheta h_t(b)/(1+c)}
               W_1^c(b;\vartheta)}{K_1(\vartheta)},
 \qquad b\in\{0,1\},
\]
and put \(u_l=\widetilde E_\vartheta L_l(N_l)\). Taking expectations and summing gives
\[
 \begin{aligned}
 l u_{l+1}&\le(l-1)u_l+C_1,\\
 (l-1)u_l&\le(l_0-1)u_{l_0}+C_1(l-l_0),\qquad l\ge l_0,\\
 u_l&\le\max\{u_{l_0},C_1\}.
 \end{aligned}
\]
The finitely many values \(l\le l_0\) are uniformly bounded. Since
\[
 L_l(s)\ge a_-\exp\{c_\phi(s-lt)^2/l\},
 \qquad
 \sup_{l\ge1}\widetilde E_\vartheta
       e^{c_\phi(N_l-lt)^2/l}\le a_-^{-1}\sup_{l\ge1}u_l<\infty,
\]
the required moment bound holds for the normalized chain.
The exact density of \(\mu_{m,\vartheta}\) relative to this chain is
\[
 D_m=\frac{K_1}{K_m}\prod_{k=1}^{m-1}R_k(N_k),\qquad
 \sup_m\widetilde E_\vartheta D_m^2<\infty.
\]
The second bound follows from the logarithmic product bound proved above
and the uniform positive bounds on \(K_m\). Taking \(\epsilon=c_\phi/2\) gives
\[
 \begin{aligned}
 E_{\mu_{m,\vartheta}}e^{\epsilon(N_m-mt)^2/m}
 &=\widetilde E_\vartheta
          [D_m e^{\epsilon(N_m-mt)^2/m}]\\
 &\le(\widetilde E_\vartheta D_m^2)^{1/2}
       \{\widetilde E_\vartheta
          e^{c_\phi(N_m-mt)^2/m}\}^{1/2}\le C.
 \end{aligned}
\]
For \(l\le m\), the exact prefix density is
\[
 \frac{d(\mu_{m,\vartheta}|_{\mathcal F_l})}{d\mu_{l,\vartheta}}
 =\frac{K_l}{K_m}\frac{H_{l,m}(N_l)}{W_l(N_l)}\le C.
\]
Applying the time-\(l\) estimate proves
\eqref{eq:rank-prefix-exponential-moment} for every \(l\le m\).
\end{proof}

\subsection{The original-rank outer scan in the strictly subcritical regime}\label{app:rank-scan-subcritical}

Throughout this subsection assume that the saddle stays in a compact
strictly subcritical set:
\begin{equation}\label{eq:rank-scan-s1}
0<\vartheta_0\le\vartheta\le\vartheta_1
 <\inf_{\eta\le t\le1-\eta} C(t)^{-1}.
\end{equation}
The upper endpoint is also strictly below $1/4$. For the high-dimensional
calibration, take $q=q_n$ and $\vartheta=\vartheta_n$ from
\eqref{eq:main-subcritical-normalization}. All constants below are uniform
over the indicated split fractions and saddle parameters. Throughout this subsection put
\[
 N=\log n,\quad R=2N,\quad m=\lfloor N^8\rfloor,\quad
 t=k/n,\quad v=t(1-t),\quad u=\operatorname{logit}t,
 \quad\theta=\vartheta/2.
\]
The parameter \(\theta\) tilts the doubled energy \(Q=2T\).

The fixed-anchor hybrid comparison retains the first and last $m$
entropy summands and couples the middle rank bridge to a Gaussian bridge
with the same endpoints. Its middle error is $o(1)$ outside events of
probability $o(n^{-A})$, where $A$ can be chosen arbitrarily. The transform
coefficient of the localized surrogate is
\begin{equation*}
\Psi_t(\vartheta)=\sqrt r\,
 K_{-1/2}(t,\vartheta)K_{+1/2}(t,\vartheta),\qquad
 r=\sqrt{1-4\vartheta}=q^{-1}.
\end{equation*}
The Gaussian middle supplies a local limit theorem and a density bound
$C/\sqrt R$, including after insertion of a sufficiently small exponential
tilt of a bounded-operator-norm quadratic observable.

Let \(\mathcal F_{n,m}^{\partial}\) be the sigma-field generated by the two
boundary label paths, and let \(\mathbb P_{\rm hyb}\) be their independent
Bernoulli law followed by the conditional Gaussian middle. With the
hybrid \(\widetilde T_n\) and retained event \(\mathcal A_n\) defined in the
fixed-split proof, set
\[
 \mathcal M_{n,t}(z)=\mathbb E_{\rm hyb}
       [e^{z\widetilde T_n}\mathbf1_{\mathcal A_n}],\qquad
 \frac{d\mathbb Q_{n,t,\vartheta}^{\mathcal A}}{d\mathbb P_{\rm hyb}}
       =\frac{e^{\vartheta\widetilde T_n}\mathbf1_{\mathcal A_n}}
                    {\mathcal M_{n,t}(\vartheta)}.
\]
For \(h>0\) and an offset \(y\), define
\[
 \begin{aligned}
 I_{n,y,h}^{G}&=\{2\widetilde T_n\in[Rq+y,Rq+y+h]\},\\
 I_{n,y,h}^{r}&=\{2T_{n,k}\in[Rq+y,Rq+y+h]\},\qquad
 \mathbb P_{n,k}^{y,h}(E)=\mathbb P(E\mid I_{n,y,h}^{r}).
 \end{aligned}
\]
The width \(h\) is fixed when \(n\to\infty\), and can subsequently tend to
zero. Expectations under \(\mathbb Q_{n,t,\vartheta}^{\mathcal A}\) are
written \(\mathbb E^{\mathcal A}_{\vartheta}\).

For $m\le l\le n-m$, set
\[
 c_{li}=\mathbf1(i\le l)-l/n,\quad
 d_l=(l-1/2)(n-l+1/2),\quad
 w_l=\frac{n^2}{d_l\,l(n-l)},\quad
 M=\sum_{l=m}^{n-m}w_l c_lc_l^\top.
\]
Let $B_i$ be the rank labels at the anchor, $X_i=B_i-t$, and $A_l=\sum_{i\le
l}X_i$. Then
\begin{equation}\label{eq:rank-scan-s3}
Q_m=\frac{X^\top MX}{v},\qquad
 W_m=\frac{X^\top M^2X}{v},\qquad
 M\mathbf1=0.
\end{equation}
Extend the coupled Gaussian cumulative path \(G_l\) linearly from its
middle endpoints to \(G_0=G_n=0\), and put
\[
 \begin{gathered}
 X_i^G=G_i-G_{i-1},\qquad E_i=X_i-X_i^G,\qquad w_q=(q-1)/(2\theta),\\
 Q_m^G=(X^G)^\top MX^G/v,\qquad
 W_m^G=(X^G)^\top M^2X^G/v.
 \end{gathered}
\]
The elementary Hardy bound gives $\|M\|\le C_\eta$; $\max_iM_{ii}\le
C_\eta/m$ and $\operatorname{tr}M=R-2\log m+O(1)$.

On the count envelope
\begin{equation}\label{eq:rank-scan-s4}
|A_l|\le C\sqrt{(l\wedge(n-l))\log n},
 \qquad m\le l\le n-m,
\end{equation}
define the actual middle entropy remainder by
\[
 \mathcal R_{n,k}^{\rm ent}
   =\sum_{l=m}^{n-m}\frac{nG_{k,l}}{d_{l,n}}-\frac{Q_m}{2}.
\]
The entropy Taylor expansion gives
\begin{equation}\label{eq:rank-scan-s5}
 |\mathcal R_{n,k}^{\rm ent}|
 \le C\{(\log n)^{3/2}/\sqrt m+(\log n)^2/m\}=o(1).
\end{equation}
This deterministic bound holds throughout every time window on which
\eqref{eq:rank-scan-s4} holds.

\begin{lemma}[Conditional middle trace and rank-transfer identities]
For anchor intervals whose offsets from $Rq$ remain bounded, and also for
the $O(\log R)$ offsets used in the scan threshold,
\begin{equation}\label{eq:rank-scan-s6}
Q_m/R\longrightarrow q,\qquad
 W_m/R\longrightarrow\frac{q-1}{2\theta}
\end{equation}
in the localized tilted anchor law and its anchor-interval versions. Let
$h_i=(MX)_i$. On \eqref{eq:rank-scan-s4},
\begin{equation}\label{eq:rank-scan-s7}
\left|\sum_i(B_i-t)h_i^2\right|
 \le C(\log n)^{3/2}/\sqrt m=o(1),
\quad
 \left|\sum_i(B_i-t)M_{ii}\right|
 \le C\sqrt{\log n/m}=o(1),
\end{equation}
and $\max_i|h_i|\le C\sqrt{\log n/m}=o(1)$.
\end{lemma}

\begin{proof}
For a fixed \(D>0\), define the boundary energy and the smaller retained event
\[
 E_\partial=T_m^{-1/2}+T_m^{+1/2},\qquad
 \mathcal B_{n,D}=\mathcal A_n\cap
 \{|Z_m^-|\vee|Z_m^+|\le\log R,\ E_\partial\le D(\log R)^2\}.
\]
Conditional on \(\mathcal F_{n,m}^{\partial}\) under
\(\mathbb Q_{n,t,\vartheta}^{\mathcal A}\), write \(\mu\) and \(\Sigma\)
for the mean and covariance of \(X^G\). For \(A=M,M^2\), set
\[
 F_A^G=(X^G)^\top AX^G/v,\qquad
 \bar F_A^G=\{\operatorname{tr}(A\Sigma)+\mu^\top A\mu\}/v,
 \qquad c_M=q,\quad c_{M^2}=w_q.
\]
The Gaussian determinant and its first derivative give
\[
 \|\Sigma\|\le C,\qquad
 |\bar F_A^G-Rc_A|\le
 C\{\log m+1+|Z_m^-|^2+|Z_m^+|^2\}.
\]
Indeed, the zero-endpoint middle has length \(R-2\log m+O(1)\), pinning
changes its covariance by rank two, and
\[
 M^2(I-2\theta M)^{-1}
   =\{M(I-2\theta M)^{-1}-M\}/(2\theta),\qquad
 \mu^\top A\mu\le C(|Z_m^-|^2+|Z_m^+|^2).
\]
For \(|s|\le s_0\), the centered conditional log transform
\[
 L_A(s)=\log\mathbb E^{\mathcal A}_{\vartheta}
    [e^{s(F_A^G-\bar F_A^G)}\mid\mathcal F_{n,m}^{\partial}]
\]
satisfies \(L_A(s)\le CRs^2\) on \(\mathcal B_{n,D}\). Define
\[
 \mathcal D_{n,a}^G=
   \{|Q_m^G/R-q|+|W_m^G/R-w_q|>a\},\qquad
 \mathcal D_{n,a}^r=
   \{|Q_m/R-q|+|W_m/R-w_q|>a\}.
\]
Inserting \(e^{sF_A^G}\), with \(s\) of the sign of the deviation,
before inversion preserves the conditional density bound \(C/\sqrt R\).
Thus, for fixed \(a,h>0\) and the stated offsets,
\[
 \begin{split}
 \mathbb Q^{\mathcal A}_{n,t,\vartheta}
       (\mathcal D_{n,a}^G\cap I_{n,y,h}^G\cap\mathcal B_{n,D})
       &\le ChR^{-1/2}e^{-c_aR},\\
 \mathbb Q^{\mathcal A}_{n,t,\vartheta}(I_{n,y,h}^G)
       &\ge chR^{-1/2}.
 \end{split}
\]
The strict boundary margin and the positive prefix transform imply
\[
 \mathbb Q^{\mathcal A}_{n,t,\vartheta}(\mathcal B_{n,D}^c)
   \le R^{C_D}e^{-c_D(\log R)^2},\qquad
 \mathbb Q^{\mathcal A}_{n,t,\vartheta}
       (\mathcal B_{n,D}^c\mid I_{n,y,h}^G)
   \le C_hR^{C_D+1/2}e^{-c_D(\log R)^2}.
\]
For every fixed \(B\), the last expression is at most \(C_{B,h}R^{-B}\).
On the coupling event \(\max_{m\le l\le n-m}|A_l-G_l|\le C\log n\),
\[
 E^\top ME\le CN^2/m,\qquad E^\top M^2E\le C E^\top ME,
\]
and Cauchy--Schwarz yields, whenever \(Q_m^G+W_m^G\le CR\),
\[
 |Q_m-Q_m^G|+|W_m-W_m^G|
 \le C\{N^{3/2}/\sqrt m+N^2/m\}.
\]
The original-probability sandwich transfers the anchor intervals with this
vanishing enlargement; its \(n^{-A}\) exception is negligible relative to
the fixed-split anchor probability. This proves \eqref{eq:rank-scan-s6}.
For \eqref{eq:rank-scan-s7}, summation by parts and $A_n=0$ give
\begin{equation}\label{eq:rank-scan-s8}
\sum_i(B_i-t)h_i^2
 =\sum_{i=1}^{n-1}A_i(h_i^2-h_{i+1}^2),\qquad
 h_i-h_{i+1}=w_iA_i,
\end{equation}
where $w_i=0$ outside the middle. Also
\[
 M_{ii}-M_{i+1,i+1}=w_i(1-2i/n).
\]
The envelope implies $|h_i|\le C\sqrt{\log n}/\sqrt{(i\wedge(n+1-i))\vee
m}$. The first sum in \eqref{eq:rank-scan-s8} is bounded by $C(\log
n)^{3/2}\sum_{i=m}^{n-m}(i\wedge(n-i))^{-3/2}$; the term from $(w_iA_i)^2$
is smaller. The diagonal identity gives the second assertion in the same
way. This proves the lemma.
\end{proof}

\begin{lemma}[The conditional label-transfer tangent]
Let \(k(z)\) be the nearest integer to
\[
 n\operatorname{logistic}(u+z/R).
\]
On
every bounded interval of $z$, under the anchor laws of \eqref{eq:rank-scan-s6},
\begin{equation}\label{eq:rank-scan-s9}
Q_m\{k(z)\}-Q_m(k)
 \Longrightarrow \sqrt{V(q)}B(z)-(q-1)|z|,
\qquad V(q)=\frac{16q^2}{q+1},
\end{equation}
with independent Brownian motions on the two half-lines.
\end{lemma}

\begin{proof}
Let \(\mathcal F_0=\sigma(B_1,\ldots,B_n)\). Conditional on \(\mathcal F_0\),
the zero and one observation orders have law
\[
 \operatorname{Unif}(\mathfrak S_{n(1-t)})
       \otimes\operatorname{Unif}(\mathfrak S_{nt}).
\]
A tilt measurable in that vector does not change this conditional law.
Thus right transfers choose a zero uniformly, and left transfers remove
a one uniformly. For a right transfer at rank \(i\), put \(v_+=v(t+1/n)\).
Then
\[
 Q_m^+-Q_m=\frac{2h_i+M_{ii}}{v_+}
                  +\left(\frac v{v_+}-1\right)Q_m,
 \qquad
 v_+=v+\frac{1-2t}{n}-\frac1{n^2}.
\]
Averaging over the zeros and converting to logit time gives
\begin{equation}\label{eq:rank-scan-s10}
-Q_m+\frac1{1-t}\sum_{B_i=0}M_{ii}+o(1)
 =-Q_m+\operatorname{tr}M+o(1).
\end{equation}
Here $\sum_{B_i=0}h_i=-X^\top MX=-vQ_m$ and the second equality uses
\eqref{eq:rank-scan-s7}. The leading predictable quadratic variation per
unit logit time is
\begin{equation}\label{eq:rank-scan-s11}
\frac4{v(1-t)}\sum_{B_i=0}h_i^2+o(R)
 =4W_m+o(R).
\end{equation}
The squared conditional mean, the diagonal increment, and the variation of
$v$ contribute $o(R)$ to \eqref{eq:rank-scan-s11}. The occupation error in
the last equality is exactly the first expression in
\eqref{eq:rank-scan-s7}.
Put
\[
 \epsilon_n=C\{\sqrt{N/m}+R/n\},\qquad
 d_n=C\left\{\frac{\log m+(\log R)^2}{R}
                     +\sqrt{N/m}+\frac Rn\right\}.
\]
All constants below are uniform for \(q\) in a compact subset of
\eqref{eq:rank-scan-s1} and for the stated anchor offsets.
For \(A=M\) or \(M^2\), write \(F_A=X^\top AX/v\).
The matrix bounds imply
\[
 A\mathbf1=0,\quad \|A\|\le C,\quad
 \operatorname{tr}A\le CR,\quad \max_iA_{ii}\le C/m,\quad
 \sum_i(AX)_i^2\le C F_A .
\]
Put \(\mathcal H_n(z)=Q_m(z)+W_m(z)\) and
\(\tau_E=\inf\{z\ge0:\mathcal H_n(z)>C_0R\}\), where \(C_0\) exceeds
twice the limiting initial value. Before \(\tau_E\),
\[
 |(AX)_i|^2\le A_{ii}X^\top AX\le CR/m.
\]
For a right transfer let \(v_+=(t+n^{-1})(1-t-n^{-1})\). Exactly,
\[
 \Delta F_A
 =\frac{2(AX)_i+A_{ii}}{v_+}
       -F_A\frac{v_+-v}{v_+}.
\]
The analogous left formula has the same absolute bounds.
Compensated jumps are consequently bounded by \(\epsilon_n\).
The number of transfers per unit logit time is
\(nv+O(1)\). Since
\[
 \sum_{B_i=0}(AX)_i=-vF_A,\qquad
 \sum_{B_i=0}A_{ii}\le \operatorname{tr}A,\qquad
 \sum_i A_{ii}^2\le CR/m,
\]
the absolute drift and predictable variance of \(F_A\), per
unit logit time, are at most \(CR\).
After the time change \(z=R(u-u_0)\), Doob's inequality and the
integrated drift bound give
\[
 \E\sup_{z\le T}|F_A(z\wedge\tau_E)-F_A(0)|^2\le C_T.
\]
For initial energy at most \(C_0R/2\), the preceding maximal second-moment
bound gives
\[
 P(\tau_E\le T)
 \le \frac{4}{C_0^2R^2}
       \E\sup_{z\le T}|\mathcal H_n(z\wedge\tau_E)-\mathcal H_n(0)|^2
 \le C_T/R^2.
\]
For a prefix containing \(a_l\) zeros, \(j\) right transfers give
\[
 Y_j\mid\mathcal F_0\sim
 \operatorname{Hypergeom}\bigl(n(1-t),a_l,j\bigr).
\]
Its centered martingale is
\[
 \frac{n(1-t)}{n(1-t)-j}
       \left\{Y_j-\frac{ja_l}{n(1-t)}\right\}.
\]
For \(j\le C_Tn/R\), its jumps are bounded and its bracket is
at most \(C_T(l\wedge(n-l))/R\).
For jumps bounded by \(c_1\), conditional centering gives
\[
 \E(e^{\lambda\Delta L}\mid\mathcal F)
 \le\exp\left\{\frac{\lambda^2
           \E((\Delta L)^2\mid\mathcal F)}
                    {2(1-c_1\lambda/3)}\right\},
 \qquad 0<\lambda<3/c_1 .
\]
If \(V_*\) bounds the bracket, stop at the first crossing and use
\(\lambda=x/(V_*+c_1x/3)<3/c_1\). The exponential bound becomes
\[
 P\{\sup L\ge x,\ \langle L\rangle\le V_*\}
 \le\exp\left\{-\lambda x+
       \frac{\lambda^2V_*}{2(1-\lambda c_1/3)}\right\}
 =\exp\left\{-\frac{x^2}{2V_*+2c_1x/3}\right\}.
\]
For \(b_l=l\wedge(n-l)\), the present values satisfy
\[
 x=C\sqrt{b_lN},\qquad V_*\le C_Tb_l/R,\qquad
 b_l\ge m=N^8+O(1)
 \ \Longrightarrow\
 \frac{x^2}{V_*+c_1x}\ge cN^2.
\]
A union over prefixes bounds the additional failure probability by
\[
 2n\exp\{-c_TN^2\}.
\]
The conditional mean of the centered prefix is
\((1-j/\{n(1-t)\})A_l(0)\). Thus a larger fixed envelope
constant contains the entire window. The left transfers satisfy
the reflected calculation.
Let \(\mathcal F_k\) contain the anchor and all transfers up to \(k\).
For a right transfer define
\[
 \begin{gathered}
 \Delta u_k=\operatorname{logit}(t+1/n)-\operatorname{logit}t,
       \qquad (\Delta u_k)^{-1}=nv+O(1),\\
 \Delta Q_m=Q_m(k+1)-Q_m(k),\\
 \mathcal D_Q=\frac{E(\Delta Q_m\mid\mathcal F_k)}{\Delta u_k},\qquad
 \mathcal V_Q=\frac{\operatorname{Var}(\Delta Q_m\mid\mathcal F_k)}{\Delta u_k}.
 \end{gathered}
\]
On the envelope, expansion of this exact increment gives
\[
 \left|\mathcal D_Q+Q_m-\operatorname{tr}M\right|
       \le C\{\sqrt{N/m}+R/n\},
\]
\[
 \left|\mathcal V_Q-4W_m\right|
       \le C\{N^{3/2}/\sqrt m+R^2/n\}.
\]
The occupation errors are those in \eqref{eq:rank-scan-s7}; the left
transfer uses the corresponding positive left logit increment.
For the variance bound, the cross and diagonal terms satisfy
\[
 \sum_i |(MX)_i|M_{ii}\le CR/\sqrt m,\qquad
 \sum_iM_{ii}^2\le CR/m.
\]
The squared conditional mean contributes at most \(CR^2/n\).
Expanding \(v_+^{-2}\) contributes at most \(CR/n\).
The same conditional determinant calculation gives, for
\(d_n\le a\le a_0\),
\[
 \mathbb Q^{\mathcal A}_{n,t,\vartheta}
  (\mathcal D_{n,Ca}^G\cap I_{n,y,h}^G\cap\mathcal B_{n,D})
       \le ChR^{-1/2}e^{-cRa^2},\qquad
 \mathbb Q^{\mathcal A}_{n,t,\vartheta}(I_{n,y,h}^G)
       \ge chR^{-1/2}.
\]
Here the exact conditional-mean corrections obey
\[
 \max_{A\in\{M,M^2\}}|\bar F_A^G-Rc_A|
       \le C\{(\log R)^2+\log m\},\qquad
 \frac{|Q_m-Q_m^G|+|W_m-W_m^G|}{R}
       \le C\{\sqrt{N/m}+N/m\}.
\]
The previously defined boundary cut and the prescribed coupling exception
therefore give
\[
 \mathbb P_{n,k}^{y,h}(\mathcal D_{n,Ca}^r)
       \le C_he^{-cRa^2}+C_{B,h}R^{-B}.
\]
The factor \(R^{1/2}/h\) from the anchor denominator has been included in
\(\mathbb Q^{\mathcal A}_{n,t,\vartheta}
 (\mathcal B_{n,D}^c\mid I_{n,y,h}^G)\), rather than in its unconditional
boundary probability.
Let \(M_n^\pm,D_n^\pm\) be the stopped compensated martingales and
integrated drifts on the two sides. The resulting bounds imply
\[
\begin{split}
 \Pp\bigg\{
 &\sup_{z\le T}\left|
      \langle M_n^\pm\rangle(z)-\frac{16q^2}{q+1}z\right|
       +\sup_{z\le T}|D_n^\pm(z)+(q-1)z|\\
 &\hspace{35mm}>C_T(a+d_n)
       \,\biggm|\,I_{n,y,h}^{r}\bigg\}
 \le C e^{-cRa^2}+\frac{C_T}{R^2a^2}+C_BR^{-B}.
\end{split}
\]
The Riemann-sum error is \(O(R/n)\). Let \(\tau_C\) be the first exit of any middle prefix from the enlarged
count envelope, and put \(\tau=\tau_E\wedge\tau_C\). The probability that
this stopping time occurs before \(T\) is
\[
 \Pp\{\tau\le T\mid I_{n,y,h}^{r}\}
       \le C_T R^{-2}+C_B R^{-B}.
\]
For \(a=R^{-1/4}\), the terms in the preceding probability bound are
\[
 a+d_n=R^{-1/4}+d_n,\qquad
 (R^2a^2)^{-1}=R^{-3/2},\qquad e^{-cRa^2}=e^{-c\sqrt R}.
\]
We apply \citet[Theorem~2.1(ii), pp.~270--271]{Whitt2007}.
That theorem requires locally square-integrable martingales
starting at zero, vanishing expected maximum squared jumps,
vanishing expected bracket jumps, and deterministic limiting
predictable brackets.
The stopped martingales satisfy
\[
 \E\sup_{z\le T}|\Delta M_n^\pm(z)|^2\le\epsilon_n^2,\qquad
 \E\sup_{z\le T}|\Delta\langle M_n^\pm\rangle(z)|
       \le\epsilon_n^2.
\]
Conditional independence of the two transfer orders gives
\[
 \langle M_n^+,M_n^-\rangle\equiv0,\qquad
 \bigl(\langle M_n^+\rangle,\langle M_n^-\rangle\bigr)
       \longrightarrow(Vz,Vz).
\]
The quantitative bracket estimate therefore verifies the remaining
martingale-limit hypothesis and yields
\[
 (M_n^+,M_n^-)\ \Rightarrow\ \sqrt V(B_+,B_-),
 \qquad B_+\ \text{independent of }B_-.
\]
The drift estimate and stopping probability remove localization.
Let \(z_k=R\{\operatorname{logit}(k/n)-u\}\), and let \(\bar Q_m\)
linearly interpolate the values \(Q_m(k)\) on this grid. Then
\[
 \sup_{|z|\le T}|\bar Q_m(z)-Q_m\{k(z)\}|\le\epsilon_n\to0,
 \qquad \max_{k}|z_{k+1}-z_k|\le C_\eta R/n\to0
\]
These bounds prove \eqref{eq:rank-scan-s9}.
\end{proof}

For one boundary, write
\[
 E_m^c(t)=\sum_{l=1}^m\frac{l}{l+c}
 h_t\{C_l(t)/l\},\qquad c\in\{-1/2,1/2\}.
\]
Lemma~\ref{lem:rank-prefix-tilt} gives, for every fixed $p<\infty$,
\begin{equation}\label{eq:rank-scan-s12}
\sup_{l\le m}E_\vartheta|C_l-lt|^p\le C_p l^{p/2},
 \qquad E_\vartheta e^{c(C_l-lt)^2/l}\le C.
\end{equation}
These are terminal-Bellman-weighted moments. Conditional thinning over a
fixed sufficiently short logit interval \(J\) preserves the bounds:
\[
 \sup_{t\in J}\E_\vartheta|C_l(t)-lt|^p\le C_p l^{p/2},
 \qquad
 \sup_{t\in J}\E_\vartheta
          e^{c(C_l(t)-lt)^2/l}\le C.
\]
One may use independent uniform times or sampling without replacement.
For a zero at rank \(i\), the influence on \(E_m^c\) is
\[
 g_i^+=\sum_{l=i}^m\frac{l}{l+c}
 \left[h_t\{(C_l+1)/l\}-h_t(C_l/l)\right].
\]
The analogous difference applies on the left. If \(D_l^\pm\) is its
\(l\)-th summand, the central Taylor bound and the remote-count
exponential tail in \eqref{eq:rank-scan-s12} give
\[
 \|D_l^\pm\|_p
 \le C_p\left\{\frac{\sqrt l+1}{l^2}
                 +\frac{\log(l+1)}l e^{-c_p l}\right\}
 \le C_p l^{-3/2}.
\]
Thus Minkowski's inequality and summation over ranks yield
\[
 \|g_i^\pm\|_p\le\sum_{l=i}^m\|D_l^\pm\|_p\le C_p i^{-1/2},
 \qquad
 E_\vartheta\sum_{i\le m}(g_i^\pm)^2
 \le C\sum_{i\le m}i^{-1}.
\]
In particular,
\begin{equation}\label{eq:rank-scan-s13}
\|g_i^\pm\|_p\le C_p i^{-1/2},\qquad
 E_\vartheta\sum_{i\le m}(g_i^\pm)^2\le C\log(m+1).
\end{equation}
The time derivative of $h_t$ must be combined with label transfers before
bounding the drift. Their first-order terms cancel:
\[
 \partial_t h_t(r)+\frac{1-r}{1-t}h_t'(r)
 =-\frac{(r-t)^2}{2v^2}+O(|r-t|^3)
\]
at $r=t$, and the discrete second-difference term is $O(1/l)$ in
expectation. Equations \eqref{eq:rank-scan-s12}--\eqref{eq:rank-scan-s13}
imply drift norm $O(\log m)$ and expected bracket $O(\log m)$ per unit logit
time. The martingale maximal inequality hence yields
\begin{equation}\label{eq:rank-scan-s14}
E_\vartheta\sup_{|z|\le T}
 |E_m^c\{k(z)\}-E_m^c(k)|^2
 \le C_T\left\{\frac{\log m}{R}
             +\frac{(\log m)^2}{R^2}\right\}=o(1).
\end{equation}
The finite-\(n\) boundary terms satisfy
\[
 \sup_{|z|\le T}|T_{\rm lower}\{k(z)\}-E_m^{-1/2}\{k(z)\}|
 +\sup_{|z|\le T}|T_{\rm upper}\{k(z)\}-E_m^{+1/2}\{k(z)\}|
        \le Cm^2/n=o(1).
\]
Together with \eqref{eq:rank-scan-s14}, this removes the boundary
increment from \eqref{eq:rank-scan-s9}.
For exponential control put \(\mathcal L=\log R\). Write \(C_l(z)\)
for the prefix count at \(k(z)\), and define
\[
 \mathcal U_{n,A,T}=\bigcap_{l=1}^m
 \{\sup_{|z|\le T}|C_l(z)-l k(z)/n|\le A\sqrt{l\mathcal L}\}.
\]
At the anchor its count bounds are
\begin{equation}\label{eq:rank-scan-s15}
|C_l-lt|\le A\sqrt{l\mathcal L},\qquad 1\le l\le m.
\end{equation}
The exponential moment in \eqref{eq:rank-scan-s12} and maximal sampling
bound give, for sufficiently large \(A=A_B\),
\[
 \Pp_\vartheta(\mathcal U_{n,A,T}^{c})\le C m^C e^{-cA^2\mathcal L}
       \le R^{-B}.
\]
This also holds under the small-interval anchor laws. The lower ranks give
\[
 \sum_{l\le C_A\mathcal L}|D_l^\pm|
 \le C\sum_{l\le C_A\mathcal L}\frac{\log(l+1)}l
 \le C_A(1+\log\mathcal L)^2.
\]
Combining this with the central estimate
\eqref{eq:rank-scan-s13} on larger ranks, on
\eqref{eq:rank-scan-s15}, yields
\begin{equation*}
\max_i|g_i^\pm|\le C_A(\log\mathcal L)^2,
 \qquad \sum_i(g_i^\pm)^2\le C_A\mathcal L^2
                  (1+\log\mathcal L)^4.
\end{equation*}
With the same right-transfer filtration and logit increment, put
\[
 D_E=\frac{E[E_m^c(k+1)-E_m^c(k)\mid\mathcal F_k]}{\Delta u_k},\qquad
 \lambda_i=\frac{\mathbf1_{\{B_i=0\}}}{n(1-t)\Delta u_k}.
\]
The left-transfer quantities are defined by reversing the labels and
the direction. On the tube,
\[
 |D_E|\le C_A\mathcal L^{C_A},\qquad
 \lambda_i\le C_\eta,\qquad
 |e^{sx}-1-sx|\le \tfrac12s^2x^2e^{|s||x|}.
\]
Combining transfer and parameter drift before taking absolute values gives
\[
 \frac{\mathcal L_{\rm logit}e^{s(E_m^c-E_m^c(t_0))}}
      {e^{s(E_m^c-E_m^c(t_0))}}
 \le |sD_E|+\tfrac12s^2e^{|s|\max_i|g_i^\pm|}
                      \sum_i\lambda_i(g_i^\pm)^2.
\]
The preceding influence bounds therefore bound this ratio by
\begin{equation}\label{eq:rank-scan-s17}
c_R\le C_s\mathcal L^{C_s}
          \exp\{C_s(\log\mathcal L)^2\}=R^{o(1)}.
\end{equation}
If \(M_E^{\rm stop}\) is the nonnegative stopped boundary maximum, the
exponential generator bound gives
\[
 \Pp_\vartheta(M_E^{\rm stop}>y)
       \le e^{c_R(s)T/R-sy},\qquad
 \E_\vartheta e^{\rho M_E^{\rm stop}}
       \le1+\frac{\rho}{s-\rho}e^{c_R(s)T/R},
 \quad 0<\rho<s.
\]
Here \(c_R(s)T/R\to0\). To remove the tube, choose
\(\vartheta<\vartheta'<\vartheta''\) satisfying
\eqref{eq:rank-scan-s1}.
Between label-change times \(E_m^c(t)\) is convex; its maximum is at an
endpoint or change time. Campbell's identity therefore gives
\begin{equation}\label{eq:rank-scan-s18}
E\exp\{s\sup_{t\in J}E_m^c(t)\}
 \le C m\sup_{t\in J}E e^{sE_m^c(t)}\le C m^{1+\kappa(s)}
\end{equation}
for compact short \(J\) and strictly subcritical \(s\).
Fixing one label costs at most \(\eta^{-1}\), while positivity gives
\[
 \begin{gathered}
 W_{m,s}=m^{-\kappa(s)}A_s e^{mg_s}\ge c m^{-\kappa(s)},\\
 \E e^{sE_m^c(t)}
 \le Cm^{\kappa(s)}\E[e^{sE_m^c(t)}W_{m,s}]
 \le Cm^{\kappa(s)},\qquad t\in J .
 \end{gathered}
\]
For the anchor weight, conditional labels at the earlier time are
binomial thinnings. The comparison uses
\[
 \|g_\vartheta'\|_\infty\le C,\qquad
 g_{\vartheta''}(x)-g_{\vartheta'}(x)\ge c(x-t)^2;
\]
the following finite-binomial calculation proves
\begin{equation}\label{eq:rank-scan-s19}
E\{W_{m,t_0,\vartheta}(C_m(t_0))\mid C_m(t)=j\}
 \le C m^{\kappa(\vartheta'')-\kappa(\vartheta)}
 W_{m,t,\vartheta''}(j)
\end{equation}
when \(J\) is short enough. Uniformly on compact sets of $t$ and saddles,
the strictly subcritical Bellman ODE gives
\begin{equation}\label{eq:rank-scan-b1}
g_{t,\vartheta}(t)=g'_{t,\vartheta}(t)=0,\quad
|g'_{t,\vartheta}(x)|\le C|x-t|,\quad
|g''_{t,\vartheta}(x)|\le C\{1+|\log(x(1-x))|\}.
\end{equation}
For \(\vartheta_1>\vartheta\) with a fixed gap, the prefix-moment argument
also proves
\begin{equation}\label{eq:rank-scan-b2}
g_{t,\vartheta_1}(x)-g_{t,\vartheta}(x)\ge c(x-t)^2.
\end{equation}
All amplitudes \(A_{t,\vartheta,c}\) are bounded above and below by positive
constants. Consequently it is enough to prove the comparison of the
exponential factors.
Suppose first \(t\ge t_0\), put \(a=t_0/t\), \(\delta=1-a\), and condition
on \(C_m(t)=mx\). Then
\[
Y=C_m(t_0)/m=m^{-1}\operatorname{Bin}(mx,a).
\]
Take \(f=g_{t_0,\vartheta}\) and define
\[
F_x(y)=f(y)-x\operatorname{KL}(y/x\Vert a),\qquad 0\le y\le x.
\]
For \(x>0\), the maximizer \(y_*\) lies in the interior and satisfies
\begin{equation}\label{eq:rank-scan-b3}
y_*/x=\operatorname{logistic}\{\operatorname{logit}(a)+f'(y_*)\}.
\end{equation}
Equation \eqref{eq:rank-scan-b1} gives
\[
 1-y_*/x\asymp\delta,\qquad
 \partial_y^2\{x\operatorname{KL}(y/x\Vert a)\}
       =\frac1{x(y/x)(1-y/x)}\asymp\frac1{x\delta}
\]
on the possible stationary-point interval. There
\[
 \frac{|f''(y)|}
      {\partial_y^2\{x\operatorname{KL}(y/x\Vert a)\}}
 \le C\delta(1+|\log\delta|)\to0.
\]
Every stationary point lies in that interval, so \(y_*\) is the unique
global maximizer. Equation \eqref{eq:rank-scan-b3} and
\(|f'(y_*)|\le C|y_*-t_0|\) give
\begin{equation}\label{eq:rank-scan-b4}
|y_*-ax|\le C\delta|x-t|,
\qquad F_x(y_*)\le f(ax)+C\delta(x-t)^2.
\end{equation}
Integrating \(F_x'\) over the interval between \(ax\) and \(y_*\) gives
\[
 F_x(y_*)-F_x(ax)
 \le C|x-t|\,|y_*-ax|
 \le C\delta(x-t)^2,
\]
as in \eqref{eq:rank-scan-b4}.
For the finite-binomial prefactor set \(z_*=y_*/x\).
The exact change from binomial parameter \(a\) to \(z_*\) gives
\begin{equation}\label{eq:rank-scan-b5}
E_a e^{mf(Y)}
=e^{mF_x(y_*)}E_{z_*}\exp\left(m\left[f(Y)-f(y_*)-f'(y_*)(Y-y_*)\right]\right).
\end{equation}
There is a number \(\beta_\delta\downarrow0\) such that, for all \(0\le y\le
x\le1\),
\begin{equation}\label{eq:rank-scan-b6}
f(y)-f(y_*)-f'(y_*)(y-y_*)
\le\beta_\delta x\operatorname{KL}(y/x\Vert z_*).
\end{equation}
When \(y/x\ge1-\sqrt\delta\), the ratio of the second derivatives is
\[
 \frac{|f''(y)|}{\partial_y^2\{x\operatorname{KL}(y/x\Vert z_*)\}}
 =\frac{y(x-y)}x|f''(y)|
 \le C\sqrt\delta\{1+|\log\delta|\}.
\]
the factor \(x|\log x|\) is uniformly bounded, so the bound also holds as
\(x\downarrow0\). For \(y/x<1-\sqrt\delta\), boundedness of \(f'\) bounds
the left side by \(C|y-y_*|\), whereas
\[
x\operatorname{KL}(y/x\Vert z_*)\ge c|y-y_*|\log(1/\delta).
\]
This proves \eqref{eq:rank-scan-b6} with
\[
\beta_\delta\le C\left\{\sqrt\delta(1+|\log\delta|)+\frac1{\log(1/\delta)}\right\}.
\]
For every binomial count \(Z\) with \(N\) trials and parameter \(p\), the
two Chernoff inequalities imply
\[
P\{N\operatorname{KL}(Z/N\Vert p)>z\}\le2e^{-z}.
\]
Thus for \(\beta_\delta<1/2\), the expectation in \eqref{eq:rank-scan-b5} is
at most \(1+2\beta_\delta/(1-\beta_\delta)\le3\), uniformly in \(m,x\),
including small or zero expected binomial counts. Combining
\eqref{eq:rank-scan-b4}--\eqref{eq:rank-scan-b5},
\begin{equation}\label{eq:rank-scan-b7}
E\exp\{m g_{t_0,\vartheta}(Y)\}
\le3\exp\{m g_{t_0,\vartheta}(ax)+Cm\delta(x-t)^2\}.
\end{equation}
Finally, smooth parameter dependence of the Bellman ODE, its common double
zero at the central point, and \eqref{eq:rank-scan-b1} give
\begin{equation}\label{eq:rank-scan-b8}
|g_{t_0,\vartheta}(ax)-g_{t,\vartheta}(x)|
\le C|t-t_0|(x-t)^2.
\end{equation}
The central double zero gives \eqref{eq:rank-scan-b8} near \(t\).
Off \(|x-t|<\epsilon\), bounded derivatives give
\[
 |g_{t_0,\vartheta}(ax)-g_{t,\vartheta}(x)|
       \le C_\epsilon|t-t_0|
       \le C_\epsilon\epsilon^{-2}|t-t_0|(x-t)^2.
\]
Choose \(J\) so short that the errors in
\eqref{eq:rank-scan-b7}--\eqref{eq:rank-scan-b8} satisfy
\[
 C(\delta+|t-t_0|)(x-t)^2
 \le\tfrac12\{g_{t,\vartheta_1}(x)-g_{t,\vartheta}(x)\},
\]
using \eqref{eq:rank-scan-b2}. This proves
\begin{equation}\label{eq:rank-scan-b9}
E\exp\{m g_{t_0,\vartheta}(Y)\}
\le3e^{m g_{t,\vartheta_1}(x)}.
\end{equation}
For \(t<t_0\), complement successes and failures.
Fixing one label changes \(x\) by at most \(m^{-1}\), so
\[
 m|g(x\pm m^{-1})-g(x)|\le\|g'\|_\infty\le C.
\]
The bounded amplitudes and \(m^{-\kappa}\) factors then give
\eqref{eq:rank-scan-s19}. For \(\vartheta<s<\vartheta_1\), use
\[
 \E[e^{\vartheta_1 E_m^c}W_{m,\vartheta_1}]\le C
\]
with Campbell's identity and \eqref{eq:rank-scan-b9}, obtaining
\begin{equation*}
E_{\mu_{m,\vartheta}}\exp\left\{s\sup_{t\in J}[E_m^c(t)-E_m^c(t_0)]\right\}
\le C m^{1+\kappa(\vartheta_1)-\kappa(\vartheta)}.
\end{equation*}
Since \(E_m^c\ge0\) and \(s>\vartheta\),
\[
 e^{(\vartheta-s)E_m^c(t_0)}\le1.
\]
The two boundary blocks obey the same bound jointly under independent
uniform times; conditioning on distinguished ranks has likelihood
\(1+O(m^2/n)\). Put
\[
 M_E=\sup_{|z|\le T}[E_m^c\{k(z)\}-E_m^c(k)].
\]
The polynomial moment and \eqref{eq:rank-scan-s18} give, for fixed
\(s>\vartheta\),
\[
 E_\vartheta[e^{\vartheta M_E};\,\mathcal U_{n,A,T}^{c}]
 \le(E_\vartheta e^{sM_E})^{\vartheta/s}
      \Pp_\vartheta(\mathcal U_{n,A,T}^{c})^{1-\vartheta/s}
 \le C m^{C\vartheta/s}R^{-B(1-\vartheta/s)}\longrightarrow0,
\]
with \(B\) chosen so that \(m^{C\vartheta/s}R^{-B(1-\vartheta/s)}\to0\).
Distinct-time conditioning has likelihood \(1+O(m^2/n)\); the higher
moment also removes collisions. On the tube,
\eqref{eq:rank-scan-s14} and \eqref{eq:rank-scan-s17} give
\[
 M_E\to0\ \text{in probability},\qquad
 \sup_n\E_\vartheta[e^{sM_E};\mathcal U_{n,A,T}]<\infty
       \quad(s>\vartheta).
\]
Uniform integrability therefore proves
\begin{equation}\label{eq:rank-scan-s20}
E_\vartheta\exp\left\{\vartheta
   \sup_{|z|\le T}[E_m^c\{k(z)\}-E_m^c(k)]\right\}\longrightarrow1.
\end{equation}

For the middle let \(\tau_E\) be the energy stopping time defined above
and put
\[
 M_{n,T}^{Q}=\max\{0,\sup_{0\le z\le T}
                    [Q_m\{k(z\wedge\tau_E)\}-Q_m(k)]\}.
\]
The bounded operator norm, vanishing jump size and
\eqref{eq:rank-scan-s10}--\eqref{eq:rank-scan-s11} give
\[
 \mathbb P_{n,k}^{y,h}(M_{n,T}^{Q}>a)
       \le C_{T,s,h}e^{-sa},\qquad s>0.
\]
For the scan, use thresholds
\[
 b_n(t)=Rq+y_n+\zeta_n(t),\qquad |y_n|\le C\log R,\qquad
 \|\zeta_n-\zeta\|_\infty\to0,\quad
 \zeta\in C[\eta,1-\eta].
\]
The studentized thresholds below have this form. Define their low-anchor
crossing contribution by
\[
 \mathfrak C_{n,T}(A)=
 \frac{P\{Q_{n,k}\le b_n(t)-A,\
          \max_{0\le z\le T}Q_{n,k(z)}>b_n(t)\}}
      {P\{Q_{n,k}>b_n(t)\}}.
\]
Split its negative anchor offsets at \(A\) and \(C_0\log R\). The
uniform anchor density supplies \(R^{-1/2}\) in each term. After division
by the anchor probability, the contribution of offsets below \(-A\) is
bounded by
\[
 \mathfrak C_{n,T}(A)
 \le C e^{-(s-\theta)A}
   +C R^{\theta C_0-B}
   +C R^{D-(\theta'-\theta)C_0},
 \qquad s>\theta,\quad\theta'>\theta.
\]
Choose
\[
 C_0>\frac D{\theta'-\theta},\qquad B>\theta C_0+1.
\]
Then \(R\to\infty\), followed by \(A\to\infty\), removes all three terms.
Positive offsets satisfy
\[
 \int_A^\infty e^{-\theta y}\,dy=\theta^{-1}e^{-\theta A}\to0.
\]
Equations \eqref{eq:rank-scan-s5}, \eqref{eq:rank-scan-s9},
\eqref{eq:rank-scan-s14}, and \eqref{eq:rank-scan-s20} consequently give
\begin{equation}\label{eq:rank-scan-s21}
\frac{P\{\max_{0\le z\le T}Q_{n,k(z)}>b_n(t)\}}
      {P\{Q_{n,k}>b_n(t)\}}
 \longrightarrow H_q(T)
 :=E\exp\left\{\sup_{0\le z\le T}
       [\sqrt{\nu(q)}B(z)-\nu(q)z/2]\right\},
\end{equation}
where
\[
 \nu(q)=\theta^2V(q)=\frac{(q-1)^2(q+1)}{4q^2}.
\]
Let $\varphi,\Phi$ be the standard normal density and distribution
function, and put $h=\nu(q)T$. Brownian scaling and reflection give
\[
 \begin{split}
 H_q(T)
 &=E\exp\{\sup_{0\le s\le h}[B(s)-s/2]\}\\
 &=(h/2+2)\Phi(\sqrt h/2)+\sqrt h\,\varphi(\sqrt h/2)
   =h/2+O(\sqrt h+1).
 \end{split}
\]
For $a>0$, define the tangent-grid coefficient
\[
 H_{q,a}(T)=E\exp\left\{\max_{z\in[0,T]\cap a\mathbb Z}
       [\sqrt{\nu(q)}B(z)-\nu(q)z/2]\right\}.
\]
The Brownian grid approximation of
\citet[Theorem~2.1(iii) and Proposition~2.2(i)]{BisewskiJasnovidov2022},
applied after the same time scaling, yields
\begin{equation}\label{eq:rank-scan-s22}
 \lim_{T\to\infty}\frac{H_q(T)}T=\frac{\nu(q)}2,
 \qquad
 \lim_{a\downarrow0}\lim_{T\to\infty}\frac{H_{q,a}(T)}T
       =\frac{\nu(q)}2.
\end{equation}
Continuity makes the boundary coefficient and studentization shift
constant to \(o(1)\) within a cell. For \(T/R\le h\le h_0\), put
\[
 \tau_{\epsilon}^{\rm tr}=\inf\{h\ge0:
 |Q_m(h)/R-q|+|W_m(h)/R-(q-1)/(2\theta)|>\epsilon\}.
\]
For any fixed anchor width \(h_1>0\) and sufficiently small fixed
\(h_0\), the conditional martingale and initial trace bounds give
\[
 \mathbb P_{n,k}^{y,h_1}(\tau_{\epsilon}^{\rm tr}\le h_0)\le Ce^{-cR}.
\]
At \(\lambda=\theta/2\), the middle exponential drift per logit time is
\begin{equation*}
R\{-\lambda(q-1)+\lambda^2V(q)/2+o(1)\}
 =-R\nu(q)/8+o(R).
\end{equation*}
Equation \eqref{eq:rank-scan-s17} changes the drift by \(o(R)\).
The higher saddle bounds discarded mass by \(R^{-A}\) times the anchor
tail, for arbitrary \(A\). The excess integral is
\[
 \int_0^\infty e^{-\theta y}e^{\lambda y}\,dy
       =\frac1{\theta-\lambda},\qquad\lambda=\theta/2.
\]
Thus
\begin{equation}\label{eq:rank-scan-s24}
P\{Q_{n,k}>b_n(t),\ Q_{n,k(h)}>b_n(t_h)\}
 \le C P\{Q_{n,k}>b_n(t)\}e^{-cRh}
       +R^{-A}P\{Q_{n,k}>b_n(t)\},
\end{equation}
Here \(A\) is arbitrary; the \(O(h)\) threshold change is absorbed in
the constant. The same stopped exponential maximum controls cells.
For \(h\ge h_0\), condition on both boundary blocks and put
\[
 n_{\rm mid}=n-2m,\qquad
 p_a=N_a/n_{\rm mid}\ge c(h_0,\eta),\quad a=1,2,3.
\]
Generate \(N_a\) independent uniform times in each category and merge them.
Every category word with these totals has probability
\[
 \frac{N_1!N_2!N_3!}{n_{\rm mid}!}.
\]
Let \(N_a(s)\) be the counting processes. Applying
\citet{KomlosMajorTusnady1975} independently gives standard bridges
\(B_1,B_2,B_3\), an exception \(o(n^{-A})\), and
\begin{equation}\label{eq:rank-scan-s25a}
\max_a\sup_{0\le s\le1}
 |N_a(s)-N_as-\sqrt{N_a}B_a(s)|\le C_A\log n.
\end{equation}
Write $\tau_j$ for the $j$th pooled order statistic and $d_j=j\wedge(n_{\mathrm{mid}}-j)$.
The pooled count equation, the Brownian bridge envelope, and inversion of
the two brackets at $j/n_{\mathrm{mid}}$ imply
\begin{equation*}
|n_{\mathrm{mid}}\tau_j-j|\le C_A\{\sqrt{d_j\log n}+\log n\}
\end{equation*}
simultaneously for all $j$. The Brownian modulus bound on intervals of these
lengths gives
\begin{equation}\label{eq:rank-scan-s25c}
\max_a\sqrt{N_a}|B_a(\tau_j)-B_a(j/n_{\mathrm{mid}})|
 \le C_A\{d_j^{1/4}(\log n)^{3/4}+\log n\}.
\end{equation}
For \(d_j\in[d,2d]\), put
\[
 \ell_d=C_A\{\sqrt{d\log n}+\log n\}/n_{\rm mid}.
\]
The Brownian reflection bound, a cover by at most \(n_{\rm mid}^2\)
fixed intervals, and \(O(\log n)\) dyadic values give
\[
 P\{\max_{|u-v|\le\ell_d}|B_a(u)-B_a(v)|
                     >C_A\sqrt{\ell_d\log n}\}
 \le Cn_{\rm mid}^2\log n\,e^{-cC_A^2\log n}=o(n^{-A}).
\]
The bridge's linear term is smaller. Multiplication by \(\sqrt{N_a}\)
gives \eqref{eq:rank-scan-s25c}. Since \(\sum_aN_a(\tau_j)=j\), the
Gaussian bridge is
\begin{equation*}
\begin{aligned}
G_a(j)&=\sqrt{N_a}B_a(j/n_{\mathrm{mid}})
       -p_a\sum_b\sqrt{N_b}B_b(j/n_{\mathrm{mid}}),\\
 \operatorname{Cov}(G_a(j),G_b(l))
 &=(p_a\mathbf1(a=b)-p_ap_b)
       \{j\wedge l-jl/n_{\mathrm{mid}}\}.
\end{aligned}
\end{equation*}
The count error is bounded by \eqref{eq:rank-scan-s25c}.
The two bridges are \(G_1\) and \(G_1+G_2\), with their displayed
covariance; add the prescribed linear endpoint means.
Let \(p^0=(t_1,t_2-t_1,1-t_2)\) be the unconditioned category
proportions. For a vector \(p\), construct \(G_{a,p}\) from the same three
standard bridges in the displayed formula, with \(N_a=n_{\rm mid}p_a\).
Let \(\ell_1(j),\ell_2(j)\) be the prescribed linear endpoint means and set
\[
 H_{1,p}(j)=G_{1,p}(j)+\ell_1(j),\quad
 H_{2,p}(j)=G_{1,p}(j)+G_{2,p}(j)+\ell_2(j),\quad
 t_1(p)=p_1,\quad t_2(p)=p_1+p_2,
\]
\[
 \mathcal Q_p=\sum_{j=1}^{n_{\rm mid}-1}\sum_{a=1}^2
       \frac{w_{m+j}H_{a,p}(j)^2}{t_a(p)(1-t_a(p))}.
\]
Here \(t_1,t_2\) are trimmed and their separation is at least \(h_0\)
in logit time; hence all three category proportions have a fixed positive
lower bound. The common bridge envelope gives
\[
 \max_a|p_a-p_a^0|\le Cm/n,\qquad
 |\mathcal Q_p-\mathcal Q_{p^0}|
       \le C(m/n)(\log n)^2=o(1),
\]
\[
 w_{m+j}\le C(m+d_j)^{-2},\qquad
 \max_{a=1,2}|H_{a,p}(j)|\le C\sqrt{(m+d_j)\log n}.
\]
Indeed, differentiating \(\sqrt{p_a}\) and the two variance divisors on
this compact simplex bounds the quadratic difference by
\(C(m/n)\log n\sum_j(m+d_j)^{-1}\).
The four sums needed for the linear and squared errors satisfy
\[
 \begin{aligned}
 \sum_j\frac{d_j^{1/4}}{(m+d_j)^{3/2}}&\le Cm^{-1/4},
 &\sum_j(m+d_j)^{-3/2}&\le Cm^{-1/2},\\
 \sum_j\frac{d_j^{1/2}}{(m+d_j)^2}&\le Cm^{-1/2},
 &\sum_j(m+d_j)^{-2}&\le Cm^{-1}.
 \end{aligned}
\]
The actual two cumulative label bridges and their quadratic sum are
\[
 \begin{split}
 A_1^{\rm cat}(j)&=N_1(\tau_j)-p_1j+\ell_1(j),\\
 A_2^{\rm cat}(j)&=N_1(\tau_j)+N_2(\tau_j)-(p_1+p_2)j+\ell_2(j),\\
 Q^{\rm cat}&=\sum_{j=1}^{n_{\rm mid}-1}\sum_{a=1}^2
        \frac{w_{m+j}\{A_a^{\rm cat}(j)\}^2}{t_a(p^0)\{1-t_a(p^0)\}}.
 \end{split}
\]
Changing the categorical variance divisors from \(p^0\) to \(p\)
costs at most \(C(m/n)(\log n)^2\). Multiplying the preceding four sums
by their powers of \(\log n\) therefore gives
\begin{equation}\label{eq:rank-scan-s25e}
 \begin{split}
 |Q^{\rm cat}-\mathcal Q_{p^0}|
 &\le |Q^{\rm cat}-\mathcal Q_p|+|\mathcal Q_p-\mathcal Q_{p^0}|\\
 &\le C_A\left\{
 \frac{(\log n)^{5/4}}{m^{1/4}}
 +\frac{(\log n)^{3/2}}{\sqrt m}
 +\frac{(\log n)^2}{m}+\frac{m(\log n)^2}{n}\right\}=o(1).
 \end{split}
\end{equation}
The squared errors and the change from \(p\) to \(p^0\) give the last
terms. The two omitted endpoint cells satisfy
\[
 \sum_{j\in\{0,n_{\rm mid}\}}\sum_{a=1}^2
      \frac{w_{m+j}\ell_a(j)^2}{t_a(p)\{1-t_a(p)\}}
       \le C\log n/m,
\]
which is absorbed by the same bound. Retaining
\[
 |z_-|+|z_+|\le2\log\log n
\]
gives the same mean envelope; a higher subcritical tilt removes its
weighted complement. Equations
\eqref{eq:rank-scan-s25a}--\eqref{eq:rank-scan-s25e} therefore give the
joint comparison. Put \(\rho=e^{-h/2}\). The sum-mode exponent is
\[
 \Lambda_h(\theta)=
 \Lambda\{\theta(1+\rho)/2\}
 +\Lambda\{\theta(1-\rho)/2\},\quad
 \Lambda(s)=(1-\sqrt{1-8s})/4.
\]
Strict convexity gives
\begin{equation}\label{eq:rank-scan-s25}
\Lambda(\theta)-\Lambda_h(\theta)\ge c(h_0)>0.
\end{equation}
The endpoint eigenfunction for this sum mode is bounded by the geometric
mean of the two one-anchor eigenfunctions. To check it, write
$a(s)=(1-\sqrt{1-4s})/2$ for the $T$-parameterization and use that $a(s)/s$
is increasing. In the sum/difference coordinates,
\begin{equation*}
\frac{a\{\vartheta(1\pm\rho)/2\}}{1\pm\rho}
 \le\frac{a(\vartheta)}2.
\end{equation*}
H\"older bounds the joint boundary factor by the one-anchor factors.
Since \(R_{n,m}=R-2\log m+O(m/n)\), its length correction satisfies
\[
 m^{-C}\le
 \exp\{(R_{n,m}-R)\Lambda_h(\theta)\}
 \le m^C=R^{8C}.
\]
Thus \eqref{eq:rank-scan-s25} gives a far-tail factor
\(R^D e^{-c(h_0)R}\), uniformly in both splits.
In \eqref{eq:rank-scan-s24} take \(A>3\), and put
\[
 \begin{gathered}
 p_n(u)=P\{Q_{n,k(u)}>b_n(t(u))\},\\
 u_i=-L_\eta+i(T+g)/R,\quad I_i=[u_i,u_i+T/R],\qquad
 \mathcal G_i=I_i\cap\{u_i+ja/R:j\ge0\}.
 \end{gathered}
\]
For the retained cells define
\[
 \mathcal C_i=\{\max_{u\in\mathcal G_i}Q_{n,k(u)}>b_n(t(u))\}.
\]
The near and far sums are therefore
\[
 \begin{aligned}
 \sum_{i<j:\,u_j-u_i\le h_0}P(\mathcal C_i\cap\mathcal C_j)
   &\le C_a e^{-cg}\sum_i p_n(u_i)
           +CR^{-A}R^2\sup_u p_n(u),\\
 \sum_{i<j:\,u_j-u_i>h_0}P(\mathcal C_i\cap\mathcal C_j)
   &\le CR^{D+2}e^{-cR}\sup_u p_n(u).
 \end{aligned}
\]
The intensity varies by \(1+o(1)\) within each fixed cell. The union and
Bonferroni bounds therefore give upper and lower Riemann sums with
respective coefficients
\[
 \frac{H_q(T)}T,\qquad
 \frac{H_{q,a}(T)-C_a e^{-cg}}{T+g}.
\]
Take \(n\to\infty\), then \(T\to\infty\) with \(g=\sqrt T\), and finally
\(a\downarrow0\). Both coefficients tend to \(\nu(q)/2\), proving
\[
 P\{\max_k Q_{n,k}>b_n(t)\}
 \sim R\frac{\nu(q)}2
             \int_{-L_\eta}^{L_\eta}p_n(u)\,du .
\]
The integration already includes both directions of logit time.

\begin{proof}[Proof of Theorem~\ref{thm:max-subcritical-gumbel}]
Write $q=q_n$ and $\vartheta=\vartheta_n$. For $h=O(\log N)$ put
\[
 z_N(h)=(q-1)\sqrt{N/2}+\frac h{\sqrt{2N}}.
\]
Take $M=\lfloor n^{1/9}\rfloor$ in
\eqref{eq:rank-mean-constant-error} and
\eqref{eq:rank-variance-constant-error}. Expanding the square root
gives uniformly for $h=O(\log N)$
\[
 \begin{split}
 ET_{n,k}&=N+m(t)+O(n^{-1/9}),\\
 \operatorname{Var}(T_{n,k})&=2N+v_0(t)+O(n^{-1/18}),\\
 ET_{n,k}+\sqrt{\operatorname{Var}(T_{n,k})}\,z_N(h)
 &=Nq+d_0(t,q)+h
       +O\{\log N/N+n^{-1/18}\},
 \end{split}
\]
where $d_0$ is defined in Appendix~\ref{app:moment-correction-definitions}. Thus
Theorem~\ref{thm:rank-fixedsplit-subcritical}, uniformly over the trimmed
splits, gives
\[
 P\{D_{n,k}>z_N(h)\}
 =\frac{\Psi_t(\vartheta)
       e^{-NJ(q)-\vartheta\{d_0(t,q)+h\}}}
        {\vartheta\sqrt{2\pi N\lambda''(\vartheta)}}\{1+o(1)\}.
\]
The threshold's spatial shift converges uniformly to a continuous function.
The preceding actual-rank cell sum applies with $R=2N$, $Q=2T$ and
\[
 \frac{\nu(q)}{\vartheta}=q-1.
\]
Consequently, for one coordinate,
\[
 \begin{split}
 P\{\mathcal M_{j,n}>z_N(h)\}
 &\sim N\nu(q)\int_\eta^{1-\eta}
 \frac{\Psi_t(\vartheta)e^{-NJ(q)-\vartheta\{d_0(t,q)+h\}}}
      {\vartheta\sqrt{2\pi N\lambda''(\vartheta)}}
                                  \frac{dt}{t(1-t)}\\
 &=\frac{(q-1)\sqrt N\,\Theta_n}
           {\sqrt{2\pi\lambda''(\vartheta)}}
                     e^{-NJ(q)-\vartheta h}
  =C_n e^{-NJ(q)-\vartheta h}.
 \end{split}
\]
The quantities $\Theta_n,C_n,A_{n,p}^{\rm S},B_{n,p}^{\rm S}$ are those
in \eqref{eq:main-subcritical-normalization}. Positivity and compact
strict-subcritical continuity imply
\[
 0<c\le\Theta_n\le C,\qquad
 \log C_n=\tfrac12\log N+O(1),\qquad
 J(q_n)=\gamma_n=\frac{\log p}{N}.
\]
For a fixed real $x$, set $h_n(x)=(\log C_n+x)/\vartheta_n$. Then
\[
 \begin{split}
 B_{n,p}^{\rm S}+A_{n,p}^{\rm S}x&=z_N\{h_n(x)\},\\
 pC_n e^{-NJ(q_n)-\vartheta_nh_n(x)}&=e^{-x},\\
 pP\{\mathcal M_{j,n}>B_{n,p}^{\rm S}+A_{n,p}^{\rm S}x\}
       &\longrightarrow e^{-x}.
 \end{split}
\]
Define the actual coordinate exceedance count
\[
 Z_{n,p}(x)=\sum_{j=1}^p\mathbf1_{
 \{\mathcal M_{j,n}>B_{n,p}^{\rm S}+A_{n,p}^{\rm S}x\}}.
\]
Each event is measurable in its latent Gaussian column. Proposition
\ref{prop:sparse-factorization} therefore gives
\[
 Z_{n,p}(x)\Rightarrow\operatorname{Poisson}(e^{-x}),\qquad
 P\{Z_{n,p}(x)=0\}\longrightarrow e^{-e^{-x}},
\]
which proves \eqref{eq:rank-subcritical-gumbel-law}.
\end{proof}

\subsection{Strict transversality and the critical Hardy profile}

Let $b_+(t)=1/C_+(t)$ denote the positive-excursion Hardy threshold. The
following deterministic result strengthens monotonicity: its derivative is
strictly positive throughout the nonquadratic phase.

It also supplies the critical profile and the envelope identity used in the
boundary scan calculation.

\begin{proposition}[Strict transversality of the positive threshold]
\label{prop:hardy-threshold-transversality}
The function $b_+(t)$ is nondecreasing. On every interval where $b_+(t)<1/4$
it is locally analytic and $b_+'(t)>0$. For the full two-sided threshold,
\[
 b(t)=b_+(t)\quad(t\le1/2),\qquad
 b(t)=b_+(1-t)\quad(t\ge1/2).
\]
In particular $b_\eta=b(\eta)=b(1-\eta)$.
\end{proposition}

\begin{proof}
In the shooting ODE set
\[
 q=\operatorname{logit}x-\operatorname{logit}t,\qquad
 p=\operatorname{logit}u-\operatorname{logit}t,\qquad A=t/(1-t).
\]
Its right-hand side has the exact simplified form
\begin{equation}\label{eq:hardy-logit-simplified}
 p'=F(q,p;b,t)
 =bq\left\{\frac1{e^{q-p}-1}+\frac1{1+Ae^q}\right\}.
\end{equation}
To verify it, put $z=e^{p-q}$ and $B=Ae^q$ and use
$(1+Bz)/\{(1+B)(1-z)\}=z/(1-z)+1/(1+B)$. The partial derivatives satisfy
\begin{equation*}
 F_p=\frac{bqe^{q-p}}{(e^{q-p}-1)^2}>0,\qquad
 F_b=F/b>0,\qquad
 F_t=-\frac{bqA'(t)e^q}{(1+Ae^q)^2}<0.
\end{equation*}
Also $F_{pp}>0$.
For $0<b<1/4$, put $a=(1-\sqrt{1-4b})/2$. The smaller-slope center-started
solution has the analytic germ
\begin{equation*}
 p_0(q;b,t)=aq+
 \frac{a(1-a)^2(1/2-t)}{2-3a}q^2+O(q^3).
\end{equation*}
Along any continuation with $p_0<q$, let $H'/H=F_p$ with a positive
normalization. At zero, $H(q)\asymp q^\beta$, $\beta=a/(1-a)<1$.
Differentiation of the ODE and the germ give
\begin{equation}\label{eq:hardy-center-sensitivities}
 \partial_b p_0(q)=H(q)\int_0^q F_b(s)H(s)^{-1}\,ds>0,
 \qquad
 \partial_t p_0(q)=H(q)\int_0^q F_t(s)H(s)^{-1}\,ds<0.
\end{equation}
The central germs and forcing terms satisfy
\[
 \frac{\partial_b p_0(q)}{H(q)}\to0,\qquad
 \frac{\partial_t p_0(q)}{H(q)}\to0,\qquad
 \frac{F_b(q)}{H(q)}=O(q^{-\beta}),\qquad
 \frac{F_t(q)}{H(q)}=O(q^{1-\beta}),\quad \beta<1.
\]
Thus the homogeneous constants vanish and both integrals converge.
Their signs give the nondecreasing threshold comparison. For strict
differentiability, construct the unique terminal branch, uniformly on
compact subsets of \((b,t)\in(0,1/4)\times(0,1)\), with
\begin{equation}\label{eq:hardy-terminal-germ}
 p_*(q;b,t)=q-\log(bq)-\frac{1+1/b}{q}+O(q^{-2}).
\end{equation}
Let $P(q)$ be the first three terms in the last display and write $p=P+v$.
Then
\[
 B(q):=F_p(q,P)=1+\frac{1/b-1}{q}+O(q^{-2}),\qquad
 R(q):=F(q,P)-P'(q)=O(q^{-2}).
\]
The terminal equation is the Volterra fixed-point equation
\begin{equation*}
 v(q)=-\int_q^\infty
 e^{-\int_q^s B(u)\,du}\{R(s)+N(s,v(s))\}\,ds,
\end{equation*}
where
\[
 |N(s,v)|\le C|v|^2,\qquad
 |N(s,v_1)-N(s,v_2)|
 \le C(|v_1|+|v_2|)|v_1-v_2|.
\]
Let \(\|v\|_Q=\sup_{q\ge Q}q^2|v(q)|\), and denote the integral map by
\(\mathcal T\). Since its kernel is at most \(e^{-(s-q)/2}\),
\[
 \|\mathcal T v\|_Q\le C_0+CQ^{-2}\|v\|_Q^2,\qquad
 \|\mathcal T v_1-\mathcal T v_2\|_Q
 \le CQ^{-2}(\|v_1\|_Q+\|v_2\|_Q)\|v_1-v_2\|_Q.
\]
Take \(\mathcal R>2C_0\) and then \(Q\) so large that
\[
 C_0+CQ^{-2}\mathcal R^2\le\mathcal R,\qquad
 2CQ^{-2}\mathcal R<1.
\]
The map preserves the closed ball and is a contraction there.
These inequalities persist on a small complex parameter neighborhood;
uniformly convergent analytic iterates give
\eqref{eq:hardy-terminal-germ}. The non-growing terminal condition is
\[
 \frac{\partial_b p_*(q)}{H_*(q)}\to0,\qquad
 \frac{\partial_t p_*(q)}{H_*(q)}\to0
 \quad(q\to\infty).
\]
Variation of constants gives
\begin{equation}\label{eq:hardy-terminal-sensitivities}
 \partial_b p_*(q)=-H_*(q)\int_q^\infty F_b(s)H_*(s)^{-1}\,ds<0,
 \qquad
 \partial_t p_*(q)=-H_*(q)\int_q^\infty F_t(s)H_*(s)^{-1}\,ds>0.
\end{equation}
Indeed the derivatives of the terminal expansion are bounded, whereas its
integrating factor satisfies
\begin{equation}\label{eq:hardy-critical-sensitivity-asymptotic}
 H_*(q)\asymp e^q q^{1/b-1}.
\end{equation}
Integrating the expansion of \(F_p\) gives \(\log H_*(q)=q+(1/b-1)\log
q+O(1)\).
The terminal branch separates finite contact from admissibility.
While two trajectories remain in a fixed critical strip, \(F_p\ge c>0\);
for their difference \(d\),
\[
 d'=F(q,p_*+d)-F(q,p_*),\qquad
 d\,d'\ge c d^2.
\]
A nonzero difference therefore reaches a fixed small magnitude. For
\(P_\pm(q)=q-\log(bq)\pm\epsilon\), sufficiently large \(q\) gives
\[
 F(q,P_+)-P_+'(q)\ge c_\epsilon,\qquad
 F(q,P_-)-P_-'(q)\le-c_\epsilon.
\]
Above \(P_+\), \((q-p)'\le-c_\epsilon\), so contact \(p=q\) occurs in
finite time. Below \(P_-\), \((q-p)'\ge c_\epsilon\), and
\[
 0\le p'(q)
 \le Cq\{e^{-c_\epsilon q}+e^{-q}\},\qquad
 \int_Q^\infty p'(q)\,dq<\infty .
\]
The two terms in \eqref{eq:hardy-logit-simplified} therefore give
\[
 \int_Q^\infty p'(q)\,dq<\infty
 \ \Longleftrightarrow\ p(\infty)<\infty
 \ \Longleftrightarrow\ u(1)<1.
\]
At \(b_0=b_+(t_0)<1/4\), match the center and terminal branches at finite
large \(Q\):
\[
 D(b,t)=p_0(Q;b,t)-p_*(Q;b,t),\qquad D(b_0,t_0)=0.
\]
The absence of finite contact makes the center branch analytic near
\((b_0,t_0)\). Equations \eqref{eq:hardy-center-sensitivities} and
\eqref{eq:hardy-terminal-sensitivities} give
\[
 D_b>0,\qquad D_t<0,\qquad
 D(b_+(t),t)=0
 \ \Longrightarrow\
 b_+'(t)=-D_t/D_b>0.
\]
The implicit-function theorem also gives analyticity. Normalize the
integrating factors on the critical orbit to obtain
\begin{equation*}
 b_+'(t)=
 \frac{\displaystyle\int_0^\infty
       \frac{bqA'(t)e^q}{(1+Ae^q)^2H(q)}\,dq}
 {\displaystyle\int_0^\infty
       \frac{F(q,p(q);b,t)}{bH(q)}\,dq}>0.
\end{equation*}
The two integrands have the bounds
\[
 O(q^{1-\beta}),\quad O(q^{-\beta})\quad(q\downarrow0),\qquad
 O(e^{-cq}q^C)\quad(q\to\infty),\qquad \beta<1,
\]
by \eqref{eq:hardy-critical-sensitivity-asymptotic}; their integrals are
finite and strictly positive.
For a positive excursion, replace \(f\) by \(f^+=\max(f,t)\). Then
\[
 h_t(f^+)\le h_t(f),\qquad
 Hf^+\ge Hf,\qquad
 Hf\ge t\ \Longrightarrow\ h_t(Hf^+)\ge h_t(Hf).
\]
It suffices to consider nonnegative perturbations. For \(t\ge1/2\) and
\(0\le x\le1-t\),
\[
 h_t(t+x)=\sum_{j=2}^\infty c_j(t)x^j,\qquad
 c_j(t)=\frac{(-1)^jt^{1-j}+(1-t)^{1-j}}{j(j-1)}\ge0
 \quad(t\ge1/2).
\]
For a nonnegative perturbation \(g\), the ordinary Hardy inequality gives
\[
 \int_0^\infty(Hg)^j\,ds
 \le\left(\frac j{j-1}\right)^j\int_0^\infty g^j\,ds
 \le4\int_0^\infty g^j\,ds,\qquad j\ge2.
\]
Multiply by \(c_j(t)\), sum, and use monotone convergence to obtain
\(J_t(t+g)\le4I_t(t+g)\). The quadratic lower bound matches.
Reflection and the global monotone comparison now give the stated form of
$b(t)$ and the trimmed minimum.
\end{proof}

\begin{proposition}[The attained critical profile and its envelope]
\label{prop:critical-hardy-profile-envelope}
Suppose $b=b_+(t)<1/4$. There is a positive critical equality profile,
unique up to dilation, with finite entropy integrals $I=bJ\in(0,\infty)$.
Writing its control as $f_t(s)$ and its average as $r_t(s)$, with
arguments suppressed only inside the following integrals, define
\[
 D_t=\int_0^\infty
       \{t(1-r_t)h_t'(r_t)-(r_t-t)\}\,ds,
\]
\[
 E_t=\int_0^\infty
       \{(f_t-t)-t(1-f_t)[\operatorname{logit}f_t-
                              \operatorname{logit}t]\}\,ds.
\]
These integrals are finite, $D_t<0$, and
\begin{equation}\label{eq:critical-profile-envelope}
 E_t+bD_t=-t(1-t)b_+'(t)J.
\end{equation}
\end{proposition}

\begin{proof}
Let $p(q;t)$ be the matched critical orbit from the preceding proposition.
Put
\[
 R_t(q)=\operatorname{logistic}(\operatorname{logit}t+q),\qquad
 F_t(q)=\operatorname{logistic}(\operatorname{logit}t+p(q)),
\]
and normalize its dilation by
\begin{equation}\label{eq:critical-profile-parametrization}
 s(q)=\exp\left\{\int_1^q
       \frac{R_t(u)(1-R_t(u))}{F_t(u)-R_t(u)}\,du\right\}.
\end{equation}
This decreases from infinity to zero. Define the physical-time functions by
\[
 r_t(s(q))=R_t(q),\qquad f_t(s(q))=F_t(q).
\] In physical rank time it satisfies
$s r_t'(s)=f_t(s)-r_t(s)$, and hence $r_t(s)=s^{-1}\int_0^s f_t(v)\,dv$, since
$s r_t(s)\to0$ at zero.
Let $k=(1+\sqrt{1-4b})/2=1-a>1/2$. The center and terminal germs give
\begin{equation}\label{eq:critical-profile-endpoint-asymptotics}
 s(q)\asymp q^{-1/k}\quad(q\downarrow0),\qquad
 s(q)\asymp q^{-1/b}\quad(q\to\infty).
\end{equation}
The physical-rank tails satisfy
\[
 |r_t(s)-t|+|f_t(s)-t|=O(s^{-k})\quad(s\to\infty),\qquad
 1-r_t(s)+1-f_t(s)\le C s^{-C}e^{-cs^{-b}}\quad(s\downarrow0).
\]
Since \(k>1/2\), boundedness of \(h_t\) and its quadratic expansion at \(t\)
give
\[
 \begin{gathered}
 0<I=\int_0^\infty h_t(f_t(s))\,ds<\infty,\qquad
 0<J=\int_0^\infty h_t(r_t(s))\,ds<\infty,\\
 \int_1^\infty\{h_t(f_t)+h_t(r_t)\}\,ds
       \le C\int_1^\infty s^{-2k}\,ds.
 \end{gathered}
\]
The Bellman equality gives
\[
 \begin{split}
 I-bJ
 &=\lim_{\epsilon\downarrow0,\,M\uparrow\infty}
                [s g(r_t(s))]_\epsilon^M=0,\\
 \epsilon g(r_t(\epsilon))
 &=\epsilon\{(1-b)\log(1/t)+o(1)\}\longrightarrow0,\qquad
 M g(r_t(M))=O(M^{1-2k})\longrightarrow0.
 \end{split}
\]
Uniqueness of the matched orbit leaves only the multiplicative constant
in \eqref{eq:critical-profile-parametrization}:
\[
 s\mapsto cs,\quad c>0.
\]
In \(q\) coordinates, the two entropy densities and their parameter
derivatives are dominated on compact nonquadratic parameter intervals by
\[
 Cq^{\,1-1/k}(1+|\log q|)^C\mathbf1_{\{q\le1\}}
 +Cq^{-1/b-1}(1+\log q)^C\mathbf1_{\{q>1\}},
\]
an integrable function. Hold the optimizing profile fixed at $t$ and,
for either sign of small $h$, put
\[
 f_h=f_t+h(1-f_t)/(1-t),\qquad r_h=r_t+h(1-r_t)/(1-t).
\]
This maps $[t,1]$ exactly to $[t+h,1]$, is admissible for either sign of
$h$, and preserves averaging. The variational definition therefore gives
\[
 b_+(t+h)\le\frac{I_{t+h}(f_h)}{J_{t+h}(f_h)},
 \qquad b_+(t)=\frac{I_t(f_t)}{J_t(f_t)}.
\]
A differentiable upper bound touching at zero for both signs has the same
derivative. The derivative of $b_+$ is already proved in
Proposition~\ref{prop:hardy-threshold-transversality}. Direct
differentiation of the comparison profile gives, with $v=t(1-t)$,
\[
 I'_{\rm dir}=-E_t/v,\qquad J'_{\rm dir}=D_t/v.
\]
By \eqref{eq:critical-profile-endpoint-asymptotics}, the derivative
integrands are bounded by \(C\mathbf1_{s\le1}+Cs^{-2k}\mathbf1_{s>1}\),
an integrable function. Thus the touching quotient is differentiable and
\[
 b_+'(t)
 =\left.\frac{d}{dh}\frac{I_{t+h}(f_h)}{J_{t+h}(f_h)}\right|_{h=0}
 =\frac{I'_{\rm dir}J-IJ'_{\rm dir}}{J^2}
 =-\frac{E_t+bD_t}{vJ}.
\]
Multiplication by \(vJ\) gives \eqref{eq:critical-profile-envelope}.
Finally the integrand of $D_t$ is
$t(1-r_t)\{\log x-(x-1)\}$ for $x=r_t(1-t)/\{t(1-r_t)\}>1$, and is strictly
negative.
\end{proof}

\subsection{The logarithmic rate throughout the parameter space}

Write $T_{n,k}=nF_{n,k}$ and $L=\log n$, and retain the deterministic Hardy
constant $C(t)$ and its reciprocal $b(t)$. Define
\[
 b_\eta=\min_{t\in[\eta,1-\eta]}b(t),\qquad C_\eta^{\rm H}=b_\eta^{-1},
 \qquad
 \lambda(\vartheta)=\frac{1-\sqrt{1-4\vartheta}}2,
\]
and, for $0<b\le1/4$, put
\begin{equation*}
 \mathcal R_b(y)=\sup_{0\le\vartheta\le b}
                  \{\vartheta y-\lambda(\vartheta)\}.
\end{equation*}
At $b=1/4$ the continuous value $\lambda(1/4)=1/2$ is used. The following is
a logarithmic tail result, not a relative-tail or fluctuation theorem in the
critical and boundary regimes.

\begin{theorem}[Global fixed-split and split-scan logarithmic rates]
\label{thm:rank-global-logarithmic-rate}
For every $y>1$ and every sequence $k_n/n\to t\in(0,1)$,
\begin{equation*}
 \frac1{\log n}\log P\{T_{n,k_n}>y\log n\}
 \longrightarrow-\mathcal R_{b(t)}(y).
\end{equation*}
For every fixed trimming fraction,
\begin{equation}\label{eq:global-split-scan-rate}
 \frac1{\log n}\log P\left\{
       \max_{k\in\mathcal K_n}T_{n,k}>y\log n\right\}
 \longrightarrow-\mathcal R_{b_\eta}(y).
\end{equation}
The logarithmic rates are zero for $y\le1$. More explicitly, with
$q_b=(1-4b)^{-1/2}$ and $q_{1/4}=\infty$,
\begin{equation}\label{eq:global-rate-piecewise}
 \mathcal R_b(y)=
 \begin{cases}
 0,&y\le1,\\
 (y-1)^2/(4y),&1<y\le q_b,\\
 by-\lambda(b),&y\ge q_b\quad(b<1/4).
 \end{cases}
\end{equation}
The upper estimates are uniform over trimmed split locations.
\end{theorem}

\begin{proof}
We first note continuity of the deterministic threshold. For $s$ near $t$
define
\[
 \alpha(s,t)=\min\{s/t,(1-s)/(1-t)\},\qquad
 f_s=s+\alpha(s,t)(f-t).
\]
This maps admissible controls at baseline $t$ into admissible controls at
baseline $s$ and commutes with averaging. Uniformly for $x\in[0,1]$,
\[
 \frac{h_s\{s+\alpha(s,t)(x-t)\}}{h_t(x)}\longrightarrow1
 \quad(s\to t).
\]
The removable central value and its Taylor expansion are
\[
 \begin{aligned}
 \left.\frac{h_s(s+\alpha(s,t)(x-t))}{h_t(x)}\right|_{x=t}
 &=\frac{\alpha(s,t)^2t(1-t)}{s(1-s)},\\
 \frac{h_s(s+\alpha(s,t)(x-t))}{h_t(x)}
 &=\frac{\alpha(s,t)^2t(1-t)}{s(1-s)}
                         \{1+O(|x-t|)\}.
 \end{aligned}
\]
Off \(|x-t|<\epsilon\), the denominator has a positive lower bound.
Uniform continuity there and the central expansion imply
\[
 \epsilon(s,t):=\sup_x
 \left|\frac{h_s(s+\alpha(s,t)(x-t))}{h_t(x)}-1\right|
 \longrightarrow0
\]
uniformly on compact interior baseline intervals. Affine averaging gives
\[
 \begin{gathered}
 (1-\epsilon)I_t(f)\le I_s(f_s)\le(1+\epsilon)I_t(f),\qquad
 (1-\epsilon)J_t(f)\le J_s(f_s)\le(1+\epsilon)J_t(f),\\
 C(s)\ge\frac{1-\epsilon(s,t)}{1+\epsilon(s,t)}C(t),\qquad
 C(t)\ge\frac{1-\epsilon(t,s)}{1+\epsilon(t,s)}C(s).
 \end{gathered}
\]
Thus \(C(s)\to C(t)\), \(b(s)\to b(t)\), and the compact trimmed interval
contains a minimizer of \(b\).
Fix $0<\vartheta<b(t)$. The probability sandwich and the real
truncated-hybrid transform in the proof of
Theorem~\ref{thm:rank-fixedsplit-subcritical} imply, for every fixed $A>0$,
\begin{equation}\label{eq:global-fixedsplit-chernoff}
 P(T_{n,k}>yL)
 \le n^{\lambda(\vartheta)-\vartheta y+o(1)}+O(n^{-A}).
\end{equation}
The bound follows from the truncated-hybrid Chernoff inequality and the
original-probability comparison. Uniformly on compact strict-subcritical
sets, taking \(A\) large gives
\[
 \limsup_{n\to\infty}L^{-1}\log P(T_{n,k}>yL)
 \le\lambda(\vartheta)-\vartheta y,\qquad \vartheta<b(t).
\]
Continuity permits \(k_n/n\to t\), and optimization gives
\[
 \inf_{0\le\vartheta<b(t)}
       \{\lambda(\vartheta)-\vartheta y\}=-\mathcal R_{b(t)}(y).
\]
For the lower bound, take a piecewise constant control with
\(f(s)=t\) for all sufficiently large \(s\), and
\[
 I=\int_0^\infty h_t(f(s))\,ds>0,\qquad
 J=\int_0^\infty h_t(Hf(s))\,ds.
\]
Choose \(S\) beyond the nonconstant part of \(f\) so that
\[
 J_S=\int_0^S h_t(Hf(s))\,ds\ge J-\epsilon,\qquad
 N=\lfloor aSL\rfloor=O(L),\quad a>0.
\]
Fix a sufficiently fine partition \(0=s_0<\cdots<s_d=S\), including all
breakpoints of \(f\), with \(f=f_\nu\) on each block, before
increasing \(n\). Put
\[
 q_\nu=\lfloor aLs_\nu\rfloor,\quad
 b_\nu=\lfloor(q_\nu-q_{\nu-1})f_\nu\rfloor,\quad
 \mathcal C_n=\bigcap_{\nu=1}^d
       \{S_{q_\nu}-S_{q_{\nu-1}}=b_\nu\}.
\]
Stirling's formula gives
\[
 \log P_t(\mathcal C_n)
 =-aL\sum_{\nu=1}^d(s_\nu-s_{\nu-1})h_t(f_\nu)+O(\log L).
\]
The conditional within-block law of large numbers costs \(e^{-o(L)}\);
thus the prescribed counts and cumulative profile give an event \(E_n\) with
\begin{equation}\label{eq:global-profile-event}
 P(E_n)=e^{-aLI+o(L)},\qquad
 \sum_{l\le N}\frac{nG_{k,l}}{d_{l,n}}
 \ge aLJ_S-o(L)\quad\hbox{on }E_n.
\end{equation}
The cumulative-profile convergence in \eqref{eq:global-profile-event} gives the entropy
limit after deleting a small initial interval. The errors satisfy
\[
 \begin{gathered}
 \int_0^\epsilon h_t(Hf(s))\,ds\le C\epsilon,\\
 \left|\sum_{l\le N}\left(\frac l{l-1/2}-1\right)
                  h_t(S_l/l)\right|\le C\sum_{l\le N}l^{-1}\le C\log L,\\
 \sum_{l\notin[1,N]\cup[m,n-m]}\frac{nG_{k,l}}{d_{l,n}}\ge0.
 \end{gathered}
\]
For a prescribed prefix \(\mathcal P\) of length \(N=O(L)\), the
full-bridge likelihood ratio is
\[
 \frac{P(\mathcal P\mid S_n=k)}{P_t(\mathcal P)}
 =1+O(L^2/n).
\]
To obtain the conditional bulk lower bound, take \(m=L^8\).
The prefix displacement and the likelihood comparison for the remaining
boundary blocks satisfy
\[
 |A_N|\le CN=O(L),\qquad
 |EA_m|/\sqrt m=O(L/\sqrt m)\longrightarrow0,\qquad
 \frac{dP_{\rm blocks\mid\mathcal P}}{dP_{\rm independent\ blocks}}
 =1+O(m^2/n).
\]
The binomial local limit formula therefore gives
\[
 \inf_{\mathcal P}P\{|Z_-|,|Z_+|\le1\mid\mathcal P\}\ge c>0 .
\]
Given these endpoints, the conditional bridge coupling and entropy
expansion give
\[
 P\{|T_{\rm mid}-Q_{\rm mid}/2|>\epsilon_n
       \mid\mathcal P,Z_-,Z_+\}\le n^{-A},\qquad \epsilon_n\to0,
\]
uniformly over the prescribed prefixes and retained endpoints.
The exact conditional OU kernel on the interval of length $R_{n,m}=2L-O(\log
L)$ gives, for every fixed $q>1$,
\begin{equation}\label{eq:global-conditional-bulk-lower}
 P\left\{\sum_{m\le l\le n-m}\frac{nG_{k,l}}{d_{l,n}}>qL
       \ \middle|\ \mathcal P\right\}
 \ge n^{-I_G(q)+o(1)},\qquad I_G(q)=\frac{(q-1)^2}{4q}.
\end{equation}
Indeed, its conditional saddle is
\(q_n=2qL/R_{n,m}\to q\), with
\(\vartheta(q_n)\to(1-q^{-2})/4<1/4\). Uniformly over endpoints in
\([-1,1]\), the exact Gaussian kernel gives
\[
 \log P_G\{Q_{\rm mid}/2>qL\}
 =-\frac{R_{n,m}}2 I_G(q_n)+O(\log L)
 =-LI_G(q)+o(L).
\]
Their conditional probability is bounded below by a positive constant.
Choosing \(A>I_G(q)+1\), the coupled lower probability is consequently
\[
 c\,e^{-LI_G(q)+o(L)}-n^{-A}
 =e^{-LI_G(q)+o(L)}.
\]
The profile and middle use disjoint ranks. Since the other entropy terms
are nonnegative, choose
\[
 1<q<y,\qquad a=\frac{y-q+\epsilon}{J_S}
 \ \Longrightarrow\ aJ_S+q=y+\epsilon.
\]
Multiplying the prefix probability in \eqref{eq:global-profile-event}
by the conditional probability in
\eqref{eq:global-conditional-bulk-lower} gives
\[
 \liminf L^{-1}\log P(T_{n,k}>yL)
 \ge-I_G(q)-\frac{y-q+\epsilon}{J_S/I}.
\]
Let $J/I\uparrow C(t)$ and then remove the truncation and slack errors. With
no rare prefix, the same Gaussian middle construction at $q>y$, followed by
$q\downarrow y$, gives the Gaussian-only lower bound. The resulting optimal
exponent is
\[
 \inf_{1\le q\le y}\{I_G(q)+b(t)(y-q)\}
 =\sup_{0\le\vartheta\le b(t)}
       \{\vartheta y-\lambda(\vartheta)\}.
\]
The minimization is explicit:
\[
 \frac{d}{dq}\{I_G(q)+b(t)(y-q)\}
 =\frac{1-q^{-2}}4-b(t),\qquad
 q_*=\min\{y,q_{b(t)}\}.
\]
The minimizing value is
\[
 I_G(y)\quad(y\le q_{b(t)}),\qquad
 I_G(q_{b(t)})+b(t)(y-q_{b(t)})
 =b(t)y-\lambda(b(t))\quad(y\ge q_{b(t)}),
\]
which proves \eqref{eq:global-rate-piecewise}. For \(k_n/n\to t\), the
affine control map preserves the limiting costs, while all bulk bounds
are locally uniform. For \(y<1\), the null moments give
\[
 P(T_{n,k}\le yL)\le\frac{C}{L(1-y)^2}\longrightarrow0.
\]
At \(y=1\), the lower bound at each \(q>1\) gives
\(\liminf L^{-1}\log P(T_{n,k}>L)\ge-I_G(q)\uparrow0\).
For the scan take \(m=L^8\), \(\delta=L^{-12}\), and a deterministic
integer grid \(\mathcal G_n\) with spacing at most \(n\delta\):
\[
 |\mathcal G_n|\le C L^{12},\qquad
 |k-k_0(k)|\le n\delta
\]
for a neighboring grid point \(k_0(k)\). We prove
\begin{equation}\label{eq:global-scan-logarithmic-modulus}
 P\left\{\max_{k\in\mathcal K_n}
 |T_{n,k}-T_{n,k_0(k)}|>\epsilon_nL\right\}\le n^{-A},
 \qquad \epsilon_n\to0.
\end{equation}
The following estimates establish this modulus for the actual entropy
statistic, including its rare adjacent-rank jumps.
Hypergeometric concentration, followed by a union bound over all split-rank
pairs, gives
\begin{equation}\label{eq:global-modulus-count-envelope}
 |A_{k,l}|\le C_A\sqrt{b_lL}
 \quad\hbox{for all trimmed }k\hbox{ and all }l,
\end{equation}
outside probability $n^{-A-2}$. Sampling-without-replacement convex
domination by a binomial law gives a Bernstein bound for each time increment
of length $r\le n\delta$. Union over all such increments and ranks, at most
$O(n^3)$ choices, gives simultaneously
\begin{equation}\label{eq:global-modulus-count-increments}
 |A_{k+r,l}-A_{k,l}|
 \le C_A\{\sqrt{\delta b_lL}+L\},
\end{equation}
outside an event of probability \(n^{-A-2}\), with Bernstein constant
larger than \(A+5\). Complementary counts give \(b_l\) at the upper end.
For the number \(Z_{\rm cell}\) of boundary ranks in a time cell,
\[
 \E Z_{\rm cell}\le Cm\delta=CL^{-4},\qquad
 j=\lceil c_A L/\log L\rceil,\qquad
 P(Z_{\rm cell}\ge j)\le(eCm\delta/j)^j.
\]
Indeed,
\[
 j\log\{j/(eCm\delta)\}=(5c_A+o(1))L,
\]
choosing \(c_A\) large gives
\[
 P\{\max_{\rm cells}Z_{\rm cell}\ge j\}
 \le CL^{12}\exp\{-(5c_A+o(1))L\}\le n^{-A-2}.
\]
On its complement, every boundary prefix of length at most \(m\)
changes by at most \(j\) labels inside a cell. Uniformly in \(k\),
\[
 \left|\sum_{l\le m\,\text{or}\,l\ge n-m}\frac{nG_{k,l}}{d_{l,n}}
         -E_m^{-1/2}(k)-E_m^{+1/2}(k)\right|
 \le Cm^2/n=o(1).
\]
For trimmed \(t\), the entropy modulus and parameter derivative are
\[
 |h_t(x+z)-h_t(x)|\le C_\eta z\log(e/z),\quad 0\le z\le1,
 \qquad |\partial_t h_t(x)|\le C_\eta.
\]
Choose a sufficiently large constant $C_0$ depending on the count envelope.
On ranks $l\le C_0L$, split at $l=j$ and use $z\le\min(1,j/l)$. Their entire
within-cell change is at most
\begin{equation*}
 Cj\{1+\log^2(C_0L/j)\}+C\delta L=o(L).
\end{equation*}
For \(C_0L\le l\le m\), the count envelope and within-cell displacement give
\[
 x_l,x_l+\Delta x_l\in[c_\eta,1-c_\eta],\qquad
 |h_t'(x_l)|\le C\sqrt{L/l},\qquad |\Delta x_l|\le j/l.
\]
Thus
\[
 |h_{t+\Delta t}(x_l+\Delta x_l)-h_t(x_l)|
 \le C\{j\sqrt L/l^{3/2}+j^2/l^2+\delta\}.
\]
For either \(c=\pm1/2\), summing the actual prefix changes gives
\begin{equation*}
 \begin{split}
 &\left|\sum_{C_0L\le l\le m}\frac l{l+c}
     \{h_{t+\Delta t}(x_l+\Delta x_l)-h_t(x_l)\}\right|\\
 &\quad\le C\sum_{l\ge C_0L}
       \{j\sqrt L/l^{3/2}+j^2/l^2\}+C\delta m
 \le C(j+j^2/L+\delta m)=o(L).
 \end{split}
\end{equation*}
The two estimates hold at both rank ends. In the middle, the count envelope
gives
\[
 \begin{gathered}
 \left|\sum_{l=m}^{n-m}\frac{nG_{k,l}}{d_{l,n}}
                 -\frac1{2v}\sum_{l=m}^{n-m}w_l A_{k,l}^2\right|
 \le CL^{3/2}\sum_{l=m}^{n-m}b_l^{-3/2}
 \le CL^{3/2}/\sqrt m=o(1),\\
 w_l=\frac{n^2}{d_{l,n}\,l(n-l)}\le Cb_l^{-2}.
 \end{gathered}
\]
Using \eqref{eq:global-modulus-count-envelope} and
\eqref{eq:global-modulus-count-increments} in the quadratic increment
gives the weighted numerator difference
\begin{equation*}
 \left|\sum_{l=m}^{n-m}w_l\{A_{k+r,l}^2-A_{k,l}^2\}\right|
 \le C\{L^2\sqrt\delta+L^{3/2}/\sqrt m
                    +\delta L^2+L^2/m\}=o(L).
\end{equation*}
The first cross term and the divisor change satisfy
\[
 \sum_{l=m}^{n-m}b_l^{-2}\sqrt{b_lL}\sqrt{\delta b_lL}
 \le CL\sqrt\delta\sum_{l=m}^{n-m}b_l^{-1}
 \le CL^2\sqrt\delta,\qquad
 |\Delta(v^{-1})|\,Q_{\rm mid}\le C\delta L^2.
\]
Combining all ranks, one may take
\[
 \begin{split}
 \epsilon_n=C_A\bigg[
 &\frac jL\{1+\log^2(C_0L/j)\}+\frac{j^2}{L^2}
       +\frac{\delta m}{L}+L\sqrt\delta+\delta L\\
 &+\frac{\sqrt L}{\sqrt m}+\frac Lm+\frac{m^2}{nL}\bigg]
 \longrightarrow0,
 \qquad
 m=L^8,\quad\delta=L^{-12},\quad j=\lceil c_A L/\log L\rceil .
 \end{split}
\]
Each concentration exception has probability at most \(n^{-A-2}\);
their union proves \eqref{eq:global-scan-logarithmic-modulus}.
Use \eqref{eq:global-fixedsplit-chernoff} at any fixed $0<\vartheta<b_\eta$,
uniformly over the grid points. The modulus gives
\[
 P\{\max_kT_{n,k}>yL\}
 \le L^{12}n^{\lambda(\vartheta)-\vartheta(y-\epsilon_n)+o(1)}
                                                 +O(n^{-A}).
\]
Taking $A$ large and then optimizing $\vartheta<b_\eta$ proves the scan
upper rate. For the lower rate choose a fixed fraction attaining $b_\eta$
and use the fixed-split result; integer rounding at a trimming endpoint is
covered by its sequential uniformity.
This proves \eqref{eq:global-split-scan-rate}.
\end{proof}

\subsection{The original MAX statistic in a bounded critical window}
\label{app:critical-core}

Throughout this subsection $N=\log n$ and the transform parameter
$\vartheta$ tilts the raw statistic $T_{n,k}=nF_{n,k}$. Retain the Bellman
functions $u_{t,\vartheta},g_{t,\vartheta}$, amplitudes $A_{t,\vartheta,c}$,
and coefficients $W_m^c,K_c$ in
\eqref{eq:bellman-u-ode}--\eqref{eq:analytic-boundary-coefficient}. In
particular
\[
 r(\vartheta)=\sqrt{1-4\vartheta},\quad
 a(\vartheta)=\frac{1-r(\vartheta)}2,\quad
 \kappa(\vartheta)=\frac{a(\vartheta)}2,\quad
 \lambda(\vartheta)=2\kappa(\vartheta).
\]
All square roots use the branch positive on the real subcritical interval.
The analysis in this subsection covers
$\{\log p-\gamma_c\log n\}/\sqrt{\log n}=O(1)$ and uses localized
transforms.

\subsubsection{Real critical boundary constants}

First fix $t<1/2$ with $b=b(t)<1/4$. The active Bellman endpoint is $x=1$;
the opposite endpoint has a strictly positive subcritical margin. This
follows from Proposition~\ref{prop:hardy-threshold-transversality},
including strict monotonicity wherever the threshold is below $1/4$.
Reflection gives the result for $t>1/2$. Write
\[
 v=t(1-t),\quad h_t(x)=\operatorname{KL}(x\|t),\quad
 h_1=\log(1/t),\quad H_t=\log\{(1-t)/t\},\quad
 a_b=a(b),\quad \kappa_b=a_b/2.
\]
For $c>-1$, and in particular $c=\pm1/2$, define
\begin{equation}\label{eq:core-boundary-exponents}
 \alpha_c=\frac12-\kappa_b-bc h_1,\qquad
 B_c=\frac{1-a_b}{2a_b}-c h_1,\qquad Q_c=B_c+1.
\end{equation}
The notation $Q_c$ here is an exponent, not the quadratic statistic. The
identity $b=a_b(1-a_b)$ gives $Q_c-\alpha_c/b=1/2$.

Put $e=1-x$, $q=h_t'(x)$, and $w_b(q)=1-u_{t,b}(1-e(q))$, where
$e(q)=\{1+\exp(q-H_t)\}^{-1}$. The following limits specify positive
deterministic constants:
\begin{equation}\label{eq:core-ode-constants}
 A_{*,c}=\lim_{q\to\infty}A_{t,b,c}(1-e(q))e(q)^{1/2}q^{-B_c},
 \qquad
 C_U=\lim_{q\to\infty}q^{-1/b}
        \left.\partial_\vartheta u_{t,\vartheta}(1-e(q))
        \right|_{\vartheta=b}.
\end{equation}
They are evaluated from the critical ODE and its linear sensitivity, with
the central normalizations already imposed on $u$ and $A$. Set, with $\psi$
denoting the digamma function,
\begin{equation}\label{eq:core-first-failure-constant}
 D_c=\frac{bA_{*,c}}{\sqrt{2\pi}}
              \exp\{bc h_1\psi(c+1)\}.
\end{equation}

\begin{lemma}[Real critical singularity]\label{lem:core-real-singularity}
The limits in \eqref{eq:core-ode-constants} exist in $(0,\infty)$. As
$\Delta\downarrow0$,
\begin{equation}\label{eq:core-real-K-singularity}
 K_c(t,b-\Delta)\sim
 \begin{cases}
 D_c\Gamma(\alpha_c)C_U^{-\alpha_c}
       \Delta^{-\alpha_c}\{\log(1/\Delta)\}^{1/2},&\alpha_c>0,\\
 (2D_c/3)\{\log(1/\Delta)\}^{3/2},&\alpha_c=0,\\
 K_c(t,b)\in(0,\infty),&\alpha_c<0.
 \end{cases}
\end{equation}
The finite critical value in the last line is a convergent first-failure
series. All ODE and continuation estimates in the proof are uniform in
compact families with the same active endpoint and a uniform margin
$b(t)<1/4$.
\end{lemma}

\begin{proof}
For $f_e(w)=w(1-w)/(w-e)$, the critical ODE has the tail form
\[
 w_b(q)=b\int_q^\infty e(s)(1-e(s))s f_{e(s)}(w_b(s))\,ds.
\]
Writing $z=w_b/(be)$, monotonicity of $w_b$ and $f_e(w)\ge1-w$ first give
$z\ge q/2$. Then $f_e(w_b)\le1+C/q$ gives $z\le q+C$. A second substitution,
using $\int_q^\infty e(s)(1-e(s))s\,ds=e(q)\{q+1+O(e(q))\}$, yields
\begin{align}\label{eq:core-critical-endpoint-expansion}
 w_b&=be\{q+1+1/b+O(q^{-1})\},\\
 g_b''(1-e)&=e^{-1}\{1-q^{-1}+O(q^{-2})\},\nonumber\\
 g_b(1)-g_b(1-e)
 &=e\{q-\log(bq)+1-(2+1/b)q^{-1}+O(q^{-2})\},\nonumber\\
 (\log A_{b,c})'(1-e)
 &=\frac1{2e}+\frac{B_c}{eq}+O((eq^2)^{-1}).\nonumber
\end{align}
Substitution in the Bellman and transport identities gives the last two
expansions. In tail coordinates the amplitude remainder satisfies
\[
 \frac d{dq}\log\{A_{b,c}(1-e)e^{1/2}q^{-B_c}\}=O(q^{-2}),
\]
and hence
\[
 A_{b,c}(1-e)=A_{*,c}e^{-1/2}q^{B_c}\{1+O(q^{-1})\},
 \qquad 0<A_{*,c}<\infty.
\]
For the parameter-gap comparison put
\[
 \delta_\Delta=1-u_{t,b-\Delta}(1),\qquad
 z(q)=w_{b-\Delta}(q)-w_b(q)\ge0.
\]
Subtracting the exact equations gives
\[
 z'=A_z z+\Delta B_z,
\]
where, writing $e=e(q)$,
\[
 A_z=be(1-e)q\left\{1+
 \frac{e(1-e)}{(w_b-e+z)(w_b-e)}\right\},\qquad
 B_z=e(1-e)q f_e(w_b+z).
\]
Both coefficients are positive and bounded above by their values at $z=0$.
The critical sensitivity $U$ solves $U'=A_0U+B_0$, with
$A_0=(bq)^{-1}+O(q^{-2})$, $B_0=O(e^{-q}q)$ and positive central germ. For
any fixed sufficiently large $q_0$, let $M(q)=\exp\{\int_{q_0}^q
A_0(s)\,ds\}$. Thus
\[
 C_U=q_0^{-1/b}
 \exp\!\left\{\int_{q_0}^\infty
            [A_0(s)-(bs)^{-1}]\,ds\right\}
 \left\{U(q_0)+\int_{q_0}^\infty B_0(s)/M(s)\,ds\right\}>0,
\]
and $U(q)=C_Uq^{1/b}\{1+O(q^{-1})\}$. At fixed $q_0$, $z(q_0)=\Delta
U(q_0)+O(\Delta^2)$, so scalar comparison gives $z(q)\le\Delta
U(q)(1+C\Delta)$. Put \(L=\log(1/\Delta)\),
\(q_\pm=L\pm M_0\log L\), with \(M_0>1/b+3\).
Variation of constants before \(q_-\), where
\(z/(w_b-e)=O(L^{-4})\), gives
\[
 z(q_-)=\Delta U(q_-)\{1+O(L^{-2})\}.
\]
On \([q_-,q_+]\), positivity and \(A_z\le A_0\) imply
\[
 \begin{split}
 z(q_-)\le z(q_+)
 &\le e^{\int_{q_-}^{q_+}A_0}
       \left\{z(q_-)+\Delta\int_{q_-}^{q_+}B_0(s)\,ds\right\},\\
 \int_{q_-}^{q_+}A_0(s)\,ds&=O(\log L/L),\qquad
 U(q_-)=C_U L^{1/b}\{1+O(\log L/L)\}.
 \end{split}
\]
Beyond \(q_+\), \(A_z z+\Delta B_z\le Ce(q)(q+1)\), so
\[
 0\le\delta_\Delta-z(q_+)
 \le C\int_{q_+}^\infty e(q)(q+1)\,dq
 \le C\Delta L^{1-M_0}.
\]
The forcing integral on the middle window is exponentially smaller than
\(U(q_-)\). Combining these inequalities proves
\begin{equation}\label{eq:core-real-gap}
 \delta_\Delta=C_U\Delta\{\log(1/\Delta)\}^{1/b}
       \left\{1+O\!\left(\frac{\log\log(1/\Delta)}
                                  {\log(1/\Delta)}\right)\right\}.
\end{equation}
For \(\vartheta<b\), define the continuation by
\[
 H_{l,\vartheta}^c(s)
 =\lim_{L\to\infty}\E_{l,s}\left[
 \exp\left\{\vartheta\sum_{k=l+1}^L
             \frac{k}{k+c}h_t(S_k/k)\right\}
 W_L^c(S_L;\vartheta)\right].
\]
The positive row construction in Lemma~\ref{lem:rank-prefix-tilt}
gives this limit at every fixed state and the harmonic identity
\begin{equation}\label{eq:core-harmonic-identity}
 \E_t[e^{\vartheta T_l^c}H_{l,\vartheta}^c(S_l)]
       =K_c(t,\vartheta).
\end{equation}
Write \(j=l-s\ge1\) and let \(R_l(s)\) be the \(W\)-row ratio.
Uniformly as \(\vartheta\uparrow b\), the endpoint bounds give
\[
 |g_\vartheta^{(3)}(x)|\le C(1-x)^{-2},\qquad
 |\partial_x^i\log A_{\vartheta,c}(x)|\le C(1-x)^{-i},
 \quad i=1,2,
\]
\[
 |\Delta x_{\rm success}|=\frac{1-x}{l+1},\qquad
 |\Delta x_{\rm failure}|\le\frac C l.
\]
The transport identity cancels the complete first-order row sum.
For \(j\) above a fixed large integer, with normalized failure hazard
\(\widetilde p_l\),
\begin{equation}\label{eq:core-hazard-row}
 |R_l(l-j)-1|\le C\{\widetilde p_l/j^2+l^{-2}\},
 \qquad \widetilde p_l\asymp w_\vartheta(1-j/l).
\end{equation}
Use the log-row quantities \(\ell_\xi,\ell_\xi^0,A_\xi^{\rm row}\) defined in the proof
of Lemma~\ref{lem:rank-prefix-tilt}, evaluated at the present
\(g_\vartheta,A_{\vartheta,c},x=1-j/l\), with \(\xi\in\{0,1\}\). The remainder keeps its transition
weight:
\[
 \rho_{l,\xi}=e^{\ell_\xi-\ell_\xi^0}-1-A_\xi^{\rm row}/l,\qquad
 \widetilde p_l|\rho_{l,0}|\le C\widetilde p_l/j^2,
 \qquad(1-\widetilde p_l)|\rho_{l,1}|\le Cl^{-2}.
\]
During residence at minority level \(j\), define the accumulated
predictable failure hazard, including the possible final step, by
\[
 J_k=k-S_k,\quad \sigma_j=\inf\{k\ge l:J_k=j\},\quad
 p_k=P(J_{k+1}=J_k+1\mid\mathcal F_k),\qquad
 V_j=\sum_{k=\sigma_j}^{\sigma_{j+1}-1}p_k .
\]
The survival product satisfies
\[
 \prod_{i<k}(1-p_i)\le e^{-\sum_{i<k}p_i},\qquad 0\le p_k\le1.
\]
Including the possible last hazard therefore gives
\[
 \begin{split}
 \Pp(V_j>x\mid\mathcal F_{\sigma_j})&\le \min(1,e^{1-x}),\\
 \E(e^{sV_j}\mid\mathcal F_{\sigma_j})
 &=1+s\int_0^\infty e^{sx}\Pp(V_j>x\mid\mathcal F_{\sigma_j})\,dx\\
 &\le1+C s\le e^{Cs},\qquad 0\le s\le s_0<1.
 \end{split}
\]
Apply this conditional bound successively with parameters \(s/j^2\).
Since \(\sum_{j\ge J}j^{-2}\le C/J\), it proves
\[
 \E\exp\!\left\{s\sum_{j\ge J}V_j/j^2\right\}
       \le\exp(Cs/J),\qquad 0\le s\le cJ^2.
\]
For \(U_J=\sum_{j\ge J}V_j/j^2\), this gives
\[
 EU_J\le C/J,\qquad
 \Pp(U_J>C_1/J)
 \le \exp\{-cJ^2(C_1/J)+C cJ\}\le e^{-c'J}
\]
when \(C_1\) is sufficiently large. Its exponential row products are
uniformly integrable. Adding the compact-interior and opposite-endpoint
occupation errors gives
\[
 \sup_{j\ge J}|H_{l,\vartheta}^c(l-j)/W_l^c(l-j;\vartheta)-1|
       \le C/J+C(1+\log l)^3/l.
\]
In a fixed central interval both starting label counts are at least
\(cl\) and subsequently do not decrease. Thus
\[
 \mathbb E_{l,s}\sum_{k\ge l}
 \left\{\frac{p_k}{J_k^2}
       +\frac{1-p_k}{S_k^2}+k^{-2}\right\}
 \le C\sum_{j\ge cl}j^{-2}+C\sum_{k\ge l}k^{-2}\le C/l,
 \qquad |H_l/W_l-1|\le C/l.
\]
For fixed minority count \(j\), let \(a_{k,j},f_{k,j}\) be the
unnormalized success and failure entries. Near criticality they satisfy
\[
 a_{k,j}\le e^{-w_\vartheta(1-j/k)/2},\qquad
 f_{k,j}\asymp w_\vartheta(1-j/k),\qquad
 w_\vartheta(1-j/k)\ge c_j\log k/k.
\]
The next-failure kernel and its total mass satisfy
\[
 \mathcal K_{l,j}(r)=
       \left(\prod_{k=l}^{l+r-1}a_{k,j}\right)f_{l+r,j},
 \qquad c_j\le\sum_{r\ge0}\mathcal K_{l,j}(r)\le C_j,
\]
\[
 \prod_{k=l}^{l+r-1}a_{k,j}
 \le C_j\exp\{-c_j[(\log(l+r))^2-(\log l)^2]\}.
\]
Finite initial times are absorbed in \(c_j,C_j\).
Backward induction from large \(J\) gives positive uniform \(H/W\)
bounds for all \(j\ge1\). On the local scale
\[
 l\to\infty,\quad\vartheta\uparrow b,\quad
 \delta_\Delta l=O(1),\qquad r=yl/\log l,
\]
the entries obey
\[
 a_{l+r,j}=1-bj\log l/l+o(\log l/l),\qquad
 f_{l+r,j}=bj L_j\log l/l+o(\log l/l),
 \quad L_j=e^{-1}(1+1/j)^{j+1/2}.
\]
The survival bound permits dominated convergence of the entire waiting
kernel. Its limiting mass is $L_j$. Since $L_j=\mathcal S(j)/\mathcal
S(j+1)$, where
\begin{equation}\label{eq:core-Gamma-continuation}
 \mathcal S(j)=\frac{\Gamma(j)e^j}{\sqrt{2\pi}\,j^{j-1/2}},
 \qquad
 \frac{H_{l,\vartheta}^c(l-j)}{W_l^c(l-j;\vartheta)}
       \longrightarrow\mathcal S(j),
\end{equation}
the finite induction has the telescoping multiplier
\[
 \prod_{k=j}^{J}L_k=\frac{\mathcal S(j)}{\mathcal S(J+1)},\qquad
 \mathcal S(J+1)=1+O(J^{-1}).
\]
The continuation beyond \(J\) has relative error \(O(J^{-1})\).
The fixed-\(J\) waiting limit followed by \(J\to\infty\) therefore gives
\[
 \lim_{J\to\infty}\frac{\mathcal S(j)}{\mathcal S(J+1)}
       \{1+O(J^{-1})\}=\mathcal S(j),\qquad
 \mathcal S(1)=\frac e{\sqrt{2\pi}},
\]
proving \eqref{eq:core-Gamma-continuation}.
Subtracting the transport equations at \(b-\Delta\) and \(b\) now gives
\begin{align}\label{eq:core-real-amplitude-comparison}
 \left|\log\frac{A_{\vartheta,c}(1-1/l)}{A_{b,c}(1-1/l)}\right|
 &\le C\Delta\log l+C\log\{1+C\delta_\Delta l/\log l\},\\
 l\left|[g_b(1)-g_b(1-1/l)]
       -[g_\vartheta(1)-g_\vartheta(1-1/l)]\right|
 &\le F(C\delta_\Delta l/\log l)+C\delta_\Delta,\nonumber
\end{align}
where
\[
 F(y)=(1+y)\log(1+y)-y\log y,\qquad
 F(y)\le C+C\log(1+y).
\]
For a fixed small \(e_0\), integration of the amplitude derivative gives
\[
 \int_{1/l}^{e_0}
 \left\{\frac{C\Delta}{e}
       +\frac{C\delta_\Delta}{e(eq+\delta_\Delta)}\right\}\,de
 \le C\Delta\log l+
       C\log\left(1+\frac{C\delta_\Delta l}{\log l}\right).
\]
The logit identity similarly bounds the action difference by
\[
 l\int_0^{1/l}
 \left\{\log\left(1+\frac{C\delta_\Delta}{e\log(1/e)}\right)
                  +C\delta_\Delta\right\}\,de
 \le F\left(\frac{C\delta_\Delta l}{\log l}\right)+C\delta_\Delta.
\]
Both bounds tend to zero when \(\delta_\Delta l=O(1)\).
If \(R\) counts initial successes, its exact \(R=r\) contribution is
\begin{equation*}
 \begin{split}
 A_{r,c}(\vartheta)
 &=t^r(1-t)\exp\left\{\vartheta h_1
  [r-c\{\psi(r+c+1)-\psi(c+1)\}]\right\}\\
 &\quad\times
 \exp\left\{\vartheta\frac{r+1}{r+1+c}h_t(r/(r+1))\right\}
 H_{r+1,\vartheta}^c(r),\qquad
 K_c(t,\vartheta)=\sum_{r\ge0}A_{r,c}(\vartheta).
 \end{split}
\end{equation*}
The all-success residual vanishes for $\vartheta<b$; positivity justifies
this unfolding. Since
$g_\vartheta(1)=(1-\vartheta)h_1+\log(1-\delta_\Delta)$,
\eqref{eq:core-critical-endpoint-expansion},
\eqref{eq:core-Gamma-continuation} and
\eqref{eq:core-real-amplitude-comparison} yield
\begin{align}\label{eq:core-first-failure-asymptotic}
 A_{r,c}(b-\Delta)
 &=D_c r^{\alpha_c-1}(\log r)^{Q_c}e^{-\delta_\Delta r}
           \{1+o(1)\},\\
 A_{r,c}(b-\Delta)
 &\le C r^{\alpha_c-1+C\Delta}(1+\log r)^{Q_c}
              (1+\delta_\Delta r)^C e^{-\delta_\Delta r}.
              \nonumber
\end{align}
The first asymptotic is uniform for bounded \(\delta_\Delta r\),
including \(\delta_\Delta r\to0\); the envelope holds for all large \(r\).
The first-failure constant is
\[
 \frac{A_{*,c}b e^{-H_t}}{\sqrt{2\pi}}\,
       (1-t)e^{b h_1+g_b(1)}
 =\frac{A_{*,c}b e^{-H_t}}{\sqrt{2\pi}}\,e^{H_t},
\]
which gives \eqref{eq:core-first-failure-constant} and retains
\(\mathcal S(1)=e/\sqrt{2\pi}\).
For \(\alpha_c>0\), put
\[
 L_\delta=\log(1/\delta_\Delta),\qquad y=\delta_\Delta r.
\]
The positive-sum envelope in
\eqref{eq:core-first-failure-asymptotic} gives
\[
 \frac{K_c(t,b-\Delta)}
      {D_c\delta_\Delta^{-\alpha_c}L_\delta^{Q_c}}
 \longrightarrow\int_0^\infty y^{\alpha_c-1}e^{-y}\,dy
 =\Gamma(\alpha_c).
\]
For \(\alpha_c=0\), \(Q_c=1/2\), and
\[
 \sum_{2\le r\le1/\delta_\Delta}\frac{(\log r)^{1/2}}r
 =\int_1^{1/\delta_\Delta}\frac{(\log x)^{1/2}}x\,dx+O(1)
 =\frac23L_\delta^{3/2}+O(1).
\]
The remaining sum is \(O(L_\delta^{1/2})\). For \(\alpha_c<0\), choose
\(\epsilon>0\) with \(\alpha_c+\epsilon<0\); the envelope is bounded by
\(C r^{\alpha_c-1+\epsilon}(1+\log r)^{Q_c}\), a summable sequence.
Dominated convergence gives the positive finite critical sum. By
\eqref{eq:core-real-gap},
\[
 \delta_\Delta^{-\alpha_c}L_\delta^{Q_c}
 \sim C_U^{-\alpha_c}\Delta^{-\alpha_c}
             \{\log(1/\Delta)\}^{Q_c-\alpha_c/b},
 \qquad Q_c-\alpha_c/b=\frac12 .
\]
This proves \eqref{eq:core-real-K-singularity}. Uniform positive margins
and the same summable envelopes give the stated compact-family
uniformity and continuity.
\end{proof}

\subsubsection{Uniform finite positive prefixes}

Let $b_0=b(\eta)<1/4$. In a sufficiently small interval
$I=[\eta,\eta+\epsilon]$, the same endpoint is active, $b_t=b(t)<1/4$ has a
uniform margin, and $b_t-b_0\asymp t-\eta$. Fix $0<c_-<c_+<\infty$ and use
the common real tilt $\vartheta_N=b_0-c_0/\sqrt N$, $c_-\le c_0\le c_+$.
Write $\Delta_t=b_t-\vartheta_N\asymp t-\eta+N^{-1/2}$.

All assertions below extend by reflection and by compact strict
subcriticality to the full trim interval.

For independent Bernoulli$(t)$ labels, put $Z_m=(S_m-mt)/\sqrt{mt(1-t)}$.
Choose a fixed $B>0$ and retain the parameter-independent event $\mathcal
A_{m,N}=\{|Z_m|\le B\sqrt N\}$. Define finite analytic functions
\begin{equation}\label{eq:core-finite-coefficients}
 \begin{split}
 B_{m,N,c}(t,z)&=m^{-\kappa(z)}
       \E_t[e^{zT_m^c}\psi_z(Z_m)\mathbf1_{\mathcal A_{m,N}}],
       \qquad \psi_z(x)=e^{a(z)x^2/2},\\
 \Psi_{N,t}(z)&=\sqrt{r(z)}B_{m,N,-1/2}(t,z)B_{m,N,+1/2}(t,z).
 \end{split}
\end{equation}

\begin{lemma}[Finite real matching]\label{lem:core-finite-real-matching}
Choose $a_1$ with $\sup_{t\in I}a(b_t)<a_1<1/2$, and fix
$K>\max\{3,2(1-a_1)/(1-2a_1)\}$, $m=\lfloor N^K\rfloor$. Uniformly in $t\in
I$ and $c_0\in[c_-,c_+]$,
\begin{equation}\label{eq:core-finite-real-match}
 \frac{B_{m,N,c}(t,\vartheta_N)}{K_c(t,\vartheta_N)}=1+o(1).
\end{equation}
Under the probability law obtained by normalizing $e^{\vartheta_N
T_m^c}H_{m,\vartheta_N}^c(S_m)$, the probabilities of $T_m^c>N^{3/4}$ and
$|Z_m|>\log N$ are smaller than every inverse power of $N$.
\end{lemma}

\begin{proof}
We first record a uniform real doubling bound. Lemma
\ref{lem:core-real-singularity} gives
\begin{equation}\label{eq:core-real-doubling}
 K_c(t,b_t-\Delta_t/2)/K_c(t,b_t-\Delta_t)\le C.
\end{equation}
For the active endpoint, write
\(\delta(t,\Delta)=1-u_{t,b(t)-\Delta}(1)\).
When an exponent vanishes at $\eta$, put
$\alpha_c(t)=\zeta$, $L=\log(1/\delta(t,\Delta_t))$. Since
$|\zeta|=O(t-\eta)$ and $\Delta_t\ge c(t-\eta)$, $|\zeta|L\to0$ whenever
$t\to\eta$. The positive first-failure sum, with $y=\log r/L$, has the
uniform form
\begin{equation}\label{eq:core-confluent-real-coefficient}
 K_c(t,b_t-\Delta_t)\sim
 D_c(t)L^{\zeta/b_t+3/2}
       \int_0^1 e^{\zeta Ly}y^{\zeta/b_t+1/2}\,dy
       \sim\frac{2D_c(\eta)}3\{\log(1/\Delta_t)\}^{3/2}.
\end{equation}
For bounded \(|\zeta|L\), put \(\delta=e^{-L}\),
\(v_r=r^{\zeta-1}(\log r)^{\zeta/b_t+1/2}\), and
\[
 J_{L,\zeta}=\int_1^{1/\delta}
              x^{\zeta-1}(\log x)^{\zeta/b_t+1/2}\,dx.
\]
The damping correction and the tail of the positive run envelope satisfy
\[
 \frac{\sum_{2\le r\le1/\delta}v_r(1-e^{-\delta r})
                  +\sum_{r>1/\delta}v_r e^{-\delta r}}
      {J_{L,\zeta}}
 \le \frac{CL^{\zeta/b_t+1/2}}{cL^{\zeta/b_t+3/2}}\le C/L.
\]
The uniform first-failure asymptotic transfers this truncation to the
actual positive sum. Its own relative approximation error remains the
one in \eqref{eq:core-first-failure-asymptotic}.
For exponents bounded away from zero, the positive Gamma integral or the
negative-exponent summable envelope applies uniformly. Together these give
\[
 \sup_{t,\Delta}\frac{K_c(t,b_t-\Delta/2)}{K_c(t,b_t-\Delta)}\le C,
 \qquad \inf_{t,\Delta}K_c(t,b_t-\Delta)>0,
\]
with the remaining compact gap range covered by positivity and continuity.
To control large prefix energy, use on \(1\le s\le m-1\)
\[
 c\le\frac{H_{m,\vartheta}^c(s)}{W_m^c(s;\vartheta)}\le C,\qquad
 m^{-C}\le A_{t,\vartheta,c}(s/m)\le m^C,\qquad
 g_{\vartheta_N}(s/m)\le g_{\vartheta_N'}(s/m),
 \quad\vartheta_N'=b_t-\Delta_t/2 .
\]
The amplitude bounds follow by integrating
\(|\partial_x\log A|\le C/[x(1-x)]\), and action monotonicity is
Lemma~\ref{lem:hardy-bellman-branch}. Consequently
\[
 \frac{H_{m,\vartheta_N}^c(s)}{H_{m,\vartheta_N'}^c(s)}
 \le Cm^C.
\]
The exact identity \eqref{eq:core-harmonic-identity} therefore gives
\begin{equation}\label{eq:core-energy-cap}
 \Pp^H_{\vartheta_N}\{T_m^c>N^{3/4},\ 1\le S_m\le m-1\}
 \le Cm^C e^{-\Delta_tN^{3/4}/2}
       \frac{K_c(t,\vartheta'_N)}{K_c(t,\vartheta_N)}
 \le N^C e^{-cN^{1/4}}.
\end{equation}
The last inequality uses \eqref{eq:core-real-doubling}.
For the omitted zero-minority state, the first-run expansion gives
\[
 P^H_{\vartheta_N}(S_m=m)
 =K_c(t,\vartheta_N)^{-1}\sum_{r\ge m}A_{r,c}(\vartheta_N)
 \le C m^C\delta_t^{-C}e^{-c\delta_t m},
 \qquad \delta_t\ge c\Delta_t\ge cN^{-1/2},
\]
by \eqref{eq:core-first-failure-asymptotic}.
The all-failure state has a uniform exponential tail at the opposite end.
For \(A_l=S_l-lt\), bounded increments and the quadratic KL bound imply
\begin{equation}\label{eq:core-deterministic-energy-tube}
 T_l^c\ge c l|A_l/l|^3-C.
\end{equation}
Indeed \(|A_{k+1}-A_k|\le1\). For
\(l-\lfloor|A_l|/4\rfloor\le k\le l\), this gives
\(|A_k|\ge |A_l|/2\) and \(k\le l\), whence
\[
 T_l^c\ge c\sum_{k=l-\lfloor|A_l|/4\rfloor}^l\frac{A_k^2}{k^2}-C
 \ge c\frac{|A_l|^3}{l^2}-C .
\]
On \(T_m^c\le N^{3/4}\), nonnegativity of the summands therefore gives
\[
 \frac{|A_l|}l\le C\left(\frac{N^{3/4}+1}{l}\right)^{1/3}
 \quad(l\le m),\qquad
 |A_l|\le\rho l/4\quad(L\le l\le m),
 \quad L=\lceil C_\rho N^{3/4}\rceil .
\]
On the larger central tube $|A_l|\le\rho l$, the true $H$-chain transition,
using $H_l/W_l=1+O(l^{-1})$, has conditional drift
$\mu_l=u_{t,\vartheta_N}(t+A_l/l)-t+O(l^{-1})$ and $A_l\mu_l\le
a_1A_l^2/l+C/l$. Its centered increment has range length one. Hoeffding's
conditional bound yields
\[
 \E_H(e^{hA_{l+1}^2}\mid\mathcal F_l)
 \le\exp\{Ch+[h(1+2a_1/l)+h^2/2]A_l^2\}.
\]
Kill at exit from the tube. Starting with $h_m=\epsilon/m$, solve
$h_l=(1+2a_1/l)h_{l+1}+h_{l+1}^2/2$ backwards. For fixed small $\epsilon$,
division by $\prod_{r=l}^{m-1}(1+2a_1/r)$ and a quadratic bootstrap give
$h_l\le C\epsilon m^{2a_1-1}l^{-2a_1}$ and $\sum_{l=L}^m h_l\le C\epsilon$.
Iteration proves
\begin{equation}\label{eq:core-killed-Riccati}
 \E_H\left[e^{\epsilon A_m^2/m};\ |A_l|\le\rho l\
                  (L\le l\le m)\mid\mathcal F_L\right]
 \le C\exp\{CA_L^2m^{2a_1-1}L^{-2a_1}\}.
\end{equation}
For every \(|A_L|\le L\),
\[
 A_L^2m^{2a_1-1}L^{-2a_1}
 \le m^{2a_1-1}L^{2-2a_1}
 \le C N^{-K(1-2a_1)+(3/2)(1-a_1)}=o(1)
\]
under the stated, stronger condition on \(K\). Markov's inequality in
\eqref{eq:core-killed-Riccati} therefore gives
\[
 \Pp^H_{\vartheta_N}
 \{T_m^c\le N^{3/4},\,|Z_m|>\log N\}
 \le C e^{-c(\log N)^2}.
\]
Adding \eqref{eq:core-energy-cap} proves the terminal-tube assertion.
Finally, throughout the whole retained event $\mathcal A_{m,N}$, central
Taylor expansion and the central continuation estimate give
\begin{equation}\label{eq:core-terminal-real-ratio}
 \frac{\psi_{\vartheta_N}(Z_m)}
 {m^{\kappa(\vartheta_N)}H_{m,\vartheta_N}^c(S_m)}
 =1+O\{N^{3/2}/\sqrt m+\sqrt{N/m}+m^{-1}\}=1+o(1).
\end{equation}
The harmonic identity now gives the exact comparison
\[
 \frac{B_{m,N,c}(t,\vartheta_N)}{K_c(t,\vartheta_N)}
 =\E^H_{\vartheta_N}\left[
 \frac{\psi_{\vartheta_N}(Z_m)}
      {m^{\kappa(\vartheta_N)}H_{m,\vartheta_N}^c(S_m)}
          \mathbf1_{\mathcal A_{m,N}}\right].
\]
The displayed ratio is uniformly \(1+O(N^{3/2}/\sqrt m+\sqrt{N/m}+m^{-1})\)
on the entire retained event. Its complement has the probability just
bounded, so \eqref{eq:core-finite-real-match} follows. A uniform error
bound is
\[
 \left|\frac{B_{m,N,c}(t,\vartheta_N)}{K_c(t,\vartheta_N)}-1\right|
 \le C N^{3/2}/\sqrt m+N^C e^{-cN^{1/4}}
       +Ce^{-c(\log N)^2}+N^C e^{-cm/\sqrt N}.
\]
For each ODE constant \(C(t)\), let \(C_Q(t)\) be its finite-\(Q\)
approximation. The endpoint remainder and finite-interval ODE continuity give
\[
 \sup_{t\in I}|C(t)-C_Q(t)|\le\omega(Q)\to0,\qquad
 |C(t)-C(t')|\le2\omega(Q)+|C_Q(t)-C_Q(t')|.
\]
First fix \(Q\), let \(t'\to t\), and then let \(Q\to\infty\).
The same compact-family hazard and next-failure envelopes dominate all
continuation estimates, proving the asserted uniformity.
\end{proof}

At $t=\eta$, combine the two lines of \eqref{eq:core-real-K-singularity}
with $\Psi_t(\vartheta)=\sqrt{r(\vartheta)}K_{-1/2}(t,\vartheta)
K_{+1/2}(t,\vartheta)$. Denote the resulting aggregate parameters by
\begin{equation*}
 \Psi_\eta(b_0-\Delta)\sim
       P_\eta\Delta^{-\alpha_\eta}\{\log(1/\Delta)\}^{\beta_\eta},
 \qquad P_\eta>0,\quad\alpha_\eta>0.
\end{equation*}
Explicitly, each positive $\alpha_c$ contributes $\alpha_c$ to $\alpha_\eta$,
$1/2$ to $\beta_\eta$, and $D_c\Gamma(\alpha_c)C_U^{-\alpha_c}$ to $P_\eta$; each
zero exponent contributes $3/2$ and $2D_c/3$ to the latter two quantities;
each negative exponent contributes its positive critical $K_c$. Multiply the
product constant by $\sqrt{r(b_0)}$. The lower-rank exponent $\alpha_{-1/2}$
is positive by \eqref{eq:core-boundary-exponents}.

The finite matching and confluence proof imply, along every $t\to\eta$,
uniformly for the stated $c_0$,
\begin{equation}\label{eq:core-spatial-real-match}
 \Psi_{N,t}(\vartheta_N)
   =\{P_\eta+o(1)\}\Delta_t^{-\alpha_\eta}
                      \{\log(1/\Delta_t)\}^{\beta_\eta},
\end{equation}
and on a fixed endpoint neighborhood
\begin{equation}\label{eq:core-spatial-envelope}
 \Psi_{N,t}(\vartheta_N)
 \le C(t-\eta+N^{-1/2})^{-\alpha_\eta}
       [1+\log\{1/(t-\eta+N^{-1/2})\}]^{\beta_\eta}.
\end{equation}
Let \(\alpha_t=\sum_{c\in\{-1/2,+1/2\}}(\alpha_c(t))_+\) and
\(\tau=t-\eta\). The power-exponent change is \(O(\tau)\), and
\(\Delta_t\ge c\tau\), so
\[
 \left|\log\Delta_t^{-[\alpha_t-\alpha_\eta]}\right|
 \le C\tau\log(1/\Delta_t)\longrightarrow0,\qquad
 \{\log(1/\Delta_t)\}^{\,O(\tau)}
 =\exp\{O(\tau\log\log(1/\Delta_t))\}\longrightarrow1.
\]
Zero individual exponents are treated by
\eqref{eq:core-confluent-real-coefficient}. Keeping the same estimates
with fixed upper constants proves \eqref{eq:core-spatial-envelope}.

Reflection gives the other endpoint, and the interior has a bounded
strict-subcritical majorant.

\subsubsection{Positive Fourier inversion and spatial integration}

Put
\begin{equation}\label{eq:core-bulk-constants}
 \begin{gathered}
 b=b_0,\qquad q_c=\lambda'(b)=(1-4b)^{-1/2},\qquad
 V=\lambda''(b)=2q_c^3,\\
 \gamma_c=bq_c-\lambda(b),\qquad
 d(t)=m(t)+\frac{q_c-1}{4}v_0(t).
 \end{gathered}
\end{equation}
Use the moment constants of Propositions~\ref{prop:rank-mean-constant}
and \ref{prop:rank-variance-constant}. Join the two retained independent
boundary paths by the OU bridge in \eqref{eq:exact-ou-bridge-transform},
with
\[
 R=2\log\{(n-m)/m\}=2N-2\log m+o(1).
\]
Write \(\widetilde T_n\) for this hybrid and
\[
 \mathcal A_n=\mathcal A_{m,N}^-\cap\mathcal A_{m,N}^+,
 \quad\mathcal M_{N,t}(z)=\mathbb E_{\rm hyb}
                    [e^{z\widetilde T_n}\mathbf1_{\mathcal A_n}],\quad
 \frac{d\mathbb Q_{n,t,\vartheta}^{\mathcal A}}{d\mathbb P_{\rm hyb}}
       =\frac{e^{\vartheta\widetilde T_n}\mathbf1_{\mathcal A_n}}
                         {\mathcal M_{N,t}(\vartheta)}.
\]
In this subsection \(\Pp_{\vartheta,t}\), \(\Pp_\vartheta\), and
\(\E_\vartheta\) mean this positive hybrid law and its expectation;
conditional middle laws are given \(\mathcal F_{n,m}^{\partial}\).
For the conditional kernel \(F_R(z;x,y)\), define
\[
 \varepsilon_R(z;x,y)=
 \frac{F_R(z;x,y)}{e^{\kappa(z)R}\sqrt{r(z)}\psi_z(x)\psi_z(y)}-1,
 \qquad
 \mathcal E_{N,t}(z)=\mathcal M_{N,t}(z)-e^{N\lambda(z)}\Psi_{N,t}(z).
\]
The exact kernel gives, on the retained endpoints and a small complex
disk about \(b\),
\[
 \sup_{|x|,|y|\le C\sqrt N}|\varepsilon_R(z;x,y)|\le CN e^{-cR}.
\]
At \(z=\vartheta+is\), \(|s|\le s_0\), the difference itself satisfies
\begin{equation*}
 \frac{|\mathcal E_{N,t}(\vartheta+is)|}{\mathcal M_{N,t}(\vartheta)}
 \le e^{-cN}N^C
       \exp\{N[\operatorname{Re}\lambda(\vartheta+is)
                                      -\lambda(\vartheta)]\}.
\end{equation*}
This follows by bounding the endpoint integrands in absolute value and
normalizing by the positive real kernel.
The same real normalization gives, for all real $s$ and
$\vartheta=b-c_0/\sqrt N$,
\begin{equation}\label{eq:core-normalized-Fourier}
 \left|\frac{\mathcal M_{N,t}(\vartheta+is)}
                   {\mathcal M_{N,t}(\vartheta)}\right|
 \le C(1+|s|)^{1/4}
   \exp\{R[\operatorname{Re}\kappa(\vartheta+is)
                                      -\kappa(\vartheta)]\}.
\end{equation}
One direct verification is to condition on the endpoints under the real
tilt. If the resulting Gaussian bridge has mean $\mu$ and covariance
operator $C_\vartheta$, its quadratic characteristic function has modulus
\[
 \det(I+s^2C_\vartheta^2)^{-1/4}
 \exp\!\left[-\frac{s^2}{2}\left\langle\mu,
 C_\vartheta(I+s^2C_\vartheta^2)^{-1}\mu\right\rangle\right]
 \le\det(I+s^2C_\vartheta^2)^{-1/4}.
\]
The zero-endpoint Mehler determinant and positive boundary averaging give
\eqref{eq:core-normalized-Fourier}. Since \(r(b)>0\), its modulus is bounded by
\[
 \left|\frac{\mathcal M_{N,t}(\vartheta+is)}{\mathcal M_{N,t}(\vartheta)}\right|
 \le C(1+|s|)^{1/4}
 \begin{cases}
 e^{-cNs^2},&|s|\le s_0,\\
 e^{-cN},&s_0<|s|\le S,\\
 e^{-cN\sqrt{|s|}},&|s|>S .
 \end{cases}
\]
Let \(f_{N,\vartheta}\) be the density of \(\widetilde T_n\) under
\(\mathbb Q_{n,t,\vartheta}^{\mathcal A}\). Its characteristic function is
\(\phi_N(s)=\mathcal M_{N,t}(\vartheta+is)/\mathcal M_{N,t}(\vartheta)\), so
\[
 \int_{\mathbb R}|\phi_N(s)|\,ds\le C/\sqrt N,\qquad
 \sup f_{N,\vartheta}\le C/\sqrt N,
\]
and the rescaled Fourier integrals at \(s=w/\sqrt N\) are uniformly
integrable. Finite positivity gives
\begin{equation}\label{eq:core-positive-complex-bound}
 |B_{m,N,c}(t,\vartheta+is)|
 \le m^{\kappa(\vartheta)-\operatorname{Re}\kappa(\vartheta+is)}
       B_{m,N,c}(t,\vartheta),
\end{equation}
Indeed \(\operatorname{Re}a(\vartheta+is)\le a(\vartheta)\).
For \(z=b-w/\sqrt N\) on a compact subset of
\(\{\operatorname{Re}w>0\}\),
\[
 m^{\kappa(\operatorname{Re}z)-\operatorname{Re}\kappa(z)}
       =\exp\{O(\log m/N)\}=1+o(1).
\]
Finite real matching and doubling therefore bound the normalized
holomorphic coefficients \(F_N(w)\): for each such compact \(K\),
\[
 \sup_N\sup_{w\in K}|F_N(w)|\le C_K,\qquad
 F_N(w)\longrightarrow F(w)\quad(w>0).
\]
Every subsequence therefore has a locally convergent further subsequence,
and the identity theorem makes all its limits equal to the analytic
continuation of \(F\). Hence \(F_N\to F\) locally uniformly.
For inversion, \eqref{eq:core-normalized-Fourier} also gives
\[
 \lim_{T\to\infty}\limsup_{N\to\infty}
 \int_{|w|>T}
 \left|\frac{\mathcal M_{N,t}(\vartheta_N+iw/\sqrt N)}
                  {\mathcal M_{N,t}(\vartheta_N)}\right|\,dw=0.
\]
Compact convergence and this tail bound justify the Fourier limit.

For a conditioning lower bound uniform in the split, consider the
finite positive edge law
\[
 \frac{d\mathbb Q_{m,N,c,\vartheta_N}^{\rm edge}}{dP_t}
 =\frac{e^{\vartheta_NT_m^c}\psi_{\vartheta_N}(Z_m)
                         \mathbf1_{\mathcal A_{m,N}}}
        {m^{\kappa(\vartheta_N)}B_{m,N,c}(t,\vartheta_N)},
\]
take \(h_N=c_-/(2\sqrt N)\). Since
$\psi_{\vartheta_N}\le\psi_{\vartheta_N+h_N}$, finite real matching and
bounded real ratios give
\begin{equation}\label{eq:core-edge-energy-moment}
 \E_{\vartheta_N,\mathrm{edge}}e^{h_NT_m^c}
 \le m^{\kappa(\vartheta_N+h_N)-\kappa(\vartheta_N)}
       \frac{B_{m,N,c}(t,\vartheta_N+h_N)}
            {B_{m,N,c}(t,\vartheta_N)}\le C.
\end{equation}
The positive hybrid boundary law is the product of the two edge laws
up to \(1+O(e^{-cN}N^C)\). Thus, for
\(B_n=T_m^{-1/2}+T_m^{+1/2}\),
\[
 \E_{\vartheta_N}e^{h_NB_n}\le C,\qquad
 \Pp_{\vartheta_N}(B_n>L\sqrt N)\le Ce^{-c_-L/2}.
\]
Choose fixed \(L\) so that the last bound is less than \(1/4\).
Lemma~\ref{lem:core-finite-real-matching} then gives
\[
 \Pp_{\vartheta_N}\{B_n\le L\sqrt N,\ |Z_-|,|Z_+|\le\log N\}
 \ge\frac12
\]
for sufficiently large \(N\). On this event the conditional middle mean
and variance are
\[
 \mu_{\rm mid}=N\lambda'(\vartheta_N)+O(\log^2N),\qquad
 \sigma_{\rm mid}^2=NV+o(N).
\]
For these boundary paths and compact \(x\),
\[
 \left|\frac{Nq_c+x\sqrt N+d(t)-B_n-\mu_{\rm mid}}{\sqrt N}\right|
       \le C_x .
\]
The conditional normal local limit then gives
\[
 \mathbb Q^{\mathcal A}_{n,t,\vartheta_N}
 \{\widetilde T_n-d(t)\in[Nq_c+x\sqrt N,Nq_c+x\sqrt N+h]
                          \mid\mathcal F_{n,m}^{\partial}\}
       \ge c_h/\sqrt N.
\]
The retained paths have probability at least \(1/2\); averaging proves
\begin{equation}\label{eq:core-anchor-width-lower}
 \Pp_{\vartheta_N,t}\{\widetilde T_n-d(t)
        \in[Nq_c+x\sqrt N,Nq_c+x\sqrt N+h]\}
        \ge c_h/\sqrt N.
\end{equation}
The conditioning bound is uniform in the split and gives an anchor
conditioning cost of at most $C\sqrt N$.

Let \(\rho_{N,t}(a)\,da=P_{\rm hyb}
 \{\widetilde T_n-d(t)\in da,\mathcal A_n\}\) and define the finite positive
mixture and its density by
\[
 \mu_N(da)=\int_\eta^{1-\eta}
    P_{\rm hyb}\{\widetilde T_n-d(t)\in da,\mathcal A_n\}
                      \frac{dt}{t(1-t)},\qquad
 \rho_N(a)=\int_\eta^{1-\eta}\rho_{N,t}(a)\frac{dt}{t(1-t)}.
\]
The associated leading transform coefficient is
\begin{equation*}
 \Theta_N(z)=\int_\eta^{1-\eta}
      e^{-z d(t)}\Psi_{N,t}(z)\frac{dt}{t(1-t)}.
\end{equation*}
It is the coefficient of a positive mixture of shifted hybrid laws.
Retain $\mathfrak J_a$ and $(\mathcal A_{\rm c},\nu_*,\ell_*)$ from
\eqref{eq:calibration-fractional-gaussian} and
\eqref{eq:calibration-core-amplitude}, and $\Theta_0,P_0,P_A$ from
Appendix~\ref{app:max-deterministic-inputs}. Put $S_\eta=e^{-bd(\eta)}$.

Equations \eqref{eq:core-spatial-real-match}--
\eqref{eq:core-spatial-envelope} give, at $z=b-c_0/\sqrt N$,
\begin{equation}\label{eq:core-integrated-real-asymptotic}
 \Theta_N(z)\sim
 \begin{cases}
 \Theta_0,&\alpha_\eta<1,\\
 P_0(\log N/2)^{\beta_\eta+1},&\alpha_\eta=1,\\
 P_A N^{(\alpha_\eta-1)/2}(\log N/2)^{\beta_\eta}
                    c_0^{-(\alpha_\eta-1)},&\alpha_\eta>1.
 \end{cases}
\end{equation}
For \(\alpha_\eta<1\), the envelope
\(\tau^{-\alpha_\eta}(1+\log(1/\tau))^{\beta_\eta}\) is integrable at zero, so dominated
convergence gives \(\Theta_0\).
For the remaining cases put \(d_N=c_0/\sqrt N\), \(\tau=t-\eta\), and use
\(b(t)-b=b_1\tau\{1+o(1)\}\). The elementary endpoint integrals are
\[
 \begin{split}
 \int_0^\epsilon\frac{\{\log(1/(d_N+b_1\tau))\}^{\beta_\eta}}
                         {d_N+b_1\tau}\,d\tau
 &=\frac{\{\log(1/d_N)\}^{\beta_\eta+1}
       -\{\log(1/(d_N+b_1\epsilon))\}^{\beta_\eta+1}}
        {b_1(\beta_\eta+1)},\\
 \int_0^\epsilon(d_N+b_1\tau)^{-\alpha_\eta}
       \{\log(1/(d_N+b_1\tau))\}^{\beta_\eta}\,d\tau
 &\sim\frac{d_N^{1-\alpha_\eta}\{\log(1/d_N)\}^{\beta_\eta}}
                 {b_1(\alpha_\eta-1)},\qquad \alpha_\eta>1.
 \end{split}
\]
For the second line, \(\tau=d_Ny/b_1\) gives
\[
 \int_0^\infty(1+y)^{-\alpha_\eta}\,dy=\frac1{\alpha_\eta-1}.
\]
Multiply each endpoint contribution by \(P_\eta S_\eta/v_\eta\).
Reflection doubles it, giving \(P_0,P_A\); the envelope controls the
complement.
Apply \eqref{eq:core-positive-complex-bound} and normal-family convergence
to the positive mixture. Fix a compact set \(K_x\subset\mathbb R\).
For its density \(\rho_N\), inversion uses
\[
 u=Nq_c+x\sqrt N,\qquad z=b-w/\sqrt N,
 \qquad x\in K_x.
\]
The exponent is
\[
 -zu+N\lambda(z)
 =-N\gamma_c-bx\sqrt N+wx+Vw^2/2+o(1).
\]
For \(a>0\) and \(\operatorname{Re}w=c_0>0\),
\[
 \begin{split}
 \frac1{2\pi i}\int_{c_0-i\infty}^{c_0+i\infty}
              w^{-a}e^{wx+Vw^2/2}\,dw
 &=\frac1{\Gamma(a)}
   \int_0^\infty y^{a-1}
   \left\{\frac1{2\pi i}\int_{c_0-i\infty}^{c_0+i\infty}
                  e^{w(x-y)+Vw^2/2}\,dw\right\}dy\\
 &=\frac1{\Gamma(a)\sqrt{2\pi V}}
           \int_0^\infty y^{a-1}e^{-(x-y)^2/(2V)}\,dy
 =\mathfrak J_a(x).
 \end{split}
\]
The Gamma representation has the integrable factor \(e^{-c_0y}\), and
the vertical Gaussian has \(e^{-V(\operatorname{Im}w)^2/2}\), which
justify the interchange. For \(a=0\) the inner Gaussian inversion alone
gives \(\mathfrak J_0\). Equations \eqref{eq:core-integrated-real-asymptotic}
and \eqref{eq:calibration-core-amplitude} consequently yield
\begin{equation}\label{eq:core-integrated-density}
 N(q_c-1)\rho_N(Nq_c+x\sqrt N)
 \sim \mathcal A_{\rm c}(x)N^{\nu_*}(\log N)^{\ell_*}
             e^{-N\gamma_c-bx\sqrt N}.
\end{equation}
For each compact $K_x\subset\mathbb R$, differentiation of
\eqref{eq:calibration-fractional-gaussian} under the Gaussian integral
and positivity of its integrand give
\[
 0<\inf_{x\in K_x}\mathcal A_{\rm c}(x)
 \le\sup_{x\in K_x}\mathcal A_{\rm c}(x)<\infty,\qquad
 \sup_{x\in K_x}|\mathcal A_{\rm c}'(x)|<\infty.
\]
For fixed-split tails the Fourier multiplier is $1/z\to1/b$; equivalently,
integrate the tilted density against $e^{-\vartheta_N y}$, $y>0$. The
density bound and compact positivity justify the same $1/b$ factor
uniformly, with the local limit approximation supplied by the Gaussian middle.

\subsubsection{Transfer to the original rank scan}

For \(\sigma\in\{-1,+1\}\), put
\[
 t_\sigma(z)=\operatorname{logistic}\{\operatorname{logit}t+\sigma z/N\},
 \qquad T_{t,\sigma}(z)=T_{n,\operatorname{round}(nt_\sigma(z))}.
\]
For the corresponding uniform observation-time label process, a function
\(F(t,B)\) has generator
\begin{equation}\label{eq:rank-local-uniform-time-generator}
 \begin{split}
 \mathcal L_{N,t}^{+}F
 &=\frac{v}{N}\partial_tF
        +\frac tN\sum_{i:B_i=0}\{F(t,B+e_i)-F(t,B)\},\\
 \mathcal L_{N,t}^{-}F
 &=-\frac{v}{N}\partial_tF
        +\frac{1-t}{N}\sum_{i:B_i=1}\{F(t,B-e_i)-F(t,B)\}.
 \end{split}
\end{equation}
These rates follow from the hazards \(dt/(1-t)\) for a zero becoming one
and \(-dt/t\) for a one being deleted. For comparison, the original
fixed-count chain has the exact discrete generator
\[
 \mathcal L_{n,N,t}^{+,d}F
   =\frac tN\sum_{i:B_i=0}\{F(t+1/n,B+e_i)-F(t,B)\}.
\]
Define the exact operator difference by
\[
 \begin{split}
 \mathfrak R_{n,N,t}^+F
 &:=(\mathcal L_{n,N,t}^{+,d}-\mathcal L_{N,t}^+)F\\
 &=\frac tN\sum_{i:B_i=0}\int_0^{1/n}
   \{\partial_tF(t+r,B+e_i)-\partial_tF(t,B)\}\,dr.
 \end{split}
\]
The reverse difference \(\mathfrak R^-\) uses \(t-1/n,B-e_i\) and
\((1-t)/N\). To bound it for entropy exponentials, put
\[
 h_t^\pm(x)=h_t(x)\mathbf1_{\{\pm(x-t)\ge0\}},\qquad
 E_f(t,B)=\sum_{l\le m}\frac l{l+c}
                  f_t\!\left(l^{-1}\sum_{i\le l}B_i\right),
 \quad f_t\in\{h_t,h_t^+,h_t^-\}.
\]
For trimmed \(t\), direct differentiation and one-label replacement give
\[
 \begin{gathered}
 |\partial_tE_f|+|\partial_{tt}E_f|\le Cm,\qquad
 |E_f(t,B\pm e_i)-E_f(t,B)|\le C\log^2(em),\\
 |\partial_tE_f(t,B\pm e_i)-\partial_tE_f(t,B)|\le C\log(em).
 \end{gathered}
\]
Derivatives at \(x=t\) are one-sided and the first derivative is continuous.
Only the first \(m\) labels affect \(E_f\). For \(F=e^{sE_f}\),
\[
 \sum_{i:B_i=0}\sup_{0\le r\le1/n}
 \frac{|\partial_tF(t+r,B+e_i)-\partial_tF(t,B)|}{F(t,B)}
 \le C_s m^2e^{C_s\log^2(em)}.
\]
The \(i>m\) terms use \(r|\partial_{tt}F|\); the other \(m\) terms
use the preceding one-label bounds. Thus
\[
 \frac{|\mathfrak R_{n,N,t}^{\pm}F|}{F}
 \le\frac{C_s}{n}\{m^2+\log(em)\}e^{C_s\log^2(em)}
 \le\frac1n e^{C_{s,K}(\log N)^2},\qquad m=N^K.
\]
For \(F_Q=e^{sQ_m/2}\), the middle count and energy caps give
\[
 \frac{|\mathfrak R_{n,N,t}^{\pm}F_Q|}{F_Q}\le N^{C_s}/n.
\]
For \(0\le H\le h_0N\), let \(\tau\)
be the first exit from those caps. Under every conditional anchor law,
\[
 E\left[\int_0^{H\wedge\tau}
       \frac{|\mathfrak R_{n,N,t_\pm(z)}^{\pm}F|}{F}\,dz
                    \ \middle|\ T_{n,k}\in[u,u+1]\right]
 \le\frac{h_0N}{n}e^{C_{s,K}(\log N)^2}\longrightarrow0.
\]
After orienting the endpoint block,
write \(\mathcal L_{\rm rev}=\mathcal L_{N,t^*}^{-}\), let \(k^*(z)\) be
the corresponding original integer split, and put
\[
 \begin{gathered}
 T_{\rm rev}(z)=T_{n,k^*(z)},\qquad T(0)=T_{n,k},\\
 p_{n,t}(u)=P\{T_{n,k}\in[u,u+1]\},\\
 \mathbb P_{n,t}^{u,1}(E)=P(E\mid T_{n,k}\in[u,u+1]).
 \end{gathered}
\]
Write \(\mathbb E_{n,t}^{u,1}\) for the corresponding expectation.
Retain the positive hybrid law \(\mathbb Q_{n,t,\vartheta_N}^{\mathcal A}\), with
\[
 m=\lfloor N^K\rfloor,\qquad
 \mathcal A_n=\mathcal A_{m,N}^-\cap\mathcal A_{m,N}^+,\qquad
 \epsilon_n^{\rm lik}=Cm^2/n,\quad \epsilon_n^{\rm bad}=C_An^{-A}.
\]
For any fixed \(A>0\), the likelihood and bridge bounds give
\begin{equation}\label{eq:core-rank-sandwich-error}
 \epsilon_n=C_A\{N^{3/2}/\sqrt m+N^2/m+m^2/n\}=o(1),
\end{equation}
and the actual probability sandwich, uniformly in real \(u\), is
\[
 \begin{split}
 &(1-\epsilon_n^{\rm lik})P_{\rm hyb}\{\widetilde T_n>u+\epsilon_n,\mathcal A_n\}-\epsilon_n^{\rm bad}\\
 &\quad\le P\{T_{n,k}>u\}\\
 &\quad\le(1+\epsilon_n^{\rm lik})P_{\rm hyb}\{\widetilde T_n>u-\epsilon_n,\mathcal A_n\}+\epsilon_n^{\rm bad}.
 \end{split}
\]
The sandwich constants come from boundary likelihoods, conditional bridge
coupling and the deterministic entropy remainder, uniformly in the gap.
Define \(P_{\rm ref}=P_{\rm hyb}\{\widetilde T_n>u,\mathcal A_n\}\).
At core levels,
\[
 u=Nq_c+O(\sqrt N),\qquad
 \log P_{\rm ref}=-N\gamma_c+O(\sqrt N+\log N).
\]
for both tails and fixed-width intervals. Thus \(A>\gamma_c+1\) gives
\[
 \frac{n^{-A}}{P_{\rm ref}}
 \le \exp\{-(A-\gamma_c)N+O(\sqrt N+\log N)\}\longrightarrow0.
\]
Compact-core inversion and tail subtraction give
\[
 P_{\rm ref}(u\pm\epsilon_n)/P_{\rm ref}(u)\to1.
\]
For a retained mark \(\mathcal B\), the same construction gives an upper
numerator bound; \eqref{eq:core-anchor-width-lower} costs at most
\(C\sqrt N\). Polynomially many marks are covered by increasing \(A\).
By \eqref{eq:core-edge-energy-moment} and the terminal-tube estimate,
\[
 \Pp_{\vartheta_N}(B_n>N^{3/4})\le CN^Ce^{-cN^{1/4}},\qquad
 \Pp_{\vartheta_N}(\max|Z_\pm|>\log N)\le Ce^{-c(\log N)^2}.
\]
After the \(C\sqrt N\) conditioning cost, the true rank anchors satisfy
\begin{equation*}
 B_n/N\longrightarrow0,\qquad
 (Z_-^2+Z_+^2)/N\longrightarrow0,
\end{equation*}
with the quantitative bound
\[
 \begin{split}
 &\mathbb P_{n,t}^{u,1}
 \{B_n>N^{3/4}\ \hbox{or}\ |Z_-|\vee|Z_+|>\log N\}\\
 &\quad\le C\sqrt N\{N^Ce^{-cN^{1/4}}+Ce^{-c(\log N)^2}\}\\
 &\qquad+\exp\{-(A-\gamma_c)N+O(\sqrt N+\log N)\}.
 \end{split}
\]
This error is smaller than every inverse power of \(N\).
For \(M\) in \eqref{eq:rank-scan-s3}, retain
\[
 Q_m=X^\top MX/v,\qquad W_m=X^\top M^2X/v.
\]
Conditional on \(\mathcal F_{n,m}^{\partial}\) under
\(\mathbb Q_{n,t,\vartheta_N}^{\mathcal A}\), let \(\mu_N,\Sigma_N\)
be the mean and covariance of the Gaussian increment vector \(X^G\).
On \(|Z_-|+|Z_+|\le2\log N\),
\[
 \|\Sigma_N\|\le C,\qquad
 v^{-1}\mu_N^\top M^i\mu_N\le C\log^2N,\qquad i=1,2.
\]
Positive and negative quadratic Chernoff perturbations give the
exponential concentration
\begin{equation}\label{eq:core-middle-critical-values}
 Q_m/(2N)=q_c+o(1),\qquad
 W_m/(2N)=(q_c-1)/b+o(1).
\end{equation}
For \(A=M,M^2\), set \(c_M=q_c\), \(c_{M^2}=(q_c-1)/b\), and
\[
 F_A^G=(X^G)^\top AX^G/v,\quad
 \bar F_A^G=\{\operatorname{tr}(A\Sigma_N)+\mu_N^\top A\mu_N\}/v,
 \quad L_{N,A}(s)=\log\mathbb E_{\vartheta_N}^{\mathcal A}
       [e^{sF_A^G}\mid\mathcal F_{n,m}^{\partial}].
\]
The determinant and its two derivatives give
\[
 |\bar F_A^G-2Nc_A|\le C(\sqrt N+\log^2N),\qquad
 \sup_{|s|\le s_0}|L_{N,A}''(s)|\le CN,
\]
so fixed normalized deviations have probability \(Ce^{-c_\epsilon N}\),
also after anchor conditioning.
For the coupling error vector,
\[
 E^\top ME\le CN^2/m,\qquad E^\top M^2E\le C E^\top ME,
\]
\[
 |Q_m-Q_m^G|+|W_m-W_m^G|
        \le C(N^{3/2}/\sqrt m+N^2/m)=o(1)
\]
by Cauchy--Schwarz.
Choose \(\epsilon_0>0\) so that
\(I_-=[\eta,\eta+2\epsilon_0]\) and
\(I_+=[1-\eta-2\epsilon_0,1-\eta]\) have the active-endpoint property.
Together with \(I_0=[\eta+\epsilon_0,1-\eta-\epsilon_0]\) they cover the
trim interval. On \(I_-\cup I_+\) use
\[
 (t^*,B_i^*,u^*)=
 \begin{cases}
 (t,B_i,-\operatorname{logit}t),&t\in I_-,\\
 (1-t,1-B_i,\operatorname{logit}t),&t\in I_+.
 \end{cases}
\]
Increasing \(u^*\) deletes a one label in the oriented coordinates.
The identities
\[
 h_{1-t}(1-x)=h_t(x),\qquad
 T_{n,k}(B)=T_{n,n-k}(1-B),\qquad
 \left|\frac{dt^*}{t^*(1-t^*)}\right|
 =\left|\frac{dt}{t(1-t)}\right|
\]
preserve the raw statistic, the two boundary energies and logit
length. The active endpoint is \(x=1\) in both endpoint blocks, with a
strict opposite-end margin; either orientation is admissible in the
compact interior. Overlapping blocks cover close pairs, and disjoint
endpoint blocks use the three-category coupling.
On the count envelope, the deterministic occupancy identities give
\[
 \left|\sum_i(B_i-t)(MX)_i^2\right|
       \le CN^{3/2}/\sqrt m,\qquad
 \max_{i:B_i=1}|Q_m(B-e_i,t-1/n)-Q_m(B,t)|
       \le C\{\sqrt{N/m}+N/n\}.
\]
Conditional on the complete oriented anchor vector,
\[
 P(B^*\mapsto B^*-e_i\mid B^*)=\frac{B_i^*}{nt^*};
\]
this remains exact under every weighting measurable in that vector.
Combining these transitions with \eqref{eq:rank-scan-s7}--
\eqref{eq:rank-scan-s11} and \eqref{eq:core-middle-critical-values}
therefore gives the reverse-logit \(z/N\) limit
\begin{equation}\label{eq:core-local-Gaussian-field}
 \widetilde Z(z)=
       \sqrt{\frac{8q_c^2}{q_c+1}}\,W(z)-(q_c-1)z,
       \qquad z\ge0.
\end{equation}
The time conversion and overshoot parameter are
\[
 T=Q_m/2,\qquad \frac zN=2\frac z{2N},\qquad
 \frac{2(q_c-1)}{8q_c^2/(q_c+1)}=b.
\]
For the boundary, put
\[
 E^\pm=\sum_{c\in\{-1/2,+1/2\}}\sum_{l=1}^m
       \frac l{l+c}\,h_t^\pm(C_l^c/l),
 \quad C_l^c=\sum_{i=1}^l B_i^c.
\]
Clip \(h_t^+\) by its tangent continuation above \(t+\rho\).
The residual is increasing, so reverse deletions have nonpositive jumps;
its parameter drift is at most \(C_\rho E^+\).
For one boundary block let \(\omega_l=l/(l+c)\), and let \(f_\rho\)
be zero on \(( -\infty,t]\), equal \(h_t\) on \((t,t+\rho]\), and
its tangent above \(t+\rho\). Put
\[
 G_i=\sum_{l=i}^m\omega_l
       \{f_\rho(C_l/l)-f_\rho((C_l-1)/l)\},\qquad
 0\le f_\rho''\le C_\rho\quad\hbox{a.e.}
\]
Discrete Hardy and Taylor's integral remainder give
\[
 \sum_iG_i^2\le CE^+,\qquad
 \sum_{i:B_i=1}\sum_{l=i}^m\omega_l
 \left\{f_\rho((C_l-1)/l)-f_\rho(C_l/l)
                         +l^{-1}f_\rho'(C_l/l)\right\}
       \le C_\rho\sum_{l=1}^m l^{-1},
\]
\[
 e^{-sG_i}-1+sG_i\le\tfrac12s^2G_i^2 .
\]
Combining these terms in the reverse generator gives, for every fixed \(s>0\),
\begin{equation}\label{eq:core-positive-boundary-generator}
 \frac{\mathcal L_{\rm rev}e^{sE^+}}{e^{sE^+}}
          \le C_s\{E^+/N+\log(m)/N\}.
\end{equation}
The generator bound holds up to every \(O(N)\) energy cap.
Starting from \(B_n\le N^{3/4}\), put
\(\tau_+=\inf\{z:E^+(z)>2N^{3/4}\}\). On \(0\le z\le H\),
\[
 \sup\frac{\mathcal L_{\rm rev}e^{sE^+}}{e^{sE^+}}
       \le C_s(N^{-1/4}+\log m/N)\to0,\qquad
 P(\tau_+\le H)\le e^{-sN^{3/4}+o(1)}.
\]
The increment vanishes, and any smaller positive exponential maximum is
uniformly integrable.
On either sufficiently small oriented endpoint block \(I\), let
\(b_-^* =\inf_{t\in I} C_+(1-t)^{-1}>b\), the threshold for the
negative excursion, and choose
\[
 b<s_1<s_2<b_-^*,\qquad \alpha_0=1-d_0,\quad 0<d_0<d_*.
\]
The constant \(d_*>0\) is chosen after \(s_1,s_2\) and the fixed
Taylor tolerance below; the compact interior uses its strict margin.
For every fixed \(\epsilon>0\),
\begin{equation}\label{eq:core-negative-thinning}
 h_{\alpha_0t}^-(y)
       \le(1+\epsilon)h_t^-(x)+C_\epsilon(y-\alpha_0x)^2,
       \qquad x,y\in[0,1].
\end{equation}
Put \(d=y-\alpha_0x\). Near \(t\), the quadratic expansion and Young's
inequality give
\[
 h_{\alpha_0t}^-(y)
 \le(1+\epsilon)h_{\alpha_0t}^-(\alpha_0x)+C_\epsilon d^2 .
\]
For \(x\le t-\rho\), uniform continuity absorbs small \(d\) into
\(\epsilon h_t^-(x)\); larger \(d\) is bounded by \(C_\epsilon d^2\).
For \(x\ge t\),
\[
 h_{\alpha_0t}^-(y)\le C(\alpha_0t-y)_+^2\le Cd^2.
\]
Let \(D_i\) be independent Bernoulli\((d_0)\) deletion indicators,
\(S_l=\sum_{i\le l}B_i\), and \(Y_l=\sum_{i\le l}(1-D_i)B_i\).
For one block define its actual energy change by
\[
 \begin{split}
 E_{c,t}^-(B)&=\sum_{l\le m}\frac l{l+c}h_t^-(S_l/l),\\
 \Delta E_c^-&=\sum_{l\le m}\frac l{l+c}
           \{h_{\alpha_0t}^-(Y_l/l)-h_t^-(S_l/l)\}.
 \end{split}
\]
The contraction \(h_{\alpha_0t}(\alpha_0x)\le h_t(x)\) gives
\[
 \Delta E_c^-\le\epsilon E_{c,t}^-(B)
       +C_\epsilon\sum_{l\le m}(Y_l-\alpha_0S_l)^2/l^2.
\]
Adding the two blocks gives the change of \(E^-\).
For a single copy \(D\) of the deletion indicator,
\[
 \log\E e^{s(D-d_0)}\le Cs^2/\log(1/d_0),\qquad s\in\mathbb R.
\]
For $|s|\le\frac12\log(1/d_0)$ use $d_0(e^s-1-s)\le\sqrt{d_0}s^2/2$; outside
this interval the linear bound suffices. Put \(Z_i=(D_i-d_0)B_i\) and
\[
 (M_H)_{ij}=\sum_{l\ge\max(i,j)}^m l^{-2},\qquad
 \mathcal Q_H(Z)=Z^\top M_HZ
       =\sum_{l\le m}l^{-2}\left(\sum_{i\le l}Z_i\right)^2.
\]
Conditional on \(B\), Gaussian linearization gives
\[
 \begin{gathered}
 E[e^{c\mathcal Q_H(Z)}\mid B]
       \le\det(I-2c\sigma_{d_0}^2M_H)^{-1/2},\\
 \sigma_{d_0}^2=C/\log(1/d_0),\qquad
 \|M_H\|\le4,\quad\operatorname{tr}M_H=O(\log m).
 \end{gathered}
\]
For fixed \(c\) and \(d_0=O(N^{-1})\),
\[
 -\tfrac12\log\det(I-2c\sigma_{d_0}^2M_H)
 \le C_c\sigma_{d_0}^2\operatorname{tr}M_H
 \le C_c\frac{\log m}{\log N}\le C_cK.
\]
For \(r\le2m\) distinguished observation times, let
\(\mathbb P_{\rm wr}\) be uniform independent sampling from
\(\{1,\ldots,n\}\), and \(\mathbb P_{\rm wor}\) its law conditional on
all times \(U_1,\ldots,U_r\) being distinct. Then
\[
 \frac{d\mathbb P_{\rm wor}}{d\mathbb P_{\rm wr}}
   =\frac{n^r}{(n)_r}\mathbf1_{\{U_a\ne U_b\ (a\ne b)\}}
   \le e^{Cr^2/n}\le e^{Cm^2/n}.
\]
Forcing a deletion at rank \(i\) changes \(Z\) by a vector \(ae_i\),
\(|a|\le1\). The full quadratic difference, including its cross term,
satisfies
\[
 \mathcal Q_H(Z+ae_i)\le2\mathcal Q_H(Z)+2\mathcal Q_H(ae_i)
       \le2\mathcal Q_H(Z)+2\sum_{l\ge i}l^{-2}
       \le2\mathcal Q_H(Z)+C/i.
\]
The preceding determinant bound applies with \(2c\) in place of \(c\);
choosing \(d_*\) smaller reserves this fixed exponential margin.
Maxima occur initially or after a deletion, so there are \(O(m)\)
distinguished terms.
Choose \(s_2\epsilon\) below the opposite-end margin and use
\[
 g_{\rm mix}(x)=
 \begin{cases}
 g_{\vartheta_N}(x),&x\ge t,\\
 g_{\vartheta_N+s_2\epsilon}(x),&x<t .
 \end{cases}
 \qquad g_{\rm mix}(t)=g_{\rm mix}'(t)=0 .
\]
Choose a fixed smooth cutoff \(\chi_-\) equal to one on \([0,t-\rho]\)
and zero on \([t-\rho/2,1]\). On \(x<t\), define
\[
 A_{{\rm mix},c}(x)
 =A_{\vartheta_N+s_2\epsilon,c}(x)^{\chi_-(x)}
           A_{\vartheta_N,c}(x)^{1-\chi_-(x)};
\]
for \(x\ge t\), set \(A_{{\rm mix},c}=A_{\vartheta_N,c}\).
Only the negative branch is used at \(\vartheta_N+s_2\epsilon\). Put
\[
 \begin{split}
 W_{{\rm mix},l}^c(s)
 &=l^{-\kappa(\vartheta_N)}A_{{\rm mix},c}(s/l)e^{lg_{\rm mix}(s/l)},\\
 V_{{\rm mix},m}^c
 &=\sum_{l=1}^m\frac l{l+c}
       \{\vartheta_Nh_t(S_l/l)+s_2\epsilon h_t^-(S_l/l)\}.
 \end{split}
\]
The central row error is \(O(l^{-1})\); the positive-end row error is
bounded by \(C/l+C\widetilde p_l/j^2\).
The accumulated central error and critical-end hazard satisfy
\[
 \sum_{l\le m}\frac Cl\le C\log m,\qquad
 E\exp\!\left\{C\sum_{l:\,j_l\ge J}
                  \frac{\widetilde p_l}{j_l^2}\right\}\le C
\]
for a fixed \(J\) large enough that \(C/J^2\le s_0\).
The finitely many levels \(j<J\) are handled by their next-failure kernels. Their row product therefore costs at most \(Cm^C\).
Let \(R_1=\min\{l\ge1:B_l=0\}-1\), with \(R_1=m\) if all
first \(m\) labels are one. Define
\[
 \begin{split}
 M_{{\rm mix},c}^{\rm ff}
 &=E_t[e^{V_{{\rm mix},m}^c}W_{{\rm mix},m}^c(S_m);R_1<m],\\
 M_{{\rm mix},c}^{\rm run}
 &=E_t[e^{V_{{\rm mix},m}^c}W_{{\rm mix},m}^c(S_m);R_1=m].
 \end{split}
\]
The same row product and the all-success endpoint identity give
\[
 M_{{\rm mix},c}^{\rm ff}\le m^CK_c(t,\vartheta_N),\qquad
 M_{{\rm mix},c}^{\rm run}\le m^C(1-\delta_\Delta)^m\le m^C.
\]
Combining both endpoints with the conditional Gaussian middle yields
\[
 E_{\rm hyb}[e^{\vartheta_N\widetilde T_n+s_2\epsilon E^-};\mathcal A_n]
       \le e^{N\lambda(\vartheta_N)}N^C,
       \qquad m=N^K.
\]
Define the negative count-tube exception before applying the ratio:
\[
 \begin{gathered}
 A_l^-(z)=\{lt^*(z)-C_l(z)\}_+,\\
 \mathcal B_-=\{\sup_{0\le z\le H}\max_{l\le m}A_l^-(z)/\sqrt l>C\log N\}.
 \end{gathered}
\]
The same piecewise Lyapunov ratio gives, under the anchor and its
retained thinnings,
\[
 \mathbb P_{n,t}^{u,1}(\mathcal B_-)\le N^Ce^{-c(\log N)^2}.
\]
On \(\mathcal B_-^c\), the small-rank influence is
\(O((\log\log N)^2)\). For a state function \(F\), write
\(\Gamma_{\rm rev}(F)=\mathcal L_{\rm rev}(F^2)-2F\mathcal L_{\rm rev}F\).
The quadratic expansion for the other ranks gives, on \(\mathcal B_-^c\),
\[
 \begin{split}
 |\mathcal L_{\rm rev}E^-|+\Gamma_{\rm rev}(E^-)
       +|e^{-sE^-}\mathcal L_{\rm rev}e^{sE^-}|&\le C_s\xi_N,\\
 \xi_N&=N^{-1}(\log N)^C e^{C(\log\log N)^2}\longrightarrow0 .
 \end{split}
\]
Thus on a fixed local window the negative increment vanishes, and a
larger fixed exponent gives uniform integrability on the tube.
Fix \(A_{\rm cap}>\gamma_c+1\), then choose \(C_{\rm cap}\) with
\(\mathcal R_{b_\eta}(C_{\rm cap})>A_{\rm cap}+1\).
For a coupling accuracy exponent \(A>0\), define
\[
 \tau_{\rm cap}=\inf\{z\ge0:T_{\rm rev}(z)>C_{\rm cap}N\},
\]
\[
 \begin{split}
 \mathcal G_{n,A,H}
 &=\{\max_{m\le l\le n-m}|A_l(0)-G_l(0)|\le C_AN\}\\
 &\quad\cap\bigcap_{l=m}^{n-m}
 \{\sup_{0\le z\le H}|A_l(z)|\le C_A\sqrt{(l\wedge(n-l))N}\}.
 \end{split}
\]
The bridge construction and conditional transfer concentration bound the
complement by the prescribed \(n^{-A}\), after increasing \(C_A\).
Theorem~\ref{thm:rank-global-logarithmic-rate} gives
\[
 \begin{gathered}
 P\{\max_kT_{n,k}>C_{\rm cap}N\}\le n^{-A_{\rm cap}},\\
 \frac{n^{-A_{\rm cap}}}{p_{n,t}(u)}
 \le e^{-(A_{\rm cap}-\gamma_c)N+O(\sqrt N+\log N)} .
 \end{gathered}
\]
The coupling exponent \(A\) in \(\mathcal G_{n,A,H}\) is chosen
independently after \(C_{\rm cap}\).
Remove the unweighted cap and coupling exceptions before dividing by the
anchor probability. Apply H\"older to the combined middle and
positive-boundary terms with
\[
 b<s<s_2,\qquad 1<p_-<s_2/s,\qquad p_+=p_-/(p_--1).
\]
For \(V=Q_m/2,E^+,E^-\), define
\[
 M_V=\max\{0,\sup_{0\le z\le H}[V(z\wedge\tau_{\rm cap})-V(0)]\},
 \quad M_{\rm safe}=M_{Q_m/2}+M_{E^+},\quad M_-=M_{E^-}.
\]
On \(\mathcal G_{n,A,H}\),
\[
 \begin{split}
 \mathbb E_{n,t}^{u,1}[e^{s(M_{\rm safe}+M_-)};\mathcal G_{n,A,H}]
 &\le\{\mathbb E_{n,t}^{u,1}[e^{sp_+M_{\rm safe}};\mathcal G_{n,A,H}]\}^{1/p_+}\\
 &\quad\times\{\mathbb E_{n,t}^{u,1}[e^{sp_-M_-};\mathcal G_{n,A,H}]\}^{1/p_-}
 \le C_H.
 \end{split}
\]
Every fixed exponent is admissible for the middle and positive-boundary
terms, so H\"older also separates these terms. If \(\mathcal B_-\) is
the negative bad-tube event, its larger moment gives
\[
 \begin{split}
 \mathbb E_{n,t}^{u,1}[e^{sp_-M_-};\mathcal B_-\cap\mathcal G_{n,A,H}]
 &\le\{\mathbb E_{n,t}^{u,1}[e^{s_2M_-};\mathcal G_{n,A,H}]\}^{sp_-/s_2}\\
 &\quad\times\mathbb P_{n,t}^{u,1}(\mathcal B_-)^{1-sp_-/s_2}\\
 &\le N^C e^{-c(\log N)^2}\longrightarrow0.
 \end{split}
\]
The constant \(C_H\) is thus supplied by the stopped good generators;
the polynomial moment is used only to discard the bad tube.
Combining the two boundary parts with the exact middle matrix martingale
gives, for some $s>b$ and fixed local length $H$,
\begin{equation}\label{eq:core-reverse-crossing}
 \mathbb P_{n,t}^{u,1}\left\{\sup_{0\le z\le H}
 [T_{\rm rev}(z\wedge\tau_{\rm cap})-T_{\rm rev}(0)]>y,
                  \ \mathcal G_{n,A,H}\right\}
            \le C_He^{-sy},\qquad 0\le y\le C_0N.
\end{equation}
On an \(O(N)\) cap, the middle matrix norm and normalized energy are
bounded and its maximum jump tends to zero. Thus every fixed middle
exponent is admissible under deletion randomness, as is every positive
boundary exponent.
On \(\{\tau_{\rm cap}>H\}\), nonnegativity of \(T_{\rm rev}(0)\) gives
\(\sup_{z\le H}[T_{\rm rev}(z)-T_{\rm rev}(0)]\le C_{\rm cap}N\).
Hence the weighted coupling exception satisfies
\[
 \begin{split}
 &\mathbb E_{n,t}^{u,1}\left[
 e^{s\max\{0,\sup_{z\le H}[T_{\rm rev}(z)-T_{\rm rev}(0)]\}};
       \mathcal G_{n,A,H}^c,\ \tau_{\rm cap}>H\right]\\
 &\qquad\le e^{sC_{\rm cap}N}\frac{n^{-A}}{p_{n,t}(u)}
 \le e^{-(A-sC_{\rm cap}-\gamma_c)N+O(\sqrt N+\log N)}\to0,
 \end{split}
\]
provided \(A>sC_{\rm cap}+\gamma_c\).
For a unit anchor interval at level \(u\), the global tilted density bound
and \eqref{eq:core-anchor-width-lower} give
\[
 \begin{split}
 \frac{P\{T(0)\in[u-y,u-y+1]\}}{p_{n,t}(u)}
       &\le Ce^{\vartheta_N y},\\
 \frac{P\{T(0)>u+y\}}{p_{n,t}(u)}
       &\le Ce^{-\vartheta_N y},\qquad y\ge0 .
 \end{split}
\]
For \(|y|\le\sqrt N\), the core argument \(x+y/\sqrt N\) remains compact.
The upper excess satisfies
\[
 P(T(0)>u+\sqrt N)/p_{n,t}(u)\le Ce^{-c\sqrt N}.
\]
For deep negative anchors apply the stopped middle/positive generators
and the averaged negative mixed transform. Conditional on the binary
anchor, apply H\"older over deletions. For some \(s>b\), with
\[
 \mathcal M_H=\max\{0,\sup_{z\le H}
            [T_{\rm rev}(z\wedge\tau_{\rm cap})-T_{\rm rev}(0)]\},
\]
this gives
\[
 \E_{\vartheta_N}[e^{s\mathcal M_H};\mathcal G_{n,A,H}]\le N^C
\]
on the stopped good set. If $T(0)\le u-\sqrt N$ and a crossing of $u$
occurs, then $\mathcal M_H\ge\sqrt N$ and $e^{-\vartheta_N T(0)}\le e^{-\vartheta_N
u}e^{\vartheta_N \mathcal M_H}$. Define
\[
 \mathcal C^-_{n,u,H}=\{T(0)\le u-\sqrt N,
  \sup_{z\le H}T_{\rm rev}(z)>u,\ \tau_{\rm cap}>H\}\cap\mathcal G_{n,A,H}.
\]
Its probability satisfies
\begin{equation}\label{eq:core-deep-negative-anchors}
 P(\mathcal C^-_{n,u,H})\le e^{-\vartheta_Nu}\mathcal M_{N,t}(\vartheta_N)N^C
             e^{-(s-\vartheta_N)\sqrt N}.
\end{equation}
Divide \eqref{eq:core-deep-negative-anchors} by the reference anchor
scale \(e^{-\vartheta_Nu}\mathcal M_{N,t}(\vartheta_N)/\sqrt N\). Then,
for every fixed polynomial cell count \(N^d\),
\[
 N^d\frac{P\{T(0)\le u-\sqrt N,\
 \sup_{0\le z\le H}T_{\rm rev}(z)>u,\ \tau_{\rm cap}>H,\ \mathcal G_{n,A,H}\}}
 {e^{-\vartheta_Nu}\mathcal M_{N,t}(\vartheta_N)/\sqrt N}
 \le N^{C+d+1/2}e^{-(s-\vartheta_N)\sqrt N}\longrightarrow0.
\]
Only the anchor is coupled; subsequent deletion follows its exact binary
law. This estimate and \eqref{eq:core-reverse-crossing} give cell uniform
integrability without extending the compact-core scalar asymptotic to
deep negative offsets.

Two pair bounds complete the scan step. For separations $H/N\le h\le h_0$,
start at the later high anchor and reverse time. At exponent $s=b/2$ the
limiting middle generator equals
\begin{equation}\label{eq:core-negative-pair-drift}
 -(q_c-1)s+\frac{4q_c^2}{q_c+1}s^2
                    =-\frac14b(q_c-1)<0.
\end{equation}
Let \(\tau_{\rm pair}\) be the first local time at which
\(B_n>\epsilon N\), a middle trace leaves a fixed small neighborhood of
\eqref{eq:core-middle-critical-values}, or the count envelope fails.
Before \(\tau_{\rm pair}\), the component exponential generators satisfy
\[
 e^{-sE^+}\mathcal L_{\rm rev}e^{sE^+}
       \le C_s\{\epsilon+\log(m)/N\},\qquad
 e^{-sE^-}\mathcal L_{\rm rev}e^{sE^-}\le C_s\xi_N=o(1)
       \quad\hbox{on }\mathcal B_-^c.
\]
For the oriented anchor parameter \(t\), put
\[
 t_h=\operatorname{logistic}(\operatorname{logit}t-h),\qquad
 d(h)=1-t_h/t,\qquad d_0=d(h_0)\le C_\eta h_0<1/2.
\]
With independent uniform deletion times \(U_i\), define
\[
 D_i(h)=B_i\mathbf1_{\{U_i\le d(h)\}},\qquad
 Y_i(h)=\frac{D_i(h)-d(h)B_i}{1-d(h)} .
\]
Conditional on \(B\), \(Y(h)\) is a vector martingale. The preceding
Bernoulli estimate and binomial conditioning give
\[
 \begin{gathered}
 \sigma^2(h_0)=\frac{C}{(1-d_0)^2\log(1/d_0)}
                      =O(1/\log(1/h_0)),\\
 \{P[\operatorname{Bin}(nt,d_0)=\lfloor ntd_0\rfloor]\}^{-1}
             \le C\sqrt{ntd_0(1-d_0)+1}\le C\sqrt n .
 \end{gathered}
\]
Gaussian linearization for \(A=M,M^2\) then gives
\[
 \log E\exp\{cY(h_0)^\top A Y(h_0)/\sigma^2(h_0)\}\le CN .
\]
The convex exponential of the martingale quadratic norm is a
submartingale. Doob's inequality yields
\[
 P\{\sup_{h\le h_0}Y(h)^\top A Y(h)>\delta N\}
 \le\exp\{-c\delta N/\sigma^2(h_0)+CN\}.
\]
The deterministic mean change is \(O(h_0N)\), and the cross term satisfies
\[
 2|X^\top AY(h)|
 \le2(X^\top AX)^{1/2}(Y(h)^\top AY(h))^{1/2}.
\]
Thus the probability of an \(O(N)\) trace deviation is \(e^{-c_\delta N}\);
its exponent can be enlarged by decreasing \(h_0\).
The positive-boundary generator \eqref{eq:core-positive-boundary-generator} gives
\[
 P\{\sup_{0\le h\le h_0}[E^+(Nh)-E^+(0)]>\epsilon N\}
 \le e^{-s\epsilon N+C_s\epsilon Nh_0}.
\]
Choose \(h_0\) so the negative quadratic exponential parameter is
admissible; its marked-prefix cost is polynomial.
Equation \eqref{eq:core-negative-pair-drift} and the preceding bounds give
\[
 \mathbb P_{n,t}^{u+z,1}
   \{T_{\rm rev}(Nh)>u,\ \tau_{\rm pair}>Nh\}
       \le Ce^{-cNh}e^{(b/2)z},
\]
\[
 \int_0^\infty e^{(b/2)z}e^{-\vartheta_Nz}\,dz
       =(\vartheta_N-b/2)^{-1}\le C .
\]
For the actual two-point event put
\[
 \mathcal P_{n,t,u}(h)=\{T_{n,\lfloor nt\rfloor}>u,
                                  T_{\rm rev}(Nh)>u\}.
\]
Summing the unit overshoot intervals and the exceptional events gives
\begin{equation}\label{eq:core-near-pair-bound}
 \frac{P\{\mathcal P_{n,t,u}(h)\}}{p_{n,t}(u)}
 \le Ce^{-cNh}+e_N,\qquad
 e_N=N^C\{e^{-cN^{1/4}}+e^{-c(\log N)^2}+e^{-c'N}\}.
\end{equation}
The bounds above control the polynomially many cells and pairs, including
overshoots beyond $\sqrt N$ and deep negative anchors.
For $h\ge h_0$, the explicit three-category conditional bridge construction
in \eqref{eq:rank-scan-s25a}-- \eqref{eq:rank-scan-s25e} gives two middle
fields with correlation $\rho\le e^{-h_0/2}<1$. Their equal-half-tilt
pressure has the strict uniform gap
\begin{equation}\label{eq:core-far-pair-pressure}
 \lambda\{b(1+\rho)/2\}+\lambda\{b(1-\rho)/2\}<\lambda(b).
\end{equation}
The energy cap and \eqref{eq:core-deterministic-energy-tube} put
the terminal proportions in a central tube. Since \(\rho<1\), the
joint Gaussian endpoint factor is bounded by the geometric mean of the
two one-anchor factors, with a strict quadratic margin.
For boundary energy \(E\le C_{\rm cap}N\),
\[
 e^{bE}\le e^{C_{\rm cap}}e^{(b-1/N)E},\qquad
 E[e^{(b-1/N)T_m^c}\psi_{b-1/N}(Z_m)
                                  \mathbf1_{\mathcal A_{m,N}}]
 \le Cm^{\kappa(b-1/N)}K_c(t,b-1/N).
\]
The second inequality follows from
\eqref{eq:core-terminal-real-ratio} and the harmonic identity with its
retained indicator.
The real coefficient and pressure gap satisfy
\[
 K_c(t,b-1/N)\le N^C,\qquad
 e^{N[\lambda\{b(1+\rho)/2\}+\lambda\{b(1-\rho)/2\}-\lambda(b)]}
       \le e^{-cN}
\]
by \eqref{eq:core-far-pair-pressure}.
H\"older for the boundary factors therefore gives, uniformly for
\(h\ge h_0\),
\[
 \frac{P\{\mathcal P_{n,t,u}(h)\}}{p_{n,t}(u)}
 \le N^Ce^{-cN}+\frac{Cn^{-A_*}}{p_{n,t}(u)},\qquad
 A_*:=\min\{A_{\rm cap},A\}>\gamma_c .
\]
The second term is \(\exp\{-(A_*-\gamma_c)N+O(\sqrt N+\log N)\}\).
The required spatial comparison is the following relative bound.
For every fixed $H$,
\begin{equation}\label{eq:core-cell-real-freezing}
 \sup_{|t'-t|\le H/N}
 \left|\frac{e^{-\vartheta_Nd(t')}\mathcal M_{N,t'}(\vartheta_N)}
              {e^{-\vartheta_Nd(t)}\mathcal M_{N,t}(\vartheta_N)}
                            -1\right|\longrightarrow0.
\end{equation}
For a sequence with an interior limit, compact subcritical matching
applies. At an endpoint, \eqref{eq:core-spatial-real-match} and
\(\Delta_t\ge c/\sqrt N\) give, uniformly for \(|t'-t|\le H/N\),
\[
 \frac{\Delta_{t'}}{\Delta_t}=1+O(H/\sqrt N),\qquad
 \frac{\Delta_{t'}^{-\alpha_\eta}\{\log(1/\Delta_{t'})\}^{\beta_\eta}}
      {\Delta_t^{-\alpha_\eta}\{\log(1/\Delta_t)\}^{\beta_\eta}}
 =1+O(H/\sqrt N).
\]
The zero-exponent case uses the same ratios in
\eqref{eq:core-confluent-real-coefficient}. Also
\[
 |d(t')-d(t)|\le\omega_d(H/N)\longrightarrow0,
\]
where \(\omega_d\) is the modulus of continuity on the trim
interval. This proves \eqref{eq:core-cell-real-freezing}, uniformly for
\(c_0\) in a compact positive interval.
To pass to anchor probabilities, put
\[
 \begin{split}
 C_{N,t}(w)
 &=e^{-(b-w/\sqrt N)d(t)-N\lambda(b-w/\sqrt N)}
                        \mathcal M_{N,t}(b-w/\sqrt N),\\
 D_N(w)&=\frac{C_{N,t'}(w)-C_{N,t}(w)}{C_{N,t}(c_0)} .
 \end{split}
\]
The real doubling bound and \eqref{eq:core-positive-complex-bound} give
\[
 \sup_N\sup_{w\in K}|D_N(w)|\le C_K
 \quad(K\Subset\{\operatorname{Re}w>0\}),\qquad
 D_N(w)\longrightarrow0\quad(w>0).
\]
Equation \eqref{eq:core-cell-real-freezing} and the identity theorem give
\(D_N\to0\) locally uniformly. To specify the inverted difference, let
\(\rho_{N,t}\) be the density of \(\widetilde T_n-d(t)\) with the retained
indicator, and put
\[
 Z_{N,t}(z)=e^{-zd(t)}\mathcal M_{N,t}(z),\quad
 f_{N,t}(a)=\frac{e^{\vartheta_N a}\rho_{N,t}(a)}{Z_{N,t}(\vartheta_N)},
 \quad
 g_{N,t'}^{(t)}(a)=\frac{e^{\vartheta_N a}\rho_{N,t'}(a)}{Z_{N,t}(\vartheta_N)}.
\]
Fourier inversion and normalized Mehler domination give
\[
 \sup_a|g_{N,t'}^{(t)}(a)-f_{N,t}(a)|
 \le\frac1{2\pi}\int_{\mathbb R}
  \left|\frac{Z_{N,t'}(\vartheta_N+is)-Z_{N,t}(\vartheta_N+is)}
                   {Z_{N,t}(\vartheta_N)}\right|ds
       =o(N^{-1/2}).
\]
Indeed, \(s=w/\sqrt N\) gives the locally vanishing factor
\(D_N(c_0-iw)\); the remaining factor is bounded by \(Ce^{-cw^2}\) on
bounded \(s\), and the outer-frequency integral is exponentially small.
For a fixed-width interval \(I\) in the core range,
\[
 \frac{|\int_I(g_{N,t'}^{(t)}-f_{N,t})(a)\,da|}
             {\int_I f_{N,t}(a)\,da}
 \le\frac{|I|\,o(N^{-1/2})}{c_{|I|}/\sqrt N}\longrightarrow0.
\]
The lower bound in \eqref{eq:core-anchor-width-lower} controls the
denominator, and the probability sandwich transfers this relative error
to rank intervals.
Put
\[
 \pi_N(h)=P\!\left\{\max_{k\in\mathcal K_n}D_{n,k}>
       (q_c-1)\sqrt{N/2}+\frac h{\sqrt{2N}}\right\},
 \qquad z_{N,h}=(q_c-1)\sqrt{N/2}+h/\sqrt{2N}.
\]
For the field in \eqref{eq:core-local-Gaussian-field}, define
\[
 v_c=b^2\frac{8q_c^2}{q_c+1}=2b(q_c-1),\qquad
 H_c(H)=E\exp\left\{\sup_{z\le H}
                    [\sqrt{v_c}W(z)-v_cz/2]\right\}.
\]
The local field, exponential integrability and relative freezing give
this same coefficient in each cell. The Brownian calculation in
\eqref{eq:rank-scan-s21}--\eqref{eq:rank-scan-s22} gives
\[
 \frac{H_c(H)}H\longrightarrow\frac{v_c}2=b(q_c-1).
\]
Put \(g=\sqrt H\), and partition the trimmed logit interval into cells
\(I_i=[u_i,u_i+H/N]\) separated by gaps of length \(g/N\).
Let \(\{J_i\}\) be these gaps and the two remaining end intervals. Define
\[
 \begin{split}
 \mathcal C(I)&=\{\max_{k:\,\operatorname{logit}(k/n)\in I}D_{n,k}>z_{N,h}\},\\
 S_{N,H}&=\sum_i P\{\mathcal C(I_i)\},\\
 G_{N,H}&=\sum_i P\{\mathcal C(J_i)\},\\
 B_{N,H}&=\sum_{i<j}P\{\mathcal C(I_i)\cap\mathcal C(I_j)\}.
 \end{split}
\]
The union and Bonferroni inequalities give
\[
 S_{N,H}-B_{N,H}\le\pi_N(h)\le S_{N,H}+G_{N,H}.
\]
Relative freezing and the single-cell limit give
\[
 \begin{gathered}
 \frac{S_{N,H}}{N\rho_N(Nq_c+h)}
       \longrightarrow\frac{H_c(H)}{b(H+g)},\\
 \limsup_{N\to\infty}\frac{G_{N,H}}{S_{N,H}}\le Cg/H.
 \end{gathered}
\]
Split each retained cell into \(O(H)\) subcells of fixed local length.
The exponential maximum bound \eqref{eq:core-reverse-crossing} extends
\eqref{eq:core-near-pair-bound} to each pair of these subcells. Summing by
separation, and then applying \eqref{eq:core-far-pair-pressure}, yields
\[
 \frac{B_{N,H}}{S_{N,H}}
 \le CH^2\sum_{r\ge0}e^{-c(g+rH)}
                    +N^{C_1}e_N+N^{C_2}e^{-cN},
\]
\[
 \limsup_{N\to\infty}
 \left|\frac{\pi_N(h)}{S_{N,H}}-1\right|
 \le Cg/H+CH^2e^{-cg}
 \longrightarrow0\qquad(H\to\infty).
\]
The gap estimate also holds in the \(N^{-1/2}\) endpoint layers by the
same relative freezing. Since \(g/H\to0\) and
\(H_c(H)/H\to b(q_c-1)\), these bounds prove
\begin{equation}\label{eq:core-scan-density-reduction}
 \pi_N(h)\sim N(q_c-1)\rho_N(Nq_c+h).
\end{equation}
The scalar-tail factor $1/b$ and cluster factor $b(q_c-1)$ combine to
$q_c-1$, and integration uses the full logit measure $dt/[t(1-t)]$.

The exact moment expansions give, uniformly in trimmed \(t\) and \(h=O(\sqrt
N)\),
\[
 \E T_{n,k}=N+m(t)+o(1),\qquad
 \operatorname{Var}(T_{n,k})=2N+v_0(t)+o(1),
\]
so the raw threshold is
\[
 Nq_c+h+d(t)+\frac{h v_0(t)}{4N}+o(1)
       =Nq_c+h+d(t)+o(1).
\]
Thus the raw threshold contains the shift \(d(t)\) in
\eqref{eq:core-bulk-constants}. Combining \eqref{eq:core-integrated-density}
and \eqref{eq:core-scan-density-reduction} proves the uniform sharp scan
formula, for every fixed compact \(K_h\subset\mathbb R\),
\begin{equation}\label{eq:core-one-coordinate-scan}
 \pi_N(h)\sim\mathcal A_{\rm c}(h/\sqrt N)
       N^{\nu_*}(\log N)^{\ell_*}e^{-N\gamma_c-bh},
 \qquad h/\sqrt N\in K_h.
\end{equation}

\subsubsection{The analytic MAX normalization}

\begin{proof}[Proof of Theorem~\ref{thm:max-critical-core}]
For bounded \(\tau_N\),
\[
 \frac{h_N+y/b}{\sqrt N}=x_N+O(\log N/\sqrt N),\qquad
 \frac{\mathcal A_{\rm c}((h_N+y/b)/\sqrt N)}
      {\mathcal A_{\rm c}(x_N)}
 =1+O(\log N/\sqrt N),
\]
by smooth positivity on compact sets. A standardized displacement
\(A_n^{\rm c}y\) changes the raw offset by \(y/b+o(1)\).
The definition of \(h_N\) gives exactly
\[
 p\,\mathcal A_{\rm c}(x_N)N^{\nu_*}(\log N)^{\ell_*}
                         e^{-N\gamma_c-bh_N}=1 .
\]
Substitution into \eqref{eq:core-one-coordinate-scan} therefore yields
\[
 p\,\pi_N(h_N+y/b)
 =e^{-y}\frac{\mathcal A_{\rm c}((h_N+y/b)/\sqrt N)}
                  {\mathcal A_{\rm c}(x_N)}\{1+o(1)\}
 \longrightarrow e^{-y}.
\]
By Proposition~\ref{prop:rank-law}, every coordinate exceedance is the same
measurable event in its latent Gaussian column. The marginal estimate and
neighborhood assumptions verify Proposition~\ref{prop:sparse-factorization}:
\[
 pP(E_{j,n})\to e^{-y}
 \ \Longrightarrow\
 \sum_{j=1}^p\mathbf1_{E_{j,n}}
       \Rightarrow\operatorname{Poisson}(e^{-y}),
 \qquad
 P\Bigl(\bigcap_{j=1}^pE_{j,n}^{\,c}\Bigr)\to e^{-e^{-y}}.
\]
\end{proof}

\subsection{Moving lower saddles and a common critical normalization}
\label{app:critical-lower-extension}

Retain $N=\log n$, $b=b_\eta<1/4$, $\lambda$, $q_c$, and $\gamma_c$
from \eqref{eq:core-bulk-constants}. Moving lower saddles require uniform
real-prefix and rank-scan estimates as the subcritical margin vanishes;
the fixed-margin theorem applies on compact strictly subcritical sets.

Let
\begin{equation}\label{eq:lower-moving-range}
 \gamma_n=\frac{\log p}{N}\to\gamma_c,\qquad
 (\gamma_c-\gamma_n)\sqrt N\to\infty.
\end{equation}
For all sufficiently large $n$, define
\begin{equation*}
 \begin{gathered}
 I(q)=\frac{(q-1)^2}{4q},\qquad
 q_n=(\sqrt{\gamma_n}+\sqrt{1+\gamma_n})^2,\\
 \vartheta_n=\frac{1-q_n^{-2}}4,\qquad
 \Delta_n=b-\vartheta_n,\qquad V_n=\lambda''(\vartheta_n)=2q_n^3 .
 \end{gathered}
\end{equation*}
Then $I(q_n)=\gamma_n$, $\Delta_n\to0$, and
$\Delta_n\sqrt N\to\infty$, since
$\gamma_c-\gamma_n=b\lambda''(b)\Delta_n+O(\Delta_n^2)$.
Put
\begin{equation*}
 \begin{split}
 d_n(t)&=m(t)+\frac{q_n-1}{4}v_0(t),\\
 \Theta_n&=\int_\eta^{1-\eta}
       \Psi_t(\vartheta_n)e^{-\vartheta_n d_n(t)}
                                    \frac{dt}{t(1-t)},\\
 C_n&=\frac{(q_n-1)\sqrt N\,\Theta_n}{\sqrt{2\pi V_n}},
 \qquad h_n=\frac{\log C_n}{\vartheta_n}.
 \end{split}
\end{equation*}
\begin{theorem}[Moving original-rank saddle below the critical point]
\label{thm:max-moving-lower}
Under the pairwise-correlation and neighborhood conditions of
Assumption~\ref{ass:copula-null} and \eqref{eq:lower-moving-range}, define
\[
 B^<_{n,p}=(q_n-1)\sqrt{N/2}+\frac{h_n}{\sqrt{2N}},
 \qquad A^<_{n,p}=\frac1{\vartheta_n\sqrt{2N}}.
\]
Then $\Pp\{(\Smax-B^<_{n,p})/A^<_{n,p}\le x\}\to\exp(-e^{-x})$.
These are the constants in \eqref{eq:main-subcritical-normalization}.
\end{theorem}

\begin{proof}
Use the two prefix events and boundary sigma-field from
\eqref{eq:core-finite-coefficients}:
\[
 \mathcal A_{m,N}^{\pm}=\{|Z_m^\pm|\le B\sqrt N\},\qquad
 \mathcal A_n=\mathcal A_{m,N}^-\cap\mathcal A_{m,N}^+,\qquad
 \mathcal F_{n,m}^{\partial}
   =\sigma(C_l^-,C_l^+:1\le l\le m).
\]
Here \(C_l^\pm\) count successes in the two revealed rank prefixes and
\(Z_m^\pm=(C_m^\pm-mt)/\sqrt{mt(1-t)}\).
Under the unretained hybrid law \(\mathbb P_{\rm hyb}\), let \(X\)
be the conditional OU bridge joining \(Z_m^-\) and \(Z_m^+\). Put
\[
 R_{n,m}=2\log\{(n-m)/m\},\qquad
 Y_\circ=\frac12\int_0^{R_{n,m}}X_v^2\,dv,\qquad
 \widetilde T_n=E_\partial+Y_\circ .
\]
The retained transform and its positive tilted law are
\[
 \mathcal M_{N,t}(\vartheta)
 :=\mathcal M_{n,t}(\vartheta)
 =\mathbb E_{\rm hyb}[e^{\vartheta\widetilde T_n};\mathcal A_n],
 \qquad
 \frac{d\mathbb Q_{n,t,\vartheta}^{\mathcal A}}
      {d\mathbb P_{\rm hyb}}
 =\frac{e^{\vartheta\widetilde T_n}\mathbf1_{\mathcal A_n}}
        {\mathcal M_{N,t}(\vartheta)}.
\]
Choose a sufficiently large fixed \(K>3\), and take
\[
 m=\lfloor N^K\rfloor,
 \qquad t\in\mathcal I_\eta,\qquad
 |\vartheta-\vartheta_n|\le\Delta_n/2,\quad\vartheta\in\mathbb R .
\]
We prove the real comparison on this neighborhood before analytic inversion:
\begin{equation}\label{eq:lower-real-matching}
 \frac{B_{m,N,c}(t,\vartheta)}{K_c(t,\vartheta)}=1+o(1),
 \qquad
 \frac{\Psi_{N,t}(\vartheta)}{\Psi_{N,t}(\vartheta_n)}\le C.
\end{equation}
For \(|\vartheta-\vartheta_n|\le\Delta_n/2\), set
\(\delta_t=b(t)-\vartheta\). The endpoint and interior ranges satisfy
\[
 \delta_t\ge\Delta_n/2,\qquad
 \delta_t\sqrt N\longrightarrow\infty,\qquad
 t-\eta=O(\delta_t)
 \ \Longrightarrow\
 (t-\eta)\log(1/\delta_t)=o(1).
\]
Thus the doubling estimate \eqref{eq:core-real-doubling} is uniform:
an endpoint exponent changes by \(o(1)\) on its logarithmic scale,
whereas an interior sequence retains a positive subcritical margin.
The real harmonic identity, action monotonicity, and polynomial
amplitude comparison therefore give
\[
 \Pp^H_\vartheta(T_m^c>N^{3/4})
 \le N^C e^{-\delta_t N^{3/4}/2}
       \frac{K_c(t,b(t)-\delta_t/2)}{K_c(t,\vartheta)}
       +N^C e^{-cm/\sqrt N}
 \le N^C e^{-cN^{1/4}}+N^C e^{-cm/\sqrt N}.
\]
The last exponential bounds the zero-minority first-run series.
By \eqref{eq:core-deterministic-energy-tube}, the cap makes all sufficiently
late prefixes central. The uniform \(a_1<1/2\) Riccati estimate
\eqref{eq:core-killed-Riccati} gives
\[
 \mathbb P_\vartheta^H
 \{|Z_m|>\log N,\ T_m^c\le N^{3/4}\}
       \le Ce^{-c(\log N)^2}.
\]
On the whole retained event, \eqref{eq:core-terminal-real-ratio} gives
\[
 \frac{\psi_\vartheta(Z_m)}
      {m^{\kappa(\vartheta)}H^c_{m,\vartheta}(S_m)}
 =1+O\{N^{3/2}/\sqrt m+\sqrt{N/m}+m^{-1}\}.
\]
Substitution in the harmonic identity proves
\eqref{eq:lower-real-matching}, with
\begin{equation}\label{eq:lower-real-error}
 \left|\frac{B_{m,N,c}(t,\vartheta)}{K_c(t,\vartheta)}-1\right|
 \le C N^{3/2}/\sqrt m+N^C e^{-cN^{1/4}}
       +Ce^{-c(\log N)^2}+N^C e^{-cm/\sqrt N}.
\end{equation}
Define
\[
 r_N^{\rm an}=c\min\{\Delta_n,(\log m)^{-1/2}\},\qquad 0<c<1/8.
\]
Then $r_N^{\rm an}\sqrt N\to\infty$. The exact positive inequality
\eqref{eq:core-positive-complex-bound} reads
\[
 |B_{m,N,c}(t,z)|
 \le m^{\kappa(\operatorname{Re}z)-\operatorname{Re}\kappa(z)}
                       B_{m,N,c}(t,\operatorname{Re}z).
\]
Analyticity of $\kappa$ on a fixed disk inside $\operatorname{Re}z<1/4$
bounds the exponent difference by $C(\operatorname{Im}z)^2$.
Thus \eqref{eq:lower-real-matching} gives
\[
 \sup_t\sup_{|z-\vartheta_n|\le2r_N^{\rm an}}
       \frac{|\Psi_{N,t}(z)|}{\Psi_{N,t}(\vartheta_n)}\le C.
\]
Cauchy's estimate yields the key quantitative limit
\begin{equation}\label{eq:lower-Cauchy-flatness}
 \sup_t\sup_{|w|\le T}
 \left|\frac{\Psi_{N,t}(\vartheta_n+iw/\sqrt N)}
                    {\Psi_{N,t}(\vartheta_n)}-1\right|
 \le\frac{C_T}{r_N^{\rm an}\sqrt N}\to0,\qquad T<\infty .
\end{equation}
The choice of \(r_N^{\rm an}\) simultaneously gives
\(r_N^{\rm an}\le c\Delta_n\) and \((r_N^{\rm an})^2\log m\le c^2\); these are,
respectively, the threshold and time-power restrictions in the preceding
complex bound. For one prefix with shift \(c\), define its probability law by
\[
 \frac{d\mathbb Q_{m,t,\vartheta}^{c}}{d\mathbb P_t}
 =\frac{e^{\vartheta T_m^c}\psi_\vartheta(Z_m)
              \mathbf1_{\{|Z_m|\le B\sqrt N\}}}
        {m^{\kappa(\vartheta)}B_{m,N,c}(t,\vartheta)} .
\]
On the real axis, the same calculation and monotonicity of
\(\psi_\vartheta\) give, for fixed \(u>0\),
\[
 \mathbb E_{\mathbb Q_{m,t,\vartheta_n}^{c}}e^{uT_m^c/\sqrt N}
 \le m^{\kappa(\vartheta_n+u/\sqrt N)-\kappa(\vartheta_n)}
 \frac{B_{m,N,c}(t,\vartheta_n+u/\sqrt N)}
      {B_{m,N,c}(t,\vartheta_n)}
 =1+o(1).
\]
Differentiating the real positive comparison also yields
\[
 \mathbb E_{\mathbb Q_{m,t,\vartheta_n}^{c}}T_m^c
 \le C\{\log m+\Delta_n^{-1}\}=o(\sqrt N).
\]
The first term is the central-prefix baseline. The earlier exponential
bound continues to control \(T_m^c>N^{3/4}\).
Define the Mehler remainder by
\[
 \mathcal E_{N,t}(z)
 :=\mathcal M_{N,t}(z)-e^{N\lambda(z)}\Psi_{N,t}(z).
\]
The exact Mehler expansion bounds this difference.
On the needed vertical line $z=\vartheta_n+is$, $|s|\le s_0$,
\[
 \frac{|\mathcal E_{N,t}(z)|}{\mathcal M_{N,t}(\vartheta_n)}
 \le N^C e^{-cN}
       e^{N\{\operatorname{Re}\lambda(z)-\lambda(\vartheta_n)\}}.
\]
All these comparisons are on the specified vertical line. The conditional
determinant bound \eqref{eq:core-normalized-Fourier}, uniform because
\(\vartheta_n\to b<1/4\), gives
\[
 \left|\frac{\mathcal M_{N,t}(\vartheta_n+is)}
                 {\mathcal M_{N,t}(\vartheta_n)}\right|
 \le
 \begin{cases}
 C e^{-cNs^2},&|s|\le s_0,\\
 C e^{-cN},&s_0\le |s|\le s_1,\\
 C e^{-cN\sqrt{|s|}},&|s|\ge s_1 .
 \end{cases}
\]
Define the unscaled tilted density and the centered Gaussian density by
\[
 \mathbb Q_{n,t,\vartheta_n}^{\mathcal A}
       \{\widetilde T_n\in dy\}=f^{(\vartheta_n)}_{N,t}(y)\,dy,
 \qquad
 \phi_V^{\rm G}(x)=\frac{e^{-x^2/(2V)}}{\sqrt{2\pi V}},
 \quad V>0 .
\]
Combine the bound with \eqref{eq:lower-Cauchy-flatness}, first on
\(|s|\sqrt N\le T\) and then on its complement:
\[
 \sup_t\int_{\mathbb R}
 \left|\mathbb E_{\mathbb Q_{n,t,\vartheta_n}^{\mathcal A}}
       e^{iw(\widetilde T_n-Nq_n)/\sqrt N}
                   -e^{-V_nw^2/2}\right|\,dw\longrightarrow0 .
\]
Fourier inversion gives
\begin{equation}\label{eq:lower-uniform-LLT}
 \sup_t\sup_{x\in\mathbb R}
 \left|\sqrt N f^{(\vartheta_n)}_{N,t}(Nq_n+x\sqrt N)
                 -\phi_{V_n}^{\rm G}(x)\right|\to0.
\end{equation}
This density belongs to the continuous retained hybrid. Undoing its tilt
uses the exact positive integral
\[
 \Pp_\mathrm{hyb}\{\widetilde T_n>u,\ \mathcal A_n\}
 =\mathcal M_{N,t}(\vartheta_n)e^{-\vartheta_n u}
    \int_0^\infty e^{-\vartheta_n v}
                  f^{(\vartheta_n)}_{N,t}(u+v)\,dv .
\]
Apply \eqref{eq:lower-uniform-LLT} on bounded \(v\), and its global
\(C/\sqrt N\) density bound to the remaining integral. Uniformly for
\(h/\sqrt N\) in compact sets and bounded \(d\), this gives
\begin{equation}\label{eq:lower-fixedsplit}
 \mathbb P_{\rm hyb}\{\widetilde T_n>Nq_n+h+d,\ \mathcal A_n\}
 =\frac{\Psi_t(\vartheta_n)}
        {\vartheta_n\sqrt{2\pi NV_n}}
   e^{-N\gamma_n-\vartheta_n(h+d)-h^2/(2NV_n)}\{1+o(1)\}.
\end{equation}
It also gives a width-one tilted interval lower bound $c_H/\sqrt N$
on every fixed compact range of $h/\sqrt N$.
Set
\[
 \epsilon_N=Cm^2/n,\qquad
 \delta_N=C\{N^{3/2}/\sqrt m+N^2/m+m^2/n\}.
\]
For any fixed \(A>0\), \eqref{eq:core-rank-sandwich-error} gives
\[
 \begin{aligned}
 &(1-\epsilon_N)\mathbb P_{\rm hyb}
       \{\widetilde T_n>u+\delta_N,\mathcal A_n\}-n^{-A}\\
 &\qquad\le\mathbb P(T_{n,k}>u)\\
 &\qquad\le(1+\epsilon_N)\mathbb P_{\rm hyb}
       \{\widetilde T_n>u-\delta_N,\mathcal A_n\}+n^{-A}.
 \end{aligned}
\]
For $E_{n,t}=\{\widetilde T_n>Nq_n+h+d\}\cap\mathcal A_n$,
\eqref{eq:lower-fixedsplit} gives
\[
 \mathbb P(E_{n,t})=n^{-\gamma_c+o(1)},\qquad
 \frac{n^{-A}}{\mathbb P(E_{n,t})}=n^{\gamma_c-A+o(1)}.
\]
Choose $A>\gamma_c+1$. The bounded-offset tail ratio then transfers
\eqref{eq:lower-fixedsplit} and its fixed-width version to the ranks.
The marked coupling also retains the prefix, $Q$, and $W$ observables.
The moment expansions give, uniformly for $h/\sqrt N$ bounded,
\begin{equation*}
 \E T_{n,k}+\sqrt{\operatorname{Var}T_{n,k}}
       \left\{(q_n-1)\sqrt{N/2}+\frac h{\sqrt{2N}}\right\}
 =Nq_n+h+d_n(t)+o(1).
\end{equation*}
The omitted product is bounded by
\[
 \left|\frac{h\,v_0(t)}{4N}\right|
 \le C|h|/N=o(1).
\]
Consequently the raw-threshold expansion is valid throughout
\(h=O(\log N)\).
Use the oriented process, generator, boundary energies and stopped
maximum from \eqref{eq:rank-local-uniform-time-generator}. In this proof
\[
 \mathcal I_{n,t}=\{T_{n,k}\in[Nq_n+h+d_n(t),Nq_n+h+d_n(t)+1]\},
 \qquad
 \mathbb P_{\rm anchor}(E)=\mathbb P(E\mid\mathcal I_{n,t}).
\]
The width-one lower bound \eqref{eq:lower-uniform-LLT} multiplies
the cap and terminal-tube error bounds in
\eqref{eq:lower-real-error} by at most $C_H\sqrt N$.
For fixed \(\delta>0\), the middle Gaussian determinant and the
preceding cap bounds give
\[
 \begin{aligned}
 &\mathbb P_{\rm anchor}
 \left\{\left|\frac Q{2N}-q_n\right|
       +\left|\frac W{2N}-\frac{q_n-1}{\vartheta_n}\right|>\delta\right\}\\
 &\qquad\le C_\delta N^C
 \{e^{-c_\delta N}+e^{-cN^{1/4}}+e^{-c(\log N)^2}\}.
 \end{aligned}
\]
The conditional bridge error in $W$ is bounded by
$CN^{3/2}/\sqrt m$. The binary occupancy summation and maximum-jump
bounds used in \eqref{eq:core-local-Gaussian-field} therefore remain valid.
Since $q_n\to q_c$, the limiting field is
\eqref{eq:core-local-Gaussian-field}, with cluster rate
\[
 b(q_c-1)=\vartheta_n(q_n-1)\{1+o(1)\}.
\]
The boundary and middle contributions satisfy
\[
 \{T_m^c\le N^{3/4}\}\ \subset\
 \{T_m^c/N\le N^{-1/4}\},
 \qquad
 \left|\frac{\mathcal L e^{sE^+}}{e^{sE^+}}\right|
 \le C_s\left\{\frac{T_m^c}{N}+\frac{\log m}{N}\right\}=o(1).
\]
Increase only the negative coefficient in \eqref{eq:core-negative-thinning}.
The interval lower bound then permits a stopped crossing moment of order
\(s>b\); its height integrals are controlled by
\[
 e^{\vartheta_n y}\mathbb P_{\rm anchor}(\mathcal M_H>y)
 \le C_H e^{-(s-\vartheta_n)y},\qquad
 s-\vartheta_n\ge s-b>0,
\]
with weight \(e^{-\vartheta_n y}\) above the anchor.
For \(y\ge\sqrt N\), conditional H\"older over deletion variables on
the stopped \(O(N)\) cap gives
\[
 \mathbb E_{\mathbb Q_{n,t,\vartheta_n}^{\mathcal A}}e^{s\mathcal M_H}\le N^C .
\]
Let
\[
 \mathcal D_{n,t}(u,H)=
 \{T_t^\sigma(0)\le u-\sqrt N,\
                 \sup_{0\le z\le H}T_t^\sigma(z)>u\}.
\]
The stopped estimate and probability sandwich give
\[
 \mathbb P\{\mathcal D_{n,t}(u,H)\}
 \le e^{-\vartheta_n u}\mathcal M_{N,t}(\vartheta_n)
       N^C e^{-(s-\vartheta_n)\sqrt N}+n^{-A}.
\]
Consequently \eqref{eq:core-deep-negative-anchors} gives
\[
 \frac{\mathbb P\{\mathcal D_{n,t}(u,H)\}}
      {e^{-\vartheta_n u}\mathcal M_{N,t}(\vartheta_n)/\sqrt N}
 \le N^{C+1/2}e^{-(s-b)\sqrt N}
   +\frac{\sqrt N\,n^{-A}}
          {e^{-\vartheta_n u}\mathcal M_{N,t}(\vartheta_n)}
 \longrightarrow0
\]
after choosing \(A\) larger than the fixed scalar rate.
The strict negative drift in \eqref{eq:core-negative-pair-drift}
persists for $q_n$ near $q_c$. Put \(t_r=\operatorname{logistic}(\operatorname{logit}t+r)\) and
\(\mathcal V_t(u)=\{\sup_{0\le z\le1}T_t^\sigma(z)>u\}\).
Equations \eqref{eq:core-near-pair-bound} and
\eqref{eq:core-far-pair-pressure} give
\[
 \begin{aligned}
 &\frac{\mathbb P\{\mathcal V_t(u)\cap\mathcal V_{t_r}(u)\}}
        {\mathbb P\{\mathcal V_t(u)\}+\mathbb P\{\mathcal V_{t_r}(u)\}}\\
 &\quad\le
 \begin{cases}
 Ce^{-cNr}+N^Ce^{-cN^{1/4}}+N^Ce^{-c(\log N)^2},
                                      &1/N\le r\le r_0,\\
 N^Ce^{-cN},&r\ge r_0 .
 \end{cases}
 \end{aligned}
\]
The second line uses the real tilt \(b-N^{-1}\).
Apply the near- and far-pair bounds after subdivision into unit cells,
then let the cell length increase. For spatial freezing, take \(|t'-t|\le H/N\):
\[
 |b(t')-b(t)|\le CH/N,\qquad
 \frac{|b(t')-b(t)|}{b(t)-\vartheta_n}
       \le\frac{CH}{N\Delta_n}=o(1).
\]
Thus \eqref{eq:lower-Cauchy-flatness} and the Fourier bound give
\[
 \frac{\Psi_{N,t'}(\vartheta_n)}{\Psi_{N,t}(\vartheta_n)}
 =1+o(1),\qquad
 d_n(t')-d_n(t)=o(1),\qquad |t'-t|\le H/N ,
\]
including confluence and the corresponding anchor-interval ratios.
The matching cell sums yield, uniformly for \(h=O(\log N)\),
\begin{equation}\label{eq:lower-moving-scan}
 \Pp\left\{\mathcal M_{j,n}>
           (q_n-1)\sqrt{N/2}+\frac h{\sqrt{2N}}\right\}
  =C_n e^{-N\gamma_n-\vartheta_n h}\{1+o(1)\}.
\end{equation}
The scalar factor $1/\vartheta_n$ cancels the cluster rate
$\vartheta_n(q_n-1)$; both trim endpoints are already included in
$\Theta_n$.
Let $L_n=\log(1/\Delta_n)$ and use the constants in
Appendix~\ref{app:max-deterministic-inputs}. The real endpoint asymptotic and
its uniform envelope give
\begin{equation}\label{eq:lower-spatial-cases}
 \Theta_n\sim
 \begin{cases}
  \Theta_0,&\alpha_\eta<1,\\
  P_0 L_n^{\beta_\eta+1},&\alpha_\eta=1,\\
  P_A\Delta_n^{1-\alpha_\eta}L_n^{\beta_\eta},&\alpha_\eta>1.
 \end{cases}
\end{equation}
The three cases follow from the same endpoint integral. For a fixed small
\(\varepsilon>0\), its singular part is proportional to
\[
 \int_0^\varepsilon
 (\Delta_n+b'(\eta)t)^{-\alpha_\eta}
 \left[\log\frac1{\Delta_n+b'(\eta)t}\right]^{\beta_\eta} dt.
\]
For \(\alpha_\eta<1\) this has an integrable limiting majorant. For
\(\alpha_\eta=1\), the substitution \(r=\Delta_n+b'(\eta)t\) gives
\[
 \frac1{b'(\eta)}
 \int_{\Delta_n}^{\Delta_n+b'(\eta)\varepsilon}
 r^{-1}\log^{\beta_\eta}(1/r)\,dr
 \sim \frac{L_n^{\beta_\eta+1}}{b'(\eta)(\beta_\eta+1)}.
\]
For \(\alpha_\eta>1\), the endpoint substitution gives
\[
 t=\frac{\Delta_n u}{b'(\eta)},\qquad
 \int_0^\infty(1+u)^{-\alpha_\eta}\,du=\frac1{\alpha_\eta-1},\qquad
 e^{-bd_n(t)}\longrightarrow e^{-bd(t)}.
\]
Reflection yields \(P_0,P_A\) in \eqref{eq:lower-spatial-cases}.
For each coordinate define
\[
 E_{j,n}(h)=
 \left\{\mathcal M_{j,n}>
 (q_n-1)\sqrt{N/2}+\frac h{\sqrt{2N}}\right\}.
\]
Moreover
\[
 L_n\le\tfrac12\log N,\qquad \log C_n=O(\log N),\qquad
 h=h_n+x/\vartheta_n
 \ \Longrightarrow\ p\mathbb P(E_{j,n}(h))\to e^{-x}
\]
by \eqref{eq:lower-moving-scan}. Proposition~\ref{prop:sparse-factorization}
gives Poisson and Gumbel convergence.
\end{proof}

\begin{proof}[Proof of Theorem~\ref{thm:max-critical-lower-unified}]
The positive kernel in \eqref{eq:main-IG-kernel} satisfies
\[
 \int_0^\infty e^{zv}g_N(v)\,dv=e^{N\lambda(z)},\qquad
 \operatorname{Re}z<1/4.
\]
Indeed,
\[
 N/2-v/4-N^2/(4v)+zv
 =N/2-(1/4-z)v-N^2/(4v),
\]
and the elementary integral
\(\int_0^\infty v^{-3/2}e^{-Av-B/v}\,dv
 =\sqrt{\pi/B}\,e^{-2\sqrt{AB}}\), with \(A=1/4-z\), \(B=N^2/4\),
gives the claimed transform.
An exact identity supplies the needed global domination. For
$u=Ny$ and $0<w<u$,
\begin{equation}\label{eq:lower-exact-kernel-ratio}
 e^{-bw}\frac{g_N(u-w)}{g_N(u)}
 =\left(\frac u{u-w}\right)^{3/2}
   \exp\left\{-[b-\vartheta(y)]w
                  -\frac{N^2w^2}{4u^2(u-w)}\right\}.
\end{equation}
First take $u=Nq_c+x\sqrt N$ with $x$ in a fixed compact set.
On writing $w=r\sqrt N$,
\[
 g_N(u-w)=
 \frac{e^{-N\gamma_c-bx\sqrt N+bw}}{\sqrt{2\pi NV}}
                 e^{-(x-r)^2/(2V)}\{1+o(1)\},
 \qquad V=\lambda''(b).
\]
For \(w=r\sqrt N\), \eqref{eq:lower-exact-kernel-ratio} gives
\[
 e^{-bw}\frac{g_N(u-w)}{g_N(u)}
 \le
 \begin{cases}
 C e^{Cr-cr^2},&0<w\le u/2,\\
 C\{u/(u-w)\}^{3/2}e^{-cN^2/(u-w)},&u/2<w<u.
 \end{cases}
\]
The second line absorbs every algebraic endpoint factor.
Dominated convergence in \eqref{eq:lower-unified-intensity}, for
\(\alpha_\eta>1\), therefore gives
\[
 \mathcal R_N^{\rm lc}(Nq_c+x\sqrt N)
 \sim(q_c-1)P_A2^{-\beta_\eta}\mathfrak J_{\alpha_\eta-1}(x)
       N^{\alpha_\eta/2}(\log N)^{\beta_\eta}
                         e^{-N\gamma_c-bx\sqrt N}.
\]
For the other two cases, \(\mathfrak J_0\) from
\eqref{eq:calibration-fractional-gaussian} gives
\[
 \begin{aligned}
 \mathcal R_N^{\rm lc}(Nq_c+x\sqrt N)
 &\sim(q_c-1)\Theta_0\mathfrak J_0(x)\sqrt N\,
                 e^{-N\gamma_c-bx\sqrt N},&&\alpha_\eta<1,\\
 &\sim(q_c-1)P_0\,2^{-\beta_\eta-1}\mathfrak J_0(x)
    \sqrt N(\log N)^{\beta_\eta+1}e^{-N\gamma_c-bx\sqrt N},
                                      &&\alpha_\eta=1,\\
 \mathcal L_N(Nq_c+x\sqrt N)&=\tfrac12\log N+O(1).
 \end{aligned}
\]
These equal \eqref{eq:calibration-core-amplitude} and
\eqref{eq:core-one-coordinate-scan}. On a lower approach, set
\[
 y_n=q_n,\qquad\Delta_n\sqrt N\to\infty,\qquad
 u=u_{0,n},\qquad w=r/\Delta_n .
\]
Then \eqref{eq:lower-exact-kernel-ratio}, for fixed \(r>0\), gives
\[
 [b-\vartheta(y_n)]w=r,\qquad
 \frac{N^2w^2}{4u_{0,n}^2(u_{0,n}-w)}
       =O\!\left(\frac{r^2}{N\Delta_n^2}\right)=o(1),\qquad
 \frac{\log(e+w)}{\log(1/\Delta_n)}\to1.
\]
For \(w\le u_{0,n}/2\), the normalized integrand is bounded by
\(Ce^{-r}\) times its integrable power and logarithm; the other
half is bounded by the preceding endpoint exponential. Thus
\[
 \alpha_\eta>1:\ \int_0^\infty e^{-r}r^{\alpha_\eta-2}\,dr=\Gamma(\alpha_\eta-1),
 \qquad
 \alpha_\eta=1:\ \mathcal L_N(u_{0,n})\sim\log(1/\Delta_n),
\]
while \(\alpha_\eta<1\) retains its finite mass. Since \(I(y_n)=\gamma_n\),
\begin{equation}\label{eq:lower-kernel-moving-match}
 p\mathcal R_N^{\mathrm{lc}}(u_{0,n})=C_n\{1+o(1)\}.
\end{equation}
In either regime, $\log\{p\mathcal R_N^{\mathrm{lc}}(u_{0,n})\}
=O(\log N)$ and, uniformly for $|h|\le C\log N$,
\begin{equation}\label{eq:lower-kernel-shift}
 \frac{\mathcal R_N^{\mathrm{lc}}(u_{0,n}+h)}
      {\mathcal R_N^{\mathrm{lc}}(u_{0,n})}
       =e^{-\zeta_n h}\{1+o(1)\}.
\end{equation}
In the core, the positive amplitude \(\mathfrak J_a\) is uniformly continuous and
bounded away from zero on the required compact set, and
\(|h|/\sqrt N\to0\). On a lower approach, the two possible errors are
\[
 \frac{|h|}{N\Delta_n}\le\frac{C\log N}{N\Delta_n}=o(1),
 \qquad
 N\!\left[I(y_n+h/N)-I(y_n)\right]
   =\vartheta(y_n)h+O((\log N)^2/N).
\]
Thus the full rate remains in the calculation even if
\(N\Delta_n^3\) does not vanish.
For a bounded negative core offset, Taylor expansion of the equation
\(I(y_n)=\gamma_c+\tau_N/\sqrt N\) gives
\[
 u_{0,n}=Nq_c+\frac{\tau_N\sqrt N}{b}+d_n^*,\qquad
 d_n^*=-\frac{\tau_N^2}{2b^3V}+o(1).
\]
Writing the core amplitude at the level without \(d_n^*\) as \(C_n^*\),
we have
\[
 \log\{p\mathcal R_N^{\rm lc}(u_{0,n})\}
   =\log C_n^*-bd_n^*+o(1),\qquad
 d_n^*-\frac{b}{\zeta_n}d_n^*=o(1).
\]
The logarithmic correction cancels this displacement. For \(|\tau_N|=O(1)\), this agrees with the core normalization
at \(Nq_c+\tau_N\sqrt N/b\); for \(\tau_N\to-\infty\), apply
\eqref{eq:lower-kernel-moving-match}.
Every subsequence in \eqref{eq:lower-unified-range} has a further
subsequence of these types, proving the common normalization.
With \eqref{eq:lower-unified-normalization}, the exact raw threshold is
\[
 \begin{aligned}
 u_n(t,x)
 &:=\mathbb ET_{n,k}
   +\sqrt{\operatorname{Var}(T_{n,k})}
       \{\mathfrak b_{n,p}^{\rm lc}+x\mathfrak a_{n,p}^{\rm lc}\}\\
 &=u_{0,n}+H_n^{\rm lc}+x/\zeta_n+d(t)+o(1).
 \end{aligned}
\]
The integrated one-coordinate intensity tends to $e^{-x}$ in
both ranges. Proposition~\ref{prop:sparse-factorization} gives
the Poisson limit and proves Theorem~\ref{thm:max-critical-lower-unified}.
\end{proof}

The divisor $\zeta_n$ in $H_n^{\rm lc}$ is retained because
$\Delta_n\log N$ need not vanish along a slow lower approach.

The positive-part logarithm in the $\alpha_\eta=1$ formula is a real
deterministic regularization, and its kink is treated by the real-variable
estimates above.

\subsection{Finite analytic scalar inversion at the nonlinear boundary}
\label{app:full-boundary-scalar}

The scalar local law at positive boundary excess is established first. The
exponential parameter tilts $T_{n,k}=nF_{n,k}$. Retain the Bellman branch,
transport amplitude and real coefficient of
Lemma~\ref{lem:hardy-bellman-branch},
\eqref{eq:boundary-transport-amplitude} and
Lemma~\ref{lem:analytic-boundary-coefficient}.

Fix a compact family on which $b=b(t)<1/4$ has a uniform strict margin and
the active Bernoulli endpoint is $x=1$; reflection treats the other active
endpoint. The opposite endpoint is strictly subcritical. Write
\[
 r(z)=\sqrt{1-4z},\quad a(z)=(1-r(z))/2,\quad
 \kappa(z)=a(z)/2,\quad \lambda(z)=2\kappa(z),\quad
 C_*=1/b,\quad V=\lambda''(b).
\]
All branches are continued from the real interval below $b$. Put
$d_c=D_cC_U^{-\alpha_c}$, with the constants in
\eqref{eq:core-boundary-exponents}--\eqref{eq:core-first-failure-constant}.
Throughout the boundary calculations, a sign subscript on an endpoint
coefficient abbreviates its rank-weight shift: for example,
\(\alpha_\pm=\alpha_{\pm1/2}(t)\), \(d_\pm=d_{\pm1/2}\), and
\(K_\pm(t,b)=K_{\pm1/2}(t,b)\); below, \(B_\pm(z)=B_{\pm1/2}(z)\).
In particular $\alpha_->0$. For the second exponent we either retain a fixed
sign margin from zero or use $|\alpha_+|\log\Lambda\le A$ with $A$ fixed.

Uniformity is restricted to these fixed-sign and bounded-confluence
families.

Let $N=\log n$, $m=\lfloor N^{K_m}\rfloor$, and use exactly the finite
positive coefficients in \eqref{eq:core-finite-coefficients}, with the
parameter-independent cutoff $|Z_m|\le B\sqrt N$. Thus
\[
 B_c(z)=m^{-\kappa(z)}\mathbb E_t
 \left[e^{zT_m^c+a(z)Z_m^2/2}
                    \mathbf1_{\{|Z_m|\le B\sqrt N\}}\right],\qquad
 \Psi_{N,t}(z)=\sqrt{r(z)}B_-(z)B_+(z).
\]
The two boundary paths are independent. Given their endpoints, let
\(Z\) be their coupled OU bridge and put
\[
 R=2\log\{(n-m)/m\},\qquad
 Y_{\partial}=T_m^{-1/2}+T_m^{+1/2},\qquad
 Y_{\circ}=\frac12\int_0^RZ(v)^2\,dv,\qquad
 \widetilde T_n=Y_{\partial}+Y_{\circ}.
\]
Write \(M_{N,t}(z)=\mathcal M_{N,t}(z)\) for its retained transform.
The function \(f_{N,t}\) is the density of
\(E\mapsto\mathbb P_{\rm hyb}\{\widetilde T_n\in E,\mathcal A_n\}\).

This is a positive subprobability measure; its mass at zero transform
parameter tends to one exponentially in $N$. The normalization factor is
therefore \(1+O(e^{-cN})\).

For \(a>0,v>0\) define
\begin{equation}\label{eq:full-scalar-Gamma-bath}
 \mathcal G_{a,v}(E)=
 \frac1{\Gamma(a)\sqrt{2\pi v}}
 \int_0^\infty x^{a-1}e^{-(E-x)^2/(2v)}\,dx,
 \qquad
 \mathcal G_{a,0}(E)=\frac{E^{a-1}}{\Gamma(a)}
 \quad(E>0).
\end{equation}
Fix positive \(c_*,C_\Lambda,e_*,e^*\), and use
\[
 c_*\sqrt N\le\Lambda\le C_\Lambda N,\qquad Q=\log\Lambda,\qquad
 y=N\lambda'(b)+\Lambda E,\qquad e_*\le E\le e^*.
\]
The coefficients in the scalar density are
\[
\begin{array}{c|c|c|l}
 &\alpha_t&\beta_t&P_t\\ \hline
 \alpha_+>0&\alpha_-+\alpha_+&1&
 \sqrt{r(b)}d_-d_+\Gamma(\alpha_-)\Gamma(\alpha_+)\\
 \alpha_+=0&\alpha_-&2&
 \sqrt{r(b)}d_-\Gamma(\alpha_-)(2d_+/3)\\
 \alpha_+<0&\alpha_-&1/2&
 \sqrt{r(b)}d_-\Gamma(\alpha_-)K_+(t,b).
\end{array}
\]
Here \(K_+(t,b)\) is the real critical coefficient from
Lemma~\ref{lem:analytic-boundary-coefficient}.
Define the finite confluent mass by
\begin{equation*}
 \mathcal H_a(Q)=\int_0^Q e^{au}u^{1/2}\,du.
\end{equation*}
For \(|\alpha_+|Q\le A\), use \(\alpha_t=\alpha_-\),
replace \(Q^{\beta_t}\) by \(Q^{1/2}\mathcal H_{\alpha_+}(Q)\),
and replace \(P_t\) by
\(\sqrt{r(b)}d_-\Gamma(\alpha_-)d_+\).
Then \(\mathcal H_{\alpha_+}(Q)\sim(2/3)Q^{3/2}\)
when \(\alpha_+Q\to0\).

\begin{theorem}[Finite-hybrid scalar law at positive boundary excess]
\label{thm:finite-hybrid-boundary-scalar}
For the stated parameter families and sufficiently large
\(K_m\), chosen in the proof, uniformly
\begin{equation}\label{eq:assembled-boundary-scalar-density}
 f_{N,t}(y)\sim e^{N\lambda(b)-by}
       P_t\Lambda^{\alpha_t-1}Q^{\beta_t}
       \mathcal G_{\alpha_t,VN/\Lambda^2}(E),
\end{equation}
The equivalent is uniform for \(t=t_0+u/\Lambda\) with bounded
\(u\), on actual grid splits, and after bounded raw shifts.
The retained upper tail is asymptotic to \(f_{N,t}(y)/b\).

For \(\theta=b-c_0/\Lambda\), with \(c_0\) in a positive compact
interval, every bounded-width interval about \(y\) has tilted
hybrid probability comparable with \(1/\Lambda\).
\end{theorem}

\subsubsection{Complex endpoint geometry used by the finite proxy}

Use \(h_t(x)=\operatorname{KL}(x\|t)\), and put \(H_t=\log\{(1-t)/t\}\) and
\[
 q=h_t'(x),\quad e(q)=1-x=\{1+e^{q-H_t}\}^{-1},\quad
 \widehat w_\Delta(q)=1-u_{b-\Delta}(1-e(q)),\quad
 \delta(\Delta)=\widehat w_\Delta(\infty).
\]
Here \(e(q)\) is the endpoint proportion. We also write
\(w_z(q)=\widehat w_{b-z}(q)=1-u_z(1-e(q))\); when its argument is \(x\),
\(w_z(x)=1-u_z(x)\). The later expansion parameter is \(1/q\).
The notation \(r(z)=\sqrt{1-4z}\) concerns a complex parameter;
the real Hardy orbit is denoted \(r_t(s)\), \(s>0\).

Put \(Q_\Delta=\log(1/|\Delta|)\).
For real-reference subtraction write
\(\theta=b-d\), \(z=\theta+is\), \(\Delta=d-is\),
\(d=c_0/\Lambda\), \(Q_\Delta\asymp\log\Lambda\),
and \(|s|/d\ge c\log\Lambda\).

\begin{lemma}[Closed-half-plane endpoint bounds]
\label{lem:full-complex-endpoint-geometry}
For sufficiently small \(|\Delta|>0\), \(\Re\Delta\ge0\),
the branch from the central germ is pole-free and analytic
in the open half-plane. Uniformly,
\begin{align}
 \delta(\Delta)&=C_U\Delta Q_\Delta^{1/b}
                \{1+O(\log Q_\Delta/Q_\Delta)\},&
 \delta'(\Delta)&=C_UQ_\Delta^{1/b}
                \{1+O(\log Q_\Delta/Q_\Delta)\},\nonumber\\
 |\delta''(\Delta)|&\le
 C|\Delta|^{-1}Q_\Delta^{1/b-1},&
 \Re\delta(\Delta)&\ge
 c\{(\Re\Delta)Q_\Delta^{1/b}
            +|\Im\Delta|Q_\Delta^{1/b-1}\}.\label{eq:full-complex-gap-bounds}
\end{align}
On the real-reference subtraction range,
\begin{equation}\label{eq:full-sharp-real-gap-subtraction}
 \Re\delta(d-is)-\delta(d)
  =\left\{\frac\pi{2b}+o(1)\right\}
                 \frac{|\Im\delta(d-is)|}{Q_\Delta}.
\end{equation}
Throughout the physical small-frequency range,
\begin{equation}\label{eq:full-absolute-Bellman-comparison}
 \Re g_z(x)\le g_\theta(x),\qquad
 |A_{z,c}(x)|\le C A_{\theta,c}(x).
\end{equation}
\end{lemma}

The amplitude comparison applies for either sign of
\(\alpha_c\).

\begin{proof}
Write $E_q=e(1-e)$, $d_b=w_b-e$, and
$f_e(w)=w(1-w)/(w-e)=(1-e)-w+E_q/(w-e)$. Subtraction of
$\widehat w_\Delta'=-(b-\Delta)E_qqf_e(\widehat w_\Delta)$ gives exactly, for
$Z=\widehat w_\Delta-w_b$,
\begin{equation}\label{eq:full-complex-Z-equation}
 Z'=bE_qq\left\{Z+\frac{E_qZ}{d_b(d_b+Z)}\right\}
                  +\Delta E_qq f_e(w_b+Z).
\end{equation}
On a fixed initial interval the analytic expansion is
$Z=U\Delta-V_2\Delta^2+O(\Delta^3)$ with $U,V_2>0$. Indeed
\[
 U'=A_0U+B_0,\quad
 V_2'=A_0V_2+\frac{bE_q^2q}{d_b^3}U^2+(A_0/b)U,
\quad
 A_0=bE_qq(1+E_q/d_b^2),\ B_0=E_qqf_e(w_b),
\]
The central germs and forcings are positive. For \(\Delta=X+iY\),
real analyticity implies
\[
 X,Y\ge0\quad\Longrightarrow\quad
 \Re Z\ge c(X+Y^2),\qquad \Im Z\ge cY .
\]
Choose a fixed large \(q_0\) with \(x>3/4\) and \(w_b\) small.
Compact analytic continuation reaches this chart, where
\[
 \Im Z=0\ \Longrightarrow\ \Im Z'\ge0,\qquad
 \Re Z=0\ \Longrightarrow\
 \Re\frac{Z}{d_b+Z}\ge0,\quad
 \Re f_e(w_b+Z)>0,\quad \Im f_e(w_b+Z)\le0.
\]
For \(X,Y\ge0\), \eqref{eq:full-complex-Z-equation} therefore preserves
the first quadrant and gives \(|d_b+Z|\ge d_b\asymp eq\).
Conjugation handles \(Y<0\). Put
\(q_\pm=Q_\Delta\pm L\log Q_\Delta\), with \(L\) fixed and large.
Variation of constants, followed by the integrable tail equation, gives
\[
 \begin{aligned}
 |Z(q)|&\le C|\Delta|q^{1/b+O(|\Delta|)},&&q\le q_+,\\
 |Z'(q)|&\le Ce^{-q}\{q|Z|+1+|\Delta|q\},&&q\ge q_+,\\
 Z(q_-)&=\Delta U(q_-)\{1+O(Q_\Delta^{-2})\},&
 U(q)&=C_Uq^{1/b}\{1+O(q^{-1})\},\\
 |Z(q_+)-Z(q_-)|&\le
 C|\Delta|Q_\Delta^{1/b-1}\log Q_\Delta .
 \end{aligned}
\]
The first bound closes the small-\(Z\) bootstrap up to \(q_+\);
the second closes it afterwards. The last two bounds yield the
asserted equivalent for \(\delta\).
Set \(D(q,\Delta)=\partial_\Delta \widehat w_\Delta(q)\) and
\(D_2(q,\Delta)=\partial_\Delta^2\widehat w_\Delta(q)\). Differentiating gives
\[
 D'=a_\Delta D+E_qqf_e(\widehat w_\Delta),\qquad
 a_\Delta=(b-\Delta)E_qq\{1+E_q/(\widehat w_\Delta-e)^2\}.
\]
The same three intervals give the bound and equivalent for
$D(\infty)=\delta'$. A second differentiation gives
\[
 D_2'=a_\Delta D_2
 -2E_qq\{1+E_q/(\widehat w_\Delta-e)^2\}D
 -2(b-\Delta)E_q^2qD^2/(\widehat w_\Delta-e)^3.
\]
For the variation-of-constants integral put
\[
 G_\Delta(\infty,q)=\exp\!\left\{\int_q^\infty a_\Delta(v)\,dv\right\},
 \qquad R(q)=|\delta|/d_b(q).
\]
On \(q\asymp Q_\Delta\), \(R\) is increasing and the quadratic forcing obeys
\[
 \begin{aligned}
 &|\Delta|Q_\Delta^{1-1/b}
 \left|G_\Delta(\infty,q)
       \frac{2(b-\Delta)E_q^2qD(q,\Delta)^2}
            {(\widehat w_\Delta(q)-e(q))^3}\right|\,dq\\
 &\qquad\le C(1+R)^{-3}\,dR,\qquad
 \int_0^\infty(1+R)^{-3}\,dR=\frac12 .
 \end{aligned}
\]
The part \(q<Q_\Delta/2\) is exponentially smaller. This proves
\(|\delta''|\le C|\Delta|^{-1}Q_\Delta^{1/b-1}\).
On the real axis all quantities are positive except the two displayed
forcings. The same substitution, with $D\sim C_UQ_\Delta^{1/b}$ and
$\widehat w_\Delta-e\sim d_b+\delta$, yields
\[
 \delta''(\Delta)\sim
   -\frac{C_U}{b\Delta}Q_\Delta^{1/b-1},\qquad \Delta>0,
\]
because $2\int_0^\infty(1+R)^{-3}\,dR=1$. Apply the absolute bound on compact subsets of the open right
half-plane to the holomorphic functions
\[
 F_d(\zeta)=d\,\delta''(d\zeta)/(\log(1/d))^{1/b-1}.
\]
The identity theorem identifies the subsequential limit as
\(-C_U/(b\zeta)\). For the normalized arc integral, split
\[
 [0,\pi/2]=[0,\pi/2-\varepsilon]\cup[\pi/2-\varepsilon,\pi/2].
\]
Local uniform convergence handles the first interval; the closed-half-plane
bound gives an error at most \(C\varepsilon\) on the second.
Let \(\varepsilon\downarrow0\) and integrate the derivative identity:
\[
 e^{-i\phi}\delta(\rho e^{i\phi})
  =\delta(\rho)-i\frac{C_U}{b}\rho
         \{\log(1/\rho)\}^{1/b-1}\phi
           +o(\rho\{\log(1/\rho)\}^{1/b-1}).
\]
The real reference is controlled by
\[
 \delta(\rho)-\rho\delta'(\rho)
 \sim\frac{C_U}{b}\rho\{\log(1/\rho)\}^{1/b-1},
 \qquad
 \frac{Cd}{|s|}\log\frac{|\Delta|}{d}=o(1).
\]
This proves \eqref{eq:full-sharp-real-gap-subtraction}.
The arc argument, or positive pretransition second variation,
also yields the lower bound in \eqref{eq:full-complex-gap-bounds}.
The quadrant comparison gives
\[
 \Re w_z\ge w_\theta,\qquad G=\Re g_z .
\]
Define
\[
 \begin{aligned}
 D_0(x)&=\log(1-t+te^{G'(x)})
       -\Re\log(1-t+te^{g_z'(x)}),\\
 \bar u(x)&=\int_0^1
 \frac{t e^{g_\theta'(x)+v(G'(x)-g_\theta'(x))}}
 {1-t+t e^{g_\theta'(x)+v(G'(x)-g_\theta'(x))}}\,dv .
 \end{aligned}
\]
The triangle inequality gives \(D_0\ge0\), and the HJB equation gives
\[
 G-xG'+\log(1-t+te^{G'})+\theta h_t(x)=D_0 .
\]
Subtracting the real equation yields
\[
 (G-g_\theta)+(\bar u-x)(G-g_\theta)'=D_0.
\]
Above \(t\), \(\bar u<x\); below \(t\), \(\bar u>x\).
The central value is nonpositive because \(\Re a(z)\le a(\theta)\).
The integrating-factor solution in the outward direction therefore
gives \(G-g_\theta\le0\).
For the amplitude use the exact simplification, with $d_z=x-u_z$,
\begin{equation*}
 (\log A_{z,c})'
  =\frac{1/2-\kappa(z)-zc h_t(x)}{d_z}
                  -\frac z2h_t'(x)-\frac12(\log d_z)'.
\end{equation*}
Integration before, across and after $Q_\Delta+O(\log Q_\Delta)$ gives
\[
 |A_{z,c}(x(q))|\asymp
 |d_z(q)|^{-1/2}\{1+\min(q,Q_\Delta)\}^{\alpha_c/b}.
\]
The analogous real expression has $d_\theta$ and $Q_d=\log(1/d)$. If
$\alpha_c\ge0$, $Q_\Delta\le Q_d$ and $|d_z|\ge d_\theta$ prove the bound.
If $\alpha_c<0$ and $q,Q_d>2Q_\Delta$, retain the square root:
\[
 d_\theta/|d_z|\le
 Ce^{-(q-Q_\Delta)}(1+q)/Q_\Delta^{1/b}
 +Ce^{-(Q_d-Q_\Delta)}(Q_d/Q_\Delta)^{1/b}.
\]
These exponentials absorb the otherwise increasing logarithmic power. The
central and opposite-end intervals use their fixed analytic margins. This
proves \eqref{eq:full-absolute-Bellman-comparison}.
\end{proof}

\subsubsection{A finite Gamma correction on the full effective-count range}

At a physical state with $j$ failures among $k$ trials put
\[
 q=h_t'(1-j/k),\quad \varepsilon=q^{-1},\quad
 \nu=\frac{k w_z(q)}{bq},\quad d=\nu-\frac{j\varepsilon}{b},\quad
 \chi=\alpha_c-\frac12,\quad \ell=\log A_{z,c}.
\]
Let
\[
 \mathcal S(\nu)=\frac{\Gamma(\nu)e^\nu}
                {\sqrt{2\pi}\nu^{\nu-1/2}},\qquad
 p(\nu)=(\log\mathcal S)'(\nu).
\]
For the finite construction below, hold $b$ and $\alpha_c$ fixed in the
reduced row. Define the exact reduced failure flow, $0\le v\le1$, by
\begin{equation}\label{eq:full-reduced-finite-flow}
 \varepsilon_v=\frac{\varepsilon}
                 {1-\varepsilon\log(1+v/j)},\qquad
 n_v'=\frac{n_v}{n_v-(j+v)\varepsilon_v/b}
                        +\frac{\varepsilon_v n_v}{j+v},\quad n_0=\nu.
\end{equation}
For $d=\nu-j\varepsilon/b$ set
\[
 G=1/d,\quad A=\nu/(2d^2)+\chi\varepsilon/(bd),\quad
 B=\chi+jA,\quad D=\nu-j\nu/d-\varepsilon\nu,
\]
and evaluate $G_v,A_v$ along \eqref{eq:full-reduced-finite-flow}. Put
$\mathcal E=\int_0^1\{(1-v)G_v-A_v\}\,dv$. For a function $P$ let
the shifted function
\(P^+(\varepsilon,j,\nu)=P(\varepsilon_1,j+1,n_1)\), and
\begin{equation*}
 \mathcal F_P=B+D\{p(\nu)+P_\nu\}-\varepsilon^2P_\varepsilon
  +\frac{b\nu}{\varepsilon}
       \left\{e^{\mathcal E}\frac{\mathcal S(n_1)}{\mathcal S(\nu)}
                       e^{P^+-P}-1\right\}.
\end{equation*}
The apparent quotient at $\varepsilon=0$ is removable.

Write $L_\nu=1+\log(1+|\nu|/j)$.

\begin{lemma}[Full-ratio finite correction]
\label{lem:full-ratio-finite-Gamma}
Fix a closed sector $|\arg\nu|\le\vartheta<\pi$ and $|\nu|\ge c j$, $j\ge1$.
For each fixed integer $M$ there are explicit analytic coefficients
$\eta_1,\ldots,\eta_M$ such that
$P_M=\sum_{h=1}^M\varepsilon^h\eta_h(j,\nu)$ and
\begin{align}
 |\partial_j^a\partial_\nu^h\eta_r|
 &\le\frac{C_{r,a,h}L_\nu^r}{j^a|\nu|^{h+1}},\nonumber\\
 |\partial_j^a\partial_\nu^h\mathcal F_{P_M}|
 &\le\frac{C_{M,a,h}|\varepsilon|^M L_\nu^M}
                  {j^{a+1}|\nu|^h}.\label{eq:full-ratio-Gamma-symbol}
\end{align}
The second bound holds on a fixed sufficiently small complex $\varepsilon$
disk, independently of $|\nu|/j$. Each prescribed finite collection of
spatial and source-parameter derivatives has these bounds.
\end{lemma}
On the physical range, these bounds give
$P_M=O_M(1/(qj))$ and the same order for its scaled two-jets.
For $U_z=W_z\mathcal S(\nu)e^{P_M}$, the reduced unscaled row defect
is $U_z\mathcal F_{P_M}/k$.

\begin{proof}
Use two finite symbol classes
\[
 \mathcal A_d:\quad
 |\partial_j^a\partial_\nu^h f|
       \le C L_\nu^d/(j^{a+1}|\nu|^h),\qquad
 \mathcal B_d:\quad
 |\partial_j^a\partial_\nu^h f|
       \le C L_\nu^d/(j^a|\nu|^{h+1}).
\]
The diagonal inverse
\begin{equation*}
 (\mathcal R_b f)(j,\nu)=
       \sum_{r\ge0}\frac{f(j+r,\nu+r)}{b(\nu+r)}
\end{equation*}
maps $\mathcal A_d$ to $\mathcal B_{d+1}$. Indeed $|\nu+r|\ge
c_\vartheta(|\nu|+r)$; summing below $r=|\nu|$ costs at most one logarithm
divided by \( |\nu| \), while the remaining sum is \(O(|\nu|^{-1})\). Put
\(R=|\nu|\) and \(L_r=1+\log(1+|\nu+r|/(j+r))\). Then
\[
 L_r\le CL_\nu,\qquad
 \sum_{r\ge0}\frac{L_r^d}{(j+r)|\nu+r|}
 \le \frac{CL_\nu^d}{R}
       \left\{1+\sum_{0\le r\le R}\frac1{j+r}\right\}
 \le \frac{CL_\nu^{d+1}}R .
\] Local uniform convergence and slightly
enlarged sector Cauchy bounds give the derivative assertions. Moreover
\[
 b\nu\{\mathcal R_b f(j+1,\nu+1)-\mathcal R_b f(j,\nu)\}=-f(j,\nu).
\]
At $\varepsilon=0$, $n_v=\nu+v$. For a sufficiently small fixed complex
$\varepsilon$ disk, comparison in the finite interval gives
\[
 |n_v|\asymp|\nu|+v,\quad
 |n_v'|\le C(1+|\varepsilon||\nu|/j),\quad
 |n_v-\nu-v|\le C|\varepsilon|(1+|\nu|/j).
\]
For each fixed \(M\), choose finitely many slightly enlarged sectors on
which the denominators remain separated. Consequently, for $\eta\in\mathcal
B_d$,
\[
 |\eta(j+1,n_1)-\eta(j,\nu)|
      \le C L_\nu^d/(j|\nu|).
\]
Along the finite shift,
\[
 \eta(j+1,n_1)-\eta(j,\nu)
 =\int_0^1\{\eta_j+n_v'\eta_\nu\}(j+v,n_v)\,dv .
\]
The fixed differentiated Stirling remainder is uniform away from
the negative real axis and gives \(\mathcal F_0\in\mathcal A_0\).
For \(r=\nu/j\), the order-one terms cancel as
\[
 \begin{aligned}
 &\chi+\frac{r}{2(r-\varepsilon/b)^2}
       +\frac{\chi\varepsilon}{b(r-\varepsilon/b)}
       -\frac{r}{2(r-\varepsilon/b)^2}
       -\frac{r\chi}{r-\varepsilon/b}\\
 &\hspace{2cm}
 =\chi\left\{1+\frac{\varepsilon/b-r}{r-\varepsilon/b}\right\}=0 .
 \end{aligned}
\]
Uniformly through \(r=\nu/j\to\infty\), including bounded \(j\),
\[
 |\log\mathcal S(\nu)|\le C|\nu|^{-1}\le Cj^{-1},
 \qquad |\nu p(\nu)|\le C|\nu|^{-1}.
\]
Analyticity covers the remaining compact region, proving
\(\mathcal F_0\in\mathcal A_0\). For the finite expansion,
\[
 P_m=\sum_{h\le m}\varepsilon^h\eta_h,\quad\eta_h\in\mathcal B_h,
 \qquad |P_m^+-P_m|\le\frac{C_m|\varepsilon|L_\nu^m}{j|\nu|}.
\]
Use the identity
\[
 \mathcal F_{P_m}-\mathcal F_0
 =DP_{m,\nu}-\varepsilon^2P_{m,\varepsilon}
   +\frac{b\nu}{\varepsilon}
       e^{\mathcal E}\frac{\mathcal S(n_1)}{\mathcal S(\nu)}
                         (e^{P_m^+-P_m}-1)
\]
and the bound \(L_\nu^m/|\nu|\le C_m/j\). On a fixed smaller disk,
\[
 \mathcal F_{P_m}-\mathcal F_0\in\mathcal A_m,\qquad
 f_{m-1}=[\varepsilon^{m-1}]\mathcal F_{P_{m-1}},\qquad
 \eta_m=\mathcal R_b f_{m-1}.
\]
The diagonal identity gives the exact cancellation
\[
 \begin{aligned}
 [\varepsilon^{m-1}]\mathcal F_{P_m}
 &=f_{m-1}+b\nu\{\eta_m(j+1,\nu+1)-\eta_m(j,\nu)\}\\
 &=f_{m-1}-f_{m-1}=0 .
 \end{aligned}
\]
After \(M\) such cancellations, the finite Cauchy--Taylor remainder
has order \(\varepsilon^M\) in \(\mathcal A_M\), proving
\eqref{eq:full-ratio-Gamma-symbol}.
\end{proof}

\subsubsection{Restoring the physical Bernoulli row}

Fix \(c\in\{-1/2,1/2\}\). Write \(z=b-\Delta\),
\(e=j/k\), \(x=1-e\), \(q=h_t'(x)\), \(w=1-u_z(x)\),
\(\nu=kw/(bq)\), and \(d=\nu-j/(bq)\).

Take the finite corrector \(P_M\) from the preceding construction.
The negative-order bounds required through order three are
\[
 |(\nu\partial_\nu)^{a_\nu}(j\partial_j)^{a_j}
           (q\partial_q)^{a_q}P_M|\leq C_M|\nu|^{-1},
 \qquad a_\nu+a_j+a_q\leq3.
\]
They hold on its physical window. Define
\[
 \begin{aligned}
 W_{k,z}(s)&=k^{-\kappa(z)}A_{z,c}(s/k)e^{kg_z(s/k)},\\
 V_l^c(s)&=\frac l{l+c}h_t(s/l),\qquad
 p_t(\xi)=t^\xi(1-t)^{1-\xi},\\
 (\mathscr P_z U)_k(s)&=
 \sum_{\xi=0}^1p_t(\xi)e^{zV_{k+1}^c(s+\xi)}U_{k+1,z}(s+\xi),\\
 U_{k,z}(s)&=W_{k,z}(s)\mathcal S(\nu)e^{P_M},\qquad
 \mathcal R_M^{\rm phys}=\frac{(\mathscr P_zU)_k(s)}{U_{k,z}(s)}.
 \end{aligned}
\]
Its reduced defect is \(\mathcal F_{P_M}/k\).

Put \(A_{kj\nu}=(kj)^{-1}+q/(k|\nu|)\), \(B_k=q^2/k^2\),
\(D_w=|\partial_z w_z(q)|\), and
\(\mathcal E=\mathcal R_M^{\rm phys}-1-\mathcal F_{P_M}/k\).

\begin{lemma}[Physical restoration with one frequency mark]
\label{lem:full-physical-row-restoration}
On the physical branch and its unit-step interpolations,
assume \(j\ge1\), \(q\ge q_0\), \(\Re\nu\ge c_0j\), and
\(|w|\le\epsilon_0\), with \(q_0\) large and
\(\epsilon_0\) small. Then
\begin{equation}
 \left|\mathcal R_M^{\rm phys}-1-\frac{\mathcal F_{P_M}}k\right|
 \leq C_M\left\{\frac{|\Delta|}{kj}
       +\frac{q|\Delta|}{k|\nu|}+\frac{q^2}{k^2}\right\}.
 \label{eq:full-physical-restoration}
\end{equation}
For \(z=\vartheta+i\omega/\Lambda\), also
\begin{equation}
 |\partial_\omega\mathcal E|
 \leq\frac{C_M}{\Lambda}
 \left\{A_{kj\nu}+B_k+
 \frac{D_w}{|w|}\bigl(|\Delta|A_{kj\nu}+B_k\bigr)\right\}.
 \label{eq:full-physical-frequency-mark}
\end{equation}
Both bounds are uniform in \(|\nu|/j\).
\end{lemma}

The domain gives \(|d|\asymp|\nu|\) and \(eq\le C|w|\).
These are row estimates; occupation and Fourier weights are
applied in the subsequent continuation argument.

\begin{proof}
Interpolate a true failure by \(k_s=k+s\), \(j_s=j+s\), \(0\leq s\leq1\).
Since \(k_s-j_s=k-j\), the exact logit and the endpoint ODE give
\begin{equation*}
 \begin{split}
 q_s&=q-\log(1+s/j),\\
 n_s'&=\frac zb(1-e_s)(1-w_s)\frac{n_s}{d_s}
          +\frac{n_s}{j_sq_s}+\frac{n_s}{k_s},
 \qquad
 n_s=\frac{k_sw_z(x_s)}{bq_s},\quad
 d_s=n_s-\frac{j_s}{bq_s}.
 \end{split}
\end{equation*}
For \(d_s(n)=n-j_s/(bq_s)\) and \(w_s(n)=bq_sn/k_s\), define
\[
 \begin{aligned}
 F_{\rm phys}(s,n)
 &=\frac zb(1-e_s)(1-w_s(n))\frac n{d_s(n)}
                         +\frac n{j_sq_s}+\frac n{k_s},\\
 F_{\rm red}(s,n)&=\frac n{d_s(n)}+\frac n{j_sq_s}.
 \end{aligned}
\]
Let \(\bar n_s'=F_{\rm red}(s,\bar n_s)\), \(\bar n_0=\nu\).
In the separated cone these functions satisfy
\[
 |F_{\rm phys}-F_{\rm red}|
 \le C(|\Delta|+e_s+|w_s|),\qquad
 |\partial_nF_{\rm red}|
 \le C\{(j_sq_s)^{-1}+j_s/(q_s|n|^2)\}.
\]
Gronwall's inequality on \(0\le s\le1\), using
\(n_s/k_s=w_s/(bq_s)\), gives
\[
 \sup_{s\le1}|n_s-\bar n_s|
       \le C(|\Delta|+|w|+eq).
\]
Let \(\ell=\log A_z\). The exact HJB and amplitude transport equations give
\[
 \frac{g_z''}{k}=\frac{z/b}{d},\qquad
 \frac{\ell'}k=\frac zb\,\frac{(1-w)\nu}{2d^2}
       +\frac{-\kappa(z)-zc h_t(x)}{bqd}-\frac{zq}{2k}.
\]
Combine the full nonlinear entropy difference with the amplitude before
expanding the row. For \(\widetilde\ell=\ell+zh_t\) this yields the exact
identity
\[
 \widetilde A:=\frac{\widetilde\ell'}k
 =\frac zb\,\frac{\nu}{2d^2}
  +\frac{-\kappa(z)-zc h_t(x)}{bqd}
  -\frac{zq}{2k}\left\{\left(\frac{\nu}{d}\right)^2-1\right\}.
\]
Since \(|(\nu/d)^2-1|\leq Cj/(q|\nu|)\), comparison with
\[
 A_{\rm red}=\frac{\nu}{2d^2}
                +\frac{\alpha_c-1/2}{bqd},
 \qquad \alpha_c=\frac12-\kappa(b)-bc h_t(1),
\]
therefore gives
\[
 |\widetilde A-A_{\rm red}|
 \le C\frac{|\Delta|+e}{|\nu|}.
\]
The exact failure exponent is
\[
 E_f=\int_0^1(1-s)\frac{x_s^2(z/b)}{d_s}\,ds
       -\int_0^1x_s\widetilde A_s\,ds
       -\kappa(z)\log(1+1/k)
       -\frac{zc h_t(x_1)}{k+1+c}.
\]
Subtract its initial integrands, and do the same for the reduced exponent.
The two initial linear row terms cancel before taking absolute values. The
true terms are
\[
 B_1=-\kappa-zch+e(\ell'+zq)+g''e^2/2,\qquad
 B_0=-\kappa-zch-x(\ell'+zq)+g''x^2/2.
\]
Their weighted sum \((1-w)B_1+wB_0=0\) is the amplitude transport equation.
The reduced terms satisfy
\[
 B_{\rm red}+kw\left\{\frac1{2d}-A_{\rm red}\right\}=0,
 \qquad B_{\rm red}=\alpha_c-\frac12+jA_{\rm red}.
\]
The second identity follows from \(kw=bq\nu\) and \(d=\nu-j/(bq)\).
Write
\[
 \begin{aligned}
 K_{\rm phys}(s)&=x_s^2(z/b)/d_s,&
 M_{\rm phys}(s)&=x_s\widetilde A_s,\\
 \bar d_s&=\bar n_s-j_s/(bq_s),&
 K_{\rm red}(s)&=\bar d_s^{-1},\\
 M_{\rm red}(s)&=\frac{\bar n_s}{2\bar d_s^2}
             +\frac{\alpha_c-1/2}{bq_s\bar d_s}.
 \end{aligned}
\]
After the preceding initial-term cancellation, the differentiated
integrands satisfy
\[
 \begin{aligned}
 |\partial_s(K_{\rm phys}-K_{\rm red})|
 +|\partial_s(M_{\rm phys}-M_{\rm red})|
 &\le C\left\{\frac{|\Delta|}{jq|\nu|}
       +\frac{|\Delta|}{|\nu|^2}+\frac q{k|\nu|}\right\}.
 \end{aligned}
\]
Indeed, differentiating \(\Delta/d\) costs
\(O((jq)^{-1}+|\nu|^{-1})\), differentiating \(e/|\nu|\)
costs \(O((k|\nu|)^{-1})\), and the flow discrepancy is multiplied
by \(|\nu|^{-2}\), where \(w/|\nu|^2=O(q/(k|\nu|))\).
The failure weight then gives the required row orders directly:
\[
 \frac{bq|\nu|}{k}
 \left\{\frac{|\Delta|}{jq|\nu|}
       +\frac{|\Delta|}{|\nu|^2}+\frac q{k|\nu|}\right\}
 \le C\left\{\frac{|\Delta|}{kj}
       +\frac{q|\Delta|}{k|\nu|}+\frac{q^2}{k^2}\right\}.
\]
The time and \(c\)-potential differences contribute \(O(q/k^2)\).
Both failure exponents are \(O(|\nu|^{-1})\), with difference
\(O((|\Delta|+e)/|\nu|+k^{-1})\), so the inequality
\[
 |(e^u-1-u)-(e^v-1-v)|
 \le C(|u|+|v|)|u-v|
\]
applies to the actual failure exponent \(E_f\) and its reduced
version \(\bar E_f=\int_0^1\{(1-s)K_{\rm red}(s)-M_{\rm red}(s)\}\,ds\).
For success put
\[
 E_s=\log\frac{t\,e^{zV_{k+1}^c(s+1)}W_{k+1,z}(s+1)}
 {(1-w)W_{k,z}(s)}.
\]
The two quadratic contributions therefore satisfy
\[
 \begin{aligned}
 |w\{(e^{E_f}-1-E_f)-(e^{\bar E_f}-1-\bar E_f)\}|
 &\le C\left\{\frac{q|\Delta|}{k|\nu|}+\frac q{k^2}\right\},\\
 |(1-w)(e^{E_s}-1-E_s)|&\le Cq^2/k^2 .
 \end{aligned}
\]
The differentiated Gamma and \(P_M\) increments have the same orders;
before multiplication by \(|w|\), their failure-end discrepancy is
\[
 O\!\left(\frac{|\Delta|+|w|+e}{|\nu|^2}\right).
\]
Summation proves \eqref{eq:full-physical-restoration}.
Differentiate with \(q,k,j,\nu\) independent to obtain
\[
 |\partial_z\mathcal E|_{\nu\ {\rm fixed}}\leq C_M(A_{kj\nu}+B_k),
 \qquad
 |\nu\partial_\nu\mathcal E|
       \leq C_M(|\Delta|A_{kj\nu}+B_k).
\]
The weighted derivative of each separated denominator has the same
order as that denominator. Since
\(\partial_z\nu/\nu=\partial_z w/w\), the physical chain rule is
\[
 \partial_\omega\mathcal E
 =\frac i\Lambda\left[
   (\partial_z\mathcal E)_{\nu\ {\rm fixed}}
   +\frac{\partial_z w}{w}\,\nu\partial_\nu\mathcal E\right].
\]
The two preceding bounds give
\eqref{eq:full-physical-frequency-mark}.
\end{proof}

\subsubsection{Actual continuations, finite terminals and short initial runs}

Let \(\xi_l\in\{0,1\}\), \(S_l=\sum_{i\le l}\xi_i\),
\(J_l=l-S_l\), and \(\mathscr F_l=\sigma(\xi_1,\ldots,\xi_l)\).
For \(k<L\) define
\[
 V_l^c(S_l)=\frac l{l+c}h_t(S_l/l),\qquad
 T_{k,L}^c=\sum_{l=k+1}^{L}V_l^c(S_l),\qquad
 p_t(\xi)=t^\xi(1-t)^{1-\xi}.
\]
For a state \((k,s)\), \(H_{k,z}(s)\) is the normalized continuation
coefficient with \(T_{k,L}^c\) as future potential and the past excluded.
It is continued from the real coefficient of
Lemma~\ref{lem:analytic-boundary-coefficient}.
Write \(j=k-s\). Its exact equation is
\[
 H_{k,z}(s)=\sum_{\xi=0}^1p_t(\xi)
 e^{z(k+1)h_t((s+\xi)/(k+1))/(k+1+c)}H_{k+1,z}(s+\xi).
\]
The analogous finite continuation $H^G_{k,z}$ uses at time $m$ the terminal
$m^{-\kappa(z)}e^{a(z)Z_m^2/2} \mathbf1_{\{|Z_m|\le B\sqrt N\}}$.
The normalizing function is the \(W_{k,z}\) defined above.

For real \(\theta<b\), define the tilted transition
\[
 p^H_{k,\theta}(\xi\mid s)
 =\frac{p_t(\xi)e^{\theta V_{k+1}^c(s+\xi)}
                  H_{k+1,\theta}(s+\xi)}
        {H_{k,\theta}(s)},\qquad \xi\in\{0,1\}.
\]
The harmonic equation makes its two probabilities sum to one.
We write \(\mathbb P_{H_\theta}\), equivalently \(\mathbb P_\theta^H\),
for this chain, with expectation \(\mathbb E_{H_\theta}\);
in this continuation proof \(\mathbb E_\theta=\mathbb E_{H_\theta}\).
Set \(p_k=p^H_{k,\theta}(0\mid S_k)\).
The row operator replaces \(H\) in the numerator by its argument and
divides the resulting sum by that argument at \((k,s)\).

For fixed \(D>0\), the continuation argument uses
\[
 \begin{gathered}
 \theta=b-c_0/\Lambda,\qquad z=\theta+i\omega/\Lambda,\qquad
 \omega\in\mathbb R,\\
 |\omega|\le\Lambda\sqrt{D\log N/N}.
 \end{gathered}
\]
At an interior physical state \(S_k\) of the \(H_\theta\) chain,
the active minority count satisfies \(j=k-S_k\ge1\).

For fixed \(A,A_0<\infty\), the finite-terminal range is
\(k\le N^A\), \(b-\theta\ge cN^{-A_0}\), and \(\Lambda\le CN\).
Choose \(K_m\) for the required polynomial precision;
\(\varepsilon_m\) denotes its selectable negative power of \(N\).
The constant \(c_B\) can be made arbitrarily large.

The separated large-count range is
\(q\le Q_\Delta+O(\log Q_\Delta)\),
\(|\delta(\Delta)|/(e(q)q)\le\varepsilon_*\), and \(J\) large.

\begin{lemma}[Uniform continuations and a separated terminal interface]
\label{lem:full-uniform-continuation-terminal}
Uniformly over interior physical states,
\(C^{-1}W_{k,\theta}\le H_{k,\theta}\le CW_{k,\theta}\), and
\begin{equation}\label{eq:full-complex-continuation-bound}
 |H_{k,z}|\le C k^{\epsilon_z}W_{k,\theta},\qquad
 \epsilon_z=\kappa(\theta)-\Re\kappa(z)=O((\Im z)^2).
\end{equation}
On the stated finite-terminal range,
\begin{align}
 \frac{|H^G_{k,z}-H_{k,z}|}{H_{k,\theta}}
 &\le C\{\varepsilon_m+e^{-c_BN}\},\nonumber\\
 \frac{|\partial_\omega(H^G_{k,z}-H_{k,z})|}{H_{k,\theta}}
 &\le C\left(1+\frac{kq^{1/b}}\Lambda\right)
                   \{\varepsilon_m+e^{-c_BN}\},
 \qquad z=\theta+i\omega/\Lambda,\label{eq:full-finite-terminal-value-mark}
\end{align}
On the separated large-count range,
\begin{equation}\label{eq:full-large-count-terminal}
 \frac{H_{k,z}}{W_{k,z}}
  =\sum_{r=0}^{\lfloor\varepsilon J\rfloor}
               \frac{f_r(q,z)}{J^r}+O(e^{-cJ}),\qquad f_0=1.
\end{equation}
\end{lemma}

The last error assertion concerns true physical integer states.
Complex parameter and state calculations use its finite analytic
sum.

\begin{proof}
Let \(p_k\) be the failure probability under the real \(H_\theta\)
chain. Equation~\eqref{eq:core-hazard-row} gives
\(p_k\ge cqj/k\) near the active endpoint and \(p_k\ge cj/k\)
until the central chart. At a count level \(j\),
\[
 \E\!\left[J_{k+1}^{-1}\mid J_k=j\right]
   =j^{-1}-\frac{p_k}{j(j+1)},\qquad
 \E\sum_{l\ge k}\frac{p_l}{J_l^2}\le \frac Cj .
\]
The corresponding recursion for \(k^{\epsilon_z}J_k^{-1/2}\)
allows the small time weight in
\eqref{eq:full-complex-continuation-bound}. For each of the finitely
many counts below a fixed \(J_0\), first unfold the next-failure
kernel; its mass is bounded above and below, and its tail satisfies
\[
 \mathbb P_{H_\theta}
 \{J_v=j\ \forall v\in\{l,\ldots,l'\}\mid J_l=j\}
 \le C\exp\{-c[(\log l')^2-(\log l)^2]\}.
\]
Backward induction over \(1\le j<J_0\) extends
\[
 C^{-1}W_{k,\theta}\le H_{k,\theta}\le CW_{k,\theta}
\]
to every \(j\ge1\). For the complex row, use the separated denominator
and transport identity of Lemma~\ref{lem:full-complex-endpoint-geometry}.
Keep the failure weight inside the row expansion. Its defect satisfies
\[
 \left|\frac{\mathscr P_zW_{k,z}}{W_{k,z}}-1\right|
 \le C\{q/(kj)+k^{-2}\}\le Cp_k/j^2+Ck^{-2}.
\]
Equation~\eqref{eq:full-absolute-Bellman-comparison} combines with
\[
 \mathbb E_\theta\sum_{l\ge k}\frac{p_l}{J_l^2}\le C/j,
 \qquad \sum_{l\ge k}l^{-2}\le C/k .
\]
Its time-weighted version makes the absolute Duhamel sum finite.
Unfolding \(j<J_0\) proves \eqref{eq:full-complex-continuation-bound}
and analyticity. For
\[
 \theta'=(\theta+b)/2,\qquad
 H_{k,\theta'}/H_{k,\theta}\le Ck^Ce^{Ck},\qquad E_N=N^{A+A_0+1},
\]
positivity gives
\[
 \Pp_{H_\theta}\{T_{k,m}^c>E_N\}
 \le Ck^Ce^{Ck-(\theta'-\theta)E_N}
 \le C e^{-cN^{A+1}}.
\]
The elementary last-half-prefix entropy bound is
\[
 \sum_{l=k+1}^{L}h_t(S_l/l)
       \ge cL|S_L/L-t|^3-Ck.
\]
It forces a central entry before a polynomial time outside that exception.
In a central chart choose \(a_1\in(a(b),1/2)\), set
\(A_l=S_l-lt\), and kill the chain on exit from a fixed larger chart.
For \(0\le v\le v_*\), with \(v_*>0\) fixed sufficiently small,
the Bernoulli conditional exponential calculation gives
\[
 \mathbb E_{H_\theta}[e^{v A_{l+1}^2}\mid A_l]
 \le\exp\{Cv+[v(1+2a_1/l)+v^2/2]A_l^2\}.
\] Solving
this Riccati recursion backwards gives a noncentrality at $m$ bounded by
$CL^{2-2a_1}m^{2a_1-1}$. Taking $K_m$ large therefore makes $\mathbb
P_{H_\theta}(|Z_m|>B\sqrt N)\le Ce^{-c_BN}$.
On the entire retained set the central transport expansion gives
\[
 \frac{H_{m,z}}{W_{m,z}}=1+O(m^{-1}),\qquad
 W_{m,z}=m^{-\kappa(z)}e^{a(z)Z_m^2/2}
       \{1+O(N^{3/2}/\sqrt m+\sqrt N/m)\}.
\]
On the complement use \eqref{eq:full-complex-continuation-bound};
the harmonic identity then proves the value bound in
\eqref{eq:full-finite-terminal-value-mark}. For its derivative take
\[
 r_z=c\min\{b-\theta,(kq^{1/b})^{-1}\}.
\]
This circle lies below the critical line and changes the initial real
action by \(O(1)\). Cauchy's formula, with that bounded reference
comparison, gives
\[
 \frac{|\partial_\omega(H^G_{k,z}-H_{k,z})|}{H_{k,\theta}}
 \le \frac{C}{\Lambda r_z}
       \{\varepsilon_m+e^{-c_BN}\}
 \le C\left(1+\frac{kq^{1/b}}{\Lambda}\right)
       \{\varepsilon_m+e^{-c_BN}\}.
\]
The analytic terminal at the count boundary is constructed from the
stated physical parameter range. The endpoint asymptotic gives
\[
 |\delta(\Delta)|\ge c|\Delta|Q_\Delta^{1/b},\qquad
 q\le C Q_\Delta,\qquad
 \frac{|\delta(\Delta)|}{e(q)q}\le\varepsilon_*
 \ \Longrightarrow\
 |\Delta|\le C\varepsilon_*e(q)q^{1-1/b}.
\]
Let \(Q_*\) be the largest initial real logit. Since
\(e(Q_*+4)\asymp e(Q_*)\), reducing \(\varepsilon_*\) by a fixed
factor places this same physical family in
\[
 |z-b|<R_{Q_*+4},\qquad R_q=c e(q)q^{1-1/b}.
\]
Thus the analytic disk covers the entire capped logit tube without
changing the lemma's physical range.
The endpoint ODE gives $|w_z-w_b|\le C|z-b|q^{1/b}$ and the integrated
characteristic perturbation
\[
 \int |a_z-a_b|\,d\tau
 \le C|z-b|e(q)^{-1}q^{1/b-2}+C|z-b|\log q=O(c/q),
 \quad a_z=\frac{w_z-e}{(1-e)w_z}.
\]
The exponential factor \(e(q)^{-1}\) localizes the perturbation at the
largest \(q\); thus the smaller disk retains the pole and inward flow.
Reduce \(\varepsilon_*\) once for the strip through \(q+4\).
With \(\epsilon_j=1/j\), the exact shifts are
\[
 \begin{aligned}
 \xi=0:&\quad
 (\epsilon_j,q)\mapsto
 \left(\frac{\epsilon_j}{1+\epsilon_j},
                  q-\log(1+\epsilon_j)\right),\\
 \xi=1:&\quad
 (\epsilon_j,q)\mapsto
 \left(\epsilon_j,q+\log\{1+e\epsilon_j/(1-e)\}\right).
 \end{aligned}
\]
For an analytic \(F(\epsilon_j,q)\), define the conjugated defect
\[
 \mathscr D_zF=\frac{\mathscr P_z(WF)}{W}-F,\qquad
 T_{v,r}f=[\epsilon_j^{r+v}]
                 \{w_z(q)^{-1}\mathscr D_z(\epsilon_j^rf(q))\}.
\]
Thus \(T_{1,r}f=-a_zf'-rf\).
Let \(\phi_\tau(q)\) solve
\[
 \partial_\tau\phi_\tau(q)=-a_z(\phi_\tau(q)),\qquad \phi_0(q)=q.
\]
Choose a fixed small \(\rho_0>0\). Pole separation gives analyticity
of \(a_z\) and the conjugated row weights on the outer tube
\(\operatorname{dist}(q,[0,Q_*+4])<2\rho_0\), with uniform row bounds.
For the flow take the inner tube
\[
 \Omega_0=\{q:\operatorname{dist}(q,[0,Q_*+4])<\rho_0\}.
\]
The outer margin keeps the row's coefficient circles uniformly inside
its analytic domain, including near \(\partial\Omega_0\).
On any compact disk in it, Picard iteration gives
\[
 \phi^{(0)}_\tau=q,\qquad
 \phi^{(m+1)}_\tau=q-\int_0^\tau a_z(\phi^{(m)}_v)\,dv,\qquad
 \|\phi^{(m+1)}-\phi^{(m)}\|_{[0,T]}
       \le M L^mT^{m+1}/(m+1)! .
\]
Here \(M=\sup|a_z|\), \(L=\sup|a_z'|\), and \(T\) is chosen to keep
the iterates in that disk. The uniform analytic limit gives joint analytic
dependence on the initial point and parameter
\citep[Theorems~4.1--4.2]{Teschl2012}.
For large real \(q\), \(c\le|a_z(q)|\le C\); on a fixed central disk,
\[
 a_z(q)=\{1-a(z)\}q+O(q^2),\qquad \Re\{1-a(z)\}\ge c>0.
\]
The real critical flow enters that disk. Its intermediate compact
segment has bounded duration and derivative. The preceding integrated
perturbation estimate gives
\[
 \int_{\phi_\tau(q)\notin\{|q|\le\rho_0/4\}}
       |a_z-a_b|\,d\tau
 \le C\varepsilon_*+C|\Delta|\log(2+Q_*) .
\]
A sufficiently small central disk is invariant, since
\[
 \Re\{\bar q\,a_z(q)\}\ge c|q|^2,\qquad
 |\phi_\tau(q)|\le Ce^{-c\tau}|q|.
\]
Outside that disk and inside \(\Omega_0\), pole separation gives
\(c_{\rho_0}\le|a_z(q)|\le C\). For every starting point in
\(\Omega_0\), up to its first exit, the scalar derivative satisfies
\[
 \partial_q\phi_\tau(q)=\frac{a_z(\phi_\tau(q))}{a_z(q)},\qquad
 |\partial_q\phi_\tau(q)|\le C.
\]
For starts in the central disk, its analytic contraction gives the same
bound, with continuous value \(e^{-\{1-a(z)\}\tau}\) at zero.
These bounds also hold for transitions from intermediate times.
Variation of constants with these bounded transition derivatives gives
\[
 \sup_{\tau\le\tau_{\rm cen}}
 |\phi_{\tau,z}(q)-\phi_{\tau,b}(q)|
 \le C\int_0^{\tau_{\rm cen}}
       |a_z-a_b|(\phi_{v,z}(q))\,dv
 \le C\varepsilon_*+C|\Delta|\log(2+Q_*).
\]
Here \(\tau_{\rm cen}\) is first entry into the central disk, with
both paths stopped at first exit from \(\Omega_0\). Choose an initial
tube width \(\rho_1>0\) and then \(\varepsilon_*\) so that
\[
 C\rho_1+C\varepsilon_*+C|\Delta|\log(2+Q_*)<\rho_0/2 .
\]
The real orbit stays in \([0,Q_*+4]\); the comparison and central contraction give
\[
 \operatorname{dist}(q,[0,Q_*+4])\le\rho_1
 \ \Longrightarrow\
 \inf_{\tau\ge0}\operatorname{dist}(\phi_{\tau,z}(q),\Omega_0^c)
       \ge\rho_0/2 .
\]

Let \(\tau_{\rm out}(q)\) be first exit from \(\Omega_0\).
For paths with \(\tau_{\rm out}(q)=\infty\), put
\[
 D_z(q)=\inf_{\tau\ge0}
       \operatorname{dist}(\phi_{\tau,z}(q),\Omega_0^c),\qquad
 \Omega_z=\{q:\tau_{\rm out}(q)=\infty,\ D_z(q)>0\}.
\]
The narrower tube surrounds the full interval \([0,Q_*+4]\).
It therefore contains every pre-exit physical logit and lies in
\(\Omega_z\), with \(D_z\ge d_\Omega>0\).
For \(q\in\Omega_z\), put
\(T_h=\inf_{0\le t\le1}\tau_{\rm out}(q+th)\).
The before-exit derivative bound gives
\[
 \sup_{\tau<T_h}\sup_{0\le t\le1}
 |\phi_{\tau,z}(q+th)-\phi_{\tau,z}(q)|
       \le C|h|,\qquad |h|\le D_z(q)/(2C).
\]
If \(T_h<\infty\), the paths would still be at distance at least
\(D_z(q)/2\) from the boundary, a contradiction. Hence \(T_h=\infty\),
all these starts are in \(\Omega_z\), and the semigroup property implies
\[
 D_z(q+h)\ge D_z(q)-C|h|,\qquad
 D_z(\phi_{s,z}(q))\ge D_z(q).
\]
Abbreviate \(D=D_z,\ \Omega=\Omega_z\) and take \(p\ge0\).
Cauchy's formula
\citep[Chapter~4, Section~2.3]{Ahlfors1979} on
\(|h|=D(q)/\{2C(p+1)\}\) yields
\[
 \|f\|_p=\sup_{q\in\Omega}D(q)^p|f(q)|,\qquad
 \|f'\|_{p+1}\le C(p+1)\|f\|_p .
\]
To check the coefficient bound uniformly in \(Q_*\), write
\[
 B_\xi=\frac{p_t(\xi)e^{zV_{k+1}^c(s+\xi)}
                         W_{k+1,z}(s+\xi)}{W_{k,z}(s)},\qquad
 B_s=B_1,\quad B_f=B_0 .
\]
As analytic functions of \(\epsilon_j\),
\[
 B_s(0)=1-w,\quad B_f(0)=w,\quad
 |B_f/w|\le C,\quad
 |B_s-(1-w)|/|w|\le C|\epsilon_j| .
\]
With the exact shifted logits \(q_s,q_f\) displayed above,
\[
 \begin{aligned}
 \frac{\mathscr D_z(\epsilon_j^rf)}{w\epsilon_j^r}
 &=\frac{B_s}{w}\{f(q_s)-f(q)\}\\
 &\quad+\frac{B_f}{w}
       \{(1+\epsilon_j)^{-r}f(q_f)-f(q)\}\\
 &\quad+\frac{B_s+B_f-1}{w}f(q),\\
 |q_s-q|&\le C|e||\epsilon_j|,\qquad
 |q_f-q|\le C|\epsilon_j|,\qquad |e|/|w|\le C .
 \end{aligned}
\]
Thus the large coefficient \(B_s/w\) multiplies a difference containing
\(e\). On \(|\epsilon_j|=cD(q)/(r+p+v)\), the shifted distances are at
least \(D(q)\{1-Cc/(r+p+v)\}\); their \(p\)-powers cost a bounded factor.
Cauchy extraction therefore gives
\[
 \|T_{v,r}f\|_{p+2(v-1)}
       \le K^v(r+p+v)^v\|f\|_p,\qquad v\ge2,
\]
with \(K\) independent of \(Q_*\).
For the operator \(a_z\partial_q+r\), the inverse is
\[
 \mathscr I_rF(q)=\int_0^\infty e^{-r\tau}F(\phi_\tau(q))\,d\tau,\qquad
 \|\mathscr I_rF\|_p
 \le\int_0^\infty e^{-r\tau}\|F\|_p\,d\tau
 =r^{-1}\|F\|_p .
\]
Uniqueness holds in each finite weighted norm. If
\((a_z\partial_q+r)g=0\) and \(\|g\|_p<\infty\), then
\[
 |g(q)|=e^{-r\tau}|g(\phi_\tau(q))|
       \le e^{-r\tau}D_z(q)^{-p}\|g\|_p
       \longrightarrow0 .
\]
Induction on \(r\) makes the constructions with different \(Q_*\)
agree on their overlaps.
Set \(f_0=1\). At expansion order \(r\ge1\), cancellation of
\(\epsilon_j^{r+1}\) gives
\[
 (a_z\partial_q+r)f_r
 =\sum_{\ell=0}^{r-1}T_{r+1-\ell,\ell}f_\ell,\qquad
 f_r=\mathscr I_r
       \left(\sum_{\ell=0}^{r-1}T_{r+1-\ell,\ell}f_\ell\right).
\]
Each summand has norm index \(2\ell+2\{r-\ell\}=2r\).
If \(\|f_\ell\|_{2\ell}\le B^\ell\ell!\) for \(\ell<r\), then
\[
 \begin{aligned}
 \|f_r\|_{2r}
 &\le\frac1r\sum_{\ell=0}^{r-1}
 K^{r+1-\ell}(r+2\ell+1)^{r+1-\ell}B^\ell\ell!\\
 &\le B^rr!\sum_{d=1}^{r}CK(CKe/B)^d
 \le B^rr!,
 \qquad \frac{r^d(r-d)!}{r!}\le e^d .
 \end{aligned}
\]
Choose \(B\) so the geometric sum is at most one. In the narrower tube,
\[
 |f_r(q)|\le(B/d_\Omega^2)^rr! .
\]
Put \(\widetilde B=B/d_\Omega^2\), \(R=\lfloor\varepsilon J\rfloor\), and
\(\rho_R=c d_\Omega/(R+1)\), choosing \(c d_\Omega\widetilde B<1/2\).
For each \(r\le R\), use the common circle
\(|\epsilon_j|=\rho_R\) in the exact analytic row.
After cancellation through order \(R+1\), the complete Cauchy--Taylor
tails of its \(R+1\) terms satisfy, for \(j\ge J\),
\[
 \begin{aligned}
 \left|w^{-1}\mathscr D_z
       \left(\sum_{r=0}^{R}j^{-r}f_r\right)\right|
 &\le C(R+1)^C\left(\frac{1/j}{\rho_R}\right)^{R+2}\\
 &\le C(R+1)^C(C\varepsilon)^{R+2}(J/j)^{R+2}
 \le e^{-cJ}(J/j)^2 .
 \end{aligned}
\]
This finite Taylor estimate controls the truncated Gevrey expansion.
In this separated large-count disk,
\[
 p_l\asymp w_{\theta,l},\qquad |w_{z,l}|\le Cw_{\theta,l},
 \qquad w_{\theta,l}=1-u_\theta(S_l/l).
\]
The harmonic occupation bound therefore yields
\[
 \mathbb E_{H_\theta}\sum_{l\ge k}
 p_l e^{-cJ}(J/J_l)^2
 \le e^{-cJ}J^2\,\frac C J
 \le Ce^{-c'J}.
\]
Thus the row error propagates to the continuation with exponential
accuracy.
The upper-\(q\) exit costs \(e^{-cqJ}\), and the lower exit lies
in the central analytic chart. The second parameter variation gives
\[
 |\partial_z^2g_z(x)|\le Ce^{-1}q^{2/b-2},\qquad
 k\{g_{\Re z}(x)-\Re g_z(x)\}
 \le Ck|\Im z|^2e^{-1}q^{2/b-2}
 \le CJ\varepsilon_*^2 .
\]
Choosing \(\varepsilon_*\) after the constants above leaves
\(e^{CJ\varepsilon_*^2}e^{-cJ}\le e^{-cJ/2}\).
This proves \eqref{eq:full-large-count-terminal} for true physical
integer states.
\end{proof}

\begin{lemma}[Short initial states under the tilted law]
\label{lem:full-short-continuation-moment}
Uniformly in the real near-critical parameter, for every initial
first-failure state with length $k$,
\begin{equation}\label{eq:full-short-continuation-moment}
 \mathbb E_{H_\theta}[T_{k,m}^c+Z_m^2]
       \le C(k^2+\log m).
\end{equation}
If $k^2+\log m=o(N)$, the same bound holds under the retained finite $G$
continuation and
\begin{equation}\label{eq:full-short-direct-mark}
 \frac{|\partial_\omega H^G_{k,z}|}{H^G_{k,\theta}}
       \le \frac{C(k^2+\log m)}\Lambda,
       \qquad z=\theta+i\omega/\Lambda.
\end{equation}
\end{lemma}

\begin{proof}
Choose a fixed \(0<\rho<\min(t,1-t)/4\). The real transition satisfies
\[
 \begin{aligned}
 p^H_{k,\theta}(1\mid kx)
 &=u_\theta(x)+O(k^{-1}),&&\rho\le x\le1-\rho,\\
 p^H_{k,\theta}(0\mid k-j)
 &\ge cqj/k,&&1-\rho<x<1,\\
 p^H_{k,\theta}(1\mid kx)&\ge c,&&0\le x<\rho .
 \end{aligned}
\]
including the pure-zero state. Choose smooth \(f>0\) off \(t\) with
\[
 f(t)=0,\qquad
 f(x)\asymp(x-t)^2\ (x\to t),\qquad
 f(x)=\frac{A_+}{1-x}\ (x\to1),\qquad
 f(x)=\frac{A_-}{x}\ (x\to0).
\]
The constants and interpolation can be chosen so that
\[
 f+(u_\theta-x)f'\le-2h_t(x).
\]
The logarithmic derivative criterion is
\[
 \frac{(x-u_b)f'}f>1+\epsilon,\qquad
 2\{1-a(b)\}>1,\qquad bq\to\infty .
\]
It holds near the center and active endpoint; interpolate where
\(x-u_b\ne0\). Choose a smooth \(\chi_\rho:[0,1]\to[0,1]\) with
\(\chi_\rho=1\) on \([0,\rho/2]\) and \(\chi_\rho=0\) on \([\rho,1]\).
The regularized Lyapunov function is
\[
 V_k=\chi_\rho(x)\frac{A_-k^2}{S_k+1}
             +(1-\chi_\rho(x))kf(x),\qquad x=S_k/k.
\]
Choose \(f(x)=A_- /x\) on \(0<x\le\rho\) and
\(f(x)=A_+/(1-x)\) on \(1-\rho\le x<1\); thus
\(V_k=A_+k^2/j\) on the upper interval.
At the upper endpoint,
\[
 \frac{\mathbb E[V_{k+1}\mid j]}{V_k}
 =(1+1/k)^2\{1-p_k/(j+1)\}.
\]
Since \(p_k/(j+1)\ge cq/(2k)\), the upper-end drift is negative
for large \(q\). The lower-end regularization has the same recursion,
with the success count plus one and a success probability bounded
below. On the intervening compact sets, Taylor expansion gives
\[
 \E[V_{l+1}-V_l\mid S_l]\le-c h_t(S_l/l)+C/l,\qquad
 l(S_l/l-t)^2\le CV_l.
\]
At a first-failure initial state \(V_k\le Ck^2\); summation yields
\[
 c\,\E\sum_{l=k}^{m-1}h_t(S_l/l)+\E V_m
 \le Ck^2+C\sum_{l=k}^{m-1}l^{-1}
 \le C(k^2+\log m).
\]
Since the next entropy is at most \(Ch_t(S_l/l)+C/l^2\) and
\(\sup_l l/(l+c)<\infty\), the preceding sum proves
\eqref{eq:full-short-continuation-moment}. Define \(\mathbb P_{G,k,\theta}^{A}\) by weighting a Bernoulli path from
\((k,S_k)\) by
\[
 (H^G_{k,\theta})^{-1}e^{\theta T_{k,m}^c}
 m^{-\kappa(\theta)}e^{a(\theta)Z_m^2/2}
 \mathbf1_{\{|Z_m|\le B\sqrt N\}}.
\]
The terminal comparison gives
\[
 \mathbb P_{H_\theta}\{|Z_m|>B\sqrt N\}
       \le \frac{C(k^2+\log m)}{B^2N}=o(1),\qquad
 \frac{d\mathbb P_{G,k,\theta}^{A}}{d\mathbb P_{H_\theta}}
       \le C\mathbf1_{\{|Z_m|\le B\sqrt N\}} .
\]
The common normalization therefore preserves the moment bound.
Differentiate this finite expectation with the event fixed:
\[
 \frac{|\partial_\omega H^G_{k,z}|}{H^G_{k,\theta}}
 \le\frac C\Lambda
   m^{\kappa(\theta)-\Re\kappa(z)}
   \left\{\E_G(T_{k,m}^c+Z_m^2)+\log m\right\}.
\]
Here the time power is \(1+o(1)\) on the Fourier window.
The retained-law moment gives \eqref{eq:full-short-direct-mark}.
\end{proof}
\subsubsection{A finite stopped residual with a direct frequency mark}

Fix \(0<a_s<1/4\). For a first-failure length \(k_0\), write
\(q_i\asymp Q\) for its exact logit and define
\[
 U_{k,z}=W_{k,z}\mathcal S(\nu)e^{P_M},\quad
 A_0=\frac{k_0q_i^{1/b}}\Lambda,\quad
 v=\frac{k_0|\Im\delta(\Delta)|}{bq_i}.
\]
Put \(W=\Lambda\sqrt{D\log N/N}\). For \(T=2^jQ\), \(j\ge0\),
and \(\sigma\in\{-1,1\}\), define
\(\mathfrak F_{T,\sigma}=\{\omega\in[-W,W]:T\le\sigma\omega\le2T\}\).
On each nonempty such interval choose
\[
 J=\lceil Q^{K_J}(1+\sup_{\omega\in\mathfrak F_{T,\sigma}}v)\rceil+1,\quad
 k_T=\lceil k_0e^{C_T}\rceil,\quad
 \tau_J=\inf\{l\ge k_0:J_l\ge J\},\quad\tau=\tau_J\wedge k_T.
\]
Along the real \(H_\theta\) chain,
\[
 L_l(z)=e^{(z-\theta)T_{k_0,l}^c}
              \frac{U_{l,z}(S_l)}{H_{l,\theta}(S_l)}.
\]
The three terms of the finite identity are defined by
\[
 \begin{aligned}
 R_{{\rm red},k_0}(z)&=
 \mathbb E_{H_\theta}\sum_{l=k_0}^{\tau-1}
 L_l(z)\,\mathcal F_{P_M}(l,S_l)/l,\\
 R_{{\rm ph},k_0}(z)&=
 \mathbb E_{H_\theta}\sum_{l=k_0}^{\tau-1}L_l(z)\mathcal E(l,S_l),\\
 R_{{\rm term},k_0}(z)&=
 \mathbb E_{H_\theta}\left[
 e^{(z-\theta)T_{k_0,\tau}^c}
 \frac{H_{\tau,z}(S_\tau)-U_{\tau,z}(S_\tau)}
      {H_{\tau,\theta}(S_\tau)}\right],\\
 \frac{H_{k_0,z}-U_{k_0,z}}{H_{k_0,\theta}}
 &=R_{{\rm red},k_0}+R_{{\rm ph},k_0}+R_{{\rm term},k_0}.
 \end{aligned}
\]
The initial state is \((k_0,k_0-1)\). The deterministic first-run
carrier is included when these values are averaged below.

\begin{lemma}[Physical propagation of the finite corrector]
\label{lem:full-stopped-boundary-residual}
Uniformly for
\(\Lambda^{a_s}\le k_0\le\Lambda Q^{B_r}\) and
\(Q\le|\omega|\le W\), the actual continuation minus \(U_{k_0,z}\)
has the stated decomposition, with
\begin{align*}
 |R_{{\rm red},k_0}|&\le
       \frac{C_M Q^{-M-1}}{1+v},&
 |\partial_\omega R_{{\rm red},k_0}|&\le
       \frac{C_M A_0 Q^{-M-1}}{1+v},\nonumber\\
 |R_{{\rm ph},k_0}|&\le
       C_M\left\{\frac{|\Delta|}{1+v}+\frac{Q^2}{k_0}\right\},&
 |\partial_\omega R_{{\rm ph},k_0}|&\le
 C_M A_0\left\{\frac{|\Delta|}{1+v}+\frac{Q^2}{k_0}\right\}
 +\frac{C_M}{\Lambda}
       \left\{\frac1{1+v}+\frac{Q^2}{k_0}\right\}.
\end{align*}
For \(|\omega|\le\min(Q,W)\), omit \((1+v)^{-1}\).
The remaining terminal errors and their frequency marks can
be made smaller than any prescribed power of \(N\).
\end{lemma}

Choose the count cutoff power to absorb the explicit count-side
mismatch in the reduced bounds. The other terms are the
true-to-analytic count-side error, time-boundary contribution,
and finite-terminal replacement.

Fix stopping boundaries on frequency dyads. Retain their
artificial integration endpoints in the subsequent inversion.

\begin{proof}
The exact integrated endpoint ODE supplies the shared tail expansion
\[
 w_z(q)=\delta(\Delta)+z e(q)(q+1)+R_z(q),\qquad
 |R_z(q)|\le C\{e+|\delta|e(q+1)+e^2(q+1)^2\}.
\]
Indeed, subtract \(ze(q)q\) from
\[
 -w_z'=ze(1-e)q(1-w_z)\{1+e/(w_z-e)\}
\]
and integrate using the separated pole and
\(\int_q^\infty e(v)v\,dv=e(q)(q+1)+O(e(q)^2(q+1))\).
Because \(\Re z=\theta\), the shared term
\(ze(q)(q+1)\) cancels in the real-reference difference.
The exact HJB identity and \eqref{eq:full-sharp-real-gap-subtraction}
then imply
\begin{align*}
 k(g_\theta-\Re g_z)
 &=k x\{\log(1-w_\theta)-\Re\log(1-w_z)\}
                         -j\log(|w_z|/w_\theta)\\
 &\ge c v_k-j\log\{C(1+v_k/j)\}-Cj-o(v_k),\qquad
 v_k=\frac{k|\Im\delta|}{bq}.
\end{align*}
Hence, for a sufficiently small fixed $\epsilon>0$,
\begin{equation*}
 \frac{|U_{k,z}|}{H_{k,\theta}}
 \le Ce^{-cv_k}\quad(j\le\epsilon v_k),\qquad
 \frac{|U_{k,z}|}{H_{k,\theta}}\le C\quad(1\le j<k,\ q\asymp Q).
\end{equation*}
The action bound is multiplied only by bounded amplitude ratios:
\[
 |A_{z,c}|/A_{\theta,c}\le C,\qquad
 |\mathcal S(\nu)e^{P_M}|\le C .
\]
Thus it applies to either sign of \(\alpha_c\).
On the fixed dyad \(\mathfrak F_{T,\sigma}\), write
\(v_T=\sup_{\omega\in\mathfrak F_{T,\sigma}}v\).
The quantities \(J,k_T,\tau_J,\tau\) retain their definitions above.
Choose \(C_T,K_J\) after the error powers. On the rectangle,
\[
 q\asymp Q,\qquad v_k\asymp v,\qquad
 p_k\ge cQj/k,\qquad
 \mathbb E_\theta\sum_{l<\tau,\,J_l=j}p_l\le1 .
\]
Dividing the last bound by the hazard lower bound yields
\begin{equation}\label{eq:full-count-occupation}
 \mathbb E_\theta\sum_{l<\tau,\ J_l=j}\frac1l
                        \le\frac C{Qj}.
\end{equation}
The Duhamel identity follows by applying the positive $H_\theta$ chain
to $U_z/H_\theta$, retaining the complex accumulated potential in each
summand. Lemma~\ref{lem:full-ratio-finite-Gamma} and
\eqref{eq:full-count-occupation} give
\[
 |R_{{\rm red},k_0}(z)|
 \le C_M Q^{-M-1}
 \left\{e^{-cv}\sum_{j\le\epsilon v}
              \frac{\log^M(2+v/j)}{j^2}
       +\sum_{j>\epsilon v}\frac{\log^M(2+v/j)}{j^2}\right\}
 \le\frac{C_M Q^{-M-1}}{1+v}.
\]
For the physical terms, use
\[
 \frac1{kj}+\frac q{k|\nu|}\le\frac{CQ}{kj},\qquad
 \sum_{k\ge k_0}\frac{q^2}{k^2}\le\frac{CQ^2}{k_0}.
\]
The same count split gives
\[
 e^{-cv}\sum_{j\le\epsilon v}j^{-2}
       +\sum_{j>\epsilon v}j^{-2}\le C/(1+v).
\]
Together with \(\sum_{k\ge k_0}q^2/k^2\le CQ^2/k_0\), this proves
both unmarked bounds. Fix the real chain and dyad boundaries before
differentiating. Endpoint ODE and amplitude variation give
\[
 |\partial_z w_z(q)|\le Cq^{1/b},\quad
 |\partial_\omega U_z|\le CA_0|U_z|,\quad
 \frac{|\partial_z w|}{\Lambda|w|}
              \le \frac{Ckq^{1/b-1}}{\Lambda|\nu|}\le CA_0.
\]
On the fixed stopping rectangle the accumulated entropy and initial
carrier are bounded by \(Ck_0\), hence their derivatives cost at most
\(CA_0\). The symbol derivatives preserve the reduced row order.
Differentiating the finite Duhamel sum before taking absolute values
therefore gives
\[
 |\partial_\omega R_{\rm red}|
 \le CA_0\,\frac{C_MQ^{-M-1}}{1+v},
\]
and, by \eqref{eq:full-physical-frequency-mark},
\[
 |\partial_\omega R_{\rm ph}|
 \le C_M A_0\left\{\frac{|\Delta|}{1+v}+\frac{Q^2}{k_0}\right\}
 +\frac{C_M}{\Lambda}
       \left\{\frac1{1+v}+\frac{Q^2}{k_0}\right\}.
\]
The reference law and stopping boundaries remain fixed under differentiation.
At the count boundary,
\[
 \frac{|\delta|}{eq}\le CQ^{-K_J},\qquad
 \sum_{r=0}^{\lfloor\varepsilon J\rfloor}f_r(q,z)J^{-r}=1+O(J^{-1}),\qquad
 \mathcal S(\nu)e^{P_M}=1+O(J^{-1}).
\]
Their explicit mismatch, including its mark, is at most
\(C(1+A_0)/J\) by \eqref{eq:full-large-count-terminal}. Choose
\[
 K_J>M+2,\qquad
 J^{-1}\le Q^{-K_J}(1+v_T)^{-1},
\]
increasing \(K_J\) for the fixed logarithms. The remaining error
\(e^{-cJ}\) has its derivative controlled by the wholly subcritical circle
in \eqref{eq:full-finite-terminal-value-mark}.
At the time boundary, the killed inverse count satisfies
\[
 \E\!\left[J_{k+1}^{-1}\mathbf1_{\{J_{k+1}<J\}}
       \mid J_k=j\right]
 \le \left(1-\frac{cQ}{2k}\right)j^{-1}.
\]
Iteration and \(J_k^{-1}\ge J^{-1}\) before the count exit give
\[
 \mathbb P_{H_\theta}\{\tau_J>k_T\}
                  \le CJ\exp(-cQ C_T).
\]
The continuation and its finite energy derivative cost fixed powers
of \(N\). Since \(Q\asymp\log N\), select the fixed \(C_T\) so that
\[
 N^{C_1}J e^{-cQC_T}\le N^{-B_0}
\]
for the prescribed \(B_0\). Choose \(K_m\) last and apply
\eqref{eq:full-finite-terminal-value-mark} to replace the infinite
continuations by retained finite ones. These choices are compatible:
\[
 \frac{J}{k_0}\le Q^{C_2}
        \left(\frac{|\delta|}{Q}+\frac1{k_0}\right)=o(1),
 \qquad q\asymp Q.
\]
They complete the finite stopped decomposition.
\end{proof}

For \(k\ge1\), with an empty sum equal to zero, put
\[
 \begin{aligned}
 \pi_{k,c}(z)&=t^{k-1}(1-t)
 \exp\left\{z\sum_{l=1}^{k-1}\frac l{l+c}h_t(1)
                  +zV_k^c(k-1)\right\},\\
 \mu_{N,c}(\{k\})&=
 \frac{\pi_{k,c}(\theta)H_{k,\theta}(k-1)}{B_c(\theta)}
 \mathbf1_{\{k\le\Lambda Q^{B_r}\}}.
 \end{aligned}
\]
Finite real matching gives \(\sum_k\mu_{N,c}(\{k\})\le C\).
With \(y_k=\delta(c_0/\Lambda)k\) and
\(V_\omega=|\omega|/Q\ge1\),
\[
 A_0\asymp y_k,\qquad v\asymp V_\omega y_k,\qquad
 \int\frac{y_k}{1+V_\omega y_k}\,\mu_{N,c}(dk)
 \le\frac C{V_\omega}.
\]
Averaging the preceding signed mark gives
\begin{equation}\label{eq:full-averaged-reduced-mark}
 \sup|R_{\rm red}|\le C_MQ^{-M-1},\qquad
 |R_{\rm red}'(\omega)|\le
                    \frac{C_MQ^{-M}}{|\omega|}
                    \quad(Q\le|\omega|\le W).
\end{equation}
This positive normalization applies to positive, zero and negative
endpoint exponents. The physical derivative is estimated before
summing by
\[
 A_0Q^2/k_0\le CQ^{1/b+2}/\Lambda,\qquad
 \frac{A_0|\Delta|}{1+v}
 \le C\frac{|\Delta|}{V_\omega}\le CQ/\Lambda .
\]
Consequently
\begin{align}
 \sup_{|\omega|\le W}|R_{\rm ph}|&\le
 C\{\Delta_{\max}+Q^2/\Lambda^{a_s}\},\nonumber\\
 \sup_{|\omega|\le\min(Q,W)}|R_{\rm ph}|&\le
 C\{Q/\Lambda+Q^2/\Lambda^{a_s}\},\nonumber\\
 \int_{Q\le|\omega|\le W}|R_{\rm ph}'(\omega)|\,d\omega
 &\le C Q^{1/b+2}W/\Lambda=o(1),
 \qquad \Delta_{\max}=C(1+W)/\Lambda.\label{eq:full-averaged-physical-mark}
\end{align}

\subsubsection{Explicit first-run summation and the finite critical mass}

Choose a fixed smooth \(\chi:[0,\infty)\to[0,1]\) with
\(\chi=1\) on \([0,1/2]\) and \(\chi=0\) on \([1,\infty)\).
With the fixed analytic proxy at the finitely many small states, define
\[
 \mathcal K_{M,\Lambda}(z)=
 \sum_{r\ge0}\chi\!\left(\frac r{\Lambda Q^{B_r}}\right)
       \pi_{r+1,c}(z)U_{r+1,z}(r).
\]
The cutoff is independent of \(z\).
For a fixed negative exponent margin, define its matched
coefficient as
\begin{equation}\label{eq:full-negative-matched-mass}
 K_c(t,b)+\{\mathcal K_{M,\Lambda}(b-\Delta)
                                  -\mathcal K_{M,\infty}(b)\}.
\end{equation}
\begin{lemma}[Finite-proxy Watson estimate and its signed mark]
\label{lem:full-explicit-first-run-Watson}
For \(\alpha_c>0\) with a fixed margin, uniformly on the
finite physical frequency window,
\begin{equation}\label{eq:full-explicit-Watson}
 \mathcal K_{M,\Lambda}(b-\Delta)
 \sim d_c\Gamma(\alpha_c)\Delta^{-\alpha_c}
                   \{\log(1/|\Delta|)\}^{1/2}.
\end{equation}
The signed derivative is bounded by
\(C|\Delta|^{-\alpha_c-1}\{\log(1/|\Delta|)\}^{1/2}\).
Removing \(P_M\) changes each bound by the relative factor
\(C_M/\log(1/|\Delta|)\).

For \(a=\alpha_c>-1\), uniformly on compact
\(\Re\zeta>0\) sets,
\begin{equation}\label{eq:full-compensated-Watson}
 \frac{\mathcal K_{M,\Lambda}(b-\zeta/\Lambda)
             -\mathcal K_{M,\Lambda}(b-c_0/\Lambda)}
                    {\Lambda^a Q^{1/2}}
 \longrightarrow d_c\int_0^\infty
           x^{a-1}(e^{-\zeta x}-e^{-c_0x})\,dx.
\end{equation}
The signed derivative bound holds with \(a\) in place of
\(\alpha_c\). For \(|a|\log(1/|\Delta|)\le A\), including zero,
\[
 \mathcal K_{M,\Lambda}(b-\Delta)
 \sim d_c\mathcal H_a(\log(1/|\Delta|)).
\]
\end{lemma}

The compensated integral equals
\(\Gamma(a)(\zeta^{-a}-c_0^{-a})\) for \(a\ne0\), with
removable value \(\log(c_0/\zeta)\) at zero.

\begin{proof}
For the real continuous run coordinate, put
\[
 k=r+1,\quad x=r/(r+1),\quad q=\log r+H_t,\quad
 T_r^c=h_1\{r-c[\psi(r+c+1)-\psi(c+1)]\},\quad h_1=h_t(1).
\]
The explicit first-failure summand is
\[
 a_M(r,z)=t^r(1-t)e^{zT_r^c}
                 e^{zk h_t(x)/(k+c)}U_{k,z}(k-1).
\]
For integration and summation use its finite truncation
\[
 a_M^\Lambda(r,z)
   =\chi\!\left(\frac r{\Lambda Q^{B_r}}\right)a_M(r,z).
\]
This uses the same parameter-independent cutoff as
\(\mathcal K_{M,\Lambda}\); the factor identities below concern \(a_M\).
Use the exact damping $\ell_\Delta=-\log(1-\delta(\Delta))$; replacing it
globally by $\delta$ would introduce an uncontrolled $r\delta^2$ term. The
HJB identity $g_z=(1-z)h_t-\operatorname{KL}(x\|1-w_z)$ gives exactly
\begin{equation}\label{eq:full-exact-first-failure-factor}
 \begin{split}
 a_M(r,z)={}&C_z e^{-zc h_1\psi(r+c+1)}
 k^{z-\kappa(z)}A_{z,c}(x)w_z(x)(1-w_z(x))^r
 x^{-zr}e^{zk h_t(x)/(k+c)}\mathcal S(\nu)e^{P_M},\\
 C_z={}&(1-t)^z\exp\{zc h_1\psi(c+1)\}.
 \end{split}
\end{equation}
In this identity \(t^r\) cancels the linear all-success energy.
The shared \(e^{-zq}\) part of \((1-w)^r\) then cancels \(k^z\).
With \(Q_c=\alpha_c/b+1/2\), collect the remaining factors as
\begin{equation*}
 a_M(r,z)=e^{-\ell_\Delta r}r^{\alpha_c-1}q^{Q_c}
                         \mathcal B_M(r,z).
\end{equation*}
The amplitude \(\mathcal B_M\) is defined by the explicit factors in
\eqref{eq:full-exact-first-failure-factor}. For each fixed derivative order
$h$, with $D_r=|\delta|r/q$,
\begin{align}
 |(r\partial_r)^h\mathcal B_M|&\le C_h(1+D_r)^{C_h},\nonumber\\
 |(r\partial_r)^h\partial_\Delta\mathcal B_M|
      &\le C_h|\ell_\Delta'|r(1+D_r)^{C_h}.\label{eq:full-real-run-symbol}
\end{align}
For fixed \(0<\rho_1<\rho_2<\infty\),
\[
 \sup_{\rho_1\le|\ell_\Delta|r\le\rho_2}
       |\mathcal B_M-D_c|\to0 .
\]
On \(r\ge|\ell_\Delta|^{-a_0}\), \(0<a_0<1\), the value and
marked bounds for \(\mathcal B_M-\mathcal B_0\) gain
\(C_M/\log(1/|\Delta|)\). For \(Z=w_z-w_b\), the exact difference equation gives
\[
 r|Z-\delta|\le C|\delta|q+
             C D_r\log(1+C/D_r).
\]
Also $rw_b=bq+b+1+O(q^{-1})$. Therefore $F=r(w_z-\delta)-zq=O(1)$ on the
finite $q$ window. Its exact logit derivative is
\[
 F_q=F-z-zq\{(1-e)^2(1-w_z)w_z/(w_z-e)-1\}.
\]
The bracket times $q$, and every fixed positive-order derivative, are
bounded by the separated-pole equation. Retaining \(\delta\) exactly and
expanding \(w_z-\delta\) gives
\[
 r\log\{(1-w_z)/(1-\delta)\}+zq
       =-F/(1-\delta)+O(|\delta|q+q^2/r)
\]
with the same differentiated bounds. Subtracting the exact transport
integrals gives
\[
 |\log(A_z/A_b)|\le C|\Delta|q+C\log(1+CD_r).
\]
The critical amplitude and the effective count satisfy
\[
 A_b=A_{*,c}e(q)^{-1/2}q^{Q_c-1}\{1+O(q^{-1})\},
 \qquad |\nu|\asymp1+D_r.
\]
Insert these expressions and the differentiated Gamma symbols in
\eqref{eq:full-exact-first-failure-factor}. They give the first
bound in \eqref{eq:full-real-run-symbol}. For the mark, direct
variation of the endpoint equation and transport integral gives
\[
 |\partial_\Delta w|\le CQ_\Delta^{1/b},\qquad
 |\partial_\Delta\log A|\le CrQ_\Delta^{1/b-1},\qquad
 |\ell_\Delta'|\asymp Q_\Delta^{1/b}.
\]
The product rule proves the marked bound; inverse-\(\nu\) differentiation
of \(P_M\) retains \(Q_\Delta^{-1}\) for
\(\mathcal B_M-\mathcal B_0\). On compact positive \(\rho=|\ell_\Delta|r\),
\[
 D_r=O(Q_\Delta^{-1}),\qquad
 r(Z-\delta)=O(\log Q_\Delta/Q_\Delta),\qquad
 \mathcal S(1)=e/\sqrt{2\pi}.
\]
The critical action therefore gives the real normalization
\[
 D_c=\frac{bA_{*,c}}{\sqrt{2\pi}}
                  e^{bc h_1\psi(c+1)}.
\]
Choose a smooth \(\chi_T\) with
\[
 \chi_T(\rho)=0\ (\rho\le T),\qquad
 \chi_T(\rho)=1\ (\rho\ge2T),\qquad
 |\chi_T^{(h)}|\le C_hT^{-h}.
\]
Put \(L_\Delta=|\ell_\Delta|^{-\alpha_c}Q_\Delta^{Q_c}\).
Integrate \(H\) times in \(\rho=|\ell_\Delta|r\), retaining both cutoffs.
On the support of derivatives of the outer cutoff,
\[
 \left|\partial_r^h\chi(r/(\Lambda Q^{B_r}))\right|\le C_hr^{-h},
 \qquad \partial_\Delta\chi(r/(\Lambda Q^{B_r}))=0 .
\]
Thus the same finite-window symbol estimates give
\[
 \begin{aligned}
 &L_\Delta^{-1}\left|
     \int_1^\infty\chi_T(|\ell_\Delta|r)a_M^\Lambda(r,z)\,dr\right|\\
 &\quad\le C\int_T^{Q_\Delta}\rho^{\alpha_c-1-H}\,d\rho
 +C Q_\Delta^{\alpha_c-H}
       \int_1^\infty (1+y)^{C_H}e^{-cy}\,dy\\
 &\quad\le C\{T^{\alpha_c-H}+Q_\Delta^{\alpha_c-H}\}.
 \end{aligned}
\]
Here the second term uses
\(\Re\ell_\Delta/|\ell_\Delta|\ge c/Q_\Delta\) and
\(\rho=Q_\Delta y\). Choose \(H>\alpha_c\) for values and
\(H>\alpha_c+1\) after a parameter derivative, which introduces
one additional \(\rho\). Taking \(Q_\Delta\to\infty\) before
\(T\to\infty\) justifies both limits, including \(\alpha_c>1\).
The sum-to-integral error obeys the elementary Euler estimate
\[
 \left|\sum_{r\ge1}a_M^\Lambda(r,z)-\int_1^\infty a_M^\Lambda(r,z)\,dr\right|
 \le |a_M^\Lambda(1,z)|+\int_1^\infty|\partial_r a_M^\Lambda(r,z)|\,dr .
\]
On its leading scale,
\[
 L_\Delta^{-1}
 \left|\sum_{r\ge1}a_M^\Lambda(r,z)-\int_1^\infty a_M^\Lambda(r,z)\,dr\right|
 \le C|\ell_\Delta|^{\min(1,\alpha_c)}Q_\Delta^C .
\]
The marked version is exponentially small in \(Q_\Delta\) on its
marked scale. After differentiation, separate
\(r<|\ell_\Delta|^{-a_0}\) and apply the same real sensitivity bound.
The outer endpoint in the scaled coordinate satisfies
\[
 |\ell_\Delta|\Lambda Q^{B_r}\ge cQ^{B_r+1/b}\longrightarrow\infty .
\]
After the preceding tail limits, the constant-amplitude integral is
\[
 \int_0^\infty e^{-\ell_\Delta r}r^{\alpha_c-1}\,dr
       =\Gamma(\alpha_c)\ell_\Delta^{-\alpha_c},\qquad
 \frac{\ell_\Delta}{\Delta}
       =C_UQ_\Delta^{1/b}\{1+o(1)\},\qquad
 Q_c-\alpha_c/b=\frac12 .
\]
They prove \eqref{eq:full-explicit-Watson} and its signed derivative.
Repeating these estimates for \(\mathcal B_M-\mathcal B_0\)
preserves its additional \(Q_\Delta^{-1}\).
For \(a\le0\), subtraction at the real reference must precede
integration:
\[
 |e^{-\zeta x}-e^{-c_0x}|\le Cx\quad(0<x\le1),\qquad
 \int_0^1x^{a-1}x\,dx=(a+1)^{-1}<\infty .
\]
Write \(\nu_z=kw_z/(bq)\), \(\nu_b=kw_b/(bq)\), and define
\[
 L_M(r,z)=\log\frac{A_{z,c}(r/(r+1))}{A_{b,c}(r/(r+1))}
 +\log\frac{\mathcal S(\nu_z)}{\mathcal S(\nu_b)}
 +P_M(z)-P_M(b).
\]
The amplitude has the same first-order cancellation:
\[
 |Z(q)|\le C|\Delta|q^{1/b},\qquad
 r|Z|\le Cx(q/Q)^{1/b},\qquad |L_M(r,z)|\le Cx/q .
\]
Since \(a/b+1/2\) may be negative, first remove
\(r<|\ell_\Delta|^{-a_0}\). Its compensated contribution is bounded by
\[
 O\bigl(|\ell_\Delta|^{(1-a_0)(a+1)}Q^C\bigr)=o(1).
\]
Put \(r_\Lambda=\Lambda/(C_UQ^{1/b})\). After the change
\(r=r_\Lambda x\), the normalized compensated integrand is
\[
 I_{\Lambda,M}(x,\zeta)=
 \frac{r_\Lambda\chi(r_\Lambda x/(\Lambda Q^{B_r}))}
      {\Lambda^aQ^{1/2}}
 \{a_M(r_\Lambda x,b-\zeta/\Lambda)
              -a_M(r_\Lambda x,b-c_0/\Lambda)\}.
\]
On the remaining block,
\[
 q/Q\ge a_0+o(1),\qquad
 |I_{\Lambda,M}(x,\zeta)|\le Cx^a\quad(0<x\le1),\qquad
 \int_0^1x^a\,dx=(a+1)^{-1}.
\]
This proves \eqref{eq:full-compensated-Watson}.
Differentiation adds the same power \(x\), reference subtraction preserves
the correction \(Q^{-1}\), and fixed runs cost \(O(\Lambda^{-1})\).
For \(-1<a<0\),
\[
 \int_0^\infty x^{a-1}(e^{-\zeta x}-1)\,dx
       =\Gamma(a)\zeta^{-a}.
\]
The finite cutoff makes the critical tail negligible; matching its constant
proves \eqref{eq:full-negative-matched-mass}. For bounded \(v\) and
\(\|F'\|_\infty<\infty\),
\[
 |F((x+v)/\Lambda)-F(x/\Lambda)|
       \le |v|\|F'\|_\infty/\Lambda .
\]
Thus a bounded-energy analytic remainder is \(O(\Lambda^{-1})\), and
\[
 \frac{\Lambda^{-1}}{\Lambda^a Q^{1/2}}
       =\Lambda^{-a-1}Q^{-1/2}\longrightarrow0 .
\]
The comparison is on the scale $\Lambda^a Q^{1/2}$. Under bounded confluence,
\[
 |a|Q_\Delta\le A,\qquad
 q_{\rm tr}=Q_\Delta-b^{-1}\log Q_\Delta+O(1),\qquad
 e^{aq}q^{a/b+1/2}\,dq
\]
is the pretransition run weight, up to its constant.
Replacing \(q_{\rm tr}\) by \(Q_\Delta\) and \(q^{a/b}\) by one gives
\[
 \begin{aligned}
 \mathcal K_{M,\Lambda}(b-\Delta)
 &=d_c\int_0^{Q_\Delta}e^{aq}q^{1/2}\,dq
       +O(Q_\Delta^{1/2}\log Q_\Delta),\\
 \mathcal H_a(Q_\Delta)&\asymp Q_\Delta^{3/2},\qquad
 |\partial_\Delta\mathcal K_{M,\Lambda}|
       \le C|\Delta|^{-1}Q_\Delta^{1/2}.
 \end{aligned}
\]
The oscillatory portion after the transition is \(O(Q_\Delta^{1/2})\)
by the preceding smooth-tail estimate. At \(a=0\), the main integral
is \((2d_c/3)Q_\Delta^{3/2}\). These formulas keep the finite mass
and its smaller varying part separate and prove uniformity through zero.
\end{proof}
\subsubsection{The two-endpoint product and local inversion}

\begin{proof}[Proof of Theorem~\ref{thm:finite-hybrid-boundary-scalar}]
Choose cutoffs independently of \(z\):
\[
 \theta=b-c_0/\Lambda,\quad z=\theta+i\omega/\Lambda,\quad
 k_s=\lfloor\Lambda^{a_s}\rfloor,\quad 0<a_s<1/4,\quad
 R_\Lambda=\Lambda Q^{B_r}.
\]
Let \(R_1=\min\{l\ge1:\xi_l=0\}-1\), truncated at \(m\), be the
initial success-run length under the one-prefix Bernoulli law.
A run \(R_1\ge R_\Lambda/2\) forces prefix entropy at least
\(cR_\Lambda\). Positivity at
\[
 \theta'=b-c_0/(2\Lambda),\qquad
 (\theta'-\theta)cR_\Lambda\ge c'Q^{B_r}
\]
therefore gives
\[
 \frac{m^{-\kappa(\theta)}
 \mathbb E_t[e^{\theta T_m^c+a(\theta)Z_m^2/2};
 |Z_m|\le B\sqrt N,\ R_1\ge R_\Lambda/2]}{B_c(\theta)}
       \le N^C e^{-cQ^{B_r}}.
\]
The discarded energy mark costs \(N^C\). Choose
\[
 B_r>1,\qquad N^C e^{-cQ^{B_r}}\le N^{-A}
\]
for the prescribed \(A\). Normalize retained weights by
\(Z_c=B_c(\theta)\). Define the two short first-failure sums by
\[
 S_c(z)=\sum_{1\le k\le k_s}\pi_{k,c}(z)H^G_{k,z}(k-1),\qquad
 S_c^{\rm app}(z)=\sum_{1\le k\le k_s}\pi_{k,c}(z)U_{k,z}(k-1).
\]
Their matched factor is
\begin{equation}\label{eq:full-matched-explicit-factor}
 P_c(\omega)=\frac{\mathcal K_{M,\Lambda}(z)
                        +S_c(\theta)-S_c^{\rm app}(\theta)}{Z_c},
 \qquad R_c(\omega)=\frac{B_c(z)}{Z_c}-P_c(\omega).
\end{equation}
Thus every endpoint exponent retains its real small-state constant. Lemma~\ref{lem:full-short-continuation-moment} gives
\[
 \sup|S_c(z)-S_c(\theta)|/Z_c+
     \int_{-W}^{W}|\partial_\omega S_c(z)|/Z_c\,d\omega
 \le C\frac{(\Lambda^{2a_s}+Q)W}{\Lambda}=o(1).
\]
The explicit short block obeys the same estimate because its
derivative is at most \(Ckq^{1/b}\le Ck^2\) and its modulus is
bounded by the true real continuation. The error is uniformly a
negative power of \(N\), since
\[
 \frac{(\Lambda^{2a_s}+Q)W}{\Lambda}
 =(\Lambda^{2a_s}+Q)\sqrt{D\log N/N}
 \le C(N^{2a_s}+Q)\sqrt{\log N/N}\longrightarrow0.
\]
For a positive endpoint exponent the normalized short mass is at most
$C(\delta(c_0/\Lambda)k_s)^{\alpha_c}$ times a fixed logarithmic factor. Its
constant correction in \eqref{eq:full-matched-explicit-factor} still obeys
$C(1+|\omega|)^{-\alpha_c}$ on the whole window, because
\[
 k_s Q^{1/b}W/\Lambda\longrightarrow0.
\]
At zero and through confluence, respectively,
\[
 \frac{S_c(\theta)}{Z_c}\sim a_s^{3/2},\qquad
 \frac{S_c(\theta)}{Z_c}
   \sim\frac{\mathcal H_a(a_sQ)}{\mathcal H_a(Q)}
       \quad(|a|Q\le A).
\]
A fixed negative endpoint retains its nonzero constant.
These terms remain in \eqref{eq:full-matched-explicit-factor}.
Measure the oscillatory error by
\begin{equation*}
 \|R_c\|_{\rm osc}=
 Q\sup_{|\omega|\le\min(Q,W)}|R_c(\omega)|
 +\sup_{|\omega|\le W}|R_c(\omega)|
 +\int_{Q\le|\omega|\le W}|R_c'(\omega)|\,d\omega.
\end{equation*}
The integral is empty if \(W<Q\). For \(W\ge Q\), put
\[
 \mathfrak D_N=\{\pm W\}\cup
 \{\pm2^jQ:0\le j\le\lfloor\log_2(W/Q)\rfloor\};
\]
for \(W<Q\), put \(\mathfrak D_N=\{-W,W\}\). Then
\[
 \sum_{\omega\in\mathfrak D_N}|R_c(\omega)|
 \le|\mathfrak D_N|\sup_{|\omega|\le W}|R_c(\omega)|
 \le CQ\sup_{|\omega|\le W}|R_c(\omega)|.
\]
The function in \eqref{eq:full-matched-explicit-factor} stays fixed;
choose each representation before differentiation.
For \(k_s<k<R_\Lambda\), apply
\eqref{eq:full-averaged-reduced-mark} and
\eqref{eq:full-averaged-physical-mark}, with count-side corrections
and fixed \(M\ge4\):
\begin{equation}\label{eq:full-final-residual-budget}
 \|R_c\|_{\rm osc}
 \le C_MQ^{1-M}
   +C\left\{\frac{Q^2}{\Lambda}
        +\frac{Q^3}{\Lambda^{a_s}}
        +Q\Delta_{\max}
        +\frac{Q^{1/b+2}W}{\Lambda}\right\}+o(1)=o(1).
\end{equation}
Here the short-block and long-run expectations are bounded above,
and \eqref{eq:full-finite-terminal-value-mark} controls the terminal
difference. Choose \(B_r>1\) and the terminal accuracy \(A_{\rm term}\)
after the fixed product order; those bounds remain \(o(1)\) after
multiplication by every prescribed fixed power of \(Q\).
Positive first-run normalization and
\[
 \frac{y}{1+Vy}\le V^{-1}\qquad(y\ge0,\ V>0)
\]
do not depend on the sign of \(\alpha_c\).
At \(\omega=0\), real matching and
\eqref{eq:full-final-residual-budget} give
\[
 \frac{B_c(\theta)}{K_c(t,\theta)}=1+o(1),\qquad P_c(0)=1+o(1).
\]
For a positive or zero exponent the short matching does not alter
the divergent coefficient. For fixed negative exponent, summability
of the critical long-run tail gives
\[
 S_c(\theta)-S_c^{\rm app}(\theta)
       \longrightarrow K_c(t,b)-\mathcal K_{M,\infty}(b),
\]
which is exactly \eqref{eq:full-negative-matched-mass}.
The single integral \(\mathcal H_a(Q)\) supplies the corresponding
uniform matching through zero.
The explicit factors have the bounded variation needed for products:
\begin{equation}\label{eq:full-explicit-factor-variation}
 \sup_{|\omega|\le W}|P_c(\omega)|+
            \int_{-W}^{W}|P_c'(\omega)|\,d\omega\le C.
\end{equation}
For $\alpha_c>0$, this follows from the signed bounds
$C(1+|\omega|)^{-\alpha_c}$ and $C(1+|\omega|)^{-\alpha_c-1}$ of
Lemma~\ref{lem:full-explicit-first-run-Watson}, together with the short
constant bound. For a fixed \(a<0\), the constant matching term has zero derivative, so
\[
 \int_{-W}^W
 \left|\partial_\omega
   \frac{\mathcal K_{M,\Lambda}(z)-\mathcal K_{M,\Lambda}(\theta)}{Z_c}
 \right|\,d\omega
 \le C\Lambda^aQ^{1/2}W^{-a}
          =C(W/\Lambda)^{-a}Q^{1/2}=o(1);
\]
its critical mass remains bounded. For the normalized zero or
bounded-confluence factor the signed bound is
\[
 |P_c'(\omega)|\le
 \frac{C}{Q(1+|\omega|)}
          \exp\{C\log(2+|\omega|)/Q\},
\]
Thus, since \(\log(2+W)\le CQ\),
\[
 \begin{aligned}
 \int_{-W}^W|P_c'(\omega)|\,d\omega
 &\le\frac C Q\int_0^W
     \frac{\exp\{C\log(2+u)/Q\}}{1+u}\,du\le C,\\
 \sup_{|\omega|\le T}|P_c(\omega)-1|
 &\le |P_c(0)-1|+\frac{C\log(2+T)}Q\longrightarrow0
       \qquad(T<\infty).
 \end{aligned}
\]
These bounds retain the zero short mass.
For the two-endpoint error write
\[
 \mathcal E_\times
 =(P_-+R_-)(P_++R_+)-P_-P_+
 =R_-P_++P_-R_++R_-R_+ .
\]
Each differentiated product is controlled by
\[
 \int |(RP)'|
 \le \|P\|_\infty\int|R'|+\|R\|_\infty\int|P'|.
\]
Use \eqref{eq:full-final-residual-budget} for \(R_\pm\) and
\eqref{eq:full-explicit-factor-variation} for \(P_\pm\); the same
bound with the two residuals controls \(R_-R_+\). Thus
\(\|\mathcal E_\times\|_{\rm osc}=o(1)\), including all dyadic
endpoint terms.
Conditional on the retained endpoints, the exact Mehler formula
\eqref{eq:exact-ou-bridge-transform} gives
\begin{equation*}
 M_{N,t}(z)=e^{N\lambda(z)}\sqrt{r(z)}B_-(z)B_+(z)
                                  +\mathcal E_N(z).
\end{equation*}
For the finite retained hybrid, expanding the endpoint quadratic and
determinant gives
\[
 R=2\log\{(n-m)/m\},\quad m=N^{O(1)},\quad |Z_m^\pm|\le B\sqrt N,
 \qquad
 \sup_{|\omega|\le W}
 \frac{|\mathcal E_N(z)|}{e^{N\lambda(\theta)}Z_-Z_+}
       \le N^Ce^{-cN}.
\]
At \(s=\omega/\Lambda\), its conditional Mehler ratio satisfies
\[
 \left|
 \frac{\mathbb E_{\rm hyb}
       [e^{(\theta+is)Y_\circ}\mid\mathcal F_{n,m}^{\partial}]}
      {\mathbb E_{\rm hyb}
       [e^{\theta Y_\circ}\mid\mathcal F_{n,m}^{\partial}]}
 \right|
 \le C(1+|s|)^C
 \begin{cases}
 e^{-cNs^2},&|s|\le s_0,\\
 e^{-cN},&s_0\le|s|\le s_1,\\
 e^{-cN\sqrt{|s|}},&|s|\ge s_1,
 \end{cases}
 \qquad 0<s_0<s_1<\infty .
\]
Choosing \(D\) in \(W\) large makes the omitted Fourier integral
\(O(N^{-A})\) for prescribed \(A\). For the remaining Gaussian factor, set
\[
 G_N(\omega)=\exp\left\{
 N[\lambda(b-(c_0-i\omega)/\Lambda)-\lambda(b)]
       +(c_0-i\omega)N\lambda'(b)/\Lambda\right\}.
\]
The exact formula and Taylor expansion near $b<1/4$ give
\[
 |G_N(\omega)|\le C e^{-cN\omega^2/\Lambda^2},\qquad
 \int_{-W}^W|G_N'(\omega)|\,d\omega\le C.
\]
Replacing it by $\exp\{VN(c_0-i\omega)^2/(2\Lambda^2)\}$ costs
$O(N^{-1/2}(\log N)^{3/2})$ in the same inversion budget. The slowly varying
factor $\sqrt{r(z)}$ has a smaller error.
On \(|\omega|\le\min(Q,W)\), absolute integration costs at most
\(CQ\|\mathcal E_\times\|_\infty=o(1)\). On each remaining interval
\([A,B]\), use \(E\ge e_*>0\) and integrate once:
\[
 \begin{aligned}
 \left|\int_A^B e^{-iE\omega}G_N(\omega)
                         \mathcal E_\times(\omega)\,d\omega\right|
 \le \frac1{e_*}\bigg\{&
 |G_N\mathcal E_\times|(A)+|G_N\mathcal E_\times|(B)\\
 &+\|\mathcal E_\times\|_\infty\int_A^B|G_N'|
  +\|G_N\|_\infty\int_A^B|\mathcal E_\times'|\bigg\}.
 \end{aligned}
\]
Summation, including dyad endpoints, gives \(o(1)\).
The positive-exponent factor, its signed derivative, and variation
of the other factor give
\[
 \left|\int_{T<|\omega|\le W}
      e^{-iE\omega}G_N(\omega)P_-(\omega)P_+(\omega)\,d\omega\right|
       \le CT^{-p}+o(1),\qquad p>0,
\]
uniformly for bounded \(VN/\Lambda^2\ge0\).
Bounded-frequency Watson convergence now proves
\eqref{eq:full-scalar-Gamma-bath} with
\eqref{eq:assembled-boundary-scalar-density}, for every \(\alpha_t>0\).
For spatial uniformity retain the actual parameters at
\[
 t=t_0+u/\Lambda,\qquad |u|\le C .
\]
At confluence, the positive mass has
\[
 \partial_a\log\mathcal H_a(Q)
 =\frac{\int_0^Q u e^{au}u^{1/2}\,du}
        {\int_0^Q e^{au}u^{1/2}\,du}\in[0,Q].
\]
Integrating its logarithmic derivative gives
\[
 e^{-|a'-a|Q}
 \le\frac{\mathcal H_{a'}(Q)}{\mathcal H_a(Q)}
 \le e^{|a'-a|Q}.
\]
Moreover \(\mathcal G_{\alpha_t,v}(E)>0\) is continuous for
\(E\) in positive compact sets and bounded \(v\ge0\).
Let \(\rho_N\) be \eqref{eq:assembled-boundary-scalar-density}.
For bounded \(h>0,v_0\), interval integration gives
\begin{equation*}
 \mathbb P_{\rm hyb}\{\widetilde T_n\in y+v_0+[0,h],\,\mathcal A_n\}
       \sim\rho_N(y)e^{-bv_0}\frac{1-e^{-bh}}b.
\end{equation*}
The density formula remains uniform for \(0\le u\le C\log N\),
because \(E+u/\Lambda\) stays in a positive compact set. For the
remaining tail, positivity at \(\theta\) gives
\[
 \frac{\mathbb P_{\rm hyb}\{\widetilde T_n>y+C\log N,\,\mathcal A_n\}}{\rho_N(y)}
 \le CN^{C_1}e^{-\theta C\log N}.
\]
Choose \(C\) so this vanishes. Integration of
\(e^{-bu}\rho_N(y)\{1+o(1)\}\) on the retained range then gives
\(\mathbb P_{\rm hyb}\{\widetilde T_n>y,\,\mathcal A_n\}\sim\rho_N(y)/b\).
Finally, tilting by the actual retained \(M_{N,t}(\theta)\) gives
\begin{equation}\label{eq:full-tilted-interval-lower}
 \mathbb Q_{n,t,\theta}^{\mathcal A}
       \{\widetilde T_n\in[y,y+h]\}
 =\frac h\Lambda c_0^{\alpha_t}
 e^{-c_0E-VNc_0^2/(2\Lambda^2)}
             \mathcal G_{\alpha_t,VN/\Lambda^2}(E)(1+o(1)).
\end{equation}
Multiplying the raw density by \(e^{\theta y}\) and dividing by the real
transform cancels the finite or confluent second-end mass. For
\(d=c_0/\Lambda\),
\[
 N\{\lambda(b)-\lambda(b-d)\}-dy
 =-c_0E-\frac{NVc_0^2}{2\Lambda^2}
          +O(N/\Lambda^3),\qquad N/\Lambda^3=O(N^{-1/2}).
\] It is uniformly at least $c h/\Lambda$.
This uniform interval lower bound is used for conditioning an original-rank
anchor.
\end{proof}
\subsubsection{Return to the original rank probabilities}

For an actual split \(t=k/n\), let \(\rho_{N,t}(y)\) denote
the expression in \eqref{eq:assembled-boundary-scalar-density},
including the finite \(\mathcal H_a\) convention through zero.
Here its argument is the raw, unshifted energy; this convention differs
from the density of \(\widetilde T_n-d(t)\) in the bounded critical window.
The studentization correction is inserted only after the raw scan.
Write \(T_{n,k}=nF_{n,k}\).

For the marked assertion, put
\[
 I=[y+v_0,y+v_0+h],\qquad
 \mathcal I^r=\{T_{n,k}\in I\},\qquad
 \mathcal I^G=\{\widetilde T_n\in I\}.
\]
Let \(\mathcal E^r,\mathcal E^G\) be measurable events on the coupled
rank and hybrid spaces. Require
\[
 \mathcal E^r\cap\mathcal C_{n,t}\subset\mathcal E^G,\qquad
 \mathbb P(\mathcal C_{n,t}^{\,c})\le Cn^{-D_{\rm prob}},\qquad
 \beta_N=\mathbb Q_{n,t,\theta}^{\mathcal A}(\mathcal E^G).
\]
Here \(\mathcal C_{n,t}\) is the intersection of the count, Gaussian
oscillation, and coupling bounds constructed in the proof. A prescribed
deterministic tolerance is incorporated in \(\mathcal E^G\).

\begin{lemma}[Original-rank scalar boundary transfer]
\label{lem:original-rank-boundary-scalar-transfer}
Under the continuous null and the parameter family
of Theorem~\ref{thm:finite-hybrid-boundary-scalar},
\begin{equation*}
 \mathbb P\{T_{n,k}>y\}\sim\frac{\rho_{N,t}(y)}{b(t)}.
\end{equation*}
For bounded \(v_0\) and \(h\) in a positive compact interval,
\begin{equation*}
 \mathbb P\{T_{n,k}\in[y+v_0,y+v_0+h]\}
  \sim\rho_{N,t}(y)e^{-b(t)v_0}
                            \frac{1-e^{-b(t)h}}{b(t)}.
\end{equation*}
Both equivalents are uniform also for rounded splits
\(t=t_0+u/\Lambda+O(n^{-1})\).
Atoms in these raw neighborhoods are \(o(\rho_{N,t}(y))\).

Under the marked-event condition above,
\begin{equation}\label{eq:full-rank-anchor-mark-transfer}
 \mathbb P(\mathcal E^r\mid\mathcal I^r)
 \le C\Lambda\beta_N+
                    C n^{-D_{\rm prob}}/\rho_{N,t}(y).
\end{equation}
\end{lemma}

The atom bound makes interval endpoint conventions immaterial.
Equation~\eqref{eq:full-rank-anchor-mark-transfer} specifies
the complete marked error.

\begin{proof}
Under the null, the labels in increasing rank order form a uniform binary
bridge with exactly $k$ successes. Write $C_l$ for its prefix count and
$A_l=C_l-lt$. The two-table entropy identity gives exactly
\begin{equation*}
 \begin{split}
 T_{n,k}&=\sum_{l=1}^{n-1}
 \frac{nG_{k,l}}{(l-1/2)(n-l+1/2)},\\
 G_{k,l}&=l h_t(C_l/l)
              +(n-l)h_t((k-C_l)/(n-l)).
 \end{split}
\end{equation*}
Choose the exceptional probability exponent before weighting any
event. Positivity of \(\mathcal G\) on the stated parameter family gives
\[
 N^{-C}e^{-N\Gamma_{N,t}}\le\rho_{N,t}(y)
       \le N^Ce^{-N\Gamma_{N,t}},\qquad
 \Gamma_{N,t}=b(t)y/N-\lambda(b(t)),\qquad
 \sup\Gamma_{N,t}=\Gamma_*<\infty.
\]
For at most \(n^d\) comparisons, choose in order
\[
 D_{\rm prob}>\Gamma_*+d+2,\qquad
 C_D,\ B,\qquad K_m>3,\qquad m=N^{K_m},
 \qquad m^2/n=N^{2K_m}e^{-N}\to0 .
\]
Write
\(\mathcal B=(C_l^-,C_l^+:1\le l\le m)\), so that
\(\sigma(\mathcal B)=\mathcal F_{n,m}^{\partial}\), and reveal this vector.
Its total success count is \(b_0=C_m^-+C_m^+\).
Their exact likelihood relative to independent Bernoulli(\(t\)) paths is
\begin{equation}\label{eq:full-boundary-relative-likelihood}
 L_n(b_0)=\frac{\mathbb P\{\operatorname{Bin}(n-2m,t)=k-b_0\}}
                   {\mathbb P\{\operatorname{Bin}(n,t)=k\}}
              =1+O_\eta(m^2/n),
 \qquad 0\le b_0\le2m.
\end{equation}
To obtain the likelihood estimate, expand each falling factorial
in the ordered sample-without-replacement probability. All its
arguments exceed \(\eta n/2\), so
\[
 |\log L_n(b_0)|
 \le C_\eta\sum_{r=0}^{2m-1}\frac rn
 \le C_\eta m^2/n .
\]
The estimate holds for every ordered pattern. Consequently, for
any nonnegative conditional event weight \(a(\mathcal B)\),
\[
 \E_{\rm bridge}a(\mathcal B)
       =\{1+O(m^2/n)\}\E_{\rm Bernoulli}a(\mathcal B),
\]
even if the expectation is a rare probability.
The two revealed boundary energies differ from $T_m^{-1/2}$ and $T_m^{+1/2}$
by at most $Cm^2/n$, deterministically. For $l\le m$,
\[
 (n-l)h_t((k-C_l)/(n-l))\le Cl^2/n,\qquad
 \frac n{(l-1/2)(n-l+1/2)}
             =\frac1{l-1/2}+\frac1{n-l+1/2}.
\]
The main cell is at most \(Cl\), and
\[
 \sum_{l\le m}O(l/n)=O(m^2/n),\qquad
 \frac n{(n-r-\tfrac12)(r+\tfrac12)}
 =\frac1{r+\tfrac12}+\frac1{n-r-\tfrac12}.
\]
Thus the upper half shift is \(+1/2\).
Given the paths, put \(s_-=C_m^--mt,\ s_+=C_m^+-mt\).
The middle bridge has length \(n-2m\) and fraction
\[
 t_{\rm mid}=t-(s_-+s_+)/(n-2m)=t+O(m/n)
\]
and centered endpoints \(s_-\) and \(-s_+\). Let \(G_l\) be the
Gaussian bridge with these endpoints, length \(n-2m\), and incremental
variance \(t_{\rm mid}(1-t_{\rm mid})\). The bridge-specific KMT coupling of
\citet[Theorem~1.2 and Remark~1.3]{DimitrovWu2021} gives, for the chosen
exponent,
\begin{equation*}
 \mathbb P\left\{\sup_{m\le l\le n-m}|A_l-G_l|>
                 C_D\log n\,\middle|\,\mathcal F_{n,m}^{\partial}\right\}
                         \le n^{-D_{\rm prob}}.
\end{equation*}
Uniformity follows from
\[
 t_{\rm mid}\in[\eta/2,1-\eta/2],\qquad
 t_{\rm mid}(1-t_{\rm mid})\ge c_\eta,\qquad
 \mathbb E_{X\sim{\rm Bernoulli}(t)}e^{a|X-t|}\le e^a
\]
for fixed \(a\). Remark~2.7 and Theorem~6.6 of the same reference
quantify this endpoint-prescribed bridge construction.
On the same Brownian bridge, let \(G_l^{(t)}\) have incremental
variance \(t(1-t)\) and the same endpoints, and put
\(\Delta G_l=G_l-G_l^{(t)}\). Then
\[
 |t_{\rm mid}(1-t_{\rm mid})-t(1-t)|=O(m/n)
 \quad\Longrightarrow\quad
 \|\Delta G\|_\infty=O(m\sqrt N/\sqrt n)
\]
outside probability \(n^{-D_{\rm prob}}\).
Hypergeometric concentration and a union bound give
\[
 |A_l|\le C_D\sqrt{(l\wedge(n-l))N}\quad(1\le l\le n-1)
\]
outside probability \(n^{-D_{\rm prob}}\).
Conditionally on the resulting \(|Z_m^\pm|\le B\sqrt N\), put
\(d_l=l\wedge(n-l)\). The affine Gaussian mean and fluctuation satisfy
\[
 |G_l|\le C_D\sqrt{d_lN},\qquad
 \sup_{0\le h\le1}|G_{l+h}-G_l|\le C_D\sqrt N
\]
outside a conditional exception of the same order.
Let \(\mathcal C_{n,t}\) be the intersection of the four preceding
bounds for \(\|A-G\|_\infty,\|\Delta G\|_\infty,|A_l|,|G_l|\) and the
unit-rank oscillations. With their constants chosen above,
\(\mathbb P(\mathcal C_{n,t}^{\,c})\le Cn^{-D_{\rm prob}}\).
On \(\mathcal C_{n,t}\),
\[
 \begin{aligned}
 \sum_{l=m}^{n-m}\frac{|A_l|^3}{d_l^3}
   &\le C N^{3/2}\sum_{l=m}^{n-m}d_l^{-3/2}
     \le C N^{3/2}/\sqrt m,\\
 \sum_{l=m}^{n-m}\frac{N|G_l|+N^2}{d_l^2}
   &\le C N^{3/2}/\sqrt m+CN^2/m .
 \end{aligned}
\]
The first line is the entropy cubic remainder; the second is the
quadratic coupling error. The half-shift correction and the
sum-to-integral error, using the \(C_D\sqrt N\) unit-rank oscillation,
are bounded by the same orders. Adding \(Cm^2/n\) for the two
boundary energies gives the following bound for \(Y_G=\widetilde T_n\):
\begin{equation}\label{eq:full-original-statistic-error}
 |T_{n,k}-Y_G|\le\varepsilon_N,
 \qquad\varepsilon_N=C_D\left\{
 \frac{N^{3/2}}{\sqrt m}+\frac{N^2}{m}+\frac{m^2}{n}\right\}=o(1).
\end{equation}
The positive hybrid is exactly
\[
 R=2\log\{(n-m)/m\},\qquad
Y_G=T_m^{-1/2}+T_m^{+1/2}
        +\frac12\int_0^RZ(s)^2\,ds,
\]
with OU endpoints
\[
 Z(0)=\frac{Z_m^-}{\sqrt{1-m/n}},\qquad
 Z(R)=-\frac{Z_m^+}{\sqrt{1-m/n}}.
\]
Keeping these in the finite Mehler formula gives the scalar theorem's
retained subprobability
\(\mathbb P_G^A(E)=\mathbb P_{\rm hyb}(E\cap\mathcal A_n)\);
endpoint replacement is within its proved error. Equations
\eqref{eq:full-boundary-relative-likelihood}--\eqref{eq:full-original-statistic-error}
then give
\begin{align}
 (1-\eta_N)\mathbb P_G^A\{Y_G>u+\varepsilon_N\}
                     -Cn^{-D_{\rm prob}}
 &\le\mathbb P\{T_{n,k}>u\}\nonumber\\
 &\le(1+\eta_N)\mathbb P_G^A\{Y_G>u-\varepsilon_N\}
                     +Cn^{-D_{\rm prob}}.\label{eq:full-original-probability-sandwich}
\end{align}
Here \(\eta_N=Cm^2/n\). On the same good event, direct interval
inclusions give, for \(h>2\varepsilon_N\),
\[
 \{Y_G\in[u+\varepsilon_N,u+h-\varepsilon_N]\}
 \subset\{T_{n,k}\in[u,u+h]\}
 \subset\{Y_G\in[u-\varepsilon_N,u+h+\varepsilon_N]\}.
\]
Apply the multiplicative likelihood bound to these inclusions,
and add \(Cn^{-D_{\rm prob}}\) for their ordinary exceptional
probabilities. Tilting is applied only to the retained hybrid
terms.
The explicit proxy satisfies
\[
 \rho_{N,t}(y+d)/\rho_{N,t}(y)=e^{-bd}(1+o(1))
\]
uniformly for bounded $d$, including $d=O(\varepsilon_N)$. Apply the scalar
tail and density theorem to the sandwich and its direct interval version.
The upper and lower bounds have the same relative limit, since
\[
 \frac{n^{-D_{\rm prob}}}{\rho_{N,t}(y)}
       \le N^C e^{-(D_{\rm prob}-\Gamma_*)N}\longrightarrow0
\]
and \(\eta_N\to0\). This proves both original tail and interval
equivalents. For an atom at a bounded offset \(v\),
\[
 \Pp\{T_{n,k}=y+v\}
 \le(1+\eta_N)\Pp_G^A\{|Y_G-y-v|\le\varepsilon_N\}
       +Cn^{-D_{\rm prob}}
 =o(\rho_{N,t}(y)).
\]
For a bad mark, keep its enlarged hybrid event in the positive comparison.
Write \(\mathbb P_\theta^A=\mathbb Q_{n,t,\theta}^{\mathcal A}\).
A width-\(h\) anchor satisfies
\[
 y\le Y_G\le y+h
 \quad\Longrightarrow\quad
 e^{-\theta h}\le e^{-\theta(Y_G-y)}\le1 .
\]
Hence untilting gives the conditional bound
\[
 \begin{aligned}
 \Pp(\mathcal E^r\mid\mathcal I^r)
 &\le
 C\frac{\Pp_\theta^A(\mathcal E^G)}
         {\Pp_\theta^A(\mathcal I^G)}
       +\frac{Cn^{-D_{\rm prob}}}{\rho_{N,t}(y)}\\
 &\le C\Lambda\beta_N
       +\frac{Cn^{-D_{\rm prob}}}{\rho_{N,t}(y)} .
 \end{aligned}
\]
This proves \eqref{eq:full-rank-anchor-mark-transfer}, including deterministic
mark enlargement. Its errors satisfy
\[
 C\Lambda N^Ce^{-c(\log N)^2}\le N^{C+1}e^{-c(\log N)^2},
 \qquad
 \frac{Cn^{-D_{\rm prob}}}{\rho_{N,t}(y)}
       \le N^Ce^{-(D_{\rm prob}-\Gamma_*)N}.
\]
For fixed \(\theta_*<\inf_t b(t)\), positivity and
\eqref{eq:full-original-probability-sandwich} give the coarse-cap bound
\[
 \Pp\{\sup_k T_{n,k}>C_{\rm cap}N\}
 \le n^{C_2-\theta_*C_{\rm cap}}+Cn^{1-D_{\rm prob}} .
\]
Choose the cap and exceptional exponents before weighting:
\[
 \theta_*C_{\rm cap}-C_2>A,\qquad D_{\rm prob}>A+1,
\]
\[
 \begin{gathered}
 0\le w_{n,i}\le e^{L_wN}\quad(i\in\mathfrak I_n),\qquad
 |\mathfrak I_n|\le n^d,\qquad
 D_{\rm prob}>\Gamma_*+L_w+d+2,\\
 \frac{n^d e^{L_wN}n^{-D_{\rm prob}}}{\rho_{N,t}(y)}
       \le N^Ce^{-2N}.
 \end{gathered}
\]
The constants depend only on the trim and error exponent.
Rounding has
\[
 |\Delta t|=O(n^{-1}),\qquad N|\Delta t|=O(N/n),\qquad
 |\Delta\log\mathcal H_a(Q)|\le Q|\Delta a|=O(Q/n),
\]
so it preserves the rate and coefficient. The deterministic
studentization map retains \(e^{-bd}\) for a bounded raw shift \(d\);
for \(d=o(1)\), this factor is \(1+o(1)\).
\end{proof}

\subsection{The complete positive critical approach}
\label{app:critical-full-positive}

Throughout this subsection \(b=b(\eta)<1/4\), \(N=\log n\), and
\(q_c,V,\gamma_c,d(t)\) have the meanings in \eqref{eq:core-bulk-constants}.
All constants are uniform on fixed trimmed split intervals. The density
notation below refers to the positive analytic expression in the hybrid
scalar local law.

We use the interval version of
Lemma~\ref{lem:original-rank-boundary-scalar-transfer} whenever conditioning
the original statistic.

\subsubsection{The deterministic positive convolution}

Write \(g_{N,b}(v)=e^{bv-N\lambda(b)}g_N(v)\). It integrates to one, has
mean \(Nq_c\) and variance \(NV\), and satisfies
\[
 g_{N,b}(Ny)=\frac{1}{2\sqrt{\pi N}\,y^{3/2}}
       e^{-N J_b(y)},\qquad
 J_b(y)=\frac{(y-q_c)^2}{4yq_c^2}.
\]
Its local Gaussian limit has variance \(V\), while this exact formula
retains all moderate-deviation terms.

For \(U=Nq_c+D\), \(D/\sqrt N\to\infty\), and \(D=o(N)\),
put \(\kappa=\vartheta(U/N)-b\) and
\[
 G_N(U)=
 \begin{cases}
  \Theta_0 g_{N,b}(U),&\alpha_\eta<1,\\
  P_0\{\log(1/\kappa)\}^{\beta_\eta+1}g_{N,b}(U),&\alpha_\eta=1,\\
  0,&\alpha_\eta>1.
 \end{cases}
\]
\begin{lemma}[Matching properties of the positive calibration kernel]
\label{lem:full-critical-kernel}
The measure in \eqref{eq:main-critical-positive-measure}
is positive. The intensity \(\mathcal R_N^{\rm fc}\) is finite
and positive near \(Nq_c\), and equivalent to
\(\mathcal R_N^{\rm lc}\) on every lower or bounded-core
critical approach.

On the positive divergent range above,
\begin{equation}\label{eq:full-critical-kernel-two-parts}
 \frac{\mathcal R_N^{\rm fc}(U)}
          {N(q_c-1)e^{N\lambda(b)-bU}}
 =
 \begin{cases}
  \{G_N(U)+w(D)\}\{1+o(1)\},&\alpha_\eta\le1,\\
  w(D)\{1+o(1)\},&\alpha_\eta>1 .
 \end{cases}
\end{equation}
For fixed \(C\), uniformly over \(|h|\le C\log N\)
and \(U/N\to q_c\),
\begin{equation*}
 \frac{\mathcal R_N^{\rm fc}(U+h)}
      {\mathcal R_N^{\rm fc}(U)}
 =e^{-\min\{\vartheta(U/N),b\}h}\{1+o(1)\}.
\end{equation*}
\end{lemma}

For \(\alpha_\eta\le1\), the error in
\eqref{eq:full-critical-kernel-two-parts} is relative to
the positive sum, including when the terms are comparable.

\begin{proof}
For \(0<\alpha_\eta<1\), the choices of \(h_0,M_{\beta_\eta},B_0\) give
\[
 \log^{\beta_\eta}(e+B)\le M_{\beta_\eta} B^{h_0}\quad(B\ge e),\qquad
 \int_{B_0}^\infty w(B)\,dB
 \le\frac{c_{\rm tail}M_{\beta_\eta}}{h_0}B_0^{-h_0}
 \le\Theta_0/2.
\]
The atom is therefore nonnegative; the other measures are locally finite.
The endpoint factor
\[
 g_N(U-B)=O\!\left((U-B)^{-3/2}
                   e^{-cN^2/(U-B)}\right),\qquad B\uparrow U,
\]
makes every convolution finite.
For \(\alpha_\eta<1\), finite mass and the exact rate give
\[
 \int e^{-bB}\frac{g_N(U-B)}{g_N(U)}\,\mu(dB)\longrightarrow\Theta_0 .
\]
Put \(R_{\rm eff}=\{(b-\vartheta(U/N))_++N^{-1/2}\}^{-1}\).
Then
\[
 R_{\rm eff}\asymp
 \begin{cases}
 \{b-\vartheta(U/N)\}^{-1},
     &(b-\vartheta(U/N))\sqrt N\to+\infty,\\
 \sqrt N,& |U-Nq_c|/\sqrt N=O(1).
 \end{cases}
\]
In both ranges \(R_{\rm eff}\to\infty\).
Convexity supplies domination there; the endpoint exponential controls
\(B\) near \(U\). For \(\alpha_\eta=1\),
\[
 \int_1^R w(B)\,dB
   \sim \frac{c_{\rm tail}}{\beta_\eta+1}(\log R)^{\beta_\eta+1}
   =P_0(\log R)^{\beta_\eta+1}.
\]
For the same effective cutoff,
\[
 \frac{\log^{\beta_\eta+1}(cR_{\rm eff})}{\log^{\beta_\eta+1}R_{\rm eff}}\to1
 \quad(c>0).
\]
This proves the logarithmic lower/core match.
For \(\alpha_\eta>1\), \(c_{\rm tail}=P_A/\Gamma(\alpha_\eta-1)\), so the kernels
coincide. On a positive approach put \(s=D/\sqrt N\) and \(v=U-B\):
\[
 \int_{D/2}^{U}g_{N,b}(U-B)w(B)\,dB
 =\int_0^{Nq_c+D/2}g_{N,b}(v)w(U-v)\,dv
 \sim w(D).
\]
Indeed, on \(|v-Nq_c|\le T\sqrt N\), regular variation gives
\(w(U-v)/w(D)\to1\). The exact Gaussian rate dominates the
complement as \(T\to\infty\). At small boundary energies,
\[
 \log\frac{g_{N,b}(U-B)}{g_{N,b}(U)}
       =\kappa B+O(B/N+B^2/N),\qquad B\le C/\kappa .
\]
For \(\alpha_\eta<1\), truncate at a fixed \(B\), then increase that
truncation to obtain the finite mass \(\Theta_0\).
For \(\alpha_\eta=1\), integration over
\(B\le\kappa^{-1+\epsilon}\), followed by the remaining interval
up to \(C/\kappa\), gives
\[
 \int_{B\le C/\kappa}g_{N,b}(U-B)\,\mu(dB)
 =P_0\log^{\beta_\eta+1}(1/\kappa)\,g_{N,b}(U)\{1+o(1)\}.
\]
The factor \(e^{\kappa B}\) changes the logarithmic integral by
only \(O(\log^{\beta_\eta}(1/\kappa))\); take
\(\epsilon\downarrow0\) afterwards.
The intervening range has the following upper bound. For \(C/\kappa\le B\le
D/2\), choose \(C\) large and differentiate \(f(B)=w(B)g_{N,b}(U-B)\):
\[
 \frac{f'(B)}{f(B)}
  =\frac{\alpha_\eta-2}{B}
    +\frac{\beta_\eta}{(e+B)\log(e+B)}
    +\vartheta((U-B)/N)-b+\frac{3}{2(U-B)}
  \ge c\kappa .
\]
The differential inequality implies
\(f(B)\le f(D/2)e^{-c\kappa(D/2-B)}\), and hence
\[
 \int_{C/\kappa}^{D/2}f(B)\,dB
 \le\frac{f(D/2)}{c\kappa}
 \le Cw(D)s^{-1}e^{-cs^2}=o(w(D)).
\]
For \(\alpha_\eta>1\), the remaining small-energy term obeys
\[
 \int_{B\le C/\kappa}w(B)\,dB
 \le C\kappa^{1-\alpha_\eta}\log^{\beta_\eta}(1/\kappa),\qquad
 \frac{g_{N,b}(U)}{w(D)}
      \int_{B\le C/\kappa}w(B)\,dB
 \le C(1+s)^C e^{-cs^2}\to0 .
\]
This proves \eqref{eq:full-critical-kernel-two-parts} relative to its
positive sum. The lower and core shift identities follow from
\eqref{eq:lower-kernel-shift} and the kernel comparisons above.
For \(s\le\log N,\ |h|\le C\log N\), put
\(\kappa_h=\vartheta((U+h)/N)-b>0\). Then
\[
 \begin{aligned}
 \left|\log\frac{g_{N,b}(U+h)}{g_{N,b}(U)}\right|
       &\le C|h|s/\sqrt N=o(1),\\
 \left|\log\frac{w(D+h)}{w(D)}\right|
       &\le C|h|/D=o(1),\\
 \left|\log\frac{\log^{\beta_\eta+1}(1/\kappa_h)}
                    {\log^{\beta_\eta+1}(1/\kappa)}\right|
       &\le\frac{C|h|}{D\log(1/\kappa)}=o(1).
 \end{aligned}
\]
For \(s>\log N\), \(G_N=o(w(D))\), leaving only the boundary bound.
Positivity preserves relative shifts:
\[
 A,B>0:\qquad \frac{A_h+B_h}{A+B}-1
 =\frac{A}{A+B}\left(\frac{A_h}{A}-1\right)
  +\frac{B}{A+B}\left(\frac{B_h}{B}-1\right)=o(1).\qedhere
\]
\end{proof}

\subsubsection{Uniform original anchors in the positive approach}

Fix \(U=Nq=Nq_c+D\), with \(D/\sqrt N\to\infty\) and \(D=o(N)\), and put
\(s=D/\sqrt N\), \(b_t=b(t)\), and
\[
 q_c(t)=
 \begin{cases}\lambda'(b_t),&b_t<1/4,\\ \infty,&b_t=1/4.\end{cases}
\]
At an individual zero endpoint retain the exponent \(\alpha_{+1/2}(t)\).
We first restrict to the split families on
which the scalar theorem is proved. For a sufficiently large fixed \(A\),
define
\begin{equation*}
 \mathcal I_N=
 \begin{cases}
 [\eta,1-\eta],&s\le\log N,\\
 \{t\in[\eta,1-\eta]:
    D\min(t-\eta,1-\eta-t)\le A\log N\},&s>\log N.
 \end{cases}
\end{equation*}
A collar of length \(H/N\) changes the scaled distance by \(HD/N=o(1)\).
For the excluded cells use
\[
 \widehat\vartheta_t=\min\{\vartheta(q),b_t-N^{-1}\},\qquad
 \Psi_{N,t}(\widehat\vartheta_t)\le N^C .
\]
The positive truncated first-run bound includes zero exponents.
Use \(t_\sigma(z)\), \(T_t^\sigma(z)\) and the stopped maximum
\(\mathcal M_H\) from \eqref{eq:rank-local-uniform-time-generator}.
Let \(C_l(z)\) count the labels in the first \(l\) pooled ranks at
\(t_\sigma(z)\), put \(A_l(z)=C_l(z)-lt_\sigma(z)\), and let \(G_l(0)\)
be its coupled conditional Gaussian bridge. As in Appendix~\ref{app:critical-core},
define
\[
 \begin{aligned}
 \mathcal C_{n,t,H}
 &=\{\max_{m\le l\le n-m}|A_l(0)-G_l(0)|\le C_{A_0}N\}\\
 &\quad\cap\bigcap_{l=m}^{n-m}
 \{\sup_{0\le z\le H}|A_l(z)|
             \le C_{A_0}\sqrt{(l\wedge(n-l))N}\},\\
 \mathcal C_t(U,H)&=\{\sup_{0\le z\le H}T_t^\sigma(z)>U\},\\
 \mathcal K_{n,t,H}
 &=\{\sup_{0\le z\le H}T_t^\sigma(z)\le C_0N\}\cap\mathcal C_{n,t,H}.
 \end{aligned}
\]
The coupling gives \(\mathbb P(\mathcal C_{n,t,H}^c)\le n^{-A_0}\).
Write \(T_0=T_t^\sigma(0)\).
On this event the stopped mixed generator gives
\[
 \begin{aligned}
 \mathbb P\{\mathcal C_t(U,H)\}
 &\le e^{-\widehat\vartheta_tU}
   \E[e^{\widehat\vartheta_tT_0+
                 \widehat\vartheta_t\mathcal M_H};\mathcal K_{n,t,H}]
       +n^{-A_0}\\
 &\le N^C
   e^{-\widehat\vartheta_tU+N\lambda(\widehat\vartheta_t)}
       +n^{-A_0}\\
 &\le N^C e^{-N\mathcal R_{b_t}(q)}+n^{-A_0}.
 \end{aligned}
\]
Here
\[
 \widehat\vartheta_tq-\lambda(\widehat\vartheta_t)
       =\mathcal R_{b_t}(q)+O(N^{-1}).
\]
Cap and coupling exceptions contribute the unweighted term \(n^{-A_0}\),
with \(A_0\) chosen after the fixed exponential weights; no scalar
interval division is used.

For \(s>\log N\), strict convexity of \(\lambda\) gives
\[
 N\{\mathcal R_{b_t}(q)-\mathcal R_b(q)\}
 \ge c\min\{D\min(t-\eta,1-\eta-t),s^2\}.
\]
Let \(\mathfrak T_N^{\rm out}\) contain the centers of mesh cells
disjoint from \(\mathcal I_N\). If
\(|\mathfrak T_N^{\rm out}|\le N^{C_1}\), then
\[
 \sum_{t\in\mathfrak T_N^{\rm out}}\mathbb P\{\mathcal C_t(U,H)\}
 \le N^{C+C_1}e^{N\lambda(b)-bU-cA\log N}
       +N^{C_1}n^{-A_0}.
\]
The reference satisfies
\[
 N^{-C}\le Nw(D)\le N^C,\qquad s^2>(\log N)^2.
\]
Choose \(A,A_0\) large; the preceding error is then
\(o(e^{N\lambda(b)-bU}Nw(D))\). Overlapping \(A,2A\) collars cover
cell rounding. Restrict all following anchors to \(\mathcal I_N\).
For \(s>\log N,\ t\in\mathcal I_N\),
\[
 \frac{N(q-q_c(t))}{D}
       =1+O\!\left(\frac{\log N}{s^2}\right),\qquad
 |\alpha_{+1/2}(t)|\log\Lambda=O((\log N)^2/D)=o(1).
\]
For \(s\le\log N\), the endpoint strip satisfies
\[
 \operatorname{dist}(t,\{\eta,1-\eta\})
       \le(\log N)^2/\sqrt N
 \quad\Longrightarrow\quad |\alpha_{+1/2}(t)|\log\Lambda=o(1).
\]
At an individual zero, put
\(\tau=\min(t-\eta,1-\eta-t)\). The lower-range logarithm obeys
\[
 b_t-\vartheta(q)\asymp\tau,\qquad
 \alpha_{+1/2}(t)\log\frac1{b_t-\vartheta(q)}
       =O\{\tau\log(1/\tau)\}\to0 .
\]
Outside the second strip the saddle is moving-lower or strictly
subcritical. Thus the fixed-sign, bounded-confluence and moving-lower
families suffice.
The next estimates concern this retained range. Use the real anchor tilt
\begin{equation}\label{eq:full-critical-adaptive-tilt}
 \vartheta_t=
 \begin{cases}
  \vartheta(q),&q\le q_c(t)-N^{-1/2},\\
  b_t-\mathcal L_t^{-1},&q>q_c(t)-N^{-1/2},
 \end{cases}
 \qquad
 \mathcal L_t=\sqrt N+[N\{q-q_c(t)\}]_+
\end{equation}
in the second line. Their common admissible range is
\[
 \vartheta_t\longrightarrow b,\qquad
 \vartheta_t<b_t,\qquad \vartheta_t\le\tfrac14-c,\qquad
 L_t=
 \begin{cases}
  \sqrt N,&q\le q_c(t)-N^{-1/2},\\
  \mathcal L_t,&q>q_c(t)-N^{-1/2},
 \end{cases}
 \qquad L_t\le\sqrt N+D .
\]

Let \(\rho_{N,t}(u)\) be the scalar density scale supplied by the strict,
moving-lower, bounded-core, or positive scalar theorem, expressed in
the raw, unshifted energy coordinate: a density stated for
\(\widetilde T_n-d(t)\) is evaluated at \(u-d(t)\).
Studentization is applied only after the scan argument. The original interval
comparison is
\begin{equation}\label{eq:full-critical-anchor-interval}
 \Pp\{T_{n,k}\in[u+v,u+v+\ell]\}
 =\rho_{N,t}(u)e^{-\vartheta_t v}
       \frac{1-e^{-\vartheta_t\ell}}{\vartheta_t}\{1+o(1)\},
\end{equation}
for fixed \(\ell>0\) and \(|v|\le C\log N\). Under the retained positive
hybrid tilt,
\[
 \mathbb Q_{n,t,\vartheta_t}^{\mathcal A}\{\widetilde T_n\in[u,u+\ell]\}
       \ge\frac{c_\ell}{L_t},\qquad
 u=U+O(1),\qquad x_t=\sqrt N\{q-q_c(t)\}.
\]
Every retained sequence has a subsequence with \(x_t\) bounded,
\(x_t\to+\infty\), or \(x_t\to-\infty\). The respective interval bounds are
\[
 \mathbb Q_{n,t,\vartheta_t}^{\mathcal A}
       \{\widetilde T_n\in[u,u+\ell]\}\ge
 \begin{cases}
 c_\ell/\sqrt N,& |x_t|\le C,\\
 c_\ell/\mathcal L_t,&x_t\to+\infty,\\
 c_\ell/\sqrt N,&x_t\to-\infty .
 \end{cases}
\]
In the second case take
\(\Lambda=\mathcal L_t\) and \(E=N(q-q_c(t))/\mathcal L_t\to1\).
The bounds use, respectively, the positive minimum of the compact core
density, \eqref{eq:full-tilted-interval-lower}, and
\eqref{eq:lower-uniform-LLT} or its strict version.
Given \(\mathcal F_{n,m}^{\partial}\), let
\(f_{\vartheta_t,Z}\) be the density of \(Y_\circ\) under
\(\mathbb Q_{n,t,\vartheta_t}^{\mathcal A}\), and write
\(f_{\rm middle}=f_{\vartheta_t,Z}\).
For \(|v|\le C\log N\), the positive-range formula replaces
\(E\) by \(E+v/\mathcal L_t\). In the core and lower ranges,
\[
 \frac{|v|\sup|f_{\rm middle}'|}{c/\sqrt N}
 \le C\frac{|v|N^{-1}}{N^{-1/2}}
 \le C\log N/\sqrt N\to0 .
\]
Use \eqref{eq:lower-Cauchy-flatness} for the moving coefficient, then
the original interval sandwich. At zero retain \(\mathcal H_a\)
with the actual lower gap or bounded-confluence positive scale;
the preceding family bounds give uniformity.

\begin{lemma}[Sublinear endpoint energy at an original high anchor]
\label{lem:full-critical-anchor-cap}
Condition on an original interval in
\eqref{eq:full-critical-anchor-interval}, with fixed width and bounded raw
shift. With conditional error smaller than every fixed negative power of
\(N\), the two retained endpoints satisfy \(|Z_-|+|Z_+|\le2\log N\), and
their total nonnegative energy \(B\) satisfies
\begin{equation}\label{eq:full-critical-sublinear-cap}
 B\le C(\sqrt N+D)+A\sqrt N\log N=o(N).
\end{equation}
The conditional quadratic middle observables \(Q,W\) have the limits
\eqref{eq:core-middle-critical-values}, with exponentially small error for
each fixed tolerance.
\end{lemma}

\begin{proof}
Theorem~\ref{thm:rank-global-logarithmic-rate} and the polynomial mesh
allow \(C_0\) and the coupling exponent to be chosen so that
\[
 \mathbb P(\sup_kT_{n,k}>C_0N)\le n^{-A_0},\qquad
 0\le B\le C_0N+o(1)
 \quad\text{on }\{\widetilde T_n\le C_0N\}.
\]
Equation~\eqref{eq:core-deterministic-energy-tube} gives a central tube for
\(l\ge C_\rho N\). Riccati propagation to \(m=N^K\) and the whole-set
comparison \eqref{eq:core-terminal-real-ratio} yield
\[
 \mathbb Q_{n,t,\vartheta_t}^{\mathcal A}
   \{|Z_-|+|Z_+|>2\log N,\ \widetilde T_n\le C_0N\}
 \le N^Ce^{-c(\log N)^2}.
\]
Division by \(c/L_t\ge c/N\) preserves this error.
For each retained boundary pattern \(\mathcal B\), write
\(\mathbb P_Z=N(m_Z,C_Z)\) for the middle field law in \(L^2[0,R]\),
without energy truncation. Its expectation is
\(\mathbb E_Z=\mathbb E_{\vartheta_t,Z}\). Put
\[
 \mu_Z=\frac12\{\operatorname{tr}C_Z+\|m_Z\|^2\},\qquad
 K_Z(a)=\log\mathbb E_{\vartheta_t,Z}e^{aY_\circ},\qquad
 \mathcal I^G=\{\widetilde T_n\in[u,u+\ell]\}.
\]
Under \(\mathbb Q_{n,t,\vartheta_t}^{\mathcal A}\),
\[
 \mathbb E[
 \mathbf1_{\{\widetilde T_n\le C_0N\}}\mathbf1_{\mathcal I^G}
 \mid\mathcal F_{n,m}^{\partial}]
 \le\mathbb P(\mathcal I^G\mid\mathcal F_{n,m}^{\partial}).
\]
Average this inequality before dividing by the single anchor probability.
In the smaller endpoint tube,
\[
 \mu_Z=N\lambda'(\vartheta_t)+O((\log N)^2),\qquad
 \|C_Z\|_{\rm op}\le C,\qquad\operatorname{tr}C_Z=O(N).
\]
Negative real retilting gives
\[
 f_{\vartheta_t,Z}(\mu_Z-a)
 \le\frac C{\sqrt N}\exp\{-c\min(a^2/N,a)\},\qquad a\ge0.
\]
On a fixed real neighborhood of zero whose translated tilt is below
\(1/4\), the Gaussian resolvent gives \(K_Z''(a)\le CN\). For
\(\xi=c\min(a/N,1)\), Taylor's theorem and a second real tilt give
\[
 f_{\vartheta_t,Z}(\mu_Z-a)
 \le\frac C{\sqrt N}
       e^{K_Z(-\xi)-K_Z(0)+\xi(\mu_Z-a)}
 \le\frac C{\sqrt N}e^{-\xi a+CN\xi^2/2}.
\]
This proves the displayed lower-tail density bound.
In the first branch of \eqref{eq:full-critical-adaptive-tilt},
\(U-N\lambda'(\vartheta_t)=0\); in its second branch,
\[
 U-N\lambda'(\vartheta_t)
 \le N\{q-q_c(t)\}+CN/\mathcal L_t
 \le C(\sqrt N+D).
\]
Exceeding \eqref{eq:full-critical-sublinear-cap} forces
\[
 \mu_Z-(U-B)\ge A\sqrt N\log N+O((\log N)^2).
\]
Set \(B_N^*=C(\sqrt N+D)+A\sqrt N\log N\).
Positive averaging and the interval lower bound give
\[
 \mathbb Q_{n,t,\vartheta_t}^{\mathcal A}
       (B>B_N^*\mid\mathcal I^G)
 \le\frac{CN^{-1/2}e^{-c(\log N)^2}}{c/L_t}
       \le CN^{1/2}e^{-c(\log N)^2}.
\]
For coupled marks \(\mathcal E^r,\mathcal E^G\) as in the original
interval sandwich, with \(I=[U,U+\ell]\), put
\(\beta_N=\mathbb Q_{n,t,\vartheta_t}^{\mathcal A}(\mathcal E^G)\).
The same argument gives
\[
 \mathbb P(\mathcal E^r\mid\mathcal I^r)
 \le CL_t\beta_N+\frac{Cn^{-A_0}}{\rho_{N,t}(U)}.
\]
For a fixed tolerance, choose
\[
 |Z_-|+|Z_+|\le\sqrt{\epsilon N},\qquad B\le\epsilon N,\qquad D=o(N).
\]
Riccati and the Gaussian lower-tail bound give excluded mass
\(Ce^{-c_\epsilon N}\). Write \(\mathbb E_0\) for the expectation with
the same centered Gaussian covariance as \(\mathbb E_Z\), and zero mean.
Recall \(Q=\|X\|^2\), \(W=\langle X,C_RX\rangle\), with \(C_R\) the
OU covariance operator. On the complement,
\[
 |\mathbb E_ZQ-\mathbb E_0Q|
   +|\mathbb E_ZW-\mathbb E_0W|
 =\|m_Z\|^2+\langle m_Z,C_Rm_Z\rangle
 \le C\epsilon N,\qquad \|C_Z\|_{\rm op}\le C .
\]
The quadratic transform consequently gives
\[
 \mathbb P_Z
  \{|Q-\mathbb E_ZQ|+|W-\mathbb E_ZW|>\delta N\}
 \le C_\delta e^{-c_\delta N}
\]
after \(\epsilon\) is chosen small relative to \(\delta\).
The polynomial anchor denominator and deterministic coupling
error preserve this fixed-tolerance exponential concentration.
\end{proof}

\subsubsection{The original scan multiplier and pair estimates}

Let \(\pi_N(U)\) be the one-coordinate probability that the studentized scan
exceeds \((U-N)/\sqrt{2N}\). We prove
\begin{equation}\label{eq:full-critical-scan-reduction}
 \pi_N(U)\sim
 N(q_c-1)\int_{\mathcal I_N}
       e^{-bd(t)}\rho_{N,t}(U)\,\frac{dt}{t(1-t)} .
\end{equation}
Choose fixed overlapping endpoint patches. On the left patch use
\(T_t^-\); on the right use \(T_t^+\), obtained by reflecting labels and
time. Either orientation is admissible on their compact interior overlap,
where both Bellman margins are strict.
Let \(\mathcal W\) be standard Brownian motion and let \(r\) denote
oriented logit time. The conditional \(Q,W\) limits,
the uniform-one deletion martingale, and
\eqref{eq:rank-scan-s6}--\eqref{eq:rank-scan-s10} give, at \(r=z/N\),
\begin{equation*}
 Z_G(z)=
 \sqrt{\frac{8q_c^2}{q_c+1}}\,\mathcal W(z)-(q_c-1)z .
\end{equation*}
The same drift and bracket identities make the largest middle
jump tend to zero. For the positive boundary part,
\[
 \left|\frac{\mathcal L e^{sE^+}}{e^{sE^+}}\right|
 \le C_s\{B/N+\log m/N\}=o(1)
\]
on the sublinear cap of Lemma~\ref{lem:full-critical-anchor-cap}.
Stop at twice this cap. The positive generator controls its upper exit;
\eqref{eq:core-negative-thinning} raises only the negative coefficient.
For deletion probability \(d_0\),
\[
 \sigma_{\rm deletion}^2=O(1/\log(1/d_0))\to0,\qquad s_2>b .
\]
All maxima below are stopped at \(C_0N\) and restricted to the coupling
event; excluded paths contribute the unweighted error \(n^{-A_0}\).
Let \(Y_\circ(z)\), \(E^+(z)\) and \(E^-(z)\) denote the coupled
local processes, stopped at this cap. Define
\[
 \begin{aligned}
 \mathcal M_H^0&=\mathbf1_{\mathcal C_{n,t,H}}
 \sup_{0\le z\le H}
 [Y_\circ(z)+E^+(z)-Y_\circ(0)-E^+(0)]_+,\\
 \mathcal M_H^-&=\mathbf1_{\mathcal C_{n,t,H}}
 \sup_{0\le z\le H}[E^-(z)-E^-(0)]_+ .
 \end{aligned}
\]
Their sum bounds \(\mathcal M_H\), up to the deterministic coupling
tolerance. The first maximum allows every fixed exponent. Choose
\[
 b<s<s_2,\qquad 1<p_-<s_2/s,\qquad p_+=p_-/(p_--1).
\]
Write
\(\mathbb P_{\rm anchor}(E)=
 \mathbb P\{E\mid T_{n,k}\in[U+v,U+v+\ell]\}\), and let
\(\mathbb E_{\rm anc}\) denote its expectation. H\"older's inequality gives
\[
 \E_{\rm anc}e^{s(\mathcal M_H^0+\mathcal M_H^-)}
 \le
 \{\E_{\rm anc}e^{sp_+\mathcal M_H^0}\}^{1/p_+}
 \{\E_{\rm anc}e^{sp_-\mathcal M_H^-}\}^{1/p_-}.
\]
The good-path generators bound both factors by constants for
fixed \(H\). Put
\[
 \mathcal T_{n,H}^-=
 \left\{\sup_{0\le z\le H}\max_{c=\pm1/2}\max_{l\le m}
 \frac{[lt_\sigma(z)-C_l^c(z)]_+}{\sqrt l}\le C\log N\right\}.
\]
On \((\mathcal T_{n,H}^-)^c\), the mixed transform
and the lower anchor bound give \(N^Ce^{-c(\log N)^2}\);
applying H\"older with a reserved part of the \(s_2\) margin
makes its contribution vanish. Hence
\begin{equation*}
 \sup_{t\in\mathcal I_N} \E\{e^{s \mathcal M_H}\mid
       T_{n,k}\in[U+v,U+v+\ell]\}\le C_{H,\ell},
 \qquad |v|\le C\log N ,
\end{equation*}
with the stopped/good convention. Denominator powers multiply only the
discarded superpolynomial sets, so the retained constant is uniform
in \(D/\sqrt N\).
For \(|v|\le C\log N\), put
\(I_v=\{T_0\in[U+v,U+v+\ell]\}\).
Equation~\eqref{eq:full-critical-anchor-interval} and the \(s>b\) moment give
\[
 \frac{\mathbb P(I_v)}{\rho_{N,t}(U)}
 \mathbb P\{\mathcal C_t(U,H)\mid I_v\}
 \le C_{H,\ell}
 \begin{cases}
 e^{-bv},&v\ge0,\\
 e^{-(s-b)|v|},&v<0.
 \end{cases}
\]
On \(C\log N<|v|\le\sqrt N\), the middle upper bound and anchor lower bound cost
\[
 \frac{C/\sqrt N}{c/L_t}\le CN^C,\qquad
 N^C e^{-c'C\log N}\to0
\]
after choosing the cutoff constant large.
The shifted anchors retain polynomial lower bounds. In the
positive range \(E+v/L_t\) belongs to a fixed compact subset of
\((0,\infty)\); in the core and Gaussian ranges \(|v|/\sqrt N\le1\).
Their adaptive tilts preserve the good generator and negligible bad sets.
For deeper negative anchors,
\[
 \{T_0\le U-\sqrt N\}\cap\mathcal C_t(U,H)\cap\mathcal K_{n,t,H}
 \subset\{\mathcal M_H\ge\sqrt N\}.
\]
The stopped generator and mixed negative transform then give
\begin{equation*}
 \mathbb P\{T_0\le U-\sqrt N,\ \mathcal C_t(U,H)\}
 \le e^{-\vartheta_t U}\mathcal M_{N,t}(\vartheta_t)
          N^C e^{-(s-\vartheta_t)\sqrt N}+n^{-A_0}.
\end{equation*}
Relative to the reference interval,
\[
 \frac{\mathbb P\{T_0\le U-\sqrt N,\mathcal C_t(U,H)\}}
      {\rho_{N,t}(U)}
 \le C L_tN^Ce^{-(s-\vartheta_t)\sqrt N}
       +\frac{Cn^{-A_0}}{\rho_{N,t}(U)}
 \longrightarrow0 .
\]
Since \(\vartheta_t\to b<s\) and \(L_t\le N\), the first bound vanishes
uniformly; positive anchors retain \(e^{-\vartheta_tv}\).
For spatial freezing, let \(T_{\rm later}\) and \(T_{\rm earlier}\)
be the statistics at the two neighboring split points. Put
\(y_{\rm later}=T_{\rm later}-U\), and condition on
\[
 \{y_{\rm later}\in[y,y+\epsilon]\},\qquad
 \{T_{\rm earlier}\in[U,U+\ell]\}.
\]
The exact increment inclusions are
\[
 [-y,\ell-y-\epsilon]\ \subset\
 [U-T_{\rm later},U+\ell-T_{\rm later}]\
 \subset\ [-y-\epsilon,\ell-y].
\]
Use \eqref{eq:full-critical-anchor-interval} on each finite bin and the
exact reverse conditional chain, then let \(\epsilon\downarrow0\). The
preceding tail bounds permit the bin range to increase. The common
coefficient is identified by
\[
 \int_{\mathbb R}e^{-by}
    \Pp\{Z_G(h)\in[-y,\ell-y]\}\,dy
 =\frac{1-e^{-b\ell}}b\,\E e^{bZ_G(h)}
 =\frac{1-e^{-b\ell}}b .
\]
The final equality follows from
\[
 \log\E e^{bZ_G(h)}
 =h\left\{-b(q_c-1)+\frac{4b^2q_c^2}{q_c+1}\right\}=0,
 \qquad b=(1-q_c^{-2})/4 .
\]
Thus the anchor coefficients have ratio \(1+o(1)\) on \(H/N\) cells.
The bin inclusions retain dependence between height and increment;
reflection handles the right patch.
Repeating the bins for the maximum gives the fixed-\(H\) coefficient
\[
 \E\exp\left\{b\sup_{0\le z\le H}Z_G(z)\right\}.
\]
Its ratio to \(H\) tends to \(b(q_c-1)\); this is the drifted Brownian
calculation in \eqref{eq:rank-scan-s21}--\eqref{eq:rank-scan-s22}. Pair bounds control the sum over distinct cells. For two unit cells
at logit distance \(h\), define
\[
 \Pi_N(t,h;U)=
 \frac{\mathbb P\{\mathcal C_t(U,1)\cap\mathcal C_{t'}(U,1)\}}
 {\mathbb P\{\mathcal C_t(U,1)\}+\mathbb P\{\mathcal C_{t'}(U,1)\}},
 \qquad |\operatorname{logit}t'-\operatorname{logit}t|=h.
\]
For separations \(H/N\le h\le
h_0\), the middle exponential generator at \(s=b/2\) has the fixed negative
value
\[
 -(q_c-1)s+\frac{4q_c^2}{q_c+1}s^2
       =-\frac14b(q_c-1).
\]
The later anchor has exponentially concentrated \(Q,W\).
On \([0,h_0]\), comparison of the exact sampling bridge with independent
deletions gives
\[
 \sigma_{\rm deletion}^2=O(1/\log(1/h_0)),\qquad
 \|M\|+\|M^2\|\le C,\qquad
 \operatorname{tr}M+\operatorname{tr}M^2=O(N).
\]
Gaussian linearization and Doob therefore give
\[
 \Pp\left\{\sup_{h\le h_0}
       (|Q(h)-Q(0)|+|W(h)-W(0)|)>\delta N
       \middle|\ \sigma(B_1(t),\ldots,B_n(t))\right\}\le e^{-c_\delta N}.
\]
Conditioning deletions on their total costs at most \(C\sqrt n\).
Choose \(h_0\) small enough that the Gaussian-linearization
exponent exceeds this \(n^{1/2}\) loss. The conditional mean
changes each energy by \(O(h_0N)\), while cross terms satisfy
\[
 |\langle X,M\Delta X\rangle|
 \le\|M^{1/2}X\|\,\|M^{1/2}\Delta X\|.
\]
The anchor norm is \(O(\sqrt N)\), so this controls the product uniformly
for \(h\le h_0\). For pairs replace
\eqref{eq:full-critical-sublinear-cap} by
\[
 B\le\epsilon N,\qquad
 |Z_-|+|Z_+|\le\sqrt{\epsilon N},\qquad
 \mathbb P_{\rm anchor}
 \{B>\epsilon N\ \text{or}\ |Z_-|+|Z_+|>\sqrt{\epsilon N}\}
 \le Ce^{-c_\epsilon N}.
\]
Riccati and the middle lower deviation give this bound because \(D=o(N)\).
Choose \(h_0\) after \(\epsilon\). Over \(Nh_0\) local units,
\[
 \frac{\mathcal L_{N,t}^{\sigma}e^{sE^+}}{e^{sE^+}}
 \le C_s\epsilon,\qquad
 \mathbb P_{\rm anchor}\{(\mathcal T_{n,Nh_0}^-)^c\}
 \le N^Ce^{-c(\log N)^2}.
\]
The summation-by-parts generator remains strictly negative.
Integrating overshoots at \(b/2<\vartheta_t\) gives
\begin{equation}\label{eq:full-critical-near-pair}
 \Pi_N(t,h;U)\le C e^{-cNh}+N^C e^{-c'(\log N)^2},
 \qquad H/N\le h\le h_0 .
\end{equation}
The leading constant is uniform in \(D/\sqrt N\).

For \(h\ge h_0\), apply the three-category conditional bridge
coupling \eqref{eq:rank-scan-s25a}--\eqref{eq:rank-scan-s25e}.
Its middle correlation satisfies \(\rho\le e^{-h_0/2}<1\).
With \(c_\pm=1\pm\rho\), strict convexity gives a fixed
\(\delta_0>0\) such that
\[
 \lambda(bc_+/2)+\lambda(bc_-/2)
       \le\lambda(b)-\delta_0 .
\]
Diagonalize the middle modes while comparing only their endpoint
quadratics:
\[
 \frac{a(bc_\pm/2)}{2c_\pm}
 =\frac b4\frac{a(bc_\pm/2)}{bc_\pm/2}
 \le\frac{a(b)}4 .
\]
Since \(a(x)/x\) increases, the endpoint weight is at most the geometric
mean of its two single-anchor weights; the bulk pressure is unchanged.
On the cap,
\[
 e^{(b-\theta)B}\le e^{C_0},\qquad
 \Pi_N(t,h;U)\le N^Ce^{-\delta_0N}\quad(h\ge h_0).
\]
Here H\"older uses their common positive boundary law.
Choose cell length \(H/N\) and lower-sum gaps
\[
 g/N,\qquad g=\sqrt H .
\]
Subdivide into unit cells before \eqref{eq:full-critical-near-pair};
its constant then stays independent of \(H\). The upper and lower
cell sums yield
\[
 \begin{aligned}
 &\limsup_{N\to\infty}
 \left|\frac{\pi_N(U)}
 {N(q_c-1)\int_{\mathcal I_N}e^{-bd(t)}
                       \rho_{N,t}(U)\,dt/[t(1-t)]}-1\right|\\
 &\quad\le Cg/H+CH^2e^{-cg}
 +\limsup_{N\to\infty}
   \{N^{C_1}e^{-c_1(\log N)^2}+N^{C_2}e^{-c_2N}\}\\
 &\quad= Cg/H+CH^2e^{-cg}\longrightarrow0
                \quad(H\to\infty,\ g=\sqrt H).
 \end{aligned}
\]
The first term is omitted spatial mass; \(N/D\to\infty\) also
permits freezing on \(1/D\) layers. The coefficient is
\[
 b^{-1}\,b(q_c-1)=q_c-1 .
\]
This proves \eqref{eq:full-critical-scan-reduction}.
The original moment expansions now map studentized levels, uniformly for
trimmed \(t\) and \(U/N\to q_c\), to
\[
 U+m(t)+\frac{U/N-1}{4}v_0(t)+o(1)=U+d(t)+o(1).
\]
This follows by expanding the exact finite-\(n\) standard deviation. The
local scalar shift is \(e^{-bd(t)}\{1+o(1)\}\). It is the factor already
included in \eqref{eq:full-critical-scan-reduction}.

\subsubsection{Spatial integration throughout the positive approach}

\begin{lemma}[The complete positive critical scalar integral]
\label{lem:full-critical-spatial}
For \(U=Nq_c+D\), \(D/\sqrt N\to\infty\), and \(D=o(N)\),
\begin{equation}\label{eq:full-critical-spatial-equivalent}
 \int_{\mathcal I_N}e^{-bd(t)}
       \rho_{N,t}(U)\frac{dt}{t(1-t)}
 \sim \int_{[0,\infty)}e^{-bB}g_N(U-B)\,\mu(dB).
\end{equation}
This includes the relative-error comparison when the Gaussian and boundary
contributions are of the same order.
\end{lemma}

\begin{proof}
Put
\[
 s=D/\sqrt N,\qquad v_\eta=\eta(1-\eta),\qquad b_1=b'(\eta),
 \qquad x_t=\sqrt N\{q-q_c(t)\},
\]
and define the normalized spatial integral
\[
 \mathscr S_N(U)=e^{bU-N\lambda(b)}
 \int_{\mathcal I_N}e^{-bd(t)}\rho_{N,t}(U)\frac{dt}{t(1-t)}.
\]
For bounded \(x_t\), use the compact core formula. For \(x_t\to+\infty\),
apply \eqref{eq:assembled-boundary-scalar-density} with
\[
 \Lambda=N(q-q_c(t)),\qquad E=1.
\]
For \(x_t\to-\infty\), use the saddle \(\vartheta(q)<b_t\) and its
actual gap \(b_t-\vartheta(q)\). Every sequence has a subsequence in one
of these ranges; the corresponding positive envelope supplies domination.
For \(s\le\log N\), set
\[
 L=(\log N)^2,\qquad 0\le\tau=t-\eta\le L/\sqrt N .
\]
Proposition~\ref{prop:hardy-threshold-transversality} gives
\[
 b_t=b+b_1\tau+O(\tau^2),\quad
 q_c(t)=q_c+Vb_1\tau+O(\tau^2),\quad
 \alpha_t-\alpha_\eta=O(\tau).
\]
Uniform real and finite matching freeze the coefficient.
At an individual zero retain \(\mathcal H_{\alpha_{+1/2}(t)}\), since
\[
 |\alpha_{+1/2}(t)|\log N\le C L\log N/\sqrt N=o(1),\qquad
 \frac{\mathcal H_{\alpha_{+1/2}(t)}(Q)}{\mathcal H_0(Q)}=1+o(1).
\]
This gives the single confluent logarithmic exponent \(\beta_\eta\).
The bulk cubic error satisfies \(L^3/\sqrt N=o(1)\).
For \(x=s-Vb_1\sqrt N\,\tau\),
\[
 N\{\lambda(b_t)-\lambda(b)\}-(b_t-b)U
       =\frac{x^2-s^2}{2V}+o(1)
\]
uniformly on the strip. Combining the three scalar ranges gives the
strip integral. Define
\[
 \phi_V(z)=\frac{e^{-z^2/(2V)}}{\sqrt{2\pi V}},\qquad
 F_{\alpha_\eta}(x)=e^{x^2/(2V)}
       \int_0^\infty r^{\alpha_\eta-1}\phi_V(x-r)\,dr.
\]
After removing \(e^{N\lambda(b)-bU}\), that integral is
\begin{equation}\label{eq:full-critical-spatial-profile}
 \frac{P_\eta e^{-bd(\eta)}}{v_\eta b_1V\Gamma(\alpha_\eta)}
 N^{(\alpha_\eta-2)/2}\left(\frac{\log N}{2}\right)^{\beta_\eta}
 e^{-s^2/(2V)}
 \int_{s-Vb_1L}^{s}F_{\alpha_\eta}(x)\,dx\ \{1+o(1)\},
\end{equation}
The error is relative to this positive integral. On divergent scalar
subranges the effective logarithms are \(\log(\sqrt N|x|)\) or \(\log(\sqrt
N/|x|)\); both equal \((\log N)/2+O(\log L)\). Around \(x=0\) use the core
formula, without inserting \(\log|x|\).
The identity
\[
 e^{x^2/(2V)}\phi_V(x-r)
       =(2\pi V)^{-1/2}e^{xr/V-r^2/(2V)}
\]
gives
\[
 F_{\alpha_\eta}(x)\sim
 \begin{cases}
 \Gamma(\alpha_\eta)V^{\alpha_\eta}/
       \{\sqrt{2\pi V}|x|^{\alpha_\eta}\},&x\to-\infty,\\
 e^{x^2/(2V)}x^{\alpha_\eta-1},&x\to+\infty.
 \end{cases}
\]
Laplace's endpoint estimate and the negative-tail asymptotic give
\[
 \int_0^s F_{\alpha_\eta}(x)\,dx
 \sim V e^{s^2/(2V)}s^{\alpha_\eta-2},\qquad
 \int_{-L}^{0}F_{\alpha_\eta}(x)\,dx
 \le C
 \begin{cases}
 L^{1-\alpha_\eta},&\alpha_\eta<1,\\
 \log L,&\alpha_\eta=1,\\
 1,&\alpha_\eta>1.
 \end{cases}
\]
The positive contribution to
\eqref{eq:full-critical-spatial-profile} is \(w(D)/2\):
its \(V\) cancels the \(V^{-1}\) in the spatial Jacobian.
Define its negative-\(x\) contribution by
\[
 I_N^-=
 \frac{P_\eta e^{-bd(\eta)}}{v_\eta b_1V\Gamma(\alpha_\eta)}
 N^{(\alpha_\eta-2)/2}\left(\frac{\log N}{2}\right)^{\beta_\eta}
 e^{-s^2/(2V)}\int_{s-Vb_1L}^{0}F_{\alpha_\eta}(x)\,dx .
\]
The preceding bounds imply
\[
 \begin{aligned}
 I_N^-/g_{N,b}(U)
 &\le CN^{(\alpha_\eta-1)/2}(\log N)^{\beta_\eta} L^{1-\alpha_\eta}=o(1),
 &&\alpha_\eta<1,\\
 \frac{I_N^-}{(\log N)^{\beta_\eta+1}g_{N,b}(U)}
 &\le C\log L/\log N=o(1),&&\alpha_\eta=1,\\
 I_N^-/w(D)&\le Cs^{2-\alpha_\eta}e^{-s^2/(2V)}=o(1),
 &&\alpha_\eta>1.
 \end{aligned}
\]
Reflection supplies the other half of the boundary coefficient.
Let \(\tau(t)=\min(t-\eta,1-\eta-t)\) and define
\[
 I_N^{\rm comp}(U)=e^{bU-N\lambda(b)}
 \int_{\mathcal I_N\cap\{\tau(t)\ge L/\sqrt N\}}
       e^{-bd(t)}\rho_{N,t}(U)\frac{dt}{t(1-t)} .
\]
On this integration range,
\(\sqrt N\{b_t-\vartheta(q)\}\to\infty\), so the moving saddle uses
\(\Psi_t(\vartheta(q))\). For \(\alpha_\eta<1\), dominated convergence gives
\(I_N^{\rm comp}(U)=\Theta_0g_{N,b}(U)\{1+o(1)\}\).
At an individual zero, the logarithmic factor depends on the local gap:
\[
 b_t-\vartheta(q)\asymp\tau,\qquad
 \alpha_{+1/2}(t)\log(1/\tau)=O\{\tau\log(1/\tau)\}\to0 .
\]
Equation~\eqref{eq:core-confluent-real-coefficient} then gives
\[
 \Psi_t(\vartheta(q))
 \le C\tau^{-\alpha_\eta}\{1+\log(1/\tau)\}^2,\qquad
 \int_0^\varepsilon\tau^{-\alpha_\eta}
                   \{1+\log(1/\tau)\}^2\,d\tau<\infty,
 \quad \alpha_\eta=\alpha_{-1/2}(\eta)<1.
\]
The confluent mass stays undivided. For \(\alpha_\eta=1\), integration gives
\[
 I_N^{\rm comp}(U)=
 P_0\left(\frac{\log N}{2}-\log L\right)^{\beta_\eta+1}
             g_{N,b}(U)\{1+o(1)\}
       =G_N(U)\{1+o(1)\}.
\]
Here \(\log s,\log L=O(\log\log N)\), so \(\log(1/\kappa)=(\log
N)/2+O(\log\log N)\). For \(\alpha_\eta>1\),
\[
 I_N^{\rm comp}(U)
 \le C N^{(\alpha_\eta-2)/2}(\log N)^{\beta_\eta}
       L^{1-\alpha_\eta}e^{-s^2/(2V)}=o(w(D)).
\]
The compact interior is smaller. Thus
\[
 \frac{|\mathscr S_N(U)-G_N(U)-w(D)|}
      {G_N(U)+w(D)}\longrightarrow0,
\]
proving \eqref{eq:full-critical-kernel-two-parts} for \(s\le\log N\),
including competition. For \(s>\log N\), put
\[
 u=D(t-\eta).
\]
At bounded \(u\), the positive scalar law gives
\[
 \frac{\rho_{N,t}(U)}{\rho_{N,\eta}(U)}
           \longrightarrow e^{-b_1u}.
\]
The logit and studentization factors can be frozen. For
\(0\le u\le A\log N\), the positive scalar formula gives
\[
 \frac{N(q-q_c(t))}{D}=1+O(u/s^2),\qquad
 \log\frac{\rho_{N,t}(U)}{\rho_{N,\eta}(U)}
       =-b_1u+O(u^2/s^2)+o(u)+O(1)
       \le-cu+C .
\]
On the retained strip,
\[
 \sup_{0\le u\le A\log N}\frac u{s^2}=o(1),\qquad
 |\alpha_{+1/2}(t)|\log D=O(\log^2N/D)=o(1)
\]
at an individual zero; fixed-sign prefactors are bounded.
The coarse cell estimate preceding \eqref{eq:full-critical-adaptive-tilt}
excludes outside crossings without a scalar denominator.
The retained envelope \(Ce^{-cu}\), dominated convergence and reflection give
\[
 \int_{\mathcal I_N}e^{-bd(t)}
       \rho_{N,t}(U)\frac{dt}{t(1-t)}
 \sim e^{N\lambda(b)-bU}
     \frac{2P_\eta e^{-bd(\eta)}}{v_\eta b_1\Gamma(\alpha_\eta)}
           D^{\alpha_\eta-2}(\log D)^{\beta_\eta} .
\]
The exact rate in the deterministic kernel gives \(G_N=o(w(D))\) in this
range. This proves \eqref{eq:full-critical-spatial-equivalent}.
\end{proof}

\subsubsection{The full critical MAX normalization}

Combining \eqref{eq:full-critical-scan-reduction} and
Lemma~\ref{lem:full-critical-spatial} proves for every positive approach
\(D/\sqrt N\to\infty\), \(D=o(N)\),
\begin{equation}\label{eq:full-critical-original-scan}
 \pi_N(Nq_c+D)\sim\mathcal R_N^{\rm fc}(Nq_c+D).
\end{equation}
For \(|h|\le C\log N\),
\[
 \left|\left(E+\frac h\Lambda\right)-E\right|
       \le\frac{C\log N}{\sqrt N}\longrightarrow0,\qquad
 \frac{|h|}{\sqrt N}\le\frac{C\log N}{\sqrt N}\longrightarrow0 .
\]
Thus the positive scalar comparison and enlarged deterministic caps remain
valid. The uniform level versions of the core and lower laws give the same
shift range.

\begin{proof}[Proof of Theorem~\ref{thm:max-full-critical}]
Lemma~\ref{lem:full-critical-kernel} and the lower theorem cover bounded
and negative offsets; \eqref{eq:full-critical-original-scan} covers positive
divergence. Every subsequence has one of these further limits.
For \(U_{0,n}=Ny_n\) in \eqref{eq:main-critical-normalization},
\[
 \begin{cases}
 I(y_n)=\gamma_n,&\gamma_n\le\gamma_c,\\
 by_n-\lambda(b)=\gamma_n,&\gamma_n>\gamma_c.
 \end{cases}
\]
Write \(U=U_{0,n}\) and choose fixed \(B_*\) beyond the cutoff in
\(\mu\), so \(\mu(dz)=w(z)\,dz\) for \(z\ge B_*\). Its explicit density gives,
for every fixed \(C_0\),
\[
 \mu([0,C_0N])\le N^{C_1},\qquad
 \inf_{B_*\le z\le C_0N}w(z)\ge N^{-C_1},\qquad
 \mu([B_*,B_*+1])\ge c>0.
\]
If \(U=Ny\le Nq_c\), then \(NI(y)=\log p\) and
\[
 \begin{aligned}
 p g_N(U)&=\frac{1}{2\sqrt\pi y^{3/2}\sqrt N},\\
 \frac{e^{-bz}g_N(U-z)}{g_N(U)}
 &=(1-z/U)^{-3/2}
   \exp\!\left\{-(b-\vartheta(y))z
          -\frac{N^2z^2}{4U^2(U-z)}\right\}\le C,
       \qquad 0\le z<U .
 \end{aligned}
\]
For \(z\le U/2\), the prefactor is bounded. For \(z>U/2\), the
last exponential is at most \(e^{-cN/(1-z/U)}\). The same ratio is
bounded below on \([B_*,B_*+1]\); integrating therefore gives
\[
 c\sqrt N\le
 N(q_c-1)p g_N(U)
       \int_{[0,U)}\frac{e^{-bz}g_N(U-z)}{g_N(U)}\,\mu(dz)
 =p\mathcal R_N^{\rm fc}(U)\le C N^{C_1+1/2}.
\]
If \(U=Nq_c+D\ge Nq_c\), then \(p e^{-bU+N\lambda(b)}=1\).
The exact density \(g_{N,b}\) satisfies
\[
 \sup_{v>0}g_{N,b}(v)\le C/\sqrt N,\qquad
 \inf_{B_*\le z-D\le B_*+1}g_{N,b}(U-z)\ge c/\sqrt N .
\]
Here \(U-z\in[Nq_c-B_*-1,Nq_c-B_*]\) in the second bound.
Integrating first over all \([0,U)\), and then over
\([D+B_*,D+B_*+1]\), gives
\[
 cN^{1/2-C_1}\le
 N(q_c-1)\int_{[0,U)}g_{N,b}(U-z)\,\mu(dz)
 =p\mathcal R_N^{\rm fc}(U)\le C N^{C_1+1/2}.
\]
Both cases prove
\begin{equation}\label{eq:full-critical-prefactor-polynomial}
 N^{-C}\le p\mathcal R_N^{\rm fc}(U_{0,n})\le N^C .
\end{equation}
and hence \(H_n=O(\log N)\). With
\(H_n=\zeta_n^{-1}\log\{p\mathcal R_N^{\rm fc}(U_{0,n})\}\),
the shift identity gives
\[
 \begin{aligned}
 p\mathcal R_N^{\rm fc}(U_{0,n}+H_n+x/\zeta_n)
 &=p\mathcal R_N^{\rm fc}(U_{0,n})
              e^{-\zeta_nH_n-x}\{1+o(1)\}\\
 &\longrightarrow e^{-x}.
 \end{aligned}
\]
For the original coordinate event \(E_{j,n}(x)\),
\[
 \mathbb P(E_{j,n}(x))=p^{-1}e^{-x}\{1+o(1)\},\qquad
 E_{j,n}(x)\in\sigma(Y_{1j},\ldots,Y_{nj})
\]
by Proposition~\ref{prop:rank-law}.
Under the stated dependence conditions,
Proposition~\ref{prop:sparse-factorization} yields
\[
 \sum_{j=1}^p\mathbf1_{E_{j,n}(x)}
       \Rightarrow\operatorname{Poisson}(e^{-x}),\qquad
 \mathbb P\!\left(\sum_j\mathbf1_{E_{j,n}(x)}=0\right)\to e^{-e^{-x}} .
\]
The threshold includes \(e^{-bd(t)}\), completing exact studentization.
\end{proof}

\subsection{The original-rank scan at fixed positive boundary excess}
\label{app:full-boundary-scan}

Write $T_{n,k}=nF_{n,k}$ and $N=\log n$. Throughout this subsection
\[
 b=b(\eta)<\tfrac14,\qquad q_c=\lambda'(b),\qquad
 y-q_c\in[E_0,E_1]\Subset(0,\infty).
\]
Every limit is uniform on this compact level range. A small fixed
neighborhood of the left trim is chosen so that $y-q_c(t)\ge E_0/2$.
We first treat its inward, increasing-time scan. Reflection gives the
right trim. The cutoff $m=N^K$ is a proof device; $K$ is fixed
sufficiently large after all error exponents are selected.

\subsubsection{Scalar input, four ends, and spatial localization}

Theorem~\ref{thm:finite-hybrid-boundary-scalar} with $\Lambda=N$, followed by
Lemma~\ref{lem:original-rank-boundary-scalar-transfer}, gives
\begin{equation}\label{eq:full-boundary-anchor}
 \begin{split}
 \Pp\{T_{n,k}\in[Ny+z,Ny+z+h]\}
  &\sim A_N(t,y)e^{-b(t)z}\frac{1-e^{-b(t)h}}{b(t)},\\
 \Pp\{T_{n,k}>Ny+z\}
  &\sim A_N(t,y)e^{-b(t)z}/b(t),\\
 A_N(t,y)
  &=\frac{P_t}{\Gamma(\alpha_t)}
       [N\{y-q_c(t)\}]^{\alpha_t-1}
       [\log(N\{y-q_c(t)\})]^{\beta_t}
       e^{-N\{b(t)y-\lambda(b(t))\}} .
 \end{split}
\end{equation}
Here \(h>0,z\) are fixed; crossings retain the positive coalescent expression.
The original interval sandwich and scalar level uniformity allow
\[
 h_N=O(\log N),\qquad
 Ny+h_N=N(y+\Delta y_N),\qquad
 \Delta y_N=h_N/N=O(\log N/N).
\]
Let \(\xi_*\) be the active binary label and \(c\) the rank-weight
shift. The four endpoint parameter triples are
\[
 (t,\xi_*,c)\in
 \bigl(\{\eta\}\times\{1\}\times\{-\tfrac12,\tfrac12\}\bigr)
 \cup
 \bigl(\{1-\eta\}\times\{0\}\times\{-\tfrac12,\tfrac12\}\bigr).
\]
Time reflection preserves \(c\). Each rank endpoint contributes
\[
 (\alpha_c^{\rm use},\beta_c,P_c)=
 \begin{cases}
 (\alpha_c,\frac12,D_c\Gamma(\alpha_c)C_U^{-\alpha_c}),
                                      &\alpha_c>0,\\
 (0,\frac32,2D_c/3),                  &\alpha_c=0,\\
 (0,0,K_c(\eta,b)),                   &\alpha_c<0 ,
 \end{cases}
\]
with the constants of \eqref{eq:core-boundary-exponents} and
\eqref{eq:core-first-failure-constant}. Consequently
\begin{equation*}
 \alpha_\eta=\sum_{c=\pm1/2}\alpha_c^{\rm use}>0,\qquad
 \beta_\eta=\sum_{c=\pm1/2}\beta_c,\qquad
 P_\eta=q_c^{-1/2}P_{-1/2}P_{+1/2}.
\end{equation*}
The middle contributes \(q_c^{-1/2}\). Positive powers combine through
\[
 \frac{x^{a_1-1}}{\Gamma(a_1)}
       *\frac{x^{a_2-1}}{\Gamma(a_2)}
       =\frac{x^{a_1+a_2-1}}{\Gamma(a_1+a_2)},\qquad a_1,a_2>0,
\]
giving the single Gamma factor in \eqref{eq:full-boundary-anchor}.
The finite factors are \(K_c\); zero factors have logarithmic power
\(3/2\). Only the two temporal trims contribute an external factor two.
Put \(u(t)=\operatorname{logit}t,\ v_t=t(1-t)\). Transversality gives
\begin{equation*}
 \frac{d}{dt}\{b(t)y-\lambda(b(t))\}
       =b_+'(t)\{y-q_c(t)\}>0 .
\end{equation*}
For $u(t_j)=u(\eta)+j/N$, bounded $j$ therefore satisfies
\begin{equation*}
 \frac{\Pp(T_{n,nt_j}>Ny)}{\Pp(T_{n,n\eta}>Ny)}
 \longrightarrow e^{-jv_\eta b_+'(\eta)(y-q_c)} .
\end{equation*}
Grid rounding has size $O(1/n)$ and causes no constant.

For a neighborhood bound, first suppose \(\alpha_+(\eta)\ne0\)
and choose the neighborhood to preserve its sign. Its scalar
coefficient ratio is at most \(Ce^{Cj\log N/N}\).
When \(\alpha_+(\eta)=0\), retain the single finite mass:
\[
 \mathcal H_{\alpha_+(t)}(Q)
       =\int_0^Qe^{\alpha_+(t)w}w^{1/2}\,dw,\qquad Q=\log N.
\]
For \(j\le N/Q\), bounded confluence gives
\[
 \frac{\mathcal H_{\alpha_+(t_j)}(Q)}{\mathcal H_0(Q)}
 \le e^{|\alpha_+(t_j)|Q}\le e^{CjQ/N}.
\]
For \(j>N/Q\), use only the finite positive transform at
\(\theta_t=b(t)-1/N\). Conditional on a retained boundary
pattern, the middle density is at most \(C/\sqrt N\); positive
averaging and untilting therefore give
\[
 \Pp(T_{n,nt}>Ny)
 \le N^C e^{-N\{b(t)y-\lambda(b(t))\}} .
\]
Because \(j>N/Q\gg\log N\), the polynomial factor is absorbed
by the increasing spatial rate. Both ranges yield
\begin{equation}\label{eq:full-boundary-space-upper}
 \frac{\Pp(T_{n,nt_j}>Ny)}{\Pp(T_{n,n\eta}>Ny)}
       \le C e^{-cj+Cj\log N/N},\qquad j\le c_0N .
\end{equation}
The mixed transform in \eqref{eq:core-negative-thinning}, together
with the stopped middle generator, gives the same coarse bound for
scan cells in the second range. Near a zero exponent the scalar
equivalents are used only on bounded-confluence families.

\subsubsection{The boundary Gibbs profile under interval conditioning}

Fix an anchor interval from \eqref{eq:full-boundary-anchor} and write
\[
 \Pp_n^{\rm anc}(B)=
 \Pp\{B\mid T_{n,k}\in[Ny+z,Ny+z+h]\},\qquad t=k/n .
\]
Here \(h>0\) and \(z\) are fixed. In the boundary-process argument
write \(\mathbb P_{\rm anchor}=\mathbb P_n^{\rm anc}\).
All probabilistic orders in this subsubsection refer to this conditional law. When a retained Gaussian
middle is used, its auxiliary coupling variables are included.

Use the earlier normalized critical orbit \(r_t\), its control
\(f_t=u_{t,b(t)}\circ r_t\), and
\(\mathcal J_t=\int_0^\infty h_t(r_t(s))\,ds\).
Its normalization fixes the dilation in
Proposition~\ref{prop:critical-hardy-profile-envelope}.
For \(i\in\{-,+\}\), define
\[
 \mathcal C_{i,n}(s)=N^{-1}C^i_{\lfloor Ns\rfloor},\qquad
 \Phi_t(x)=\log s_t(x),\qquad J_*=\lfloor N^d\rfloor,
 \quad 0<d<1/4,
\]
and let \(\tau_i=\inf\{l:l-C_l^i=J_*\}\). Put
\[
 A_{i,n}=
 \frac{\tau_i}{N}\exp\{-\Phi_t(C^i_{\tau_i}/\tau_i)\}
 \quad\text{if }\tau_i\le m,\ C^i_{\tau_i}/\tau_i>t;
 \qquad A_{i,n}=0\quad\text{otherwise}.
\]
Under \(\mathbb P_n^{\rm anc}\), \(Y_\circ/N\to q_c(t)\) in
probability, and for each \(0<\epsilon<S<\infty\),
\begin{equation}\label{eq:full-boundary-Gibbs}
 \begin{gathered}
 \sup_{\epsilon\le s\le S}
 |\mathcal C_{i,n}(s)-s r_t(s/A_{i,n})|=o_P(1),\\
 A_{-,n}+A_{+,n}=\frac{y-q_c(t)}{\mathcal J_t}+o_P(1).
 \end{gathered}
\end{equation}
When \(A_{i,n}=0\), the expression \(s r_t(s/A_{i,n})\) means \(ts\).
Only the sum of allocations is used below.
Under the true retained tilt,
\[
 \theta_N=b(t)-c/N,\qquad
 \mathcal M_{N,t}(\theta_N)=e^{N\lambda(\theta_N)}N^{O(1)},\qquad
 \mathbb Q_{n,t,\theta_N}^{\mathcal A}
 \{\widetilde T_n\in[Ny+z,Ny+z+h]\}\ge N^{-C}.
\]
Theorem~\ref{thm:finite-hybrid-boundary-scalar} gives success probability
\(u_{\theta_N}(S_l/l)+O(l^{-1})\) on compact states.
Near the active endpoint, with \(j=l-S_l\), its failure hazard satisfies
\[
 \frac{p_l}{w_{\theta_N}(S_l/l)}
 =1+O\{j^{-1}+(1+\log l)^3/l\}.
\]
This is obtained from the two neighboring positive continuation
ratios and the finite row expansion.

Work with one prefix, so \(S_l=C_l^i\) and \(\tau=\tau_i\).
For the normalized orbit,
\[
 \Phi_t'(x)=\{u_{t,b(t)}(x)-x\}^{-1},\qquad
 P_l=\log l-\Phi_t(S_l/l).
\]
Write \(x_l=S_l/l\), \(j_l=l-S_l\),
\(u_l=\mathbb E_{H_{\theta_N}}(\xi_{l+1}\mid\mathscr F_l)\), and define
\[
 \begin{aligned}
 \Delta\mathscr M_{l+1}
   &=P_{l+1}-P_l-
           \mathbb E_{H_{\theta_N}}(P_{l+1}-P_l\mid\mathscr F_l),\\
 R_l&=\mathbb E_{H_{\theta_N}}(P_{l+1}-P_l\mid\mathscr F_l)
       -\frac{1-\Phi_t'(x_l)(u_l-x_l)}l .
 \end{aligned}
\]
The critical first-order term vanishes when \(u_l=u_{t,b(t)}(x_l)\).
The finite Taylor expansion gives
\[
 |\Delta\mathscr M_{l+1}|\le C/j_l,\qquad
 \mathbb E[(\Delta\mathscr M_{l+1})^2\mid\mathscr F_l]
 \le C\{p_l/j_l^2+l^{-2}\},\qquad
 |R_l|\le C\{p_l/j_l^2+(1+\log l)^3/l^2\}.
\]
For \(\tau_j=\inf\{l:J_l=j\}\), put
\(A_j=\sum_{\tau_j\le l<\tau_{j+1}}p_l\). The survival product gives
\[
 \mathbb P_{H_{\theta_N}}(A_j>x\mid\mathscr F_{\tau_j})
       \le\min(1,e^{1-x}).
\]
Let \(\tau_{\rm c}\) be the first time after \(\tau\) at which
the cumulative energy exceeds \(C_{\rm cap}N\), or \(x_l\le t+\rho\),
or \(l=m\). Multiply the hazard bounds with weights \(j^{-2}\),
then apply the bounded-jump inequality:
\[
 \Pp\!\left\{\sum_{l\ge\tau}\frac{p_l}{j_l^2}>
                         \frac C{J_*}\right\}\le e^{-cJ_*},
 \qquad
 \mathbb P_{H_{\theta_N}}\left\{
 \sup_{\tau\le v\le\tau_{\rm c}}
 \left|\sum_{l=\tau}^{v-1}(\Delta\mathscr M_{l+1}+R_l)\right|
       >J_*^{-1/4}\right\}\le e^{-cJ_*^{1/2}},
\]
after absorbing logarithmic factors. Randomness before \(\tau\) remains
the orbit dilation. The lower entropy bound, apart from \(O(J_*)\) ranks, gives
\[
 \tau\le CN,\qquad l_{\rm entry}\le C_\rho N .
\]
These justify stopping before any orbit approximation.
Moreover,
\[
 1-u_{\theta_N}(1)=O\{N^{-1}(\log N)^{1/b}\},
\]
so the tilt changes phase by \(N^{-d}\) times logarithms.
A positive allocation has post-\(\tau\) energy comparable to
\(\tau\{\log(\tau/J_*)\}^{1/b}\). Hence
\[
 \tau\{\log(\tau/J_*)\}^{1/b}\le CN
 \quad\Longrightarrow\quad
 \tau=O(N/(\log N)^{1/b}),\qquad
 \sum_{l=1}^{\tau}\frac{l}{l+c}h_t(S_l/l)=o(N).
\]
In the central tube let $a=a(b(t))<1/2$ and $A_l=S_l-lt$.
For small fixed $\epsilon>0$ and a central start $L$,
\begin{equation*}
 E(A_l^2\mid A_L)\le
 Cl+C A_L^2(l/L)^{2a+\epsilon}+Cl^2e^{-cL},
 \qquad L\le l\le m .
\end{equation*}
The exit term is retained on this polynomial horizon.
The exponential tube version gives a $\sqrt l\log N$ noise envelope.
Put \(k_*=1-a>1/2\), \(a_l=l/(l+c)\), and
\begin{align*}
 d_t^+(x)&=t(1-x)h_t'(x)-(x-t),\\*
 d_t^-(x)&=(x-t)-(1-t)x h_t'(x),\displaybreak[1]\\
 B_{n,S}&=\frac1N\sum_{SN<l\le m}a_lh_t(x_l),\displaybreak[1]\\
 I_{n,S}&=\frac1N\sum_{SN<l\le m}a_l
       \{|d_t^+(x_l)|+|d_t^-(x_l)|\}\\*
 &\quad+\frac1N\sum_{i=1}^m
 \left\{\sum_{\max(i,\lfloor SN\rfloor+1)\le l\le m}
              \frac{a_lh_t'(x_l)}l\right\}^2 .
\end{align*}
The central moment bound and the adjoint Hardy inequality give
\begin{equation}\label{eq:full-boundary-profile-tail}
 \max(B_{n,S},I_{n,S})
 \le CS^{1-2k_*+\epsilon}+o_P(1),\qquad
 0<\epsilon<2k_*-1 .
\end{equation}
Precisely, for every fixed \(S,\zeta>0\),
\[
 \mathbb P_n^{\rm anc}
 \{\max(B_{n,S},I_{n,S})>CS^{1-2k_*+\epsilon}+\zeta\}\to0 .
\]
Take \(n\to\infty\) before \(S\to\infty\).
Put \(E_l=\sum_{v=1}^l v(v+c)^{-1}h_t(S_v/v)\).
Independently of phase, the terminal tube follows from
\[
 E_l\ge cl|S_l/l-t|^3-C .
\]
Late ranks enter the central chart, where the killed recursion gives
\[
 \E e^{\epsilon A_m^2/m}\le C,\qquad
 \exp\{C|S_m/m-t|A_m^2/m\}
       \le \exp\{\epsilon A_m^2/m\}
\]
after increasing fixed \(K\); the second bound absorbs terminal matching.
Put \(\mathbb Q=\mathbb Q_{n,t,\theta_N}^{\mathcal A}\).
Consequently, for the middle energy \(Y_\circ\),
\[
 \begin{aligned}
 \mathbb Q\{|Z_m^-|+|Z_m^+|>2\log N\}
       &\le Ce^{-c(\log N)^2},\\
 \mathbb E_{\mathbb Q}(Y_\circ\mid\mathcal F_{n,m}^{\partial})
       &=Nq_c(t)+O((\log N)^2),\\
 \operatorname{Var}_{\mathbb Q}(Y_\circ\mid\mathcal F_{n,m}^{\partial})
       &=O(N).
 \end{aligned}
\]
The last two estimates hold on \(\max_\pm|Z_m^\pm|\le\log N\).
Continuation and anchor division cost only \(N^C\), transferring the phase,
tail and energy limits to \eqref{eq:full-boundary-Gibbs}.
Allocations below \(\delta\) carry at most \(C\delta N\) energy;
let \(\delta\downarrow0\).

\subsubsection{Conditional rank flips and the two L\'evy processes}

Fix a left anchor and write outer logit time as \(z/N\).
In this local calculation abbreviate \(b=b(t)\), \(q_c=q_c(t)\).
Let \(\Pi_i\) be the chronological row index of pooled rank \(i\), and
\(B_i(t)=\mathbf1_{\{\Pi_i\le k\}}\), \(t=k/n\).
Given \(B(t)=(B_1(t),\ldots,B_n(t))\), zero locations are uniform
among future times and one locations among past times. Hence
the respective selection probabilities are
\[
 \begin{aligned}
 \mathbb P\{k<\Pi_i\le k_t^+(dz)\mid B(t)\}
   &=t\,dz/N+o(N^{-1}),&& B_i(t)=0,\\
 \mathbb P\{k_t^-(dz)<\Pi_i\le k\mid B(t)\}
   &=(1-t)\,dz/N+o(N^{-1}),&& B_i(t)=1 .
 \end{aligned}
\]
For disjoint scaled blocks, sampling without replacement gives convergence
of all fixed joint factorial moments to their Poisson products.
Conditioning on the label-measurable anchor preserves temporal ordering:
\[
 \mathcal L(\Pi\mid B(t),\,T_{n,k}\in[Ny+z,Ny+z+h])
       =\mathcal L(\Pi\mid B(t)).
\]
A zero at scaled rank \(s\) changes entropy by
\[
 g(s)=\int_s^\infty h_t'(r_t(w))\,\frac{dw}{w}
     =\{\operatorname{logit}f_t(s)-\operatorname{logit}t\}/b(t)>0.
\]
On each \([\epsilon,S]\), the entropy Taylor remainder is
uniformly small; the half shifts \(c=\pm1/2\) change the
limiting jump and drift by \(o(1)\). Along a subsequence with \(A_{i,n}\Rightarrow A_i\), the two
conditional Lévy measures are
\[
 \begin{aligned}
 \nu_{i,+}(dx)&=A_i\int_0^\infty
          t(1-f_t(s))\,\delta_{g(s)}(dx)\,ds,\\
 \nu_{i,-}(dx)&=A_i\int_0^\infty
          (1-t)f_t(s)\,\delta_{-g(s)}(dx)\,ds .
 \end{aligned}
\]
Here \(\delta_a\) denotes unit point mass at \(a\).
Independent superposition sums \(A_-\) and \(A_+\), so only
their total in \eqref{eq:full-boundary-Gibbs} enters the limit.

Compensation precedes the infinite-rank limit. For
$g_S(s)=\int_s^S h_t'(r_t(w))\,dw/w$, Fubini gives
\[
 \int_0^S t(1-f_t)g_S\,ds=\int_0^S t(1-r_t)h_t'(r_t)\,ds .
\]
Combining this with the explicit parameter derivative gives $D_t^+$.
Backward compensation gives
\[
 D_t^-=\int_0^\infty\{(r_t(s)-t)-(1-t)r_t(s)h_t'(r_t(s))\}\,ds<0.
\]
Here
\[
 D_t^+=\int_0^\infty
 \{t(1-r_t(s))h_t'(r_t(s))-(r_t(s)-t)\}\,ds .
\]
Both compensated drift integrands are quadratic at infinity.
The profile tail gives
\[
 g(s)=O(s^{-k_*}),\qquad
 \int_S^\infty g(s)^2\,ds=O(S^{1-2k_*}),
\]
and \eqref{eq:full-boundary-profile-tail} transfers this
control to finite ranks. Near zero, \(g(s)=O(s^{-b(t)})\).
Choose \(\epsilon_N=(\log N)^{-M}\), \(1<M<1/b(t)\), and
\(L_0=\lfloor\epsilon_NN\rfloor\). For a zero label at rank \(i\), its finite entropy increment is
\[
 g_{i,n}=\sum_{l=i}^{m}\frac l{l+c}
       \left\{h_t\!\left(\frac{C_l+1}{l}\right)-h_t(C_l/l)\right\}.
\]
Write \(g_i=g_{i,n}\). This difference obeys
\[
 g_i\le C\log L_0\{1+\log(L_0/i)\}+g_{L_0},\qquad
 g_{L_0}\le C\epsilon_N^{-b(t)}.
\]
For every fixed \(p<M\),
\[
 \frac1N\sum_{i\le L_0}|g_i|^p
 \le C_p\epsilon_N(\log N)^p
          +C_p\epsilon_N^{1-b(t)p}\longrightarrow0 .
\]
Because \(b(t)<1/4\), choose \(p>3\). The early selections satisfy
\[
 \mathbb P_n^{\rm anc}
 \{\exists\,i\le\lfloor\epsilon N\rfloor:
             k<\Pi_i\le k_t^+(H)\}\le C_H\epsilon .
\]
Thus truncation and polynomial moments control large jumps without
increasing the forward critical exponent.
At the strict Gaussian tilt \(b(t)<1/4\),
Appendix~\ref{app:rank-scan-subcritical} gives the middle limit
\[
 Z_0(z)=\mathscr M_0(z)-(q_c(t)-1)z,\qquad
 \mathscr M_0(z)=\sqrt{\frac{8q_c(t)^2}{q_c(t)+1}}\,\mathcal W(z),
 \qquad
 \langle\mathscr M_0\rangle_z=\frac{8q_c(t)^2}{q_c(t)+1}z .
\]
The retained middle starts at \(m\gg N\), so a fixed boundary core
contributes \(o(1)\) to it. Its conditional characteristic function
factors from the finite Poisson cores. Let \(J_n^\sigma\) be the boundary increment process. Its
compensated jump representation is
\[
 J_n^\sigma(z)=
 \sum_i\int_0^z g_{i,n}^\sigma(v)
        \{dN_i^\sigma(v)-\lambda_i^\sigma(v)\,dv\}
        +\int_0^z d_n^\sigma(v)\,dv ,
\]
where \(N_i^\sigma\) counts the oriented flip at rank \(i\),
\(\lambda_i^\sigma\) is its conditional rate from
\eqref{eq:rank-local-uniform-time-generator}, and
\(d_n^\sigma=\mathcal L_{N,t}^\sigma E_\partial\).
Define \(J_n^{\sigma;\epsilon,S}\) by restricting these finite rank sums,
including the corresponding drift sums, to
\(\epsilon N\le i,l\le SN\).
For every fixed \(H,\zeta>0\), the preceding bounds give
\[
 \lim_{\epsilon\downarrow0}\lim_{S\uparrow\infty}\limsup_n
 \mathbb P_n^{\rm anc}
 \{\sup_{0\le z\le H}|J_n^\sigma(z)-J_n^{\sigma;\epsilon,S}(z)|
                                        >\zeta\}=0 .
\]
Hence the full Brownian and jump processes factor. Let \(Z_\pm\)
be their two oriented Lévy limits, started at zero, and define their
Laplace exponents by
\(\mathbb E e^{sZ_\pm(z)}=\exp\{z\psi_\pm(s)\}\).
They are
\begin{align*}
 \psi_+(s)&=-(q_c-1)s+\frac{4q_c^2}{q_c+1}s^2\nonumber\\
 &\quad+\frac{y-q_c}{\mathcal J_t}\left[
 D_t^+s+\int_0^\infty t(1-f_t(w))
          (e^{s g(w)}-1-sg(w))\,dw\right],
          &&0\le s\le b,                            \\
 \psi_-(s)&=-(q_c-1)s+\frac{4q_c^2}{q_c+1}s^2\nonumber\\
 &\quad+\frac{y-q_c}{\mathcal J_t}\left[
 D_t^-s+\int_0^\infty(1-t)f_t(w)
          (e^{-s g(w)}-1+sg(w))\,dw\right],
          &&s\ge0 .                                
\end{align*}
The profile envelope gives
\begin{equation}\label{eq:full-boundary-Levy-duality}
 \begin{gathered}
 t(1-f_t(w))e^{b g(w)}=(1-t)f_t(w)\quad(w>0),\qquad
 \psi_-(s)=\psi_+(b-s)-\psi_+(b)\quad(0\le s\le b),\\
 \psi_+(b)=-v_t b_+'(t)(y-q_c)<0,\qquad
 \psi_+'(0)=-(q_c-1)+(y-q_c)D_t^+/\mathcal J_t<0.
 \end{gathered}
\end{equation}
For $s>b$ the reverse exponent is defined directly by its displayed integral.

\subsubsection{Reverse crossing, finite windows, and the analytic supremum}

Choose a fixed interval \(I=[\eta,\eta+2\epsilon_0]\) so that the
active endpoint and the strict opposite-end margin persist on \(I\).
Fix \(t_0=k_0/n\in[\eta,\eta+\epsilon_0]\), \(H>0\), and define the
inward process on its trimmed domain by
\[
 T(z)=T_{n,\operatorname{round}
       [n\operatorname{logistic}\{\operatorname{logit}t_0+z/N\}]}.
\]
Use the actual integer endpoints
\[
 \begin{gathered}
 k_H=\operatorname{round}\!\left[
 n\operatorname{logistic}\{\operatorname{logit}t_0+H/N\}\right],
 \qquad t_H=k_H/n,\\
 H_n=N(\operatorname{logit}t_H-\operatorname{logit}t_0)
       =H+O(N/n),\\
 z_k=N\{\operatorname{logit}(k/n)-\operatorname{logit}t_0\},
 \quad r_k=H_n-z_k,\quad k_0\le k\le k_H,\\
 T_H=T_{n,k_H},\qquad
 T_H^{\rm R}(r_k)=T_{n,k},\qquad
 R_H(r_k)=T_H^{\rm R}(r_k)-T_H .
 \end{gathered}
\]
Suprema over \(r\) below mean maxima over this grid, with a constant
interpolation between its points. In particular,
\[
 \max_{k_0\le k\le k_H}T_{n,k}
       =T_H+\max_{0\le r\le H_n}R_H(r).
\]
The anchor is always the terminal statistic. For \(h>0\), define
\[
 I_H(u,h)=\{T_H\in[u,u+h)\},\qquad
 \mathbb P_H^{u,h}(B)=\mathbb P(B\mid I_H(u,h)).
\]
Write \(\mathbb E_H^{u,h}\) for expectation under this conditional law.
Fix a compact interval
\[
 \mathcal Y\Subset(\sup_{t\in I}q_c(t),\infty)
\]
containing the level range above and its sufficiently small fixed
enlargement. Scalar equivalents and conditional bounds are used
on fixed-sign compact intervals, or along sequences with
\[
 |\alpha_c(t_0)|\log N\le C_0\quad\bigl(c=\pm\tfrac12,\ \alpha_c(\eta)=0\bigr),
 \qquad C_0<\infty.
\]
The unnormalized bounds are uniform on \(I\).
The closed-interval scalar law also gives, for fixed
\(0<a\le1\),
\[
 \limsup_n\sup_{u/N\in\mathcal Y}
 \frac{\mathbb P(T_H=u)}{A_N(t_H,u/N)}
 \le
 \limsup_n\sup_{u/N\in\mathcal Y}
 \frac{\mathbb P(T_H\in[u-a,u+a])}{A_N(t_H,u/N)}
 \le C a.
\]
Letting \(a\downarrow0\) shows that half-open intervals have the same
scalar equivalent. Their probabilities are positive for all
sufficiently large \(n\).

Let \(B^H\) be the complete label vector at \(k_H\). Conditional on
\(B^H\), the chronological locations of its one labels are a uniform
permutation of \(\{1,\ldots,k_H\}\). The terminal interval is
\(B^H\)-measurable, so the original backward transition is exactly
\[
 \mathbb E\{F((k-1)/n,B_{k-1})\mid B_k\}
   =\frac1k\sum_{i:B_{k,i}=1}F((k-1)/n,B_k-e_i).
\]
The same identity holds after conditioning on \(I_H(u,h)\).
Write \(\mathbb E_{\rm del}\) for expectation over this conditional
uniform permutation. Thus increasing \(r\) deletes a uniformly
selected existing one.
The scaled finite-count generator is
\[
 \mathcal L_{n,N,t}^{-,d}F
 =\frac{1-t}{N}\sum_{i:B_i=1}
       \{F(t-1/n,B-e_i)-F(t,B)\},\qquad t=k/n.
\]
Its continuous counterpart is \(\mathcal L_{N,t}^{-}\) in
\eqref{eq:rank-local-uniform-time-generator}.

Put \(t_r=k/n\) at \(r=r_k\). Let \(C_l^c(r)\) be the prefix counts
and \(B_i^{H,c}\) the terminal labels at rank end \(c=\pm1/2\).
With the middle matrix \(M\) from \eqref{eq:rank-scan-s3}, define
\[
 \begin{gathered}
 X=B_k-t_r\mathbf1,\qquad
 Q_m(r)=\frac{X^\top MX}{t_r(1-t_r)},\\
 C_l^c(0)=\sum_{i\le l}B_i^{H,c},\qquad
 A_l^{\rm R}(r)=\sum_{i\le l}B_{k,i}-lt_r.
 \end{gathered}
\]
The three energy components are
\[
 E_H^\pm(r)=
 \sum_{c=\pm1/2}\sum_{l\le m}\frac l{l+c}
          h_{t_r}^{\pm}(C_l^c(r)/l),\qquad
 V_H(r)=\tfrac12Q_m(r),\qquad
 \overline T_H(r)=V_H(r)+E_H^+(r)+E_H^-(r).
\]
Couple the terminal middle \(A_l^{\rm R}(0)\) to \(G_l^H\), sampling
the auxiliary variables conditionally on \(B^H\) independently of its
chronological permutation. For a prescribed \(A>0\), use
\[
 \begin{split}
 \mathcal C_{n,H}^{\rm R}
 &=\left\{\max_{m\le l\le n-m}
                   |A_l^{\rm R}(0)-G_l^H|\le C_AN\right\}\\
 &\quad\cap\bigcap_{l=m}^{n-m}
 \left\{\sup_{0\le r\le H_n}|A_l^{\rm R}(r)|
                     \le C_A\sqrt{(l\wedge(n-l))N}\right\},\\
 \mathcal A_{n,H}
 &=\{|Z_m^-(t_H)|\vee|Z_m^+(t_H)|\le B\sqrt N\},\\
 \mathcal K_{n,H}^{\rm R}
 &=\mathcal C_{n,H}^{\rm R}\cap\mathcal A_{n,H}
     \cap\left\{\max_{k_0\le k\le k_H}T_{n,k}\le C_{\rm cap}N\right\}.
 \end{split}
\]
Choose \(B\) to include the terminal count envelope. The coupling and
deterministic entropy estimates give
\[
 \sup_{0\le r\le H_n}
       |T_H^{\rm R}(r)-\overline T_H(r)|
 \le \epsilon_n,\qquad
 \epsilon_n=C_A\{N^{3/2}/\sqrt m+N^2/m+m^2/n\}=o(1)
\]
on \(\mathcal C_{n,H}^{\rm R}\). The global rate and the coupling
bound imply, for any prescribed \(A_0\),
\[
 \mathbb P((\mathcal K_{n,H}^{\rm R})^c)\le n^{-A_0},
\]
after increasing \(C_{\rm cap},A,B\). These exceptions are removed as
unweighted probabilities. The retained Gaussian middle is untruncated
conditional on each prefix pattern; an energy-cap indicator is placed
only in a numerator and can be removed for an upper bound.

Let \(\tau_H\) be the first exit from the energy cap or the middle
count envelope. For a component \(W\), put
\[
 M_W=\sup_{0\le r\le H_n}
           [W(r\wedge\tau_H)-W(0)]_+,\qquad
 M_H^0=M_{V_H}+M_{E_H^+},\qquad M_H^-=M_{E_H^-},
\]
and define the actual reverse maximum by
\[
 M_H^{\rm R}=\sup_{0\le r\le H_n}
           [T_H^{\rm R}(r\wedge\tau_H)-T_H]_+.
\]
On \(\mathcal K_{n,H}^{\rm R}\), the stopping time exceeds \(H_n\) and
\[
 M_H^{\rm R}\le M_H^0+M_H^-+2\epsilon_n.
\]

The proof of \eqref{eq:core-positive-boundary-generator} applies to
this actual reverse orientation throughout the \(O(N)\) cap.
For \(F=e^{aE_H^+}\), its exact finite-count correction is
\[
 \begin{split}
 \mathfrak R_{n,N,t}^-F
 &=-\frac{1-t}{N}\sum_{i:B_i=1}\int_0^{1/n}
       [\partial_tF(t-v,B-e_i)-\partial_tF(t,B)]\,dv,\\
 \frac{|\mathfrak R_{n,N,t}^-F|}{F}
 &\le \eta_{n,a}:=\frac1n e^{C_{a,K}(\log N)^2}.
 \end{split}
\]
Consequently the exact scaled generator satisfies
\[
 \frac{\mathcal L_{n,N,t}^{-,d}e^{aE_H^+}}{e^{aE_H^+}}
 \le C_a\{E_H^+/N+\log m/N\}+\eta_{n,a}\le C_a
\]
before the cap, for every fixed \(a>0\).

For the middle put \(h_i=(MX)_i\),
\(v_t=t(1-t)\), and
\(\Gamma^-(W)=\mathcal L^-_{N,t}(W^2)-2W\mathcal L^-_{N,t}W\).
The matrix satisfies \(M\mathbf1=0\), \(\|M\|\le C\), and
\(\operatorname{tr}M\le CN\). The exact continuous-time identities are
\[
 \begin{split}
 \mathcal L^-_{N,t}V_H
   &=\frac1{2N}\left\{-Q_m+\frac1t\sum_{i:B_i=1}M_{ii}\right\},\\
 \Gamma^-(V_H)
   &=\frac{1-t}{4Nv_t^2}
             \sum_{i:B_i=1}(-2h_i+M_{ii})^2
     \le \frac C N\{W_m+\operatorname{tr}(M^2)\}\le C .
 \end{split}
\]
Here \(W_m=X^\top M^2X/v_t\le\|M\|Q_m\le CN\) on the cap.
The occupancy estimate and the exact variance change give
\[
 \max|\Delta V_H|\le C\{\sqrt{N/m}+N/n\},\qquad
 \frac{|\mathfrak R_{n,N,t}^-e^{aV_H}|}{e^{aV_H}}
       \le \eta_{n,a}.
\]
Using \(e^{ax}-1-ax\le a^2x^2e^{a|x|}/2\) proves the same bounded
exponential generator for \(V_H\), for every fixed \(a>0\).

To convert the generator bound into a maximum bound, distinguish
the deletion index from its logit clock:
\[
 \begin{gathered}
 J_H=k_H-k_0,\qquad \rho_j=r_{k_H-j}\quad(0\le j\le J_H),\\
 \nu_H=\min\bigl(\{j:\rho_j\ge\tau_H\}\cup\{J_H\}\bigr),\qquad
 W_j=W(\rho_j),\qquad \Delta W_j=W_{j+1}-W_j,\\
 \rho_{j+1}-\rho_j
   =\frac{N}{nt_{\rho_j}(1-t_{\rho_j})}+O(N/n^2).
 \end{gathered}
\]
Let \(\mathscr F_j\) contain \(B^H\), the first \(j\) deleted labels,
and the terminal auxiliary variables. The exact scaled generator gives
nonnegative bounds \(b_{a,j}\) satisfying, before \(\nu_H\),
\[
 \mathbb E(e^{a\Delta W_j}\mid\mathscr F_j)
       \le e^{b_{a,j}},\qquad
 b_{a,j}\le C_a(\rho_{j+1}-\rho_j),\qquad
 \sum_{j<\nu_H}b_{a,j}\le C_a(H+1).
\]
Hence
\[
 L_j=\exp\!\left\{
 a(W_{j\wedge\nu_H}-W_0)
       -\sum_{i<j\wedge\nu_H}b_{a,i}\right\}
\]
is a nonnegative supermartingale. Stopping at its first level
crossing, for \(0<s<a\),
\[
 \begin{split}
 \mathbb P(M_W>x\mid B^H)
       &\le e^{C_a(H+1)-a x},\\
 \mathbb E(e^{sM_W}\mid B^H)
       &=1+s\int_0^\infty e^{sx}
                   \mathbb P(M_W>x\mid B^H)\,dx\\
       &\le1+\frac{s e^{C_a(H+1)}}{a-s}.
 \end{split}
\]
The component \(M_H^0\) thus admits every fixed exponent with
a constant independent of the terminal label vector on the cap.
H\"older's inequality separates its two components.

The opposite-end threshold is
\[
 b_t^-=
 \left\{\sup_{\substack{f:[0,\infty)\to[0,t]\\0<I_t(f)<\infty}}
                \frac{J_t(f)}{I_t(f)}\right\}^{-1}.
\]
It exceeds \(b(t)\) uniformly on the chosen interval. Shrink \(I\)
if necessary and choose
\[
 b_{\max}:=\sup_{t\in I}b(t)<s<s_2<\inf_{t\in I}b_t^-,
 \qquad 1<p_-<s_2/s,\quad p_+=p_-/(p_--1).
\]
Define the negative good tube at the terminal-oriented path by
\[
 \mathcal T_{n,H}^{-,{\rm R}}
 =\left\{\sup_{0\le r\le H_n}
       \max_{c=\pm1/2}\max_{l\le m}
       \frac{[lt_r-C_l^c(r)]_+}{\sqrt l}\le C\log N\right\}.
\]
To bound this tube, put \(\ell_N=\log N\),
\(L_0=\lceil C\ell_N^2\rceil\), and
\(B_N=C(1+\log\ell_N)^2\). For a deletion let \(\Delta_iE_H^-\) be its
energy change with \(t\) held fixed. The finite \(x\log x\) difference,
including counts equal to one, and Hardy's inequality give
\[
 \begin{gathered}
 E_H^-\le C_K(1+\ell_N)^3,\qquad
 \max_{i:B_i=1}|\Delta_iE_H^-|\le B_N,\\
 \sum_{i:B_i=1}(\Delta_iE_H^-)^2
       \le C\{L_0B_N^2+\ell_N^2\log m\}
       \le C_K(1+\ell_N)^3 .
 \end{gathered}
\]
The convex Taylor trace is at most \(C\log m\): for each affected
count \(C_l\ge1\), multiplication by its \(C_l\) possible deleted
labels cancels the \(x^{-1}\) singularity in the integral remainder.
After summing \(l^{-1}\), the drift and jump inequalities yield
\[
 \begin{split}
 &|\mathcal L^-_{N,t}E_H^-|+\Gamma^-(E_H^-)
       +|e^{-aE_H^-}\mathcal L^-_{N,t}e^{aE_H^-}|\\
 &\quad\le \frac{C_{a,K}}N(1+\log N)^3
             e^{C_{a,K}(1+\log\log N)^2},\\
 \xi_{n,a}
 &:=
 \frac{C_{a,K}}N(1+\log N)^3
             e^{C_{a,K}(1+\log\log N)^2}
       +\frac1n e^{C_{a,K}(\log N)^2}\longrightarrow0 .
 \end{split}
\]
The second term is the exact finite-count correction. Stopping also
at the first tube exit and applying the preceding supermartingale
argument bounds its exponential maximum by a constant.

The retained mixed transform controls the complementary tube.
At the terminal split put
\[
 b_H=b(t_H),\qquad \theta_H=b_H-c_0/N,\qquad c_0>0.
\]
For \(r\le H_n\), the deletion fraction is
\(d(r)=1-t_r/t_H=O_H(N^{-1})\). Define \(D_i^c(r)\) to be the
indicator that terminal one label \(i\) in end \(c\) has been deleted
by time \(r\), and put it equal to zero when \(B_i^{H,c}=0\).
Equation
\eqref{eq:core-negative-thinning}, with
\(\alpha_0=t_r/t_H\), implies
\[
 E_H^-(r)-E_H^-(0)
 \le\varepsilon E_H^-(0)
   +C_\varepsilon\sum_{c=\pm1/2}
      \sum_{l\le m}l^{-2}
       \left(\sum_{i\le l}(D_i^c(r)-d(r))B_i^{H,c}\right)^2 .
\]
For comparison, take independent Bernoulli\((d(r))\) indicators
on the terminal one-label locations. Put
\[
 (\mathsf H_m)_{ij}
       =\sum_{\max(i,j)\le l\le m}l^{-2},\qquad
 \xi_i^c(r)=B_i^{H,c}\{D_i^c(r)-d(r)\}.
\]
Then each quadratic form is
\((\xi^c(r))^\top\mathsf H_m\xi^c(r)\), with
\[
 \|\mathsf H_m\|\le4,\qquad
 \operatorname{tr}\mathsf H_m=O(\log m),\qquad
 \log\mathbb E e^{u(D_i-d)}
       \le \sigma_d^2u^2/2,\quad \sigma_d^2=C/\log(1/d).
\]
For each fixed \(a>0\), independent Gaussian linearization gives
\[
 \begin{split}
 \mathbb E e^{a\xi^\top\mathsf H_m\xi}
 &\le\det(I-2a\sigma_d^2\mathsf H_m)^{-1/2}\\
 &\le\exp\!\left\{
 \frac{a\sigma_d^2\operatorname{tr}\mathsf H_m}
       {1-2a\sigma_d^2\|\mathsf H_m\|}\right\}\le C_{a,H}.
 \end{split}
\]
At \(d=0\), \(\xi=0\). With
\(P_c=\operatorname{diag}(B_1^{H,c},\ldots,B_m^{H,c})\),
\[
 \|P_c\mathsf H_mP_c\|\le4,\qquad
 \operatorname{tr}(P_c\mathsf H_mP_c)
       \le\operatorname{tr}\mathsf H_m.
\]
Condition on \(B^H\). Index its distinguished one labels by
\[
 \mathcal I_H=\{(c,i):c=\pm\tfrac12,\ 1\le i\le m,
                                  \ B_i^{H,c}=1\},\qquad
 \mathfrak m_H=|\mathcal I_H|\le2m=o(k_H).
\]
Let \(J_\iota\in\{1,\ldots,k_H\}\) be the deletion index of label
\(\iota\in\mathcal I_H\). Under \(\mathbb P_{\rm ind}\), these indices
are independent and uniform; under \(\mathbb P_{\rm del}\), they are distinct and uniform. Thus
\[
 \begin{gathered}
 \mathbf J=(J_\iota)_{\iota\in\mathcal I_H},\qquad
 (k_H)_{\mathfrak m_H}
       =\prod_{v=0}^{\mathfrak m_H-1}(k_H-v),\\
 \frac{d\mathbb P_{\rm del}}{d\mathbb P_{\rm ind}}(\mathbf J\mid B^H)
 =\frac{k_H^{\mathfrak m_H}}{(k_H)_{\mathfrak m_H}}
       \mathbf1_{\{|\{J_\iota:\iota\in\mathcal I_H\}|=\mathfrak m_H\}},\\
 \log\frac{k_H^{\mathfrak m_H}}{(k_H)_{\mathfrak m_H}}
 =\sum_{v=0}^{\mathfrak m_H-1}-\log(1-v/k_H)
 \le\frac{2}{k_H}\sum_{v=0}^{\mathfrak m_H-1}v
 \le Cm^2/n .
 \end{gathered}
\]
Set \(J_{(c,i)}=k_H+1\) when \(B_i^{H,c}=0\). For either mark law,
put
\[
 \begin{gathered}
 d_j=j/k_H,\qquad
 \xi_{j,i}^c=B_i^{H,c}
              \{\mathbf1_{\{J_{(c,i)}\le j\}}-d_j\},\qquad
 \boldsymbol\xi_j^c=(\xi_{j,i}^c)_{i=1}^m,\\
 \mathcal Q_j=\sum_{c=\pm1/2}
                 (\boldsymbol\xi_j^c)^\top\mathsf H_m\boldsymbol\xi_j^c,\\
 \mathcal E_j^-=
 \sum_{c=\pm1/2}\sum_{l\le m}\frac l{l+c}
 h_{(k_H-j)/n}^-
 \!\left(\frac1l\sum_{i\le l}B_i^{H,c}
                         \mathbf1_{\{J_{(c,i)}>j\}}\right),
 \qquad 0\le j\le J_H .
 \end{gathered}
\]
Under the deletion law, \(\mathcal E_j^-=E_H^-(\rho_j)\).
The thinning inequality gives
\[
 \mathcal E_j^--E_H^-(0)
       \le\varepsilon E_H^-(0)+C_\varepsilon\mathcal Q_j,
 \qquad d_j\le J_H/k_H=O_H(N^{-1}).
\]
When no distinguished label is deleted, its counts are constant and
\(t\) decreases. Since
\[
 \partial_t h_t^-(x)=\frac{(t-x)_+}{t(1-t)}\ge0,
\]
only the initial value and the distinguished deletion indices can
attain a new maximum. In particular,
\[
 \begin{split}
 M_H^-&\le\max_{0\le j\le J_H}
                   [\mathcal E_j^--E_H^-(0)]_+,\\
 e^{aM_H^-}
 &\le e^{a\varepsilon E_H^-(0)}
   \left[1+\sum_{\iota\in\mathcal I_H}
       \mathbf1_{\{J_\iota\le J_H\}}
                         e^{aC_\varepsilon\mathcal Q_{J_\iota}}\right].
 \end{split}
\]
For \(\iota=(c_\iota,i_\iota)\) and \(1\le j\le J_H\), force its
indicator at index \(j\) to equal one in the independent model:
\[
 \begin{gathered}
 \zeta_{\iota,j}=1-\mathbf1_{\{J_\iota\le j\}}\in\{0,1\},\qquad
 \boldsymbol\xi_j^{c,\iota}
 =\boldsymbol\xi_j^c+
   \mathbf1_{\{c=c_\iota\}}\zeta_{\iota,j}e_{i_\iota},\\
 \mathcal L_{\rm ind}
       ((\boldsymbol\xi_j^{c,\iota})_{c=\pm1/2}\mid B^H)
 =\mathcal L_{\rm ind}
       ((\boldsymbol\xi_j^c)_{c=\pm1/2}\mid J_\iota=j,B^H).
 \end{gathered}
\]
Positivity of \(\mathsf H_m\) implies
\[
 \begin{split}
 (\xi+\zeta e_i)^\top\mathsf H_m(\xi+\zeta e_i)
 &\le2\xi^\top\mathsf H_m\xi
                       +2\zeta^2(\mathsf H_m)_{ii}\\
 &\le2\xi^\top\mathsf H_m\xi+C/i,\qquad |\zeta|\le1,\\
 \sum_{c=\pm1/2}(\boldsymbol\xi_j^{c,\iota})^\top
                 \mathsf H_m\boldsymbol\xi_j^{c,\iota}
 &\le2\mathcal Q_j+C/i_\iota .
 \end{split}
\]
Apply the preceding determinant estimate to the two independent
blocks with exponent \(2aC_\varepsilon\). Uniformly in \(j\) and \(\iota\),
\[
 \begin{split}
 \mathcal M_{\iota,j}
 &:=\mathbb E_{\rm ind}
      [e^{aC_\varepsilon\mathcal Q_j}\mid J_\iota=j,B^H]\\
 &\le e^{CaC_\varepsilon/i_\iota}
      \mathbb E_{\rm ind}[e^{2aC_\varepsilon\mathcal Q_j}\mid B^H]
       \le C_{a,\varepsilon,H,K}.
 \end{split}
\]
The likelihood comparison is applied to this nonnegative marked
upper bound, which no longer depends on the stopping rule. It yields
\[
 \begin{split}
 \mathbb E_{\rm del}[e^{aM_H^-}\mid B^H]
 &\le e^{Cm^2/n+a\varepsilon E_H^-(0)}
  \left[1+\sum_{\iota\in\mathcal I_H}\sum_{j=1}^{J_H}
                      \frac{\mathcal M_{\iota,j}}{k_H}\right]\\
 &\le e^{Cm^2/n+a\varepsilon E_H^-(0)}
          \left[1+C_{a,\varepsilon,H,K}
                         \mathfrak m_H\frac{J_H}{k_H}\right]\\
 &\le N^C e^{a\varepsilon E_H^-(0)} .
 \end{split}
\]
Here \(a,\varepsilon,H,K\) are fixed; \(\varepsilon\) is subsequently
chosen within the opposite-end margin.

Choose \(s_2\varepsilon\) below the uniform opposite-end margin.
The finite positive continuation with mixed potential
\(\theta_Hh_{t_H}^++(\theta_H+s_2\varepsilon)h_{t_H}^-\),
proved in Appendix~\ref{app:critical-core}, applies with the local
gap \(b_H-\theta_H=c_0/N\). Let \(\widetilde T_H\) be the retained
hybrid at this terminal split. It yields
\[
 \mathbb E_{\rm hyb}
 \left[e^{\theta_H\widetilde T_H
                  +s_2\varepsilon E_H^-(0)};\mathcal A_{n,H}\right]
       \le N^C e^{N\lambda(\theta_H)}.
\]
For the moment transfer, let
\[
 \begin{split}
 \mathcal C_H^0
 &=\left\{\max_{m\le l\le n-m}|A_l^{\rm R}(0)-G_l^H|\le C_AN\right\}\\
 &\quad\cap\bigcap_{l=m}^{n-m}
 \{|A_l^{\rm R}(0)|\le C_A\sqrt{(l\wedge(n-l))N}\}
 \cap\{T_H\le C_{\rm cap}N\}.
 \end{split}
\]
Conditioning first on the terminal labels and their auxiliary
variables, for \(0<a\le s_2\), gives
\[
 \begin{split}
 &\mathbb E[e^{\theta_HT_H+aM_H^-};I_H(u,h)\cap\mathcal K_{n,H}^{\rm R}]\\
 &\quad\le N^C\mathbb E[e^{\theta_HT_H+a\varepsilon E_H^-(0)};
                                     \mathcal C_H^0\cap\mathcal A_{n,H}]\\
 &\quad\le N^Ce^{\theta_H\epsilon_n}
       \mathbb E_{\rm hyb}[e^{\theta_H\widetilde T_H+a\varepsilon E_H^-(0)};
                                     \mathcal A_{n,H}].
 \end{split}
\]
Removal of the coupling indicator uses the pointwise comparison on \(\mathcal C_H^0\) and leaves its complement unweighted.
For \(u/N\in\mathcal Y\) and fixed \(h>0\), the scalar lower bound is
\[
 \inf_{u/N\in\mathcal Y}
 e^{\theta_Hu-N\lambda(\theta_H)}
       \mathbb P(I_H(u,h))\ge c_hN^{-C}.
\]
Consequently,
\[
 \begin{split}
 \mathbb E_H^{u,h}[e^{s_2M_H^-};\mathcal K_{n,H}^{\rm R}]
 &\le\frac{e^{-\theta_Hu}
       \mathbb E[e^{\theta_HT_H+s_2M_H^-};I_H(u,h)\cap\mathcal K_{n,H}^{\rm R}]}
             {\mathbb P(I_H(u,h))}
 \le N^C.
 \end{split}
\]
The negative count-tube estimate proved with
\eqref{eq:core-negative-thinning}, divided by this same scalar lower
bound, gives
\[
 \mathbb P_H^{u,h}((\mathcal T_{n,H}^{-,{\rm R}})^c)
       \le N^Ce^{-c(\log N)^2}.
\]
The reserved exponent now yields
\[
 \begin{split}
 &\mathbb E_H^{u,h}
 [e^{sp_-M_H^-};
       (\mathcal T_{n,H}^{-,{\rm R}})^c\cap\mathcal K_{n,H}^{\rm R}]\\
 &\quad\le
 \{\mathbb E_H^{u,h}
       [e^{s_2M_H^-};\mathcal K_{n,H}^{\rm R}]\}^{sp_-/s_2}
 \{\mathbb P_H^{u,h}
       ((\mathcal T_{n,H}^{-,{\rm R}})^c)\}^{1-sp_-/s_2}
 \le N^Ce^{-c'(\log N)^2}.
 \end{split}
\]
On the good tube the stopped generator supplies a fixed constant.
Combining \(M_H^0\) and the negative component by H\"older's inequality gives
\begin{equation}\label{eq:full-boundary-reverse-crossing}
 \sup_{u/N\in\mathcal Y}
 \mathbb E_H^{u,h}
       [e^{sM_H^{\rm R}};\mathcal K_{n,H}^{\rm R}]
       \le C_{H,h},\qquad s>b_{\max}.
\end{equation}
The constant may depend on the fixed bin width \(h\).

Conditional H\"older and the terminal comparison also give
\[
 \begin{split}
 &\mathbb E[e^{\theta_HT_H+sM_H^{\rm R}};\mathcal K_{n,H}^{\rm R}]\\
 &\quad\le N^C e^{2s\epsilon_n}
       \mathbb E[e^{\theta_HT_H+s\varepsilon E_H^-(0)};
                            \mathcal C_H^0\cap\mathcal A_{n,H}]\\
 &\quad\le N^C e^{(\theta_H+2s)\epsilon_n}
       \mathbb E_{\rm hyb}[e^{\theta_H\widetilde T_H+s\varepsilon E_H^-(0)};
                            \mathcal A_{n,H}]
 \le N^C e^{N\lambda(\theta_H)}.
 \end{split}
\]

Let \(U=Ny\), \(p_0(U)=\mathbb P(T_{n,k_0}>U)\), and
\(\mathcal C_H(U)=\{\max_{k_0\le k\le k_H}T_{n,k}>U\}\).
Choose \(0<\delta<E_0/4\). For unit bins
\(I_H(U-\ell-1,1)\), \(R\le\ell\le\delta N\), the scalar law at
level \(y-(\ell+1)/N\) and compact positivity of its prefactor give
\[
 \frac{\mathbb P(I_H(U-\ell-1,1))}{p_0(U)}
       \le C_H e^{b_H(\ell+1)}.
\]
The prefactor comparison here is between two strict positive excesses
in a fixed compact interval. At a zero individual exponent the
positive coalescent finite mass is retained. On such a bin a crossing
requires \(M_H^{\rm R}\ge\ell\); \eqref{eq:full-boundary-reverse-crossing} gives
\[
 \begin{split}
 &\frac{\mathbb P\{\mathcal C_H(U),\
               U-\delta N\le T_H\le U-R,\
               \mathcal K_{n,H}^{\rm R}\}}{p_0(U)}\\
 &\qquad\le C_H\sum_{\ell\ge R-1}
           e^{b_H(\ell+1)-s\ell}
       \le C_H'e^{-(s-b_{\max})R}.
 \end{split}
\]
For deeper negative offsets, conditioning is on the terminal statistic
\(T_H\). On
\[
 \mathcal D_{N,H}
 =\{T_H\le U-\delta N,\ \mathcal C_H(U)\}
       \cap\mathcal K_{n,H}^{\rm R},
\]
we have \(M_H^{\rm R}\ge U-T_H\), whence
\[
 \begin{split}
 \theta_HT_H+sM_H^{\rm R}
 &\ge \theta_HU+(s-\theta_H)(U-T_H)\\
 &\ge \theta_HU+(s-\theta_H)\delta N,\\
 \mathbb P(\mathcal D_{N,H})
 &\le N^C e^{N\lambda(\theta_H)-\theta_HU
                          -(s-\theta_H)\delta N}.
 \end{split}
\]
Since \(t_H-t_0=O_H(N^{-1})\),
\[
 N\{\lambda(\theta_H)-\lambda(b(t_0))
                  -(\theta_H-b(t_0))y\}=O_H(1).
\]
After division by \(p_0(U)\), the last bound is
\(N^Ce^{-(s-\theta_H)\delta N+O_H(1)}=o(1)\).
For the positive offsets no crossing estimate is needed:
\[
 \limsup_n\frac{\mathbb P(T_H>U+R)}{p_0(U)}
       \le C_H e^{-b(t_0)R}.
\]
The discarded cap and coupling probabilities divided by \(p_0(U)\)
are at most
\[
 N^C\exp\{-N[A_0-\sup_{(t,y)\in I\times\mathcal Y}
                              \{b(t)y-\lambda(b(t))\}]\}=o(1),
\]
by choosing \(A_0\) larger than the displayed compact supremum.
These bounds establish uniform integrability of the terminal-bin
crossing probabilities.

Along a convergent admissible parameter sequence, let \(Z_\pm\)
denote the two L\'evy processes at its limiting parameters and put
\[
 \begin{gathered}
 t_0\longrightarrow t_*,\qquad y\longrightarrow y_*,\qquad b=b(t_*),\\
 M_-=\sup_{0\le r\le H}Z_-(r),\qquad
 \mathscr G_H(v)=\mathbb P(M_->-v).
 \end{gathered}
\]
The function \(\mathscr G_H\) is nondecreasing. Its continuity set
\[
 \mathcal D=\{v\in\mathbb R:\mathbb P(M_-=-v)=0\}
\]
is dense, because a probability distribution has at most countably
many atoms. Choose a finite partition
\[
 \begin{gathered}
 R>0,\qquad \{-R,R\}\subset\mathcal D,\qquad
 \Pi=\{-R=v_0<v_1<\cdots<v_J=R\}\subset\mathcal D,\\
 h_j=v_{j+1}-v_j,\qquad
 |\Pi|=\max_{0\le j<J}h_j,\qquad
 w_j=\int_{v_j}^{v_{j+1}}e^{-bv}\,dv .
 \end{gathered}
\]
Define its terminal bins and the truncated crossing probability by
\[
 \mathcal I_{j,n}=I_H(U+v_j,h_j),\qquad
 P_{n,R}=
 \frac{\mathbb P\{\mathcal C_H(U)\cap\mathcal K_{n,H}^{\rm R},
                         \ -R\le T_H-U<R\}}
      {A_N(t_H,y)}.
\]
For each \(j\), the exact bin inclusions are
\[
 \begin{split}
 \mathcal I_{j,n}\cap\mathcal K_{n,H}^{\rm R}
                   \cap\{M_H^{\rm R}>-v_j\}
 &\subseteq\mathcal I_{j,n}\cap\mathcal K_{n,H}^{\rm R}
                   \cap\mathcal C_H(U)\\
 &\subseteq\mathcal I_{j,n}\cap\mathcal K_{n,H}^{\rm R}
                   \cap\{M_H^{\rm R}>-v_{j+1}\}.
 \end{split}
\]
For this fixed partition, the scalar interval law and the terminal
Gibbs-profile and reverse-label limits give
\[
 \begin{gathered}
 \frac{\mathbb P(\mathcal I_{j,n})}{A_N(t_H,y)}
       \longrightarrow w_j,\\
 \mathbb P_H^{U+v_j,h_j}
   \{\mathcal K_{n,H}^{\rm R},\,M_H^{\rm R}>-a\}
       \longrightarrow \mathscr G_H(a),
       \qquad a\in\{v_j,v_{j+1}\}.
 \end{gathered}
\]
Every width \(h_j>0\) is fixed when \(n\to\infty\).
The terminal intervals are disjoint, so, with
\[
 L_\Pi=\sum_{j=0}^{J-1}w_j\mathscr G_H(v_j),\qquad
 U_\Pi=\sum_{j=0}^{J-1}w_j\mathscr G_H(v_{j+1}),
\]
the inclusions yield
\[
 L_\Pi\le\liminf_nP_{n,R}
       \le\limsup_nP_{n,R}\le U_\Pi .
\]
Monotonicity gives a quantitative bound on the difference of the sums:
\[
 \begin{split}
 L_\Pi&\le\int_{-R}^{R}e^{-bv}\mathscr G_H(v)\,dv\le U_\Pi,\\
 0\le U_\Pi-L_\Pi
 &=\sum_{j=0}^{J-1}w_j\{\mathscr G_H(v_{j+1})-\mathscr G_H(v_j)\}\\
 &\le e^{bR}|\Pi|\{\mathscr G_H(R)-\mathscr G_H(-R)\}
 \le e^{bR}|\Pi|.
 \end{split}
\]
Taking partitions with \(|\Pi|\downarrow0\) therefore proves
\[
 \lim_nP_{n,R}=\int_{-R}^{R}e^{-bv}\mathscr G_H(v)\,dv .
\]
Bounded truncation under a unit terminal interval and
\eqref{eq:full-boundary-reverse-crossing} also give
\[
 \begin{split}
 \mathbb E e^{sM_-}
 &=\sup_{L\ge1}\lim_n
   \mathbb E_H^{U,1}
    [(e^{sM_H^{\rm R}}\wedge L);\mathcal K_{n,H}^{\rm R}]\\
 &\le C_{H,1},\qquad s>b.
 \end{split}
\]

The comparison of the terminal density with the initial tail uses the
rate derivative and endpoint displacement:
\[
 \begin{gathered}
 \partial_t\{b(t)y-\lambda(b(t))\}
       =b_+'(t)\{y-q_c(t)\},\qquad
 t_H-t_0=\frac{H_n}{N}v_{t_0}+O_H(N^{-2}),\\
 \frac{A_N(t_H,y)}{A_N(t_0,y)}
       =\exp\{-H v_{t_0}b_+'(t_0)(y-q_c(t_0))+o(1)\}
       =e^{H\psi_+(b)+o(1)}.
 \end{gathered}
\]
Here \(A_N\) uses the positive coalescent scalar expression when
an individual exponent approaches zero. In that case
\[
 \begin{gathered}
 |\alpha_c(t_H)-\alpha_c(t_0)|\ell_N=O_H(\log N/N),\\
 e^{-|\alpha_c(t_H)-\alpha_c(t_0)|\ell_N}
 \le\frac{\mathcal H_{\alpha_c(t_H)}(\ell_N)}
          {\mathcal H_{\alpha_c(t_0)}(\ell_N)}
 \le e^{|\alpha_c(t_H)-\alpha_c(t_0)|\ell_N}
       =1+O_H(\log N/N).
 \end{gathered}
\]
The other polynomial and logarithmic prefactors have ratio \(1+o(1)\).
The scalar tail then gives
\[
 \frac{p_0(U)}{A_N(t_0,y)}\longrightarrow\frac1b,\qquad
 \frac{A_N(t_H,y)}{p_0(U)}
       \longrightarrow b e^{H\psi_+(b)}.
\]
The preceding unit-bin, deep-offset, and positive-tail bounds permit
removal of the offset cutoff:
\[
 \begin{split}
 0&\le\limsup_n\left\{
 \frac{\mathbb P(\mathcal C_H(U))}{p_0(U)}
       -\frac{A_N(t_H,y)}{p_0(U)}P_{n,R}\right\}\\
 &\le C_H e^{-(s-b_{\max})R}+C_H e^{-bR}.
 \end{split}
\]
Let \(R\uparrow\infty\) through continuity endpoints.
The integrand is nonnegative, and
\[
 \begin{split}
 \int_{\mathbb R}e^{-bv}\mathscr G_H(v)\,dv
 &=\mathbb E\int_{-M_-}^{\infty}e^{-bv}\,dv\\
 &=\frac1b\mathbb E e^{bM_-}<\infty .
 \end{split}
\]
Consequently,
\[
 \begin{split}
 \frac{\mathbb P(\mathcal C_H(U))}{p_0(U)}
 &\longrightarrow
 b e^{H\psi_+(b)}\int_{\mathbb R}e^{-bv}\mathscr G_H(v)\,dv\\
 &=e^{H\psi_+(b)}
              \mathbb E e^{b\sup_{0\le r\le H}Z_-(r)} .
 \end{split}
\]
The argument also permits raw shifts \(O(\log N)\), by absorbing
them into \(y_N=y+O(\log N/N)\) throughout the compact scalar range.

A finite-dimensional change of measure gives its forward expression.
On a finite grid
\(0=r_0<\cdots<r_m=H\), put
\[
 \frac{d\mathbb Q_b}{d\mathbb P}
       =e^{bZ_+(H)-H\psi_+(b)},\qquad
 W_i=Z_+(H-r_i)-Z_+(H).
\]
The \(b\)-moment is finite by
\eqref{eq:full-boundary-Levy-duality}. Under \(\mathbb Q_b\),
reversed disjoint increments remain independent, and, for
\(0\le a\le b\),
\[
 \begin{split}
 \mathbb E_{\mathbb Q_b}e^{a(W_i-W_{i-1})}
 &=\exp\{(r_i-r_{i-1})
               [\psi_+(b-a)-\psi_+(b)]\}\\
 &=e^{(r_i-r_{i-1})\psi_-(a)}.
 \end{split}
\]
Thus \((W_i)_i\) has the law of \((Z_-(r_i))_i\), and
\[
 \begin{split}
 \mathbb E e^{b\max_iZ_+(H-r_i)}
 &=e^{H\psi_+(b)}
       \mathbb E_{\mathbb Q_b}e^{b\max_iW_i}\\
 &=e^{H\psi_+(b)}
       \mathbb E e^{b\max_iZ_-(r_i)}.
 \end{split}
\]
Monotone refinement gives
\[
 \frac{\mathbb P(\mathcal C_H(U))}{p_0(U)}
       \longrightarrow
       \mathbb E e^{b\sup_{0\le z\le H}Z_+(z)} .
\]
Only the \(b\)-moment of \(Z_+\) enters this identity.

Set \(t_0=\lceil n\eta\rceil/n\) and
\[
 t_{j,n}=\frac1n\operatorname{round}
       [n\operatorname{logistic}\{\operatorname{logit}t_0+j/N\}].
\]
For the unit cell \([j,j+1]\), use \(t_{j+1,n}\) as its terminal
anchor in the fixed-\(H=1\) crossing estimate. Equation
\eqref{eq:full-boundary-space-upper} then gives
\[
 \frac{\Pp\{\sup_{H\le z\le N[\operatorname{logit}(\eta+\epsilon_0)
                    -\operatorname{logit}t_0]}T(z)>Ny\}}
      {\Pp\{T(0)>Ny\}}
 \le C\sum_{j\ge\lfloor H\rfloor}e^{-cj}\le Ce^{-cH}.
\]
For \(j\le N/\log N\),
\[
 |t_{j+1,n}-\eta|\le C(j+1)/N,\qquad
 |\alpha_c(t_{j+1,n})|\log N\le C
       \quad\text{if }\alpha_c(\eta)=0.
\]
Thus the bounded-confluence estimates apply.
For \(j>N/\log N\), the spatial exponential absorbs the polynomial
unnormalized bound. Outside the fixed neighborhood, the global rate has a gap.
Thus take \(N\to\infty\), then \(H\to\infty\).
The limit is spectrally positive with
\[
 \mathbb EZ_+(1)<0,\qquad \psi_+(b)<0 .
\]
Its supremum identity gives
\begin{equation*}
 E e^{b\sup_{z\ge0}Z_+(z)}
   =\frac{b\psi_+'(0)}{\psi_+(b)}
   ={\cal H}^{\rm bd}_t(y).
\end{equation*}
Kella (2012), \emph{Journal of Applied Probability} 49, 883--887,
equation (5), gives this transform
(\url{https://arxiv.org/pdf/1111.7099}).
Small bounded jumps admit every positive moment. For large jumps,
\[
 t(1-f_t)e^{bg}=(1-t)f_t ,
\]
so the \(b\)-moment is finite. With
\(M_j=\sup_{0\le v\le1}\{Z_+(j+v)-Z_+(j)\}\),
\[
 \E e^{b\sup_{z\ge0}Z_+(z)}
 \le\sum_{j\ge0}\E e^{b(Z_+(j)+M_j)}
 =\E e^{bM_0}\sum_{j\ge0}e^{j\psi_+(b)}<\infty.
\]
Since \(\psi_+(b)<0\), analytic continuation and then monotone convergence
as the exponent increases to \(b\) prove the positive-moment identity.
Let \(Z_{\rm L}\) and \(Z_{\rm R}\) be the limiting inward increments
from the left and right temporal trims. Reflecting every label and the
chronological time gives
\[
 t_{\rm R}(z)=1-t_{\rm L}(z),\qquad
 \mathcal L(Z_{\rm R})=\mathcal L(Z_{\rm L}).
\]
At raw height \(Ny+z\), either endpoint has one-sided tail
\begin{equation}\label{eq:full-boundary-one-side}
 \frac{P_\eta{\cal H}^{\rm bd}_\eta(y)}
      {b\Gamma(\alpha_\eta)}
 [N(y-q_c)]^{\alpha_\eta-1}[\log(N(y-q_c))]^{\beta_\eta}
 e^{-N\{by-\lambda(b)\}-bz}\{1+o(1)\}.
\end{equation}

\subsubsection{Removing the intersection of the two temporal endpoints}

Choose disjoint fixed neighborhoods of the two trims:
\[
 |u(t_1)-u(t_2)|\ge h_0>0,\qquad |\rho|\le\rho_{\rm time}<1 .
\]
Here \(\rho_{\rm time}\) is a fixed upper bound for the two split-time
bridge correlations on these separated neighborhoods.
Reveal their common three-category boundary labels.
All category proportions stay bounded below. Let \(T_i=T_{n,nt_i}\) and let \(Y_i\) be their jointly coupled
hybrid statistics. The conditional bridge construction in
Appendix~\ref{app:rank-scan-subcritical} gives an event
\(\mathcal C_{n,t_1,t_2}\) such that
\[
 \max_{i=1,2}|T_i-Y_i|
 \le C_D\{N^{3/2}/\sqrt m+N^2/m+m^2/n\}=o(1),\qquad
 \mathbb P(\mathcal C_{n,t_1,t_2}^{\,c})\le n^{-D},
\]
while retaining each boundary energy of order \(O(N)\).
At \(\theta_N=b-1/N\), the exact modes have pressure
\begin{equation*}
 L_\rho(\theta_N)=
 \lambda(\theta_N(1+\rho)/2)+\lambda(\theta_N(1-\rho)/2),
 \qquad \lambda(b)-L_\rho(b)\ge\delta_0>0.
\end{equation*}
Compare only their endpoint eigenfunctions.
For $c_\pm=1\pm\rho$, monotonicity of $a(s)/s$ gives
\[
 \frac{a(\theta_Nc_\pm/2)}{2c_\pm}
       \le\frac{a(\theta_N)}4.
\]
The endpoint quadratic is bounded by the geometric mean of its
single-anchor versions, while the bulk remains \(NL_\rho(\theta_N)\).
For split \(i\), write \(E_{\partial,i}\) for its two-prefix energy
and \(Z_{i,\pm}\) for its standardized terminal counts, and put
\[
 W_i=m^{-2\kappa(\theta_N)}
 e^{\theta_NE_{\partial,i}}
 \psi_{\theta_N}(Z_{i,-})\psi_{\theta_N}(Z_{i,+})
 \mathbf1_{\{|Z_{i,-}|\vee|Z_{i,+}|\le B\sqrt N\}}.
\]
On the common positive boundary law, H\"older gives
\[
 \mathbb E(W_1W_2)^{1/2}
       \le(\mathbb EW_1)^{1/2}(\mathbb EW_2)^{1/2}\le N^C
\]
after cancellation of the \(m\)-powers. Moreover,
\[
 e^{(b-\theta_N)E_{\rm bd}}\le e^{C_{\rm cap}}
 \quad(E_{\rm bd}\le C_{\rm cap}N),\qquad
 (b-\theta_N)|Z_\pm|^2\le B^2 .
\]
Thus changing \(b\) to \(\theta_N\) costs only a constant
both in the boundary energy and in the Gaussian terminal.
Put
\[
 \mathcal A_n^{(2)}
 =\bigcap_{i=1}^2\{|Z_{i,-}|\vee|Z_{i,+}|\le B\sqrt N\},\qquad
 \mathcal K_n^{(2)}
 =\mathcal C_{n,t_1,t_2}\cap
   \{E_{\partial,1}\vee E_{\partial,2}\le C_{\rm cap}N\}.
\]
The conditional middle comparison yields
\[
 \mathbb E_{\rm hyb}
 [e^{\theta_N(Y_1+Y_2)/2};\mathcal A_n^{(2)}\cap\mathcal K_n^{(2)}]
 \le N^Ce^{NL_\rho(\theta_N)} .
\]
This is a positive retained-hybrid estimate and uses the
\(O(N)\) boundary cap throughout.

Choose $\epsilon<\delta_0/(4b)$. Chernoff gives
\[
 \Pp\{T_1>N(y-\epsilon),T_2>N(y-\epsilon)\}
 \le N^C e^{-N\{by-\lambda(b)+\delta_0/2\}}+n^{-D}.
\]
Use the polynomial mesh and the \(o(N)\) modulus in
\eqref{eq:global-scan-logarithmic-modulus}. Let \(\mathfrak P_N\)
be its far pairs of split points, with \(|\mathfrak P_N|\le N^{C_1}\).
Then
\[
 \frac{\displaystyle\sum_{(k_1,k_2)\in\mathfrak P_N}
       \mathbb P(T_{n,k_1}>Ny,T_{n,k_2}>Ny)}
      {\mathbb P(T_{n,n\eta}>Ny)}
 \le N^{C_2}e^{-\delta_0N/2}
       +N^{C_1}\frac{n^{-D}}{\Pp(T_{n,n\eta}>Ny)}
 \longrightarrow0
\]
after \(D\) is chosen larger than the fixed scalar rate.
The two temporal endpoint probabilities can consequently be
added.

\subsubsection{Original studentization and the coordinate maximum}

Propositions~\ref{prop:rank-mean-constant} and
\ref{prop:rank-variance-constant} give
\[
 ET_{n,k}=N+m(t)+o(1),\qquad
 \operatorname{Var}(T_{n,k})=2N+v_0(t)+o(1).
\]
A standardized height $(y-1)\sqrt{N/2}$ has raw threshold
$Ny+d_0(t,y)+o(1)$. The reflected temporal endpoint has the same
moment constants. Equation~\eqref{eq:full-boundary-one-side} and
the far intersection bound prove
\begin{equation}\label{eq:full-boundary-coordinate-tail}
 \Pp\{{\cal M}_{j,n}>(y-1)\sqrt{N/2}\}
 \sim{\cal C}^{\rm bd}_\eta(y)
       N^{\alpha_\eta-1}[\log(N(y-q_c))]^{\beta_\eta}
       e^{-N\{by-\lambda(b)\}}.
\end{equation}
The one-sided supremum contributes exactly
$\mathcal H^{\rm bd}_\eta(y)$, which includes the temporal scan multiplier.

For \(y_n\) in Theorem~\ref{thm:max-full-boundary},
\(pe^{-N\{by_n-\lambda(b)\}}=1\).
Let
\[
 c_n=\log\{\mathcal C_\eta^{\rm bd}(y_n)
              N^{\alpha_\eta-1}[\log(N(y_n-q_c))]^{\beta_\eta}\}.
\]
The center adds the raw shift \(c_n/b+o(1)\), and
\(A_{n,p}^{\rm B}x\) adds \(x/b+o(1)\).
Since \(c_n=O(\log N)\), both are covered by the uniform
scalar and studentization expansions. Equation
\eqref{eq:full-boundary-coordinate-tail} then gives
\[
 p\,\Pp\{\mathcal M_{j,n}>
       B_{n,p}^{\rm B}
          +A_{n,p}^{\rm B}x\}
 =e^{c_n}e^{-c_n-x}\{1+o(1)\}\longrightarrow e^{-x}.
\]
The coordinate events are measurable in their latent Gaussian
columns. Proposition~\ref{prop:sparse-factorization} gives
the Poisson limit under the stated dependence assumptions,
and its zero-count probability proves
Theorem~\ref{thm:max-full-boundary}.

\begin{proof}[Proof of Theorem~\ref{thm:max-full-analytic}]
On a subsequence with \(\gamma_n\to\gamma<\gamma_c\), apply
Theorem~\ref{thm:max-subcritical-gumbel} and
\eqref{eq:main-subcritical-normalization}. For \(\gamma>\gamma_c\), apply
Theorem~\ref{thm:max-full-boundary}.
When \(\gamma=\gamma_c\), Theorem~\ref{thm:max-full-critical} treats
\(|\gamma_n-\gamma_c|\le N^{-1/4}\). Outside this strip,
\[
 \begin{aligned}
 \gamma_n<\gamma_c-N^{-1/4}
 &\ \Longrightarrow\ (b-\vartheta_n)\sqrt N\ge cN^{1/4},\\
 \gamma_n>\gamma_c+N^{-1/4}
 &\ \Longrightarrow\ NE_n\ge cN^{3/4}.
 \end{aligned}
\]
The first implication permits the moving-lower saddle; the second
permits the boundary term of the positive convolution.
In the last range,
\[
 \mathcal H_\eta^{\rm bd}(q_c+E)
 \sim\frac{b(q_c-1)}
          {\eta(1-\eta)b'(\eta)E},\qquad E\downarrow0,
\]
This matches \eqref{eq:main-boundary-normalization}, retaining
the factor \(\{\log(NE_n)\}^{\beta_\eta}\).
All selected formulas agree on every convergent further subsequence;
compactness proves the assertion without a dimension-ratio limit.
If \(b_\eta=1/4\), compact dimension ratios instead give
\[
 \sup_n\vartheta_n\le\tfrac14-\epsilon
\]
for some \(\epsilon>0\), so the strict theorem applies.
\end{proof}
\section{The joint limit and Cauchy calibration}\label{app:joint-null}

Put \(N=\log n\), \(\kappa_p=C(1+\mathfrak r_p)\), and
\(a_n=1/n\).  All time vectors below have a fixed number of
distinct points in \(\mathcal I_\eta\).

\begin{lemma}\label{lem:joint-gaussian-projection}
Let \(X,Z\) be jointly standard Gaussian vectors with canonical
correlation at most \(h\).  If \(\E f(X)=0\), \(f\in L^2\), and
\(q\ge2\) satisfies \(h\sqrt{q-1}\le1\), then
\begin{equation}\label{eq:joint-gaussian-projection}
 \|\E\{f(X)\mid Z\}\|_q
 \le h\sqrt{q-1}\,\|f(X)\|_2.
\end{equation}
For \(A\in\sigma(Z)\) with \(\Pp(A)>0\),
\[
 \E\bigl[|\E\{f(X)\mid Z\}|\mid A\bigr]
 \le \Pp(A)^{-1/q}h\sqrt{q-1}\,\|f(X)\|_2.
\]
\end{lemma}

\begin{proof}
Write \(C=\operatorname{Cov}(Z,X)\), and let \(f_d\) be the
orthogonal projection of \(f\) onto degree \(d\) Gaussian Hermite
chaos. For a contraction matrix \(B\), define \(\Gamma(B)\) by
the degree-\(d\) action \(B^{\otimes d}\); thus
\[
 f=\sum_{d\ge1}f_d,\qquad
 \E\{f(X)\mid Z\}=\{\Gamma(C)f\}(Z),\qquad \|C\|_{\rm op}\le h.
\]
For a standard normal vector \(Z_0\) of the argument dimension, put
\[
 (T_\rho g)(z)=\E g(\rho z+\sqrt{1-\rho^2}Z_0).
\]
Gaussian hypercontractivity gives
\[
 \|T_\rho g\|_q\le\|g\|_2,\qquad
 0\le\rho\le(q-1)^{-1/2}.
\]
This dimension-free form follows from the Gaussian logarithmic
Sobolev inequality; see \citet[pp.~445--447]{Ledoux1992}.
Factor the conditional contraction into \(T_{1/\sqrt{q-1}}\)
and the contraction with first-chaos matrix \(\sqrt{q-1}\,C\).
Since \(f\) has no constant chaos,
\[
 \|\Gamma(\sqrt{q-1}\,C)f\|_2^2
 \le h^2(q-1)\sum_{d\ge1}\|f_d\|_2^2.
\]
Writing \(Q(Z)=\E\{f(X)\mid Z\}\), the factorization gives
\[
 \|Q\|_q
 \le\|\Gamma(\sqrt{q-1}\,C)f\|_2
 \le h\sqrt{q-1}\,\|f\|_2,
\]
which proves \eqref{eq:joint-gaussian-projection}. H\"older's inequality gives
\[
 \E(|Q|\mid A)
 =\frac{\E(|Q|\mathbf1_A)}{\Pp(A)}
 \le\frac{\|Q\|_q\Pp(A)^{1-1/q}}{\Pp(A)}
 \le\Pp(A)^{-1/q}h\sqrt{q-1}\,\|f\|_2 .
\]
\end{proof}

\begin{proof}[Proof of Theorem~\ref{thm:joint-independence}]
Fix \(r\ge1\), \(x\in\mathbb R\), and a separated set
\(A=\{j_1,\ldots,j_r\}\).  Write
\[
 E_{j,n}(x)=
 \left\{\mathcal M_{j,n}>
       \mathfrak b_{n,p}+\mathfrak a_{n,p}x\right\},
 \qquad E_A=\bigcap_{j\in A}E_{j,n}(x).
\]
Theorem~\ref{thm:max-full-analytic} and
\eqref{eq:tail-separated-factorization} give, uniformly in \(A\),
\begin{equation}\label{eq:joint-rare-probability}
 \pi_n:=\Pp\{E_{j,n}(x)\}\sim e^{-x}/p,\qquad
 \Pp(E_A)=\pi_n^r\{1+O_r(\delta_p\log p)\}.
\end{equation}
Delete \(\mathcal B(A)=\bigcup_{j\in A}\mathcal B_{j,p}\), and put
\(J=\{1,\ldots,p\}\setminus\mathcal B(A)\).
For \(j\in J\) define \(h_j=\|R_{jA}R_A^{-1/2}\|_2\).
Separation, symmetry of the neighborhoods, and fixed \(r\) give
\[
 R_A=I_r+O_r(\delta_p),\qquad
 \sum_{j\in J}h_j\le C_r\delta_p,\qquad
 |\mathcal B(A)|\le rD_p.
\]
For distinct \(j,k\in J\), put
\[
 R_{jk}^{(2)}=
 \begin{pmatrix}1&\rho_{jk}\\ \rho_{jk}&1\end{pmatrix}.
\]
Define its canonical correlation with \(Y_A\) by
\[
 h_{\{j,k\},A}
 =\left\|(R_{jk}^{(2)})^{-1/2}
       R_{\{j,k\},A}R_A^{-1/2}\right\|_{\rm op}.
\]
It satisfies
\[
 \begin{split}
 h_{\{j,k\},A}
 &\le(1-\rho_\star)^{-1/2}
       \left\|R_{\{j,k\},A}R_A^{-1/2}\right\|_{\rm HS}\\
 &\le(1-\rho_\star)^{-1/2}(h_j+h_k).
 \end{split}
\]
For every matrix \(B\), tensoring with \(I_n\) preserves its
operator norm. Applied to the covariance blocks, this gives
\[
 \|B\otimes I_n\|_{\rm op}=\|B\|_{\rm op},\qquad
 h_{\{j,k\},A}\le C(h_j+h_k).
\]
For centered errors with \(\max_j\|R_j\|_4\le L_N\), set
\(f_{jk}=R_jR_k-\E(R_jR_k)\) and \(q=C_r\log p\). Then
\[
 \|f_{jk}\|_2\le2L_N^2,\qquad
 \Pp(E_A)^{-1/q}\le C_r,\qquad
 h_{\{j,k\},A}\sqrt{q-1}\le C_r\delta_p\sqrt{\log p}\le1.
\]
Lemma~\ref{lem:joint-gaussian-projection}, applied to the complete
columns, therefore gives
\[
 \E\!\left[
 \left|\E(R_jR_k\mid Y_A)-\E(R_jR_k)\right|
 \,\middle|\,E_A\right]
 \le C(h_j+h_k)\sqrt{\log p}\,L_N^2.
\]
For \(j=k\), the same calculation uses the single-column block.
Since
\[
 \sum_{j,k\in J}(h_j+h_k)
 =2|J|\sum_{j\in J}h_j\le C_rp\delta_p,
\]
expansion of the conditional square gives
\[
 \begin{split}
 &\E\{(\sum_{j\in J}R_j)^2\mid E_A\}
       -\E(\sum_{j\in J}R_j)^2\\
 &\quad=\sum_{j,k\in J}
   \E[\E(R_jR_k\mid Y_A)-\E(R_jR_k)\mid E_A]\\
 &\quad\le Cp\delta_p\sqrt{\log p}\,L_N^2.
 \end{split}
\]
Thus
\begin{equation}\label{eq:joint-rare-replacement}
 \frac{\E\{(\sum_{j\in J}R_j)^2\mid E_A\}}{pN}
 \le
 \frac{\E(\sum_{j\in J}R_j)^2}{pN}
 +C\frac{\delta_p\sqrt{\log p}\,L_N^2}{N}.
\end{equation}
For deleted columns take a fixed \(q_r>2r\).
The fixed moments in \eqref{eq:sum-full-coordinate-moments}
and \eqref{eq:joint-rare-probability} give
\[
 \begin{split}
 \E\!\left[
 \left|p^{-1/2}\sum_{j\in\mathcal B(A)}D_{j,n}(k)\right|
 \,\middle|\,E_A\right]
 &\le p^{-1/2}\sum_{j\in\mathcal B(A)}
       \|D_{j,n}(k)\|_{q_r}\Pp(E_A)^{-1/q_r}\\
 &\le C_{r,q_r}p^{-1/2}rD_p\,p^{r/q_r}.
 \end{split}
\]
Absorbing the fixed factor \(r\), we obtain
\begin{equation}\label{eq:joint-rare-deletion}
 \E\!\left[
 \left|p^{-1/2}\sum_{j\in\mathcal B(A)}D_{j,n}(k)\right|
 \,\middle|\,E_A\right]
 \le C_{r,q_r}D_p p^{-1/2+r/q_r}.
\end{equation}
The deletion error vanishes since
\[
 D_p p^{-1/2+r/q_r}=p^{-1/2+r/q_r+o(1)}\longrightarrow0.
\]
Fix \(t=k/n\). With the quadratics of
Lemma~\ref{lem:sum-full-population-comparison}, define
\[
 \begin{aligned}
 R_j^{(1)}&=T_j(k)-Q_j(k)-\E\{T_j(k)-Q_j(k)\},\\
 R_j^{(2)}&=Q_j(k)-Q_j^{\rm pop}(k)
                    -\E\{Q_j(k)-Q_j^{\rm pop}(k)\},\\
 R_j^{(3)}&=\frac12\sum_i a_i(t)^2
       \{\chi_{1/n}(Y_{ij},Y_{ij})
                    -\E\chi_{1/n}(Y_{ij},Y_{ij})\}.
 \end{aligned}
\]
Lemma~\ref{lem:sum-full-population-comparison} and
\eqref{eq:joint-rare-replacement} give
\[
 \begin{aligned}
 \max_{a,j}\|R_j^{(a)}\|_4&\le C,\\
 \sum_{a=1}^3\frac{\E\{(\sum_{j\in J}R_j^{(a)})^2\mid E_A\}}{pN}
 &\le C\left\{\frac{\kappa_p}{N}
                 +\frac{\delta_p\sqrt{\log p}}N\right\}
 \le C\left\{\frac{\kappa_p}{N}+\frac{\delta_p}{\sqrt N}\right\}.
 \end{aligned}
\]
After exact studentization, the conditional squared error at a
fixed time is therefore bounded by
\begin{equation}\label{eq:joint-rank-row-error}
 \varepsilon_{\rm row}
 =C\left\{\frac{\kappa_p}{N}
          +\frac{\delta_p}{\sqrt N}\right\}.
\end{equation}
Here \(\kappa_p/N\to0\) and \(\delta_p/\sqrt N\to0\).
It remains to treat the row statistic under \(E_A\).
Whiten the selected rows and regress each remote column:
\[
 G_i=R_A^{-1/2}Y_{i,A},\qquad
 Y_{ij}=c_j^\top G_i+\sigma_jV_{ij},\qquad
 \|c_j\|=h_j,\quad \sigma_j^2=1-h_j^2.
\]
The array \(V\) is independent of \(G\), and its rows are independent.
The conditional covariance and standardized correlation are
\[
 \operatorname{Cov}(Y_{i,J}\mid G_i)
   =R_{JJ}-(c_j^\top c_k)_{j,k\in J},
 \qquad
 \rho^V_{jk}=\frac{\rho_{jk}-c_j^\top c_k}{\sigma_j\sigma_k}.
\]
Because \(\max_jh_j\le C_r\delta_p\) and \(\sum_jh_j\le C_r\delta_p\),
\[
 \begin{split}
 \sum_{k\in J}|\rho^V_{jk}|
 &\le(1-C_r\delta_p^2)^{-1}
       \left\{\sum_{k\in J}|\rho_{jk}|+h_j\sum_{k\in J}h_k\right\}
 \le C(1+\mathfrak r_p)=\kappa_p,\\
 \max_{j\ne k}|\rho^V_{jk}|
 &\le\frac{\rho_\star+C_r\delta_p^2}{1-C_r\delta_p^2}
 \le\rho_1<1,\qquad
 \|R^V\|_{\rm op}\le\max_j\sum_k|\rho^V_{jk}|\le\kappa_p .
 \end{split}
\]
Here \(\rho_1\in(\rho_\star,1)\) is fixed and \(n\) is sufficiently large.
For fixed \(B>0\), choose \(C_B\) sufficiently large and define
\[
 \mathcal A_B=\{\max_{i\le n}\|G_i\|\le C_B\sqrt N\}
       \in\sigma(G).
\]
The Gaussian union bound gives \(\Pp(\mathcal A_B^c)\le n^{-B}\).
If \(\overline\gamma\) bounds \(\log p/\log n\), then
\(\Pp(\mathcal A_B^c\mid E_A)\le C n^{-B+r\overline\gamma}\).
Let \(F_{ij}(u)=
\Phi\{(\Phi^{-1}(u)-c_j^\top G_i)/\sigma_j\}\).
On \(\mathcal A_B\), Gaussian density and tail ratios satisfy
\begin{equation}\label{eq:joint-shifted-marginals}
 \sup_{\substack{i,j\in J\\a_n\le u\le1-a_n}}
 \left\{
 \left|\frac{F_{ij}(u)}u-1\right|
 +\left|\frac{1-F_{ij}(u)}{1-u}-1\right|
 +|F_{ij}'(u)-1|
 \right\}\le C_B\delta_pN.
\end{equation}
To verify the bound, write \(y=\Phi^{-1}u\) and
\(\zeta_{ij}=c_j^\top G_i\). On \(\mathcal A_B\),
\[
 |\zeta_{ij}|\le C_Bh_j\sqrt N,\qquad |y|\le C\sqrt N,\qquad
 F_{ij}'(u)=\sigma_j^{-1}
       \exp\!\left\{\frac{y^2}{2}
                    -\frac{(y-\zeta_{ij})^2}{2\sigma_j^2}\right\}.
\]
Consequently,
\[
 \left|\log F_{ij}'(u)\right|
 \le C\{h_j^2(1+y^2)+|\zeta_{ij}y|+\zeta_{ij}^2\}
 \le C_B\delta_pN .
\]
For a normal tail, the logarithmic derivative is bounded by
\(C(1+|y|)\). Applying the mean-value formula to its location and
scale arguments therefore gives
\[
 \left|\log\frac{F_{ij}(u)}u\right|
 +\left|\log\frac{1-F_{ij}(u)}{1-u}\right|
 \le C\{|\zeta_{ij}|(1+|y|)+h_j^2(1+y^2)+\zeta_{ij}^2\}
 \le C_B\delta_pN .
\]
Since \(\delta_pN\to0\), \(|e^x-1|\le e^{|x|}|x|\) proves
\eqref{eq:joint-shifted-marginals}.
Introduce the Hilbert space
\[
 \mathcal H=\bigoplus_{j\in J}
 L^2\!\left([a_n,1-a_n],\frac{du}{u^2(1-u)^2}\right).
\]
For \(w(u)=u(1-u)\), write
\[
 \Phi_i=(\mathbf1\{Y_{ij}\le\Phi^{-1}u\}-u)_{j,u},
 \quad\mu_i=\E(\Phi_i\mid G),\quad
 \xi_i=\Phi_i-\mu_i,\quad C_i=\E(\xi_i\otimes\xi_i\mid G).
\]
Conditional on \(G\), the \(\xi_i\) are independent centered rows.
For \(f=(f_j)_{j\in J}\in\mathcal H\), set
\(\zeta_j=\langle f_j,\xi_{i,j}\rangle\).
The one-column Hardy bound in Lemma~\ref{lem:sum-full-row-kernel-bounds},
after the change of variable \(u\mapsto F_{ij}(u)\), gives
\(\E(\zeta_j^2\mid G)\le C\|f_j\|^2\).
Indeed \eqref{eq:joint-shifted-marginals} gives
\[
 \frac{du}{w(u)^2}
 \le C\,\frac{dF_{ij}(u)}{w(F_{ij}(u))^2}.
\]
Conditional maximal correlation and \(2ab\le a^2+b^2\) now imply
\[
 \begin{split}
 \langle f,C_if\rangle
 &=\E\{(\sum_j\zeta_j)^2\mid G\}\\
 &\le C\sum_{j,k}|\rho^V_{jk}|\|f_j\|\|f_k\|
 \le C\max_j\sum_k|\rho^V_{jk}|\sum_j\|f_j\|^2
 \le\kappa_p\|f\|^2,
 \end{split}
\]
after fixing the constant in \(\kappa_p=C(1+\mathfrak r_p)\).
Also,
\[
 \|\Phi_i\|^2
 \le C\sum_{j\in J}
 \left\{\frac1{\Phi(Y_{ij})\vee a_n}
       +\frac1{(1-\Phi(Y_{ij}))\vee a_n}\right\}.
\]
For \(\mathcal V_{ij}=\|\xi_{i,j}\|^2\), integration with
\eqref{eq:joint-shifted-marginals} gives
\[
 \E(\mathcal V_{ij}\mid G)\le CN,\qquad
 \E(\mathcal V_{ij}^2\mid G)\le Cn.
\]
These bounds include both exterior endpoint intervals.
Gaussian maximal correlation applies to the centered functions
\(\mathcal V_{ij}\) and \(\mathcal V_{ik}\).  Their marginal means can differ;
their canonical correlation remains \(|\rho^V_{jk}|\).
Therefore
\[
 \left|\operatorname{Cov}(\mathcal V_{ij},\mathcal V_{ik}\mid G)\right|
 \le C|\rho^V_{jk}|n,
 \qquad \sum_{j\ne k}|\rho^V_{jk}|\le Cp\kappa_p.
\]
The covariance expansion is
\[
 \begin{split}
 \E\{(\sum_j\mathcal V_{ij})^2\mid G\}
 &=\left\{\sum_j\E(\mathcal V_{ij}\mid G)\right\}^2
   +\sum_j\operatorname{Var}(\mathcal V_{ij}\mid G)
   +\sum_{j\ne k}\operatorname{Cov}(\mathcal V_{ij},\mathcal V_{ik}\mid G)\\
 &\le Cp^2N^2+Cpn+Cn\sum_{j\ne k}|\rho^V_{jk}|\\
 &\le C\{p^2N^2+p\kappa_p n\}.
 \end{split}
\]
Together with the operator bound, this proves
\begin{equation}\label{eq:joint-feature-moments}
 \|C_i\|_{\rm op}\le\kappa_p,\qquad
 \E(\|\xi_i\|^4\mid G)
 \le C\{p\kappa_p n+p^2N^2\}.
\end{equation}
The Gaussian Poincar\'e argument in the proof of
Lemma~\ref{lem:sum-full-row-kernel-bounds}, followed by a linear
change of variables, gives, for a centered function \(f\) of a
Gaussian vector with covariance matrix \(\Sigma\),
\[
 \|f\|_4\le C\|\Sigma\|_{\rm op}^{1/2}\,\||\nabla f|\|_4 .
\]
For \(f=\langle\xi_i,\xi_h\rangle\), one derivative equals
\[
 -\frac{\phi(y)}{w(\Phi y)^2}
       \{\mathbf1(Y_{hj}\le y)-F_{hj}(\Phi y)\},
 \qquad a_n<\Phi y<1-a_n.
\]
It is zero elsewhere; smooth approximation justifies the
Poincar\'e calculation for this absolutely continuous kernel.
The centered indicator has fourth moment at most \(Cw(u)\).
The Gaussian tail bound gives
\[
 \phi(\Phi^{-1}u)^4\le Cw(u)^4\log^2(e/w(u)).
\]
Consequently,
\[
 \E|\partial_{y_j}f|^4
 \le C\int_{a_n}^{1-a_n}
          \frac{\phi(\Phi^{-1}u)^4}{w(u)^7}\,du
 \le Ca_n^{-2}N^2.
\]
The second derivative moment uses the same integral with
\(\phi(\Phi^{-1}u)^2\le Cw(u)^2\log(e/w(u))\):
\[
 \E|\partial_{y_j}f|^2
 \le C\int_{1/n}^{1-1/n}\frac{\log(e/w(u))}{w(u)}\,du
 \le CN^2.
\]
Let \(\Gamma_j\) be the sum of the squared derivatives in the
first- and second-row coordinates of column \(j\).  Then
\[
 \E(\Gamma_j\mid G)\le CN^2,\qquad
 \E(\Gamma_j^2\mid G)\le Cn^2N^2.
\]
After marginal standardization, the two-row Gaussian column blocks
have cross-covariance \(\rho^V_{jk}I_2\).  Maximal correlation gives
\[
 |\operatorname{Cov}(\Gamma_j,\Gamma_k\mid G)|
 \le C|\rho^V_{jk}|n^2N^2.
\]
Thus
\[
 \begin{split}
 \E(|\nabla f|^4\mid G)
 &=\left\{\sum_j\E(\Gamma_j\mid G)\right\}^2
   +\sum_j\operatorname{Var}(\Gamma_j\mid G)
   +\sum_{j\ne k}\operatorname{Cov}(\Gamma_j,\Gamma_k\mid G)\\
 &\le Cp^2N^4+Cpn^2N^2
       +Cn^2N^2\sum_{j\ne k}|\rho^V_{jk}|\\
 &\le C\{p\kappa_p n^2N^2+p^2N^4\}.
 \end{split}
\]
The preceding Poincar\'e bound consequently proves
\begin{equation}\label{eq:joint-inner-fourth}
 \E\{|\langle\xi_i,\xi_h\rangle|^4\mid G\}
 \le C\kappa_p^2\{p\kappa_p n^2N^2+p^2N^4\},\qquad i\ne h.
\end{equation}
For \(z\in\mathbb R^r\), define the conditional-mean map
\(\mu(z)=(\mu_j(u;z))_{j,u}\in\mathcal H\) by
\[
 \mu_j(u;z)
 =\Phi\{(\Phi^{-1}u-c_j^\top z)/\sigma_j\}-u.
\]
Put \(H_A^2=\sum_{j\in J}h_j^2\). On
\(\|z\|\le C_B\sqrt N\), its derivative satisfies
\[
 \|d_z\mu\|_{\rm op}^2
 \le C\sum_{j\in J}h_j^2
       \int_0^1\frac{\phi(\Phi^{-1}u)^2}{w(u)^2}\,du
 \le C H_A^2.
\]
For \(R=C_B\sqrt N\), define
\[
 \begin{aligned}
 \Pi_R(0)&=0,\qquad
 \Pi_R(z)=z\min\{1,R/\|z\|\}\quad(z\ne0),\\
 \widetilde\mu&=\mu\circ\Pi_R,\qquad
 \operatorname{Lip}(\Pi_R)\le1,\qquad
 \operatorname{Lip}(\widetilde\mu)\le CH_A.
 \end{aligned}
\]
Thus \(\widetilde\mu=\mu\) on \(\mathcal A_B\).
For a nonnegative integrable variable \(V\), set
\(\operatorname{Ent}(V)=\E(V\log V)-(\E V)\log(\E V)\),
with \(0\log0=0\). The Gaussian logarithmic Sobolev inequality
\citep[p.~445, equation~(1)]{Ledoux1992} states
\[
 \operatorname{Ent}(g^2)\le2\E|\nabla g|^2,\qquad
 g\in W^{1,2}(\gamma),
\]
where \(\gamma\) is standard Gaussian measure.
For an \(L\)-Lipschitz real function \(f\), put
\(H(s)=\log\E e^{s(f-\E f)}\) and \(g_s=e^{s(f-\E f)/2}\).
Bounded smooth approximation gives
\[
 \frac{\operatorname{Ent}(g_s^2)}{\E g_s^2}
       =sH'(s)-H(s),\qquad
 \frac{2\E|\nabla g_s|^2}{\E g_s^2}
       \le\tfrac12s^2L^2.
\]
Hence
\[
 sH'(s)-H(s)\le\tfrac12s^2L^2,\qquad
 \left\{\frac{H(s)}s\right\}'\le\tfrac12L^2\quad(s>0).
\]
Since \(H(s)/s\to0\) at zero, integration and replacement of \(f\)
by \(-f\) give
\[
 \E e^{s(f-\E f)}\le e^{s^2L^2/2},\qquad s\in\mathbb R.
\]
Apply this bound to the norm of each weighted Hilbert sum.
Write \(\widetilde\mu\) for the projected map.
The variance change \(\sigma_j-1=O(h_j^2)\) gives
\[
 \|\mu(0)\|^2\le C\sum_{j\in J}h_j^4\le C H_A^4,\qquad
 \|\mu(z)\|\le C H_A(\|z\|+H_A)
 \quad(\|z\|\le C_B\sqrt N).
\]
For example, the first bound follows by integrating
\(h_j^4y^2\phi(y)^2/w(\Phi y)^2\) with respect to \(d\Phi(y)\).
Independence of \(G_i\), their fixed dimension \(r\), and the
Lipschitz bound imply
\[
 \E\left\|\sum_i a_i(t)
    \{\widetilde\mu(G_i)-\E\widetilde\mu(G_i)\}\right\|^2
       =\sum_i a_i(t)^2
          \E\|\widetilde\mu(G_i)-\E\widetilde\mu(G_i)\|^2
 \le C_r H_A^2.
\]
Here one may bound the Hilbert variance by one half of
\(\E\|\widetilde\mu(G)-\widetilde\mu(G')\|^2\) for independent copies.
The map
\[
 (z_1,\ldots,z_n)\longmapsto
 \left\|\sum_i a_i(t)
       \{\widetilde\mu(z_i)-\E\widetilde\mu(G_i)\}\right\|
\]
is \(CH_A(\sum_i a_i(t)^2)^{1/2}=CH_A\)-Lipschitz and has mean at
most \(C_rH_A\). The preceding concentration inequality gives
\[
 \Pp\!\left\{
 \left\|\sum_i a_i(t)
       \{\widetilde\mu(G_i)-\E\widetilde\mu(G_i)\}\right\|
       >C_rH_A+xH_A\right\}\le e^{-cx^2}.
\]
Take \(x=C_B\sqrt N\). On \(\mathcal A_B\),
\(\widetilde\mu(G_i)=\mu_i\), while \(\sum_i a_i(t)=0\) cancels
the common mean. Together with
\(\max_i\|\mu_i\|^2\le C_BNH_A^2\), this gives, outside another
event of probability \(Cn^{-B}\),
\begin{equation}\label{eq:joint-mean-good}
 \left\|\sum_i a_i(t)\mu_i\right\|^2
 +\max_i\|\mu_i\|^2
 \le C_BNH_A^2\le C_BN\delta_p^2.
\end{equation}
Fix distinct \(t_1,\ldots,t_d\in\mathcal I_\eta\), and define
\[
 \mathcal A_B^*=\mathcal A_B\cap
 \bigcap_{\ell=1}^d
 \left\{\left\|\sum_i a_i(t_\ell)\mu_i\right\|^2
             +\max_i\|\mu_i\|^2\le C_BNH_A^2\right\}.
\]
The preceding bounds and a finite union give
\[
 \Pp((\mathcal A_B^*)^c)\le C_{B,d}n^{-B},\qquad
 \Pp((\mathcal A_B^*)^c\mid E_A)
             \le C_{B,d}n^{-B+r\overline\gamma}.
\]
Put \(M_t=\sum_i a_i(t)\mu_i\) and define
\[
 \begin{split}
 U_{\rm c}(t)&=\sum_{i<h}a_i(t)a_h(t)\langle\xi_i,\xi_h\rangle,\\
 U_{\rm L}(t)&=\sum_i a_i(t)
                  \langle\xi_i,M_t-a_i(t)\mu_i\rangle,\\
 U_{\rm M}(t)&=\tfrac12\{\|M_t\|^2-\sum_i a_i(t)^2\|\mu_i\|^2\}.
 \end{split}
\]
The quadratic statistic then decomposes exactly as
\[
 \sum_{i<h}a_i(t)a_h(t)\langle\Phi_i,\Phi_h\rangle
 =U_{\rm c}(t)+U_{\rm L}(t)+U_{\rm M}(t).
\]
Independence of the centered rows and
\eqref{eq:joint-feature-moments} give
\[
 \begin{split}
 |U_{\rm M}(t)|
 &\le\tfrac12\{\|M_t\|^2+\max_i\|\mu_i\|^2\},\\
 \E\{U_{\rm L}(t)^2\mid G\}
 &=\sum_i a_i(t)^2
   \langle M_t-a_i(t)\mu_i,C_i(M_t-a_i(t)\mu_i)\rangle\\
 &\le2\kappa_p
      \{\|M_t\|^2+\max_i a_i(t)^2\max_i\|\mu_i\|^2\}.
 \end{split}
\]
Substitution of \eqref{eq:joint-mean-good} yields
\begin{equation}\label{eq:joint-first-projection-error}
 \frac{|U_{\rm M}(t)|}{\sqrt{2pN}}
 \le C\delta_p^2\sqrt{\frac Np},\qquad
 \frac{\E\{U_{\rm L}(t)^2\mid G\}}{2pN}
 \le C\frac{\kappa_p\delta_p^2}{p}.
\end{equation}
Thus the preceding bounds control both terms due to the row-dependent
regression means under conditional centering.
For one column, its conditional covariance kernel is
\[
 K_{i,j}(u,v)
 =F_{ij}(u\wedge v)-F_{ij}(u)F_{ij}(v)
 =F_{ij}(u\wedge v)\{1-F_{ij}(u\vee v)\}.
\]
Writing \(K_0(u,v)=(u\wedge v)\{1-(u\vee v)\}\),
\eqref{eq:joint-shifted-marginals} gives
\[
 K_{i,j}(u,v)=K_0(u,v)\{1+O(\delta_pN)\}
\]
uniformly on the integration square. Therefore
\[
 \begin{split}
 \operatorname{tr}(C_{i,jj}C_{h,jj})
 &=\int_{a_n}^{1-a_n}\int_{a_n}^{1-a_n}
   \frac{K_{i,j}(u,v)K_{h,j}(u,v)}
        {w(u)^2w(v)^2}\,du\,dv\\
 &=\{1+O(\delta_pN)\}
   \int_{a_n}^{1-a_n}\int_{a_n}^{1-a_n}
       \frac{K_0(u,v)^2}{w(u)^2w(v)^2}\,du\,dv\\
 &=4N+O(\delta_pN^2+1),
 \end{split}
\]
where the last integral is evaluated in
\eqref{eq:sum-full-operator-bounds}.
For \(j\ne k\), put
\(\mathcal C_\rho(u,v)=\Phi_\rho(\Phi^{-1}u,\Phi^{-1}v)-uv\),
where \(\Phi_\rho\) is the standard bivariate normal distribution
function with correlation \(\rho\). The conditional cross-covariance kernel is
\[
 K_{i,jk}(u,v)=
 \mathcal C_{\rho^V_{jk}}\{F_{ij}(u),F_{ik}(v)\}.
\]
The preceding change-of-variable inequality and the Gaussian
copula estimate in Lemma~\ref{lem:sum-full-row-kernel-bounds} give
\[
 \|C_{i,jk}\|_{\rm HS}^2
 \le C\int_0^1\int_0^1
       \frac{\mathcal C_{\rho^V_{jk}}(u,v)^2}
            {w(u)^2w(v)^2}\,du\,dv
 \le C|\rho^V_{jk}|.
\]
Cauchy--Schwarz therefore gives
\[
 |\operatorname{tr}(C_{i,jk}C_{h,kj})|
 \le\|C_{i,jk}\|_{\rm HS}\|C_{h,jk}\|_{\rm HS}
 \le C|\rho^V_{jk}|,
\]
uniformly in \(i,h\).
Since \(|J|=p+O_r(D_p)\), summing the diagonal and off-diagonal
blocks gives
\[
 \begin{split}
 \operatorname{tr}(C_iC_h)
 &=\sum_{j\in J}\operatorname{tr}(C_{i,jj}C_{h,jj})
   +\sum_{\substack{j,k\in J\\j\ne k}}
          \operatorname{tr}(C_{i,jk}C_{h,kj})\\
 &=4|J|N+O\{p\delta_pN^2+p+p\kappa_p\}\\
 &=4pN\left[1+O\left\{\delta_pN+\frac{\kappa_p}{N}
                                      +\frac{D_p}{p}\right\}\right].
 \end{split}
\]
Thus, uniformly on \(\mathcal A_B^*\),
\begin{equation}\label{eq:joint-conditional-trace}
 \operatorname{tr}(C_iC_h)=4pN(1+O(\varepsilon_{\rm cov})),
 \quad
 \varepsilon_{\rm cov}
 =\delta_pN+\frac{\kappa_p}{N}+\frac{D_p}{p},
 \quad i\ne h.
\end{equation}
Fix \(z\in\mathbb R^d\).
Write \(\mathcal C=(\mathcal C(t_a,t_b))_{a,b=1}^d\) for the
matrix of the previously defined covariance kernel, and set
\[
 W_{ih}=\sum_{\ell}z_\ell a_i(t_\ell)a_h(t_\ell),\qquad
 U_{\rm c}=\sum_{i<h}W_{ih}\langle\xi_i,\xi_h\rangle,\qquad
 U_{\rm L}=\sum_\ell z_\ell U_{\rm L}(t_\ell),\qquad
 U_{\rm M}=\sum_\ell z_\ell U_{\rm M}(t_\ell).
\]
Then \(|W_{ih}|\le C_z/n\), its absolute row sums are bounded,
and
\[
 \sum_{i<h}W_{ih}^2
 =\tfrac12 z^\top\mathcal C z+O_z(n^{-1}).
\]
If \(z=0\), the characteristic-function assertion is immediate.
Otherwise distinct time points give \(z^\top\mathcal C z>0\),
and \eqref{eq:joint-conditional-trace} implies
\[
 s_G^2:=\E(U_{\rm c}^2\mid G)
 =2pN\{z^\top\mathcal C z+O_z(\varepsilon_{\rm cov})\}.
\]
Order the independent conditional rows and define
\(\mathcal F_h=\sigma(G,\xi_1,\ldots,\xi_h)\). The variables
\[
 d_h=\sum_{i<h}W_{ih}\langle\xi_i,\xi_h\rangle
\]
are martingale differences. Their predictable variance and fourth
moments are bounded as follows.
The diagonal part of their predictable variance uses
\(A_i=\sum_{h>i}W_{ih}^2C_h\), for which
\(\|A_i\|_{\rm op}\le C\kappa_p/n\).
Independence and \eqref{eq:joint-feature-moments} give
\[
 \begin{split}
 \operatorname{Var}\left(
    \sum_i\langle\xi_i,A_i\xi_i\rangle\,\middle|\,G\right)
 &=\sum_i\operatorname{Var}
       (\langle\xi_i,A_i\xi_i\rangle\mid G)\\
 &\le\sum_i\|A_i\|_{\rm op}^2\E(\|\xi_i\|^4\mid G)\\
 &\le n\,\frac{C\kappa_p^2}{n^2}
             \{p\kappa_p n+p^2N^2\}\\
 &=C\kappa_p^2\{p\kappa_p+p^2N^2/n\}.
 \end{split}
\]
For the off-diagonal part let
\(L_{ih}=W_{ih}\mathbf1(i<h)\), and let \(C_{\rm diag}\)
have blocks \(C_i\).  Then
\[
 T=C_{\rm diag}^{1/2}(L\otimes I_{\mathcal H})
                         C_{\rm diag}^{1/2},\qquad
 \|T\|_{\rm op}\le C\kappa_p,\quad
 \|T\|_{\rm HS}^2\le CpN.
\]
The off-diagonal bracket has block matrix \(TT^*\), and
\[
 \|TT^*\|_{\rm HS}^2
 =\operatorname{tr}\{(T^*T)^2\}
 \le\|T\|_{\rm op}^2\|T\|_{\rm HS}^2
 \le C\kappa_p^2pN .
\]
We obtain
\begin{equation}\label{eq:joint-martingale-bracket}
 \E\left[
 \left\{\frac{\sum_h\E(d_h^2\mid\mathcal F_{h-1})}{s_G^2}
                         -1\right\}^2\,\middle|\,G\right]
 \le C\left\{
 \frac{\kappa_p^3}{pN^2}
 +\frac{\kappa_p^2}{n}
 +\frac{\kappa_p^2}{pN}\right\}.
\end{equation}
Conditioning first on \(\xi_h\), the fourth-moment expansion
of a sum of independent centered variables gives
\[
 \begin{split}
 \sum_h\E(d_h^4\mid G)
 &\le C\sum_{i<h}W_{ih}^4
                   \E\langle\xi_i,\xi_h\rangle^4\\
 &\quad+C\sum_h
 \E\left\langle\xi_h,
             \left(\sum_{i<h}W_{ih}^2C_i\right)\xi_h
       \right\rangle^2.
 \end{split}
\]
The first sum is bounded using \(\sum_{i<h}W_{ih}^4\le Cn^{-2}\)
and \eqref{eq:joint-inner-fourth}; the second uses
\(\|\sum_{i<h}W_{ih}^2C_i\|_{\rm op}\le C\kappa_p/n\) and
\eqref{eq:joint-feature-moments}. Hence
\[
 \begin{split}
 \sum_h\E(d_h^4\mid G)
 &\le\frac{C\kappa_p^2}{n^2}
          \{p\kappa_p n^2N^2+p^2N^4\}
       +n\,\frac{C\kappa_p^2}{n^2}
          \{p\kappa_p n+p^2N^2\}\\
 &\le C\left\{
     p\kappa_p^3N^2+\frac{p^2\kappa_p^2N^4}{n^2}
      +p\kappa_p^3+\frac{p^2\kappa_p^2N^2}{n}\right\}.
 \end{split}
\]
Dividing by \(s_G^4\asymp(pN)^2\) gives
\begin{equation}\label{eq:joint-martingale-fourth}
 s_G^{-4}\sum_h\E(d_h^4\mid G)
 \le C\left\{
 \frac{\kappa_p^3}{p}
 +\frac{\kappa_p^2N^2}{n^2}
 +\frac{\kappa_p^3}{pN^2}
 +\frac{\kappa_p^2}{n}\right\}.
\end{equation}
Apply the fourth-moment martingale normal-approximation
inequality stated in Appendix~\ref{app:sum-full}
\citep[Theorem~1.1]{Mourrat2013}.
Conditional on each fixed \(G\in\mathcal A_B^*\), the \(d_h\)
are martingale differences with finite fourth moments.
Equations \eqref{eq:joint-martingale-bracket} and
\eqref{eq:joint-martingale-fourth} verify every term of this bound.
Put
\[
 b_N=\frac{\kappa_p^3}{p}
 +\frac{\kappa_p^2N^2}{n^2}
 +\frac{\kappa_p^3}{pN^2}
 +\frac{\kappa_p^2}{n}
 +\frac{\kappa_p^2}{pN}.
\]
The assumptions imply
\[
 \kappa_p/N\longrightarrow0,\qquad
 \kappa_p^4/p\longrightarrow0,\qquad b_N\longrightarrow0.
\]
Let \(F_G\) be the conditional distribution of \(U_{\rm c}/s_G\),
and let \(Z\sim N(0,1)\) have distribution function \(\Phi\).
For \(R\ge1\), integration by parts and the unit second moments give
\[
 \begin{split}
 \left|\int_{\mathbb R}e^{isy}\,d(F_G-\Phi)(y)\right|
 &\le(2+2|s|R)\|F_G-\Phi\|_\infty
       +\Pp(|U_{\rm c}/s_G|>R\mid G)+\Pp(|Z|>R)\\
 &\le C_sRb_N^{1/5}+2R^{-2}.
 \end{split}
\]
The choice \(R=b_N^{-1/15}\) makes both terms \(O_s(b_N^{2/15})\).
Apply this estimate at \(s=s_G/\sqrt{2pN}\), which is uniformly bounded
by \eqref{eq:joint-conditional-trace}. Moreover,
\[
 \left|e^{-s_G^2/(4pN)}-e^{-z^\top\mathcal C z/2}\right|
 \le C_z\varepsilon_{\rm cov},\qquad
 \E\!\left[\left|\frac{U_{\rm L}+U_{\rm M}}{\sqrt{2pN}}\right|
                    \,\middle|\,G\right]
 \le C\left\{\delta_p^2\sqrt{\frac Np}
             +\delta_p\sqrt{\frac{\kappa_p}p}\right\}
\]
on the retained Gaussian event, by
\eqref{eq:joint-first-projection-error}. Define
\[
 U=\sum_{i<h}W_{ih}\langle\Phi_i,\Phi_h\rangle
       =U_{\rm c}+U_{\rm L}+U_{\rm M}.
\]
Using
\(|e^{ix}-e^{iy}|\le|x-y|\) and
\(\Pp((\mathcal A_B^*)^c\mid E_A)\le Cn^{-B+r\overline\gamma}\)
therefore gives
\begin{equation}\label{eq:joint-conditional-row-cf}
 \begin{split}
 &\left|\E\!\left[
 e^{iU/\sqrt{2pN}}\,\middle|\,E_A\right]
       -e^{-z^\top\mathcal C z/2}\right|\\
 &\quad\le C_z\left\{
 b_N^{2/15}+\varepsilon_{\rm cov}
 +\delta_p^2\sqrt{N/p}
 +\delta_p\sqrt{\kappa_p/p}
 +n^{-B+r\overline\gamma}\right\}.
 \end{split}
\end{equation}
Take \(B>r\overline\gamma+2\).
Equations \eqref{eq:joint-rare-deletion},
\eqref{eq:joint-rank-row-error}, and
\eqref{eq:joint-conditional-row-cf} prove
\begin{equation}\label{eq:joint-original-conditional-cf}
 \begin{split}
 &\left|\E\!\left[
 e^{i\sum_\ell z_\ell
 D_{\mathrm{sum},n,p}(k_n(t_\ell))}\,\middle|\,E_A\right]
       -e^{-z^\top\mathcal C z/2}\right|\\
 &\quad\le C_z\left\{
 D_pp^{-1/2+r/q_r}+\sqrt{\varepsilon_{\rm row}}
 +b_N^{2/15}+\varepsilon_{\rm cov}
 +\delta_p^2\sqrt{N/p}
 +\delta_p\sqrt{\kappa_p/p}+n^{-2}\right\}.
 \end{split}
\end{equation}
Each displayed term tends to zero under the stated assumptions.
Define
\[
 N_n(x)=\sum_j\mathbf1\{E_{j,n}(x)\},\qquad
 W_n=(D_{\mathrm{sum},n,p}(k_n(t_\ell)))_{\ell=1}^d,
\]
and let \(\mathcal S_r\) be the ordered tuples of \(r\) distinct
indices satisfying the separation condition of
Proposition~\ref{prop:sparse-factorization}.
Write \(\epsilon_{n,r,z}\) for the explicit right-hand side of
\eqref{eq:joint-original-conditional-cf}. Then
\[
 \begin{split}
 &\left|\sum_{A\in\mathcal S_r}
 \E(e^{iz^\top W_n}\mathbf1_{E_A})
 -e^{-z^\top\mathcal C z/2}
       \sum_{A\in\mathcal S_r}\Pp(E_A)\right|\\
 &\qquad\le\epsilon_{n,r,z}\sum_{A\in\mathcal S_r}\Pp(E_A)
       =O_{r,z}(\epsilon_{n,r,z}),\\
 &\left|\sum_{A\notin\mathcal S_r}
           \E(e^{iz^\top W_n}\mathbf1_{E_A})\right|
       \le\sum_{A\notin\mathcal S_r}\Pp(E_A)\longrightarrow0.
 \end{split}
\]
The last limit and
\(\sum_{A\in\mathcal S_r}\Pp(E_A)\to e^{-rx}\)
follow from Proposition~\ref{prop:sparse-factorization}.
Consequently,
\begin{equation}\label{eq:joint-mixed-factorial}
 \E\{e^{iz^\top W_n}(N_n(x))_r\}
 \longrightarrow e^{-z^\top\mathcal C z/2}e^{-rx},
 \qquad r=0,1,\ldots.
\end{equation}
For \(r=0\), this is the SUM finite-dimensional limit.
The marginals are tight. For each fixed \(r\),
\[
 ((k)_r)^2\le C_r\{1+(k)_{2r}\},\qquad k\in\mathbb N_0,
\]
and the factorial-moment limits imply
\[
 \sup_n\E\{(N_n(x))_r^2\}<\infty,\qquad
 \sup_n\E\!\left[(N_n(x))_r
             \mathbf1\{(N_n(x))_r>K\}\right]\le C_r/K .
\]
Thus \eqref{eq:joint-mixed-factorial} passes to every subsequential
joint limit \((W,N_x)\).
Its marginals are Gaussian and Poisson with
\(\lambda_x=e^{-x}\).  For fixed \(z\), put
\[
 g_z(k)=\E(e^{iz^\top W}\mid N_x=k)-e^{-z^\top\mathcal C z/2},
 \qquad
 H_z(t)=\sum_{k\ge0}g_z(k)\Pp(N_x=k)t^k.
\]
For every complex \(t\),
\[
 \sum_{k\ge0}|g_z(k)|\Pp(N_x=k)|t|^k
 \le2e^{\lambda_x(|t|-1)}<\infty.
\]
Hence \(H_z\) is entire, and \eqref{eq:joint-mixed-factorial} gives
\[
 H_z^{(r)}(1)
 =\E\{e^{iz^\top W}(N_x)_r\}
       -e^{-z^\top\mathcal C z/2}\E(N_x)_r=0,
 \qquad r\ge0.
\]
Its Taylor series at \(1\) is identically zero. Comparing the
coefficients at \(0\) yields
\[
 \E(e^{iz^\top W}\mid N_x=k)
 =e^{-z^\top\mathcal C z/2},\qquad k\in\mathbb N_0,
\]
and hence
\[
 \E\{e^{iz^\top W}\mathbf1(N_x=0)\}
       =e^{-z^\top\mathcal C z/2}e^{-e^{-x}}.
\]
At finite \(n\),
\[
 \left\{(\Smax-\mathfrak b_{n,p})/\mathfrak a_{n,p}\le x\right\}
       =\{N_n(x)=0\}.
\]
Together with \eqref{eq:sum-full-functional-limit}, this yields
joint tightness. Every subsequential limit
\((\mathbb G,G_{\max})\), with \(\mathbb G\) continuous, satisfies
\[
 \Pp\{(\mathbb G(t_1),\ldots,\mathbb G(t_d))\in B,\,
                    G_{\max}\le x\}
 =\Pp\{(\mathbb G(t_1),\ldots,\mathbb G(t_d))\in B\}
                    e^{-e^{-x}} .
\]
Finite cylinders on a countable dense time set generate the Borel
sigma field on continuous paths. Thus \(\mathbb G\) and \(G_{\max}\)
are independent. Finally,
\[
 \left|\sup_t f(t)-\sup_t g(t)\right|\le\|f-g\|_\infty
\]
proves \eqref{eq:joint-independent-limit}.
\end{proof}

\Needspace{4\baselineskip}
\begin{proof}[Proof of Corollary~\ref{cor:cauchy-null}]
The OU supremum has a continuous distribution by the argument
following \eqref{eq:sum-full-functional-limit}. The Gumbel
distribution is also continuous.
The two continuous survival transforms in
\eqref{eq:component-analytic-pvalues} therefore converge to
independent \(U_1,U_2\sim\operatorname{Unif}(0,1)\).
Their endpoints have probability zero.
Writing \(C_j=\cot(\pi U_j)\), independence gives
\[
 \E e^{it\{\omega C_1+(1-\omega)C_2\}}
 =e^{-\omega|t|}e^{-(1-\omega)|t|}
 =e^{-|t|}.
\]
Continuous mapping proves \eqref{eq:cauchy-null-limit} and,
for every fixed \(\alpha\in(0,1)\),
\[
 \Pp\{\Pcauchy(\omega)\le\alpha\}\longrightarrow\alpha.
\]
\end{proof}

\section{Population discrepancy, power, and localization}\label{app:alternative}

\subsection{Population identification}

Let $F,G$ be continuous distribution functions, let
$0<\lambda<1$, and put $H=\lambda F+(1-\lambda)G$.

\begin{lemma}[Weighted Jensen gap]\label{lem:jensen}
The population statistic and gap satisfy
\[
 \cZ_\lambda(F,G)=\frac{\pi^2}{3}-\cD_\lambda(F,G),
 \qquad \cD_\lambda(F,G)\ge0.
\]
The gap vanishes if $F=G$ and is strictly positive if $F\ne G$
on a set of positive $H$-measure.
\end{lemma}

\begin{proof}
If $F=G=H$, then
\begin{align*}
\cZ_\lambda(F,F)
&=-\int_0^1\frac{u\log u+(1-u)\log(1-u)}{u(1-u)}\,du\\
&=-\int_0^1\frac{\log u}{1-u}\,du
  -\int_0^1\frac{\log(1-u)}{u}\,du\\
&=2\sum_{r=1}^{\infty}\frac1{r^2}
=\frac{\pi^2}{3}.
\end{align*}
For general \(F,G\), subtraction in \cref{eq:population-Z} gives
\[
 \frac{\pi^2}{3}-\cZ_\lambda(F,G)
 =\int
 \frac{\lambda\phifun(F)+(1-\lambda)\phifun(G)-\phifun(H)}
      {H(1-H)}\,dH
 =\cD_\lambda(F,G),
\]
which is \cref{eq:gap}. Since
\[
\phifun''(u)=\frac1u+\frac1{1-u}>0,\qquad u\in(0,1),
\]
strict convexity gives
\[
\lambda\phifun\{F(x)\}+(1-\lambda)\phifun\{G(x)\}
-\phifun\{H(x)\}\ge0.
\]
The inequality is strict whenever $F(x)\ne G(x)$ and $H(x)\in(0,1)$. Integration with respect to $H$ proves the result.
\end{proof}

\begin{proposition}[Population identification]\label{prop:identification}
For every coordinate $j$,
\[
 \Delta_j(t)\le\Delta_j(\lambda^\star),\qquad 0<t<1.
\]
If $F_j^{(1)}\ne F_j^{(2)}$ on a set of positive pooled-marginal
measure, the inequality is strict for $t\ne\lambda^\star$.

If at least one coordinate changes in this sense, then
\[
 \operatorname*{arg\,max}_{0<t<1}\Delta_{\mathrm{sum}}(t)
 =\operatorname*{arg\,max}_{0<t<1}\Delta_{\mathrm{max}}(t)
 =\{\lambda^\star\},
\]
\[
 \operatorname*{arg\,min}_{0<t<1}Q_{\mathrm{sum}}(t)
 =\operatorname*{arg\,min}_{0<t<1}Q_{\mathrm{max}}(t)
 =\{\lambda^\star\}.
\]
\end{proposition}

\begin{proof}
Fix coordinate $j$ and suppress the subscript. Let
$H=\lambda^\star F^{(1)}+(1-\lambda^\star)F^{(2)}$.
For $t<\lambda^\star$, define
\[
a_t=\frac{\lambda^\star-t}{1-t},
\qquad
F^R_t=a_tF^{(1)}+(1-a_t)F^{(2)}.
\]
Convexity of $\phifun$ gives, pointwise,
\begin{align*}
t\phifun(F^{(1)})+(1-t)\phifun(F^R_t)
&\le t\phifun(F^{(1)})
 +(1-t)a_t\phifun(F^{(1)})
 +(1-t)(1-a_t)\phifun(F^{(2)})\\
&=\lambda^\star\phifun(F^{(1)})+(1-\lambda^\star)\phifun(F^{(2)}).
\end{align*}
The pooled distribution is unchanged, so the difference is exactly
\[
 \begin{split}
 \Delta(\lambda^\star)-\Delta(t)
 &=(1-t)\int
 \frac{a_t\phifun(F^{(1)})+(1-a_t)\phifun(F^{(2)})
               -\phifun(a_tF^{(1)}+(1-a_t)F^{(2)})}
      {H(1-H)}\,dH\\
 &\ge0.
 \end{split}
\]
Here \(0<a_t<1\); strict convexity makes the integral positive
when the marginals differ on a set of positive \(H\)-measure.
For $t>\lambda^\star$, let $b_t=\lambda^\star/t$ and
$F^L_t=b_tF^{(1)}+(1-b_t)F^{(2)}$. Then
\begin{align*}
t\phifun(F^L_t)+(1-t)\phifun(F^{(2)})
&\le tb_t\phifun(F^{(1)})
+t(1-b_t)\phifun(F^{(2)})
+(1-t)\phifun(F^{(2)})\\
&=\lambda^\star\phifun(F^{(1)})+(1-\lambda^\star)\phifun(F^{(2)}),
\end{align*}
Thus
\[
 \begin{split}
 \Delta(\lambda^\star)-\Delta(t)
 &=t\int
 \frac{b_t\phifun(F^{(1)})+(1-b_t)\phifun(F^{(2)})
                 -\phifun(b_tF^{(1)}+(1-b_t)F^{(2)})}
      {H(1-H)}\,dH\ge0 .
 \end{split}
\]
For \(t>\lambda^\star\), \(0<b_t<1\), so the same strictness
criterion applies. At every interior \(t\),
\[
 F^L_t-F^R_t=
 \begin{cases}
 \dfrac{1-\lambda^\star}{1-t}(F^{(1)}-F^{(2)}),
       &t<\lambda^\star,\\[1mm]
 F^{(1)}-F^{(2)},&t=\lambda^\star,\\[1mm]
 \dfrac{\lambda^\star}{t}(F^{(1)}-F^{(2)}),
       &t>\lambda^\star .
 \end{cases}
\]
Lemma~\ref{lem:jensen} gives a positive gap for every changed
coordinate and zero for every unchanged one. Hence a maximizing
coordinate is changed whenever at least one coordinate changes.
If
$\mathcal J=\{j:F_j^{(1)}\ne F_j^{(2)}\}$, then for every
$t\ne\lambda^\star$,
\begin{align*}
 \Delta_{\mathrm{sum}}(\lambda^\star)-\Delta_{\mathrm{sum}}(t)
 &=p^{-1}\sum_{j\in\mathcal J}
 \{\Delta_j(\lambda^\star)-\Delta_j(t)\}>0,\\
 \Delta_{\mathrm{max}}(t)
 &=\Delta_{j_t}(t)
 <\Delta_{j_t}(\lambda^\star)
 \le\Delta_{\mathrm{max}}(\lambda^\star),
 \quad j_t\in\operatorname*{arg\,max}_j\Delta_j(t).
\end{align*}
Therefore both gaps have the unique maximizer $\lambda^\star$, and the two
population criteria have the unique minimizer $\lambda^\star$.
\end{proof}

\subsection{Auxiliary bounds for the uniform approximation}

In this subsection retain the entropy notation
\[
 h(u)=-u\log u-(1-u)\log(1-u),\qquad
 h(0)=h(1)=0,\qquad
 \varpi(d)=d\log(e/d),\quad \varpi(0)=0.
\]

\begin{lemma}[Elementary entropy bounds]\label{lem:entropy-bounds}
For $u,v\in[0,1]$ and $d=|u-v|\le1/2$,
\[
|h(u)-h(v)|\le\varpi(d).
\]
Moreover,
\[
h(u)\le u\log(e/u),\qquad 0\le u\le\frac12,
\]
\[
h(u)\le 2u(1-u)\log\!\left\{\frac{e}{u(1-u)}\right\},
\qquad 0<u<1.
\]
If $a,b\in[0,1]$ and $t\in[0,1]$, then
\[
t h(a)+(1-t)h(b)\le h\{ta+(1-t)b\}.
\]
\end{lemma}

\begin{proof}
For $0<u<1$,
\[
h'(u)=\log\!\left(\frac{1-u}{u}\right),
\qquad
h''(u)=-\frac1u-\frac1{1-u}<0.
\]
 Hence $h$ is concave, and Jensen's inequality gives
\[
 h\{ta+(1-t)b\}-th(a)-(1-t)h(b)\ge0,
 \qquad a,b,t\in[0,1].
\]
Suppose $0\le v\le u\le1/2$ and $d=u-v$. Since $s\mapsto\log(1/s)$ is decreasing,
\begin{align*}
h(u)-h(v)
&=\int_v^u\log\!\left(\frac{1-s}{s}\right)ds\\
&\le\int_v^{v+d}\log(1/s)\,ds
\le\int_0^d\log(1/s)\,ds
=d\log(e/d).
\end{align*}
The same bound holds for $1/2\le v\le u\le1$ because $h(u)=h(1-u)$. If $v<1/2<u$, then
\[
|h(u)-h(v)|=|h(1-u)-h(v)|,
\qquad
|(1-u)-v|\le u-v=d,
\]
and the preceding bound applies because $\varpi$ is increasing on $[0,1]$.
For $0\le u\le1/2$,
\[
-(1-u)\log(1-u)
\le (1-u)\frac{u}{1-u}=u,
\]
so
\[
h(u)\le u\log(1/u)+u=u\log(e/u).
\]
For $0<u\le1/2$,
\[
u(1-u)\ge\frac u2,
\qquad
\log\!\left\{\frac{e}{u(1-u)}\right\}\ge\log(e/u),
\]
which gives
\[
h(u)\le2u(1-u)\log\!\left\{\frac{e}{u(1-u)}\right\}.
\]
The case $1/2<u<1$ follows from symmetry.
\end{proof}

Fix $\eta\in(0,1/2)$ and $t\in[\eta,1-\eta]$.
Let $F,G$ be continuous distribution functions and
$H=tF+(1-t)G$.

Let $Y_1,\ldots,Y_n$ be distinct, and suppose the distribution
functions $\widehat F,\widehat G$ satisfy
\[
 \widehat H=t\widehat F+(1-t)\widehat G
 =\frac1n\sum_{i=1}^n\mathbf1(Y_i\le\cdot).
\]
For $Y_{(1)}<\cdots<Y_{(n)}$ and
$v_{\ell,n}=(\ell-1/2)/n$, set
\[
 \widehat{\cZ}_{n,t}(\widehat F,\widehat G)
 =\frac1n\sum_{\ell=1}^n
 \frac{t h\{\widehat F(Y_{(\ell)})\}
       +(1-t)h\{\widehat G(Y_{(\ell)})\}}
 {v_{\ell,n}(1-v_{\ell,n})}.
\]

\begin{lemma}[Deterministic stability of the coordinate statistic]\label{lem:stability}
There are constants $c_\eta\in(0,1)$ and $C_\eta<\infty$ such that, whenever
\[
\max\{\|\widehat F-F\|_\infty,\|\widehat G-G\|_\infty,n^{-1}\}
\le\varepsilon\le c_\eta,
\]
we have
\[
\left|\widehat{\cZ}_{n,t}(\widehat F,\widehat G)-\cZ_t(F,G)\right|
\le C_\eta\varepsilon^{1/2}\log(e/\varepsilon).
\]
\end{lemma}

\begin{proof}
Let $H^{-1}(u)=\inf\{x:H(x)\ge u\}$ and define, for $0<u<1$,
\[
f(u)=F\{H^{-1}(u)\},
\qquad
g(u)=G\{H^{-1}(u)\}.
\]
Extend both functions to the endpoints by
$f(0)=g(0)=0$ and $f(1)=g(1)=1$.
Continuity of $H$ gives $H\{H^{-1}(u)\}=u$. For $0<u<v<1$,
\begin{align*}
v-u
&=H\{H^{-1}(v)\}-H\{H^{-1}(u)\}\\
&=t\bigl[f(v)-f(u)\bigr]+(1-t)\bigl[g(v)-g(u)\bigr],
\end{align*}
so
\[
0\le f(v)-f(u)\le\frac{v-u}{t}\le\frac{v-u}{\eta},
\qquad
0\le g(v)-g(u)\le\frac{v-u}{1-t}\le\frac{v-u}{\eta}.
\]
If $H(x)=H(y)$ for $x<y$, then
\[
0=t\{F(y)-F(x)\}+(1-t)\{G(y)-G(x)\},
\]
and both increments are nonnegative. Hence $F(y)=F(x)$ and $G(y)=G(x)$, so
\[
f\{H(x)\}=F(x),
\qquad
g\{H(x)\}=G(x).
\]
Set
\[
\widetilde h_t(u)=t h\{f(u)\}+(1-t)h\{g(u)\},
\qquad
r_t(u)=\frac{\widetilde h_t(u)}{u(1-u)}.
\]
The probability integral transform and $t f(u)+(1-t)g(u)=u$ give
\[
\cZ_t(F,G)=\int_0^1r_t(u)\,du,
\qquad
0\le\widetilde h_t(u)\le h(u).
\]
If $d=|u-v|\le\eta/2$, Lemma~\ref{lem:entropy-bounds} yields
\begin{align*}
|\widetilde h_t(u)-\widetilde h_t(v)|
&\le t\varpi\!\left(\frac d t\right)
 +(1-t)\varpi\!\left(\frac d{1-t}\right)\\
&=d\log(et/d)+d\log\{e(1-t)/d\}\\
&\le2\varpi(d).
\end{align*}
Write
\[
q_\ell=\frac\ell n,
\qquad
\widetilde h_{\ell,n}=t h\{\widehat F(Y_{(\ell)})\}
 +(1-t)h\{\widehat G(Y_{(\ell)})\},
\qquad
\widetilde h_\ell=\widetilde h_t\{H(Y_{(\ell)})\}.
\]
Because $\widehat H(Y_{(\ell)})=q_\ell$ and
\[
\|\widehat H-H\|_\infty
\le t\|\widehat F-F\|_\infty
 +(1-t)\|\widehat G-G\|_\infty
\le\varepsilon,
\]
we have
\[
|H(Y_{(\ell)})-q_\ell|\le\varepsilon,
\qquad
|H(Y_{(\ell)})-v_{\ell,n}|\le\frac32\varepsilon.
\]
Take $c_\eta$ small enough that
\[
b=4\sqrt\varepsilon\le\frac14,
\qquad
\frac32\varepsilon\le\frac\eta2.
\]
Let
\[
\mathcal I_b=\{\ell:b\le v_{\ell,n}\le1-b\}.
\]
Concavity of $h$ gives
\[
0\le\widetilde h_{\ell,n}
\le h\{t\widehat F(Y_{(\ell)})+(1-t)\widehat G(Y_{(\ell)})\}
=h(q_\ell).
\]
For $v_{\ell,n}<b$, $q_\ell\le2v_{\ell,n}$ and $q_\ell\le1/2$, hence
\[
\frac{\widetilde h_{\ell,n}}{v_{\ell,n}(1-v_{\ell,n})}
\le4\log\!\left(\frac e{v_{\ell,n}}\right).
\]
Let $N_b=\#\{\ell:v_{\ell,n}<b\}$. Since $n^{-1}\le\varepsilon$,
\[
nb=4n\sqrt\varepsilon\ge4\sqrt n,
\qquad
N_b\le2nb.
\]
Using $v_{\ell,n}\ge\ell/(2n)$ and
\[
\log(N_b!)=\sum_{\ell=1}^{N_b}\log\ell
\ge\int_1^{N_b}\log x\,dx
=N_b\log N_b-N_b+1,
\]
we obtain
\begin{align*}
\frac1n\sum_{v_{\ell,n}<b}
\frac{\widetilde h_{\ell,n}}{v_{\ell,n}(1-v_{\ell,n})}
&\le\frac4n\sum_{\ell=1}^{N_b}\log\!\left(\frac{2en}{\ell}\right)\\
&\le\frac{4N_b}{n}\log\!\left(\frac{2e^2n}{N_b}\right)
\le Cb\log(e/b).
\end{align*}
 For the upper boundary set $r=n-\ell+1$. Then
\[
 1-q_\ell=(r-1)/n,\qquad
 v_{\ell,n}(1-v_{\ell,n})=v_{r,n}(1-v_{r,n}).
\]
On this boundary $r/n<1/2$, so monotonicity of $h$ gives
\[
 h(q_\ell)=h((r-1)/n)\le h(r/n)=h(q_r).
\]
The lower-boundary upper bound therefore also gives
\[
\frac1n\sum_{v_{\ell,n}>1-b}
\frac{\widetilde h_{\ell,n}}{v_{\ell,n}(1-v_{\ell,n})}
\le Cb\log(e/b).
\]
Moreover,
\begin{align*}
\int_0^br_t(u)\,du+\int_{1-b}^1r_t(u)\,du
&\le2\int_0^b\frac{h(u)}{u(1-u)}\,du\\
&\le4\int_0^b\log(e/u)\,du
\le Cb\log(e/b).
\end{align*}
For $\ell\in\mathcal I_b$, Lemma~\ref{lem:entropy-bounds} and the modulus bound for $\widetilde h_t$ imply
\begin{align*}
|\widetilde h_{\ell,n}-\widetilde h_t(v_{\ell,n})|
&\le|\widetilde h_{\ell,n}-\widetilde h_\ell|
 +|\widetilde h_t\{H(Y_{(\ell)})\}-\widetilde h_t(v_{\ell,n})|\\
&\le\varpi(\varepsilon)+2\varpi(3\varepsilon/2)
\le C\varpi(\varepsilon).
\end{align*}
Therefore
\[
\left|
\frac1n\sum_{\ell\in\mathcal I_b}
\frac{\widetilde h_{\ell,n}-\widetilde h_t(v_{\ell,n})}
{v_{\ell,n}(1-v_{\ell,n})}
\right|
\le\frac{C\varpi(\varepsilon)}b.
\]
For $u,v\in[b/2,1-b/2]$ and $d=|u-v|\le1/n$, the exact identity
\[
\left|\frac1{u(1-u)}-\frac1{v(1-v)}\right|
=\frac{d|1-u-v|}{u(1-u)v(1-v)}
\]
and Lemma~\ref{lem:entropy-bounds} give
\begin{align*}
|r_t(u)-r_t(v)|
&\le\frac{|\widetilde h_t(u)-\widetilde h_t(v)|}{u(1-u)}
 +\widetilde h_t(v)
 \left|\frac1{u(1-u)}-\frac1{v(1-v)}\right|\\
&\le\frac{C_\eta}{b}
\left\{\varpi(d)+d\log(e/b)\right\}.
\end{align*}
Let $I_{\ell,n}=((\ell-1)/n,\ell/n]$. Since $b\ge4/\sqrt n$, every $I_{\ell,n}$ with $\ell\in\mathcal I_b$ is contained in $[b/2,1-b/2]$. Hence
\begin{align*}
\left|
\frac1n\sum_{\ell\in\mathcal I_b}r_t(v_{\ell,n})
-\int_{\cup_{\ell\in\mathcal I_b}I_{\ell,n}}r_t(u)\,du
\right|
&\le\frac{C_\eta}{b}
\left\{\varpi(1/n)+\frac{\log(e/b)}n\right\}\\
&\le\frac{C_\eta\varpi(\varepsilon)}b.
\end{align*}
The symmetric difference between
$\cup_{\ell\in\mathcal I_b}I_{\ell,n}$ and $[b,1-b]$ has length at most $2/n$. Since
\[
r_t(u)\le\frac{h(u)}{u(1-u)}\le C\log(e/b),
\qquad b/2\le u\le1-b/2,
\]
we also have
\[
\left|
\int_{\cup_{\ell\in\mathcal I_b}I_{\ell,n}}r_t(u)\,du
-\int_b^{1-b}r_t(u)\,du
\right|
\le\frac{C\log(e/b)}n
\le\frac{C\varpi(\varepsilon)}b.
\]
Combining the boundary and interior bounds yields
\begin{align*}
\left|\widehat{\cZ}_{n,t}(\widehat F,\widehat G)-\cZ_t(F,G)\right|
&\le C_\eta\left\{b\log(e/b)+\frac{\varpi(\varepsilon)}b\right\}\\
&\le C_\eta\sqrt\varepsilon\log(e/\varepsilon).
\end{align*}
\end{proof}

\begin{lemma}[Uniform empirical-cdf bound]\label{lem:empirical-cdf}
Under Assumption~\ref{ass:regularity}, with $L_{n,p}=\log(16pn^4)$,
\[
\Pp\!\left[
\max_{k\in\cK_n}\max_{1\le j\le p}
\max\left\{
\|\widehat F^{(k)}_{1j}-F^L_{j,k/n}\|_\infty,
\|\widehat F^{(k)}_{2j}-F^R_{j,k/n}\|_\infty
\right\}
>\sqrt{\frac{L_{n,p}}{m_n}}
\right]
\le\frac{1}{2n^3}.
\]
\end{lemma}

\begin{proof}
For a nonempty prefix or suffix $I$ contained entirely in regime $q\in\{1,2\}$, let $r=|I|$ and
\[
\widehat F_{j,I}(x)=\frac1r\sum_{i\in I}\mathbf1(X_{ij}\le x).
\]
The observations in $I$ are independent and share the continuous
distribution function $F^{(q)}_j$.
The two-sided Dvoretzky--Kiefer--Wolfowitz--Massart inequality
\citep{Massart1990} applies for every sample size $r$ and threshold
$x>0$, with bound $2e^{-2rx^2}$. Taking
$x=\sqrt{L_{n,p}/(2r)}$ gives
\[
\Pp\!\left(
\|\widehat F_{j,I}-F^{(q)}_j\|_\infty
>\sqrt{\frac{L_{n,p}}{2r}}
\right)
\le2e^{-L_{n,p}}.
\]
Let \(\mathscr P_q\) be the nonempty prefixes and suffixes of
regime \(q\). Define
\[
 \mathcal E_{n,p}=
 \bigcap_{q=1}^2\bigcap_{I\in\mathscr P_q}\bigcap_{j=1}^p
 \left\{\|\widehat F_{j,I}-F_j^{(q)}\|_\infty
                   \le\sqrt{\frac{L_{n,p}}{2|I|}}\right\}.
\]
There are at most \(2n\) such prefixes and suffixes. Therefore
\[
\Pp(\mathcal E_{n,p}^{c})
\le4pn e^{-L_{n,p}}
=\frac{1}{4n^3}
\le\frac{1}{2n^3},
\]
For either segment
\(S=\{1,\ldots,k\}\) or \(S=\{k+1,\ldots,n\}\), let
\[
 S_q=S\cap I_q,\qquad r_q=|S_q|,\qquad r=|S|=r_1+r_2,
 \quad I_1=\{1,\ldots,\tau^\star\},\quad
 I_2=\{\tau^\star+1,\ldots,n\}.
\]
Every nonempty \(S_q\) is one of the regime prefixes or suffixes
included in \(\mathcal E_{n,p}\). With zero-length terms omitted,
\[
 \widehat F_{j,S}=\sum_{q:r_q>0}\frac{r_q}{r}\widehat F_{j,S_q},
 \qquad
 F_{j,S}=\sum_{q:r_q>0}\frac{r_q}{r}F_j^{(q)} .
\]
Consequently, on \(\mathcal E_{n,p}\),
\[
 \begin{split}
 \|\widehat F_{j,S}-F_{j,S}\|_\infty
 &\le\sum_{q:r_q>0}\frac{r_q}{r}
                  \sqrt{\frac{L_{n,p}}{2r_q}}\\
 &=\sqrt{\frac{L_{n,p}}2}\,
                  \frac{\sqrt{r_1}+\sqrt{r_2}}{r}
 \le\sqrt{\frac{L_{n,p}}r}
 \le\sqrt{\frac{L_{n,p}}{m_n}},
 \end{split}
\]
where the third inequality uses
\((\sqrt{r_1}+\sqrt{r_2})^2\le2(r_1+r_2)\).
For the left segment \(F_{j,S}=F^L_{j,k/n}\); for the right
segment \(F_{j,S}=F^R_{j,k/n}\). This includes \(k=\tau^\star\),
where each segment has only one nonzero term.
\end{proof}

Suppose the pooled observations have no ties within each coordinate.
For a uniformly random permutation $\pi$, define
\[
 \mathcal D_{\pi,n,p}
 =\max_{\substack{k\in\cK_n,\ 1\le j\le p\\1\le\ell\le n}}
   \max_{q\in\{1,2\}}
 \left|\widehat F^{\pi,(k)}_{qj}(X_{(\ell)j})-\frac\ell n\right|.
\]

\begin{lemma}[Conditional permutation empirical-distribution bound]\label{lem:permutation-cdf}
For every such pooled sample,
\[
 \Pp_\pi\!\left(
 \mathcal D_{\pi,n,p}>\sqrt{\frac{L_{n,p}}{2m_n}}
 \,\middle|\,\bX_1,\ldots,\bX_n\right)\le\frac1{4n^2}.
\]
\end{lemma}

\begin{proof}
Fix $(j,k,\ell)$. Among the $n$ pooled observations for coordinate $j$, exactly $\ell$ have rank at most $\ell$. Therefore
\[
\widehat F^{\pi,(k)}_{1j}(X_{(\ell)j})
=\frac1k\sum_{i=1}^k\mathbf1\{R_{\pi(i)j}\le\ell\}
\]
has conditional mean \(u_\ell=\ell/n\). Let
\(\xi_1,\ldots,\xi_k\) be independent Bernoulli\((u_\ell)\)
replacement labels. Theorem~4 of \citet{Hoeffding1963} gives
\[
 \begin{aligned}
 &\E_\pi e^{s\sum_{i=1}^k\{\mathbf1(R_{\pi(i)j}\le\ell)-u_\ell\}}
       \le\prod_{i=1}^k\E e^{s(\xi_i-u_\ell)}
       =e^{k\psi_{u_\ell}(s)},\\
 &\psi_u(s)=\log(1-u+ue^s)-us,\qquad
 \psi_u(0)=\psi_u'(0)=0,\qquad
 \psi_u''(s)=\frac{u(1-u)e^s}{(1-u+ue^s)^2}\le\tfrac14 .
 \end{aligned}
\]
Integrating the last bound twice gives
\(\psi_u(s)\le s^2/8\), so, conditionally on the pooled values,
\[
 \E_\pi\exp\left[
 s\sum_{i=1}^k\{\mathbf1(R_{\pi(i)j}\le\ell)-\ell/n\}
 \,\middle|\,\bX_1,\ldots,\bX_n\right]\le e^{ks^2/8}.
\]
For \(x>0\), the upper-tail Chernoff exponent is
\[
 \inf_{s>0}\{-skx+ks^2/8\}
 =-2kx^2,\qquad s_{\rm opt}=4x .
\]
Applying the same calculation to the negative sum gives
\[
\Pp_\pi\!\left(
\left|\widehat F^{\pi,(k)}_{1j}(X_{(\ell)j})-\frac\ell n\right|>x
\,\middle|\,\bX_1,\ldots,\bX_n
\right)
\le2e^{-2kx^2}.
\]
The right segment is also a simple random sample without replacement, so
\[
\Pp_\pi\!\left(
\left|\widehat F^{\pi,(k)}_{2j}(X_{(\ell)j})-\frac\ell n\right|>x
\,\middle|\,\bX_1,\ldots,\bX_n
\right)
\le2e^{-2(n-k)x^2}.
\]
Taking
\[
x=\sqrt{\frac{L_{n,p}}{2m_n}}
\]
and summing the bounds over at most $2pn^2$ triples and segments gives
\[
4pn^2e^{-L_{n,p}}=\frac{1}{4n^2}.
\]
\end{proof}

\subsection{Uniform approximation}

\begin{proof}[Proof of Proposition~\ref{prop:uniform}]
On the event in Lemma~\ref{lem:empirical-cdf}, fix $(j,k)$ and put
\[
t=\frac kn,
\qquad
F=F^L_{j,k/n},
\qquad
G=F^R_{j,k/n},
\qquad
\widehat F=\widehat F^{(k)}_{1j},
\qquad
\widehat G=\widehat F^{(k)}_{2j}.
\]
Then $t\in[\eta,1-\eta]$ and
\[
tF+(1-t)G
=\lambda^\star F^{(1)}_j+(1-\lambda^\star)F^{(2)}_j.
\]
The statistic in Lemma~\ref{lem:stability} is exactly $Z_j(k)$, while
\[
\cZ_t(F,G)=\frac{\pi^2}{3}-\Delta_j(k/n).
\]
Since
\[
\max\{\|\widehat F-F\|_\infty,
\|\widehat G-G\|_\infty,n^{-1}\}
\le\varepsilon_{n,p},
\]
Lemma~\ref{lem:stability} gives the first assertion simultaneously for all $(j,k)$.
For the permutation assertion, let $R_{ij}$ be the pooled rank of $X_{ij}$ and set $Y_{ij}=R_{ij}/n$. For each coordinate, the pooled transformed observations are
\[
\left\{\frac1n,\frac2n,\ldots,1\right\}.
\]
On the event in Lemma~\ref{lem:permutation-cdf}, the two segment empirical distribution functions of the transformed ranks satisfy, at every grid point $\ell/n$,
\[
\left|\widehat U^{\pi,(k)}_{qj}(\ell/n)-\frac\ell n\right|
\le\sqrt{\frac{L_{n,p}}{2m_n}},
\qquad q\in\{1,2\}.
\]
Between adjacent grid points the empirical distribution functions are constant, hence
\[
\|\widehat U^{\pi,(k)}_{qj}-U\|_\infty
\le\sqrt{\frac{L_{n,p}}{2m_n}}+\frac1n
\le\varepsilon_{n,p},
\]
where $U(u)=u$ on $[0,1]$. Lemma~\ref{lem:stability}, applied with $F=G=U$, gives
\[
\left|Z_j^\pi(k)-\cZ_{k/n}(U,U)\right|
\le C_\eta a_{n,p}.
\]
Lemma~\ref{lem:jensen} gives $\cZ_{k/n}(U,U)=\pi^2/3$. The continuity assumption and independence imply
\[
\Pp(X_{ij}=X_{rj})=0,
\qquad i\ne r,
\]
so the pooled no-tie event has probability one after taking a finite union over $(i,r,j)$.
\end{proof}

\begin{proof}[Proof of Proposition~\ref{prop:studentized-separation}]
Fix $\comp\in\{\mathrm{sum},\mathrm{max}\}$ and write
\[
\delta_{\comp}(t)=\Delta_{\comp}(\lambda^\star)-\Delta_{\comp}(t)\ge0,
\qquad
\delta_{\mu,n,k}=\mu_{n,k}-\frac{\pi^2}{3}.
\]
For $t=k/n$ and $k\in\cK_n$, \cref{eq:population-standardized-location-score} gives
\begin{align*}
\frac{\Psi_{\comp,n,p}(\lambda^\star)-\Psi_{\comp,n,p}(t)}{w_{\comp,p}}
&=\frac{\Delta_{\comp}(\lambda^\star)+\delta_{\mu,n,\tau^\star}}{s_{n,\tau^\star}}
-\frac{\Delta_{\comp}(t)+\delta_{\mu,n,k}}{s_{n,k}}\\
&=\frac{\delta_{\comp}(t)}{s_{n,\tau^\star}}
+\Delta_{\comp}(t)\left(\frac1{s_{n,\tau^\star}}-\frac1{s_{n,k}}\right)
+\frac{\delta_{\mu,n,\tau^\star}}{s_{n,\tau^\star}}
-\frac{\delta_{\mu,n,k}}{s_{n,k}}.
\end{align*}
Because $0\le\Delta_{\comp}(t)\le\Delta_{\comp}(\lambda^\star)$,
\[
\left|\frac1{s_{n,\tau^\star}}-\frac1{s_{n,k}}\right|
\le\frac{s_{n,\max}-s_{n,\min}}{s_{n,\min}^2},
\qquad
\left|\frac{\delta_{\mu,n,\tau^\star}}{s_{n,\tau^\star}}
-\frac{\delta_{\mu,n,k}}{s_{n,k}}\right|
\le\frac{2\delta_{\mu,n}}{s_{n,\min}}.
\]
Therefore
\begin{align*}
\frac{s_{n,\min}}{w_{\comp,p}}
\{\Psi_{\comp,n,p}(\lambda^\star)-\Psi_{\comp,n,p}(t)\}
&\ge
\frac{s_{n,\min}}{s_{n,\max}}\delta_{\comp}(t)
-\Delta_{\comp}(\lambda^\star)
\frac{s_{n,\max}-s_{n,\min}}{s_{n,\min}}
-2\delta_{\mu,n}\\
&=\frac{\delta_{\comp}(t)}{\kappa_n}
-\Delta_{\comp}(\lambda^\star)(\kappa_n-1)-2\delta_{\mu,n}.
\end{align*}
Taking the minimum over the grid points satisfying
$|k/n-\lambda^\star|\ge\epsilon$ proves the first assertion.
The null-mean assertion is \eqref{eq:null-mean-uniform-bound}.
The last assertion is Proposition~\ref{prop:rank-covariance}.
\end{proof}

Define
\[
\mathcal E_0=\left\{
\max_{k\in\cK_n}\max_{1\le j\le p}
\left|Z_j(k)-\left\{\frac{\pi^2}{3}-\Delta_j(k/n)\right\}\right|
\le C_\eta a_{n,p}\right\}.
\]
Proposition~\ref{prop:uniform} gives
$\Pp(\mathcal E_0^c)\le(2n^3)^{-1}$.

\begin{proof}[Proof of Theorem~\ref{thm:power}]
On $\mathcal E_0$, evaluation at the true split yields
\begin{align*}
\Ssum
&\ge
\frac{\sqrt p\{\mu_{n,\tau^\star}-A_n(\tau^\star)\}}
{s_{n,\tau^\star}}\\
&\ge
\frac{\sqrt p}{s_{n,\tau^\star}}
\{\overline\Delta_n-\varrho_{n,p}\}.
\end{align*}
The dense signal condition makes the last expression exceed the fixed
critical value $c_{\mathrm{sum},1-\alpha}(\eta)$ for all sufficiently large \(n\), so
\[
\Pp_{H_1}\{\phi_{\mathrm{sum},\alpha}=1\}
\ge\Pp(\mathcal E_0)\ge1-\frac1{2n^3}.
\]
Choose $j_\star$ such that $\Delta_{j_\star}=\Delta_{\max,n}$ and put
\[
L_{n,p}^{\mathrm{max}}
=\frac{\Delta_{\max,n}-\varrho_{n,p}}{s_{n,\tau^\star}}.
\]
On $\mathcal E_0$,
\[
\Smax\ge\mathcal M_{j_\star,n}
\ge
\frac{\mu_{n,\tau^\star}-Z_{j_\star}(\tau^\star)}
{s_{n,\tau^\star}}
\ge L_{n,p}^{\mathrm{max}}.
\]
Under \cref{eq:sparse-power-ld},
\[
\frac{\Smax-\mathfrak b_{n,p}}{\mathfrak a_{n,p}}
\ge
\frac{L_{n,p}^{\mathrm{max}}-\mathfrak b_{n,p}}{\mathfrak a_{n,p}}
\longrightarrow+\infty.
\]
For all sufficiently large \(n\),
\[
\Pp_{H_1}\{\phi_{\mathrm{max},\alpha}=1\}
\ge\Pp(\mathcal E_0)\ge1-\frac1{2n^3}.
\]
\end{proof}

\begin{proof}[Proof of Corollary~\ref{cor:cauchy-power}]
For $u\downarrow0$,
\[
\tan\{\pi(1/2-u)\}=\cot(\pi u)
=\frac1{\pi u}\{1+O(u^2)\}\longrightarrow+\infty.
\]
Let
\[
 C_{r,n}=\tan\{\pi(1/2-P_{r,n,p})\},
 \qquad r\in\{\mathrm{sum},\mathrm{max}\}.
\]
If $P_{\mathrm{sum},n,p}\to0$ and $C_{\mathrm{max},n}=O_{\Pp}(1)$,
fix $L<\infty$ and $\epsilon>0$.  Tightness supplies a constant
$B<\infty$, independent of $n$, such that
\[
 \limsup_{n\to\infty}
 \Pp\{|(1-\omega)C_{\mathrm{max},n}|>B\}<\epsilon.
\]
On the complementary event, $\Tcauchy(\omega)\le L$ implies
$\omega C_{\mathrm{sum},n}\le L+B$.  Hence
\[
 \limsup_{n\to\infty}\Pp\{\Tcauchy(\omega)\le L\}
 \le
 \limsup_{n\to\infty}
 \Pp\{\omega C_{\mathrm{sum},n}\le L+B\}+\epsilon
 =\epsilon.
\]
Letting \(\epsilon\downarrow0\) proves divergence in probability.
The same argument applies after interchanging the components.
If both component \(p\)-values tend to zero, then, for every fixed \(L\),
\[
 \{\omega C_{\mathrm{sum},n}
       +(1-\omega)C_{\mathrm{max},n}\le L\}
 \subseteq
 \{C_{\mathrm{sum},n}\le L\}\cup\{C_{\mathrm{max},n}\le L\},
\]
whose probability tends to zero. Thus
\[
 \Tcauchy(\omega)\xrightarrow{\Pp}+\infty,
 \qquad
 \Pcauchy(\omega)
 =\frac12-\frac1\pi\arctan\{\Tcauchy(\omega)\}\xrightarrow{\Pp}0.
\]
\end{proof}

\begin{proof}[Proof of Theorem~\ref{thm:location}]
For the dense score, on $\mathcal E_0$,
\begin{align*}
\max_{k\in\cK_n}
\left|D_{\mathrm{sum},n,p}(k)-\Psi_{\mathrm{sum},n,p}(k/n)\right|
&=\max_{k\in\cK_n}
\frac{\sqrt p\,|A_n(k)-Q_{\mathrm{sum}}(k/n)|}{s_{n,k}}\\
&\le\frac{C_\eta\sqrt p\,a_{n,p}}{s_{n,\min}}.
\end{align*}
For the sparse score, the elementary inequality
\[
\left|\max_j x_j-\max_j y_j\right|
\le\max_j|x_j-y_j|
\]
gives, on the same event,
\begin{align*}
&\max_{k\in\cK_n}
\left|D_{\mathrm{max},n,p}(k)-\Psi_{\mathrm{max},n,p}(k/n)\right|\\
&\quad=\max_{k\in\cK_n}
\left|
\max_{1\le j\le p}\frac{\mu_{n,k}-Z_j(k)}{s_{n,k}}
-\max_{1\le j\le p}
\frac{\mu_{n,k}-\pi^2/3+\Delta_j(k/n)}{s_{n,k}}
\right|\\
&\quad\le\frac{C_\eta a_{n,p}}{s_{n,\min}}.
\end{align*}
Consequently, for either $\comp\in\{\mathrm{sum},\mathrm{max}\}$,
\begin{equation}\label{eq:proof-location-uniform}
\max_{k\in\cK_n}
\left|D_{\comp,n,p}(k)-\Psi_{\comp,n,p}(k/n)\right|
\le \mathfrak e_{\comp,n,p},
\qquad
\mathfrak e_{\comp,n,p}=\frac{C_\eta w_{\comp,p}a_{n,p}}{s_{n,\min}}.
\end{equation}
Because $\widehat\tau_{\comp}$ maximizes $D_{\comp,n,p}$,
\[
D_{\comp,n,p}(\widehat\tau_{\comp})\ge D_{\comp,n,p}(\tau^\star).
\]
On $\mathcal E_0$, \cref{eq:proof-location-uniform} yields
\begin{align*}
\Psi_{\comp,n,p}(\lambda^\star)
-\Psi_{\comp,n,p}(\widehat\tau_{\comp}/n)
&\le
\left|\Psi_{\comp,n,p}(\lambda^\star)-D_{\comp,n,p}(\tau^\star)\right|\\
&\quad+D_{\comp,n,p}(\tau^\star)-D_{\comp,n,p}(\widehat\tau_{\comp})\\
&\quad+
\left|D_{\comp,n,p}(\widehat\tau_{\comp})
-\Psi_{\comp,n,p}(\widehat\tau_{\comp}/n)\right|\\
&\le2\mathfrak e_{\comp,n,p}.
\end{align*}
If $|\widehat\tau_{\comp}/n-\lambda^\star|\ge\epsilon$, the left-hand side is at least $g_{\comp,n,p}(\epsilon)$. Under \cref{eq:separation-location},
\[
g_{\comp,n,p}(\epsilon)>2\mathfrak e_{\comp,n,p}
\]
for all sufficiently large $n$. Hence
\[
\Pp\!\left(
\left|\frac{\widehat\tau_{\comp}}{n}-\lambda^\star\right|\ge\epsilon
\right)
\le\Pp(\mathcal E_0^c)
\le\frac1{2n^3}\longrightarrow0.
\]
Under \cref{eq:local-separation-rate}, the preceding score inequality gives
\[
\underline c_{\comp,n,p}\min\!\left
\{
|\widehat\tau_{\comp}/n-\lambda^\star|^{\beta_{\comp}},1
\right\}
\le2\mathfrak e_{\comp,n,p}.
\]
Because $r_{\comp,n,p}\to0$, for all sufficiently large $n$,
\[
\frac{2\mathfrak e_{\comp,n,p}}{\underline c_{\comp,n,p}}
<r_{\comp,n,p}^{\beta_{\comp}}<1.
\]
Therefore, on $\mathcal E_0$,
\[
\left|\frac{\widehat\tau_{\comp}}{n}-\lambda^\star\right|
\le\left(\frac{2C_\eta w_{\comp,p}a_{n,p}}
{s_{n,\min}\underline c_{\comp,n,p}}\right)^{1/\beta_{\comp}}
<r_{\comp,n,p}.
\]
The probability bound and stochastic order follow from
$\Pp(\mathcal E_0^c)\le(2n^3)^{-1}$.
\end{proof}

\section{Theory for componentwise and Cauchy-adaptive wild binary segmentation}
\label{app:wbs}

Throughout this appendix, \(N=N_n^{\mathrm W}=\log(en)\).
The probabilistic bounds below use Assumption~\ref{ass:wbs-explicit};
the aggregate SUM bounds and the resulting Cauchy bounds additionally
use \eqref{eq:wbs-sum-copula}. The reference functions
\(F_\eta,F_{\mathrm G}\) and normalizers
\(\mathfrak a_{L,p},\mathfrak b_{L,p}\) are used through their
deterministic definitions and bounds. Constants may depend on the
fixed model and trimming constants, but not on \(n,p\), the signal
strengths, or the active coordinate sets.

\subsection{Analytic scales and homogeneous-interval tails}

For an interval of length $L$, put $N_L=\log L$,
$\gamma_L=(\log p)/N_L$, and define
\[
 q(\gamma)=(\sqrt\gamma+\sqrt{1+\gamma})^2,\qquad
 \zeta(\gamma)=\min\{(1-q(\gamma)^{-2})/4,b_\eta\},\qquad
 y(\gamma)=\frac{\gamma+\lambda\{\zeta(\gamma)\}}{\zeta(\gamma)}.
\]
Here $\lambda$ and $b_\eta$ are the analytic quantities defined in
Section~\ref{sec:null}; the original normalizers
$\mathfrak a_{L,p},\mathfrak b_{L,p}$ use its three-branch rule.

\begin{lemma}[Uniform analytic scales]\label{lem:wbs-weak-analytic-scales}
Fix $0<\eta<1/2$. As $L\to\infty$, uniformly over $0<\gamma_-\le\gamma_L\le\gamma_+<\infty$,
\begin{equation}\label{eq:wbs-weak-analytic-scales}
 \mathfrak a_{L,p}^{-1}=\zeta(\gamma_L)\sqrt{2N_L},\qquad
 \mathfrak b_{L,p}=(y(\gamma_L)-1)\sqrt{N_L/2}
                         +O\{\log N_L/\sqrt{N_L}\}.
\end{equation}
In particular, $\mathfrak a_{L,p}\asymp N_L^{-1/2}$ and
$|\mathfrak b_{L,p}|\le C\sqrt{N_L}$. If $c_\ell n\le L\le n$
and $\zeta_n=\zeta\{(\log p)/(\log n)\}$, then
$|\zeta(\gamma_L)-\zeta_n|\le C/\log n$ and $c\le\zeta_n\le C$.
\end{lemma}
\begin{proof}
The common formula \eqref{eq:full-max-branch-selection} gives the
scale identity and the center correction
$\mathfrak a_{L,p}\log\mathcal C_{L,p}$. Write $C_L$ for the
coefficient in \eqref{eq:main-subcritical-normalization} with $n=L$.
In the subcritical branch,
the switching width and the endpoint bounds
\eqref{eq:core-real-K-singularity} and
\eqref{eq:core-confluent-real-coefficient} give
\[
 b_\eta<1/4\ \Longrightarrow\
 b_\eta-\frac{1-q(\gamma_L)^{-2}}4\ge cN_L^{-1/4},\qquad
 N_L^{-C}\le C_L\le N_L^C.
\]
For $b_\eta=1/4$, the compact range of $\gamma_L$ keeps the
subcritical tilt uniformly below $1/4$, giving the same bound.
In the boundary branch, its excess $E_L=y(\gamma_L)-q_c$ satisfies
$cN_L^{-1/4}\le E_L\le C$; the explicit positive coefficient in
\eqref{eq:main-boundary-coefficients} and
\eqref{eq:main-boundary-normalization} therefore give
\[
 |\log\mathcal C_{L,p}|
 \le C\{\log N_L+|\log E_L|+|\log\log(N_LE_L)|\}
 \le C\log N_L.
\]
The critical bound follows from
\eqref{eq:full-critical-prefactor-polynomial}. Thus all three branches
satisfy $|\log\mathcal C_{L,p}|\le C\log N_L$.
On the stated compact dimension range,
$c\le\zeta(\gamma_L)\le C$ and $1+c\le y(\gamma_L)\le C$,
which proves \eqref{eq:wbs-weak-analytic-scales}. Finally,
\[
 \left|\frac{\log p}{\log L}-\frac{\log p}{\log n}\right|
 =\frac{\log p\,|\log(n/L)|}{\log L\log n}\le\frac C{\log n};
\]
the function $\zeta$ is Lipschitz on the compact dimension range.
\end{proof}

Write $\overline F_\eta=1-F_\eta$, where $F_\eta$ is the
distribution function of $\mathcal M_\eta$ in Section~\ref{sec:null}.
The following bounds concern that reference distribution, independently
of the distribution of the observed statistic.

\begin{lemma}[Reference tails]\label{lem:wbs-weak-reference-tails}
Fix $0<\eta<1/2$. For every fixed $h>0$ there are positive constants $c,C_{\eta,h}$
such that
\begin{equation}\label{eq:wbs-weak-reference-tails}
 \frac{c}{1+x}e^{-x^2/2}\le\overline F_\eta(x)
       \le C_{\eta,h}e^{-e^{-2h}x^2/2},\qquad x\ge0.
\end{equation}
Consequently, as $u\downarrow0$,
\[
 c_{\mathrm{sum},1-u}(\eta)\asymp\sqrt{\log(1/u)},\qquad
 g_{1-u}=\log(1/u)+O(u).
\]
\end{lemma}
\begin{proof}
Retaining a single stationary normal value gives
\[
 \overline F_\eta(x)\ge\int_x^{x+(1+x)^{-1}}
                   \frac{e^{-v^2/2}}{\sqrt{2\pi}}\,dv
 \ge\frac{c}{1+x}e^{-x^2/2}.
\]
Let $L_\eta=\log\{(1-\eta)/\eta\}$. By stationarity, the OU
representation of $\mathcal M_\eta$ is the supremum on $[0,2L_\eta]$ of
\[
 U_t=e^{-t}(U_0+B_{e^{2t}-1}),\qquad
 U_0\sim N(0,1),\quad U_0\perp B.
\]
The process $U_0+B_v$ has the same law as
$\widetilde B_{1+v}$ for a Brownian motion $\widetilde B$.
The exponential martingale
$\exp\{\theta\widetilde B_v-\theta^2v/2\}$, stopped at the
first crossing on a finite grid and then with $\theta=z/T$, gives
\[
 \Pp\{\max_{v\in\mathcal D}\widetilde B_v\ge z\}
 \le e^{-\theta z+\theta^2T/2}=e^{-z^2/(2T)},\qquad
 \mathcal D\subset[0,T],\quad z\ge0.
\]
Increasing finite grids and path continuity give the same bound
for the supremum. For a block $ih\le t\le(i+1)h$,
\[
 \{\sup U_t\ge x\}
 \subseteq\{\sup_{v\le e^{2(i+1)h}}\widetilde B_v\ge xe^{ih}\},
\]
so its probability is at most $\exp\{-e^{-2h}x^2/2\}$.
There are at most $1+\lceil2L_\eta/h\rceil$ blocks, proving
\eqref{eq:wbs-weak-reference-tails}. Its two sides bracket the
upper $u$-quantile between constant multiples of
$\sqrt{\log(1/u)}$. The remaining expansion follows from
$g_{1-u}=-\log\{-\log(1-u)\}$ and
$-\log(1-u)=u\{1+O(u)\}$.
\end{proof}

\begin{proof}[Proof of Lemma~\ref{lem:wbs-cauchy-dominance}]
Put \(q=q_{\mathrm C,n}\). Stabilization bounds either cotangent
below by \(-\cot(\pi q)\). For \(0<x\le\pi/4\),
\(x^{-1}-x\le\cot x\le x^{-1}\). Thus a component value at most
\(q/4\) gives
\[
 T_{\mathrm C}(I)\ge\tfrac12\cot(\pi q/4)-\tfrac12\cot(\pi q)
 \ge\frac{3}{2\pi q}-\frac{\pi q}{8}
 \ge\frac1{\pi q}\ge\cot(\pi q),
\]
where the third inequality uses \(q\le1/4\).
Applying the decreasing inverse cotangent proves
\(P_{\mathrm C}(I)\le q\). Conversely, if both original component
values exceed \(q\), both clipped values do as well, since
\(1-q>q\). Their average cotangent is then less than \(\cot(\pi q)\),
so \(P_{\mathrm C}(I)>q\). Taking the contrapositive proves the converse.
\end{proof}

Recall the stopping levels in \eqref{eq:wbs-weak-stopping}.
The independently sampled interval pool has at most \(M_n+1\) members.

\begin{lemma}[Stopping on homogeneous intervals]\label{lem:wbs-weak-null-stopping}
Under Assumption~\ref{ass:wbs-explicit}, fix the trimming fraction
\(0<\eta<1/2\) and use the exact local rank constants
\(\mu_{L,r},s_{L,r}\). Let \(\mathcal R_n^0\) consist of the intervals
in the sampled pool that are contained in a single true regime.
For the stopping levels in \eqref{eq:wbs-weak-stopping},
\begin{equation}\label{eq:wbs-max-null-stopping}
 \mathbb P\left\{\exists I\in\mathcal R_n^0:
       P_{\mathrm{max}}(I)\le q_{\mathrm{max},n}
       \,\middle|\,\mathcal R_n\right\}
 \le CpM_n n e^{-c\Lambda_n}=o(1).
\end{equation}
This MAX bound imposes no restriction on dependence among coordinates.
If condition~\eqref{eq:wbs-sum-copula} also holds, then
\begin{equation}\label{eq:wbs-weak-null-stopping}
 \begin{split}
 &\mathbb P\left\{\exists I\in\mathcal R_n^0,\
       g\in\{\mathrm{sum},\mathrm{max},\mathrm C\}:
       P_g(I)\le q_{g,n}
       \,\middle|\,\mathcal R_n\right\}\\
 &\hspace{20mm}\le CM_n n e^{-cN^2}
                       +CpM_n n e^{-c\Lambda_n}=o(1).
 \end{split}
\end{equation}
Both bounds remain valid when a data-dependent subset of
\(\mathcal R_n^0\) is examined.
\end{lemma}
\begin{proof}
Condition on the data-independent pool and fix \(I=(s,e]\in\mathcal R_n^0\),
with \(L=e-s\). At a trimmed split \(k\), put \(r=k-s\) and
\[
 D_{j,I}(k)=\frac{\mu_{L,r}-Z_{j,I}(k)}{s_{L,r}},\qquad
 \sigma^0_{L,r}=Ls_{L,r}.
\]
For each coordinate, the observations on \(I\) are independent with
one continuous marginal law. Their chronological ranks are therefore
uniform. Exact studentization and the deterministic null-moment
expansion \eqref{eq:main-null-moment-orders} give
\[
 (\sigma^0_{L,r})^2=2\log L+O_\eta(1)\asymp N.
\]
The centered raw deficit \(L\{\mu_{L,r}-Z_{j,I}(k)\}
=\sigma^0_{L,r}D_{j,I}(k)\) obeys the scalar uniform-rank bound
\eqref{eq:rank-raw-bernstein}. Consequently,
\[
 \mathbb P\{|D_{j,I}(k)|\ge x\mid\mathcal R_n\}
 \le C\exp\{-c\min(x^2,x\sqrt N)\},\qquad x>0.
\]
The non-strict event is covered by applying that bound at half its
positive threshold. No joint coordinate law has been used.

For \(u\in\{q_{\mathrm{max},n},q_{\mathrm C,n}\}\),
\(\log(1/u)=\Lambda_n+O(\log M_n+1)\sim\Lambda_n\).
The deterministic analytic scales and reference quantiles in
Lemmas~\ref{lem:wbs-weak-analytic-scales} and
\ref{lem:wbs-weak-reference-tails} give
\[
 c_{L,p}(u):=\mathfrak b_{L,p}+\mathfrak a_{L,p}g_{1-u}
       \asymp\Lambda_n/\sqrt N>0.
\]
Monotonicity of the Gumbel reference distribution and the union over
coordinates and trimmed splits imply
\[
 \begin{split}
 \mathbb P\{P_{\mathrm{max}}(I)\le u\mid\mathcal R_n\}
 &\le CpL\exp[-c\min\{c_{L,p}(u)^2,c_{L,p}(u)\sqrt N\}]\\
 &\le CpL e^{-c\Lambda_n},
 \end{split}
\]
since \(\Lambda_n/N\to\infty\). There are at most \(M_n+1\le2M_n\)
pool members for large \(n\). Summing this bound proves
\eqref{eq:wbs-max-null-stopping}.

Now impose \eqref{eq:wbs-sum-copula}. The margins are constant on
this homogeneous interval, so its transformed rows are iid
\(N_p(0,\mathbf R_p)\), with \(B_p\le\overline B_n\).
Apply \eqref{eq:sum-pointwise-bernstein} in
Lemma~\ref{lem:sum-rank-exponential-controls} with sample size \(L\)
and local split \(r=k-s\). Since \(\log L\asymp N\), applying the
strict-tail bound at \(x/2\) gives
\[
 \mathbb P\{|D_{\mathrm{sum},I}(k)|\ge x\mid\mathcal R_n\}
 \le C\exp\left[-c\min\left\{
       \frac{x^2}{B_p},\frac{x\sqrt{pN}}{B_p}\right\}\right],\qquad x>0.
\]
The lemma allows singular \(\mathbf R_p\) and arbitrary finite
\(B_p\); conditioning on the data-independent pool does not change
the law on a fixed interval.

For \(u\in\{q_{\mathrm{sum},n},q_{\mathrm C,n}\}\), the OU reference
quantile satisfies \(c_{\mathrm{sum},1-u}(\eta)\asymp\sqrt{\Lambda_n}\).
Since \(p\ge n^{a_-}\) and \(\overline B_n=N^{\beta_{\mathrm W}}\),
\[
 \min\left\{\frac{\Lambda_n}{B_p},
       \frac{\sqrt{\Lambda_n pN}}{B_p}\right\}
 \ge\min\{N^2,\sqrt p\,N^{(3-\beta_{\mathrm W})/2}\}
 \ge cN^2.
\]
Another union over the trimmed splits gives
\[
 \mathbb P\{P_{\mathrm{sum}}(I)\le u\mid\mathcal R_n\}
       \le CL e^{-cN^2}.
\]
The converse in Lemma~\ref{lem:wbs-cauchy-dominance} gives
\[
 \{P_{\mathrm C}(I)\le q_{\mathrm C,n}\}
 \subseteq\{P_{\mathrm{sum}}(I)\le q_{\mathrm C,n}\}
       \cup\{P_{\mathrm{max}}(I)\le q_{\mathrm C,n}\}.
\]
Combining this inclusion with the two component bounds and summing
over the fixed pool proves \eqref{eq:wbs-weak-null-stopping}.
The polynomial bound on \(p\), \(M_n=o(N)\), and
\(\Lambda_n=N^{\beta_{\mathrm W}+2}\) make both displayed error
bounds tend to zero. A subset can only decrease each rejection event.
Both results follow from finite-sample concentration and deterministic
reference tails.
\end{proof}

\subsection{Relative population geometry on an interval}
\label{app:wbs-relative-geometry}

Put \(K_0=\lceil d_*^{-1}\rceil\), a fixed upper bound for \(K_n+1\), and write
\[
 \mathcal S_{q,n}=\{j:F_j^{(q)}\ne F_j^{(q+1)}\},\qquad
 s_{q,n}=|\mathcal S_{q,n}|.
\]
For an interval \(J=(a,b]\) with integer endpoints, put
\[
 \mathcal Q(J)=\{q:a<\tau_q<b\},\qquad
 \mathcal S_J=\bigcup_{q\in\mathcal Q(J)}\mathcal S_{q,n},
 \qquad s_J=|\mathcal S_J|,
\]
\[
 A_\Sigma(J)=\max_{q\in\mathcal Q(J)}\Delta_{\Sigma,q,n},
 \qquad
 A_\infty(J)=\max_{q\in\mathcal Q(J)}\Delta_{\infty,q,n},
 \qquad E_\Sigma(J)=nA_\Sigma(J),\quad E_\infty(J)=nA_\infty(J).
\]
An empty maximum and an empty sum are zero. These quantities concern
all adjacent changes in \(J\); the sets \(\mathcal S_{q,n}\) need not coincide.
For \(I=(a_I,b_I]\subseteq J\), \(L=|I|\), use continuous mixtures between
integer splits and define
\[
 f_{j,I}(t)=\Delta_{j,I}(a_I+Lt),\qquad
 f_{\Sigma,I}(t)=\sum_{j=1}^p f_{j,I}(t),\qquad
 f_{\infty,I}(t)=\max_{j\le p}f_{j,I}(t).
\]
The actual trimmed endpoints are
\(\eta_L=\lceil\eta L\rceil/L\) and \(1-\eta_L\).

\begin{lemma}[Relative population bounds]\label{lem:wbs-relative-population}
Under Assumption~\ref{ass:wbs-explicit}, there is a constant \(C_{\rm var}\),
depending only on the density-ratio and spacing constants, such that,
for every such interval \(J\) and every subinterval,
\[
 0\le f_{\Sigma,I}(t)\le C_{\rm var}t(1-t)A_\Sigma(J),\qquad
 0\le f_{j,I}(t)\le C_{\rm var}t(1-t)A_\infty(J),\quad 0\le t\le1.
\]
On each open cell between consecutive true changes, for
\(f=f_{\Sigma,I}\) or \(f=f_{j,I}\),
\[
 f''(t)\ge f(t)\ge0.
\]
If an interval \(I_q\) has endpoints
\[
 a_{I_q}\in[\tau_q-\mathfrak d_n/2,\tau_q-\mathfrak d_n/3],
 \qquad
 b_{I_q}\in[\tau_q+\mathfrak d_n/3,\tau_q+\mathfrak d_n/2],
\]
then it isolates change \(q\), its change fraction lies in \([2/5,3/5]\),
and
\[
 \frac{|I_q|}{n}f_{\Sigma,I_q}(t_{I_q,\tau_q})
       \ge c_{\rm iso}\Delta_{\Sigma,q,n},\qquad
 \frac{|I_q|}{n}f_{\infty,I_q}(t_{I_q,\tau_q})
       \ge c_{\rm iso}\Delta_{\infty,q,n},
 \qquad c_{\rm iso}=d_*/2
\]
for all sufficiently large \(n\), including integer rounding.
In particular, define
\[
 c_{\rm sel}=c_{\rm iso}/16,\qquad
 \eta_0=\min\{1/8,\ c_{\rm iso}/(512C_{\rm var})\}.
\]
For every fixed \(0<\eta<\eta_0\),
\[
 \sup_{t\in[0,2\eta]\cup[1-2\eta,1]}f_{\Sigma,I}(t)
       \le(c_{\rm sel}/16)A_\Sigma(J),\qquad
 \sup_{t\in[0,2\eta]\cup[1-2\eta,1]}f_{\infty,I}(t)
       \le(c_{\rm sel}/16)A_\infty(J).
\]
\end{lemma}
\begin{proof}
For the reference distribution \(H_j\) in the density-ratio condition, put
\[
 \|F-G\|_j^2=
 \int\frac{\{F(x)-G(x)\}^2}{H_j(x)^2\{1-H_j(x)\}^2}\,dH_j(x).
\]
The bounds on the densities also bound CDFs and upper tails:
\[
 c_*H_j\le F\le C_*H_j,\qquad
 c_*(1-H_j)\le1-F\le C_*(1-H_j).
\]
They hold for every regime mixture. Thus the norm is finite for a
difference of two such mixtures. With \(U=tF+(1-t)G\), Taylor's integral
formula and \(-h''(v)=\{v(1-v)\}^{-1}\) give
\[
 c\,\frac{t(1-t)(F-G)^2}{H_j(1-H_j)}
 \le h(U)-th(F)-(1-t)h(G)
 \le C\,\frac{t(1-t)(F-G)^2}{H_j(1-H_j)} .
\]
Also
\[
 c\,\frac{dH_j}{H_j(1-H_j)}
 \le\frac{dU}{U(1-U)}
 \le C\,\frac{dH_j}{H_j(1-H_j)} .
\]
Integration proves
\[
 c_Dt(1-t)\|F-G\|_j^2
 \le\mathcal D_t(F,G)
 \le C_Dt(1-t)\|F-G\|_j^2 .
\]
In particular, for the adjacent quantities used in the signal conditions,
\[
 \frac{6c_D}{25}
       \|F_j^{(q)}-F_j^{(q+1)}\|_j^2
 \le d_{qj,n}\le
 \frac{C_D}{4}\|F_j^{(q)}-F_j^{(q+1)}\|_j^2 .
\]
The spacing condition gives \(K_n+1\le K_0\).
For regimes \(u<v\) in \(J\), Cauchy--Schwarz gives
\[
 \|F_j^{(u)}-F_j^{(v)}\|_j^2
 \le K_0\sum_{q\in\mathcal Q(J)}
       \|F_j^{(q)}-F_j^{(q+1)}\|_j^2 .
\]
The same inequality holds for two mixtures: express their difference as
a convex combination of regime differences and apply convexity of
the squared norm. Consequently
\[
 \sum_j\|F^L_{j,I,k}-F^R_{j,I,k}\|_j^2
 \le C\sum_{q\in\mathcal Q(J)}\Delta_{\Sigma,q,n}
 \le CK_0A_\Sigma(J),
\]
\[
 \max_j\|F^L_{j,I,k}-F^R_{j,I,k}\|_j^2
 \le C\max_j\sum_{q\in\mathcal Q(J)}d_{qj,n}
 \le CK_0A_\infty(J).
\]
These bounds and the preceding comparison prove the first assertions.
They include any short regime fragments produced by the endpoints of \(I\).

On a homogeneous split cell fix a threshold \(x\), and write
\(P=F^L_{j,I,t}(x)\), \(R=F^R_{j,I,t}(x)\), \(Q=F_j^{(q)}(x)\)
for its current regime, and \(u=tP+(1-t)R\).
The pooled distribution and hence \(u\) do not depend on the split. Put
\[
 J(t,x)=h(u)-th(P)-(1-t)h(R).
\]
The identities
\[
 P'=(Q-P)/t,\qquad R'=(R-Q)/(1-t)
\]
give
\[
 J'=-\operatorname{KL}(Q\Vert P)+\operatorname{KL}(Q\Vert R),
 \qquad
 J''=\frac{(Q-P)^2}{tP(1-P)}
             +\frac{(Q-R)^2}{(1-t)R(1-R)} .
\]
Bernoulli variance decomposition yields
\[
 u(1-u)=tP(1-P)+(1-t)R(1-R)+t(1-t)(P-R)^2.
\]
Thus
\[
 J''\ge\frac{(Q-P)^2+(Q-R)^2}{u(1-u)}
       \ge\frac{(P-R)^2}{2u(1-u)},\qquad
 J\le\frac{t(1-t)(P-R)^2}{u(1-u)},
\]
so \(J''\ge J/\{2t(1-t)\}\ge2J\).
Integrating against the fixed pooled measure divided by \(u(1-u)\)
proves \(f''\ge f\). Summation preserves this inequality.

The density-ratio bounds and the mixture-distance estimates above
provide integrable bounds for these derivatives on the trimmed cells,
justifying differentiation under the integral. Finally, the isolating rectangles give
\[
 |I_q|\ge2\lceil\mathfrak d_n/3\rceil\ge2\mathfrak d_n/3,\qquad
 t_{I_q,\tau_q}\in[2/5,3/5],
\]
with a possible inward rounding of the rectangles. Substitution in
the definitions of \(d_{qj,n}\), followed by a sum or maximum,
proves the witness bounds. The last assertion follows from
\(t(1-t)\le2\eta\) on the two outer bands and
\(2C_{\rm var}\eta\le c_{\rm iso}/256=c_{\rm sel}/16\).
\end{proof}

\subsection{Uniform rank bounds under weak changes}

For $I=(s,e]$ of length $L$ and $k\in\mathcal K(I)$, write
$r=k-s$ and $t=r/L$. Let $F_{ij}$ be the population marginal cdf
of row $i$, and define the population mixtures
\[
 P=F^L_{j,I,k}=\frac1r\sum_{i=s+1}^kF_{ij},\qquad
 R=F^R_{j,I,k}=\frac1{L-r}\sum_{i=k+1}^eF_{ij},\qquad
 H_{j,I}=tP+(1-t)R=\frac1L\sum_{i=s+1}^eF_{ij}.
\]
Define the raw statistic and its exact null constants by
\[
 \begin{aligned}
 \overline Z_L&=\sum_{l=1}^{L-1}
       \frac{Lh(l/L)}{(l-1/2)(L-l+1/2)},&
 T_{j,I}(k)&=L\{\overline Z_L-Z_{j,I}(k)\},\\
 m^0_{L,r}&=\E_0T_{j,I}(k),&
 \sigma^0_{L,r}&=Ls_{L,r},\qquad
 D_{j,I}(k)=\frac{T_{j,I}(k)-m^0_{L,r}}{\sigma^0_{L,r}}.
 \end{aligned}
\]
Here $\E_0$ denotes uniform chronological ordering of the pooled
ranks; these constants are unchanged under an alternative.
With $w(u)=u(1-u)$, put
\[
 \begin{aligned}
 P_0(u)&=P\{H_{j,I}^{-1}(u)\},&
 R_0(u)&=R\{H_{j,I}^{-1}(u)\},\\
 J_t(v_1,v_2;u)&=t\varphi(v_1)+(1-t)\varphi(v_2)-\varphi(u),&
 K_{j,I,k}(u)&=\frac{J_t(P_0(u),R_0(u);u)}{w(u)}.
 \end{aligned}
\]
Thus $\Delta_{j,I}(k)=\int_0^1K_{j,I,k}(u)\,du$.
For comparison across interval lengths, also put
\[
 X_{j,I}(k)=\frac{D_{j,I}(k)-\mathfrak b_{L,p}}{\mathfrak a_{L,p}},
 \qquad \zeta_n=\zeta\{(\log p)/(\log n)\},\qquad
 Y_{j,I}(k)=X_{j,I}(k)/\zeta_n.
\]

\begin{lemma}[A signal-dependent uniform rank bound]\label{lem:wbs-weak-rank}
Fix $0<\eta<1/2$. Suppose the rows are independent and all regime margins satisfy
$c_*\le dF_j^{(q)}/dH_j\le C_*$ for continuous $H_j$, with
fixed positive constants $c_*,C_*$. Let the data-independent pool
have at most $M_n+1$ intervals of lengths at least $\ell_n$.
For sufficiently large $\ell_n$ and $1\le z\le c\ell_n$,
conditionally on the pool, an event of
probability at least $1-Cp(M_n+1)n^3e^{-z}$ satisfies
\begin{equation}\label{eq:wbs-weak-raw-bound}
 |T_{j,I}(k)-L\Delta_{j,I}(k)|
 \le C\{\sqrt{L\Delta_{j,I}(k)\,z\log(eL)}+z\log(eL)\}
\end{equation}
simultaneously over all its intervals, trimmed splits and coordinates.
If $\ell_n\ge c_\ell n$, $n^{a_-}\le p\le n^{a_+}$ and
$M_n=o(N)$, where $N=\log(en)$ and $0<a_-<a_+<\infty$ are fixed,
then $z=C_zN$ gives failure probability at most $Cn^{-3}$ for
sufficiently large fixed $C_z$. On the same event, the original
analytic MAX normalization satisfies
\begin{equation}\label{eq:wbs-weak-studentized-bound}
 |Y_{j,I}(k)-L\Delta_{j,I}(k)|
 \le C\{N\sqrt{L\Delta_{j,I}(k)}+N^2+L\Delta_{j,I}(k)/N\}.
\end{equation}
The constants do not require independence across coordinates.
\end{lemma}
\begin{proof}
Condition on the pool and first fix $I,j,k$. Every regime cdf
and every mixture has density between fixed positive constants
relative to $H=H_{j,I}$. For $A=(s,k]$ or $(k,e]$, write
$m=|A|$, $F_A=m^{-1}\sum_{i\in A}F_{ij}$ and
$\widehat F_A=m^{-1}\sum_{i\in A}\mathbf1\{X_{ij}\le x\}$.
If $H(x)=v$, the centered indicators
$\xi_i=\mathbf1\{X_{ij}\le x\}-F_{ij}(x)$ satisfy
\[
 |\xi_i|\le1,\qquad \sum_{i\in A}\E\xi_i^2\le Cm w(v),\qquad
 \log\E\exp\left(\theta\sum_{i\in A}\xi_i\right)
 \le\frac{Cm w(v)\theta^2}{2(1-|\theta|/3)},\quad |\theta|<3.
\]
The last inequality follows by expanding the exponential,
using $\E\xi_i=0$, $\E|\xi_i|^q\le\E\xi_i^2$ for $q\ge2$,
and $q!\ge2\cdot3^{q-2}$. Exponential Markov inequality, applied
with each sign of $\theta$, gives
\[
 \Pp\left\{|\widehat F_A(x)-F_A(x)|
       >C\left(\sqrt{w(v)z/L}+z/L\right)\right\}\le2e^{-z},
 \qquad m\ge\eta L.
\]
Apply this at $x_v=H^{-1}(v/L^2)$, $1\le v<L^2$, with the
two infinite endpoints. Between consecutive grid points,
$F_A(x_{v+1})-F_A(x_v)\le C/L^2$; monotonicity of both cdfs
and $|w(u)-w(v)|\le|u-v|$ therefore give
\begin{equation}\label{eq:wbs-weak-weighted-cdf}
 |\widehat F_A(x)-F_A(x)|
 \le C\{\sqrt{H(x)(1-H(x))z/L}+z/L\},\qquad x\in\mathbb R.
\end{equation}
There are at most $2p(M_n+1)n$ side--split--coordinate choices
and at most $n^2+1$ grid points for each. Their exceptional
probabilities sum to at most $Cp(M_n+1)n^3e^{-z}$.
The pooled empirical cdf satisfies the same inequality by taking
the convex combination of the two side inequalities.

Work on this common event. Write $u_l=l/L$, let $x_l$ be the
$l$th pooled order statistic, and put
$\widehat P_l=\widehat F^L_{j,I,k}(x_l)$ and
$\widehat R_l=\widehat F^R_{j,I,k}(x_l)$. Continuity and row
independence imply that there are almost surely no ties, so
\[
 t\widehat P_l+(1-t)\widehat R_l=u_l
       =tP_0(u_l)+(1-t)R_0(u_l).
\]
Choose $m_*=\lceil C_*'z\rceil\le L/4$, increasing $C_*'$
and decreasing the constant in $z\le c\ell_n$ if necessary.
Inverting the pooled version of \eqref{eq:wbs-weak-weighted-cdf}
and using the bounded derivatives of $P_0,R_0$ gives, for
$m_*\le l\le L-m_*$,
\[
 \begin{aligned}
 |H(x_l)-u_l|&\le C\{\sqrt{w(u_l)z/L}+z/L\},\\
 |\widehat P_l-P_0(u_l)|+|\widehat R_l-R_0(u_l)|
     &\le e_l:=C\{\sqrt{w(u_l)z/L}+z/L\}.
 \end{aligned}
\]
The inversion follows from
$|v-u_l|\le C\{\sqrt{w(v)z/L}+z/L\}$ with $v=H(x_l)$;
$w(v)\le w(u_l)+|v-u_l|$ and
$C\sqrt{|v-u_l|z/L}\le |v-u_l|/2+C'z/L$
give the first displayed bound.
Every probability on the segment joining the empirical and
population pairs is then between fixed positive multiples of
$u_l$, and its complement between such multiples of $1-u_l$.

Suppress the subscripts on $K$, set $d_l=P_0(u_l)-R_0(u_l)$,
and write $\operatorname{logit}v=\log\{v/(1-v)\}$ for $0<v<1$.
Taylor's formula for the entropy, with the common-mixture
constraint in the preceding display, gives
\[
 K(u_l)\asymp d_l^2/w(u_l)^2,\qquad
 |\operatorname{logit}P_0(u_l)-\operatorname{logit}R_0(u_l)|
                         \le C|d_l|/w(u_l).
\]
Indeed the directional first derivative of $J_t$ is
$t\{\operatorname{logit}P_0-\operatorname{logit}R_0\}$ times
the increment of $P_0$, whereas its second directional derivative
has absolute value at most $C/w(u_l)$. Consequently
\begin{equation}\label{eq:wbs-weak-entropy-taylor}
 \left|\frac{J_t(\widehat P_l,\widehat R_l;u_l)}{w(u_l)}-K(u_l)\right|
 \le C\left\{\sqrt{K(u_l)}\frac{e_l}{w(u_l)}
                       +\frac{e_l^2}{w(u_l)^2}\right\}.
\end{equation}
The same density-ratio bounds imply, almost everywhere,
$|P_0'|+|R_0'|\le C$ and
\begin{align*}
 0\le K(u)&\le C,\displaybreak[1]\\
 \frac d{du}J_t(P_0,R_0;u)
 &=t\{\operatorname{logit}P_0-\operatorname{logit}u\}P_0'
 +(1-t)\{\operatorname{logit}R_0-\operatorname{logit}u\}R_0',\displaybreak[1]\\
 \left|\frac d{du}J_t(P_0,R_0;u)\right|&\le C,
 \qquad |K'(u)|\le C/w(u).
\end{align*}
Integration over the mesh cells, with the two endpoint cells
bounded by $K\le C$, yields
\[
 \left|\sum_{l=1}^{L-1}K(u_l)-L\Delta_{j,I}(k)\right|
 \le C+C\int_{1/L}^{1-1/L}\frac{du}{w(u)}\le C\log(eL).
\]
The elementary harmonic sums give
\[
 \begin{aligned}
 \sum_{l=m_*}^{L-m_*}w(u_l)^{-1}&\le CL\log(eL),&
 \sum_{l=m_*}^{L-m_*}w(u_l)^{-2}&\le CL^2/m_*,\\
 \sum_{l=m_*}^{L-m_*}\frac{e_l^2}{w(u_l)^2}
 &\le C\left\{\frac zL\sum_lw(u_l)^{-1}
            +\frac{z^2}{L^2}\sum_lw(u_l)^{-2}\right\}
 \le Cz\log(eL).
 \end{aligned}
\]
Summing \eqref{eq:wbs-weak-entropy-taylor} and applying
Cauchy--Schwarz thus bounds its left-hand side sum by
\[
 C\sqrt{\{L\Delta_{j,I}(k)+\log(eL)\}z\log(eL)}+Cz\log(eL)
 \le C\{\sqrt{L\Delta_{j,I}(k)z\log(eL)}+z\log(eL)\}.
\]
The exact raw identity is
\[
 T_{j,I}(k)=\sum_{l=1}^{L-1}
 \frac{L^2J_t(\widehat P_l,\widehat R_l;u_l)}{(l-1/2)(L-l+1/2)}.
\]
For any probabilities with $tv_1+(1-t)v_2=u$, including zero
empirical cells, the chi-square bound for Bernoulli relative
entropy gives
\[
 0\le J_t(v_1,v_2;u)
 \le\frac{t(v_1-u)^2+(1-t)(v_2-u)^2}{w(u)}\le C_\eta w(u).
\]
Each omitted rank therefore contributes at most a constant, so
the two rank tails cost at most $Cm_*$. Moreover
\[
 \left|\frac{L^2}{(l-1/2)(L-l+1/2)}-\frac1{w(u_l)}\right|
                       \le\frac{C}{Lw(u_l)^2};
\]
the half-shift weight replacement costs at most
$C\sum_{l=1}^{L-1}\{Lw(u_l)\}^{-1}\le C\log(eL)$.
These bounds prove \eqref{eq:wbs-weak-raw-bound}.

Now take $z=C_zN$. The failure bound is at most $Cn^{-3}$ by
$p\le n^{a_+}$ and $M_n=o(N)$. The deterministic null moment
expansions \eqref{eq:main-null-moment-orders} give
\[
 m^0_{L,r}=\log L+O(1),\qquad
 \sigma^0_{L,r}=\sqrt{2\log L}\{1+O(N^{-1})\},
\]
uniformly on the trimmed grid, under either sampling law for
the observed statistic. Lemma~\ref{lem:wbs-weak-analytic-scales}
therefore implies
\[
 \frac1{\zeta_n\mathfrak a_{L,p}\sigma^0_{L,r}}
                  =1+O(N^{-1}),\qquad
 \frac{|\mathfrak b_{L,p}|}{\zeta_n\mathfrak a_{L,p}}=O(N).
\]
Insert these expressions into
\[
 Y_{j,I}(k)=\frac{T_{j,I}(k)-m^0_{L,r}}
                   {\zeta_n\mathfrak a_{L,p}\sigma^0_{L,r}}
                -\frac{\mathfrak b_{L,p}}{\zeta_n\mathfrak a_{L,p}}.
\]
Equation~\eqref{eq:wbs-weak-raw-bound} then yields
\eqref{eq:wbs-weak-studentized-bound}, including the relative
studentization term $L\Delta_{j,I}(k)/N$.
\end{proof}

\begin{lemma}[Aggregation on changed and unchanged coordinates]
\label{lem:wbs-weak-sum}
Suppose Assumption~\ref{ass:wbs-explicit} and
\eqref{eq:wbs-sum-copula} hold, and fix \(0<\eta<1/2\). Define
\[
 Y_{\Sigma,I}(k)=\sqrt{2pN}\,D_{\mathrm{sum},I}(k).
\]
There is an event whose conditional probability, given the interval
pool, is at least \(1-Cn^{-2}\), on which, simultaneously for every
integer-endpoint interval \(J\), every \(I\in\mathcal R_n\) contained
in \(J\), and every \(k\in\mathcal K(I)\),
\[
 \begin{split}
 \left|Y_{\Sigma,I}(k)
       -|I|\sum_{j=1}^p\Delta_{j,I}(k)\right|
 \le C\bigg\{&
       \frac{E_\Sigma(J)}N
       +N\sqrt{s_JE_\Sigma(J)}+s_JN^2\\
       &+\sqrt{p\overline B_n}\,N+\overline B_nN
       \bigg\}.
 \end{split}
\]
In particular, the same event satisfies the bound
\[
 \left|Y_{\Sigma,I}(k)
       -|I|\sum_{j=1}^p\Delta_{j,I}(k)\right|
 \le C\left\{
 \sqrt{pN\Lambda_n}
 +N\sqrt{s_JE_\Sigma(J)}
 +s_JN^2+\frac{E_\Sigma(J)}N
 \right\}.
\]
\end{lemma}

\begin{proof}
Write \(I=(s,e]\), \(L=e-s\), and \(r=k-s\), and retain the exact
rank deficit and its null moments:
\[
 T_{j,I}(k)=L\{\overline Z_L-Z_{j,I}(k)\},\qquad
 m^0_{L,r}=L(\overline Z_L-\mu_{L,r}),\qquad
 \sigma^0_{L,r}=Ls_{L,r}.
\]
The deterministic null expansions in
\eqref{eq:main-null-moment-orders} imply, uniformly for
\(c_\ell n\le L\le n\) and trimmed \(r\),
\[
 |m^0_{L,r}|\le CN,\qquad
 (\sigma^0_{L,r})^2=2\log L+O(1),\qquad
 \left|\frac{\sqrt{2N}}{\sigma^0_{L,r}}-1\right|
       \le \frac CN.
\]
These expansions concern the deterministic calibration constants and
hold under heterogeneous observations as well.
Set \(b_{L,r}=\sqrt{2N}/\sigma^0_{L,r}\). The exact identity
\[
 Y_{\Sigma,I}(k)
 =b_{L,r}\sum_{j=1}^p\{T_{j,I}(k)-m^0_{L,r}\}
\]
retains the exact null mean.

Choose \(z=C_1N\) in Lemma~\ref{lem:wbs-weak-rank}, with \(C_1\)
sufficiently large. On its uniform event, Cauchy--Schwarz and
\(L\le n\) give
\[
 \begin{split}
 &\left|
 \sum_{j\in\mathcal S_J}\{T_{j,I}(k)-m^0_{L,r}\}
       -L\sum_{j\in\mathcal S_J}\Delta_{j,I}(k)
 \right|\\
 &\quad\le
 C N\sum_{j\in\mathcal S_J}\sqrt{L\Delta_{j,I}(k)}
       +Cs_JN^2\\
 &\quad\le
 C N\sqrt{Ls_J\sum_{j\in\mathcal S_J}\Delta_{j,I}(k)}
       +Cs_JN^2\\
 &\quad\le C\{N\sqrt{s_JE_\Sigma(J)}+s_JN^2\}.
 \end{split}
\]
The last line uses the population comparison in
Lemma~\ref{lem:wbs-relative-population},
\[
 \sup_{\substack{I\subseteq J,\ |I|\ge c_\ell n\\
                 k\in\mathcal K(I)}}
       \sum_{j=1}^p\Delta_{j,I}(k)\le CA_\Sigma(J).
\]
Coordinates outside \(\mathcal S_J\) have identical marginal
distributions throughout \(J\). Their common Gaussian copula is
unchanged as well, so their joint row vectors are iid within each
such \(I\). In particular \(\Delta_{j,I}(k)=0\) for these coordinates.

Define their partial dense score by
\[
 D_{0,J,I}(k)=\frac1{\sqrt p}
       \sum_{j\notin\mathcal S_J}
       \frac{T_{j,I}(k)-m^0_{L,r}}{\sigma^0_{L,r}}.
\]
Put \(d_J=|\mathcal S_J^c|\). If \(d_J=0\), this score is
identically zero. Otherwise, the unchanged margins and
\eqref{eq:wbs-sum-copula} give an iid Gaussian subvector with
correlation matrix equal to the corresponding principal submatrix
of \(\mathbf R_p\). Write \(B_{0,J}\) for the maximum absolute row
sum of this submatrix; then \(B_{0,J}\le B_p\le\overline B_n\). Since
\[
 D_{0,J,I}(k)=\sqrt{\frac{d_J}{p}}
       \left\{\frac1{\sqrt{d_J}}
                    \sum_{j\notin\mathcal S_J}D_{j,I}(k)\right\},
\]
\eqref{eq:sum-pointwise-bernstein}, applied to this \(d_J\)-coordinate
sample of length \(L\), gives, conditionally on the pool,
\[
 \begin{split}
 \mathbb P\{|D_{0,J,I}(k)|>x\mid\mathcal R_n\}
 &\le2\exp\left[-c\min\left\{
       \frac{px^2}{d_JB_{0,J}},
       \frac{x\sqrt{pN}}{B_{0,J}}\right\}\right]\\
 &\le2\exp\left[-c\min\left\{
       \frac{x^2}{\overline B_n},
       \frac{x\sqrt{pN}}{\overline B_n}\right\}\right],
 \qquad x>0.
 \end{split}
\]
Here \(d_J\le p\) and \(\log L\asymp N\); the cited lemma permits
singular covariance matrices. Taking
\(x=C\{\sqrt{\overline B_nz}+\overline B_nz/\sqrt{pN}\}\)
therefore gives, also in the case \(d_J=0\),
\[
 \Pp\left\{
 |D_{0,J,I}(k)|
 >C\left(\sqrt{\overline B_nz}
          +\frac{\overline B_nz}{\sqrt{pN}}\right)
 \,\middle|\,\mathcal R_n\right\}\le2e^{-z}.
\]
There are at most \(C M_n n^3\) triples \((J,I,k)\). With
\(z=C_2N\) and \(C_2\) sufficiently large, a union bound therefore
gives
\[
 \sup_{J,I,k}\sqrt{2pN}\,|D_{0,J,I}(k)|
 \le C\{\sqrt{p\overline B_n}\,N+\overline B_nN\}
\]
except on an event of conditional probability at most \(Cn^{-2}\).
The corresponding union bound in Lemma~\ref{lem:wbs-weak-rank}
has a further factor \(p\); it has the same stated probability
after increasing \(C_1\), since \(p\le n^{a_+}\).

Combining the active and inactive parts, using
\(|b_{L,r}|\le C\), and keeping the exact inactive centering gives
\[
 \begin{split}
 Y_{\Sigma,I}(k)-L\sum_j\Delta_{j,I}(k)
 ={}&(b_{L,r}-1)L\sum_j\Delta_{j,I}(k)\\
 &+b_{L,r}
   \left[\sum_{j\in\mathcal S_J}(T_{j,I}(k)-m^0_{L,r})
                     -L\sum_j\Delta_{j,I}(k)\right]\\
 &+\sqrt{2pN}\,D_{0,J,I}(k).
 \end{split}
\]
The first term is bounded by \(CE_\Sigma(J)/N\), proving the
first assertion. Finally,
\[
 \sqrt{p\overline B_n}\,N
       \le \sqrt{pN\Lambda_n},\qquad
 \frac{\overline B_nN}{\sqrt{pN\Lambda_n}}
       =\sqrt{\frac{\overline B_n}{pN}}\longrightarrow0,
\]
because \(p\ge n^{a_-}\) and \(\overline B_n\) is a fixed power
of \(N\). This proves the second assertion.
\end{proof}

\subsection{Small Cauchy values and the sampled interval pool}
\label{app:wbs-weak-pool}

Throughout this subsection put \(N=N_n^{\mathrm W}\),
\(\Lambda_n=\overline B_nN^2\), and
\[
 q_{\mathrm C,n}=\frac{e^{-\Lambda_n}}{4M_n}.
\]
The symbols \(F_\eta\) and \(F_{\mathrm G}\) denote the original OU-supremum
and standard Gumbel reference distributions.

We use Lemma~\ref{lem:wbs-weak-reference-tails} with \(h=\log2\),
so its OU upper bound is \(C_\eta e^{-x^2/8}\).
The elementary Gumbel bound is
\[
 \tfrac12e^{-x}\le1-F_{\mathrm G}(x)\le e^{-x},\qquad x\ge0,
\]
because \(y/2\le1-e^{-y}\le y\) for \(0\le y\le1\).

\begin{lemma}[A small component in the stabilized Cauchy value]
\label{lem:wbs-small-cauchy-value}
Suppose \(0<q_{\mathrm C,n}\le1/4\) and use the upper clipping in
Algorithm~\ref{alg:cauchy-wbs}. If \(u=\min\{P_{\mathrm{sum}}(I),P_{\mathrm{max}}(I)\}\le q_{\mathrm C,n}/4\),
then
\[
 u\le P_{\mathrm C}(I)\le8u.
\]
\end{lemma}
\begin{proof}
For \(0<v\le1/4\),
\[
 \frac1{2\pi v}\le\cot(\pi v)\le\frac1{\pi v}.
\]
The larger clipped component contributes at least
\(-\tfrac12\cot(\pi q_{\mathrm C,n})\), whereas the smaller component is unaltered.
Consequently
\[
 T_{\mathrm C}(I)\ge\frac1{4\pi u}-\frac1{2\pi q_{\mathrm C,n}}
                 \ge\frac1{8\pi u}>0.
\]
Since \(T_{\mathrm C}(I)>0\),
\[
 P_{\mathrm C}(I)=\pi^{-1}\arctan\{T_{\mathrm C}(I)^{-1}\}
          \le\{\pi T_{\mathrm C}(I)\}^{-1}\le8u.
\]
This proves the upper bound. Both clipped component
values are at least \(u\); monotonicity of the cotangent and its inverse
proves the lower bound.
\end{proof}

Let \(\mathfrak J_q\) be the isolating rectangle of
Lemma~\ref{lem:wbs-relative-population}, interpreted with inward integer
rounding. Let \(\mathcal E_{\rm iso}\) mean that the sampled pool meets
every \(\mathfrak J_q\).
For a deterministic radius \(r\), let \(\mathcal E_{\rm end}(r)\) mean
that every sampled endpoint has distance greater than \(2r+2\) from every
internal true change.
For \(g\in\{\mathrm{sum},\mathrm{max},\mathrm C\}\), define
the method-specific homogeneous stopping event
\[
 \mathcal E_{\rm null,g}
 =\bigcap_{I\in\mathcal R_n^0}\{P_g(I)>q_{g,n}\},
\]
where \(\mathcal R_n^0\) is the homogeneous subfamily in
Lemma~\ref{lem:wbs-weak-null-stopping}.

\begin{lemma}[Isolation and endpoint avoidance]
\label{lem:wbs-pool-events}
Under Assumption~\ref{ass:wbs-explicit}, for \(r=o(n)\),
\[
 \mathbb P(\mathcal E_{\rm iso}^c)\le K_n e^{-cM_n},\qquad
 \mathbb P\{\mathcal E_{\rm end}(r)^c\}
       \le CK_nM_n(r+1)/n.
\]
\end{lemma}
\begin{proof}
The eligible integer interval family has order \(n^2\) members,
uniformly under
\(c_\ell n\le\ell_n\le2\mathfrak d_n/3\).
Each isolating rectangle has at least \(c\mathfrak d_n^2\ge cn^2\)
members for large \(n\), giving its fixed positive sampling probability.
Independence of the interval draws and a union over \(q\) prove the
first bound. At most \(n\) eligible intervals have any prescribed
endpoint. The union of the endpoint neighborhoods has at most
\(CK_n(r+1)\) integer points, so a union over both endpoints and
the \(M_n\) intervals proves the second bound. If the full interval is
also included deterministically, its endpoints \(0,n\) do not enter
these neighborhoods for \(r=o(n)\).
\end{proof}
With fixed \(0<\eta<1/2\) and levels \eqref{eq:wbs-weak-stopping},
Lemma~\ref{lem:wbs-weak-null-stopping} gives
\(\mathbb P(\mathcal E_{\rm null,max}^c)\to0\) under
Assumption~\ref{ass:wbs-explicit}. For \(g=\mathrm{sum},\mathrm C\),
the same conclusion for \(\mathcal E_{\rm null,g}\) additionally uses
\eqref{eq:wbs-sum-copula}. The MAX event places no condition on SUM values.

\subsection{Selection and recursive consistency}
\label{app:wbs-weak-recursion}

\begin{proof}[Proofs of Theorems~\ref{thm:wbs-sum-consistency},
\ref{thm:wbs-max-consistency}, and \ref{thm:wbs-cauchy-consistency}]
Fix one of the three model sequences, with method
\(g\in\{\mathrm{sum},\mathrm{max},\mathrm C\}\), and define its minimum
signal ratio from
\eqref{eq:wbs-component-signal-ratios} and
\eqref{eq:wbs-cauchy-signal-ratio} by
\[
 \kappa=
 \begin{cases}
  \min_{q\le K_n}R_{\Sigma,q,n},&g=\mathrm{sum},\\
  \min_{q\le K_n}R_{\infty,q,n},&g=\mathrm{max},\\
  \min_{q\le K_n}R_{\mathrm C,q,n},&g=\mathrm C.
 \end{cases}
\]
Put
\[
 \delta_n=\kappa^{-1/4}+N^{-1},\qquad
 r_n=\lceil C_rn\delta_n\rceil .
\]
The constant \(C_r\) is chosen below and depends only on the common
model constants. The assumptions give
\[
 \delta_n\longrightarrow0,\qquad
 M_n\delta_n
 =\left(\frac{M_n^4}{\kappa}\right)^{1/4}+\frac{M_n}{N}
 \longrightarrow0.
\]
In particular \(r_n=o(n)\), \(M_n(r_n+1)/n\to0\), and
\(4r_n<\mathfrak d_n\), \(2r_n<\eta\ell_n\) for large \(n\).

For all three methods, Assumption~\ref{ass:wbs-explicit} permits
Lemma~\ref{lem:wbs-weak-rank} with a sufficiently large fixed multiple
of \(N\) as the logarithmic deviation parameter. For
\(g=\mathrm{sum},\mathrm C\), the corresponding theorem additionally
imposes \eqref{eq:wbs-sum-copula}, so Lemma~\ref{lem:wbs-weak-sum}
applies as well. Denote the rank event alone by
\(\mathcal E_{\rm weak,max}\), and its intersection with the SUM event
by \(\mathcal E_{\rm weak,g}\) for \(g=\mathrm{sum},\mathrm C\).
Then \(\mathbb P(\mathcal E_{\rm weak,g}^c)\to0\) under the
respective theorem's assumptions, uniformly over every true span
\(J=(\tau_a,\tau_b]\), every sampled \(I\subseteq J\), and every split.

For selection, express the bounds on this event in the following scales.
Put
\[
 Y_{\Sigma,I}(k)=\sqrt{2pN}\,D_{\mathrm{sum},I}(k),\qquad
 X_{j,I}(k)=\frac{D_{j,I}(k)-\mathfrak b_{L,p}}{\mathfrak a_{L,p}},
 \qquad Y_{\infty,j,I}(k)=X_{j,I}(k)/\zeta_n,
\]
where
\[
 \gamma_n=\frac{\log p}{\log n},\qquad \zeta_n=\zeta(\gamma_n),
 \qquad 0<c\le\zeta_n\le C.
\]
For component labels write \(A_{\mathrm{sum}}=A_\Sigma\),
\(A_{\mathrm{max}}=A_\infty\), \(f_{\mathrm{sum},I}=f_{\Sigma,I}\), and
\(f_{\mathrm{max},I}=f_{\infty,I}\).
With the quantities of Lemma~\ref{lem:wbs-relative-population}, the weak
bounds and \(L\le n\) yield the following inequalities; the SUM
inequality is asserted only for \(g=\mathrm{sum},\mathrm C\):
\[
 \begin{aligned}
 |Y_{\Sigma,I}(k)-Lf_{\Sigma,I}(t_{I,k})|
 &\le C\left\{\sqrt{pN\Lambda_n}
       +N\sqrt{s_JE_\Sigma(J)}+s_JN^2+E_\Sigma(J)/N\right\},\\
 |Y_{\infty,j,I}(k)-Lf_{j,I}(t_{I,k})|
 &\le C\left\{N\sqrt{E_\infty(J)}+N^2+E_\infty(J)/N\right\}.
 \end{aligned}
\]
The SUM bound controls the joint null fluctuation of the unchanged
subvector. The \(E/N\) terms account for exact studentization and
the common scale across interval lengths.

Consider a nonempty \(\mathcal Q(J)\), and suppress \(J\) in the next
displays. For SUM define \(H=E_\Sigma^2/(pN)\), for MAX define
\(H=E_\infty\), and for Cauchy define
\[
 H=\max\{E_\Sigma^2/(pN),E_\infty\}.
\]
Under the SUM condition, for every \(q\in\mathcal Q(J)\),
\[
 \frac{(n\Delta_{\Sigma,q,n})^2}{pN}
 \ge\kappa^2\left(\Lambda_n+\frac{s_{q,n}^2N^3}{p}\right).
\]
Under the Cauchy condition, at each \(q\) at least one of this
SUM inequality and
\[
 n\Delta_{\infty,q,n}
 \ge\kappa\left(\Lambda_n+\frac{s_{q,n}^2N^3}{p}\right)
\]
holds, with \(\kappa^2\) in the SUM inequality. Since \(\kappa\ge1\)
eventually and
\(s_J\le K_0\max_{q\in\mathcal Q(J)}s_{q,n}\), these inequalities imply
\[
 H\ge c\kappa^2(\Lambda_n+s_J^2N^3/p)
       \quad\hbox{for SUM},\qquad
 H\ge c\kappa(\Lambda_n+s_J^2N^3/p)
       \quad\hbox{for Cauchy}.
\]
For MAX alone, \(H\ge\kappa\Lambda_n\), without a support penalty.

In the SUM and Cauchy cases put \(t_J=s_J^2N^3/(pH)\).
Since \(E_\Sigma\le\sqrt{pNH}\),
\[
 \left\{\frac{N\sqrt{s_JE_\Sigma}}{\sqrt{pNH}}\right\}^2
 =\frac{s_JE_\Sigma N}{pH}
 \le\frac{s_JN^{3/2}}{\sqrt{pH}}=t_J^{1/2},
 \qquad
 \frac{s_JN^2}{\sqrt{pNH}}=t_J^{1/2}.
\]
It follows that, for the SUM bound when needed,
\[
 |Y_{\Sigma,I}(k)-Lf_{\Sigma,I}(t_{I,k})|
       \le C\delta_n\sqrt{pNH}.
\]
For \(g=\mathrm{max},\mathrm C\), \(E_\infty\le H\) and
\(H\ge c\kappa\Lambda_n\ge c\kappa N^2\), so
\[
 |Y_{\infty,j,I}(k)-Lf_{j,I}(t_{I,k})|
       \le C\{N/\sqrt H+N^2/H+N^{-1}\}H
       \le C\delta_n H.
\]
The constants are uniform over the bounded family of active true spans.

Work now on
\[
 \mathcal E=\mathcal E_{\rm weak,g}\cap\mathcal E_{\rm iso}
            \cap\mathcal E_{\rm end}(r_n)\cap\mathcal E_{\rm null,g}.
\]
An admissible recursive node \(J'=(a',b']\) has two labels \(a<b\) with
\[
 |a'-\tau_a|\le r_n,\qquad |b'-\tau_b|\le r_n;
 \qquad a=0\Longrightarrow a'=0,\quad
 b=K_n+1\Longrightarrow b'=n.
\]
Every sampled \(I=(s,e]\subseteq J'\) lies inside
\(J=(\tau_a,\tau_b]\). Indeed, if \(\tau_a\) is internal and
\(s\le\tau_a\), then \(a'\le s\le\tau_a\) implies
\(|s-\tau_a|\le r_n\), contrary to \(\mathcal E_{\rm end}(r_n)\).
The right endpoint is identical. Thus every actual candidate excludes
the previously assigned boundary changes.

If \(\mathcal Q(J)\ne\varnothing\), the isolating witness for each
\(q\in\mathcal Q(J)\) is contained in \(J'\), because
\[
 s_q\ge\tau_q-\mathfrak d_n/2
       \ge\tau_a+\mathfrak d_n/2>a',\qquad
 e_q\le\tau_q+\mathfrak d_n/2
       \le\tau_b-\mathfrak d_n/2<b'.
\]
For a componentwise procedure, choose the witness for its strongest
adjacent signal in this span. Its population raw value is at least
\(c_{\rm iso}E_\Sigma\) for SUM or \(c_{\rm iso}E_\infty\) for MAX.
The error bounds are \(o(E_\Sigma)\) and \(o(E_\infty)\), respectively.
Reference-tail monotonicity therefore proves both acceptance and
\[
 \max_t\frac Ln f_{\Sigma,\widehat I}(t)
       \ge(c_{\rm iso}/2)A_\Sigma(J)
 \quad\hbox{or}\quad
 \max_t\frac Ln f_{\infty,\widehat I}(t)
       \ge(c_{\rm iso}/2)A_\infty(J).
\]
Its witness negative log p-value is at least \(cH\), and
\(H/\Lambda_n\to\infty\), so its p-value is below the applicable
stopping level.

For Cauchy, consider the component attaining
\(H=\max\{E_\Sigma^2/(pN),E_\infty\}\), and let \(I_0\) be its
strongest-boundary witness. Lemma~\ref{lem:wbs-weak-reference-tails} and the error bounds give
\[
 u_0:=\min\{P_{\mathrm{sum}}(I_0),P_{\mathrm{max}}(I_0)\}
       \le Ce^{-cH},\qquad u_0/q_{\mathrm C,n}\longrightarrow0.
\]
Let \(\widehat I\) be the interval selected by the Cauchy procedure and
\(\widehat g\) its component with the smaller p-value.
Lemma~\ref{lem:wbs-small-cauchy-value} and the Cauchy interval selection rule give
\[
 P_{\mathrm C}(\widehat I)\le P_{\mathrm C}(I_0)\le8u_0=o(q_{\mathrm C,n}).
\]
This proves acceptance. Lemma~\ref{lem:wbs-cauchy-dominance} then gives
\(P_{\widehat g}(\widehat I)\le q_{\mathrm C,n}\), so clipping does not alter the
smaller component. Monotonicity of the cotangent and its inverse yields
\[
 P_{\widehat g}(\widehat I)\le P_{\mathrm C}(\widehat I)\le8u_0 .
\]
Thus the small-tail comparison bounds the p-value of the component
selected by the algorithm.

The lower reference-tail bounds show that
\[
 D_{\mathrm{sum},\widehat I}(\widehat k)\ge c\sqrt H
       \quad\hbox{if }\widehat g=\mathrm{sum},\qquad
 X_{\widehat j,\widehat I}(\widehat k)\ge cH
       \quad\hbox{if }\widehat g=\mathrm{max}.
\]
The term \(\log(1+D)\) in the OU lower tail is negligible: rank scores
are polynomially bounded in \(n,p\), whereas \(H\gg\Lambda_n\gg N\).
The population upper bound of
Lemma~\ref{lem:wbs-relative-population} and the error estimates now give
\[
 E_\Sigma\ge c\sqrt{pNH}\quad\hbox{if }\widehat g=\mathrm{sum},
 \qquad
 E_\infty\ge cH\quad\hbox{if }\widehat g=\mathrm{max}.
\]
Thus the selected component has evidence comparable with \(H\), and
its approximation error is \(O(\delta_n)\) relative to its
population scale.

Put \(c_{\rm sel}=c_{\rm iso}/16\).
If the actual component had population maximum
\[
 \max_t\frac Ln f_{\widehat g,\widehat I}(t)
       <c_{\rm sel}A_{\widehat g}(J),
\]
its score would be at most twice \(c_{\rm sel}\) times its common
node scale. Its own strongest-boundary witness has score at least
\(c_{\rm iso}/2\) times the same scale. If \(u\) is its actual p-value
and \(v\) this same-component witness p-value, the reference bounds yield
\[
 v/u\longrightarrow0,\qquad v/q_{\mathrm C,n}\longrightarrow0.
\]
For SUM the positive difference between the two exponential
coefficients is
\[
 \tfrac18(c_{\rm iso}/2)^2-\tfrac12(2c_{\rm sel})^2
       =3c_{\rm iso}^2/128>0;
\]
for MAX the difference is \(c_{\rm iso}/2-2c_{\rm sel}>0\).
The exact stabilized transform also satisfies
\[
 T_{\mathrm C}(\widehat I)\le C/u,\qquad
 T_{\mathrm C}(I_q)\ge c/v-C/q_{\mathrm C,n},
\]
contradicting the Cauchy selection rule. Hence the selected component
has population maximum at least \(c_{\rm sel}A_{\widehat g}(J)\).

For each of the three procedures take
\[
 f(t)=\frac Ln f_{\Sigma,\widehat I}(t)
       \quad\hbox{if SUM is selected},\qquad
 f(t)=\frac Ln f_{\widehat j,\widehat I}(t)
       \quad\hbox{if MAX is selected}.
\]
In the latter case \((\widehat j,\widehat k)\) is an actual maximizing
coordinate--split pair. The preceding comparisons and the uniform
error bounds imply
\[
 M_f:=\max_{\eta_L\le t\le1-\eta_L}f(t)
       \ge(c_{\rm sel}/2)A,\qquad
 f(\widehat t)\ge M_f-C\delta_n A,
\]
where \(A=A_\Sigma(J)\) or \(A_\infty(J)\) for the selected component.
Since \(f''\ge f\ge0\) on every true-regime cell and its cell endpoints,
including \(\eta_L,1-\eta_L\), are integer split fractions, the
population maximum is attained on the split grid.
Choose the fixed trimming constant sufficiently small that
\[
 0<\eta<1/8,\qquad 2C_{\rm var}\eta\le c_{\rm sel}/16 .
\]
Then both actual trim endpoint values, and all values in the two
outer \(2\eta\) bands, are at most \(c_{\rm sel}A/16\).

Partition the trimmed interval by its true change fractions, and let
\([v,w]\) contain \(\widehat t\). If both endpoints are true changes,
\(w-v\ge d_*\). Comparison with the solution of \(y''=y\) having both
endpoint values \(M_f\) gives
\[
 f(t)\le M_f\frac{\cosh\{t-(v+w)/2\}}{\cosh\{(w-v)/2\}},
 \qquad M_f-f(t)\ge cM_f(t-v)(w-t).
\]
For the comparison, a positive interior maximum of \(f-y\) would
contradict \((f-y)''-(f-y)\ge0\).
The last inequality follows from
\[
 \cosh\{(w-v)/2\}-\cosh\{t-(v+w)/2\}
 =2\sinh\{(t-v)/2\}\sinh\{(w-t)/2\}
\]
and \(\sinh x\ge x\). Since \(M_f\ge cA\), this proves
\[
 \min(\widehat t-v,w-\widehat t)\le C\delta_n .
\]
If \(v=\eta_L\) is a trim endpoint, convexity gives instead
\[
 f(\widehat t)\le
       \frac{w-\widehat t}{w-v}\frac{M_f}{4}
       +\frac{\widehat t-v}{w-v}M_f,\qquad
 w-\widehat t\le C\delta_n.
\]
The right trim cell is identical. A cell with both endpoints at the
trims contradicts \(f(\widehat t)\ge M_f-C\delta_n A\).
Thus a true change \(\tau_q\in I\) satisfies
\[
 |\widehat k-\tau_q|\le Cn\delta_n\le r_n
\]
after \(C_r\) is fixed sufficiently large.

Since \(I\subseteq(\tau_a,\tau_b]\), endpoint avoidance excludes
both assigned boundary changes, and hence \(a<q<b\).

The two children have label pairs \((a,q)\) and \((q,b)\).
Their endpoints remain within \(r_n\) of the corresponding true
boundaries, their lengths are at least \(\mathfrak d_n-2r_n>0\), and
their unassigned sets partition \(\mathcal Q(J)\setminus\{q\}\).
If \(b=a+1\), every eligible interval lies entirely in the single
true regime \((\tau_a,\tau_{a+1}]\), by endpoint avoidance.
The event \(\mathcal E_{\rm null,g}\) makes the node stop;
an empty interval pool also stops by the algorithm's definition.
Induction from labels \((0,K_n+1)\) therefore gives exactly \(K_n\)
accepted splits, one within \(r_n\) of each true change.
The disjoint neighborhoods, \(2r_n<\mathfrak d_n\), preserve their order.

Finally,
\[
 \begin{aligned}
 \mathbb P(\mathcal E^c)
 &\le\mathbb P(\mathcal E_{\rm weak,g}^c)
      +K_ne^{-cM_n}+CK_nM_n(r_n+1)/n
      +\mathbb P(\mathcal E_{\rm null,g}^c)
 \longrightarrow0.
 \end{aligned}
\]
Since \(r_n/n\to0\), for every fixed \(\epsilon>0\) the required
count-and-location event holds on \(\mathcal E\) for all large \(n\).
The three conclusions follow under their respective signal conditions,
with active coordinate sets and signal branches allowed to vary across
changes.
\end{proof}

\section{Finite-sample assessment of the asymptotic null calibration}
\label{app:analytic-size}

We assess the finite-sample accuracy of the null reference distributions
in Theorems~\ref{thm:sum-null} and \ref{thm:max-full-analytic} and
Corollary~\ref{cor:cauchy-null}. The original rank statistics are
calibrated directly using the Gaussian-process and analytic Gumbel
references, and their component $p$-values form the equal-weight Cauchy
combination. This experiment complements the permutation-calibrated
comparisons in Section~\ref{sec:simulation}.

We generate independent observation vectors
$\bX_1,\ldots,\bX_n\sim N_p(\bm0,\mathbf R_p)$ under three
cross-coordinate dependence structures. The independent design has
$\mathbf R_p=\mathbf I_p$. For the block design, let
$\mathcal B_g=\{3(g-1)+1,\ldots,\min(3g,p)\}$,
$g=1,\ldots,\lceil p/3\rceil$, and set
\[
 (\mathbf R_p)_{j\ell}=
 \begin{cases}
 1,&j=\ell,\\
 0.3,&j\ne\ell,\quad j,\ell\in\mathcal B_g
                      \text{ for some }g,\\
 0,&\text{otherwise}.
 \end{cases}
\]
The AR(1) design has
\[
 (\mathbf R_p)_{j\ell}=0.3^{|j-\ell|}.
\]
Both dependent designs satisfy the copula conditions of
Assumption~\ref{ass:copula-null}. Their off-diagonal row sums are
uniformly bounded, and the neighborhood condition follows from the
fixed block size or geometric correlation decay.

The sample-size and dimension pairs are
\[
 (n,p)\in\{(100,100),(200,200),(200,500),
          (500,200),(500,500),(1000,500)\}.
\]
Combining these pairs with the three dependence structures gives
18 configurations. The trimming fraction is $\eta=0.1$, and all
integer splits in
$\cK_n=\{\lceil0.1n\rceil,\ldots,n-\lceil0.1n\rceil\}$ are scanned.
Each configuration uses $R_{\mathrm{MC}}=2{,}000$ independent
replications at nominal levels $\alpha\in\{0.01,0.05,0.10\}$.
Coordinatewise strictly increasing transformations preserve the ranks,
so the results are invariant under these marginal transformations.

For each sample, we compute the original standardized rank deficits
$D_{j,n}(k)$, the SUM and MAX scans in \eqref{eq:global-stats}, and the
Cauchy combination. The split-specific quantities $\mu_{n,k}$ and
$s_{n,k}$ are evaluated by the exact hypergeometric rank-path
recursions in Appendix~\ref{app:rank}. The same null samples are used
for all three procedures and nominal levels. For the calibrated
$p$-value $P_{\nu}^{(r)}$ in replication $r$, the empirical size is
\[
 \widehat s_{\nu}(\alpha)
 =\frac1{R_{\mathrm{MC}}}\sum_{r=1}^{R_{\mathrm{MC}}}
             \mathbf1\{P_{\nu}^{(r)}\le\alpha\},
 \qquad \nu\in\{\mathrm{sum},\mathrm{max},\mathrm C\}.
\]
At the nominal rejection probability, the Monte Carlo standard error
is $\{\alpha(1-\alpha)/R_{\mathrm{MC}}\}^{1/2}$.
Pointwise $95\%$ Wilson intervals quantify binomial sampling
uncertainty conditional on the computed reference distributions.

For SUM, Theorem~\ref{thm:sum-null} gives the reference variable
$\sup_{|u|\le\log9}\mathbb U(u)$, where $\mathbb U$ is stationary
with covariance $e^{-|u-v|}$. We simulate 200,000 independent
reference paths, independently of the null samples, on a grid of
32,768 equal subintervals. With $\Delta=2\log9/32768$, the grid
values follow
\[
 U_0\sim N(0,1),\qquad
 U_m=e^{-\Delta}U_{m-1}
       +\sqrt{1-e^{-2\Delta}}\,\xi_m,\qquad
 \xi_m\overset{\mathrm{iid}}{\sim}N(0,1).
\]
These transitions are exact at the grid points, and the grid maximum
approximates the continuous supremum. Upper-tail probabilities are
evaluated from the empirical reference distribution with one added
to the numerator and denominator. Comparison with the nested grid
of 16,384 subintervals indicates limited discretization sensitivity
in the reference quantiles. Reference-path Monte Carlo error and
grid approximation error are separate from the uncertainty measured
by the Wilson intervals.

For MAX, we use the analytic normalizing pair in
\eqref{eq:full-max-branch-selection}. Numerical solution of the
entropy Hardy problem at $\eta=0.1$ gives
$b_\eta\approx0.2499825$ and $\gamma_c\approx29.38$.
All six values of $\log p/\log n$ select the subcritical branch, so
$(\mathfrak a_{n,p},\mathfrak b_{n,p})
=(A_{n,p}^{\rm S},B_{n,p}^{\rm S})$ from
\eqref{eq:main-subcritical-normalization}.
The Bellman and transport equations are solved with tolerance
$10^{-10}$. The Bernoulli endpoint recursion is truncated at
$1,024$, and the moment series at index $4,096$ in each summation
index, with the first-order tail correction for $m(t)$ and
extrapolation for $v_0(t)$. Symmetry reduces the spatial integral
to $[\eta,1/2]$, where composite Simpson integration uses
17 equally spaced nodes.

Numerical sensitivity is assessed by halving the spatial resolution
and the endpoint and moment truncation levels, and by increasing
the spatial grid to 33 nodes for $(n,p)=(100,100)$. The resulting
changes in the MAX center are small relative to the observed
quantile discrepancies. The exact rank moments are verified by
complete subset enumeration at $n=8$, the scan by an independent
direct implementation, and the Bernoulli endpoint recursion by
enumeration of short binary sequences. Comparisons at
$n=500,1000,2000$ also show decreasing differences between the
exact rank moments and the moment-series limits.

The MAX $p$-value is
$1-F_{\mathrm G}\{(\Smax-\mathfrak b_{n,p})/\mathfrak a_{n,p}\}$,
and the Cauchy $p$-value uses $\omega=1/2$ as in
Corollary~\ref{cor:cauchy-null}. Thus both component calibrations
are numerical evaluations of their limiting reference distributions.

\begin{figure}[!htbp]
\centering
\includegraphics[width=\textwidth]{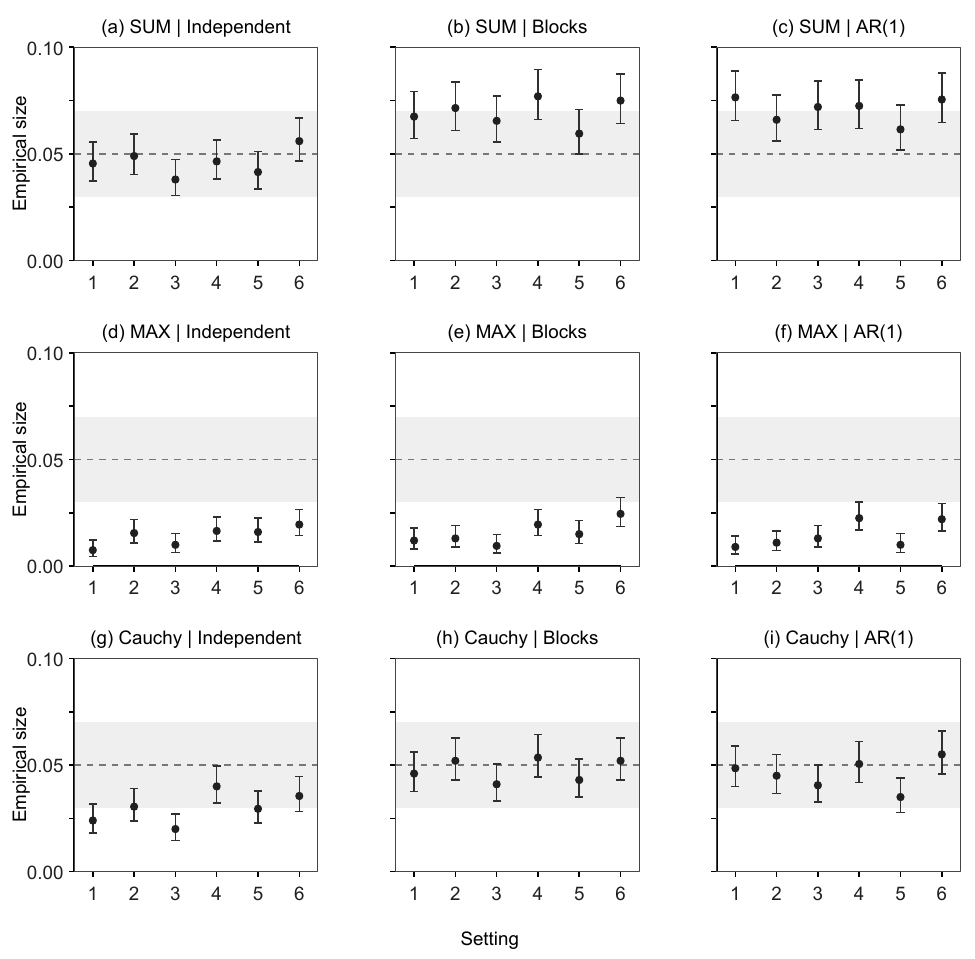}
\caption{Empirical size at level $0.05$ for the direct asymptotic
calibrations. Columns correspond to independent coordinates,
block dependence, and AR(1) dependence; rows correspond to
SUM, MAX, and Cauchy. Settings 1--6 on the horizontal axes
are $(n,p)=(100,100),(200,200),(200,500),(500,200),
(500,500),(1000,500)$, respectively. Each point uses
2,000 null samples and each error bar is a pointwise $95\%$
Wilson interval. The dashed line marks $0.05$ and the
shaded band marks the descriptive range $[0.03,0.07]$.}
\label{fig:analytic-size}
\end{figure}

Table~\ref{tab:analytic-size-full} reports the rejection frequencies,
and Figure~\ref{fig:analytic-size} displays empirical size at
level $0.05$ with pointwise Wilson intervals. The shaded band
$[0.03,0.07]$ provides a descriptive comparison range.

\begin{table}[!htbp]
\centering
\begin{threeparttable}
\caption{Empirical rejection frequencies for direct asymptotic calibration.
Each entry uses 2,000 independent null samples, with trimming fraction
$\eta=0.1$. Block and AR(1) correlations are $0.3$.}
\label{tab:analytic-size-full}
\small
\setlength{\tabcolsep}{3.8pt}
\renewcommand{\arraystretch}{1.12}
\begin{tabular}{rr*{9}{r}}
\toprule
& & \multicolumn{3}{c}{$\alpha=0.01$}
  & \multicolumn{3}{c}{$\alpha=0.05$}
  & \multicolumn{3}{c}{$\alpha=0.10$}\\
\cmidrule(lr){3-5}\cmidrule(lr){6-8}\cmidrule(lr){9-11}
$n$ & $p$ & SUM & MAX & Cauchy & SUM & MAX & Cauchy & SUM & MAX & Cauchy\\
\midrule
\multicolumn{11}{l}{\textit{Independent coordinates}}\\
100 & 100 & 0.0090 & 0.0010 & 0.0075 & 0.0455 & 0.0075 & 0.0240 & 0.0875 & 0.0225 & 0.0530\\
200 & 200 & 0.0135 & 0.0010 & 0.0065 & 0.0490 & 0.0155 & 0.0305 & 0.0965 & 0.0350 & 0.0640\\
200 & 500 & 0.0080 & 0.0025 & 0.0065 & 0.0380 & 0.0100 & 0.0200 & 0.0715 & 0.0250 & 0.0450\\
500 & 200 & 0.0180 & 0.0025 & 0.0140 & 0.0465 & 0.0165 & 0.0400 & 0.0895 & 0.0415 & 0.0600\\
500 & 500 & 0.0060 & 0.0020 & 0.0050 & 0.0415 & 0.0160 & 0.0295 & 0.0930 & 0.0435 & 0.0595\\
1000 & 500 & 0.0100 & 0.0015 & 0.0060 & 0.0560 & 0.0195 & 0.0355 & 0.1050 & 0.0460 & 0.0755\\
\midrule
\multicolumn{11}{l}{\textit{Independent blocks of at most three coordinates}}\\
100 & 100 & 0.0215 & 0.0015 & 0.0165 & 0.0675 & 0.0120 & 0.0460 & 0.1175 & 0.0305 & 0.0760\\
200 & 200 & 0.0210 & 0.0030 & 0.0125 & 0.0715 & 0.0130 & 0.0520 & 0.1225 & 0.0320 & 0.0820\\
200 & 500 & 0.0190 & 0.0010 & 0.0110 & 0.0655 & 0.0095 & 0.0410 & 0.1175 & 0.0235 & 0.0690\\
500 & 200 & 0.0235 & 0.0020 & 0.0140 & 0.0770 & 0.0195 & 0.0535 & 0.1290 & 0.0510 & 0.0985\\
500 & 500 & 0.0165 & 0.0020 & 0.0125 & 0.0595 & 0.0150 & 0.0430 & 0.1170 & 0.0335 & 0.0715\\
1000 & 500 & 0.0195 & 0.0035 & 0.0115 & 0.0750 & 0.0245 & 0.0520 & 0.1295 & 0.0620 & 0.0960\\
\midrule
\multicolumn{11}{l}{\textit{AR(1) dependence across coordinates}}\\
100 & 100 & 0.0225 & 0.0015 & 0.0160 & 0.0765 & 0.0090 & 0.0485 & 0.1215 & 0.0200 & 0.0815\\
200 & 200 & 0.0150 & 0.0010 & 0.0100 & 0.0660 & 0.0110 & 0.0450 & 0.1240 & 0.0270 & 0.0785\\
200 & 500 & 0.0180 & 0.0020 & 0.0090 & 0.0720 & 0.0130 & 0.0405 & 0.1220 & 0.0305 & 0.0740\\
500 & 200 & 0.0235 & 0.0025 & 0.0170 & 0.0725 & 0.0225 & 0.0505 & 0.1295 & 0.0535 & 0.0905\\
500 & 500 & 0.0160 & 0.0025 & 0.0125 & 0.0615 & 0.0100 & 0.0350 & 0.1195 & 0.0275 & 0.0685\\
1000 & 500 & 0.0220 & 0.0025 & 0.0150 & 0.0755 & 0.0220 & 0.0550 & 0.1215 & 0.0440 & 0.1005\\
\bottomrule
\end{tabular}
\begin{tablenotes}[flushleft]
\footnotesize
\item SUM uses the simulated Ornstein--Uhlenbeck reference distribution;
MAX uses the analytic Gumbel normalization; Cauchy combines their
theoretical $p$-values with equal weights. No permutation calibration
or fitted finite-sample critical values are used. All sample vectors
are independent over the observation index.
\end{tablenotes}
\end{threeparttable}
\end{table}

Under coordinate independence, SUM is generally close to the nominal
level, with moderate variation across sample sizes and dimensions.
Cross-coordinate dependence produces systematically higher rejection
frequencies, particularly in the more extreme tail. The inflation
persists across the sample sizes considered and is consistent with
the larger empirical variance of the dense score under dependence.
The accuracy of the Gaussian-process approximation therefore depends
on the coordinate dependence as well as the sample size.

MAX is conservative across all configurations and nominal levels.
The Wilson intervals indicate that this under-rejection exceeds
Monte Carlo sampling variation. Table~\ref{tab:analytic-max-quantiles}
compares the analytic critical values with empirical upper quantiles
under coordinate independence.

\begin{figure}[!htbp]
\centering
\includegraphics[width=\textwidth]{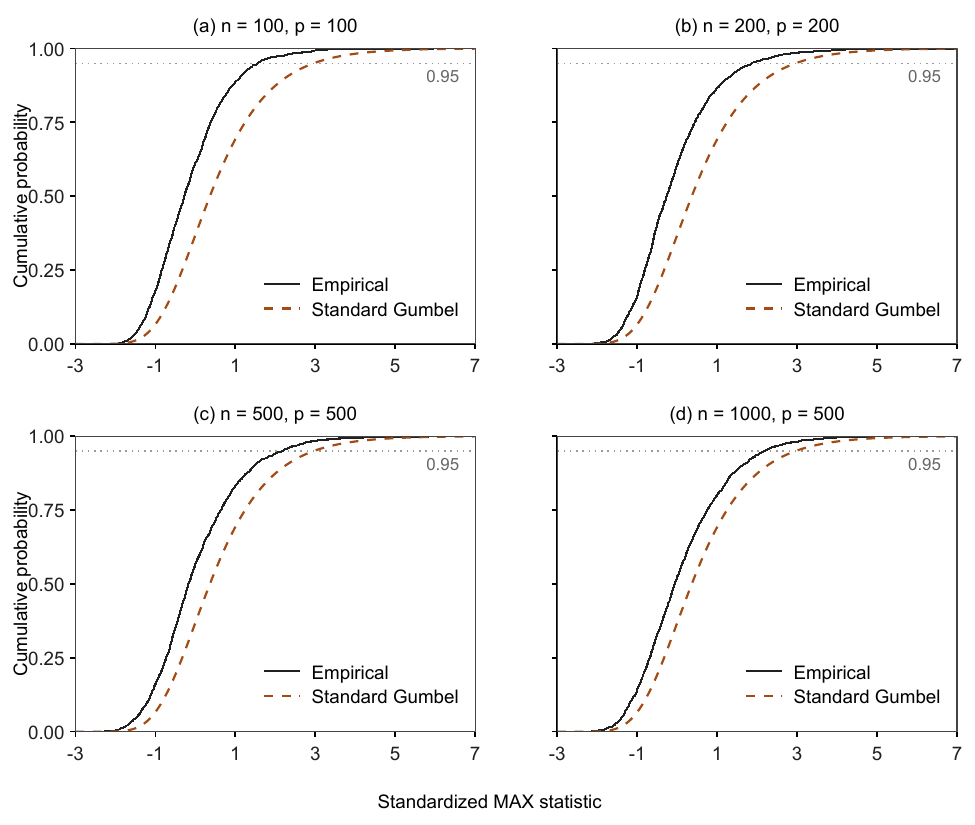}
\caption{Empirical distribution functions of
$(\Smax-\mathfrak b_{n,p})/\mathfrak a_{n,p}$ under independent
coordinates, compared with the standard Gumbel distribution.
Each empirical curve uses 2,000 samples; the horizontal
reference line marks cumulative probability $0.95$.
The displayed range $[-3,7]$ emphasizes the central and
upper-tail comparison. The empirical upper quantiles remain
below the analytic Gumbel quantiles.}
\label{fig:analytic-max-cdf}
\end{figure}

\begin{table}[!htbp]
\centering
\begin{threeparttable}
\caption{Analytic MAX normalization and the upper null quantile under
coordinate independence. The empirical quantile uses the same 2,000
null samples as Table~\ref{tab:analytic-size-full}.}
\label{tab:analytic-max-quantiles}
\small
\setlength{\tabcolsep}{5pt}
\begin{tabular}{rrrrrrr}
\toprule
$n$ & $p$ & $\mathfrak a_{n,p}$ & $\mathfrak b_{n,p}$ &
$c_{\mathrm{max},0.95}$ & Empirical $0.95$ quantile & Difference\\
\midrule
100 & 100 & 1.3580 & 6.9590 & 10.9925 & 9.0734 & 1.9191\\
200 & 200 & 1.2661 & 7.6047 & 11.3652 & 9.9624 & 1.4028\\
200 & 500 & 1.2582 & 8.7019 & 12.4391 & 10.9638 & 1.4753\\
500 & 200 & 1.1778 & 7.3469 & 10.8453 & 9.7778 & 1.0675\\
500 & 500 & 1.1690 & 8.3699 & 11.8421 & 10.8993 & 0.9428\\
1000 & 500 & 1.1142 & 8.1609 & 11.4702 & 10.6016 & 0.8686\\
\bottomrule
\end{tabular}
\begin{tablenotes}[flushleft]
\footnotesize
\item $c_{\mathrm{max},0.95}=\mathfrak b_{n,p}+
\mathfrak a_{n,p}\{-\log[-\log(0.95)]\}$.
The last column is the analytic critical value minus the empirical
quantile; empirical quantiles are descriptive and are not used for testing.
\end{tablenotes}
\end{threeparttable}
\end{table}

The analytic MAX critical values lie above the empirical upper
quantiles. At fixed dimension, this discrepancy decreases as the
sample size increases, although an appreciable gap remains.
Figure~\ref{fig:analytic-max-cdf} shows the same pattern in the
normalized distribution functions. The quantile differences are
substantially larger than the changes produced by numerical
refinement, indicating that numerical evaluation alone does not
account for the conservative calibration.

The Cauchy combination tends to be conservative under coordinate independence
and closer to the nominal level under the dependent designs,
especially at the intermediate nominal level. Its accuracy varies
across significance levels, with some over-rejection in the more
extreme tail. The opposing calibration errors of SUM and MAX,
together with residual finite-sample dependence, may contribute to
this pattern.

\enlargethispage{\baselineskip}
These results show that direct asymptotic calibration has uneven
finite-sample accuracy: SUM is sensitive to coordinate dependence,
whereas the analytic MAX thresholds remain conservative over the
range examined. The Cauchy combination partly moderates these
discrepancies but does not eliminate them uniformly across nominal
levels. Whole-vector permutation provides the exchangeability-based
calibration established in Appendix~\ref{app:rank} and used in
Section~\ref{sec:simulation}.

\clearpage

\end{document}